\documentclass[
  paper=A4, 		
  pagesize, 		
  DIV=calc, 		
  headings=small,	
  english,  		
  12pt, 			
  listof=totoc, 
  index=totoc, 
  BCOR5mm, 			
]{scrbook}
\usepackage[T1]{fontenc}	
\usepackage[utf8]{inputenc}	
\usepackage{fixltx2e}  		
\usepackage[english]{babel} 	
\usepackage[english]{translator}
\usepackage{graphicx} 		
\usepackage{subcaption}
\usepackage[acronym,toc,numberedsection,nonumberlist]{glossaries}
\usepackage[hyphens]{url} 	
\usepackage[draft,a4paper,breaklinks,bookmarksnumbered]{hyperref}  %
\usepackage{comment}

\usepackage{graphicx} 
\usepackage{caption}
\usepackage{cite}
\usepackage{amsfonts,amssymb,amsmath,mathrsfs,bm}
\usepackage{multirow}
\usepackage[usenames,dvipsnames]{xcolor}
\usepackage{xspace} 
\usepackage{bm} 
\usepackage[utf8]{inputenc} 
\usepackage{eucal}
\usepackage{float}
\usepackage[normalem]{ulem}
\usepackage{tabularx}
\usepackage{framed} 

\newcommand{\Ham}{\hat{H}}
\newcommand{\nnm}{\nonumber}
\newcommand{\doe}{\partial}
\newcommand{\be}{\begin{equation}}
\newcommand{\ee}{\end{equation}}
\newcommand{\bse}{\begin{subequations}}
	\newcommand{\ese}{\end{subequations}}

\newcommand{\mr}{\mathrm}
\newcommand{\tr}{\textrm}
\newcommand{\mc}{\mathcal}

\newcommand{\bs}{\boldsymbol}
\newcommand{\ms}{\mathsf}

\newcommand{\ve}{\varepsilon}

\usepackage{
  ellipsis, 
  ragged2e, 
  marginnote,
}
\usepackage[tracking=true]{microtype}%
\DeclareMicrotypeSet*[tracking]{my}%
  { font = */*/*/sc/* }%
\SetTracking{ encoding = *, shape = sc }{ 45 }

\usepackage{%
  lmodern, 
}

\makeglossaries

\newacronym[\glslongpluralkey={Public-Key-Infrasturkturen}]{PKI}{PKI}{Public-Key-Infrasturktur}
\newacronym{IDM}{IDM}{Identity-Management}
\newacronym{ESB}{ESB}{Enterprise-Service-Bus}
\newacronym{ID}{ID}{Identifikator} 
\newacronym{SigG}{SigG}{Signaturgesetz}
\newacronym{CCC}{CCC}{Chaos Computer Club e. V.}
\newacronym{BSI}{BSI}{Bundesamt für Sicherheit in der Informationstechnik}

\newglossaryentry{glos:PKI}{
name=Public-Key-Infrasturktur,
description={Eine Public-Key-Infrasturktur ist eine...}
}
\newglossaryentry{glos:SC}{
name=Smart-Card,
description={EineSmart-Card ist eine...}
}
\newglossaryentry{glos:SigG}{
name=Signaturgesetz,
description={EineSmart-Card ist eine...}
}
\graphicspath{{./gfx/}}

\begin{document}
\pdfinfo{                               
    /Author (Dein Name)
    /CreationDate (D:20130310111111)    
                                        %
                                        %
    /ModDate (D:20130310111111)         
    /Creator (TeX \& TXC)               
    /Producer (pdfTeX)                  
    /Title () 													
    /Subject ()                         
    /Keywords ()                        
}
\frontmatter
\pagenumbering{Roman}
\thispagestyle{empty}
\begin{center}
		
    \includegraphics[width=3cm]{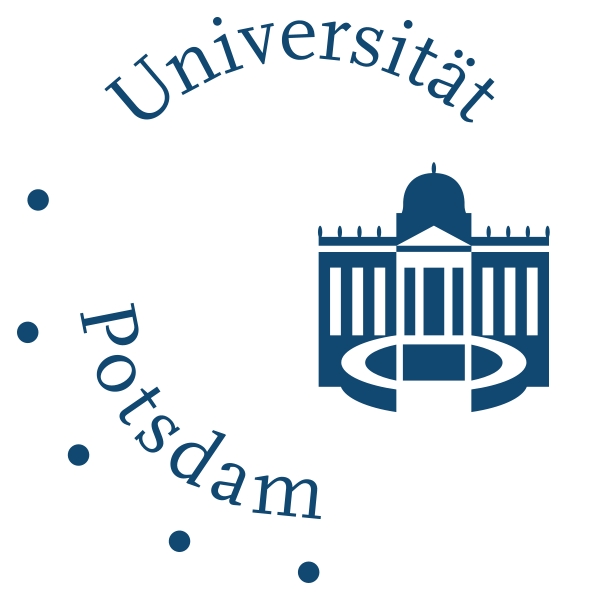}\\
    \vspace{.5cm}
    
    {\Huge Universität Potsdam}\\
    \vspace{1cm}
    
    {\Large Institut für Physik und Astronomie\\[1mm]}
    
    \vspace{2cm}
    {\Large \textbf{Accurate waveform models for gravitational-wave astrophysics: synergetic approaches from analytical relativity}}\\
    \vspace*{3mm}
    
    \vspace{1.0cm}
    {\normalsize \textbf{Dissertation}}\\ \vspace{0.5cm}
    {\normalsize zur Erlangung des wissenschaftlichen Grades}\\
    {\normalsize doctor rerum naturalium}\\
    {\normalsize (Dr. rer. nat.)}\\
    \vspace{1cm}
    
    \vspace{0.5cm}

    \parbox{1cm}{
      \begin{large}
        \begin{tabbing}
	        Kandidat: \hspace{1.5cm} \=Andrea Antonelli\\[2mm]
	    	  Gutachterin: \>Prof. Dr. Alessandra Buonanno\\
	    		Abgabedatum: \> \today\\
        \end{tabbing}
      \end{large}
    }\\
    
    \vspace{5cm} 
\end{center}
 
\thispagestyle{empty}
\cleardoublepage
    
\thispagestyle{empty}
\cleardoublepage


\begin{center}
    \textbf{Abstract}
\end{center}
\vspace*{1cm}
\noindent Gravitational-wave (GW) astrophysics is a field in full blossom. Since the landmark detection of GWs from a binary black hole on September 14th 2015, fifty-two compact-object binaries have been reported by the LIGO-Virgo collaboration. Such events carry astrophysical and cosmological information ranging from an understanding of how black holes and neutron stars are formed, what neutron stars are composed of, how the Universe expands, and allow testing general relativity in the highly-dynamical strong-field regime. It is the goal of GW astrophysics to extract such information as accurately as possible. Yet, this is only possible if the tools and technology used to detect and analyze GWs are advanced enough. A key aspect of GW searches are waveform models, which encapsulate our best predictions for the gravitational radiation under a certain set of parameters, and that need to be cross-correlated with data to extract GW signals. Waveforms  must be very accurate to avoid missing important physics in the data, which might be the key to answer the fundamental questions of GW astrophysics. The continuous improvements of the current LIGO-Virgo detectors, the development of next-generation ground-based detectors such as the Einstein Telescope or the Cosmic Explorer, as well as the development of the Laser Interferometer Space Antenna (LISA), demand accurate waveform models. While available models are enough to capture the low spins, comparable-mass binaries routinely detected in LIGO-Virgo searches,  those for sources from both current and next-generation ground-based and spaceborne detectors must be accurate enough to detect binaries with large spins and asymmetry in the masses. 
Moreover, the thousands of sources that we expect to detect with future detectors demand accurate waveforms to mitigate biases in the estimation of signals' parameters due to the presence of a foreground of many sources that overlap in the frequency band. This is recognized as one of the biggest challenges for the analysis of future-detectors' data, since biases might hinder the extraction of important astrophysical and cosmological information from future detectors' data. In the first part of this thesis, we discuss how to improve waveform models for binaries with high spins and asymmetry in the masses. In the second, we present the first generic metrics that have been proposed to predict biases in the presence of a foreground of many overlapping signals in GW data.

For the first task, we will focus on several classes of \emph{analytical} techniques. Current models for LIGO and Virgo studies are mostly based on appropriate resummations of the post-Newtonian (PN, weak-field, small velocities) approximation. 
However, two other approximations have risen in prominence, the post-Minkowskian (PM, weak-field only) approximation natural for unbound (scattering) orbits and the small-mass-ratio (SMR) approximation typical of binaries in which the mass of one body is much bigger than the other. These are most appropriate to binaries with high asymmetry in the masses that challenge current waveform models.
Moreover, they allow one to ``cover'' regions of the parameter space of coalescing binaries, thereby improving the interpolation (and faithfulness) of waveform models.
The analytical approximations to the relativistic two-body problem can synergically be included within the effective-one-body (EOB) formalism, in which the two-body information from each approximation can be recast into an effective problem of a mass orbiting a deformed Schwarzschild (or Kerr) black hole. The hope is that the resultant models can cover both the low-spin comparable-mass binaries that are routinely detected, and the ones that challenge current models.
The first part of this thesis is dedicated to a study about how to best incorporate information from the PN, PM, SMR and EOB approaches in a synergistic way. We also discuss how accurate the resulting waveforms are, as compared against numerical-relativity (NR) simulations. 
We begin by comparing PM models, whether alone or recast in the EOB framework, against PN models and NR simulations. We will show that PM information has the potential to improve currently-employed models for LIGO and Virgo, especially if recast within the EOB formalism. This is very important, as the PM approximation comes with a host of new computational techniques from particle physics to exploit. 
Then,  we show how a combination of PM and SMR approximations can be employed to access previously-unknown PN orders, deriving the third subleading PN dynamics for spin-orbit and (aligned) spin$_1$-spin$_2$ couplings. Such new results can then be included in the EOB models currently used in GW searches and parameter estimation studies, thereby improving them when the binaries have high spins. 
Finally, we build an EOB model for quasi-circular nonspinning binaries based on the SMR approximation (rather than the PN one as usually done). We show how this is done in detail without incurring in the divergences that had affected previous attempts, and compare the resultant model against NR simulations. We find that the SMR approximation is an excellent approximation for \emph{all} (quasi-circular nonspinning) binaries, including both the comparable-mass binaries that are routinely detected in GW searches and the ones with highly asymmetric masses. In particular, the SMR-based models compare much better than the PN models, suggesting that SMR-informed EOB models might be the key to model binaries in the future. 

In the second task of this thesis, we work within the linear-signal approximation and describe generic metrics to predict inference biases on the parameters of a GW source of interest in the presence of confusion noise from unfitted foregrounds and from residuals of other signals that have been incorrectly fitted out. We illustrate the formalism with simple (yet realistic) LISA sources, and demonstrate its validity against Monte-Carlo simulations. The metrics we describe pave the way for more realistic studies to quantify the biases with future ground-based and spaceborne detectors.


\tableofcontents 
\mainmatter 
\pagenumbering{arabic}


\colorlet{shadecolor}{orange!15}
\newcommand{\dwmunu}{_{\mu\nu}}
\newcommand{\upmunu}{^{\mu\nu}}
\newcommand{\Adown}{_{\text{A}}}
\newcommand{\subreal}{_{\text{EOB}}}
\newcommand{\bind}{_{\text{bind}}}
\newcommand{\R}{r}
\newcommand{\Pp}{p}
\newcommand{\nNNLOSO}{N$^3$LO SO }
\newcommand{\Spinonespintwo}{spin$_1$--spin$_2$ }
\newcommand{\Sonestwo}{S$_1$S$_2$ }
\newcommand{\refs}{\, {\color{red}[Refs]\, }}
\newcommand{\hicsuntdracones}{{
		\color{red} \underline{Checked up to here}.
}}

\newcommand{\zzero}{\zzz^{(0)}}
\newcommand{\zone}{\zzz^{(1)}}
\newcommand{\ztwo}{\zzz^{(2)}}
\newcommand{\GSF}{GSF }
\newcommand{\mm}{Gm}
\newcommand{\chap}{Chapter}
\newcommand{\chaps}{Chapters}
\newcommand{\Lbar}{\frac{L}{\mu}}
\newcommand{\Lbarbyrsquared}{\frac{L^2}{\mu^2r^2}}
\newcommand{\prbar}{\frac{p_r}{\mu}}
\newcommand{\rtest}{R}
\newcommand{\invrad}{\Upsilon}

\chapter{Introduction}
\label{chap:introduction}

\section{Preface and organization of the material}

\label{sec:GWhist}

The detection of 52 black-hole (BH), neutron-star, and black-hole--neutron-star binaries in the span of five years mark gravitational-wave (GW) astrophysics as a young and vibrant field~\cite{Abbott:2016blz,Abbott:2017oio,TheLIGOScientific:2017qsa,Abbott:2020niy,LIGOScientific:2021qlt}, whose ultimate goals are to unravel what drives the birth, life and death of neutron stars and BHs, present independent ways to measure the Universe expansion, and test general relativity in hitherto poorly explored regions of strong gravitational fields. 
Yet, one could say that gravitational-wave theory already has a long history.
Soon after he penned down his equations in 1915~\cite{Einstein:1915ca}, Einstein theorized the presence of gravitational waves using a linearized approximation, deriving for the first time the quadrupole formula that is at the core of GW astrophysics~\cite{Einstein:1916cc}.
While critics of the derivation and validity of quadrupole formula existed (see the ``quadrupole controversy'', as recounted for instance in Ref.~\cite{Kennefick:2007zz}), it became clear that GWs do  exist and carry away energy with them. One of the most convincing arguments came from Richard Feynman at the Chapel Hill conference in 1957~\cite{Kennefick:2007zz} (later expounded upon by Hermann Bondi~\cite{Bondi:1957dt}), where he used an analogy with ''beeding stick'' where beads are moved as a byproduct of incoming GWs to explain that GWs do work and thus carry energy. In particular, this implied that GWs can in principle be detected, bringing about the possibility of opening up a completely new branch of astrophysics. The possibility of detecting GWs suggested by Feynman and Bondi's calculations prompted Joseph Weber to search for GWs for the first time ever~\cite{PhysRev.117.306}, and indeed make the first claims of their detection~\cite{PhysRevLett.22.1320}. While his claims failed to convince the wider scientific community, the observation of the Hulse-Taylor pulsar PSR B1913+16 a few years later marked the first confirmed detection of GWs~\cite{Hulse:1974eb}.  This pulsar and its companion neutron star continuously lose orbital energy with an observable decay in their orbital period. Within general relativity (GR), this decay is due to GWs and one can predict it using post-Newtonian (PN) techniques, which fully exploit the large separation between the compact objects in the binary and their small velocity relative to the speed of light. The decay was found to match the prediction from emission due to GWs, confirming their existence~\cite{Weisberg:1981mt,Taylor:1982zz}.

The interest generated by Weber's work had meanwhile set up in motion the development of the technical machinery needed to measure GWs on Earth. The challenge was considerable,
manly because gravity only couples weakly to matter and thus leaves only a small imprint on a detector. Still, already in 1972 Rainer Weiss proposed a concept for a laser interferometer and was able to predict the major sources of noise that instrumentalists had to deal with~\cite{Weiss_detector}.  Weiss' analysis forms the basis for the two detectors composing the Laser Interferometer Gravitational-wave Observatory (LIGO) (Hanford in the state of Washington and Livingston in Lousiana), the Virgo detector in Italy and the KAGRA detector in Japan~\cite{Somiya:2011np}. Along with the instrumental challenge, another thread of research initiated by Weber~\cite{PhysRev.117.306} was that to develop robust statistical techniques to dig GW signals out of data. This thread was picked up by Kip Thorne's group and associates at Caltech~\cite{Cutler:1992tc,Finn:1992wt,Finn:1992xs,Flanagan:1997fn,Flanagan:1997kp} (see also important early work by Sanjeev Dhurandhar and Bangalore Sathyaprakash~\cite{Sathyaprakash:1991mt,Dhurandhar:1992mw}).
A crucial technique to disentangle signals from data is matched filtering, which cross-correlates the detector's noise with a bank of waveform templates, namely predictions for the shape of the signal from binaries of compact objects in motion around each other. Waveforms need to be very accurate to disentangle signals from detector noise. The need for such accurate templates sparked interest in providing both numerical and
(approximate) analytical solutions to the Einstein equations, 
and the synergy between numerical and analytical techniques forms the basis of the templates used in current GW searches.

The long process that has led to the development of such advanced waveform models, and instrumental and data analysis tools bore its fruits on September 14th, 2015, when the LIGO detectors picked up the GWs from a binary BH~\cite{Abbott:2016blz}. This event marks a turning point in GW astronomy (rightfully vindicating the ``astronomy'' in its name), as well as for astrophysics in general. Almost two years after the first detection, in August 2017, the Virgo detector joined the network~\cite{Abbott:2017oio}, playing a crucial role in the first detection of an electromagnetic counterpart of a binary neutron star in 2017~\cite{TheLIGOScientific:2017qsa}. This event is credited for the birth of multimessenger astrophysics, as the same binary was detected with GWs and all bands across the electromagnetic spectrum. As the sensitivity of the network increased between observation runs, the detectors were also able to detect 50 more events~\cite{Abbott:2020niy,LIGOScientific:2021qlt}. This marks another important paradigm shift in which the full set of GW events can be used as a whole to begin answering questions regarding the formation channels leading to the observed BHs, measure quantities of interest for cosmology (such as the Hubble constant~\cite{Abbott:2019yzh}), and test strong gravity \cite{LIGOScientific:2016lio,LIGOScientific:2019fpa,LIGOScientific:2020tif}.

The future of GW astrophysics is promising.
While the current network of detectors reshapes the landscape of astrophysics, preparations are being made for upgrades of the current network with KAGRA and LIGO-India \cite{Unnikrishnan:2013qwa}, and for future detectors both on ground (see the Einstein Telescope~\cite{Punturo:2010zz} and the Cosmic Explorer~\cite{Reitze:2019iox}) and in space, with the Laser Interferometer Space Antenna (LISA)~\cite{Audley:2017drz}. Future detectors promise to detect thousands (if not millions) of sources. This presents both an exciting astrophysical and fundamental physics prospect, since the ``loudness'' of the expected sources will allow the most precise astrophysical and cosmological inference from GWs, as well as a unique set of challenges, mainly related to dealing with such ``loudness''. Among the challenges, it has been estimated that waveform models will not be accurate enough to infer from GWs without introducing systematic biases already in the next LIGO and Virgo observing runs~\cite{Purrer:2019jcp,Relton:2021cax}. In particular binaries with high spins and unequal masses present the biggest problems, as they lie in a region of the parameter space that is not well covered by the templates and numerical simulations in use nowadays (which prompts us to focus on these binaries in the present thesis). On top of the instrumental and waveform modelling challenges, future data analyses will have to deal with the simultaneous inference with inaccurate models from an unkwown number of sources (the so-called ``global fit'') and the inference of the parameters of a source of interest in the presence of a foreground of other signals. The challenge lies in our ability to accurately extract the signals' parameters without incurring in biases, which may hinder our ability to infer astrophysical and cosmological data out of the detected GW sources. In this thesis, we extend metrics~\cite{Flanagan:1997kp,PhysRevD.71.104016,Cutler:2007mi} that were devised to set (individual) waveform accuracy requirements for LIGO and Virgo to provide a way of assessing the biases in both the above situations.

The manuscript is organized as follows. The introductory \chap~\ref{chap:introduction} gives an overview of the original work in this thesis and its context. 
Section~\ref{sec:approx} gives an overview of the main \emph{analytical} techniques that allow one to predict the waveform of GW signals with great accuracy. We discuss the theoretical foundations of gravitational radiation in Sec.~\ref{sec:quad}, and the post-Newtonian (PN), post-Minkowskian (PM) and small-mass-ratio (SMR) approximations, and the effective-one-body (EOB) framework that forms the basis of some of the current state-of-the-art waveforms used in LIGO-Virgo studies in sections~\ref{sec:PN},~\ref{sec:PM},~\ref{sec:SMR} and~\ref{sec:EOBform}, respectively. We also discuss numerical simulations of coalescing binaries in Sec.~\ref{sec:NR}, and synergies between all the above in Sec.~\ref{sec:interplay}. Given the broad range of topics, a detailed review of each topic is not possible. Instead, we have opted to include only the information that is directly relevant to the subsequent sections. 
In Sec.~\ref{sec:improvingweak}, we discuss how to improve models using the PM approximation: Sec.~\ref{sec:1paper} contains a research summarizes \chap~\ref{chap:two}, where we compare PM and PN models against numerical solutions to the Einstein equations; Sec.~\ref{sec:3&4paper} is an alternative derivation of the main results in \chaps~\ref{chap:four} and~\ref{chap:five}, where we combine PM and SMR information synergistically to derive the third-subleading PN order for spinning binaries at the spin-orbit and (aligned) spin$_1$-spin$_2$ order. Both are needed to improve models for the highly spinning binaries for which models may not be sufficient already in the next observing runs.
In Sec.~\ref{sec:2paper} we discuss how to improve waveforms in the strong-field regime, presenting an EOB model based on the SMR approximation (rather than the PN one as all others). We also discuss how this can be used as a basis to improve models with both equal and unequal masses. 
In Section~\ref{sec:PE}, we describe the main ideas behind the detection and characterization of GW signals, including a summary of \chap~\ref{chap:six}, where we provide generic metrics to assess biases in the presence of inaccurately removed signals and from the presence of an unfitted foreground of signals. Finally, \chap~\ref{chap:seven} reports the main conclusions of the thesis. Appendices \ref{app:DJStoISO@4PN} and \ref{app:chipred} complement Sec. \ref{sec:3&4paper}, while the rest are the appendices in Refs.~\cite{Antonelli:2019fmq,Antonelli:2019ytb,Antonelli:2020aeb,Antonelli:2020ybz,Antonelli:2021vwg}. Because of stylistic preference, ``we'' is used instead of ``I''. The exception is the subsection below, where the research contributions towards the publications in this thesis are reported.

\subsection*{Research contributions}
\label{sec:overview}

Sections~\ref{sec:1paper},~\ref{sec:3&4paper},~\ref{sec:2paper} and~\ref{sec:5paper} include some of the original  research work I have carried out during my PhD. The published articles can be found in \chaps~\ref{chap:two} to~\ref{chap:six}. Here I summarize my contributions to these.

\begin{itemize}
	\item Chapter~\ref{chap:two} is
	\newline
	
	\textbf{Andrea Antonelli}, Alessandra Buonanno, Jan Steinhoff, Maarten van de Meent, Justin Vines. \emph{Energetics of two-body Hamiltonians in post- Minkowskian gravity}. Phys. Rev., D99(10):104004, 2019.
	\newline
	
	The main goal of the paper is to compare various PM models and the PN models that form the basis of current LIGO-Virgo waveforms against NR simulations.
	My main contribution towards this publication was the development of a reliable code to estimate the binding energies of a given Hamiltonian. I have used these to perform all the analyses in the paper, producing the results in Figs.~\ref{fig:energywPM},~\ref{fig:energyJPM},~\ref{fig:energymixed},~\ref{fig:PMvsclassicEOB},~\ref{fig:evolq1},~\ref{fig:vanPMvsPNq1} (which were plotted by collaborators). 
	I have also contributed to the interpretation of the results. 
	The author list is in alphabetical order.
	
	\item Chapter~\ref{chap:three} is
	\newline
	
	\textbf{Andrea Antonelli}, Maarten van de Meent, Alessandra Buonanno, Jan Steinhoff, Justin Vines. \emph{Quasicircular inspirals and plunges from nonspinning effective-one-body Hamiltonians with gravitational self-force information}. Phys. Rev., D101(2):024024, 2020.
	\newline
	
	In this paper, which I have led, I have analyzed the ``light-ring divergence problem'' and found an EOB Hamiltonian that does not contain unphysical divergences at the light ring (see Sec.~\ref{sec:2paper} for more details). I have performed the analytical calculations in the paper. I have showed the absence of the divergence more explicitly with plunges of the effective mass through this potentially pathological strong-field region, producing Fig.~\ref{fig:orbitsep}. I have carried out the binding-energy and dephasing comparisons of the model against NR simulations in Figs.~\ref{fig:compEJ},~\ref{fig:compEO},~\ref{fig:wavephase}, and~\ref{fig:dphivsq}. I have written all the sections in the paper, with the exception of Appendix~\ref{sec:appred}.
	
	\item Chapter~\ref{chap:four} is
	\newline
	
	\textbf{Andrea Antonelli}, Chris Kavanagh, Mohammed Khalil, Jan Steinhoff, Justin Vines. \emph{Gravitational spin-orbit coupling through third- subleading post-Newtonian order: from first-order self-force to arbitrary mass ratios}. Phys. Rev. Lett., 125(1):011103, 2020.
	\newline
	
	I was part of a collaboration that calculated the third-subleading PN dynamics for spin-orbit couplings. 
	The main result I contributed towards are the new terms in the scattering-angle at third-subleading order [first line of Eq.~\eqref{chiNNNLOSO}].
	My results provide independent checks for calculations performed also by M.~Khalil and J.~Vines. I have contributed to the interpretations of the results. The author list is in alphabetical order.
	
	\item Chapter~\ref{chap:five} is
	\newline
	
	\textbf{Andrea Antonelli}, Chris Kavanagh, Mohammed Khalil, Jan Steinhoff, Justin Vines. \emph{Gravitational spin-orbit and aligned spin1-spin2 couplings through third-subleading post-Newtonian orders}. Phys. Rev. D, 102:124024, 2020.
	\newline
	
	The present publications details the derivation in the letter reported in \chap~\ref{chap:four}, and it extends the analysis to bilinear \Sonestwo couplings for spin-aligned configurations. Accordingly, I have extended the calculation performed in \chap~\ref{chap:four} to the bilinear order in the spins, and matched it with indipendent calculations by M.~Khalil and J.~Vines. I have written Sec.~\ref{sec:N3LOSO}. The author list is in alphabetical order.
	
	\item Chapter~\ref{chap:six} is the manuscript
	\newline
	
	\textbf{Andrea Antonelli}, Ollie Burke, Jonathan R. Gair. \emph{Noisy neighbours: inference biases from overlapping gravitational-wave signals}. e-print:2104.01897. Accepted for publication in the Monthly Notices of the Royal Astronomical Society.
	\newline
	
	I have led the paper together with Ollie Burke.
	I have proposed and carried out the extension of the Cutler-Vallisneri formalism to retrieve biases when both unmodelled and incorrectly-modelled foregrounds are present in the data. I have produced (and given the interpretation behind) Figs.~\ref{fig:conf_noise_LISA},~\ref{fig:conf_noise_bias},~\ref{fig:R_theta_correlations}, and~\ref{fig:inacc_rem_bias}. I have also checked the results presented in Figs.~\ref{fig:LISA_Corner}, and~\ref{fig:posteriors_LISA}. I have played a major role in the writing of all the sections in the paper, with the exception of Sec.~\ref{sec:Source_Confusion_Bias} (where my role was minor) and Sec.~\ref{sec:GF} (where I played no role).
    The codes that I have developed for this project, which were used to plot the figures mentioned above (and Figs.~\ref{fig:conf_noise_LISAnew}, \ref{fig:conf_noise_biasnew}, and~\ref{fig:inacc_rem_biasnew}), can be found at the link \text{\url{https://github.com/aantonelli94/GWOP}}.
\end{itemize}

\subsection*{Notation}

\begin{figure}
	\centering
	\includegraphics[width=\linewidth]{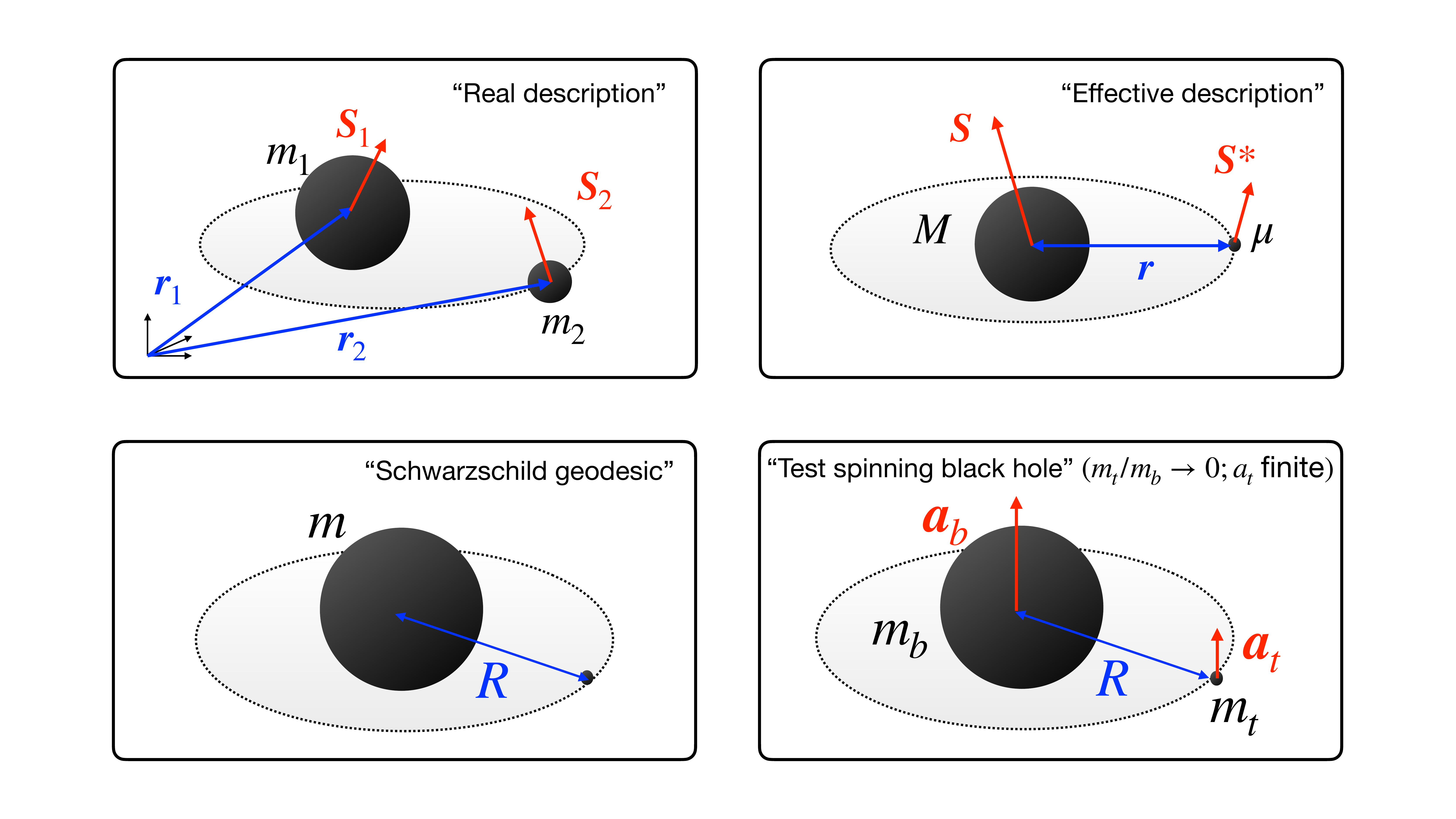}
	\caption{Configurations of BH binaries considered in the thesis' introduction. \emph{Top left}: the ``real description'' is used to report PN results in Sec. \ref{sec:PN}, and in the derivations of  new third-subleading PN terms with spins in Sec.\ref{sec:3&4paper} (in which we restrict our attention to spin-aligned configurations in the equatorial plane). \emph{Top right}:  the ``effective description'' is used in Sec.\ref{sec:EOBform}, Sections \ref{sec:1paper}, and \ref{sec:2paper} for nonspinning configurations, and Sec. \ref{sec:3&4paper_SO} to report the EOB Hamiltonians for generic spin-orbit and (aligned) spin$_1$-spin$_2$ couplings. \emph{Bottom left:} the configuration for Schwarzschild geodesics used in the calculation of the 2PM test-body scattering angle for nonspinning systems in Sec.\ref{sec:PM}.  \emph{Bottom right:} the ``test spinning'' configuration  used in the calculation of the test-body scattering angle for spinning systems in Sec.\ref{sec:3&4paper_SO}, in which we take the spins to be aligned. All the quantities are introduced in the ``Notation'' section.
	}
	\label{fig:notation}
\end{figure}

This section unifies the notation used in the introductory chapter.
We consider binaries in either bound or unbound orbits, which we assume to consist of point particles in an approximately flat space. In the former case, we consider the four domains reported in Fig.~\ref{fig:notation}. In the generic two-body problem (sometimes referred to as the ``real description''), we define the masses $m_1$ and $m_2$ such that $m_1<m_2$. The two bodies are located at separations $\boldsymbol{r}_1$ and $\boldsymbol{r}_2$ (with $r_i\equiv |\boldsymbol{r}_i|$). 
We use the following combinations of the masses:
\begin{align}
& M=m_1+m_2&\text{(total mass)},\nnm\\
& \mu =\frac{m_1m_2}{m_1+m_2}&\text{(reduced mass)},\nnm\\
& q\equiv \frac{m_1}{m_2}\leq 1&\text{(mass ratio)},\nnm\\
& \nu =\frac{m_1m_2}{(m_1+m_2)^2}&\text{(symmetric mass ratio)},\nnm\\
& \delta = \frac{m_2 - m_1}{M}&\text{(mass difference)}.\nnm
\end{align}
In the centre-of-mass frame, we describe the binary with masses $M$ and $\mu$, and with relative orbital separation $\boldsymbol{r}=\boldsymbol{r}_2 - \boldsymbol{r}_1$.
The angular momentum $\boldsymbol{L}$ is related to the binary's relative momentum 3-vector $\bm{p}$, in such a way that
\begin{equation}
p^2 = p_r^2 + \frac{L^2}{r^2}, \quad
p_r= \bm{p}\cdot\frac{\bm r}{r}, \quad
L=|\bm{L}|=|\bm{r}\times\bm{p}|.
\end{equation}
We work in the equatorial plane ($\theta=\pi/2$), in which $p_\theta=0$.
We also use the reduced versions $l$, $\hat p_\phi$ and $p_\phi$ of the angular momentum,
\begin{equation}
l \equiv \hat p_\phi \equiv \frac{p_\phi}{\mu} = \frac{L}{GM\mu}\,.
\end{equation}
For spinning two-body systems in the real description, we consider both generic and \emph{aligned-spin} configurations (Sections \ref{sec:PN} and \ref{sec:3&4paper} respectively). In aligned-spin binaries, the canonical spins $\boldsymbol{S}_i$ are parallel to the orbital angular momentum $\boldsymbol{L}$. We also sometimes find it useful to employ the mass-reduced spins
\begin{gather}
\bm{a}_1 = \frac{\bm{S}_1}{m_1}, \qquad 
\bm{a}_2 = \frac{\bm{S}_2}{m_2},
\end{gather}

In the ``effective description'' of Fig.~\ref{fig:notation}, we describe the central body with $M$ and the effective mass with $\mu$. The relative distance between them is $\bm r$. In this domain, we employ effective spins $\boldsymbol{S}$ and $\boldsymbol{S}^*$, such that
\begin{gather}  \label{eq:sstar}
\boldsymbol{S}=\boldsymbol{S}_1+\boldsymbol{S}_2, \quad \boldsymbol{S}_* = \frac{m_2}{m_1}\boldsymbol{S}_1+\frac{m_1}{m_2}\boldsymbol{S}_2.
\end{gather}
While the spins are aligned in the starting configurations of Sec.\ref{sec:3&4paper}, some of the results obtained in the effective description are valid for precessing spins. 

In the lower panels of Fig.~\ref{fig:notation} we move away from the two point-particle description and consider a Schwarzschild geometry, in which the central mass is $m$ and relative distance between BH and test body is $\rtest$, and a ``test spinning BH'' geometry, defined such that the test mass $m_t$ is negligible with respect to the background field generated by $m_b=m$ (in the sense that $m_t/m_b\rightarrow0$). The spins are defined as
\begin{equation}
\boldsymbol{a}_b =\frac{\boldsymbol{S}}{M}, \quad \boldsymbol{a}_t =\frac{\boldsymbol{S}_*}{M}.
\end{equation}
The test spin $a_t =|\bm{a}_\mr t|$ is kept finite as the mass ratio between test mass and central mass vanishes. In the ``test spinning BH'' scenario we consider aligned-spin configurations.
In the remainder of the manuscript, we set $c=1$, with the exception of sections~\ref{sec:PN} and~\ref{sec:3&4paper}  where we reintroduce it to keep track of PN ordering. The metric signature is mostly plus (-,+,+,+). 

\section{Waveform modelling: analytical, semi-analytical and numerical approaches}
\label{sec:approx}

The aim of waveform models is to predict the gravitational radiation of compact-object binaries under a certain set of phenomenologically-relevant parameters (such as the bodies' spins, masses, and the system's eccentricity).
Information from the compact objects' spins may unravel their astrophysical origin, whether it is from a field binary or dynamical scenario~\cite{Abbott:2020niy}.
In the field binary scenario BHs form through stellar collapse~\cite{Oppenheimer:1939ue,May:1966zz,Penrose:1969pc}, and their rates can be calculated with population-synthesis codes~\cite{Belczynski:2001uc,Belczynski:2016obo}. In the dynamical formation scenario, binaries are the results of the dynamical interaction of BHs in dense regions such as globular clusters~\cite{Sadowski:2007dz,OLeary:2007iqe,OLeary:2008myb,Miller:2008yw, Tsang:2013mca,Rodriguez:2015oxa,Rodriguez:2016kxx}. Field binaries tend to align their individual spins~\cite{Campanelli:2006fg}, while the spins of those formed in clusters tend to avoid any preferential direction~\cite{Rodriguez:2016vmx}, and in general spins are predicted to be an important discriminant to distinguish compact objects formed in either scenario~\cite{Vitale:2015tea,Rodriguez:2016vmx,Farr:2017uvj,Fishbach:2017dwv}.  
Likewise, BHs formed in dynamical clusters tend to be more massive and have mass ratios $q< 1$, while binaries formed of components out of field formation scenarios tend to reach more comparable masses $q\sim 1$. 
A similar idea holds for the eccentricity. Stellar-collapse BH binaries have spent long enough in such a configuration for their orbits to have circularized by the time they reach LIGO frequencies ($f\sim 10$ Hz)~\cite{Peters:1964zz}. Still, binaries formed dynamically could have residual eccentricities at these frequencies~\cite{Zevin:2017evb,Zevin:2018kzq,Gondan:2018khr,Samsing:2018ykz,Rodriguez:2018pss,Romero-Shaw:2019itr}.
Moreover, tests of strong gravity also require the most accurate waveforms available, be it to model specific features of a theory that is directly targeted, or in agnostic tests to avoid systematic biases, while an accurate modelling of precession effects~\cite{Ossokine:2020kjp} and higher modes~\cite{Cotesta:2018fcv} is  needed to avoid misinterpreting the source, or to resolve the binary's sky localization (with important consequences in the extraction of cosmological parameters of interest such as the Hubble constant~\cite{Schutz:1986gp,Abbott:2019yzh}). Accurate waveforms also allow one to infer the equation-of-state of neutron stars through measurements of the tidal deformability~\cite{Abbott:2018wiz}.
For all these reasons, such models are crucial to infer astrophysical and cosmological information from GW observations. 
While improvements on waveforms with spins, eccentricity, precession or higher modes are needed in the future, we focus on spins and mass ratios in this thesis. Our focus on spins is motivated by the fact that current models of spinning binaries may not be sufficient in the next LIGO-Virgo-KAGRA runs~\cite{Purrer:2019jcp,Relton:2021cax}. 
Likewise, sources with disparate mass ratios are difficult to model. However, as discussed, they allow to distinguish BH formation scenarios as well as give the possibility, through the observations of extreme mass-ratio inspirals, to test gravity to high levels of accuracy.

Waveforms are conventionally subdivided in three phases: the \emph{inspiral} (with plunges sometime included in this phase), \emph{merger} and \emph{ringdown}.  
The inspiral corresponds to the phase in which the compact objects of the binary are far apart from each other, and are slowly approaching one another. Roughly around the binary's innermost stable circular orbit (ISCO, $r_\text{ISCO}=6 \mm$ for a Schwarzschild BH), the slow inspiral phase smoothly transitions into a quick plunge (inward radial motion) that roughly ends at the light ring (LR, $r_\text{LR}=3 \mm$ for a Schwarzschild BH) of the system. Here, one enters the merger phase. Finally the compact objects merge into a single remnant and enter the ringdown phase, where the perturbations of the remnant object are quickly dissipated away.
There are two main strategies to model binary coalescences. One could solve the relativistic two-body problem \emph{numerically}, obtaining accurate but slow-to-reproduce waveforms, or \emph{analytically}, obtaining fast-to-evaluate waveforms that however are limited to their respective domains of approximation. The former are prerogative of the field of numerical relativity (NR), briefly discussed in Sec.~\ref{sec:NR}.
The latter is prerogative of the field of \emph{analytical relativity}, which aims to include ever more physics in the inspiral and plunge portions of the waveform using approximations to the Einstein equations.
The three approximations that are currently employed are
the PN, PM and SMR ones, described in more detail in the sections below. The ringdown necessitates analytical techniques from BH perturbation theory, and is modelled as a superposition of quasinormal modes.
All these analytical and numerical techniques can be combined in frameworks such as the EOB one (also discussed below). This is a semi-analytical method that synergetically combines information from each approximation, NR simulations, and BH perturbation theory to generate waveforms that cover all the phases of the binary's coalescence, from inspiral through merger to ringdown. Before delving into a description of each approach, we present the basic feature of GWs that underlies all of them, the wavelike propagation of gravity very far from the source.

\subsection{Fundamentals of gravitational waves}
\label{sec:quad}

Gravitational waves are perturbations $h_{\mu\nu}$ of a given background spacetime $g_{\mu\nu}$ that must be a solution to the Einstein field equations,
\begin{equation}\label{EE}
R_{\mu\nu}-\frac{1}{2}g_{\mu\nu}R=8\pi G T_{\mu\nu}\,,
\end{equation}
where $R_{\mu\nu}$ is the Ricci tensor, $R$ the Ricci scalar and $T_{\mu\nu}$ the stress-energy tensor. 
In the present thesis we also need the concept of geodesic equations,
	\begin{equation}\label{geodgbold}
	\frac{d^2 x^{\alpha}}{d\tau^2}+\Gamma^{\alpha}_{\mu\nu}\frac{dx^\mu}{d\tau}\frac{dx^\nu}{d\tau}=0\,,
	\end{equation}
where $\Gamma^{\alpha}_{\mu\nu}$ are the Christoffel symbols, $x^\alpha$ the worldline, and $\tau$ an affine parameter associated to its proper time. Such equations describe the motion of test particles in some smooth spacetime $g_{\mu\nu}$. In terms of the worldlines $x_i^\alpha$ and covariant components of the velocities $u_{i\mu} = g_{\mu\nu}u_i^{\mu}$ (for particle $i$ with respect to $\tau$) they can be also written as
\begin{equation}\label{dudtau}
\frac{dx_i^\alpha}{d\tau}=g^{\mu\nu} u_{i\nu}\,, \quad
\frac{du_{i\mu}}{d\tau}=-\frac{1}{2}\partial_\mu g^{\alpha\beta}(x_i) u_{i\alpha}u_{i\beta}\,.
\end{equation}

\begin{figure}[t]
	\centering
	\includegraphics[width=\linewidth]{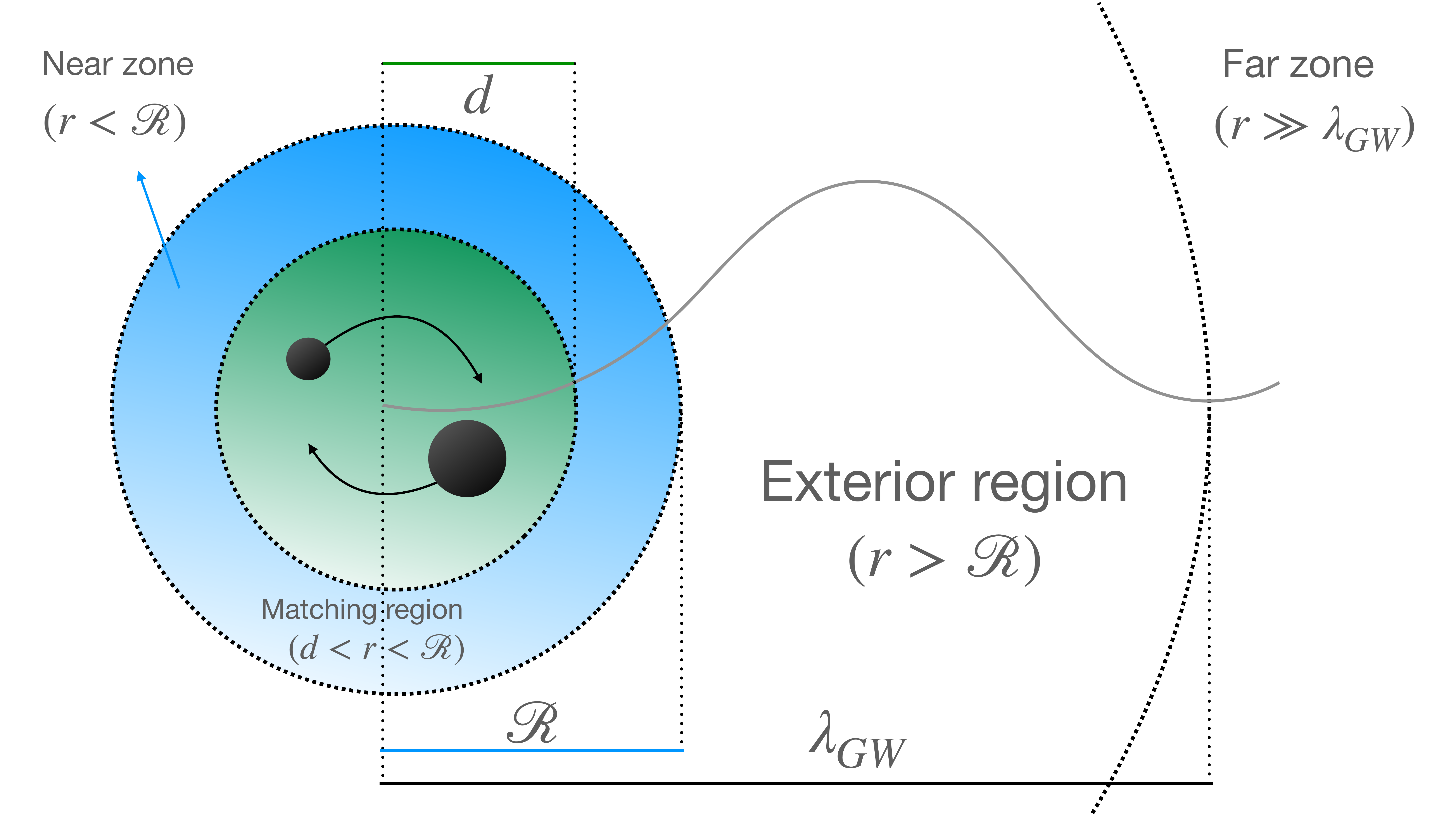}
	\caption{Schematic representation of the GW zones around and away from a source. The definitions are found in the main text. Not to scale.}
	\label{fig:PN_zones}
\end{figure}

We ignore momentarily what generates the waves in Eq.\eqref{EE} and focus instead on the perturbations' behaviour far away from the source. First of all, one needs to rigorously define what is ``far''. 
We define the \emph{far zone} as the set of field points $r$ such that $r\gg \lambda_\text{GW}$, where $\lambda_\text{GW}$ is the wavelength of the GWs (see Fig.~\ref{fig:PN_zones}). At such distances we can neglect the effect of the presence of sources in the field equations ($T_{\mu\nu}=0$). The perturbation in this domain is about the limit of a flat (Minkowski) background
\begin{equation}\label{eq:g_linearized}
	g_{\mu\nu}=\eta_{\mu\nu}+h_{\mu\nu}+\mathcal{O}(G^2)\,\qquad \text{with } h_{\mu\nu}\ll 1 \,.
\end{equation}
The perturbation $h_{\mu\nu}$ scales as $G$, which we also use as an expansion parameter.
The equation governing the propagation of the perturbation $h_{\mu\nu}$ in the far zone is a linearised version of Eq. \eqref{EE}, which must be obtained expanding the Christoffel symbols $\Gamma^{\mu}_{\alpha\beta}[h]$ 
and Riemann curvature tensor $R_{\nu\alpha\beta}^{\mu}[h]$ 
in the perturbation $h\dwmunu$. 
Introducing the D'Alembertian symbol $\Box=\partial_{\alpha}\partial^{\alpha}$ and redefining the perturbations as,
\begin{equation}
\bar{h}_{\mu\nu}=h_{\mu\nu}-\frac{1}{2}\eta_{\mu\nu}h, 
\qquad  (\text{with trace }h=\eta^{\mu\nu}h_{\mu\nu})
\end{equation}
the linearized Einstein equations are
\begin{equation}
\Box\bar{h}_{\mu\nu}-\partial^{\alpha}\partial_{\mu}\bar{h}_{\nu\alpha}-\partial^{\alpha}\partial_{\nu}\bar{h}_{\mu\alpha}+\eta_{\mu\nu}\partial^{\alpha}\partial^{\beta}\bar{h}_{\alpha\beta}=-16\pi G T_{\mu\nu}+\mathcal{O}(G^2)\,.\label{lineq}
\end{equation}
Details can be found, e.g., in Maggiore's textbook~\cite{Maggiore:1900zz} and Ref.~\cite{Flanagan:2005yc}.

The equations can be simplified further by making specific gauge choices. To see the gauge freedom present in the equations, one can consider a gauge function $\xi_{\mu}$, such that the transformed metric is $h'_{\mu\nu}=h_{\mu\nu}+\partial_{\mu}\xi_{\nu}+\partial_{\nu}\xi_{\mu}$.
One can then check that the Riemann tensor is invariant under these changes at linear order in the perturbation. 
Part of the gauge freedom can be fixed setting up the Lorenz gauge by imposing $\partial^{\nu}\bar{h}_{\mu\nu}=0$ (under which, the transformed perturbation $\bar{h}'_{\mu\nu}$ satisfies $\partial^{\nu}\bar{h}_{\mu\nu}=\partial^{\nu}\bar{h}'_{\mu\nu}+\Box\xi_{\mu}=0$).
Equation \eqref{lineq} in the far zone is then 
\begin{equation}\label{eq:lineq_T0}
\Box \bar{h}_{\mu\nu}=0\,, \qquad \text{with } \partial^{\nu}\bar{h}_{\mu\nu}=0\,,
\end{equation}
which is an equation that admits \emph{wave-like} solutions of the form
\begin{equation}\label{sol}
\bar{h}_{\mu\nu}=\mathfrak{R}[a_{\mu\nu}e^{i k_{\alpha}x^{\alpha}}]\,.
\end{equation}
One only needs the propagation four-vector $k_{\alpha}$ to satisfy some conditions. The equation $\Box \bar{h}_{\mu\nu}=0$ implies a dispersion relation $k_{\alpha}k^{\alpha}=0$ (recall to this end that $\partial_{\alpha}\rightarrow ik_{\alpha}$), while the Lorenz gauge implies that $\partial^{\nu}\bar{h}_{\mu\nu} =-ik^{\nu}a_{\mu\nu}e^{i k_{\alpha}x^{\alpha}}= 0$, and therefore a transversality condition $k^{\nu}a_{\mu\nu}=0$.
To make further progress, one could pick a gauge choice of $\xi_{\mu}$ such that $\Box\xi_{\mu}=0$ (corresponding to the case in which the Lorenz gauge is valid even for the transformed perturbation, e.g. $\partial^{\nu}\bar{h}'_{\mu\nu}=0$). Such a choice leads to a \emph{transverse-traceless} (TT) perturbation $h^{TT}_{\mu\nu}$ that still satisfies a wave equation and the Lorenz condition. 
The transverse and traceless nomenclature can be argued in the following way. The Einstein's equations have 10 independent variables, which are reduced to 6 independent ones  when the Lorenz condition is imposed. If we further choose $\Box\xi_{\mu}=0$, we are free to choose 4 conditions and reduce the number of independent variables for the linearised Einstein's equations to only 2. We impose 
four standard conditions on the perturbation: $\bar h^{TT}=0$, $\bar h^{TT}_{0i}=0$, $\bar h^{TT}_{00}=0$ and $\bar h_{ij}^{TT}n^{j}=0$ (for $n^j$ the components of the spatial basis vector $\textbf{n}$), which imply that the waves are both traceless and transverse. Indeed, with this choice we can picture gravitational waves as being represented by a matrix which is traceless, transverse and that contains 2 degrees of freedom (or polarizations) with amplitudes $a_{+}$ and $a_{\times}$. 
For motion along the $z$-direction with frequency $\omega$ (and ignoring constant phase shifts), the plane-wave solution reads
\begin{equation}\label{HxH+}
\bar h_{ij}^{TT}(x^{\mu})=\begin{pmatrix} 
a_{+} & a_{\times} & 0 \\
a_{\times} & -a_{+}&  0 \\
0 & 0 & 0
\end{pmatrix}_{ij} \cos\left[\omega\left(t-\frac{z}{c}\right)\right]\,.
\end{equation}
One can also define a projection operator $\Lambda_{ij}^{kl}=(P_{i}^{k}P_{j}^{l}+P_{i}^{l}P_{j}^{k}-P_{ij}P^{kl})/2$
(with $P_{ik}=\delta_{ik}-n_{i}n_{k}$ a combination of the spatial unit vectors and Kronecker delta)~\cite{Maggiore:1900zz}, such to transform a perturbation $\bar h_{\mu\nu}$ into a TT-gauge perturbation. That is, $
\bar h^{TT}_{\mu\nu}=\Lambda_{ij}^{\mu\nu}\bar h_{ij}$.

\subsection{Post-Newtonian dynamics}
\label{sec:PN} 

The field equations~\eqref{eq:lineq_T0} at the basis of the far-zone GW propagations are not valid ``close'' to the binary, where the wavezone assumption $r\gg \lambda_\text{GW}$ breaks down.
One of the leading analytical schemes that are used to describe fields and motion in the presence of sources is the post-Newtonian approach.
As the name suggests, the leading order in this framework is given by Newtonian theory (which is treated as exact), with successive corrections given in powers of the inverse of the speed of light $c$. To keep track of PN ordering, we insert the PN parameter $c$ in all expressions of the present section. Conventionally, a $n$PN term is a correction of order $\mc O(c^{-2n})$.
The Newtonian metric, with time and spatial coordinates split as $x^\mu = (ct,\boldsymbol{x})$, is 
\begin{equation}
ds^2 = g\dwmunu dx^\mu dx^\nu =  - (c^2 - 2U) dt^2+\delta_{ij}dx^idx^j+\mc O(c^{-2})\,, 
\end{equation}
where one notices the presence of the Newtonian gravitational potential $U$. The field equations for $U$ are
\begin{equation}
\nabla U = - 4\pi G\rho +\mc O(c^{-2})\,,
\end{equation}
where $\rho$ is the energy density found from the stress-energy tensor via $T^{00} = \rho c^2$.
In fact, PN corrections assume that the binary system has small velocities, which implies $v/c\ll 1$ and, for virialized binary systems, also weak fields $G\ll 1$. 
Relativistic corrections to the equations of motion can also be recast about the limit of Newtonian gravity, more precisely in the form of relativistic corrections to the Newtonian acceleration.
This requires thinking of the binary's compact objects as point particles with well specified positions and momenta.
In harmonic coordinates, we define $r_{12}=|\boldsymbol{x}_1 - \boldsymbol{x}_2|$ as the distance between the bodies, $\boldsymbol{n}_{12} = (\boldsymbol{x}_1 - \boldsymbol{x}_2)/r_{12}$ as the unit vector, $\boldsymbol{v}_1=d \boldsymbol{x}_1/dt$ and $\boldsymbol{a}_1=d \boldsymbol{v}_1/dt$ as the coordinate velocity and acceleration of body 1, and $\boldsymbol{v}_{12}= \boldsymbol{v}_1- \boldsymbol{v}_2$ as the relative velocity. For the first body, we can write
\begin{equation}\label{eq:aS0}
\boldsymbol{a}_1 = -\frac{Gm_2}{r_{12}^2} +\frac{1}{c^2} \boldsymbol{a}_1^\text{1PN} +\frac{1}{c^3} \boldsymbol{a}_1^\text{1.5PN}  +\frac{1}{c^4} \boldsymbol{a}_1^\text{2PN}  +\frac{1}{c^5} \boldsymbol{a}_1^\text{2.5PN} + \dots
\end{equation}
The first relativistic corrections $\boldsymbol{a}_1^\text{1PN}$ were first calculated in by Lorentz and Droste~\cite{Droste} (and extended to $N$-body systems by Einstein, Infeld, and Hoffman~\cite{Einstein:1938yz}), and can be written down in terms of harmonic coordinates (see Eq. (203) in Ref. \cite{Blanchet:2013haa})
\begin{align}
 \boldsymbol{a}_1^\text{1PN}=\bigg[&\frac{5 G^2 m_1 m_2}{r_{12}^3}+\frac{4 G^2 m_2^2}{r_{12}^3}\nnm\\
 &+\frac{G m_2}{r_{12}^2}\bigg(\frac{3}{2}(\boldsymbol{n}_{12}\cdot\boldsymbol{v}_1)^2-v_1^2+4 (\boldsymbol{v}_1\cdot\boldsymbol{v}_2) -2 v_2^2\bigg)\bigg]\boldsymbol{n}_{12}\nnm\\
 +&\frac{G m_2}{r_{12}^2}[4(\boldsymbol{n}_{12}\cdot\boldsymbol{v}_1) - 3(\boldsymbol{n}_{12}\cdot\boldsymbol{v}_2)]\boldsymbol{v}_{12}.
\end{align}
These are (time-symmetric) \emph{conservative} corrections that admit a Hamiltonian (or Lagrangian) formulation.
Corrections to Eq.~\eqref{eq:aS0}, either in the form of acceleration corrections, Hamiltonians or Lagrangians, have been calculated at 2PN~\cite{Ohta:1973je,Ohta:1974kp,Ohta:1974pq,Damour:1981bh,Damour:1985mt}, 3PN~\cite{Jaranowski:1997ky,Jaranowski:1999ye,Jaranowski:1999qd,Damour:2000we,Blanchet:2000nv,Blanchet:2000ub,Itoh:2003fy,Foffa:2011ub} and 4PN~\cite{Jaranowski:2015lha,Damour:2014jta,Damour:2016abl,Bernard:2015njp,Bernard:2016wrg,Bernard:2017bvn,Marchand:2017pir,Foffa:2019rdf,Foffa:2019yfl} orders. The interested reader can find more details about the techniques used in Refs.~\cite{Blanchet:2013haa,Goldberger:2006bd} (and references therein). In \chaps~\ref{chap:two} and~\ref{chap:three} (and sections~\ref{sec:1paper} and~\ref{sec:2paper}) we make use of Hamiltonians for nonspinning binaries encoding information up to 4PN order.

The knowledge of the conservative dynamics is not enough to fully describe binaries' dynamics, as it lacks information about the emission of GWs out of the system. 
The PN formalism is however not suited to describe this feature, since solutions in this scheme diverge at large separations. The divergence arises from insisting on treating time and spatial coordinates in $x^\mu$ on a separate footing, which is required to describe the istantaneous action at a distance in the leading order Newtonian gravity. In this scenario, the time derivatives of the field $h\dwmunu$ are subdominant with respect to spatial derivatives. In other words, in PN theory retardation effects are considered small perturbation of istantaneous contributions, for instance when expanding the metric at retarded time,
\begin{equation}
	h\dwmunu\bigg(t-\frac{r}{c}\bigg) = h\dwmunu(t) +\mc O\bigg(\frac{r}{c}\bigg)\,.
\end{equation}
At distances in the far zone, $r \gg \lambda_\text{GW}\sim (c/v)d$ (with $d$ the typical source's size), retardation effects diverge in an unphysical way.
This motivates us to introduce a certain distance $\mathcal{R}\ll \lambda_\text{GW}$ from the source, below which PN solutions are valid. We define $r< \mathcal{R}$ as the \emph{near zone}, and the region $\mathcal{R}<r<+\infty$ outside of it as the \emph{exterior region} (see Fig.~\ref{fig:PN_zones}). To understand the imprint of the PN dynamics on wave propagation in the exterior region (and therefore on the waveform detected by observatories), we need other approximation schemes as the PN one.

The leading method was proposed by Luc Blanchet, Thibault Damour, and Bala Iyer~\cite{Blanchet:1985sp,Blanchet:1989ki,Damour:1990gj}, and uses matched-asymptotic expansions to connect the PN-expanded field solutions in the near zone to solutions that are only expanded in weak fields $G$ only, i.e., PM solutions. The latter are valid in the far zone (and in fact they give the linearized field equations~\eqref{eq:lineq_T0} at leading order). By systematically matching multipolar expansions of both PN and PM solutions in a region where both are valid, the matching region of field points $d < r < \mathcal{R}$ (with $d$ the characteristic size of the source, see Fig.~\ref{fig:PN_zones}), wave solutions that are valid for any separation $r$ can be constructed. Due to the multipolar structure of the matching, the method also goes by the name of ``multipolar PM-PN'' approach.
Through the matching, one transfers the information from the near zone gathered by the PN solutions (which account for source effects from multipoles of $T\dwmunu$) to the far zone. At lowest order in this scheme, the information propagation is described by the \emph{quadrupole} formula.
This is one of the fundamental equations to understand GWs, as it tells us which kind of systems generate waves that can be detected in the far zone.
One can derive an expression for the quadrupole formula from the linearized Einstein equations~\eqref{lineq} in the Lorenz gauge,
\begin{equation}\label{waveq}
\Box \bar{h}_{\mu\nu}=-\frac{16 \pi G}{c^4} T_{\mu\nu}+\mc O(G^2)\,.
\end{equation}
An equation of this sort is solved with Green's function methods~\cite{Flanagan:2005yc}, using a retarded Green's function that tells us how the field at a point $x^\mu=(ct,\textbf{x})$ is connected to the source location $x'^\mu=(ct',\textbf{x}')$ and that satisfies $\Box G(t,\textbf{x}; t',\textbf{x}')=\delta(\textbf{x}- \textbf{x}')\delta(t-t')$, 
\begin{equation}\label{eq:greensfunction}
G(t,\textbf{x}; t',\textbf{x}')=-\frac{1}{4\pi}\frac{1}{|\textbf{x}-\textbf{x}'|}\delta \bigg(t-\frac{|\textbf{x}-\textbf{x}'|}{c}-t'\bigg)\,,
\end{equation}
which gives the following solution
\begin{align}\label{hnonexp}
\bar{h}_{ij}(t,\textbf{x})=&-\frac{16\pi G}{c^4}\int d^4 x' G(t,\textbf{x}; t',\textbf{x}') T_{ij}(x')+\mc O(G^2)\nonumber\\
=&\frac{4 G}{c^4}\int d^3 \text{x}' \frac{1}{|\textbf{x}-\textbf{x}'|} T_{ij}\bigg(t-\frac{|\textbf{x}-\textbf{x}'|}{c}, \textbf{x}'\bigg)+\mc O(G^2)\,.
\end{align}
We evaluate the perturbation at a distance $r\gg d$  away from the characteristic size of the source. As a result, $|\textbf{x}-\textbf{x}'|=r-\textbf{x}'\cdot\textbf{n}+\mathcal{O}(d^2/r)$~\cite{Maggiore:1900zz}, under which
\begin{equation}\label{eq:sol_h_1PM}
\bar{h}^{\text{TT}}_{ij}(t,\textbf{x})=\frac{4 G}{r c^4}\Lambda_{ij}^{kl}(\textbf{n})\int d^3\text{x}' \,  T_{kl}\bigg(t-\frac{r}{c}+\frac{\textbf{x}'\cdot\textbf{n}}{c}, \textbf{x}'\bigg)+\mc O(G^2)\,.
\end{equation}
Requiring that the source moves slowly, is weakly stressed and weakly self gravitating, we have that $|T^{0i}/T^{00}|\sim\sqrt{|T^{ij}/T^{00}|}\sim \sqrt{|U/c^2|}\sim v/c\ll 1$. This approximation is natural for virialized binaries in a bound orbit. We expand $T_{kl}$,
\begin{equation}\label{texp}
T_{kl}\bigg(t-\frac{r}{c}+\frac{\textbf{x}'\cdot\textbf{n}}{c}, \textbf{x}'\bigg) = T_{kl}\bigg(t-\frac{r}{c}\bigg)
+\mathcal{O}\bigg(\frac{1}{c}\bigg)\,.
\end{equation}
Inserting \eqref{texp} in \eqref{hnonexp}, we get a TT perturbation evaluated at retarded time $t'=t-r/c$,
\begin{align}
\bar{h}^{\text{TT}}_{ij}(t,\textbf{x})=\frac{4 G}{r c^4} \Lambda_{ij}^{kl}(\textbf{n})\bigg[&\int d^3\text{x}' \,  T_{kl}(t',\textbf{x}')
\bigg]\bigg|_{t'=t-r/c}+\mathcal{O}\bigg(\frac{1}{c^5},G^2\bigg)\,.\label{eq:quad_full}
\end{align}
Equation~\eqref{eq:quad_full} can be seen as a first formulation of the quadrupole formula. To understand why the expression is fundamentally quadrupolar, and to reduce it to the standard formula often quoted in the literature~\cite{Maggiore:1900zz,Flanagan:2005yc}, a few more manipulations of the stress-energy tensor are in order. We first spell out useful definitions for the mass monopole $M$, dipole $J^i$ and quadrupole $I^{ij}$ as the moments of the mass density $\rho(t,\textbf{x})=c^{-2}T^{00}(t,\textbf{x})$,
\begin{align}
&M\equiv\int d^3 \text{x} \rho(t,\textbf{x})\,,\quad J^i\equiv\int d^3 \text{x} \rho(t,\textbf{x}) \text{x}^i\,,\quad I^{ij}\equiv\int d^3 \text{x} \rho(t,\text{x}) \text{x}^i  \text{x}^j\,.\nonumber
\end{align}
Using Gauss law, and enforcing the $\nu=0$ component of the conservation of energy and momentum $\partial_\mu T^{\mu\nu}=0+\mc O(G^2)$,
one can readily show that the time-variation of the monopole vanishes, $\dot M=0$.
This can be ascribed to a statement of conservation of mass, and it implies that monopoles do not generate GWs. 
It can be similarly shown
that the the lowest-order of the stress-energy tensor implies that linear momentum vanishes, $\dot P = 0$,
and that therefore dipolar terms do not contribute to gravitational perturbations~\cite{Maggiore:1900zz}.
The first multipole moment that does contribute is the mass quadrupole, as it does so through its nonvanishing second derivative $\ddot I_{kl}\propto T_{kl}$.
Finally, we can express the TT-gauge gravitational perturbation with the standard quadrupole formula,
\begin{equation}\label{quad}
\bar h^{\text{TT}}_{ij}(t,\textbf{x})=\frac{2G}{c^4r}\Lambda_{ij}^{kl}(\textbf{n})\ddot{I}_{kl}\bigg(t-\frac{r}{c}\bigg)+\mathcal{O}\left(\frac{1}{c^5},G^2\right)\,.
\end{equation}
The quadrupole formula tells us two important facts. First, only objects that have time-varying mass quadrupoles generate GWs. Isolated stars, for instance, do not emit GWs unless they have some asymmetry and are rotating (such as pulsars), or they go through gravitational collapse (such as supernovae). On the other hand, coalescing binaries do emit GWs. Second, the quadrupole formula tells us that spacetime is stiff and hardly perturbed, since the small numerical factor $G/c^4$ on the RHS of Eq.~\eqref{quad} implies $h\sim\mathcal{O}(10^{-21})$ for typical binaries~\cite{Sathyaprakash:2009xs}. It is also important to note that the same physical information from the quadrupole formula can be recast in terms of a ``radiation-reaction'' (RR) force that backreacts on the (conservative) orbital motion.
For a correct representation of the (local) equations of motion, one must also account for this correction.
From the matching of near-zone PN solutions to far-zone fluxes related to GW emission from the system, the leading RR acceleration related to quadrupolar emission of GWs appears at 2.5PN order in $\boldsymbol a_1^\text{2.5PN}$ from Eq.~\eqref{eq:aS0}, and it reads~\cite{Damour:1981bh,Iyer:1993xi,Iyer:1995rn} (see also Eq. (203) of Ref. \cite{Blanchet:2013haa})
\begin{align}
	\boldsymbol{a}_1^\text{RR}=\bigg[&\frac{208 G^3 m_1 m_2^2}{15r_{12}^4}(\boldsymbol{n}_{12}\cdot\boldsymbol{v}_{12}) - \frac{24 G^3 m_1^2 m_2}{5r_{12}^4}(\boldsymbol{n}_{12}\cdot\boldsymbol{v}_{12})\nnm\\
	&+\frac{12 G^2 m_1 m_2}{5r_{12}^3}(\boldsymbol{n}_{12}\cdot\boldsymbol{v}_{12})v_{12}^2\bigg]\boldsymbol{n}_{12}\nnm\\
	+&\bigg[\frac{8 G^3 m_1^2 m_2}{5r_{12}^4}-\frac{32 G^3 m_1 m_2^2}{5r_{12}^4}-\frac{4 G^2 m_1 m_2}{5r_{12}^3}v_{12}^2\bigg]\boldsymbol{v}_{12}.
\end{align}
This acceleration correction is an example of a \emph{dissipative} term associated to gravitational radiation. These are known up to 4.5PN~\cite{Iyer:1993xi,Iyer:1995rn,Gopakumar:1997ng,Marchand:2020fpt}. 

Other types of corrections to the acceleration~\eqref{eq:aS0} arise when endowing the point particles in the binary with internal structure, e.g., with spins. 
Spinning binaries have a rich phenomenology. For instance, if the spins are not parallel to the angular momentum of the binary,
the system will experience precession effects from the torques the misalignment induces on the spins (which may lead to misinterpreting GW sources if not properly modelled). The first linear-in-spin (\emph{spin-orbit}, SO) corrections appear in the equations of motion at 1.5PN order. The related acceleration $\boldsymbol{a}_\text{SO}$ that appears in $\boldsymbol{a}_1^\text{1.5PN}$ in Eq.~\eqref{eq:aS0} can be found, e.g., in Ref.~\cite{Blanchet:2013haa} [see Eq.~(390) there].
Alternatively, one can write down a Hamiltonian at SO order~\cite{Damour:2007nc}, 
\begin{equation}\label{eq:HSO}
H_{\text{SO}}(\boldsymbol{r}_i,\boldsymbol{p}_i,\boldsymbol{S}_i)=\sum_i  \boldsymbol{\Omega}_{s,i} \cdot \boldsymbol{S}_i\,,
\end{equation}
where $\boldsymbol{r}_i$ and $\boldsymbol{p}_i$ are the bodies' position and momentum, and $\boldsymbol{\Omega}_{s,i}$ is the angular velocity at which the $i^\text{th}$ canonical spin $\textbf S_i$ precesses. Using Arnowitt-Deser-Misner (ADM) coordinates~\cite{Arnowitt:1962hi}, Eq.~\eqref{eq:HSO} can be rewritten as
\begin{equation}\label{eq:Hspinorbit}
H_{\text{SO}}(\boldsymbol{r},\boldsymbol{p},\boldsymbol{S},\boldsymbol{S}^*) = \frac{G}{c^2} \frac{\boldsymbol{L}}{r^3}\cdot(g_S^\text{ADM}\boldsymbol{S} + g_{S^*}^\text{ADM}\boldsymbol{S}^*)\,,
\end{equation}
where all vectorial quantities are defined in the notation section. The coefficients $g_{S,S^*}^\text{ADM}$ are known as ``gyro-gravitomagnetic'' coefficients, as they parametrize the gravitomagnetic field $\sim \boldsymbol{L}/r^3$~\cite{Damour:2008qf}, and at leading 1.5PN order in ADM coordinates simply read $g_{S}^\text{ADM}=2$ and $g_{S^*}^\text{ADM}=3/2$.
Using the Poisson brackets $\{r^a,p_b\}=\delta^a_b$ and $\{S_i^a,S_i^b\}= \epsilon^{ab}_c S^c_i$, one derives from Eq.~\eqref{eq:HSO} a Newtonian-looking (but relativistic) precession equation for the spins,
\begin{equation}\label{eq:Stot_evol}
\frac{d\textbf{S}_i}{dt}=\{ \textbf{S}_i, H_{\text{SO}}(\boldsymbol{r},\boldsymbol{p},\boldsymbol{S}_i)\} = \boldsymbol{\Omega}_{s,i}\times \textbf{S}_i\,.
\end{equation}
For weak fields and small velocities, and restricting the attention to the case of a test gyroscope with negligible spin of the secondary, the equation reduces to the de Sitter's precession formula~\cite{deSitter:1916zza} that has been tested with Gravity Probe B~\cite{Everitt:2011hp}.
In Sec.~\ref{sec:interplay}, we provide more details as to how $\boldsymbol{\Omega}_{s,i}$ can be used to compare approximations to the relativistic two-body problem.

Finally, the spinning PN dynamics is known well beyond the leading 1.5PN SO order. The relative 1PN SO corrections to the equations of motion were calculated by Tagoshi \emph{et al.}~\cite{Tagoshi:2000zg}, Faye \emph{et al.}~\cite{Faye:2006gx}, Damour, Jaranowski and Sch\"afer in ADM coordinates~\cite{Damour:2007nc}, and by Porto~\cite{Porto:2010tr} and Levi~\cite{Levi:2010zu} using effective-field-theory (EFT) methods. In the literature, these are often referred to as next-to-leading SO PN corrections, abbreviated to NLO. The 3.5PN order (next-to-next-to leading correction, N$^2$LO) was reached with Hamiltonian and EFT techniques~\cite{Hartung:2011te,Hartung:2013dza,Marsat:2012fn,Levi:2015uxa}. The current state-of-the-art conservative PN SO dynamics is the 4.5PN (N$^3$LO) order, as we have calculated with the method of Ref.~\cite{Bini:2019nra} in \chaps~\ref{chap:four} and~\ref{chap:five} (see also Ref.~\cite{Levi:2020kvb} for partial results in the EFT scheme). The SO dynamics has some radiation-raction (dissipative) force terms and fluxes as well contributing to 3PN~\cite{Blanchet:2011zv} and 4PN order~\cite{Marsat:2013caa}, see Sec.~11.3 of Ref.~\cite{Blanchet:2013haa}.
One could push the spin dynamics to higher orders in the spin,
gaining with each term a half PN order with respect to the corresponding relative SO coupling (since spins scale as $c^{-1}$). The \emph{spin-spin} couplings contain both spin-squared terms $\boldsymbol{S}_i^2$ and bilinear spin$_1$-spin$_2$ terms. They have been calculated at 2PN order in~\cite{Barker:1975ae,DEath:1975wqz,Thorne:1984mz,Poisson:1997ha}, 3PN order in Refs.~\cite{Steinhoff:2007mb,Porto:2008jj,Porto:2008tb,Levi:2008nh,Steinhoff:2008ji,Hergt:2010pa,Hergt:2011ik,Bohe:2015ana}, and 4PN order in Refs.~\cite{Hergt:2010pa,Levi:2011eq,Hartung:2011ea,Hartung:2013dza,Levi:2011eq,Levi:2015ixa,Levi:2014sba}. In \chap~\ref{chap:five} of this thesis, we extend the knowledge of spin-spin couplings by retrieving the spin$_1$-spin$_2$ (henceforth $S_1S_2$) terms
at 5PN order for spin-aligned configurations with the method of Ref.~\cite{Bini:2019nra} (see also Ref.~\cite{Levi:2020uwu} for partial results). Cubic or quartic-in-spin terms have also been calculated~\cite{Levi:2016ofk,Khalil:2020mmr,Levi:2019kgk,Levi:2020lfn}.

\subsection{Post-Minkowskian dynamics}\label{sec:PM}

Another scheme that can be made applicable to the (near-zone) orbital dynamics of binaries with compact objects at large separations is the PM approximation.
In this approach, one expands the metric $g\dwmunu$ in weak fields $G \ll1$ and about the limit of the flat (Minkowski) metric $\eta\dwmunu$,
\begin{align}\label{eq:gmunu_gexp}
g\dwmunu &= \eta\dwmunu + G h^{(1)}\dwmunu + G^2 h^{(2)}\dwmunu +\mathcal{O}(G^3)\,.
\end{align}
The expanded metric is coupled to the geodesic equations for particle $i$, Eq.~\eqref{dudtau}, where the worldline and four-velocity $u_{i\alpha}$ are also expanded in weak fields,
\begin{align}
x_{i}^\alpha(\tau) &= x^\alpha_{i,0}(\tau) + G x^\alpha_{i,1}(\tau) + G^2 x^\alpha_{i,2}(\tau)+\mathcal{O}(G^3)\,,\nnm\\
u_{i}^\alpha(\tau) &= u^\alpha_{i,0}(\tau) + G u^\alpha_{i,1}(\tau) + G^2 u^\alpha_{i,2}(\tau)+\mathcal{O}(G^3)\label{puexpPM}\,.
\end{align}
The expansion of the metric is the same as the one used in Eq.~\eqref{eq:g_linearized} in the context of linearized gravity. We have added superscripts to indicate the PM order of the expanded quantities, and to help the presentation in this section. The perturbation $h\dwmunu$ in the previous sections corresponds to $h^{(1)}\dwmunu$ here, and the field equations~\eqref{lineq} and~\eqref{eq:lineq_T0} there are valid at 1PM order. Conventionally $n$PM expressions correspond to corrections of order $\mathcal{O}(G^n)$. 

Not only can the PM approximation describe the far-zone propagation of gravitationally-bound systems at low velocities,  as was the case in the virialized binaries in the previous section where the expansion was in the parameter $v/c\sim G$, but also the near-zone dynamics of gravitationally-unbound systems.
In the analysis below, we consider the point-particle stress-energy tensor
\begin{equation}\label{Tmunu_PMexp}
T^{\mu\nu} = \sum_{i=1,2} m_i \int d\tau_i u_i^\mu u_i^\nu \frac{\delta^4 (x-x_i(\tau_i))}{\sqrt{-g}}\,,
\end{equation}
where $g = \det g\dwmunu$, which can be used to study the unbound motion of a scattering source due to the presence of another. This is the natural example of a system in which the near-zone orbital dynamics can be described by the PM approximation. 
Scattering information is encoded in changes to the momenta $p_{i\mu} \equiv m_i g\dwmunu u^\nu_i$ of the scattering particle. In the centre-of-mass frame, the net change from incoming momenta from past infinity to momenta at future infinity (also known as ``impulse''~\cite{Kosower:2018adc}) is found from Eq.~\eqref{dudtau},
\begin{equation}\label{eq:impulse}
\Delta p_{i\mu} = -\frac{m_i}{2}\int_{-\infty}^\infty d\tau_i \partial_\mu g^{\alpha\beta}(x_i) u_{i\alpha}u_{i\beta}\,.
\end{equation}
 One should be aware that, in principle, the metric $g_{\mu\nu}$ diverges at the location of the particle. Within the PM scheme in mind here, one can show that the integral above is finite after an appropriate regularization of the metric (see, e.g., the use of the Hadamard regularization in this context in Sec.II of Ref.\cite{Damour:2017zjx}). The regularization amounts to only retaining Feynman-like diagrams corresponding to the exchange of gravitons, while neglecting those corresponding to graviton self-interactions along the same wordline \cite{Damour:2017zjx}.

The magnitude of the impulse $\ms Q\equiv \sqrt{\Delta p_{1\mu}\Delta p_{1}^\mu}$ (also ``momentum mass transfer'') is a Lorentz-invariant quantity, and it is related to the \emph{scattering angle} $\chi$ via
\begin{equation}\label{chidefinition}
\sin \frac{\chi}{2} = \frac{\ms Q}{2p_\infty}\,.
\end{equation}
Here, $p_\infty$ is the magnitude at infinity of the equal (and opposite) momenta of the scattering particles
\begin{equation}\label{pinfdef}
p_\infty =\frac{m_1 m_2}{E}\sqrt{\gamma^2-1}\,,
\end{equation}
where $\gamma$ is the relative Lorentz factor and $E$ the total energy, which we divide by the rest mass $M$ (defining $\Gamma$),
\begin{equation}\label{gammaGammadefs}
\gamma\equiv u_1 \cdot u_2 = \frac{1}{\sqrt{1-v^2}},\qquad \left(\frac{E}{M}\right)^2 \equiv \Gamma^2 = 1+2\nu (\gamma-1)\,.
\end{equation}

An intriguing feature of the momentum mass transfer $\ms Q$ is its dependence on the symmetric mass ratio $\nu$, which we leverage in Sec.~\ref{sec:N3LOSO}, as well as \chaps~\ref{chap:four} and~\ref{chap:five}. 
More in detail, $\ms Q$ (and relatedly a combination of $\chi$) does not depend on $\nu$ up to 2PM order, and are only \emph{linear} in $\nu$ through 4PM~\cite{Vines:2018gqi,Bini:2019nra,Damour:2019lcq}. For spinning systems the same mass-ratio dependence holds (at least) for spin-aligned configurations~\cite{Antonelli:2020ybz}. This is quite a simple mass dependence at such high (and nonlinear) orders in the PM approximation. Specifically, it allows for the possibility of
deriving the scattering two-body dynamics from geodesic calculations up to 2PM order, the 4PM dynamics from linear-in-$\nu$ gravitational self force (GSF) calculations for hyperbolic orbits (see Ref.~\cite{Long:2021ufh} for work along this direction), as well as the two-body PN scattering dynamics using doubly expanded PN-SMR analytical results at linear order in the mass ratio. 
Furthermore, as we argue in this section, the two-body scattering information can be used to describe binaries in \emph{bound} orbits, with potential applications to GW searches and inference studies.

The simple dependence on $\nu$ of $\ms Q$ is a consequence of the iterative structure of PM calculations, Poincar\'e symmetry and symmetry under exchange of mass (and spin) indices. The derivations can be found in Ref.\cite{Damour:2019lcq} and Sec. \ref{sec:massdep} (which we follow here). The iterative structure is as follows. The $\mathcal{O}(G^n)$ corrections in the worldline $x^\alpha_{i,n}(\tau)$ and velocity $u^\alpha_{i,n}(\tau)$ from Eq.~\eqref{puexpPM} ``correct'' the stress-energy tensor $T\dwmunu$ from Eq.~\eqref{Tmunu_PMexp}, which in turn is inserted in the field equations to obtain the $h^{n+1}\dwmunu$ correction to the metric~\eqref{eq:gmunu_gexp}. Finally, $h^{n+1}\dwmunu$ determines the $\mathcal{O}(G^{n+1})$ corrections in the worldline $x^\alpha_{i,n+1}(\tau)$ and velocity $u^\alpha_{i,n+1}(\tau)$, and the procedure starts over.
At leading order, the system's degrees of freedom in~\eqref{puexpPM} are
\begin{align}
& x_{i,0}^\mu (\tau_i) = y_i^\mu + u_i^\mu \tau_i\,,\\
& p^\mu_{i,0} (\tau_i) = m_i u_i^\mu\,,
\end{align}
and the stress-energy tensor~\eqref{Tmunu_PMexp} (with $g = \eta$) sources the (trace-reversed) metric perturbation 
\begin{equation}
\bar h^{(1)}\dwmunu = 4 G \sum_i m_i\frac{u_{i\mu}u_{i\nu}}{r_i(x)},
\end{equation}
which depends on the Minkowski distance $r_i (x) = \sqrt{(x - y_i)^2 + [u_i\cdot (x-y_i)]^2}$ for field point $x$ from the geodesic in the rest frame, $y_i$. We note the linear dependence on the bodies' masses $m_i$. Using Poincar\'e invariance, the impulse~\eqref{eq:impulse} depends on the bodies' worldlines only through the impact parameter $b^\mu = y_1^\mu - y_2^\mu$, which are orthogonal to the four velocity $u_1\cdot b = u_2\cdot b$ and have magnitude $b\equiv \sqrt{b_\mu b^\mu}$. The 1PM impulse generated by the leading order worldlines is in fact~\cite{Bini:2017xzy}
\begin{equation}
\Delta p_{1\mu} = - \Delta p_{2\mu} = - \frac{2 G m_1 m_2}{\sqrt{\gamma^2-1}} \left[(2\gamma^2 -1)\frac{b^\mu}{b^2}+\mathcal{O}(G)\right] \,.
\end{equation}
The momentum transfer is readily found to be
\begin{equation}\label{Qat1PM}
\ms Q =  \frac{2 G m_1 m_2}{b} \left[\frac{(2\gamma^2 -1)}{\sqrt{\gamma^2-1}}+\mathcal{O}(G)\right]\equiv \frac{2 G m_1 m_2}{b} \left[\ms Q^\text{1PM}+\mathcal{O}(G)\right]\,.
\end{equation}
The leading order of $x^\alpha_{i,0}(\tau)$ and $u^\alpha_{i,0}(\tau)$ eventually leads to the 1PM correction to $\ms Q$. Therefore, through the iterations in the PM scheme, the $x^\alpha_{i,n}(\tau)$ and $u^\alpha_{i,n}(\tau)$ corrections determine the functions $\ms Q^{(n+1)\text{PM}}$ that correct Eq.~\eqref{Qat1PM}. Now, the crucial point to understand the $\nu$ dependence of $\ms Q$ is the mass structure of the metric perturbations $h^{(n)}\dwmunu$. Continuing the iterative procedure, the $h^{(n)}\dwmunu$ perturbations are polynomials of degree $n$
\begin{alignat}{3}\label{eq:hmassstructure}
h_{\mu\nu}^{(1)}(x)&=m_1h_{\mu\nu,m_1}^{(1)}(x)+m_2h_{\mu\nu,m_2}^{(1)}(x),
\nnm\\
h_{\mu\nu}^{(2)}(x)&=m_1^2h_{\mu\nu,m_1^2}^{(2)}(x)+m_2^2h_{\mu\nu,m_2^2}^{(2)}(x)+m_1m_2h_{\mu\nu,m_1m_2}^{(2)}(x),
\nnm\\
&\vdots
\end{alignat}
where $h_{\mu\nu,m^im^j}^{(n)}(x)$ are functions at field point $x$ of degree $n$ and multiplied by $m_1^i m_2^j$ (with $i+j= n$). As the momenta and four velocities at order $\mathcal{O}(G^n)$ lead to the $h^{(n+1)}\dwmunu$ perturbation, and coupling this to the important fact that they are parametrized by the mass-independent leading order values $x^\alpha_{i,0}(\tau)$ and $u^\alpha_{i,0}(\tau)$, the terms $x^\alpha_{i,n}(\tau)$ and $u^\alpha_{i,n}(\tau)$ must also have the same polynomial-in-$n$ mass dependence of Eq.~\eqref{eq:hmassstructure}. Through these, one can see that the polynomial structure carries over to the $\ms Q^{(n+1)\text{PM}}$ corrections of Eq.~\eqref{Qat1PM}. Defining $\ms Q^{n\text{PM}}_{m^im^j}$ to have the same index structure as Eq.~\eqref{eq:hmassstructure}, the mass transfer is found to be of the following form
\begin{alignat}{3}
\ms Q&=\frac{2Gm_1m_2}{b}\bigg[\ms Q^\mr{1PM}
\\\nnm
&\quad+\frac{G}{b}\bigg(m_1\ms Q^\mr{2PM}_{m_1}+m_2\ms Q^\mr{2PM}_{m_2}\bigg)
\\\nnm
&\quad+\frac{G^2}{b^2}\bigg(m_1^2\ms Q^\mr{3PM}_{m_1^2}+m_2^2\ms Q^\mr{3PM}_{m_2^2}+m_1m_2 \ms Q_{m_1m_2}^\mr{3PM}\bigg)
\\\nnm
&\quad+\frac{G^3}{b^3}\bigg(m_1^3\ms Q^\mr{4PM}_{m_1^3}+m_2^3\ms Q^\mr{4PM}_{m_2^3}+m_1^2m_2 \ms Q_{m_1^2m_2}^\mr{4PM}+m_1m_2^2 \ms Q_{m_1m_2^2}^\mr{4PM}\bigg)\bigg]+\mc O(G^5).
\end{alignat}
The other symmetry we make use of is the invariance under $i\leftrightarrow j$, which leads to $\ms Q^{n\text{PM}}_{m_1^im_2^j} = \ms Q^{n\text{PM}}_{m_1^jm_2^i}$ and therefore to the following expression through 4PM order
\begin{alignat}{3}
&\ms Q=\frac{2Gm_1m_2}{b}\bigg[\ms Q^\mr{1PM}
+\frac{G}{b}(m_1+m_2)\ms Q^\mr{2PM}_{m_2}
\nnm\\
&\quad+\frac{G^2}{b^2}\bigg((m_1^2+m_2^2)\ms Q^\mr{3PM}_{m_2^2}+m_1m_2 \ms Q_{m_1m_2}^\mr{3PM}\bigg)
\\\nnm
&+\frac{G^3}{b^3}\bigg((m_1^3+m_2^3)\ms Q^\mr{4PM}_{m_2^3}
+m_1m_2(m_1+m_2) \ms Q_{m_1m_2^2}^\mr{4PM}\bigg)\bigg]+\mc O(G^5).
\end{alignat}
Noticing that the highest mass dependence is still linear in $\nu$, $m_1^3 + m_2^3 = M^3 (1-3\nu)$, as well as that the 3PM functions are also $\nu$-dependent through $m_1^2 + m_2^2 = M^2 (1-2\nu)$, we can rewrite the impulse magnitude at 4PM order as
\begin{alignat}{3}\label{Qtilde}
\ms Q&=\frac{2Gm_1m_2}{b}\bigg[\ms Q^\mr{1PM}
+\frac{GM}{b}\ms Q^\mr{2PM}_{m_2}+\Big(\frac{GM}{b}\Big)^2\bigg(\ms Q^\mr{3PM}_{m_2^2}+\nu \tilde{\ms Q}_{m_1m_2}^\mr{3PM}\bigg)
\\\nnm
&\quad+\Big(\frac{GM}{b}\Big)^3\bigg(\ms Q^\mr{4PM}_{m_2^3}+\nu \tilde{\ms Q}_{m_1m_2^2}^\mr{4PM}\bigg)+\mc O(G^4)\bigg]\equiv \frac{2Gm_1m_2}{b} \ms{\tilde Q}(b,\nu),
\end{alignat}
with $\ms{\tilde Q}(b,\nu)$ collecting the $\nu$-dependence of $\ms Q$. If we were to continue to higher orders, because of the mass structure of the metric perturbations we would find that the 5PM and 6PM corrections $Q^{5,6\mr{PM}}$ are quadratic in $\nu$, $Q^{7,8\mr{PM}}$ are cubic, and so on. Finally, it is easy to show that the mass dependence is the same for the combination $\chi/\Gamma$, where $\Gamma$ is defined to be the (mass reduced) energy of the system in Eq.~\eqref{gammaGammadefs}. Starting from the definition Eq.~\eqref{chidefinition}, and using $p_\infty$ from~\eqref{pinfdef} and $\ms{\tilde Q}$ from~\eqref{Qtilde}, we have
\begin{align}\label{eq:whereGammafrom}
\sin\frac{\chi}{2} &= \frac{\ms Q}{2p_\infty} = \frac{\ms Q E}{2 m_1 m_2\sqrt{\gamma^2-1}} \nnm\\
&= \frac{\ms Q M\Gamma}{2 m_1 m_2\sqrt{\gamma^2-1}} = \frac{G M\Gamma}{ b\sqrt{\gamma^2-1}} \ms{\tilde Q}(b,\nu)\,.
\end{align}
When $\chi$ is expanded out of the argument of the sine, the numerical coefficients in front of the $n$PM corrections in $\ms{\tilde Q}(b,\nu)$ are altered, but its overall dependence on $\nu$ is not. Dividing $\chi$ by the only other symmetric-mass-ratio dependence in $\Gamma=\Gamma(\nu)$, one can see that $\chi/\Gamma$ has the same mass structure as Eq.~\eqref{Qtilde} (and can be parametrized to be linear in $\nu$ at 4PM order).\footnote{The mass-ratio dependence of $\ms Q$ holds for spin-aligned systems as well. In this scenario, the antisymmetric spin tensor $S_i^{\mu\nu}$ is another degree of freedom alongside $x_i^{\mu}$ and $u_i^{\mu}$, which we also expand in $G$,
\begin{equation}
S_i^{\mu\nu}(\tau_i) = S_{i0}^{\mu\nu}(\tau_i) + G S_{i1}^{\mu\nu}(\tau_i) +\mc{O}(G^2),\qquad \text{with }\quad  S_{i0}^{\mu\nu}(\tau_i) = m_i \epsilon^{\mu\nu}_{\rho\sigma} u_i^\rho u_i^\sigma\,.
\end{equation}
The solution for the trace-reversed $\bar h_1^{\mu\nu}$, in harmonic gauge ($\doe_\mu \bar h_1^{\mu\nu}=0$), is
\begin{alignat}{3}
\bar h_1^{\mu\nu}(x)&=4 \sum_{\mr i}m_\mr i \Big(u_\mr i^\mu u_\mr i^\nu+u_\mr i^{(\mu}\epsilon^{\nu)}{}_{\rho\sigma\lambda}u_\mr i^\rho a_{\mr i}^\sigma\doe^\lambda\Big)\frac{1}{r_\mr i},
\end{alignat}
and the field perturbation depends on the spin through the mass-reduced spin vector
\begin{equation}
a_\mr i^\mu=-\frac{1}{2m_\mr i}\epsilon^\mu{}_{\nu\rho\sigma}u_{\mr i}^\nu S_{\mr i0}^{\rho\sigma}\,.
\end{equation}
The (mass-independent) spin vector parametrizes the $h^{(n+1)}\dwmunu$ alongside $x^\alpha_{i,0}$ and $u^\alpha_{i,0}$, implying it does not affect its mass structure. For spin-aligned configurations, one can show that the only Lorentz-invariants the impulse depends on are $b$, $\gamma$, and the spin magnitudes $a_1$ and $a_2$ (with $a^\mu = a_i n^\mu$)~\cite{Vines:2017hyw}. The PM corrections to $\ms Q$ in Eq.~\eqref{Qtilde} for spinning systems up to bilinear $a_1 a_2$ spin corrections can be expanded to be
\begin{alignat}{3}
\ms Q^{n\mr{PM}}_{m_1^im_2^j}&=\ms Q^{n\mr{PM}}_{m_1^im_2^ja^0}(\gamma)+\frac{a_1}{b}\ms Q^{n\mr{PM}}_{m_1^im_2^ja_1}(\gamma)
+\frac{a_2}{b}\ms Q^{n\mr{PM}}_{m_1^im_2^ja_2}(\gamma)+\frac{a_1a_2}{b^2}\ms Q^{n\mr{PM}}_{m_1^im_2^ja_1a_2}(\gamma)\,,\nonumber
\end{alignat}
in such a way that the mass-ratio dependence of the nonspinning case is preserved.}

As mentioned, one of the immediate corollaries of the discussion above is that, up to 2PM order $\mc O(G^2)$, the test-body limit ($\nu\rightarrow0$) completely specifies the scattering dynamics of two-body systems at arbitrary mass ratios. Crucially, the unbound motion one determines in this way can be made applicable to bound orbits, of the type detected by GW interferometers, due to the fact that the scattering angle is a gauge-invariant quantity that encapsulates the near-zone dynamics for both unbound and bound orbits~\cite{Damour:2016gwp,Damour:2014afa,Bini:2017wfr,Bini:2017xzy,Damour:2017zjx,Vines:2017hyw}. The way we do this in the present thesis is through the mediation of (coordinate-dependent) Hamiltonians.
Using the Hamilton's equations for the conservative dynamics with spherical coordinates $\{r,\phi\}$ and conjugate momenta $\{p_r,L\}$,
\begin{align}\label{Ham_eqs}
\dot r &= \frac{\partial H}{\partial p_r}\,,&\qquad \dot \phi = \frac{\partial H}{\partial L}\,,\nnm\\
\dot p_r &= -\frac{\partial H}{\partial r}\,, &\qquad \dot p_\phi = -\frac{\partial H}{\partial \phi}\,,
\end{align}
we have that 
\begin{equation}
\dot \phi = \frac{\partial H}{\partial L} =\frac{\partial p_r}{\partial L} \left(\frac{\partial H}{\partial p_r}\right)= \frac{\partial p_r}{\partial L} \dot r.
\end{equation}
The expression is integrated for the scattering angle $\chi$
\begin{equation}\label{chi_workingdef}
\Delta\phi = \int_{r_\text{min}}^\infty \frac{\dot \phi}{\dot r} dr= \int_{r_\text{min}}^\infty  \frac{\partial p_r}{\partial L} dr= \chi + \pi\,.
\end{equation}
Notice that $p_r$ can be found from Hamiltonians by inverting $H=H(r,p_r,L)\rightarrow p_r(r,L,H)$.\footnote{In general, anything related to $p_r$ (such as its integral over bound orbits, also known as the radial action) is an equally-valid starting point. For instance, Refs.~\cite{Kalin:2019rwq,Kalin:2019inp} propose an alternative approach that leverages the radial action and analytic continuation to make scattering data valid for bound orbits without using Hamiltonians.}
We illustrate this Hamiltonian approach by first deriving the scattering dynamics at 2PM order, and then using the information to specify a Hamiltonian in a generic gauge (that contains the first 2PM Hamiltonian found in the literature~\cite{Damour:2017zjx}). The important takeaway point is that such a Hamiltonian, while specified with scattering data from unbound motion, is valid for both bound and unbound orbits and could thus be used in principle to describe GWs from LIGO-Virgo sources.

The first task is to obtain the test-body scattering angle $\chi_t$ at 2PM order, which, due to the lack of mass-ratio structure through this order, can be made valid for binaries with arbitrary masses. This exercise is also useful to understand how scattering angles can be calculated in practice, as the method presented here is the same underlying all the scattering angle calculations in this thesis. We begin from the Schwarzschild metric (in the equatorial plane with $\theta=\pi/2$),
\begin{equation}
-d\tau^2 = -\frac{\rtest-2\mm}{\rtest}dt^2 + \frac{\rtest}{\rtest-2\mm} d\rtest^2 + \rtest^2 d\phi^2\,,
\end{equation} 
From the mass-shell constraint, $g\dwmunu p^\mu p^\nu = - \mu^2$, we have that
\begin{equation}\label{massshell_schw}
-1 = -\frac{r-2\mm}{\rtest}\dot t^2 + \frac{\rtest}{\rtest-2\mm} \dot \rtest^2 + \rtest^2 \dot\phi^2\,.
\end{equation}
We identify the ($\mu$-rescaled) conserved energy and angular momentum
\begin{equation}\label{cons_quantitites_Schwarz}
\gamma = \frac{\rtest-2\mm}{\rtest}\dot t \,, \qquad \ell = \rtest^2\dot \phi\,.
\end{equation}
We have parametrized the test-mass energy per unit mass with the Lorentz factor $\gamma$ from Eq.~\eqref{gammaGammadefs}~\cite{Vines:2018gqi}. From the conserved quantities~\eqref{cons_quantitites_Schwarz} and the definition of $\gamma$~\eqref{gammaGammadefs}, as well as relating $\ell$ to the impact parameter via 
\begin{equation}\label{ellofb}
\ell = \gamma v b
\end{equation}
and rearranging~\eqref{massshell_schw} to give $\dot \rtest$, we have that the following integrand for the scattering angle
\begin{align}\label{eq:phidotrdot_schw}
 \frac{\dot\phi}{\dot \rtest} &= \frac{\ell }{\rtest^2} \left[\gamma^2 - \frac{\rtest-2\mm}{\rtest} \frac{\ell^2 + \rtest^2}{\rtest^2}\right]^{-1/2} = \frac{\gamma v b}{\sqrt{\gamma^2\rtest^4 - \rtest(\rtest-2\mm)(\rtest^2+\ell^2)}}
\nnm\\
&= \frac{v b}{\sqrt{\rtest^2 v^2(\rtest^2-b^2)+2\mm \rtest[v^2b^2 + \rtest^2 (1-v^2)]}}\,.
\end{align}
We solve the scattering angle~\eqref{chi_workingdef} with (test-body) integrand $\dot\phi/\dot \rtest$. We introduce an auxiliary inverse radius $\invrad\equiv 1/\rtest$, such that
\begin{equation}
\chi_t+\pi = \text{Fp} \int_0^{\invrad_\text{min}} \frac{2bv}{\sqrt{v^2(1-b^2\invrad^2) + 2 \mm \invrad (1-v^2 +b^2\invrad^2v^2)}}d\invrad
\end{equation}
is solved with the method of finite parts (``Fp'') order by order in an expansion in $\mm$.\footnote{We note that the test-body scattering angle in this setting is finite and does not necessarily require the method of finite parts to be solved. The method presented here is the one used throughout the thesis (including for integrals involving Hamiltonians that cannot be solved in closed form). The motivation behind discussing it for the test-body scattering case is that, in this setting, the expressions are not too cumbersome.} The minimum inverse radius $\invrad_\text{min}$ is given by the largest root of the leading PM order of the integrand's denominator. 
We PM expand the test-body scattering angle,
\begin{equation}\label{eq:chiSchw}
\chi_t (b,v)= \sum_{i\geq 0} \chi_t^{i\text{PM}}(b,v)
\end{equation}
The leading order vanishes, as obtained expanding both integrand and integration limit $\invrad_\text{min} = b^{-1}+\mc{O}(G)$,
\begin{equation}
\chi_t^{0\text{PM}} +\pi = \text{Fp} \int_0^{\invrad_\text{min}} \left(\frac{2b}{\sqrt{1-b^2\invrad^2}}\right)d\invrad = 2 \arcsin(b\invrad)\bigg|^{b^{-1}+\mc{O}(G)}_0=\pi +\mc{O}(G)\,.
\end{equation}
At higher orders, one needs to expand both integrand and integration limit in $G$ to avoid infinities. The method of finite parts allows us to bypass the expansion of the latter by simply dropping the infinities that arise when keeping $\invrad_\text{min} = b^{-1}+\mc{O}(G)$ as an integration limit.\footnote{These infinities are artifacts of our choice of excluding PM corrections to $\invrad_\text{min}$. One would find that all the infinite terms cancel out in a systematic PM expansion of both integrand and integration limits, hence why these are simply neglected.} For example, at the next-to-leading 1PM order in the test-body scattering angle we have
\begin{align}
\chi^{1\text{PM}}_t &= -\text{Fp} \int_0^{b^{-1}}
\frac{2b\, \mm\, v\, \invrad\, (1-v^2+b^2\invrad^2v^2)}{[v^2(1-b^2\invrad^2)]^{3/2}}d\invrad \nonumber\\
& = \frac{2\mm}{bv^2} \frac{[1+v^2(1-b^2\invrad^2)]}{\sqrt{1-b^2\invrad^2}}\bigg|_{b^{-1}+\mc{O}(G)}^0 = \frac{2\mm}{b} \frac{1+v^2}{v^2}-(\infty)\,,
\end{align}
where the last infinity is dropped and only the finite part of $\chi^{1\text{PM}}_t$ is kept. With a similar technique, we can carry out the integrals for the subsequent PM orders. We gather the leading finite parts,
\begin{align}\label{eq:chi_t_from2PM}
\chi^{1\text{PM}}_t&=\frac{2\mm}{b} \frac{1+v^2}{v^2},\\
\chi^{2\text{PM}}_t&=\frac{3\pi}{4}\frac{(\mm)^2}{b^2}\frac{4+v^2}{v^2},\\
\chi^{3\text{PM}}_t&=\frac{2(\mm)^3}{3b^3}\frac{(-1+15 v^2 +45 v^4 + 5 v^6)}{v^6},\\
\chi^{4\text{PM}}_t&=\frac{105\pi}{64}\frac{(\mm)^4}{b^4}\frac{16+16v^2+v^4}{v^4},\\
\vdots\nonumber
\end{align}
The 1PM and 2PM expressions can be expressed in terms of $\ell$ and $\gamma$ instead of $b$ and $v$ with the coordinate mappings~\eqref{gammaGammadefs} and~\eqref{ellofb},
\begin{align}
\chi^{1\text{PM}}_t&=\frac{2\mm}{\ell}\frac{(2\gamma^2-1)}{\sqrt{\gamma^2-1}},\\
\chi^{2\text{PM}}_t&=\frac{3(\mm)^2\pi}{4\ell^2}(5\gamma^2-1)\,.
\end{align}
As expected from the discussion of the mass-ratio dependence of the scattering angle above, the arbitrary-mass-ratio scattering dynamics can be determined from these two test-body results. Indeed, noticing that the test-body angular momentum $\ell$ is linked to the canonical angular momentum via $L = \mu\ell/\Gamma$~\cite{Vines:2018gqi}, we have
\begin{align}\label{chi_West}
\chi_t  &=  \frac{2\mm}{\ell}\frac{(2\gamma^2-1)}{\sqrt{\gamma^2-1}}+ 
\frac{2(\mm)^2\pi}{4\ell^2}(5\gamma^2-1) +\mc{O}(G^3)\\
&= \frac{1}{\Gamma}\left[
\frac{2\mm\mu}{ L}\frac{(2\gamma^2-1)}{\sqrt{\gamma^2-1}}+ 
\frac{3\pi}{4\Gamma}\frac{(\mm\mu)^2}{ L^2}(5\gamma^2-1)\right]+\mc{O}(G^3)\equiv \frac{\chi_{\text{2-body}}}{\Gamma}+\mc{O}(G^3)\nonumber\,.
\end{align}
In the last line we have isolated the arbitrary-mass scattering angle $\chi_{\text{2-body}}$ of the binary, which agrees with the expression found by Westpfahl~\cite{Westpfahl:1985}.

The second and final task in this investigation into the 2PM dynamics is the specification of a Hamiltonian valid for bound orbits from the arbitrary-mass scattering angle~\eqref{chi_West}.
In the parametrization used above, we pick a Hamiltonian $H=M\Gamma$ with $\Gamma$ from Eq.~\eqref{gammaGammadefs}. Notice that the parametrization is such that the total energy of the system $\Gamma$ is linked to the test-body's energy $\gamma$ (in units of $\mu$) with an expression that is reminiscent of the EOB ``energy map'' (see Sec.~\ref{sec:EOBform}). We can say that $\Gamma$ is written in a EOB gauge. The Hamiltonian ansatz is written down through the expansion of ``effective'' energy $\gamma$. Here we choose a quasi-isotropic ansatz in which the Hamiltonian depends on the angular momentum only through the combination $L^2/(\mu^2r^2)$. That is,
\begin{align}\label{gamma_ans}
\gamma^2 = 1+\left(\prbar\right)^2 + \Lbarbyrsquared &+ \left[f_1(\gamma_0^2) + g_1(\gamma_0^2) \Lbarbyrsquared \right]\frac{G M}{r}\\
&+ \left[f_2(\gamma_0^2) + g_2(\gamma_0^2) \Lbarbyrsquared \right]\left(\frac{G M}{r}\right)^2+\mc O (G^3)\,,\nnm
\end{align}
where $g_i$ and $f_i$ are arbitrary functions of $\gamma_0^2\equiv1+ p_r^2/\mu^2 + L^2/(\mu^2r^2)$, the 0PM portion of the Hamiltonian. The ansatz is inverted to obtain an expression for $p_r$ [where we set $df_1/d(\gamma_0^2)=0$ and $dg_1/d(\gamma_0^2)=0$ for simplicity]
\begin{align}
\left(\prbar\right)^2 &= \gamma^2  - \left[f_1 + g_1 \Lbarbyrsquared \right]\frac{G M}{r}
- \left[f_2 + g_2 \Lbarbyrsquared \right]
\left(\frac{G M}{r}\right)^2+\mc O(G^3)\nonumber.
\end{align}
Taking the derivative with respect to $L$, the expression for the scattering angle's integrand is
\begin{align}
\frac{\partial p_r}{\partial L} &= \frac{L}{\mu p_{r,0}} + \frac{GM}{r}\frac{L}{\mu}\left(\frac{g_1}{p_{r,0}}+
\frac{f_1+ \Lbarbyrsquared g_1}{2p^3_{r,0}}
\right) +\nonumber\\
&\left(\frac{GM}{r}\right)^2\frac{L}{\mu} \bigg[\frac{g_1}{p_{r,0}}+ \frac{f_2+f_1g_1+\Lbarbyrsquared(g_1^2+g_2)}{2p^3_{r,0}}+ \frac{3\left(f_1+ \Lbarbyrsquared g_1\right)^2}{8p^5_{r,0}}\bigg]+\mc O(G^3)\nonumber,
\end{align}
with $p_{r,0}\equiv\mu\sqrt{\gamma^2 - 1 - L^2/(\mu^2r^2)}$.
Integrating Eq.~\eqref{chi_workingdef} order-by-order in $G$ with the method of finite parts, one obtains the following prediction for the scattering angle from the chosen Hamiltonian ansatz through 2PM order
\begin{align}\label{chi2PM_pred}
\chi_{\text{2-body}}^\text{pred} =& -\frac{GM\mu}{L}\frac{f_1}{\gamma^2-1}\\
&+\frac{\pi(GM\mu)^2}{16L^2}
\left[4 f_1 g_1-8 f_2+\left(\gamma
^2-1\right) \left(3 g_1^2-4
g_2\right)\right]+\mc O (G^3)\nonumber\,.
\end{align}
The prediction can be matched with Westpfahl's scattering angle~\eqref{chi_West}. At 1PM order, this completely specifies the gauge freedom $f_1$ in the prediction, 
\begin{equation}
f_1 = -2 (2\gamma^2 -1)\,.
\end{equation}
The matching at 2PM order~\eqref{chi_West} does not specify the remaining degrees of freedom $g_1$, $g_2$ and $f_2$, which shows the gauge freedom that is present in Hamiltonians. Different choices for the unspecified functions lead to various resummations. One could choose to work with completely isotropic Hamiltonian that does not depend on $L$. Setting $g_1=g_2=0$ in Eq.~\eqref{gamma_ans}, or equivalently in Eq.~\eqref{chi2PM_pred}, one would find that $f_2 = 3(5\gamma^2-1)/\Gamma$ with this resummation.
Other choices can lead us to the 1PM and 2PM Hamiltonians in the EOB gauge~\cite{Damour:2016gwp,Damour:2017zjx}. Matching those Hamiltonians with the ansatz~\eqref{gamma_ans}, one gets
\begin{align}
g_1 = 2\,, \qquad g_2 = -4,,\qquad f_2 = \frac{3-11\Gamma-15\gamma_0^2+23 \gamma_0^2\Gamma}{2\Gamma}\,,
\end{align}
with $\gamma_0=\sqrt{1 + p_r^2/\mu^2  +L^2/(\mu^2r^2)}$.
The important point to make here is that the same 2PM scattering data from Westpfahl's scattering angle is contained in these Hamiltonians. 
Even if such Hamiltonians are expressed in different gauges, they are equivalent in this sense. In Sec.~\ref{sec:1paper}, we compare Hamiltonians with the same PM information to understand which resummation leads to a better base for future waveform models for LIGO-Virgo-KAGRA studies.
The possibility of using scattering data organized in a PM approximation to directly model ready-to-be-used Hamiltonians has been one of the catalysts for the increase in interest in the PM approximation and the binary scattering dynamics in recent years. The other catalyst is related
to the \emph{scattering-amplitudes} revolution, namely the increasing availability of efficient computational techniques from (quantum) amplitude considerations that can be applied to the (classical) scattering two-body problem~\cite{Cheung:2018wkq,Kosower:2018adc,Cheung:2020gyp,Bern:2019crd,Bern:2019nnu,Bern:2020buy,Bern:2021dqo,Bjerrum-Bohr:2018xdl,Bjerrum-Bohr:2019kec,Kalin:2019inp,Kalin:2019rwq,Kalin:2020fhe,Mogull:2020sak,Jakobsen:2021smu}. Such on-shell scattering amplitude techniques allow us to push the knowledge of the nonspinning PM conservative dynamics up to 4PM~\cite{Bern:2021dqo}, of the leading dissipative dynamics terms at 3PM order~\cite{Damour:2020tta}, of the (spin-aligned) dynamics up to 2PM order for spin-spin couplings~\cite{Kosmopoulos:2021zoq} (which was corroborated by effective-field-theory techniques in Ref.~\cite{Liu:2021zxr}), and of tidal interactions at 2PM order~\cite{Kalin:2020lmz}. In the early works to push the state of the art~\cite{Cheung:2018wkq,Bern:2019nnu,Bern:2021dqo}, the framework to calculate PM expansions starts from the ``classical limit'' of scattering amplitude of massive scalar particles minimally coupled to gravity~\cite{Kosower:2018adc,Damour:2019lcq}, from which a match with the amplitude from a nonrelativistic effective-field-theory of massive scalar particles is performed to determine its potential $V(p,r)=\sum_{n=1}(G/r)^n c_n (p^2)$ in terms of momenta $p^2=p_r^2+L^2/r^2$ and separation $r$ at 4PM order~\cite{Bern:2021dqo} (e.g. obtaining coefficients up to $c_4$).\footnote{The 4PM potential has been derived for the local-in-time dynamics only. The complete dynamics also contains tail terms that have not been calculated in Ref.~\cite{Bern:2021dqo}} Crucial ingredients of the calculations are generalized unitarity~\cite{Bern:1994cg,Bern:1994zx,Bern:2008qj} and double-copy constructions~\cite{Bern:2010ue}.
The potential $V(p,r)$, and the associated Hamiltonian
\begin{equation}\label{eq:vanillaHmPM}
H_{\text{mPM}}(r,p)=\sqrt{m_1^2+p^2}+\sqrt{m_2^2+ p^2} 
+\sum_{n=1}^m\frac{G^n}{r^n}c_n(p^2),
\end{equation}
(here in terms of centre-of-mass quantity and using an isotropic gauge for the 0PM kinetic part) can be applied to the case of binaries in a bound orbit as discussed above.

All in all, the insight from relativistic scattering results, whether obtained through quantum or classical techniques, hold the promise to improve the waveform models used in LIGO-Virgo analyses. There are two ways in which this can be achieved, along the lines of what has been discussed so far. One could improve models \emph{directly} by basing waveform models on the PM-expanded potentials $V(p,r)$, or one could use the PM approximation to \emph{indirectly} yield information about the PN expressions. In \chap~\ref{chap:two} (see also Sec.~\ref{sec:1paper}), we study whether PM Hamiltonians lead to better waveform models than those currently employed in LIGO-Virgo template banks and inference studies. The nonspinning PM Hamiltonians found from scattering-amplitude calculations, and containing (up to) the 3PM conservative dynamics, form the backbone for the model comparison analysis. In \chaps~\ref{chap:four} and~\ref{chap:five} (see also Sec.~\ref{sec:3&4paper}), we make use of the mass dependence of the scattering angle discussed at great length here to improve PN models with spins.

\subsection{Small-mass-ratio approximation}
\label{sec:SMR}

The PM and PN approximation break down when the gravitational fields become too strong. 
One approach that allows us to reach this interesting region of spacetime is the SMR approximation, in which the Einstein equations are perturbed about small mass ratios $q\ll1$.
The theoretical ground to model the inspiral and plunge of binaries with the  SMR approximation is the notion of gravitational \emph{self force}, i.e., the effect of a small compact object's own gravitational field on its motion around a more massive companion. At the zeroth order of this approximation, the motion can be derived to be the geodesic of the chosen background spacetime. Subsequent corrections can be thought of as a force on the small compact object due to the perturbation of the background it creates.
The main motivation to pursue SMR corrections to a binary's motion has been (and continues to be) the possibility of using them to model extreme-mass-ratio inspirals (EMRIs)~\cite{Amaro-Seoane:2014ela,Babak:2017tow}, potential LISA sources of stellar-mass BHs of $m\sim 10 M_\odot$ orbiting supermassive BHs with $M\sim 10^6 M_\odot$. 
Recently, \GSF theory has also started to become an important ingredient for numerical and analytical approaches to model binaries with more comparable masses, potentially including intermediate-mass-ratio inspirals~\cite{LeTiec:2011bk,Barausse:2011dq,LeTiec:2011dp,Rifat:2019ltp,vandeMeent:2020xgc,Warburton:2021kwk}.

The central tenets of \GSF theory are the \emph{MiSaTaQuWa} equations of motion~\cite{Quinn:1996am,Mino:1996nk}, which describe the perturbing force on a test mass
around its massive companion (see also Refs.~\cite{Barack:2009ux,Pound:2015tma} for pedagogical discussions). These equations must deal with two concepts that are problematic in GR, ``forces'' and ``test masses''.
Test masses are problematic in GR because they imply the use of the point-particle stress-energy tensor~\eqref{Tmunu_PMexp}, which is ill-defined as a source for the nonlinear Einstein equations~\cite{Pound:2015tma}. To make sense of the equations of motion one needs to work in approximation schemes (such as the previous example of linearized gravity). In \GSF theory, the metric of the system $\boldsymbol{g}\dwmunu=g\dwmunu+ \tilde h\dwmunu +\mathcal{O}(q^2)$ is then expanded around a background $g_{\mu\nu}$ (that must satisfy the vacuum Einstein equations globally~\cite{Poisson:2011nh,Pound:2015tma}) with a small perturbation $\tilde h\dwmunu$ (which we take here to carry over a small parameter, such as the mass ratio $q$). 
To understand what is meant by gravitational ``force'' in \GSF theory,
we assume for the moment that the perturbation is smooth and that it
admits a generalized equivalence principle (GEP),
according to which $\boldsymbol{g}\dwmunu$ is the local metric and test bodies move alongside it in geodesic motion~\cite{DEath:1975jps,Kates:1980zz,Thorne:1984mz}. 
From the perspective of $\boldsymbol{g}\dwmunu$, the small object therefore satisfies the geodesic equation~\eqref{geodgbold},
where 
the Christoffel symbols are calculated with respect to the full metric $\boldsymbol{g}\dwmunu$. If we insist on treating the test body as moving in the geodesic of $g\dwmunu$, which is usually easier to deal with in pratical calculations, the motion is noninertial and 
the test body is accelerated by~\cite{Barack:2009ux}
\begin{equation}\label{geod}
F^\alpha_{\text{grav}}=\mu a^\alpha=\mu\left[ \frac{d^2x^\alpha}{d\tau^2}+\Gamma^{\alpha}_{\mu\nu}\frac{dx^\mu}{d\tau} \frac{dx^\nu}{d\tau}\right]\,,
\end{equation}
in which the Christoffel symbols $^{g}\Gamma^{\alpha}_{\mu\nu}$ are found from the background metric $g\dwmunu$.
From Eq.\eqref{geod} and geodesic motion, one can find the following equation for the gravitational force~\cite{Barack:2009ux,Barack:2018yvs}
\begin{align}\label{eq:misataquwa}
F^\alpha_{\text{grav}}[\tilde h]=-\frac{\mu}{2}\left(g^{\alpha\sigma}+\frac{dx^\alpha}{d\tau}\frac{dx^\sigma}{d\tau}\right)(2\tilde h_{\sigma(\mu;\nu)}-\tilde h_{\mu\nu;\sigma}) \frac{dx^\mu}{d\tau}\frac{dx^\nu}{d\tau}\,.
\end{align}
In this sense, the appearance of such a (fictitious) force is an artifact of our insistence on treating the test body as moving along geodesics in  $g\dwmunu$, when the full metric in fact includes an extra perturbation $\tilde h\dwmunu$. 

While useful to understand in what sense to interpret the GSF force, the above argument does not apply to point particles in GR, which would imply a nonsmooth perturbation at the particle's location. 
For this reason, rigurous derivations of the \GSF do not start from the concept of a geodesic of a point particle, but rather from a rigorous application of matched asymptotic expansions of the Einstein equations. With such analyses one can justify the picture of a point particle moving in geodesic motion around an effective metric, implying that geodesic motion is a derived notion in GSF theory (see, e.g. Sec.~4.1 of Ref.~\cite{Pound:2015tma}).
The remaining issue is to find a suitable metric perturbation $\tilde h\dwmunu$ that solves the field equations coupled to the geodesic motion from which Eq.~\eqref{eq:misataquwa} is derived  (and that must also satisfy the GEP).
In the seminal papers~\cite{Mino:1996nk,Quinn:1996am}, 
a split of the retarded metric perturbation $h^\text{ret}\dwmunu(x)$ at point $x$ on the worldline was defined as $h^\text{ret}\dwmunu (x)= h\dwmunu^{\text{(dir)}}(x) + h\dwmunu^{\text{(tails)}}(x)$. The two contributions are a direct piece $h\dwmunu^{\text{(dir)}}(x)$, which is the portion of the field that propagates along null curves to the point mass location $x$, and a tail contribution $h\dwmunu^{\text{(tails)}}(x)$, which instead is the portion propagating within the past light cone (and backreacting off the spacetime curvature). It can be shown~\cite{Pound:2015tma} that the direct piece diverges at the location of the particle with a (Coulomb-like) divergence $r^{-1}$, where $r$ is the geodesic distance from the particle, and exerts no force on the test body. On the other hand, the tail piece exerts a gravitational force of the form of Eq.~\eqref{eq:misataquwa}. References~\cite{Quinn:1996am,Mino:1996nk} then proposed to consider the (effective) metric $\boldsymbol{g}\dwmunu=g\dwmunu+h\dwmunu^{\text{(tails)}}$, implying that the original \emph{MiSaTaQuWa} equation is $F_\text{grav}[h\dwmunu^{\text{(tails)}}]$.
On a more practical level, the tail piece of the perturbation is hard to implement in calculations, as it does not satisfy a field equation and is not generally twice differentiable at the particle location~\cite{Poisson:2011nh}. Another procedure for the regularization of the metric perturbation that circumvents these problems was proposed by Detweiler and Whithing~\cite{Detweiler:2002mi}.
Here, the retarded metric perturbation can be split into a singular (S) and a regular (R) contribution, $h\dwmunu^{\text{ret}} = h\dwmunu^{\text{(S)}} + h\dwmunu^{\text{(R)}}$. The singular field exhibits the same Coulomb-like $r^{-1}$ divergence as $h\dwmunu^{\text{(dir)}}$, but unlike $h\dwmunu^{\text{(dir)}}$ it solves the (inhomogeneous) linearized Einstein equations. That is, $h\dwmunu^{\text{(S)}}$ has the characteristics of a ``self-field''. The latter contribution $h\dwmunu^{\text{(R)}}$ is a regular field that contains the same information as the tail piece $h\dwmunu^{\text{(tails)}}$, but is also a smooth solution of the (homogeneous) linearized Einstein equations. That is, it behaves as an external field that is independent of the particle. 
The effective metric $\boldsymbol{g}\dwmunu=g\dwmunu+h\dwmunu^{\text{(R)}}$ (which solves the GEP, see Eq.~(10) in Ref.~\cite{Pound:2015tma}) can therefore be used as a way to disentangle the contributions of the external field from the self-field, in a way that is not possible in the original direct-tail split of the metric. The regular-singular split is often the preferred choice for practical calculations.

Numerical impementations of the self force employ either the mode-sum regularization~\cite{Barack:1999wf,Barack:2001bw} or puncture schemes~\cite{Barack:2007jh,Barack:2007we}. 
The reader is referred to \chap 4 in Ref.~\cite{Barack:2018yvs} (and references therein) for more details on either scheme.
To first order in the mass ratio, the \GSF program has been successful in the calculation of the force in a circular~\cite{Barack:2007tm,Shah:2010bi} and eccentric orbit~\cite{Barack:2010tm} around a Schwarzschild background, as well as for circular~\cite{Shah:2012gu}, eccentric~\cite{vandeMeent:2016pee}, and generic~\cite{vandeMeent:2017bcc} orbits in Kerr.  
The \GSF results used in the present thesis are based on state-of-the-art codes by Van de Meent, which use a metric reconstruction procedure~\cite{Cohen:1974cm,Chrzanowski:1975wv,Wald:1978vm,Ori:2002uv,Lousto:2002em,Keidl:2010pm,Pound:2013faa,Merlin:2014qda,vandeMeent:2015lxa,Merlin:2016boc} and the mode-sum regularization to reach the first-order \GSF in Kerr for generic orbits (though we will only make use of results for circular orbits in Schwarzschild). 
To second order in the mass ratio, the puncture method has been used to calculate the \GSF for circular orbits in a Schwarzschild background~\cite{Pound:2019lzj} (see also Refs.~\cite{Pound:2012nt,Pound:2012dk,Pound:2014koa,Pound:2015wva,Pound:2014xva,Miller:2016hjv,Pound:2017psq} for the groundwork of this result).
The resultant self-force information can then be used to evolve EMRIs (see Refs.~\cite{Warburton:2011fk,Osburn:2015duj} for evolutions along eccentric orbits in Schwarzschild), or it can be included in frameworks to generate EMRI waveforms. The latter use a two-timescale analysis in which secular and orbital effects can be separated by virtue of their different timescales~\cite{Miller:2020bft}. Within this approach, one can use near-identity transformation to speed up calculations of the inspiral~\cite{vandeMeent:2018rms}, ``stitch together'' geodesics to recreate the full structure of the waveform~\cite{Hughes:2021exa}, and use GPUs to speed the full waveforms' generation for LISA data-analysis applications~\cite{Chua:2020stf,Katz:2021yft}. At present such waveforms are only available at adiabatic order, though the development of second-order \GSF codes opens the gate to obtain post-adiabatic waveforms, which are needed for an accurate modelling of EMRIs~\cite{Miller:2020bft}.

Along with the numerical \GSF studies, one can also pursue \emph{analytical} calculations of gauge-invariant expressions. 
These are useful for two reasons:  they can be used to cross-check different numerical schemes, or to transfer knowledge from \GSF theories onto other frameworks~\cite{Tiec:2014lba} (such as the EOB one~\cite{Bini:2014zxa,Akcay:2015pjz,Bini:2015bfb,Bini:2017xzy}). 
The expressions can be obtained through the analytical solutions to the Regge-Wheeler-Zerilli~\cite{Regge:1957td,Zerilli:1971wd} and Teukolsky equations~\cite{Teukolsky:1973ha} found by Mano, Suzuki and Takasugi~\cite{Mano:1996mf}: the method is called and ``MST'' after the latter three authors. The MST method yields results in a double PN-SMR expansion, and can be used to push analytical calculations of gauge-invariant quantities to very high PN orders (much higher than currently-known complete PN orders, see Sec.~\ref{sec:PN}), but about the limit of small mass ratios $q$.

\subsection{The effective-one-body formalism}
\label{sec:EOBform}

The EOB formalism provides a relativistic analog of the two-body to one-body reduction in the solution to the Newtonian's two-body problem. 
The main idea of this approach is to map the real two-body motion onto the  motion of an effective particle around a \emph{deformed} Schwarzschild or Kerr background~\cite{Buonanno:1998gg,Buonanno:2000ef}. 
The overarching goal of the EOB program is to describe the relativistic coalescence of two compact objects and the GW emission out of the system. It aims to do so through a combination of analytical and numerical results and by providing a physically-motivated resummation that, starting from such results, can extend their domains of validity to cover the parameter space of coalescing binaries. 
For instance, an EOB model based on PN theory is expected to be valid in regions of stronger gravity than models solely described by the PN approximation \cite{Damour:2000we}; an EOB model based on SMR-expanded quantities is expected to be valid for more comparable-mass binaries than EMRI models~\cite{vandeMeent:2018rms,Miller:2020bft,Katz:2021yft}; an EOB model with both PN and SMR information is expected to be valid in regions of strong gravity and for comparable-mass binaries.
The inspirals and plunges described by EOB Hamiltonians can be further calibrated to NR simulations and complemented by a superposition of quasi-normal modes for the ringdown to provide full inspiral-merger-ringdown models for LIGO-Virgo analyses~\cite{Pan:2011gk,Taracchini:2012ig,Taracchini:2013rva,Bohe:2016gbl,Cotesta:2018fcv,Ossokine:2020kjp}. We focus here on inspirals and plunges, which necessitate two ingredients to be described: a Hamiltonian encapsulating the conservative dynamics, and a prescription for the GW emission through GW fluxes. We discuss both in the following.

The crux of the EOB formalism is the interplay between the real motion of the binary's bodies, described by a Hamiltonian $H_\text{EOB}$ that encapsulates the conservative dynamics of the approximation one starts with (PN, PM or SMR), and the effective motion. This effective motion is itself described by a Hamiltonian $H_\text{eff}$ onto which information from the two-body problem is resummed. 
The effective Hamiltonian is expressed in the centre-of-mass orbital separation $r$ and total and reduced masses $M$ and $\mu$. If the system is spinning, spins are expressed as  $\boldsymbol S$ and $\boldsymbol S_*$ [see Eq. \eqref{eq:sstar} for the definitions].
A dictionary between real and effective domains is then imposed through several mappings. First, an equality between the real and effective action integrals is imposed~\cite{Buonanno:1998gg,Damour:2017zjx}. The standard choice in the nonspinning case, in which we can limit the analysis to a fixed plane described by polar coordinates, is to set
$I_\text{\R}^{\text{real}}=(2\pi)^{-1}\oint
{\Pp}_{\R}\, d\R= I_\text{\R}^{\text{eff}}$ and
$I_\phi^{\text{real}}= (2\pi)^{-1}\oint
{\Pp}_{\phi}\, d\phi= I_\phi^{\text{eff}}$ (where $\Pp_r$ and $\Pp_\phi$ are radial and angular momenta). In this case, the EOB approach further requires a simple functional relation $H_{\text{EOB}} = f(H_{\text{eff}})$ between the real
$H_{\text{EOB}}(r,p_r,p_\phi,\nu)$ and effective
$H_{\text{eff}}(r,p_r,p_\phi,\nu)$ Hamiltonians, the \emph{energy map}~\cite{Buonanno:1998gg,Damour:2016gwp}
\begin{equation}\label{eq:enmap}
H_{\text{EOB}} =M\sqrt{1+2\nu\bigg(\frac{H_{\text{eff}}}{\mu }-1\bigg)}\,.
\end{equation}

The above is the main resummation in EOB theory.
In its original incarnation, $H_{\text{eff}}$ describes the motion of a effective body with mass $\mu$, and is determined by a mass-shell constraint~\cite{Damour:2000we}
\begin{equation}\label{massshell}
g^{\mu\nu}_\text{eff}\Pp_\mu\Pp_\nu+\mu^2 +Q(r,p_r,p_\phi,\nu)
=0,
\end{equation}
where $\Pp_\mu = (\Pp_t,\Pp_r,\Pp_\theta=0,\Pp_\phi)$ is the momentum four-vector and $Q(r,p_r,p_\phi,\nu)$ is a function that quantifies any potential deviations from geodesic motion. The effective metric $g^{\mu\nu}_\text{eff}$ is chosen to be a deformation of Schwarzschild (with parameter $\nu$),
\begin{align}
ds^{2}=-A(\R,\nu)dt^{2} + D(\R,\nu)[A(\R,\nu)]^{-1} d\R^{2}+\R^{2} (d\theta^2 + \sin^2\theta d\phi^2)\,,
\end{align}
in which the potentials reduce to the Schwarzschild metric $A_0(r)=1-2GM/r$ and $D_0(r)=1$ in the test particle limit, $\nu\rightarrow0$. This choice for the effective metric implies that the geodesic limit is included in the EOB framework by construction, and is based on the assumption that the transition from extreme ($\nu\rightarrow0$) to comparable ($\nu=1/4$) mass ratios is smooth. 
The inclusion of the exact test-body limit in the EOB framework, together with resummations, also introduces nonperturbative features (such as the ISCO and light ring) that complement the perturbative two-body information of the real two-body motion.
The effective Hamiltonian is found from the mass-shell constraint noticing that $-\Pp_t = H_\text{eff}$. In the equatorial plane, it is easily found to be
\begin{equation}\label{Ham_eff}
\left(\frac{ H_{\text{eff}}}{\mu}\right)^2=A(r,\nu)\left[1+
\frac{G M}{ r^2}\frac{\Pp_\phi^{2}}{\mu^2}+\frac{A(r,\nu)}{D(\R,\nu)}\frac{\Pp_{r}^{2}}{\mu^2}
+\frac{Q(r,\Pp_r,\Pp_\phi,\nu)}{\mu^2}\right].
\end{equation}
The 1PN conservative dynamics is wholly encapsulated by the above Hamiltonian with potentials taken at their Schwarzschild values.
At 2PN, one can start from a Hamiltonian in a PN Taylor-expanded form, $H_\text{PN}(r,p_r,p_\phi)$, and map \emph{all} of the 2PN information into strikingly compact corrections to these Schwarzschild values. Indeed, the information from the complicated $H_\text{PN}(r,p_r,p_\phi)$ expression (see Eqs.~2.6 in Ref.~\cite{Buonanno:1998gg}) at this order read $A_{2\text{PN}}(r,\nu)=1-2GM/r+2\nu (GM/r)^3$ and $ D_{2\text{PN}}(r,\nu)=1-6\nu (GM/r)^2$. 
At 3PN order, the complete PN dynamics cannot be resummed into the $g_\text{eff}$ metric only, and partial information must be included in the phase-space function $Q(r,p_r,p_\phi,\nu)$.
To uniquely specify the dynamics at this order with the above effective Hamiltonian, one needs to make a choice for the form of $Q(r,\Pp_r,\Pp_\phi,\nu)$. One of the standard choices is the Damour-Jaranowski-Sch\"afer (DJS) gauge, which has been widely applied to the mapping of PN information into the EOB Hamiltonian (see~\cite{Damour:2000we,Damour:2014jta,Damour:2015isa} and references therein). This gauge can be defined by requiring that $Q(r,\Pp_r,\Pp_\phi,\nu)$ is proportional to $\Pp_r$
(without dependence on $\Pp_\phi$).
The DJS EOB potentials have been extended to the 4PN order in Ref.~\cite{Damour:2015isa} and to partial higher-PN orders in Refs.~\cite{Bini:2019nra,Bini:2020nsb}. Moreover, the EOB Hamiltonian in the DJS gauge has also been used to resum SMR information in the potentials through the Detweiler redshift~\cite{Barausse:2011dq,Akcay:2012ea} (even though this leads to unphysical divergences at the binary's LR, as discussed in detail in Sec.~\ref{sec:2paper}). See also Ref.~\cite{Akcay:2015pjz} for the inclusion of SMR information in the $Q$ function.

Another gauge that is used in the EOB literature (and that we use in the next sections) is the \emph{post-Schwarzschild} (PS) gauge. Here, one abandons the idea of mapping two-body information to deformations of the effective metric. Instead, the effective motion is kept to be the test-body limit, while relativistic two-body information gets mapped to the phase-space function $Q$ in its entirety. For a nonspinning system, the Schwarzschild-geodesic Hamiltonian is given in the equatorial plane by
\begin{equation}\label{eq:Hs}
\left(\frac{H_\mr S}{\mu}\right)^2 \equiv \hat H_\mr S^2 =
\left(1- \frac{2GM}{r}\right)\,\left [1+\frac{G M}{ r^2}\frac{\Pp_\phi^{2}}{\mu^2 }+\left(1- \frac{2GM}{r}\right)\frac{\Pp_{r}^{2}}{\mu^2}\right ]\,,
\end{equation} 
while the effective Hamiltonian takes the form 
\begin{equation}\label{eq:Heff_PS}
H^\text{PS}_\text{eff} = \sqrt{H_\mr S^2 (r,p_r,p_\phi) + Q(r, H_\mr S, \nu)}\,.
\end{equation}
This gauge choice has been first proposed by Damour in the context of PM-expanded EOB Hamiltonians~\cite{Damour:2017zjx}, where the $Q$ function is expanded in the weak field parameter $G$ and fixed through a matching with the scattering angle. This procedure is performed at 2PM order by Damour~\cite{Damour:2017zjx} using Westpfahl's scattering angle~\cite{Westpfahl:1985}, and at 3PM order in \chap~\ref{chap:two} using the 3PM scattering angle from Ref.~\cite{Bern:2019nnu}.
We also use this gauge in \chap~\ref{chap:three}  in the context of a fully SMR-based Hamiltonian (to avoid the light-ring divergence, see discussion in Sec.~\ref{sec:2paper}).

To conclude our discussion on EOB Hamiltonians, we discuss \emph{spinning} systems. The EOB Hamiltonian's spin sector has mostly been studied in the context of the PN approximation. One has to choose  whether to resum PN information around the test-spin motion in a Kerr spacetime~\cite{Barausse:2009aa,Barausse:2009xi}, or around the test-mass motion~\cite{Balmelli:2013zna,Balmelli:2015zsa} (each with a different resummation of the spins' information, see Ref.~\cite{Khalil:2020mmr} for a detailed comparison of the two). The spin Hamiltonian we use in \chaps~\ref{chap:four} and~\ref{chap:five} uses the latter approach, and reads~\cite{Damour:2001tu,Balmelli:2013zna,Balmelli:2015zsa}
\begin{align}\label{HSO}
{H}_\text{eff} &= \bigg[
A
\left( \mu^2+ p^2  + B_{p_r} p_r^2  + B_{L} \frac{L^2a^2}{r^2} + \mu^2Q  
\Big)\right]^{1/2}  
\nonumber\\&\quad  
+ \frac{GMr}{\Lambda}  \boldsymbol{L}\cdot \left(g_S \boldsymbol{S} + g_{S^*} \boldsymbol{S}^*\right),
\end{align}
where $\Lambda = (r^2+a^2)^2 - \Delta a^2$ with $\Delta = r^2 - 2GMr + a^2$, $p^2 = p_r^2 + L^2/r^2$, and $a=|\boldsymbol S|/M $ the magnitude of the reduced total spin. The second line of the Hamiltonian is the EOB version of the (ADM) SO Hamiltonian~\eqref{eq:Hspinorbit}.
The potentials read~\cite{Balmelli:2013zna,Balmelli:2015zsa}
\begin{subequations}
	\begin{align}
	A &= \frac{\Delta r^2}{\Lambda} \left(A^0 +  A^\text{even}\right), \\
	B_{p_r} &= \left(1 - \frac{2GM}{r} + \frac{a^2}{r^2}\right) \left(A^0 D^0 +  B_{p_r}^\text{even}\right)  - 1, \\
	B_L &= - \frac{r^2 + 2GM r}{\Lambda}, \\
	Q &= Q^0 + Q^\text{even}.
	\end{align}
\end{subequations}
The zero-spin corrections $A^0(r),~D^0(r)$ and $Q^0(r)$ are given, e.g., by Eq.~(28) of Ref.~\cite{Khalil:2020mmr} and are based on the 4PN nonspinning Hamiltonian derived in Ref.~\cite{Damour:2015isa}. The corrections in odd powers of $S^{1,3,5,\dots}$ are encoded in the gyro-gravitomagnetic factors $g_S$, and $g_{S^*}$~\cite{Damour:2008qf,Nagar:2011fx,Barausse:2011ys}, 
while those in even powers are added through extensions of the $A$, $B$ and $Q$ potentials, $A^\text{even}, \, B_{p_r}^\text{even}$, and $Q^\text{even}$. The state-of-the art expressions with this gauge for the spin sector can be found in \chap~\ref{chap:five} for the spin-orbit and \Sonestwo (aligned-spin) sector, and in Ref.~\cite{Khalil:2020mmr} for higher spin orders.

The second ingredient to build inspiralling waveforms are the fluxes describing the radiation reaction on the binary's orbit in the form of a backreacting force. The basic set of equations for inspiralling orbits in the EOB framework are the Hamilton equations augmented with a radiation-reaction force $\mathcal{F}_\text{RR}$.
In terms of a nonspinning EOB Hamiltonian $H_{\text{EOB}}( r,p_r,p_\phi)$, the equations are a modification of the Hamilton's equations~\eqref{Ham_eqs} that takes into account the radiation reaction,
	\begin{align*}
	& \frac{d r}{d t}=\frac{\partial H^{\text{EOB}} }{\partial p_{r}}\,, \qquad& \Omega\equiv\frac{d\phi}{d t}=\frac{\partial H^{\text{EOB}}}{\partial p_{\phi}}\,, \\
	&\frac{dp_{ r}}{d t}=-\frac{\partial H^{\text{EOB}}}{\partial r}+\mathcal{F}_{\text{RR}}\frac{p_{r}}{p_{\phi}}\,, \qquad& \frac{dp_{\phi}}{d t}=\mathcal{F}_{\text{RR}}\,,
	\end{align*}
The radiation reaction force $\mathcal{F}_\text{RR}$ drives the inspiral of the system and it contains semi-analytical two-body information \cite{Damour:2007xr,Damour:2008gu,Pan:2010hz}. 
The GW flux for quasi-circular (nonspinning) orbits can be written as~\cite{Damour:2008gu}
\begin{equation}
\mathcal{F}_\text{RR} = \frac{M\Omega}{8\pi}
\sum_{l=2}^{l_\text{max}}\sum_{m=l-2}^{l} m^2 \big|\frac{rh_{lm}}{GM}\big|^2\,,
\end{equation}
where $r$ is the binary's orbital separation.
The modes $h_{lm}$ are built from PN theory, and can be resummed multiplicatively (see e.g., Ref.~\cite{Damour:2008gu}),
\begin{equation}
h_{lm}^{(\epsilon)}=
h_{lm}^{(N,\epsilon)}S^{(\epsilon)}_{\text{eff}}T_{lm}e^{i\delta_{lm}}\rho_{lm}^{(l)}f_{lm}^{\text{NQC}}.
\end{equation}
Here, $h_{lm}^{(N,\epsilon)}$ is the Newtonian contribution (with $\epsilon$ representing the parity of the mode). The second term, $S^{(\epsilon)}_{\text{eff}}$, is inspired by the source term present in the right hand side of the Regge-Wheeler-Zerilli equation~\cite{Regge:1957td,Zerilli:1971wd}, which is used to describe the test particle limit in BH perturbation theory's calculations. The third term, $T_{lm}$, contains ``tail'' information from the nonlinear backreaction of GWs on the perturbed spacetime. Further, $\delta_{lm}$ and $\rho_{lm}$ are phase and amplitude corrections that are needed to recover the expected PN expansion of $h_{lm}^{(\epsilon)}$, while $f_{lm}^{\text{NQC}}$ contains calibrated terms from numerical calculations. For a detailed analysis of each term, see Ref.~\cite{Damour:2008gu}. We use the flux in the way described above (but omitting $f_{lm}^{\text{NQC}}$) to drive the inspirals of the models presented in Sec. \ref{sec:2paper} and \chap~\ref{chap:three}.
 
\subsection{Numerical relativity}
\label{sec:NR}
 
 While the work of the present thesis is mostly analytical, we will make use of NR simulations for coalescing BH binaries.
 These simulations solve the full vacuum Einstein equations (rewritten in the 3+1 decomposition~\cite{Arnowitt:1962hi}) and provide the most accurate predictions of the nonlinear radiative dynamics to date, which we can then use as a ``golden standard'' to compare various (semi-)analytical waveform models. 
 The first NR simulations of the last orbits, merger and ringdown of a binary BHs were produced in 2005 by Pretorius~\cite{Pretorius:2005gq} (followed at close quarters by Campanelli \emph{et al.}~\cite{Campanelli:2005dd} and Baker \emph{et al.}~\cite{Baker:2005vv}). Since then, various groups have carried out simulations for equal~\cite{Baker:2006ha,Buonanno:2006ui,Hannam:2007ik} and unequal-mass~\cite{Berti:2007fi,Gonzalez:2008bi,Buchman:2012dw} nonspinning binaries, binaries with aligned spins~\cite{Berti:2007nw,Hannam:2007wf,Chu:2009md}, and precessing binaries~\cite{Szilagyi:2009qz,Campanelli:2008nk}. These works use a variety of techniques that are reviewed, e.g., in Refs.~\cite{Centrella:2010mx,Baumgarte:2010ndz,Sperhake:2011xk}.
 In this thesis, we employ simulations for nonspinning and aligned-spin BH binaries from the Simulating eXtreme Spacetimes (SXS) collaboration~\cite{SXS} (originally carried out for Ref.~\cite{Ossokine:2017dge}). 
 The advantage of using waveform predictions from NR simulations is their unmatched accuracy, especially in the strong-field regime and through the (otherwise-inaccessible) merger. The price to pay is in computational time, since a typical NR simulation requires months on supercomputers to be run.
 As GW data analyses for current detectors require one to fill in banks of hundreds of thousands of templates [as needed for matched filtering, see Sec. \ref{sec:detection}], it is not feasible to use NR waveforms only, and a synergy with analytical approximation schemes is required.

\subsection{Synergy between approximation schemes}\label{sec:interplay}

\begin{figure}
	\centering
	\includegraphics[width=\linewidth]{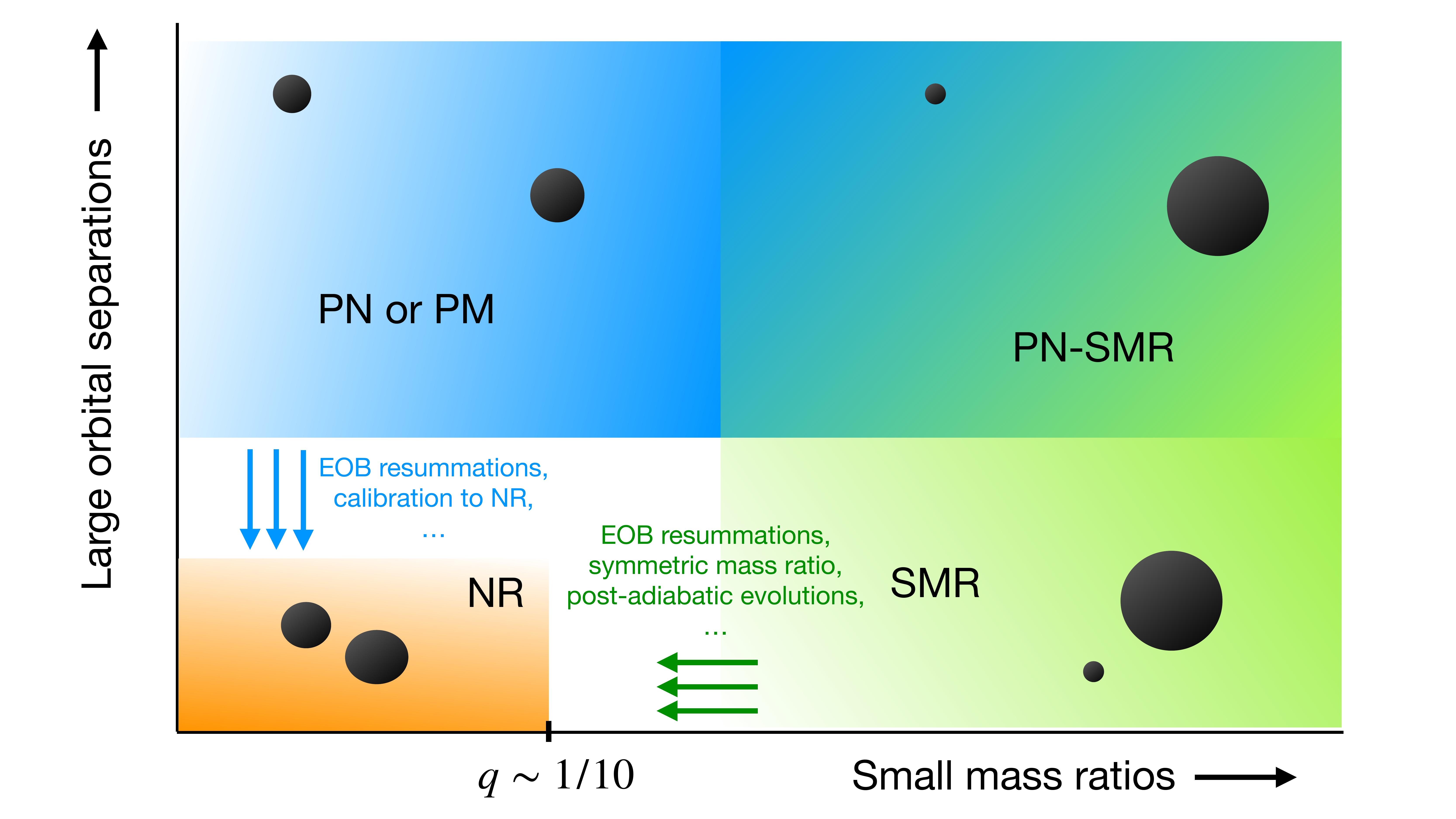}
	\caption{
	Domains of validity for the approaches to the relativistic two-body problem. At large orbital separations, we have the PN and PM approximations (the latter both for bound and unbound orbits). For small mass ratios and large separations, they both overlap with the SMR approximation, which however extends to smaller separations (where gravity is stronger). NR simulations are valid for comparable masses (currently up to $q\sim 1/10$) and near merger. 
    The ``gap'' between PN/PM and NR can be filled with EOB resummations or calibration campaigns; that between the SMR and NR domains with (EOB) resummations  (see \chap~\ref{chap:three}), by using the symmetric mass ratio $\nu$~\cite{LeTiec:2011bk,LeTiec:2011dp,Barausse:2011dq}, or with post-adiabatic waveforms~\cite{vandeMeent:2020xgc}.
}
	\label{fig:paramspace}
\end{figure}

The synergy between the analytical and numerical approaches is crucial for the success of GW astrophysics. There are two complementary levels at which this is manifest. On one hand, waveforms can be generated from  analytical and numerical approaches and combined to accurately cover the inspiral, merger and ringdown. This approach is at the base of the \texttt{SEOBNR}~\cite{Bohe:2016gbl,Babak:2016tgq,Cotesta:2018fcv,Ossokine:2020kjp}, \texttt{TEOBResumS}~\cite{Nagar:2018plt,Nagar:2018zoe} and \texttt{IMRPhenom}~\cite{Hannam:2013oca,Husa:2015iqa,Khan:2015jqa,Khan:2019kot,Garcia-Quiros:2020qpx,Pratten:2020fqn,Pratten:2020ceb} families of models used in LIGO-Virgo studies~\cite{LIGOScientific:2018mvr}, which ``calibrate'' resummations of PN waveforms with NR results.
On the other hand, one can consider combining analytical approaches \emph{before} generating a waveform. The rationale here is to obtain waveforms that cover a larger region of parameter space, and that are less reliant on calibrations to numerical simulations. 

\begin{figure}
	\centering
	\includegraphics[width=\linewidth]{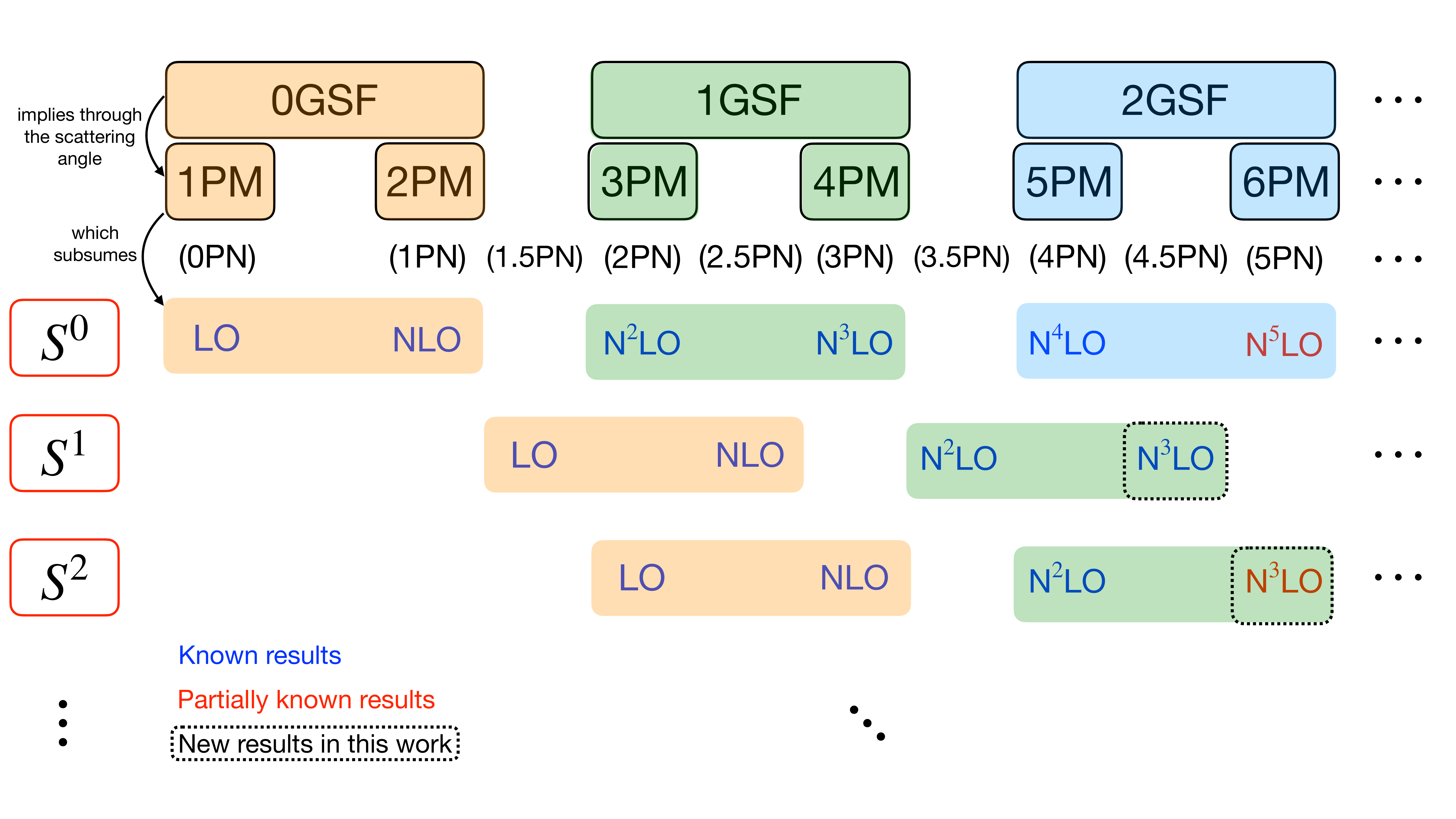}
	\caption{Relation between approximations for the (local) conservative binary dynamics.  The 0\GSF (test-body) limit specifies the 1PM and 2PM order through the scattering angle's mass dependence, which in turn subsume the leading-order (LO) and next-to-leading order (NLO) in the PN approximation (all terms in orange). 
	For nonspinning systems (``$S^0$'') the LO and NLO correspond to 0PN and 1PN terms, for spin-orbit (``$S^1$'') couplings to 1.5PN and 2.5PN, for the spin-spin (``$S^2$'') ones to 2PN and 3PN, and so on. From linear-in-$\nu$ terms (``1\GSF'') one can retrieve the next-to-next-to LO (N$^2$LO) and  next-to-next-to-next-to LO (N$^3$LO) PN terms in green; from the $\mathcal{O}(\nu^2)$ terms, the partially-shown higher-PN order terms in blue.
	 Known results in the PN expansion are reported in dark blue; partial results in red (see Sec.~\ref{sec:PN} for more details). The new results of this thesis are the third-subleading PN orders for $S^1$ and the spin$_1$-spin$_2$ couplings [that complete part of $S^2$], as highligted by the dashed boxes.
	}
	\label{fig:PM_SF_PN}
\end{figure}

\begin{figure}
	\centering
	\includegraphics[width=\linewidth]{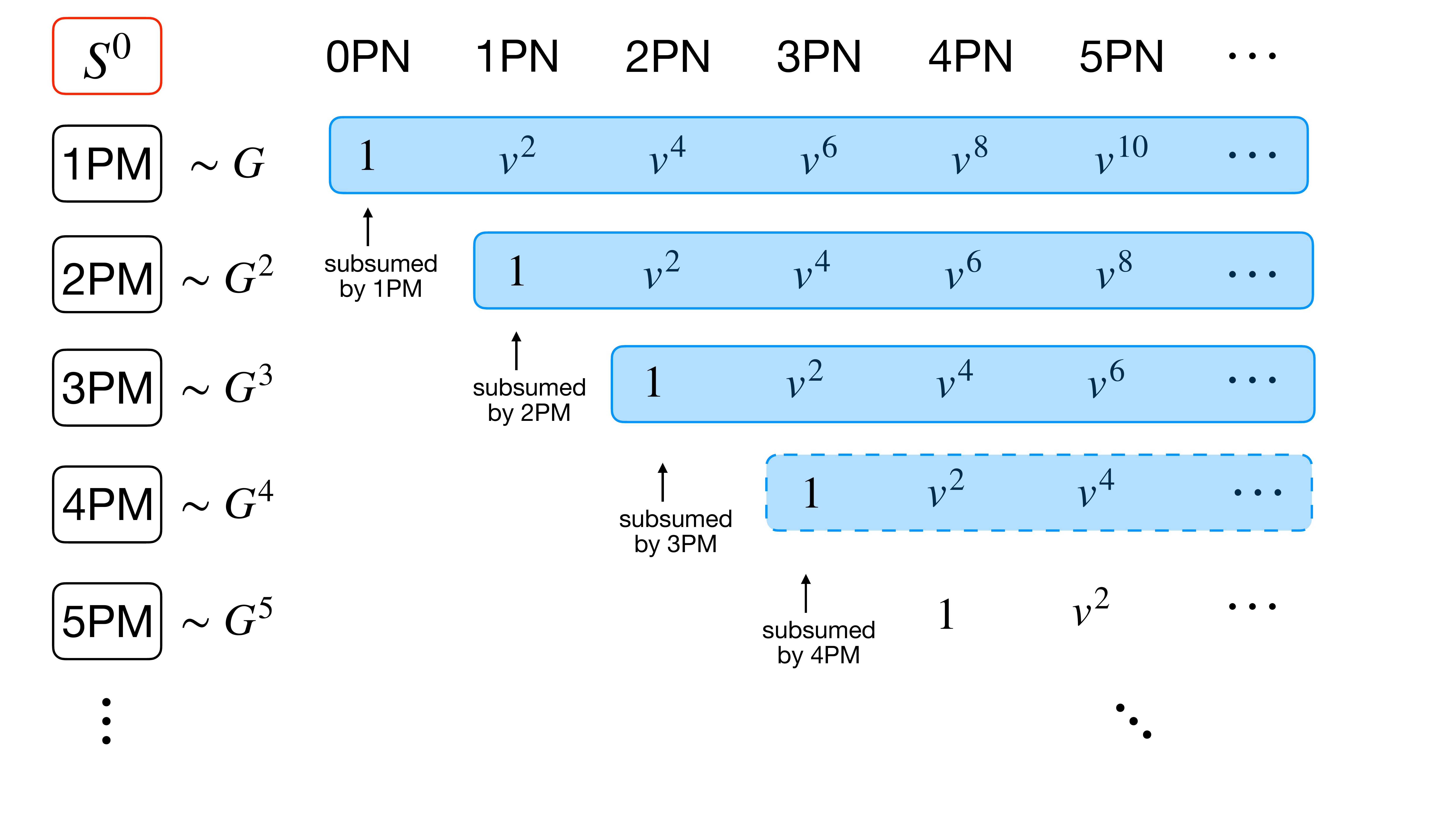}
	\caption{
		The figure shows the relation between PM and PN results for the conservative dynamics of nonspinning (``$S^0$'') systems. In blue we show the known PM results from scattering-amplitude and EFT results~\cite{Cheung:2018wkq,Bern:2019nnu,Kalin:2020fhe}: solid borders indicate complete knowledge of the dynamics, the dashed border at 4PM indicates that only the local conservative dynamics is available. The important point to notice here is that obtaining PM expansions at the $n^\text{th}$ order automatically retrieves the $(n-1)$PN (local) conservative dynamics, while yielding an infinite number of terms at high velocities.
	}
	\label{fig:PMvsPN_S0}
\end{figure}

It is instructive to consider the domains of validity of the approaches discussed thus far. In Fig.~\ref{fig:paramspace}, we show the coverage of the parameter space composed of orbital separation $r$ and mass ratio $q$.\footnote{The cartoon in Fig.~\ref{fig:paramspace} does not take into account other important parameters such as spins or eccentricity, and is only really a good indicator of the respective domains for nonspinning systems. In this nonspinning case, the reach of analytical approximations has been quantified in Fig.(4) of Ref.~\cite{vandeMeent:2020xgc}.} 
The PN and PM approximations -- the latter when recast for bound orbits, see Sec.~\ref{sec:PM} -- are complementary in a region that broadly coincides with the inspiral of the binary. In a sense the PM approximation is more general, since with the same expansion in $G$ it does not assume small velocities. In fact, the N$^n$LO PN order is \emph{completely} subsumed by the $(n+1)$PM conservative dynamics. For instance, the 1PM order subsumes the leading PN order, which includes the 0PN nonspinning dynamics, the 1.5PN spin-orbit terms, the 2PN spin-spin couplings, and so on at higher orders in spins. The relation between PM and PN results is shown for higher orders in Fig.~\ref{fig:PM_SF_PN}, where we also include the relation between PM and \GSF results that the scattering angle's mass dependence imposes (recall the discussion in Sec.~\ref{sec:PM}). \footnote{
	The interdependence between approaches to the relativistic two-body problem in Fig.~\ref{fig:PM_SF_PN} has been checked comparing results from independent calculations for the (local) conservative dynamics. It holds up to N$^4$LO (4PN) for nonspinning systems, and up to the N$^2$LO SO and \Sonestwo dynamics. The N$^3$LO results we obtain in this thesis rely on the conjecture that they can be obtained from \GSF results at linear order, though they still need to be checked with independent calculations that do not depend on the assumption of this relation.
}
Each PM order resums  an infinite number of terms in the PN velocity expansion, as seen for nonspinning binaries in Fig.~\ref{fig:PMvsPN_S0}.
The intertwinement between PM and PN dynamics brings forth self-evident synergy opportunities, as retrieving expansions in one approximation provides part of the other's without further rearrangements. Less clearly, but just as importantly, both can be linked to (and compared against) the SMR and NR schemes valid in the strong-field regime. In this case one needs the mediation of \emph{gauge-invariant} quantities, namely expressions that encapsulate the conservative and dissipative two-body dynamics, and that can be used as benchmarks for comparisons. 

The increased availability of gauge-invariant quantities has led to fervent activity linked both to the possibility of checking the inner workings of each formalisms and to begin quantifying the limits of the respective domains of validity in Fig.~\ref{fig:paramspace} (see Ref.~\cite{Tiec:2014lba} for more details). 
One of the most intriguing results comes from the comparison of gauge-invariant quantities from \GSF codes against those extracted from NR simulations. Using either the periastron advance~\cite{LeTiec:2011bk} or the binding energy~\cite{LeTiec:2011dp,Barausse:2011dq} it was found that the SMR corrections to both quantities encapsulate two-body information remarkably well for all the mass ratio considered in the comparisons. Surprisingly, this includes the \emph{equal-mass} case that one would not expect to be modelled by the \GSF formalism at all. 
The very encouraging agreement between \GSF and NR results for comparable-mass binaries is reached when the SMR corrections are expressed in terms of an expansion in the symmetric mass ratio $\nu$ (rather than $q$). The presence (or better, lack) of a ``gap'' between the SMR and NR approaches for comparable masses in Fig.~\ref{fig:paramspace} has been quantified for nonspinning systems in Ref.~\cite{vandeMeent:2020xgc}. There, it has been estimated that, for systems in a quasi-circular orbit, post-adiabatic corrections to the GW phase that include dissipative effects at second order in $\nu$ could reach an accuracy similar to that of current NR simulations in the regime of intermediate-mass-ratio inspirals, and even potentially for more comparable-mass systems.
These results open up the possibility of using the SMR approximation as a base to accurately model all binaries throughout the plunge in the strong-field regime. Such a notion is at the base of \chap~\ref{chap:three}, where we include SMR information into a EOB Hamiltonian, obtain inspiralling and plunging waveforms from it, and quantify the improvement from PN waveforms brought forth by SMR corrections via the gauge-invariant binding energy. 

To transfer two-body information between schemes, we employ the \emph{spin-precession frequencies} and \emph{Detweiler redshift}, and the formalism of the ``first law'' of binary BH (BBH) mechanics. At this stage, we focus on circular orbits.
The spin-precession invariant is defined as $\psi_i=\Omega_{s,i}/\Omega$, where $\Omega^2_{s,i}\equiv\boldsymbol{\Omega}_{s,i}\cdot \boldsymbol{\Omega}_{s,i} $ are the magnitudes of the bodies' precession frequencies from Eq.~\eqref{eq:Stot_evol}, while $\Omega$ is the binary's orbital frequency. One can also recast the latter as the PN parameter $x \equiv (GM\Omega)^{2/3}$. 
Within the PN formalism, $\psi_i$ has been calculated, e.g., using the near-zone metric in harmonic coordinates~\cite{Bohe:2012mr} and the ADM Hamiltonian~\cite{Dolan:2014pja}. 
In \GSF theory, it has been calculated as an expansion in the mass ratio, $\psi_i = \psi_i^{(0)}(x) + q\, \Delta\psi_i(x)+\mathcal{O}(q^2)$, where $\psi_i^{(0)}(x)$ is the background (geodesic) value and
$\Delta\psi_i(x)$ the secular change to the spin precession. This change can be thought of as the result of a fictitious self-torque that appears from the failing of the spins to be parallel transported along the background metric $g\dwmunu$ (while they are along $\boldsymbol{g}\dwmunu=g\dwmunu+ h^{\text{(R)}}\dwmunu$~\cite{Harte:2011ku}).
The self torque can be calculated numerically or analytically~\cite{Dolan:2013roa,Akcay:2016dku,Kavanagh:2017wot,Akcay:2017azq,Bini:2018ylh}. In the work presented in \chaps~\ref{chap:four} and~\ref{chap:five}, we use the analytical results in a double PN-SMR approximation obtained by Kavanagh \emph{et al.}~\cite{Kavanagh:2017wot} and Bini \emph{et al.}~\cite{Bini:2018ylh,Bini:2019lkm} using MST techniques. 
The other quantity that we employ is the Detweiler redshift~\cite{Detweiler:2008ft}, defined as $z_i\equiv (u_i^t)^{-1}=d\tau_i/dt$ for a particle $i$ in the \GSF effective metric $\boldsymbol{g}\dwmunu$ with four-velocity $u_i^\alpha=dx^\alpha/d\tau_i$ moving with proper time $\tau_i$ with respect to coordinate time $t$. As for the spin-precession invariant, the Detweiler redshift can be calculated wholly within PN theory~\cite{Detweiler:2008ft,Blanchet:2009sd,Blanchet:2010zd}, or expanded in the mass ratio as $z_i(x)=z_i^{(0)}(x) + q\, \Delta z_i(x) + \mathcal{O}(q^2)$, where $z_i^{(0)}(x)$ is the background value of the redshift and $\Delta z_i(x)$ the conservative SMR effect. The Detweiler redshift has been the first quantity to provide a benchmark between approximations to the two-body problem, and it has been calculated to very high accuracy numerically in a Schwarzschild background (by Van de Meent, see app.~\ref{sec:appred}) and Kerr~\cite{Shah:2012gu}, and to high-PN orders in analytical double PN-SMR expansions up to eccentric orbits in Kerr~\cite{Bini:2019lcd}. The latter two calculations in a Kerr background employ a notion of generalized redshift  
$z_i\equiv \langle(u_i^t)^{-1} \rangle =\langle d\tau_i/dt \rangle$, where $\langle\cdot\rangle$ is the orbit average~\cite{Barack:2011ed}.
The definition of the Detweiler redshift suggests that one may think of it as the gravitational redshift of the particle in the curved spacetime in which it moves. However, it is calculated through a particular combination of the regularized metric perturbation $h_{uu}\equiv h^{(R)}\dwmunu u_i^\mu u_i^\nu$~\cite{Detweiler:2008ft} that does not correspond to the physical retarded perturbation. A better interpretation has been proposed by Le Tiec and collaborators~\cite{LeTiec:2011ab}, which used an analogy between the ``first law of BBH mechanics'' and the laws of BH mechanics to establish a link between the redshift and the surface gravity $\kappa$ of the small companion (later confirmed with NR simulations~\cite{Zimmerman:2016ajr}).

The first law of BBH mechanics relates \emph{global} quantities (such as Hamiltonians) to \emph{local} quantities (such as the Detweiler redshift or spin-precession frequencies). Somewhat restrictively, it can be considered a bridge that connects gauge-invariant information to Hamiltonians.
It has been derived for nonspinning point particles in circular orbits in Ref.~\cite{LeTiec:2011ab}, spinning particles on circular orbits in Ref.~\cite{Blanchet:2012at}, nonspinning particles in eccentric orbits in Refs.~\cite{Tiec:2015cxa,Blanchet:2017rcn}, and precessing eccentric orbits of a point mass in the small mass-ratio approximation~\cite{Fujita:2016igj}.
In Sec.~\ref{sec:2paper}, we employ the first law for circular orbits valid for nonspinning systems, which reads
\begin{equation}\label{eq:firstlaw_circ}
\delta M -\Omega \delta L = \sum_i z_i\delta m_i\,,
\end{equation}
and gives the variation in the ADM mass $M$ and angular momentum $L$ as a response to the variation in the individual masses.\footnote{In this version of the first law, the identification of the redshift with the surface gravity $\kappa$ of the individual BHs is manifest, since the analogous first law of BH thermodynamics reads
\begin{equation}
\delta M_\text{BH} -\Omega_\text{H} \delta J_\text{BH} = 4\kappa m_\text{irr}\delta m_\text{irr}\,,
\end{equation}
with $m_\text{irr}^2 = A/(16\pi)$ the irreducible mass, $A$ the surface area, $\Omega_\text{H}$ the BH's horizon frequency, and $M$ and $J$ its mass and angular momentum~\cite{Bardeen:1973gs}.
}
Equation \eqref{eq:firstlaw_circ} is derived for conservative spacetimes with helical symmetry, that is, with a helical Killing vector. However, it requires the use of (ADM) quantities that are not defined in a global sense, in this setting, at the order in mass ratios required in this work. One can still define a formally-equivalent (but Hamiltonian-based) version of the first law that uses a local notion of energy along the orbit of the particle, which would suggest an identification of such local energy to the Bondi mass at any retarded time \cite{Fujita:2016igj}. The identification has been shown to agree well, though with some differences, at second order in a perturbation in the mass ratio \cite{Pound:2019lzj}.

In Sec.~\ref{sec:3&4paper} we employ quantities calculated via a first law for spin-aligned eccentric systems at arbitrary mass ratios, which can be argued to be of the form
\begin{equation}\label{eq:firstlawtemp}
\delta E = \Omega_r \delta I_r + \Omega \delta I_\phi +\sum_i (z_i\delta m_i + \Omega_{s,i}\delta S_i)\,,
\end{equation}
where we have introduced the ADM energy $E\equiv H$, the radial $I_r$ and angular $I_\phi$ action variables
\begin{equation}
I_r = \frac{1}{2\pi}\oint dr p_r\,, \qquad I_r = \frac{1}{2\pi}\oint d\phi p_\phi \equiv L\,,
\end{equation}
and the radial $\Omega_r$ and angular $\Omega$ frequencies, 
\begin{equation}
\Omega_r = \frac{\partial H}{\partial I_r}\,, \qquad \Omega = \frac{\partial H}{\partial I_\phi} = \frac{\partial H}{\partial L}\,.
\end{equation}
As the names suggest, the $z_i$ and $\Omega_{s,i}$ in Eq.~\eqref{eq:firstlawtemp} can be identified with the generalized redshift and (averaged) spin precession frequencies in Eq.~\eqref{eq:Stot_evol}. We refer the reader to Sec.~\ref{firstlaw} for an argument as to why this identification is possible for spin-aligned, eccentric, and arbitrary-mass-ratio binaries.
In terms of a Hamiltonian, they can be found to be
\begin{equation}\label{1lawnew}
z_i=\bigg\langle \frac{\partial H}{\partial m_i} \bigg\rangle, \qquad
\Omega_{s,i} =\bigg\langle \frac{\partial H}{\partial S_i}\bigg\rangle.
\end{equation}
As an alternative to the use of Hamiltonians, one could use the radial action $I_r$. The first law in terms of $I_r$ implies that
\begin{align}
&\frac{T_r}{2\pi}\equiv \frac{1}{2\pi}\oint dt = \bigg(\frac{\partial I_r}{\partial E}\bigg)_{L,m_i,S_i}\,, \qquad &
\frac{\Phi}{2\pi}\equiv \frac{1}{2\pi}\oint d\phi=-\bigg(\frac{\partial I_r}{\partial L}\bigg)_{E,m_i,S_i}\,,\nnm\\
&\frac{\mathcal{T}_i}{2\pi} \equiv \frac{1}{2\pi}\oint d\tau_i=-\bigg(\frac{\partial I_r}{\partial m_i}\bigg)_{E,L,m_j,S_i}\,, \qquad &
\frac{\Omega_{s,i}}{2\pi}=-\frac{1}{T_r}\bigg(\frac{\partial I_r}{\partial S_i}\bigg)_{E,L,m_i,S_j}\nnm\,.
\end{align}
The subscripts indicate quantities that must be kept fixed in the radial action. The generalized redshift and spin-precession invariant are calculated through radial actions as
\begin{equation}\label{redgennew}
z_i= \frac{\mathcal{T}_i}{T_r}\,,\qquad \psi_i = \frac{\Omega_{s,i}}{\Phi}\,.
\end{equation}
In \chaps~\ref{chap:four} and~\ref{chap:five}, the redshift and precession frequencies are found using both the Hamiltonian and radial action approaches.

\section{Improving the orbital dynamics with insight from black-hole scattering dynamics}
\label{sec:improvingweak}

Having argued the synergy opportunities between approximation schemes, we now focus our attention to the relations between the PM and PN dynamics. In principle, both can be used to describe binaries in a bound orbit at separations relevant for current GW detectors. However, the availability of results in the PN approximation implies this is used as a basis to model coalescing binaries in GW searches and inference studies. It is therefore natural to ask whether one can use available insight from relativistic scattering, and the associated PM expansions, to improve the models used for LIGO-Virgo analyses. The most direct approach is to compare various models with PM information between themselves and with PN models used in current GW studies; a more indirect approach employs insight from relativistic scattering (in the form of the mass-dependence of $\chi/\Gamma$ discussed in Sec.~\ref{sec:PM}) to improve PN models.

\subsection{Direct approach: the energetics of post-Minkowskian gravity}
\label{sec:1paper}

The breakthroughs in the calculations of the PM dynamics at high orders imply it is now possible to check how well the PM and PN approximations compare with each other against NR simulations.
In \chap~\ref{chap:two} we do this, using the binding energy as a benchmark to establish which approximation constitutes a better basis for models. 
Potential improvements from the PM results are most easily addressed with two key comparisons: i) between different classes of PM models, to check how well PM information is resummed in Hamiltonians, ii) and between the best performing PM models and the PN ones used in GW searches and inference studies.

A first class of purely PM models that is available to us can be obtained from the Hamiltonian in Eq.~\eqref{eq:vanillaHmPM}. This form of the Hamiltonian has been chosen to present the first results at 2PM and 3PM order for nonspinning systems from scattering-amplitude calculations~\cite{Cheung:2018wkq,Bern:2019nnu,Bern:2019crd}. In the comparison studies we employ models obtained inserting $c_1$, $c_2$ and $c_3$ from Eq.~(10) of Ref.~\cite{Bern:2019nnu} into Eq.~\eqref{eq:vanillaHmPM}.\footnote{
In principle $c_4$ is known from the 4PM local dynamics \cite{Bern:2021dqo}. We leave the corresponding Hamiltonian out of the comparison analysis, as it subject of current work.} Another class of PM models can be obtained employing the EOB Hamiltonian in the PS gauge, in which the effective Hamiltonian is~\eqref{eq:Heff_PS}.
In terms of the mass-reduced quantities
\begin{gather}\label{defs_massreduced}
\hat H^\text{PS}_\text{eff}\equiv\frac{H^\text{PS}_\text{eff}}{\mu},\quad \hat H_\mr S\equiv \frac{H_\mr S}{\mu} = \gamma,\quad u\equiv\frac{GM}{r },
\end{gather}
(as well as those from the notation section), the effective Hamiltonian can be found PM-expanding the phase space function $Q(u,\hat H_\mr S^2,\nu)$ in the parameter $u\sim G$,
\begin{align}\label{HeffPS}
(\hat H^\text{PS}_\text{eff})^2&=
\hat H_\mr S^2
+(1-2u)\Big[u^2q_\mr{2PM}(\hat H_\mr S^2,\nu)+u^3q_\mr{3PM}(\hat H_\mr S^2,\nu)
+\mc O(G^4)\Big],\nnm\\
&\text{with }\hat H_\mr S^2 = (1-2u) \left[1 + l^2 u^2 + (1-2u) \hat{p}_r^2 \right].
\end{align}
The functions $q_\mr{n\text{PM}}$ can be specified by requiring that they reproduce the known scattering angle for two-body systems. One first notices that the $q_\mr{1\text{PM}}(\hat H_\mr S,\nu)$ term is missing in Eq.~\eqref{HeffPS}, as a result of the fact that the energy map applied to $\hat H^\text{PS}_\text{eff}= \hat H_\mr S+\mathcal{O}(G^2)$ is enough to reproduce the two-body dynamics at 1PM order \cite{Damour:2016gwp,Damour:2017zjx}. At 2PM and 3PM order, the functions are easily obtained with results from the literature. Damour gives in (5.6) and (5.9) of Ref.~\cite{Damour:2017zjx} expressions for these functions in terms of $ \hat{\mathcal{E}}_\text{eff}\equiv\gamma \equiv \hat H_\mr S $, the coefficients of the Schwarzschild test-body scattering angle $\chi_t$ in our Eq.~\eqref{eq:chi_t_from2PM}, and the generic (two-body) scattering-angle coefficients. Bern \emph{et al.}~\cite{Bern:2019nnu} provide these last coefficients, as calculated with scattering-ampitude techniques, see their Eq.~(12). In the notation we are using in this introduction, the 3PM scattering angle from Bern \emph{et al.}~\cite{Bern:2019nnu} reads $\chi_\text{3PM}(\Gamma,l)=2\sum_{n=1}^3l^{-n}\chi_n(\Gamma)+\mc O(G^4)$, with
\begin{subequations}\label{eq:chi2PM3PM}
	\begin{align}
	\chi_1&=\frac{2\gamma^2-1}{\sqrt{\gamma^2-1}},\\
	\chi_2&=\frac{3\pi}{8}\frac{5\gamma^2-1}{\Gamma},\\
	\chi_3&=\frac{64\gamma^6-120\gamma^4+60\gamma^2-5}{3(\gamma^2-1)^{3/2}}-\frac{4}{3}\frac{\nu}{\Gamma^2}\gamma\sqrt{\gamma^2-1}(14\gamma^2+25)
	\nnm\\
	&\quad-8\frac{\nu}{\Gamma^2}(4\gamma^4-12\gamma^2-3)\sinh^{-1}\sqrt{\frac{\gamma-1}{2}}.
	\end{align}
\end{subequations}
This implies that the effective Hamiltonian's coefficients at the same orders are
\begin{subequations}
	\begin{align}
	q_\mr{2PM}&=\frac{3}{2}(5\hat H_\mr S^2-1)\Bigg(1-\frac{1}{\sqrt{1+2\nu(\hat H_\mr S-1)}}\Bigg),\label{qtwo}
	\\
	q_\mr{3PM}&=-\frac{2\hat H_\mr S^2-1}{\hat H_\mr S^2-1}q_\mr{2PM}
	+\frac{4}{3}\nu\hat H_\mr S\frac{14\hat H_\mr S^2+25}{1+2\nu(\hat H_\mr S-1)}\nnm\\
	&\quad+\frac{8\nu}{\sqrt{\hat H_\mr S^2-1}}\frac{4\hat H_\mr S^4-12\hat H_\mr S^2-3}{1+2\nu(\hat H_\mr S-1)}\sinh^{-1}\sqrt{\frac{\hat H_\mr S-1}{2}}. \label{qthree}
	\end{align}
\end{subequations}
Henceforth, we denote $H^{\text{EOB,PS}}_{n\text{PM}}$ the PM EOB Hamiltonians in the PS gauge up to $n$PM order, obtained inserting the effective Hamiltonians at the desired order in the energy map. The amount of two-body information the $H^{\text{EOB,PS}}_{n\text{PM}}$ Hamiltonians encapsulate is equivalent to that of $H_{n\text{PM}}$, since both are constructed from the same gauge-invariant coefficients~\eqref{eq:chi2PM3PM}. In the comparisons, we employ both classes of Hamiltonians to check the importance of properly resumming the scattering angle's PM information. 
A reasonable benchmark for what consititutes a ``good'' resummation of the relativistic conservative dynamics are the well-studied EOB models in the DJS gauge~\cite{Damour:2000we,Damour:2015isa}, see Sec.~\ref{sec:EOBform}, namely the gauge choice used in EOB models for LIGO-Virgo analyses.
In the comparisons, we use the nonspinning EOB Hamiltonian with effective energy given by Eq.~\eqref{Ham_eff} and circular-orbit conservative-dynamics information encapsulated in the $A(u,\nu)$ potential,\footnote{For circular orbits $p_r=0$, and both $D(u,\nu)$ and $Q$ can be safely neglected.} which we expand up to 4PN order,
\begin{align}\label{eq:A4PN}
&A(u,\nu)= 1-2u + u^3 a_\text{2PN}(\nu) + u^4 a_\text{3PN}(\nu)+ u^5 a_\text{4PN}(\ln  u,\nu) + \mc O(G^6)\,,\nnm\\
&\, \nnm\\
&\qquad a_\text{2PN}(\nu)= 2\nu \,, \qquad a_\text{3PN}(\nu) = \frac{94 \nu }{3}-\frac{41}{32} \pi ^2\nu,\\
&a_\text{4PN}(\ln u,\nu)= -\frac{4237\nu}{60}  +\frac{128\nu }{5}\gamma_\text{E}+\frac{2275 \pi ^2 \nu
}{512}\nonumber\\
&\qquad \qquad \qquad \quad
+\frac{256}{5} \nu  \ln 2 +\nu ^2\left(\frac{41 \pi^2}{32}-\frac{221}{6}\right)+\frac{64}{5} \nu  \ln u.\nnm
\end{align}
Here, $\gamma_\text{E}$ is Euler's gamma function.
We denote with $H^{\text{EOB}}_{m\text{PN}}$ the EOB DJS gauge models up to $m$PN.

From a given model, we calculate the (mass-reduced) binding energy as $E_\text{mod} \equiv (H_\text{mod}(r,p_r,L)-M)/\mu$, and compare it against those extracted from NR simulations $E_\text{NR}$. As discussed, Hamiltonians depend on gauge choices. It is possible to remove all gauge information by reducing the expressions to analytical expansions for the binding energy in the circular orbit limit. However, this is an approach we do not follow here. Instead, we trace out parametric curves 
of binding energy as a function of a gauge-invariant quantity, such as the orbital frequency $\Omega$, treating the Hamiltonians as exact (and therefore retaining gauge information). The rationale for seeking such parametric relations is that they allow one to assess how good a Hamiltonian is in resumming information from the conservative dynamics of the system ~\cite{Damour:2011fu,LeTiec:2011dp,Nagar:2015xqa,Ossokine:2017dge}.
When GW emission is present, as in the case of NR simulations, these curves still depend most sensitively on the conservative dynamics~\cite{Ossokine:2017dge}.
In the following, we use the binding energy $E_\text{NR}(\Omega)$ from an NR simulation of a nonspinning binary in a circular orbit, as extracted in Ref.~\cite{Ossokine:2017dge}. The simulation (SXS ID: 0180~\cite{SXS}) is for a system with comparable masses $q=1$ and spans about $\sim 28$ orbital cycles in a region of strong fields.
The analogous circular-orbit binding energy $E_\text{mod}(\Omega)$ from Hamiltonians is retrieved setting $H_\text{mod}(r,p_r=0,L=L(\Omega))$ in the definition of $E_\text{mod}$. The relation $L=L(\Omega)$ is obtained inverting $\Omega \equiv \dot \phi = \partial H_\text{mod}(r,p_r=0,L)/\partial L$ in the Hamilton's equations~\eqref{Ham_eqs}. 
The inversion is performed numerically treating the Hamiltonian as exact.  Alternative analytical PM expressions for the binding energy are presented in Refs.~\cite{Kalin:2019rwq,Kalin:2019inp}.
For the models' binding energies, we choose to neglect radiation-reaction effects. While this means we should not expect exact agreement with the NR results in the region of strong fields past the innermost stable circular orbit (ISCO), any discrepancy between the models we obtain in this case is solely due to the resummation provided from the Hamiltonian.

\begin{figure*}
	\centering
	\includegraphics[width=.9\linewidth]{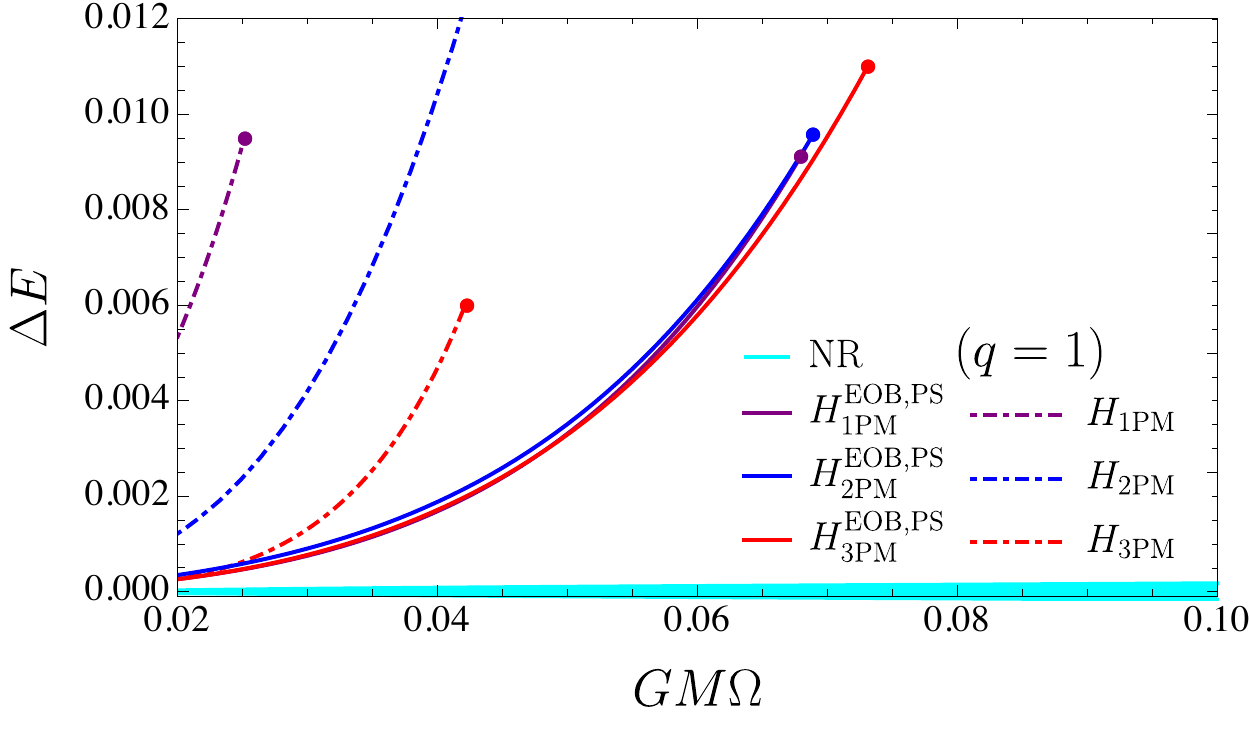}
	\caption{Binding-energy comparison between PM Hamiltonians in the isotropic and EOB PS gauge. The shaded area in cyan is an estimate of the NR error~\cite{Ossokine:2017dge}. The curves stop at the ISCO of the binary, whenever that is present in the orbital dynamics. The plot has been obtained with the code I developed to get the results in \chap~\ref{chap:two}. }
		\label{fig:energywPMnew}
\end{figure*}

\begin{figure}
	\centering
	\includegraphics[width=.9\linewidth]{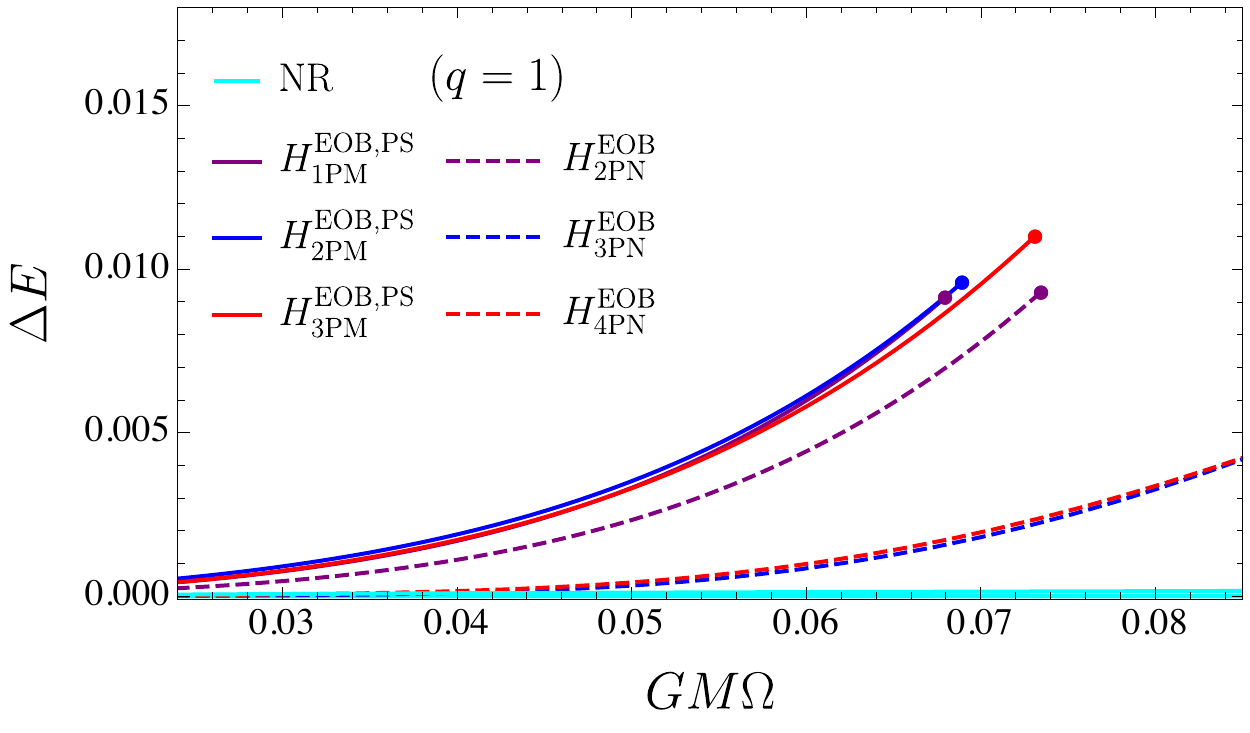}
	\caption{Same as Fig.~\ref{fig:energywPMnew}, but comparing PM Hamiltonians in the EOB PS gauge with PN models in  the EOB DJS gauge. }
	\label{fig:PMvsclassicnew}
\end{figure}

The first comparison between PM models, $H_\text{nPM}$ and $H^{\text{EOB,PS}}_\text{nPM}$ (for $n= 1,2,3$), is found in Fig.~\ref{fig:energywPMnew}, where we plot the difference $\Delta E \equiv |E_\text{mod} - E_\text{NR}|$ between the binding energy from the model and from NR.
An important point to notice is the improvement in weak fields ($GM\Omega\lesssim 0.04$) when more PM orders are included in the $H_\text{nPM}$ models. This first result points to the usefulness of the PM approximation in encoding information about binaries in a bound orbit.
Moreover, we see that the EOB resummation improves on those models, by extending the validity of the PM approximation to regions of stronger gravity. Here, however, it is noticed that the PS-gauge EOB Hamiltonian is not efficient at resumming two-body information, since increasing the PM order one only slightly improves the geodesic limit (which in this gauge is equivalent to the 1PM Hamiltonian~\cite{Damour:2016gwp,Damour:2017zjx}). In Fig.~\ref{fig:PMvsclassicnew} we compare the EOB models based on the PM approximation with $H^{\text{EOB}}_{n\text{PN}}$ (with $n=2,3,4$), namely those based on the PN approximation and used in LIGO-Virgo analyses. 
We notice that the latter still produce $E(\Omega)$-curves that are substantially closer to NR result than the PM approximants, especially for weak fields $GM\Omega\lesssim 0.05$. From this result, we gather that, while we saw in Fig.~\ref{fig:energywPMnew} that the pursuit of higher-order PM corrections is useful, the third-subleading PM order for nonspinning systems is not yet enough to give models that perform better than those currently employed.
This is in fact expected (though not guaranteed \emph{a priori}), since the 3PM dynamics only contains complete PN information up to the second order, while in fact the 3PN or 4PN orders are needed to reach accurate waveforms \cite{Bohe:2016gbl,Nagar:2018zoe}.

The above conclusions are based on a number of assumptions that have been made to retrieve the binding energies, and that could be lifted in future studies.  One could consider more eccentric or unbound orbits, which are closer to the natural domain of PM calculations. For the latter purpose, one could for instance compare scattering-angle calculations obtained either numerically~\cite{Damour:2014afa} or analytically from PN, PM and EOB approximations (work along these lines has recently been carried out~\cite{Nagar:2020xsk}). 

\subsection{Indirect approach: novel post-Newtonian terms from relativistic scattering and self-force theory}
\label{sec:3&4paper}

The analysis of the previous section leaves out the possibility of using new insights from scattering to improve models indirectly. Here we use the scattering angle's mass dependence and analytical \GSF results to obtain new terms in the PN expressions used to model spinning binaries, following a procedure first outlined in Ref.~\cite{Bini:2019nra} for nonspinning systems. In the process, we derive the \nNNLOSO and \Sonestwo PN dynamics, the former for generic and the latter for spin-aligned configurations. The derivations are very technical and best divided in smaller steps. The  goal  is to specify the conservative dynamics encapsulated in the scattering angle at the desired orders. We briefly state the methodology as a roadmap for the section ahead.

We must first obtain the scattering angle $\chi_t$ for a test spinning BH [Eq.~\eqref{eq:chi_t_N3LOa2}], which is needed to fix the spinning two-body dynamics at 2PM (NLO) and half of the dynamics through 4PM (N$^3$LO). From the test-body scattering we propose a parametrized form of the two-body angle  $\chi^{\text{N}^3\text{LO}}$ [Eq.~\eqref{chiN3LO_gen_tomatch}], which includes unknown coefficients at linear order in the mass ratio [Eq.~\eqref{eq:all_unknowns}]. 
We introduce parametrized Hamiltonians with unknown coefficients at the orders we want to specify [Eq.~\eqref{SOHam} for SO and~\eqref{S1S2Ham} for \Sonestwo couplings], and obtain predictions for the scattering angle from them [Eqs.~\eqref{eq:chipredSO} and~\eqref{eq:S1S2chipred}]. The predictions are matched to $\chi^{\text{N}^3\text{LO}}$ to specify some of the Hamiltonian's unknown coefficients, as well as to introduce the unkwnown coefficients from $\chi^{\text{N}^3\text{LO}}$ into the Hamiltonian.
From these Hamiltonians, one calculates predictions for the Detweiler redshift and spin-precession invariants [Eqs.~\eqref{UpredGSF_SO} and~\eqref{psipredGSF_SO} for the SO, and~\eqref{psipredGSF_s1s2} for the \Sonestwo dynamics] and matches them with results from MST techniques. If enough constraints are imposed, the unknowns from $\chi^{\text{N}^3\text{LO}}$ can be retrieved.
The method we use in the present section is alternative to the one in \chap~\ref{chap:five}, which does not rely on Hamiltonians, and has been used to check the results obtained both in the literature (for lower orders in the PN expansion) and with the method reported in  \chap~\ref{chap:five}. The results presented here agree with those in the published work.


\subsubsection{Scattering angle in the extended test-body limit}
\label{sec:chi_test4PM}

To obtain the scattering angle in the test-body limit with spins we use the extended test-body configuration in Fig.~\ref{fig:notation}. That is, we endow (aligned) spins $a_b$ and $a_t$ to the bodies described by background (``b'') and test (``t'') masses $m_b$ and $m_t$. We set the mass ratio $m_t/m_b \rightarrow 0$ and allow the spin $a_t$ to retain a finite extension.
Physically, the choice of a vanishing mass ratio implies that the secondary body has negligible effect on the background gravitational field, while retaining a finite size. The ``test BH'' (which would best be identified with a naked singularity~\cite{Vines:2018gqi,Siemonsen:2019dsu}) moves in the Kerr background described by ($m_b$,$a_b$) according to the Mathisson-Papapetrou-Dixon (MPD) equations~\cite{Mathisson:1937zz,Papapetrou:1951pa,Dixon:1970zza,Dixon:1979}. Reference~\cite{Bini:2017pee} solves the set of MPD equations (with the Tulczyjew spin supplementary condition~\cite{Tulczyjew:1959}) to obtain the test-spin version of the scattering angle's integrand~\eqref{eq:phidotrdot_schw}, which is their Eq.~(66). We expand this to 4PM order and linear order in each spin\footnote{
For ease of presentation, we do not report the rather cumbersome integrand. In our notation, however, it can be easily obtained noticing that $\hat s$ and $\hat a$ in Ref.~\cite{Bini:2017pee} are, respectively, $a_t$ and $a_b$, while their $u$ is the inverse of the radial separation in Kerr, $u = 1/\rtest$. One must then simply take the square root of (the inverse of) their Eq.~(66), expand $u\sim G$ to fourth order, and the whole expression to linear order in $a_t$ and $a_b$. The result can be expressed in terms of $(b,v)$ using the fact that their $\hat E= \gamma$ and $\hat L = \ell$ are linked to $(b,v)$ via Eqs.~\eqref{gammaGammadefs} and~\eqref{ellofb}.
}
and solve for the scattering with the method of finite parts reported in Sec.~\ref{sec:PM}, namely integrating each PM order and discarding the resultant infinities. The test-body scattering angle through these orders reads,
\begin{align}\label{eq:chi_t_N3LOa2}
\chi^t(b,v)= \chi^{t}_{a^0}(b,v) &+ a_t\, \chi^{t}_{a_t}(b,v) + a_b\, \chi^{t}_{a_b}(b,v) \nonumber\\
&+ a_t a_b\, \chi^{t}_{a_ta_b}(b,v)+\mc O(G^5,a_t^2,a_b^2)\,,
\end{align}
where $\chi^{t}_{a^0}(b,v)$ matches the nonspinning scattering angle through 4PM order given by Eq.~\eqref{eq:chiSchw} with coefficients~\eqref{eq:chi_t_from2PM} (and $m\rightarrow m_b$). The primary and secondary spin's angle  $\chi^{t}_{a_t}(b,v)$ are expanded as in Eq.~\eqref{eq:chiSchw}, with
\begin{align}\label{eq:chitab}
\chi^{t,\text{1PM}}_{a_b}(b,v)&=-\frac{4 Gm_b}{b^2 v}\,,\\
\chi^{t,\text{2PM}}_{a_b}(b,v)&=-\frac{2 \pi(Gm_b)^2}{b^3 v^3}(2+3 v^2)\,,\nnm\\
\chi^{t,\text{3PM}}_{a_b}(b,v)&=-\frac{12 (Gm_b)^3}{b^4 v^5}(1+10 v^2+5 v^4)\,,\nnm\\
\chi^{t,\text{4PM}}_{a_b}(b,v)&=-\frac{21 \pi}{2}\frac{  (Gm_b)^4}{b^5 v^5}(8+20 v^2+5 v^4)\,,\nnm
\end{align}
and
\begin{align}\label{eq:chitat}
\chi^{t,\text{1PM}}_{a_t}(b,v)&=-\frac{4 Gm_b}{b^2 v}\,,\\
\chi^{t,\text{2PM}}_{a_t}(b,v)&=-\frac{3 \pi}{2}\frac{(Gm_b)^2}{b^3 v^3}(2+3 v^2)\,,\nnm\\
\chi^{t,\text{3PM}}_{a_t}(b,v)&=-\frac{8 (Gm_b)^3}{b^4 v^5}(1+10 v^2+5 v^4)\,,\nnm\\
\chi^{t,\text{4PM}}_{a_t}(b,v)&=-\frac{105 \pi}{16}\frac{  (Gm_b)^4}{b^5 v^5}(8+20 v^2+5 v^4)\,.\nnm
\end{align}
Similarly, the biliear-in-spin portion of the test-spin scattering angle reads
\begin{align}\label{eq:chitabt}
\chi^{t,\text{1PM}}_{a_ta_b}(b,v)&=\frac{4 Gm_b (1+v^2)}{b^3 v^2}\,,\\
\chi^{t,\text{2PM}}_{a_ta_b}(b,v)&=\frac{3 \pi}{2}\frac{  (Gm_b)^2}{b^4 v^4}(2+15 v^2+3 v^4)\,,\nnm\\
\chi^{t,\text{3PM}}_{a_ta_b}(b,v)&=\frac{8 (Gm_b)^3}{b^5 v^6}(1+35v^2+55v^4+5 v^6)\,,\nnm\\
\chi^{t,\text{4PM}}_{a_ta_b}(b,v)&=\frac{105 \pi }{16}\frac{ (Gm_b)^4}{b^6v^6}(24+140 v^2+ 95 v^4+ 5 v^6)\,.\nnm
\end{align}

From the test-body scattering angle we can specify the dynamics through NLO, as well as the geodesic portion through N$^3$LO in a PM expansion. 
In principle, with knowledge of linear-in-$\nu$ (analytical) \GSF expressions with spins through 4PM order, we would be able to complete the specification of the PM two-body dynamics at N$^3$LO. However, such results are not yet available. Our approach is then to determine the scattering angle in a PN expansion through the same order, since it still contains unknown PN terms and because in this case  results in a double SMR-PN expansion for spinning binaries are available. 
We write the PN-expanded two-body scattering angle through N$^3$LO\footnote{In the remainder of this chapter, the "order'' in N$^n$LO is understood to refer to orders in a PN expansion.} as follows
\begin{equation}\label{chiN3LO_gen_tomatch}
\chi^{\text{N}^3\text{LO}}(b,v) =\chi_{a^0}(b,v) +  \chi_{a^1}(b,v)+\chi_{a_1a_2}(b,v)+\mc O(G^5,a_1^2, a_2^2)\,.
\end{equation}
The nonspinning part $\chi_{a^0}$ can be found in Eq.~(4.32a) of Ref.~\cite{Vines:2018gqi}, and it reads
\begin{align}\label{eq:chi_a0}
\frac{\chi_{a^0}}{\Gamma}=&
\frac{GM}{v^2b} \left[2+2v^2+\mc O(v^8)\right]+\pi\frac{(GM)^2}{b^2
	v^4}\left[3 v^2+\frac{3}{4}v^4+\mc O(v^8)\right]\\
&+\frac{(GM)^3}{b^3 v^6}\bigg[-\frac{2}{3}+2 \frac{15-\nu}{3} v^2
+ \frac{60-13\nu}{2} v^4 + \frac{40-227\nu}{12} v^6+\mc O(v^8)\bigg]
\nonumber\\
& +\frac{\pi(GM)^4}{b^4 v^8}
\bigg[15\frac{7-2\nu}{4} v^4 + \left(\frac{105}{4}-\frac{437}{8}\nu+\frac{123}{128}\pi^2\nu\right) v^6+\mc O(v^8)\bigg]\nonumber\,.
\end{align}
The geodesic part is easily obtained expanding Eqs.~\eqref{eq:chiSchw} and~\eqref{eq:chi_t_from2PM} in the velocity $v$, with the identification of the masses $m_b \rightarrow M$ (which is consistent with the fact that in the secondary mass in the test spin limit has a negligible effect on the total gravitational field). The part linear in the mass ratio is obtained from two-body results (e.g., from the known Hamiltonian through 3PN order in Ref.~\cite{Vines:2018gqi}).
Notice the presence of a $\Gamma$ in the denominator of the left-hand side of the expression, which comes from the momentum dependence in the definition of the scattering angle~\eqref{eq:whereGammafrom}. Since $\Gamma\rightarrow1$ in the test-body limit, the coefficients of $\chi_t$ are the same as those in $\chi_{a^0}$.

In the following sections, we are interested in specifying $\chi_{a^1}(b,v)$ and $\chi_{a_1a_2}(b,v)$ through N$^3$LO. In fact, the expressions are known in a PN expansion up to next-to-next-to leading order from Eqs.~(4.32) of Ref.~\cite{Vines:2018gqi}. However, we pretend not to know them in order to present an independent derivation. We can write the SO contribution as
\begin{align}\label{eq:chi_tomatch}
\frac{\chi_{a^1}}{\Gamma}=
\frac{\delta a_-}{b}v\bigg\{-&\frac{\pi  (GM)^2}{b^2
	v^4}\left[\frac{1}{2}+\frac{3}{4} v^2+\mathcal{O}(v^8)\right]\\
&-\frac{(GM)^3}{b^3 v^6}\bigg[2+20 v^2+10v^4-\nu \chi^{\nu^1}_{3-}(v) +\mathcal{O}(v^8)\bigg]
\nonumber\\
& +\frac{\pi(GM)^4}{b^4 v^8}\bigg[-\frac{63}{4} v^2-\frac{315}{8} v^4-\frac{315}{32} v^6+\nu \chi^{\nu^1}_{4-}(v)+\mathcal{O}(v^8)\bigg]\bigg\}
\nonumber\\
+\frac{a_+}{b}v\bigg\{&-\frac{4 GM}{b
	v^2}\left[1+\mathcal{O}(v^8)\right]-\frac{\pi  (GM)^2}{b^2
	v^4}\left[\frac{7}{2}+\frac{21}{4} v^2+\mathcal{O}(v^8)\right]\nonumber\\
&+\frac{(GM)^3}{b^3 v^6}\bigg[-10-100 v^2-50v^4+\nu\chi^{\nu^1}_{3+}(v)+\mathcal{O}(v^8)\bigg]\nonumber\\
+\pi&\frac{(GM)^4}{b^4 v^8}\bigg[-\frac{273}{4} v^2-\frac{1365}{8} v^4-\frac{1365}{32} v^6+\nu \chi^{\nu^1}_{4+}(v)+\mathcal{O}(v^8)\bigg] \bigg\}\nnm\,,
\end{align}
where we have defined
\begin{equation}\label{def:apam}
a_+ \equiv a_1 + a_2, \qquad a_- \equiv a_2 - a_1.
\end{equation}
The geodesic part of Eq.~\eqref{eq:chi_tomatch} in this parmetrization is fixed by the PN expansion of the functions~\eqref{eq:chitab} and~\eqref{eq:chitat}, and requiring that $m_b \rightarrow M$, $a_b \rightarrow a_2$ and $a_t \rightarrow a_1$.
At 3PM-N$^2$LO, it matches (the geodesic part of) Eq.~(4.32) in Ref.~\cite{Vines:2018gqi}. The 4PM and (vanishing) 3PM-$\mc O(v^6)$ terms are specified here for the first time. Notice the appearance of $\delta = (m_2-m_1)/M$ in front of the $a_-$ parametrization, as required to match the two-body results from the literature~\cite{Vines:2018gqi}. As $\nu \rightarrow 0$, we also have that $\delta\rightarrow1$.
The functions $\chi^{\nu^1}_{3\pm}(v)$ and $\chi^{\nu^1}_{4\pm}(v)$ are what is left to specify in the \nNNLOSO dynamics.

Similarly to the SO contributions, the \Sonestwo can be written as
\begin{align}\label{S1S2chi}
\chi_{a_1 a_2}=\frac{a_1 a_2}{b^2} \bigg\{&\frac{G M }{b
	v^2}\left[4+4 v^2+\mathcal{O}(v^8)\right]+\\
&\frac{\pi \left(G M \right)^2}{b^2 v^4}\left[3+\frac{45 }{2}v^2+\frac{9
}{2}v^4+\mathcal{O}(v^8)\right]\nnm\\
&\frac{(G M)^3}{b^3
	v^6} \bigg[8+280 v^2+440 v^4+40 v^6+\nu  \chi^{\nu^1}_{3\times}(v)+\mathcal{O}(v^8)\bigg]+\nonumber\\
&\frac{\pi\left(G M \right)^4}{b^4 v^8} \bigg[\frac{315 }{2}v^2+\frac{3675
}{4}v^4+\frac{9975 }{16}v^6+\nu  \chi^{\nu^1}_{4\times}(v)+\mathcal{O}(v^8)\bigg]\bigg\}\nonumber,
\end{align}
with geodesic terms obtained PN expanding the functions~\eqref{eq:chitabt}, and using the same mass and spin identifications between two-body and test spin configurations. 
The 4PM and 3PM-$\mc O(v^6)$ terms are also specified here for the first time. The rest match Eq.(4.32c) in Ref.~\cite{Vines:2018gqi} after taking into account the different spin parametrizations, linked via Eq.~\eqref{def:apam}.
The functions we want to specify to complete the knowledge of the dynamics at N$^3$LO with bilinear terms in the spins are $\chi^{\nu^1}_{3\times}(v)$ and $\chi^{\nu^1}_{4\times}(v)$.
Both these and the SO functions must be PN expanded, consistently with the way the geodesic part has been. We can therefore expand them to $\mc O(v^6)$ order, and collect the 24 unknown coefficients as follows
\begin{equation}\label{eq:all_unknowns}
\left(\begin{array}{c}
\chi^{\nu^1}_{3\pm}(v)\\
\chi^{\nu^1}_{4\pm}(v)\\
\chi^{\nu^1}_{3\times}(v)\\
\chi^{\nu^1}_{4\times}(v)
\end{array}\right)
=
\left(\begin{array}{cccc}
\chi^{\nu^1}_{31\pm} & \chi^{\nu^1}_{33\pm} & \chi^{\nu^1}_{35\pm} & \chi^{\nu^1}_{37\pm}	\\
\chi^{\nu^1}_{41\pm} & \chi^{\nu^1}_{43\pm} & \chi^{\nu^1}_{45\pm} & \chi^{\nu^1}_{47\pm}	\\
\chi^{\nu^1}_{30\times} & \chi^{\nu^1}_{32\times} & \chi^{\nu^1}_{34\times} & \chi^{\nu^1}_{36\times}	\\
\chi^{\nu^1}_{40\times} & \chi^{\nu^1}_{42\times} & \chi^{\nu^1}_{44\times} & \chi^{\nu^1}_{46\times}
\end{array}\right)
\left(\begin{array}{c}
1\\
v^2\\
v^4\\
v^6
\end{array}\right)\,.
\end{equation}

\subsubsection{Derivation of the third-subleading spin-orbit PN dynamics}\label{sec:3&4paper_SO}

In this section, we aim at specifying the 16 coefficients gathered in $\chi^{\nu^1}_{3\pm}(v)$ and $\chi^{\nu^1}_{4\pm}(v)$, thereby deriving the \nNNLOSO PN dynamics. We associate an isotropic-gauge Hamiltonian to such dynamics, 
\begin{equation}\label{HN3LOSO}
H^{\text{N}^3\text{LO}}_\text{SO}(r,p^2)\equiv H_{a^0}(r,p^2)+H_{a^1}(r,p^2)+\mc O(|a|^2)\,.
\end{equation}
The nonspinning part is taken to be complete through \emph{fourth}-subleading order for reasons that will become clear below. The 4PN isotropic-gauge nonspinning Hamiltonian can be found through a canonical transformation from the DJS-gauge one~\cite{Damour:2015isa}, see Appendix~\ref{app:DJStoISO@4PN}. A unique Hamiltonian can be specified by making a choice that fixes the 0PM contributions. We use the ``real gauge'', defined by requiring that these contributions match the PN expansion of 
\begin{equation}\label{eq:realgauge}
H_{a^0}^\text{0PM} = \sqrt{m_1^2+p^2} +  \sqrt{m_2^2+p^2} \,.
\end{equation} 
The (aligned-)spin orbit part $H_{a^1}(p^2)$ is complete through third-subleading order
\begin{alignat}{2} \label{SOHam}
&H_{a^1}=\frac{L}{c r^2}\bigg(\frac{7}{4}a_+ +\frac{\delta}{4}a_-\bigg)&\\
&+\frac{L}{c^3 r^2}(a_+\text{ }\delta a_-)\bigg[\begin{pmatrix}\alpha_{13+}\\\alpha_{13-}\end{pmatrix}\frac{p^2}{\mu^2}\frac{GM}{r}+\begin{pmatrix}\alpha_{23+}\\\alpha_{23-}\end{pmatrix}\frac{(GM)^2}{r^2}\bigg]&\nonumber\\
&+\frac{L}{c^5 r^2}(a_+\text{ }\delta a_-)\bigg[\begin{pmatrix}\alpha_{15+}\\\alpha_{15-}\end{pmatrix}\frac{p^4}{\mu^4}\frac{GM}{r}+\begin{pmatrix}\alpha_{25+}\\\alpha_{25-}\end{pmatrix}\frac{p^2}{\mu^2}\frac{(GM)^2}{r^2}+\begin{pmatrix}\alpha_{35+}\\\alpha_{35-}\end{pmatrix}\frac{(GM)^3}{r^3}\bigg]&\nonumber\\
&+\frac{L}{c^7 r^2}(a_+\text{ }\delta a_-)\bigg[\begin{pmatrix}\alpha_{17+}\\\alpha_{17-}\end{pmatrix}\frac{p^6}{\mu^6}\frac{GM}{r}+\begin{pmatrix}\alpha_{27+}\\\alpha_{27-}\end{pmatrix}\frac{p^4}{\mu^4}\frac{(GM)^2}{r^2}+\begin{pmatrix}\alpha_{37+}\\\alpha_{37-}\end{pmatrix}\frac{p^2}{\mu^2}\frac{(GM)^3}{r^3} &\nonumber\\
&\qquad\qquad\qquad\qquad\qquad\qquad\qquad\qquad\qquad\qquad
\qquad\qquad
+\begin{pmatrix}\alpha_{47+}\\\alpha_{47-}\end{pmatrix}\frac{(GM)^4}{r^4}\bigg]\,. \nonumber
\end{alignat}
Notice that the coefficients $\alpha_{ij\pm}$ are such that $i$ stands for the power of $G$ and $j$ for the power of $1/c$ (reintroduced as parameter in the Hamiltonian to keep track of the PN order), while $\pm$ refers to whether they multiply $a_+$ or $a_-$ in the vector multiplication with $(a_+\text{ }\delta a_-)$.
In principle, the coefficients are known through N$^2$LO SO order~\cite{Vines:2018gqi}. As in the case of the scattering angle, we keep them unknown to check if the procedure indipendently reproduces results in the literature.\footnote{The expression in Ref.~\cite{Vines:2018gqi} is specified using an EOB gauge for the 0PM contributions, see their Eq.~(4.21), so that additional calculations are in order to compare results with the literature.}

From $H^{\text{N}^3\text{LO}}_\text{SO}(r,p^2,\{\alpha_{ij\pm}\})$, we calculate a prediction for the scattering angle by inverting the expression to get $p_r(H^{\text{N}^3\text{LO}}_\text{SO}\equiv \Gamma M,L,\{\alpha_{ij\pm}\})$, taking the derivative with respect to $L$, and solving order by order in the PM expansion with the method of finite parts. To make contact with Eq.~\eqref{eq:chi_tomatch}, the expression is further expanded to third-subleading PN order. We express the results in terms of $(b,v)$ instead of $\Gamma$ and $L$ using, respectively, Eq.~\eqref{gammaGammadefs} and~\cite{Vines:2018gqi}
\begin{equation}
L= \frac{b\, v\, \gamma}{\Gamma}+\frac{\Gamma-1}{2\nu} \left(a_+ + \frac{\delta a_-}{\Gamma}\right)\,.
\end{equation}
The prediction reads 
\begin{equation}
\chi^\text{(pred)}_{\text{N}^3\text{LO}}(b,v) = \chi_{a^0}(b,v)+\chi_{a^1}^\text{(pred)}(b,v)+\mc O(G^5, a_1^2, a_1a_2, a_2^2),
\end{equation}
with $\chi_{a^0}(b,v)$ given by Eq. \eqref{eq:chi_a0}, and SO correction
\begin{align}\label{eq:chipredSO}
\frac{\chi_{a^1}^\text{(pred)}}{\Gamma} = 
\frac{\delta a_-}{b}v\bigg[&\frac{GM}{bv^2}\chi^\text{(pred)}_{1-}(v)+\frac{\pi  (GM)^2}{b^2
	v^4}\chi^\text{(pred)}_{2-}(v)\\
&+\frac{(GM)^3}{b^3 v^6}\chi^\text{(pred)}_{3-}(v)+\frac{\pi(GM)^4}{b^4 v^8}\chi^\text{(pred)}_{4-}(v)+\mc O(v^8)\bigg]
\nonumber\\
+\frac{a_+}{b}v\bigg[&\frac{GM}{bv^2}\chi^\text{(pred)}_{1+}(v)+\frac{\pi  (GM)^2}{b^2
	v^4}\chi^\text{(pred)}_{2+}(v)\nonumber\\
&+\frac{(GM)^3}{b^3 v^6}\chi^\text{(pred)}_{3+}(v)+\frac{\pi(GM)^4}{b^4 v^8}\chi^\text{(pred)}_{4+}(v)+\mc O(v^8)\bigg]\,.\nnm
\end{align}
The functions are too long to be reported here, but are included in Appendix~\ref{app:chipred} for completeness. Matching the functions in Eq.~\eqref{eq:chipredSO} with the ansatz~\eqref{eq:chi_tomatch} [and unknowns~\eqref{eq:all_unknowns}] order-by-order in the velocity $v$ and mass ratio $\nu$, we obtain constraints for the Hamiltonian and scattering angle coefficients. Solving them consistently gives the following 1PM
\begin{align}\label{1PMalphacoeff}
&\alpha_{13-}=\frac{5}{16}(1+\nu )\,, \, \, \, \, \, \alpha_{13+}=\frac{1}{16} (-5+41 \nu )\,,\nonumber\\
&\alpha_{15-}=\frac{1}{32} \left(-7+18 \nu +11 \nu ^2\right)\,, \quad \alpha_{15+}=\frac{1}{32} \left(7-62 \nu +101 \nu ^2\right)\,,\nonumber\\
&\alpha_{17-}=\frac{1}{256} \left(45-221 \nu +174 \nu ^2+93 \nu ^3\right)\,,\nonumber\\
&\alpha_{17+}=\frac{1}{256} \left(-45+479 \nu -1366 \nu ^2+933 \nu ^3\right)\,,\nonumber\\
\end{align}
and 2PM coefficients
\begin{align}\label{2PMalphacoeff}
&\alpha_{23-}=-\frac{1}{2} (1+\nu )\,, \, \, \, \, \, \alpha_{23+}=-\frac{11}{2}-5 \nu\,,\nonumber\\
&\alpha_{25-}=-\frac{3}{16} \left(9+23 \nu +4 \nu ^2\right)\,, \quad\alpha_{25+}=-\frac{3}{16} \left(-9+121 \nu +46 \nu ^2\right)\,,\nonumber\\
&\alpha_{27-}=\frac{1}{16} \left(26-79 \nu -146 \nu ^2-15 \nu ^3\right)\,, \nonumber\\
&\alpha_{27+}=\frac{1}{16} \left(-26+299 \nu -680 \nu ^2-192 \nu ^3\right)\,.\nonumber\\
\end{align}
The Hamiltonian is therefore fully specified at 2PM-N$^3$LO order, which is expected since the geodesic motion included in Eq.~\eqref{eq:chi_tomatch}, is enough to specify the two-body dynamics.
At 3PM order, the constraints allow us to determine some of the coefficients in~\eqref{eq:all_unknowns},
\begin{align}
& \chi^{\nu^1}_{31-}=0\,,\, \, \chi^{\nu^1}_{33-}=0\,,\, \,\chi^{\nu^1}_{31+}=0\,,\, \, \chi^{\nu^1}_{33+}=10\,,\label{3PMv2chi}\\
& \chi^{\nu^1}_{41-}=0\,,\, \, \chi^{\nu^1}_{43-}=0\,,\, \,\chi^{\nu^1}_{41+}=\frac{3}{4}\,,\, \, \chi^{\nu^1}_{43+}=\frac{39}{4}\,.
\end{align}
The first line matches results for the scattering angle through next-to-next to leading PN order in Ref.~\cite{Vines:2018gqi}, which is an important check for the calculation; the second line is composed of new N$^3$LO terms. One also gets additional constraints between 3PM and 4PM scattering-angle coefficients,
\begin{equation}\label{chicon}
\chi^{\nu^1}_{45-}=\frac{3}{8} (4 \chi^{\nu^1}_{35-}+15)\,,\, \, \, \chi^{\nu^1}_{45+}=\frac{3}{8} (4 \chi^{\nu^1}_{35+}+105)\,,
\end{equation}
as well as between coefficients of the Hamiltonian and of the parametrized scattering angle. At 3PM order, these are
\begin{align}\label{3PMalphacoeff}
\alpha_{35-}=&\frac{9}{16}+\frac{\nu}{16}  (39-4 \chi^{\nu^1}_{35-})+\frac{5 \nu ^2}{16}\,,\\
\alpha_{35+}=&\frac{159}{16}+\frac{\nu}{16} (565-4 \chi^{\nu^1}_{35+})+\frac{95 \nu ^2}{16}\,,\nonumber\\
\alpha_{37-}=&\frac{315}{64}+\nu  \left(\frac{1331}{64}-\frac{\chi^{\nu^1}_{37-}}{4}\right)+\nu
^2 \left(\frac{539}{32}-\frac{3 \chi^{\nu^1}_{35-}}{8}\right)+\frac{7 \nu
	^3}{16}\,,\nonumber\\
\alpha_{37+}=&-\frac{315}{64}+\nu  \left(\frac{6775}{64}-\frac{\chi^{\nu^1}_{37+}}{4}\right)+\nu
^2 \left(\frac{4589}{32}-\frac{3 \chi^{\nu^1}_{35+}}{8}\right)+\frac{175 \nu
	^3}{16}\,,\nonumber
\end{align}
while at 4PM they are
\begin{align}\label{4PMalphacoeff}
\alpha_{47-}=&-\frac{7}{16}+\nu ^2 \left(-\frac{39}{8}+\chi^{\nu^1}_{35-}\right)+\frac{\nu
	^3}{8}+\nonumber\\
&\nu  \left(-\frac{185}{48}+\frac{41 \pi ^2}{64}+2 \chi^{\nu^1}_{35-}+\frac{3 \chi^{\nu^1}_{37-}}{2}-\frac{2 \chi^{\nu^1}_{47-}}{3}\right)\,,\nonumber\\
\alpha_{47+}=&-\frac{217}{16}+\nu ^2 \left(-\frac{189}{2}+\chi^{\nu^1}_{35+}\right)-\frac{5 \nu
	^3}{2}+\nonumber\\
&\nu  \left(-\frac{2599}{48}-\frac{41 \pi ^2}{64}+2 \chi^{\nu^1}_{35+}+\frac{3\chi^{\nu^1}_{37+}}{2}-\frac{2 \chi^{\nu^1}_{47+}}{3}\right)\,.
\end{align}

The Hamiltonian now depends on the remaining six unconstrained coefficients, $H^{\text{N}^3\text{LO}}_\text{SO}=H^{\text{N}^3\text{LO}}_\text{SO}(r,p_r, L; \{\chi^{\nu^1}_{35\pm},\chi^{\nu^1}_{37\pm},\chi^{\nu^1}_{47\pm}\})$, which can be fixed employing \GSF results.
One way is to directly integrate the generalized redshift and spin precession frequency~\eqref{1lawnew}. The procedure (see e.g., Refs.~\cite{Akcay:2015pza,Bini:2019lcd,Bini:2019lkm}) involves parametrizing the Hamiltonian with semi-latus rectum $\bar p$, eccentricity $\bar e$, and Darwin's relativistic anomaly $\chi$~\cite{Darwin1961}. From the definition of the latter, we have that
\begin{equation}
r = \frac{\bar p\, M}{\bar e(1+\cos \chi)}\,,
\end{equation}
while $p_r(\bar p,\bar e)$ and $L(\bar p,\bar e)$ in the Hamiltonian are obtained inverting $p_r(H^{\text{N}^3\text{LO}}_\text{SO}\equiv E,L,r)$, solving $p_r=0$ at periastron ($\chi =0$) and apastron ($\chi = \pi$) for $L(\bar p,\bar e)$ and $E(\bar p,\bar e)$, and inserting the results back into $p_r$. The initial inversion is performed retaining terms up to $\mc O(\bar p^{-9/2})$, namely N$^3$LO, and $\mc O(\bar e^6)$. 
Predictions for generalized redshift and spin-precession invariant (say for ``body 1'') can be calculated in terms of $(\bar p,\bar e)$ solving four integrals, see Sec.~\ref{sec:interplay}. These are related to the Hamiltonian as follows
\begin{align}\label{eq:integrals}
T_r&= \oint dt = \oint \left(\frac{\partial H^{\text{N}^3\text{LO}}_\text{SO}}{\partial p_r}\right)^{-1} dr = 2 \int_{0}^\pi \left(\frac{\partial H^{\text{N}^3\text{LO}}_\text{SO}}{\partial p_r}\right)^{-1}\frac{dr}{d\chi}d\chi\,,\\
\Phi&= \oint d\phi 
= \oint \left(\frac{\partial H^{\text{N}^3\text{LO}}_\text{SO}}{\partial L}\right)dt = 2 \int_{0}^\pi \left(\frac{\partial H^{\text{N}^3\text{LO}}_\text{SO}}{\partial L}\right)\left(\frac{\partial H^{\text{N}^3\text{LO}}_\text{SO}}{\partial p_r}\right)^{-1}\frac{dr}{d\chi}d\chi\,,\nnm\\
z_1^{\text{(pred)}}&=\frac{1}{T_r}\oint \left(\frac{\partial H^{\text{N}^3\text{LO}}_\text{SO}}{\partial m_1}\right) dt = \frac{2}{T_r} \int_{0}^\pi \left(\frac{\partial H^{\text{N}^3\text{LO}}_\text{SO}}{\partial m_1}\right)\left(\frac{\partial H^{\text{N}^3\text{LO}}_\text{SO}}{\partial p_r}\right)^{-1}\frac{dr}{d\chi}d\chi\,,\nnm\\
\psi^{\text{(pred)}}_1&=\frac{1}{\Phi}\oint \left(\frac{\partial H^{\text{N}^3\text{LO}}_\text{SO}}{\partial S_1}\right)dt=
\frac{2}{\Phi} \int_{0}^\pi \left(\frac{\partial H^{\text{N}^3\text{LO}}_\text{SO}}{\partial S_1}\right)\left(\frac{\partial H^{\text{N}^3\text{LO}}_\text{SO}}{\partial p_r}\right)^{-1}\frac{dr}{d\chi}d\chi\,.\nnm
\end{align}
Our goal is to match the predictions $z^{\text{(pred)}}_1(\bar p,\bar e)$ and $\psi^{\text{(pred)}}_1(\bar p,\bar e)$ to results obtained from MST techniques. However the set of coordinates $(\bar p,\bar e)$ is gauge dependent. For this reason, we introduce gauge-independent variables $(x,\iota)$ that can be used between PN and \GSF schemes~\cite{Akcay:2015pza,Bini:2019lcd}
\footnote{
	The denominator of $\iota$ scales as 1PN. Because of this, the overall PN ordering of $\psi_1$ is scaled down in such a way that manifestly nonlocal-in-time (4PN nonspinning) terms appear in the N$^3$LO corrections in the literature~\cite{Kavanagh:2017wot,Bini:2019lkm}. To reproduce these terms, we have included the 4PN nonspinning tail terms in the Hamiltonian.
}
\begin{equation}
x \equiv (GM\Omega)^{2/3} = \left(\frac{\Phi}{T_r}\right)^{2/3},\quad
\iota\equiv \frac{3x}{\Phi/2\pi-1}.
\end{equation}
The expressions $z_1(x,\iota)$ and $\psi_1 (x,\iota)$, which are valid for arbitrary masses at this stage, are expanded to N$^3$LO [$\mc O(x^{11/2})$ and $\mc O(x^4)$ respectively].
When including known terms up to N$^2$LO, they agree with Eq.~(50) of Ref.~\cite{Bini:2019lcd} and Eq.~(83) of Ref.~\cite{Bini:2019lkm} up to that order. The full expressions up to N$^3$LO are lengthy, but provided as a \texttt{Mathematica} file in the Supplemental Materials of Ref.~\cite{Antonelli:2020ybz}.
Through the auxiliary gauge-independent variables $y$ and $\lambda$, which isolate the mass ratio dependence of $x$ and $\iota$ as
\begin{align}\label{ylambdanew}
y&= (Gm_2 \Omega_\phi)^{2/3} = \frac{x}{(1 + q)^{2/3}} \,,\\
\lambda &=\frac{3y}{\Phi/(2\pi)-1} = \frac{\iota}{(1 + q)^{2/3}} \,,
\end{align}
we expand $U_1\equiv z_1^{-1}$ and $\psi_1$ to first order in the mass ratio $q$, first order in the massive body's  spin $ a_2 = m_2\hat a$, and zeroth order in the spin of the smaller companion $a_1$,
\begin{align}
&U_1=  U^\text{(0)}_{1a^0} +\hat{a}\, U^\text{(0)}_{1a}+q\left(\delta U^\text{GSF}_{1a^0} +\hat{a}\, \delta U^\text{GSF}_{1a}\right)+\mathcal{O}(q^2,\hat{a}^2)\,,\\
&\nonumber\\
&\psi_1=  \psi^\text{(0)}_{1a^0} +\hat{a}\, \psi^\text{(0)}_{1a}+q\left(\delta \psi^\text{GSF}_{1a^0} +\hat{a}\, \delta \psi^\text{GSF}_{1a}\right)+\mathcal{O}(q^2,\hat{a}^2)\label{eq:psia}\,.
\end{align}
To compare the 1SF corrections $\delta U^{\text{GSF}}_{1\cdots}$ and $\delta \psi^{\text{GSF}}_{1\cdots}$ with those derived in the literature, we link the $(y,\lambda)$ variables to another set $(p,e)$ corresponding to semi-latus rectum and eccentricity associated to the frequencies of an unperturbed Kerr spacetime (see Appendix~\ref{Kerrvar} for details).
The terms needed to solve for the N$^3$LO SO unknowns are $\delta U^\text{GSF}_{1a}$ and  $\delta\psi_{1\, a^0}^\text{GSF}$, for which we obtain
\begin{align}\label{UpredGSF_SO}
\delta U^\text{GSF}_{1a}&=\left(3-\frac{7 }{2}e^2-\frac{1}{8}e^4\right)
p^{-5/2}+\left(18-4 e^2-\frac{117}{4}e^4\right)
p^{-7/2}\nnm\\
+\bigg[&\frac{271}{4}+\frac{\chi^{\nu^1}_{35+}+\chi^{\nu^1}_{35-}}{2}+\frac{287 e^2}{2}+e^4
\left(-\frac{11399}{32}-\frac{15 (\chi^{\nu^1}_{35+}+\chi^{\nu^1}_{35-})}{16}\right)\bigg]p^{-9/2}\nnm\\
+\bigg\{&181+\frac{3}{4}(\chi^{\nu^1}_{35+}+\chi^{\nu^1}_{35-})-\frac{5 }{2}(\chi^{\nu^1}_{37+}+\chi^{\nu^1}_{37-})+\frac{4
}{3}(\chi^{\nu^1}_{47+}+\chi^{\nu^1}_{47-})\nnm\\
+e^2& \bigg[\frac{37445}{24}-\frac{41
	\pi ^2}{8}+\frac{1}{4}(\chi^{\nu^1}_{37+}+\chi^{\nu^1}_{37-})-\frac{5}{2}(\chi^{\nu^1}_{37+}+\chi^{\nu^1}_{37-})+2 (\chi^{\nu^1}_{47+}+\chi^{\nu^1}_{47-})\bigg]\nnm\\
+e^4 &\bigg[\frac{1}{96} \left(-251287+615 \pi
^2\right)-\frac{45}{32}(\chi^{\nu^1}_{35+}+\chi^{\nu^1}_{35-})+\nnm\\
&\quad+\frac{135}{16}(\chi^{\nu^1}_{37+}+\chi^{\nu^1}_{37-})-5 (\chi^{\nu^1}_{47+}+\chi^{\nu^1}_{47-})\bigg]\bigg\} p^{-11/2}+\mc O(p^{-13/2}),
\end{align}
and
\begin{align}\label{psipredGSF_SO}
\delta \psi^\text{GSF}_{1a^0}&=-p^{-1}+\left(\frac{9}{4}+e^2\right)
p^{-2}\\
&+\bigg\{\frac{893}{16}-\frac{123 \pi ^2}{64}-\frac{1}{4}(\chi^{\nu^1}_{35+}-\chi^{\nu^1}_{35-})\nnm\\
&\qquad+e^2 \left[\frac{143}{4}-\frac{123 \pi
	^2}{256}-\frac{3}{8}(\chi^{\nu^1}_{35+}-\chi^{\nu^1}_{35-})\right]\bigg\} p^{-3}\nnm\\
&+
\bigg\{-\frac{319511}{2880}+\frac{1256 \gamma_\text{E} }{15}+\frac{15953 \pi
	^2}{6144}+\frac{3 }{8}(\chi^{\nu^1}_{35+}-\chi^{\nu^1}_{35-})\nnm\\
&+\frac{5}{4}(\chi^{\nu^1}_{37+}-\chi^{\nu^1}_{37-})-\frac{2}{3}(\chi^{\nu^1}_{47+}-\chi^{\nu^1}_{47-})+\frac{296 }{15}\ln 2+\frac{729}{5}\ln 3+\frac{628}{15}\ln p^{-1}\nnm\\
&\qquad+e^2 \bigg[-\frac{3983}{480}+\frac{536
	\gamma_\text{E} }{5}-\frac{55217 \pi ^2}{4096}+\frac{7 }{16}(\chi^{\nu^1}_{35+}-\chi^{\nu^1}_{35-})\nnm\\
&\qquad\qquad+\frac{25}{8}(\chi^{\nu^1}_{37+}-\chi^{\nu^1}_{37-})-2 (\chi^{\nu^1}_{47+}-\chi^{\nu^1}_{47-})\nnm\\
&\qquad\qquad+\frac{11720}{3}\ln 2-\frac{10206}{5}\ln 3+\frac{268 }{5}\ln p^{-1}\bigg]\bigg\}p^{-4}+\mc O(p^{-5}).\nnm
\end{align}

These results are compared with GSF results (obtained using MST techniques) in Eq.~(4.1) of Ref.~\cite{Kavanagh:2016idg}, Eq.~(23) of Ref.~\cite{Bini:2016dvs} and Eq.~(20) of Ref.~\cite{Bini:2019lcd} for the redshift, here gathered into
\begin{align}
\delta U^\text{MST}_{1a}&=\left(3-\frac{7 }{2}e^2-\frac{1}{8}e^4\right)
p^{-5/2}+\left(18-4 e^2-\frac{117}{4}e^4\right)
p^{-7/2}\\
+\bigg(&87+\frac{287}{2}e^2-\frac{6277}{16}e^4\bigg)p^{-9/2}\nnm\\
+\bigg[&\frac{3890}{9}-\frac{241 \pi
	^2}{96}+e^2 \left(\frac{5876}{3}-\frac{569 \pi
	^2}{64}\right) +e^4 \left(-3547+\frac{2025 \pi
	^2}{128}\right)\bigg] p^{-11/2}\nnm\\
+&\mc O(p^{-13/2}),\nnm
\end{align}
and Eq.~(3.33) of Ref.~\cite{Kavanagh:2017wot} for the precession frequency, 
\begin{align}
\delta \psi^\text{MST}_{1a^0}&=-p^{-1}+\left(\frac{9}{4}+e^2\right)
p^{-2}\\
&+\left[\frac{1}{64} \left(2956-123 \pi
^2\right)+\frac{1}{256} e^2 \left(5456-123 \pi
^2\right)\right] p^{-3}\nnm\\
&+
\bigg[-\frac{587831}{2880}+\frac{1256 \gamma_\text{E}
}{15}+\frac{31697 \pi ^2}{6144}+\frac{296}{15}\ln 2+\frac{729}{5}\ln 3+\frac{628}{15}\ln p^{-1}\nnm\\
&\quad+e^2 \bigg(-\frac{164123}{480}+\frac{536
	\gamma_\text{E} }{5}-\frac{23729 \pi ^2}{4096}\nnm\\
&\qquad\qquad+\frac{11720}{3}\ln 2-\frac{10206}{5}\ln 3+\frac{268 }{5}\ln p^{-1}\bigg)\bigg]p^{-4}+\mc O(p^{-5})\nnm.
\end{align}
The first lines of each set of variables (the NLO) agree. At N$^2$LO, the expressions depend on the remaining unknown coefficients. Matching $\delta \psi^\text{MST}_{1a^0}=\delta \psi^\text{GSF}_{1a^0}$ and $\delta U^\text{MST}_{1a}=\delta U^\text{GSF}_{1a}$ order by order in $p$ and $e$, the N$^2$LO constraints can be consistently solved for
\begin{equation}\label{NNLOsolnew}
\chi^{\nu^1}_{35-}=0\,, \, \, \, \, \, \, \, \, \,  \chi^{\nu^1}_{35+}=\frac{77}{2}\,.
\end{equation}
This is another important double check in our work, since these parameters were the last to retrieve to fully confirm the known scattering-angle values at N$^2$LO SO order from Ref.~\cite{Vines:2018gqi}.
From the earlier constraint~\eqref{chicon}, the coefficients~\eqref{NNLOsolnew} also imply
\begin{equation} 
\chi^{\nu^1}_{45-}=\frac{45}{8}\,, \, \, \, \, \, \, \, \, \,  \chi^{\nu^1}_{45+}=\frac{777}{8}\,.
\end{equation}
In light of the N$^2$LO SO solutions, the \nNNLOSO constraints at each order in $p$ and $e$ can be consistently solved for the remaining four unknowns,
\begin{align}
& \chi^{\nu^1}_{37-}=0\,, \, \, \, \, \,  \, \, \, \, \, \,  \, \, \, \, 
\chi^{\nu^1}_{47-}=\frac{257}{96}+\frac{251}{256}\pi ^2\,, \\
& \chi^{\nu^1}_{37+}=\frac{177}{4}\,,  \, \, \, \,  \, \, \, \,  \chi^{\nu^1}_{47+}=\frac{23717}{96}-\frac{733}{256}\pi ^2\,.
\end{align}
The \nNNLOSO scattering angle~\eqref{eq:chi_tomatch} is finally
\begin{align}\label{chiNNNLOSOnew}
\frac{\chi_{a^1}}{\Gamma}=
\frac{\delta a_-}{b}v\bigg\{&-\frac{\pi  (GM)^2}{b^2
	v^4}\left[\frac{1}{2}+\frac{3}{4} v^2+\mathcal{O}(v^8)\right]\\
&+\frac{(GM)^3}{b^3 v^6}\bigg[-2-20 v^2-10v^4+\mathcal{O}+\mathcal{O}(v^8)\bigg]\nonumber\\
& +\frac{\pi(GM)^4}{b^4 v^8}\bigg[-\frac{63}{4} v^2-\frac{315}{8} v^4-\frac{315}{32} v^6\nonumber\\
&+\nu	\bigg(\frac{3 }{4}v^2+\frac{45 }{8}v^4+\left(\frac{257}{96}+\frac{251 \pi
	^2}{256}\right) v^6\bigg)+\mathcal{O}+\mathcal{O}(v^8)\bigg]\bigg\}
\nonumber\\
+\frac{a_+}{b}v&\bigg\{-\frac{4 GM}{b
	v^2}\bigg[1+\mathcal{O}(v^8)\bigg]-\frac{\pi  (GM)^2}{b^2
	v^4}\left[\frac{7}{2}+\frac{21}{4} v^2+\mathcal{O}+\mathcal{O}(v^8)\right]\nonumber\\
&+\frac{(GM)^3}{b^3 v^6}\bigg[-10-100 v^2-50v^4\nonumber\\
&+\nu 
\bigg(10 v^2+\frac{77 }{2}v^4+\frac{177 }{4}v^6\bigg)+\mathcal{O}(v^8)\bigg]\nonumber\\
& +\frac{\pi(GM)^4}{b^4 v^8}\bigg[-\frac{273}{4} v^2-\frac{1365}{8} v^4-\frac{1365}{32} v^6\nonumber\\
&+\nu\bigg(\frac{39 }{4}v^2+\frac{777 }{8}v^4+\left(\frac{23717}{96}-\frac{733 \pi
	^2}{256}\right)v^6\bigg)+\mathcal{O}(v^8)\bigg] \bigg\}\,.\nonumber
\end{align}

We conclude the derivation of the conservative \nNNLOSO PN dynamics specifying the remaining 3PM and 4PM coefficients of the N$^3$LO-SO Hamiltonian. From the constraints~\eqref{3PMalphacoeff}-\eqref{4PMalphacoeff} and the scattering-angle coefficients above, we get
\begin{align}\label{alphaspec}
\alpha_{35-}=&\frac{9}{16}+\frac{39 \nu }{16}+\frac{5 \nu ^2}{16}\,,\nonumber\\
\alpha_{35+}=&\frac{159}{16}+\frac{411 \nu }{16}+\frac{95 \nu ^2}{16}\,, \nonumber\\
\alpha_{37-}=&\frac{315}{64}+\frac{1331 \nu }{64}+\frac{539 \nu ^2}{32}+\frac{7 \nu
	^3}{16}\,,\nonumber\\
\alpha_{37+}=&-\frac{315}{64}+\frac{6067 \nu }{64}+\frac{4127 \nu ^2}{32}+\frac{175
	\nu ^3}{16}\,,\nonumber\\
\alpha_{47-}=&-\frac{7}{16}+\left(-\frac{203}{36}-\frac{5 \pi ^2}{384}\right) \nu
-\frac{39 \nu ^2}{8}+\frac{\nu ^3}{8}\,,\nonumber\\
\alpha_{47+}=&-\frac{217}{16}+\left(-\frac{2717}{36}+\frac{487 \pi ^2}{384}\right)
\nu -56 \nu ^2-\frac{5 \nu ^3}{2}\,.
\end{align}
That is, the fully-specified \nNNLOSO Hamiltonian is given by Eq.~\eqref{SOHam} with coefficients at 1PM and 2PM orders given, respectively, by Eqs.~\eqref{1PMalphacoeff} and~\eqref{2PMalphacoeff}, and 3PM and 4PM coefficients given by Eq.~\eqref{alphaspec}.

\subsubsection{Derivation of the third-subleading \Sonestwo PN dynamics}
\label{sec:S1S2}

The derivation of the third-subleading \Sonestwo PN dynamics is very similar to the one already presented above for the spin-orbit couplings. The goal is to specify the coefficients in $\chi^{\nu^1}_{3\times}(v)$ and $\chi^{\nu^1}_{4\times}(v)$ from~\eqref{eq:all_unknowns}. We extend the Hamiltonian~\eqref{HN3LOSO} to include a contribution that is bilinear in the spins, $H^{\text{N}^3\text{LO}}_\text{\Sonestwo}(r,p^2)\equiv H_{a^0}(r,p^2)+H_{a^1}(r,p^2)+H_{a_{1}a_{2}}(r,p^2)+\mc O(a_1^2,a_2^2)$, where $H_{a^0}(p^2)$ and $H_{a^1}(r,p^2)$ are the Hamiltonians fully specified at \nNNLOSO order from Appendix~\ref{app:DJStoISO@4PN} and Sec.~\ref{sec:3&4paper_SO},\footnote{
	The $H_{a^0}(p^2)$ and $H_{a^1}(r,p^2)$ contributions are specified for a smoother presentation. This is by no means a requirement.
} and $H_{a_{1}a_{2}}(r,p^2)$ is the new contribution with ansatz
\begin{alignat}{2} \label{S1S2Ham}
H_{a_{1}a_{2}}&=\frac{\mu}{ r^2}a_1 a_2\bigg[\alpha_{10\times}\frac{GM}{r}\bigg]+\, \, \, \, & \\
\frac{\mu}{c^2 r^2}a_1 a_2\bigg[&\alpha_{12\times}\frac{p^2}{\mu^2}\frac{GM}{r}+\alpha_{22\times}\frac{(GM)^2}{r^2}\bigg]+\, \, \, \, &\nonumber\\
\frac{\mu}{c^4 r^2}a_1 a_2\bigg[&\alpha_{14\times}\frac{p^4}{\mu^4}\frac{GM}{r}+\alpha_{24\times}\frac{p^2}{\mu^2}\frac{(GM)^2}{r^2}+\alpha_{34\times}\frac{(GM)^3}{r^3}\bigg]+\, \, \, \, &\nonumber\\
\frac{\mu}{c^6 r^2}a_1 a_2\bigg[&\alpha_{16\times}\frac{p^6}{\mu^6}\frac{GM}{r}+\alpha_{26\times}\frac{p^4}{\mu^4}\frac{(GM)^2}{r^2}+\alpha_{36\times}\frac{p^2}{\mu^2}\frac{(GM)^3}{r^3}+\alpha_{46\times}\frac{(GM)^4}{r^4}\bigg]. &\nonumber
\end{alignat}
The terms up to N$^2$LO, $\mathcal{O}(c^{-4})$, are easily retrievable from known terms in the same gauge~\cite{Vines:2018gqi}. Again, they are assumed not to be to known to perform useful checks along the way.
Solving for the scattering angle through N$^3$LO and including cross terms in the spins, the prediction reads $\chi^\text{(pred)}_{\text{N}^3\text{LO}}(b,v) = \chi_{a_0}(b,v)+\chi_{a_1}(b,v)+\chi_{a_1a_2}^\text{(pred)}(b,v)+\mc O(G^5, a_1^2, a_2^2)$. The angles $\chi_{a_0}(b,v)$ and $\chi_{a_1}(b,v)$ are fully specified as discussed in Sec.~\ref{sec:3&4paper_SO}, while $\chi_{a_1a_2}^\text{(pred)}(b,v)$ is the prediction at bilinear-order in the spins
\begin{align}
\chi_{a_1 a_2}^\text{(pred)}=\frac{a_1 a_2}{b^2} \bigg\{&\frac{G M }{bv^2}\chi_{1\times}^\text{(pred)}(v)+\frac{\pi \left(G M \right)^2}{b^2 v^4}\chi_{2\times}^\text{(pred)}(v)+\frac{(G M)^3}{b^3
	v^6} \chi_{3\times}^\text{(pred)}(v)\nonumber\\
&+ \frac{\pi\left(G M \right)^4}{b^4 v^8} \chi_{4\times}^\text{(pred)}(v)+\mathcal{O}(v^8)\bigg\}\,.
\label{eq:S1S2chipred}
\end{align}
The functions are reported in Appendix~\ref{app:chipred} for completeness. Matching the prediction~\eqref{eq:S1S2chipred} to~\eqref{S1S2chi} with~\eqref{eq:all_unknowns} order by order in $G$, $v$, and $\nu$, the mass dependence of the scattering angle implies the following 1PM and 2PM coefficients,
\begin{align}\label{2PMalphaS1S2}
&\alpha_{10\times}=-1\,, \, \, \, \, \, \alpha_{12\times}=-\frac{1}{2}-\frac{5 \nu }{4}\,, \, \, \, \, \,\alpha_{14\times}=\frac{3}{8}-\frac{9 \nu }{16}-\frac{11 \nu ^2}{8}\,, \nonumber\\
&\alpha_{16\times}=-\frac{5}{16}+\frac{39 \nu }{32}-\frac{93 \nu ^3}{64}\,, \, \, \, \, \,\alpha_{22\times}=5+\frac{7 \nu }{4}\,, \\
&\alpha_{24\times}=\frac{21}{8}+\frac{837 \nu }{32}+\frac{27 \nu ^2}{8}\,, \, \, \, \, \,\alpha_{26\times}=-\frac{11}{4}-\frac{25 \nu }{16}+\frac{435 \nu ^2}{8}+\frac{145 \nu ^3}{32}\,.\nonumber
\end{align}
It also implies
\begin{equation}\label{chi44cross}
\chi^{\nu^1}_{30\times}=0, \quad \chi^{\nu^1}_{32\times}=8, \quad\chi^{\nu^1}_{40\times}=0, \quad \chi^{\nu^1}_{42\times}=\frac{15}{2}, \quad \chi^{\nu^1}_{44\times}=-\frac{495}{16}+\frac{45 }{32}\chi^{\nu^1}_{34\times}\,.
\end{equation}
At 3PM and 4PM we also get the following constraints on the Hamiltonian coefficients,
\begin{align}\label{acrosscon}
\alpha_{34\times}=&-\frac{101}{8}-\frac{835 \nu }{32}-\frac{5 \nu ^2}{8}-\frac{3 \nu  }{16}\chi^{\nu^1}_{34\times}\,,\nonumber\\
\alpha_{36\times}=&-\frac{77}{10}+\nu  \left(-\frac{6737}{32}-\frac{3
}{32}\chi^{\nu^1}_{34\times}-\frac{3 }{16}\chi^{\nu^1}_{36\times}\right)\nonumber\\
&+\nu
^2 \left(-\frac{37111}{320}-\frac{3 }{16}\chi^{\nu^1}_{34\times}\right)-\frac{819 \nu ^3}{320}\,,\nonumber\\
\alpha_{46\times}=&\frac{111}{5}+\nu  \left(-\frac{1255}{16}+\frac{27
}{16}\chi^{\nu^1}_{34\times}+\frac{9 }{8}\chi^{\nu^1}_{36\times}-\frac{8
}{15}\chi^{\nu^1}_{46\times}\right)\nonumber\\
&+\nu ^2 \left(\frac{4019}{80}-\frac{143 \nu ^3}{160}+\frac{9
}{16}\chi^{\nu^1}_{34\times}\right)-\frac{143 \nu ^3}{160}\,.
\end{align}
These results give a Hamiltonian $H^{\text{N}^3\text{LO}}_\text{\Sonestwo}(r,p_r,L;\{\chi^{\nu^1}_{34\times},\chi^{\nu^1}_{36\times},\chi^{\nu^1}_{46\times}\})$ that depends on the remaining \Sonestwo coefficients to be fixed. 
We find that the $\delta \psi^\text{GSF}_{1a}$ spin correction in Eq.~\eqref{eq:psia} contains the right amount of (gauge-invariant) information needed to extend the domain of validity of the \GSF results to arbitrary-mass binaries where both bodies are spinning. This is because it contains information about both the massive spin $a_2=m_2 \hat a$, through the power of $\hat a$ it is expanded to first order in, and the secondary spin $a_1$, since the derivative with respect to $S_1$ that defines $\psi_1$ brings linear-in-$a_1$ information down to zeroth order.\footnote{Alternatively, we could have used the redshift and expanded in the secondary spin. However, analytical results in a double PN-SMR expansion are not available for the secondary spin. The unavailability of these results is also the reason why we have restricted the attention to the \Sonestwo dynamics and have not sought to find the complete spin-spin couplings, for which more constraints from the linear-in-$a_1$ correction to $\psi_1$ are needed to fix the $S_1^2$ coefficients.}
From this, using the procedure outlined in the previous section, consistently working through N$^3$LO with terms bilinear in spin, and reducing the arbitrary-mass-ratio results (also available as Supplemental Material in Ref.~\cite{Antonelli:2020ybz}) to first order in $q$, zeroth order in $a_1$ and first order in $\hat a$, we find
\begin{align}\label{psipredGSF_s1s2}
\delta \psi^\text{\GSF}_{1a}=&-\frac{1}{2}p^{-3/2}+\left(-\frac{41}{8}-\frac{e^2}{8}\right)p^{-5/2}\\
+&\bigg[
-\frac{159}{32}-\frac{123 \pi ^2}{64}-\frac{3}{16}\chi^{\nu^1}_{34\times}+e^2 \left(-\frac{89}{4}-\frac{123 \pi
	^2}{256}-\frac{9}{32}\chi^{\nu^1}_{34\times}\right)
\bigg] p^{-7/2}\nnm\\
+&
\bigg[
-\frac{4220237}{5760}+\frac{1256 \gamma_\text{E} }{15}+\frac{75841 \pi
	^2}{6144}+\frac{21}{32}\chi^{\nu^1}_{34\times}+\frac{15
	}{16}\chi^{\nu^1}_{36\times}-\frac{8}{15}\chi^{\nu^1}_{46\times}\nnm\\
&\quad+\frac{296}{15}\ln 2+\frac{729}{5}\ln 3+\frac{628 }{15}\ln p^{-1}+e^2
\bigg(-\frac{932729}{640}+\frac{536 \gamma_\text{E} }{5}\nnm\\
&\quad+\frac{7703
	\pi ^2}{4096}+\frac{57}{64}\chi^{\nu^1}_{34\times}+\frac{75 }{32}\chi^{\nu^1}_{36\times}-\frac{8
	}{5}\chi^{\nu^1}_{46\times}\nnm\\
&\quad+\frac{11720 }{3}\ln 2-\frac{10206 }{5}\ln 3+\frac{268}{5}\ln p^{-1}\bigg)
\bigg] p^{-9/2} +\mc O(p^{-11/2})\nnm.
\end{align}
The expression matches known results in the \GSF literature through NLO $\mathcal{O}(p^{-5/2})$ (see Eq.(52) and (56) of Ref.~\cite{Bini:2019lkm} for the $e^0$ and $e^2$ contributions, respectively). The results in the literature are here reported for convenience:
\begin{align}
\delta \psi^\text{MST}_{1a}=&-\frac{1}{2}p^{-3/2}+\left(-\frac{41}{8}-\frac{e^2}{8}\right)p^{-5/2}\\
+&\left[\frac{237}{32}-\frac{123 \pi ^2}{64}+e^2
\left(-\frac{59}{16}-\frac{123 \pi ^2}{256}\right)\right] p^{-7/2}\nnm\\
+&\bigg\{
-\frac{2580077}{5760}+\frac{1256 \gamma_\text{E} }{15}+\frac{52225 \pi ^2}{6144}+\frac{296}{15}\ln 2+\frac{729 }{5}\ln 3+\frac{628}{15}\ln p^{-1}\nnm\\
&\quad+ e^2\bigg[-\frac{274889}{640}+\frac{536 \gamma_\text{E} }{5}-\frac{39529 \pi ^2}{4096}\nnm\\
&\qquad\quad+\frac{11720}{3}\ln 2-\frac{10206 }{5}\ln 3+\frac{268}{5}\ln p^{-1}
\bigg]
\bigg\} p^{-9/2} +\mc O(p^{-11/2})\nnm.
\end{align}
At N$^2$LO, the \GSF results constrain our prediction. At each order in $e$, the N$^2$LO portions of $\delta \psi^\text{MST}_{1a}$ and $\delta \psi^\text{\GSF}_{1a}$ can be consistently matched with
\begin{equation}
\chi^{\nu^1}_{34\times}=-66\,,
\end{equation}
which implies, through Eq.~\eqref{chi44cross}, that
\begin{equation}
\chi^{\nu^1}_{44\times}=-\frac{495}{4}\,.
\end{equation}
At N$^3$LO, the constraints are solved at each order in the eccentricity for
\begin{equation}
\chi^{\nu^1}_{36\times}=-\frac{1093}{5}\,, \, \, \, \, \chi^{\nu^1}_{46\times}=-\frac{7995}{8}+\frac{1845 \pi ^2}{256}\,.
\end{equation}
We conclude that the spin-precession frequency provides enough information to fully specify the scattering angle through third subleading \Sonestwo PN order. For completeness, this reads
\begin{align}\label{S1S2chispec}
\chi_{a_1 a_2}=\frac{a_1 a_2}{b^2} \bigg\{&\frac{G M }{b
	v^2}\left[4+4 v^2+\mathcal{O}(v^8)\right]+\frac{\pi \left(G M \right)^2}{b^2 v^4}\left[3+\frac{45 }{2}v^2+\frac{9
}{2}v^4+\mathcal{O}(v^8)\right]\nonumber\\
+&\frac{(G M)^3}{b^3
	v^6} \bigg[8+280 v^2+440 v^4+40 v^6\nnm\\
&+\nu 
\bigg(8 v^2-66 v^4-\frac{1093 }{5}v^6\bigg)+\mathcal{O}(v^8)\bigg]\\
+&\frac{\pi\left(G M \right)^4}{b^4 v^8} \bigg[\frac{315 }{2}v^2+\frac{3675
}{4}v^4+\frac{9975 }{16}v^6\nonumber\\
&+\nu  \bigg[\frac{15 }{2}v^2-\frac{495 }{4}v^4+\left(-\frac{7995}{8}+\frac{1845 \pi
	^2}{256}\right) v^6\bigg]+\mathcal{O}(v^8)\bigg]\bigg\}\,. \nnm
\end{align}
From the new coefficients, we get the 3PM and 4PM Hamiltonian coefficients using Eq.~\eqref{acrosscon},
\begin{align}
\alpha_{34\times}=&-\frac{101}{8}-\frac{439 \nu }{32}-\frac{5 \nu ^2}{8}\,,\nonumber\\
\alpha_{36\times}=&-\frac{77}{10}-\frac{26137 \nu }{160}-\frac{33151 \nu ^2}{320}-\frac{819 \nu
	^3}{320}\,,\nonumber\\
\alpha_{46\times}=&\frac{111}{5}+\frac{7781 \nu }{80}-\frac{123 \pi ^2 \nu }{32}+\frac{1049 \nu
	^2}{80}-\frac{143 \nu ^3}{160}\,,
\end{align}
completing the specification of the Hamiltonian.

\subsubsection{The N$^3$LO dynamics in waveform models with spins}
\label{sec:comps}

Once the new PN information has been encoded in the scattering-angle dynamics or Hamiltonian, it can be translated into expressions used to model binaries in a bound orbit. One can for instance obtain the \nNNLOSO corrections to the gyro-gravitomagnetic ratios $g_S$ and $g_{S^*}$ in the EOB effective Hamiltonian for (aligned-)spin binaries, 
\begin{equation}\label{Heff_gyros}
H^\text{SO}_\text{eff} = \frac{1}{c^2 r^3} L [g_S(r ,p) S + g_{S^*}( r , p)S^*].
\end{equation}
Through a canonical transformation from the quasi-isotropic Hamiltonians we have specified in the previous section, the new terms can be found to be those in Eq.~\eqref{chiNNNLOSO} (see also Eqs.~(55) and (56) of Ref.~\cite{Nagar:2011fx} for lower-order corrections).
Since $g_{S,S^*}$ can be made independent of the spins~\cite{Damour:2008qf,Nagar:2011fx,Barausse:2011ys}, the \nNNLOSO terms derived here from a spin-aligned configurations are indeed valid for \emph{generic} spin orientations. Noticing that the spins enter Hamiltonians only through the scalar $\boldsymbol L\cdot \boldsymbol S_i$ at this spin order,\footnote{For instance, see Eq.~\eqref{eq:Hspinorbit} for the generic SO Hamiltonian in ADM coordinates, from which one can reproduce the precession equation for the spins~\eqref{eq:Stot_evol}.} one simply needs to rewrite the effective Hamiltonian~\eqref{Heff_gyros} as 
\begin{equation}\label{Heff_gyrosgen}
H^\text{SO}_\text{eff} = \frac{1}{c^2 r^3} \boldsymbol L\cdot [g_S(\boldsymbol r ,\boldsymbol p) \boldsymbol S + g_{S^*}(\boldsymbol r , \boldsymbol p)\boldsymbol S^*].
\end{equation}
In this sense, the \nNNLOSO PN corrections we have found can be employed to model generic (and possibly precessing) spinning binaries.
In the case of the \Sonestwo dynamics, one can translate the (partial) spin-spin results into higher-order-in-spin corrections to the EOB potentials, as described in Sec.~\eqref{sec:EOBform}. The results of the corrections to the potentials can be found in Sec.~\ref{sec:NRcomp} of \chap~\ref{chap:five}, see Eqs.~\eqref{SSpots}. These corrections are only valid for aligned spins, since the generic-spins case has additional contributions proportional to $(\bm{n}\cdot \bm{S}_1)(\bm{n}\cdot \bm{S}_2)$~\cite{Khalil:2020mmr}, which vanish for aligned spins and cannot be fixed from aligned-spin self-force results or be removed by canonical transformations.

\begin{figure}
	\centering
	\includegraphics[width=.8\linewidth]{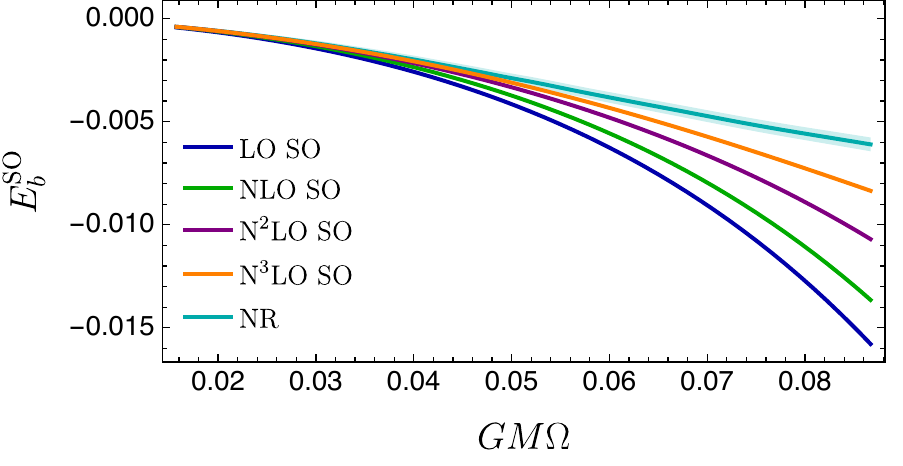}
	\caption{
		Comparison of the gauge-invariant relation between the circular-orbit aligned-spin spin-orbit binding energy $E_b$ and $\Omega$.
		The figure shows results obtained numerically from the EOB Hamiltonians with effective SO Hamiltonian~\eqref{Heff_gyros} and increasing PN orders in $g_{S,S^*}$, and NR results with errors from Refs.~\cite{Ossokine:2017dge,SXS}.
		The linear-in-spin contribution is isolated for equal masses ($q=1$) as discussed in the main text. Adapted from Ref.\cite{Antonelli:2020aeb}.}
	\label{fig:bindingenergynew}
\end{figure}

We can assess the improvements from the novel terms by comparing binding energies in a circular orbit and aligned spins to numerical simulations~\cite{Dietrich:2016lyp,Ossokine:2017dge}.
Spin-dependent contributions in this configuration can be isolated approximating the binding energy as a (PN-like) expansion in the spins, $E_\text{b} (S_1,S_2) = E_b^{S^0} + E_b^\text{SO} + E_b^\text{SS}+\dots$, and rearranging the $E_\text{b} (S_1,S_2)$ energies into linear combinations of the spins. For instance, we can isolate SO binding energy contributions for equal masses using the linear combination in Eq.~(14) of Ref.~\cite{Ossokine:2017dge}.
The comparison between the SO contributions from Hamiltonians and NR results is shown in Fig.~\ref{fig:bindingenergynew}, and allows us to assess the improvements from increasing orders in the gyro-gravitomagnetic corrections as a function of the binary's orbital frequency $\Omega$. The new corrections (included in the orange line) lead to better comparisons against NR results (cyan, with error estimated as in Ref.~\cite{Ossokine:2017dge}). This result is important, as it illustrates that the procedure discussed here can potentially improve waveform models through a combination of all the available analytical techniques to solve Einstein's equations. In Sec.~\ref{sec:NRcomp}, we present more details and additional comparisons, including at \Sonestwo order.

\section[Improving the strong-field binary dynamics with the small-mass-ratio$\dots$]{Improving the strong-field binary dynamics with the small-mass-ratio approximation}
\label{sec:2paper}

Successive orders in the PN and PM approximations allow us to obtain ever more accurate waveform models. Historically, the PN has been used as the theoretical ground to model waveforms, though we have discussed that the PM approximation could soon be a valid addition. Both however lose accuracy in the strong-field regime. One can see this, for instance, in Fig.~\ref{fig:PMvsclassicnew}, where the binding-energy curves of both PN and PM models considerably deviate from NR for strong fields $GM\Omega\gtrapprox 0.05$. Accurate waveform models for future data analyses require a handle both on the strong-field regime -- as needed, e.g., for the high-SNR sources in LISA such as supermassive BH binaries or for the BH binaries to be detected with next-generation detectors -- and on binaries with a large asymmetry in the masses -- like the \texttt{GW190412} event~\cite{LIGOScientific:2020stg}, or the EMRIs one expects to detect with LISA~\cite{vandeMeent:2018rms,Chua:2020stf,Hughes:2021exa,Katz:2021yft}. 
The SMR approximation is an excellent theoretical basis in both cases, as \GSF results are not limited to weak fields and are most naturally defined when the masses are highly asymmetric.

It was discussed in Sec.~\ref{sec:interplay} that SMR results can be extended towards the regime of comparable masses, see Fig.~\ref{fig:paramspace}. One possibility is to base EOB Hamiltonians also on SMR information (henceforth ``EOBSMR'' Hamiltonians), and use them to model both equal and unequal mass binaries. This is one of the main motivations driving the analysis of \chap~\ref{chap:three}, where we assess the above proposition in the case of nonspinning binaries in a quasi-circular orbit. The interplay between the EOB formalism and the \GSF framework was first proposed in Ref.~\cite{Damour:2009sf}. References~\cite{Barausse:2011dq,LeTiec:2011bk,Akcay:2012ea} later focussed on obtaining a Hamiltonian for nonspinning circular-orbit binaries in an SMR expansion to first order in the mass ratio, resumming the complete PN information in the process. Rearranging Eq.~\eqref{eq:firstlaw_circ} to give an expression for the circular-orbit binding energy as a function of the gauge-independent radius $x=(GM\Omega)^{2/3}$,  Ref.~\cite{LeTiec:2011dp} finds
\begin{align}\label{eq:Ebcirc_GSF}
E_{b,\text{circ}} &= \left(\frac{1-2x}{\sqrt{1-3x}}-1\right)+\nu\, E_{b,\text{circ}}^\text{\GSF} + \mc O(\nu^2),\qquad \text{with}\nnm\\
E_{b,\text{circ}}^\text{\GSF}&= \frac{1}{2}\Delta z-\frac{x}{3}\frac{d\Delta z}{dx}-1+\sqrt{1-3x}+\frac{x}{6}\frac{7-24x}{(1-3x)^{3/2}}, 
\end{align} 
where $\Delta z(x)$ is the SMR expansion of the redshift about the Schwarzschild limit, $z_1 =\sqrt{1-3x}+\nu \Delta z(x)+\mc O(\nu^2)$. Matching this binding energy for the one obtained from an ansatz Hamiltonian in the EOB DJS gauge~\eqref{Ham_eff} in the circular-orbit limit, the companion paper~\cite{Barausse:2011dq} obtains the following correction to the conservative dynamics contained in the EOB $A(u,\nu)$ potential,
\begin{align}\label{eq:LReq}
A(u, \nu)&=1-2u+\nu a(u)+\mathcal{O}(\nu^2),\nnm\\
\text{with}\quad a(u)&= \Delta z(u)\sqrt{1-3u}-u\bigg(1+\frac{1-4u}{\sqrt{1-3u}}\bigg).
\end{align}
Self-force data at linear order in a Schwarzschild background can then be included in the EOB Hamiltonian through appropriate fits to $\Delta z(u)$. A fit that extends just below the ISCO ($u\sim 1/5$) was used in Refs.~\cite{Barausse:2011dq,LeTiec:2011dp}, while one extending to the light ring (LR, $u=1/3$) in Akcay \emph{et al.}~\cite{Akcay:2012ea}. In the latter work, a fundamental problem of the EOBSMR description in the DJS gauge was uncovered, namely a pole at the LR. Given that the binding energy diverges at the LR for a massless particle in a circular orbit around a Schwarzschild background, the presence of a pole there might not sound surprising. The problem in Eq.~\eqref{eq:LReq} is that, taking this expected pole into account in the circular-orbit limit, and including numerical \GSF information, the EOB $A$-potential \emph{still} diverges at the LR with a power $a(u)\propto (1-3u)^{-1/2}$~\cite{Akcay:2012ea}.
In fact, one obtains $a(u)$ through the matching between binding energies from the EOB Hamiltonian with $a(u)$ and Eq.~\eqref{eq:Ebcirc_GSF}. At the LR, and for circular orbits, the former scales as $a(u)L_\text{circ}^2\propto a(u)(1-3u)^{-1}$, with $L_\text{circ}$ the angular momentum for circular orbits, whereas the latter as $E_{b,\text{circ}}^\text{\GSF}\propto (1-3u)^{-3/2}$. This implies that $a(u)\propto (1-3u)^{-1/2}$ in order to reproduce the expected divergence from $E_{b,\text{circ}}^\text{\GSF}$. 
Through the $A$-potential, the pole therefore makes its way into the EOBSMR Hamiltonian in the DJS gauge (henceforth $H^{\text{EOB}}_\text{SMR}$), which is supposed to be valid for \emph{generic} orbits. The presence of this pole in the generic Hamiltonian, when we would expect it only for circular orbits, is referred to here as the ``LR divergence problem''.

The first step to obtain a viable EOBSMR Hamiltonian without poles at the LR for generic orbits is to consider another gauge for the effective Hamiltonian. The problem of the DJS description is that the angular momentum $L_\text{circ}$ appears with a power that is too low to reproduce the full divergent behaviour of $E_{b,\text{circ}}^\text{\GSF}$ in the circular-orbit limit. The strategy to avoid the LR divergence problem is then to find a gauge in which high enough powers of $L_\text{circ}$ (or an analogous quantity) can be accommodated. 
In \chap~\ref{chap:three}, we employ an EOB Hamiltonian in the PS gauge~\eqref{eq:Heff_PS} that is expanded around the Schwarzschild limit with a phase-space function $ Q^{\text{PS}}_\text{SMR} = \mu^2\hat{Q}^{\text{PS}}_\text{SMR}$ containing SMR information, 
\begin{equation}\label{HPSSMR}
\hat H_{\text{eff}}^{\text{PS}}=\sqrt{\hat H_{\text{S}}^{2}+(1-2u)\hat{Q}^{\text{PS}}_\text{SMR}(u,\nu,\hat H_{\text{S}})}\,.
\end{equation}
Here  $\hat H^2_{\text{S}}$, whose expression appears in Eq.~\eqref{eq:Hs}, is smooth at the LR for generic orbits, but scales as $\hat H_{\text{S}}\propto (1-3u)^{-1/2}$ (and thus analogously to $L_\text{circ}$) for circular orbits. In this gauge, the pole at the LR is to be ``absorbed'' by appropriate powers of $\hat H_{\text{S}}$.
First, one needs to understand what potential problems arise in the fitting of the Detweiler redshift. In Appendix~\ref{sec:appred}, Van de Meent fits a particular polynomial form of the redshift that contains both the expected divergence at the LR and a (milder) logarithmic divergence [see Eq.~\eqref{eq:zfit}]. We can single out these problematic terms rewriting the redshift as
\begin{equation}\label{zsmrnew}
\Delta z(x)=\frac{\Delta z^{(0)}(x)}{1-3x}+\frac{\Delta z^{(1)}(x)}{\sqrt{1-3x}}+\frac{\Delta z^{(2)}(x)}{1-3x}\log E_S^{-2}
\,,
\end{equation}
where $E_S \equiv (1-2x)/\sqrt{1-3x}$ is the circular-orbit limit of $\hat H_S$, and the functions $\Delta z^{(i)}(x)$ can be fitted to be smooth at the LR in terms of the gauge-invariant radius $x$.
In terms of these functions, Eq.~\eqref{eq:Ebcirc_GSF} reads
	\begin{align}\label{eq:Ebcirc_GSF_123}
E_{b,\text{circ}}^\text{\GSF}&=\sqrt{1-3 x}-1+\frac{(7-24 x) x}{6 (1-3 x)^{3/2}}\\
&
+\frac{1}{2(1-3x)}
\bigg[\Delta z^{(0)}(x)+\Delta z^{(1)}(x)\sqrt{1-3 x}
+\Delta z^{(2)}(x)\ln E_S(x)\bigg]
\nonumber\\
&
-\frac{x}{3(1-3x)}\bigg\{\frac{3\Delta z^{(0)}(x)}{1-3 x}+\frac{3 \Delta z^{(1)}(x)}{2 \sqrt{1-3 x}}
\nonumber\\
&
\qquad\qquad\qquad+\bigg[\frac{1-6 x}{(1-2x)(1-3x)}+\frac{3\ln E_S(x)}{(1-3x)}\bigg]\Delta z^{(2)}(x)
\nonumber\\
&
\qquad\qquad\qquad+\frac{\Delta z^{(0)}}{dx}+ \sqrt{1-3 x}\,\frac{\Delta z^{(1)}}{dx}+ \frac{\Delta z^{(2)}}{dx}\ln E_S(x)\bigg\}\,.\nonumber
\end{align}
An ansatz for $\hat{Q}^{\text{PS}}_\text{SMR}$ that could capture all the LR poles in Eq.~\eqref{zsmrnew} is
\begin{equation}\label{Qans}
\hat{Q}^{\text{PS}}_{\text{SMR}}(u,\nu,\hat H_{\text{S}})=\nu\big[f_{0}(u)\hat H_{\text{S}}^{5}+f_{1}(u)\hat H_{\text{S}}^{2}+f_{2}(u)\hat H_{\text{S}}^{3}\ln \hat H^{-2}_{\text{S}}\big]\,.
\end{equation}
The role of the $\Ham_{\text{S}}^5$ term is to capture the global divergence $(1-3x)^{-2}$ of Eq.~\eqref{eq:Ebcirc_GSF_123}.\footnote{In principle, a $\Ham_{\text{S}}^3$ term will suffice to capture the divergence~\cite{Damour:2017zjx}. However, we find that this minimal choice leads to evolutions that are not well behaved for systems with comparable masses.} The second term $\Ham_{\text{S}}^2$ is devised to incorporate the $\sqrt{1-3x}$ terms appearing in the numerator of the same equation, which would make the Hamiltonian imaginary after the light ring. The term proportional to $\ln\Ham_{\text{S}}^{-2}$ incorporates the logs in the fit that would make the Hamiltonian non-smooth at the light ring. In order to fix the functions $f_i$ in Eq.~\eqref{Qans}, a prediction for the binding energy from the ansatz Hamiltonian~\eqref{HPSSMR} is obtained. For this, one needs to solve the circular-orbit condition
\begin{equation}
\dot p_r=-\frac{\partial  H_\text{EOB}}{\partial r}(r,p_r=0,p_\phi^{\text{circ}},\nu)=0\,,
\end{equation}
and obtain an expression for the angular momentum $p_\phi^{\text{circ}}(u)$ at order $\mc O(\nu)$ in the circular-orbit limit. The result is in terms of the gauge-dependent radius $u$, and must be expressed in terms of the gauge-independent $x$, which is done inverting $u(x)$ in a perturbative expansion in $\nu$ using the definition of $x=(GM\Omega)^{2/3}$ and the orbital frequency
\begin{equation}
\Omega = \frac{\partial H_\text{EOB}}{\partial p_\phi}(r,p_r=0,p_{\phi}^{\text{circ}},\nu)\,.
\end{equation}
The final expression for the binding energy [see Eq.~\eqref{Eeob} in \chap~\ref{chap:three} for the result and for more details] can be matched to Eq.~\eqref{eq:Ebcirc_GSF_123}.
Further imposing that the functions $f_i$ can be split into
\begin{align}
&f_i(x)=\tilde{f}_i(x)+\sum_{j=0}^{2}f_i^{(j)}(x)\Delta z^{(j)}(x)\,,
\end{align}
in such a way that the functions $\tilde f_i$ and $f_i^{(j)}$ are smooth at the light ring, we can solve the constraints from the binding-energy matching to get the following nonvanishing functions,
\begin{subequations}
	\begin{align}
	&\tilde{f}_0(x)=-\frac{x (1-3x)\left(1-4x\right)}{(1-2x)^5}\,,\quad\tilde{f}_1(x)=-\frac{x}{(1-2 x)^2}\,,\\
	&f_0^{(0)}(x)=\frac{1-3x}{(1-2x)^5}\,,\quad
	f_1^{(1)}(x)=\frac{1}{(1-2x)^2}\,,\quad
	f_2^{(2)}(x)=\frac{1}{(1-2x)^3}\,.\nnm
	\end{align}
\end{subequations}
We therefore obtain the following PS phase space function in the Hamiltonian,
\begin{align}\label{eq:QSMR_fin}
\frac{\hat{Q}^{\text{PS}}_{\text{SMR}}}{\nu}(u,\nu,\hat H_{\text{S}})=&(1-3u)\bigg[\frac{\Delta z^{(0)}(u)}{(1-2u)^5}-\frac{(1-4 u)\text{ }u}{(1-2 u)^5}\bigg]\hat H_{\text{S}}^5\\
&+\bigg[\frac{\Delta z^{(1)}(u)}{(1-2 u)^2}-\frac{u}{(1-2
	u)^2}\bigg]\hat H_{\text{S}}^2+\frac{\Delta z^{(2)}(u)}{(1-2u)^3}\hat H_{\text{S}}^{3}\ln\hat H_{\text{S}}^{-2}\,.\nonumber
\end{align}
The resultant EOBSMR Hamiltonian~\eqref{HPSSMR} with~\eqref{eq:QSMR_fin}, henceforth $H^{\text{EOB,PS}}_\text{SMR}$, does not contain poles in its generic-orbit form, as expected by construction.

The Hamiltonian specified by~\eqref{eq:QSMR_fin} resums circular-orbit PN information at linear order in the mass ratio. However, generic-orbit information is known in the PS gauge at least through 3PN information~\cite{Damour:2017zjx}. In this case, the phase-space function is
\begin{align}
\label{QPScoeffnew}
\hat Q^{\text{PS}}_{\text{3PN}}=&3\nu u^{2}Y+5\nu u^{3}+\bigg(3\nu-\frac{9}{4}\nu^{2}\bigg)u^{2}Y^{2}
\\
&+\bigg(27\nu-\frac{23}{4}\nu^{2}\bigg)u^{3}Y+\bigg(\frac{175}{3}\nu-\frac{41\pi^{2}}{32}\nu-\frac{7}{2}\nu^{2}\bigg)u^{4}\,,\nonumber
\end{align}
with PN parameter $Y \equiv (\hat H^{2}_{\text{S}}-1)\sim \mathcal{O}(1/c^2)$. We refer to Hamiltonians with $n$PN information in the EOB PS gauge as $H_{n{\text{PN}}}^{\text{EOB,PS}}$ to distinguish them from those in the DJS gauge (namely $H_{n{\text{PN}}}^{\text{EOB}}$, as in previous sections). We can add additional (non-circular) information from the PN expansion onto the SMR phase-space function~\eqref{eq:QSMR_fin}. In Sec.~\ref{sec:noncirc}, we propose a ``correction'' $\Delta\hat Q^{\text{PS}}$, such that an EOBSMR model with complete PN information at linear order in $\nu$ and additional generic-orbit information up to 3PN order is
\begin{equation}\label{eq:QSMR3PN}
\hat Q^{\text{PS}}_{\text{SMR-3PN}}=\hat Q^{\text{PS}}_{\text{SMR}}+\Delta\hat Q^{\text{PS}}\,,
\end{equation}
with $\Delta\hat Q^{\text{PS}}\equiv \Delta\hat Q^{\text{PS}}_{\text{extra}}-\Delta\hat Q^{\text{PS}}_{\text{count}}$. The first term is obtained as the difference between Eq.~\eqref{QPScoeffnew} and the 3PN expansion of Eq.~\eqref{eq:QSMR_fin}, and is found to be 
\begin{align}\label{eq:Qextra}
\Delta\hat Q^{\text{PS}}_{\text{extra}}=& 3\nu u^2 Y+\bigg(3\nu-\frac{9}{4} \nu^2 \bigg) u^2Y^2+3 \nu u^3 \nonumber\\
&+\bigg(22\nu-\frac{23}{4} \nu^2 \bigg) u^3Y+\bigg(16 \nu-\frac{7}{2} \nu^2 \bigg)  u^4\,.
\end{align}
The second term is $\Delta\hat Q^{\text{PS}}_{\text{count}}=\nu (9\, u^3Y^2+96\, u^4Y+112\, u^5)$, and it ensures the addition of Eq.~\eqref{eq:Qextra} to~\eqref{eq:QSMR_fin} does not spoil the matching between the circular-orbit binding energy from the SMR-3PN Hamiltonian and the one from Eq.~\eqref{eq:Ebcirc_GSF_123} at all PN orders and linear order in $\nu$. We refer to the EOBSMR Hamiltonian in the PS gauge with phase-space function~\eqref{eq:QSMR3PN} as the $H^{\text{EOB,PS}}_\text{SMR-3PN}$ Hamiltonian. The models we have defined in this section, and that are used in the analysis to come, can be found in Table~\ref{table:modelsnew}.

\begin{table}
	\caption{{\textbf{Two-body EOB Hamiltonians.} All the EOB Hamiltonians connect the ``real'' to the ``effective'' descriptions through the energy map~\eqref{eq:enmap} connecting real (EOB) and effective Hamiltonians. The EOB Hamiltonians in the DJS gauge below have effective Hamiltonian~\eqref{Ham_eff}, with varying $A$-potentials in the circular-orbit limit. The PS-gauge Hamiltonians have effective energy~\eqref{HeffPS}, with varying phase-space functions.}
		\label{table:modelsnew}}
	\centering
	\begin{tabular}{lp{7cm}p{2.5cm}}
		\hline
		
		$H_{n\text{PN}}^{\text{EOB}}$	&	$n$PN  Hamiltonian in DJS gauge& Eq.~\eqref{eq:A4PN}	\\
		
		$H_{\text{SMR}}^{\text{EOB}}$
		& SMR Hamiltonian in the DJS gauge (with LR divergence)	& Eq.~\eqref{eq:LReq} \\
		
		$H_{\text{SMR}}^{\text{EOB,PS}}$
		& SMR Hamiltonian in PS gauge	& Eq.~\eqref{eq:QSMR_fin} \\

		$H_{n{\text{PN}}}^{\text{EOB,PS}}$
		&   $n$PN  Hamiltonian in PS gauge & 
		Eq.~\eqref{QPScoeffnew}	\\

		$H_{\text{SMR-3PN}}^{\text{EOB,PS}}$					
		& SMR-3PN Hamiltonian in PS gauge & Eq.~\eqref{eq:QSMR3PN} \\

	\end{tabular}
\end{table}

\begin{figure}[t]
	\includegraphics[width=\columnwidth]{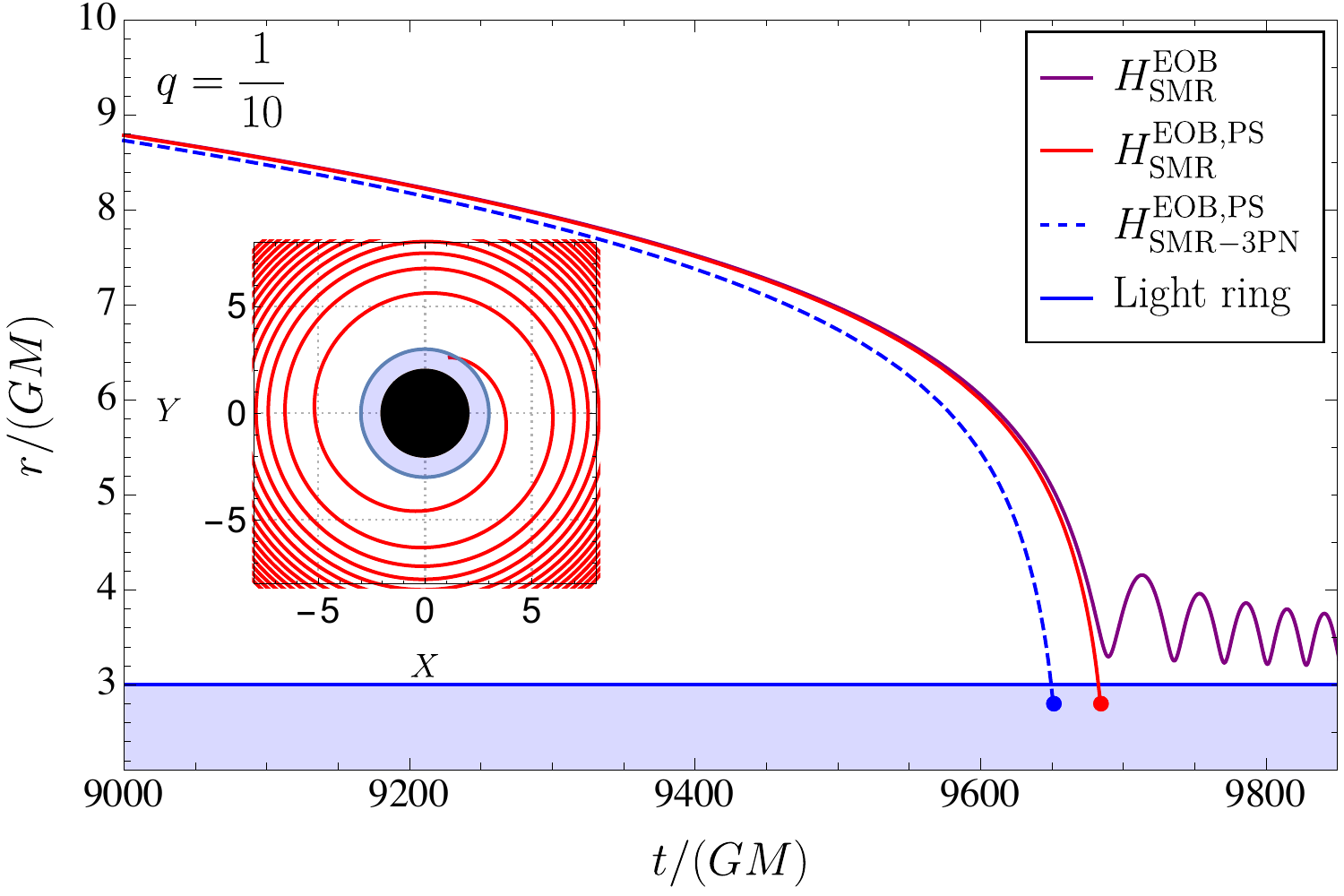}
	\caption{The evolved orbital separation for the SMR Hamiltonians is presented. The effective masses of models $H_{\text{SMR}}^{\text{EOB,PS}}$ and $H_{\text{SMR-3PN}}^{\text{EOB,PS}}$ plunge through the LR radius. Conversely, the plunge of the effective mass of $H_{\text{SMR}}^{\text{EOB}}$ presents unphysical features associated to the LR-divergence. Adapted from Ref.~\cite{Antonelli:2019fmq}.
	}
	\label{fig:orbitsepnew}
\end{figure}

\begin{figure*}
	\centering
	\includegraphics[width=.75\columnwidth]{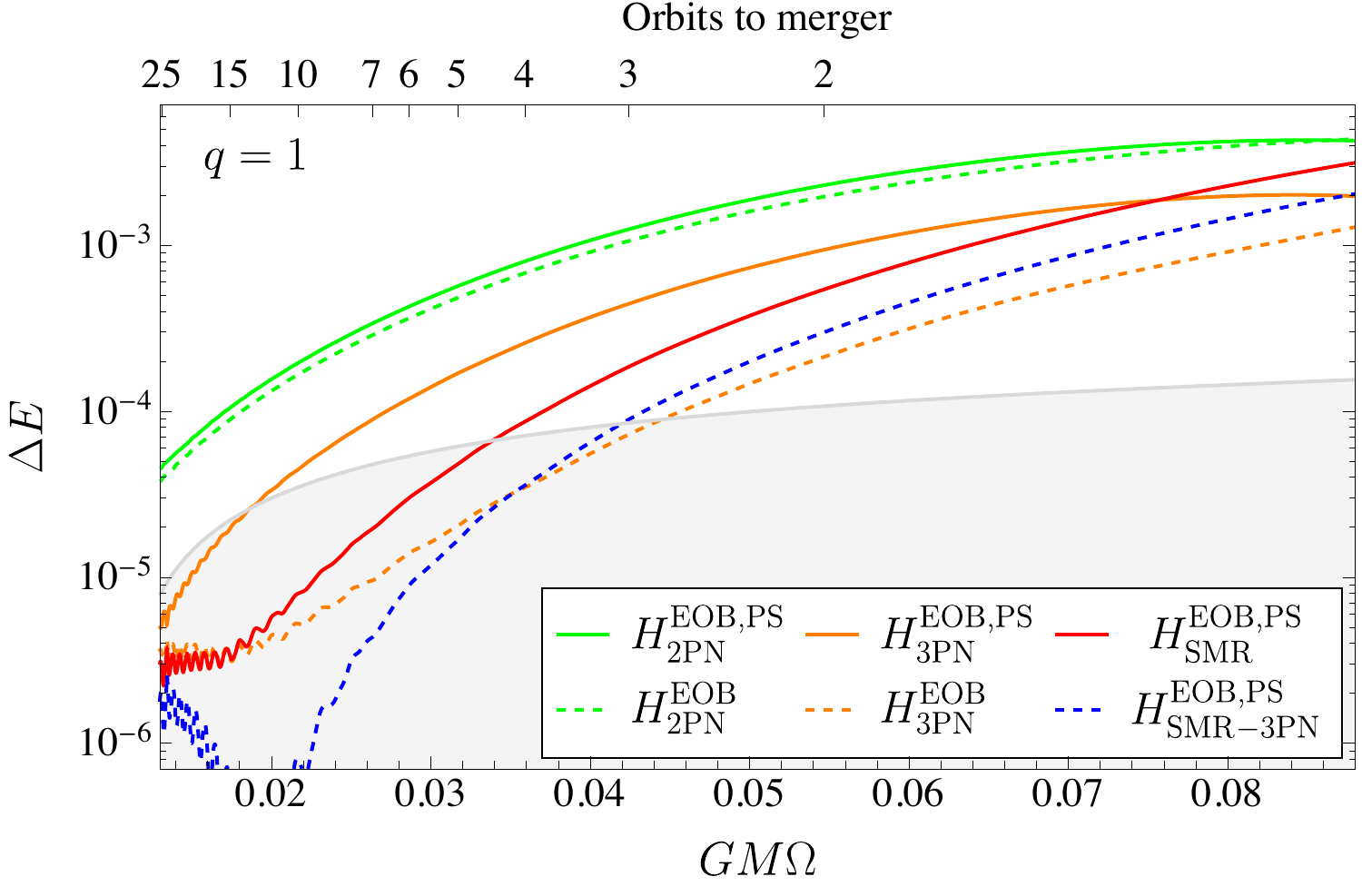}
	\caption{Differences $\Delta E$ in binding energy from NR for our SMR Hamiltonians versus frequency $GM\Omega$.
	We compare these to similar results for PN models up to third order, in both PS and DJS gauges. The estimated NR error is shown in grey. Adapted from Ref.~\cite{Antonelli:2019fmq}.}
	\label{fig:compEOnew}
\end{figure*}

The most direct way to check that the $H^{\text{EOB,PS}}_\text{SMR}$ and $H^{\text{EOB,PS}}_\text{SMR-3PN}$ models are free of poles at the LR is to evolve \emph{quasi}-circular inspirals from each Hamiltonian, for which we use the Hamilton equations and the EOB flux prescription of Sec.~\ref{sec:EOBform}. 
With a GW-driven inspiral, we can track the time evolution of the orbital separation $r(t)$, phase $\phi(t)$, angular $p_\phi(t)$, and radial momenta $p_r(t)$ of the binary. The orbital separation evolutions of the EOBSMR models, with and without PN information, in both DJS and PS gauges, are seen in Fig.~\ref{fig:orbitsepnew} for a binary with $q=1/10$. One can inspect the presence of unphysical effects from the LR divergence in the DJS gauge,\footnote{The pole at the LR modifies the potential of the binary through the Hamiltonian, making it diverge there. The effective mass $\mu$ ``bounces off'' this divergent potential, while GW emission pushes it back towards it. This gives rise to the oscillatory behaviour in Fig.~\ref{fig:orbitsepnew}.} and the absence in both models in the PS gauge. These time evolutions allow us to extract a host of information from the binary in the presence of gravitational radiation. For instance, they get us access to binding-energy curves that are not calculated assuming no GW radiation out of the system (as in Sec.~\ref{sec:1paper}), but rather evolved smoothly throughout the inspiral and plunge. This also implies that we can avoid restricting our attention to weak fields, and focus instead on comparing models in the challenging region of strong gravitational fields, where we expect the SMR information to be especially valuable. 
The binding energy is extracted from (``evolutions'' of) EOB Hamiltonians $E_\text{EOB}(t)\equiv H_\text{EOB}[r(t),p_\phi(t),p_r(t),\phi(t)]$, and again compared against the NR simulations from the SXS collaboration~\cite{Ossokine:2017dge}.
We plot the binding energy difference $\Delta E$ for $q=1$ between NR binding energy and EOB prediction as a function of frequency $\Omega\equiv\dot \phi(t)$ in Fig.\ref{fig:compEOnew}. In the weak-field regime, $H^{\text{EOB,PS}}_\text{SMR}$ and $H^{\text{EOB,PS}}_\text{SMR-3PN}$ resum two-body information very well, with differences $\Delta E$ that are consistent with the error from different resolutions of the NR simulations used, in gray (the same holds when we consider $q=1/10$, see Fig.~\ref{fig:compEO}). We compare the models with SMR information with PN models in both DJS and PS at the 2PN and 3PN orders. The differences between EOBSMR models and NR results are generally smaller than those with EOB models based on the PN approximation. 
The fact that the SMR approximation resums two-body information for comparable masses this well for GW-driven evolutions is not a trivial statement, and is a first confirmation that the results in Ref.~\cite{Barausse:2011dq,Tiec:2013twa} (obtained considering the conservative dynamics only) hold more generally when the binaries radiate GWs.


The results are encouraging and hold for a range of comparisons (presented in Sec.~\ref{sec:energetics}), and point to the utility of EOBSMR models as a base for future families of waveform templates. To crystallise this thought, we compare the \emph{dephasing} $\Delta\phi_{22}=\phi_\text{NR}-\phi_\text{EOB}$ of the (2,2) mode phase from NR simulations and EOB prediction.
For a proper comparison, the EOB and NR waveforms must be aligned~\cite{Pan:2011gk}. Usually, this amounts to minimizing a certain function
\begin{equation}
\Xi(\Delta t,\Delta \phi)=\int_{t^\text{alig}_1}^{t^\text{alig}_2}[\phi_\text{NR}(t)-\phi_\text{EOB}(t+\Delta t)-\Delta\phi]^2 dt\\,
\end{equation} 
over the time and phase shifts, $\Delta t$ and $\Delta\phi$. 
The integrating interval [$t^\text{alig}_1,t^\text{alig}_2$] is chosen in the inspiral of the NR simulation, and large enough to average out the numerical noise, which is needed to avoid junk radiation at the beginning of NR simulations. From the alignment procedure described above, one can obtain the phase and amplitude time-shift to be applied to the EOB model to align it with the NR waveforms, i.e., the aligned waveforms are

\begin{figure*}[t]
	\centering
	\includegraphics[width=.7\columnwidth]{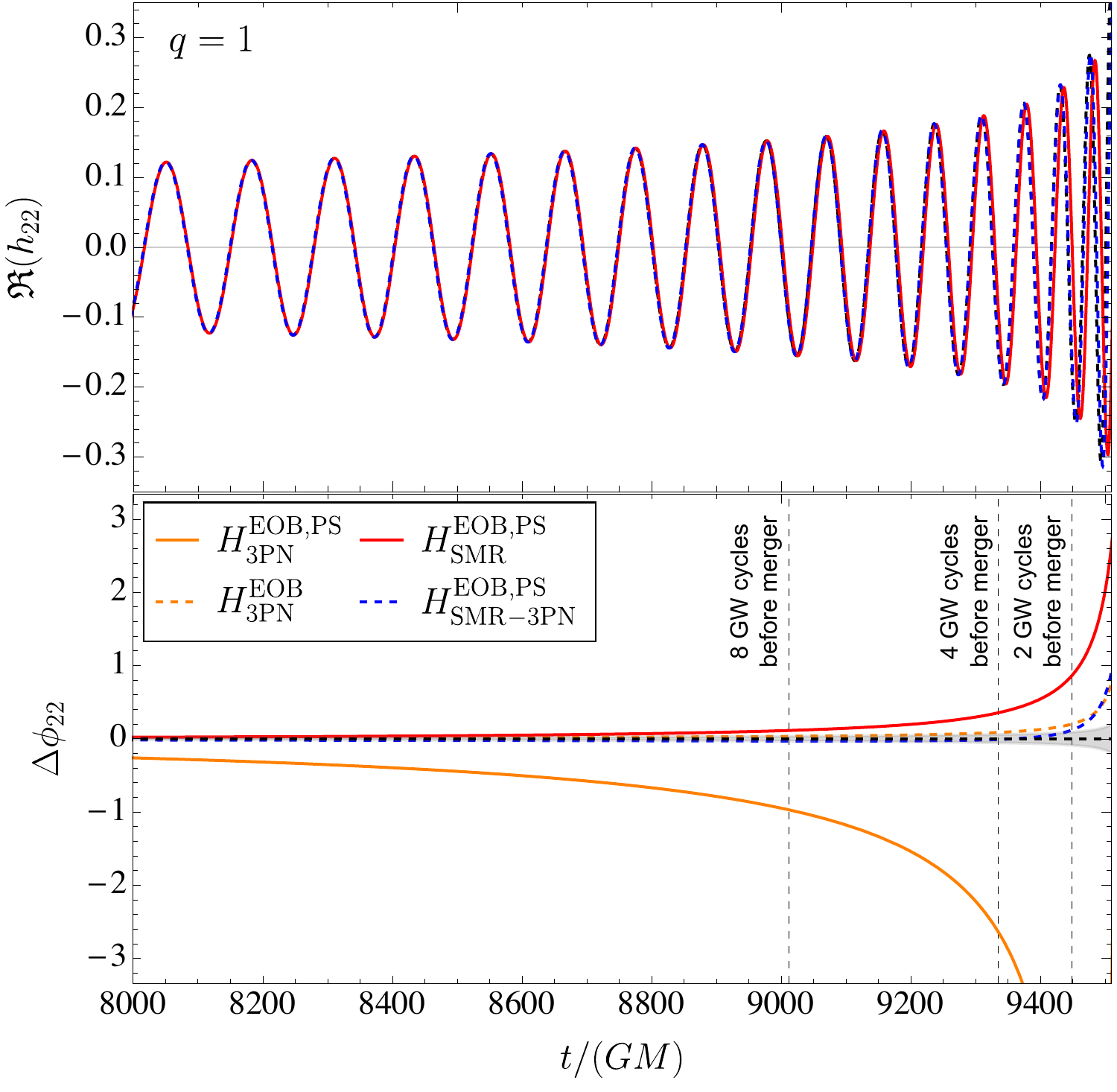}
	\hspace{12pt}
	\caption{In the top panels, the real parts $\mathfrak{R}(h_{22})$ of the ($\ell,m$)=(2,2) mode EOB waveform for the SMR, SMR-3PN models are shown and compared to the NR waveforms (in dashed-black, overlapping with the EOB waveforms up to few GW cycles to merger). In the lower panels, the dephasing of SMR and PN EOB models from the NR simulations is calculated. Also shown are the times corresponding to 8, 4 and 2 GW cycles before NR merger. Adapted from Ref.~\cite{Antonelli:2019fmq}. }
	\label{fig:wavephasenew}
\end{figure*}
\begin{figure*}[t]
	\includegraphics[width=.5\columnwidth]{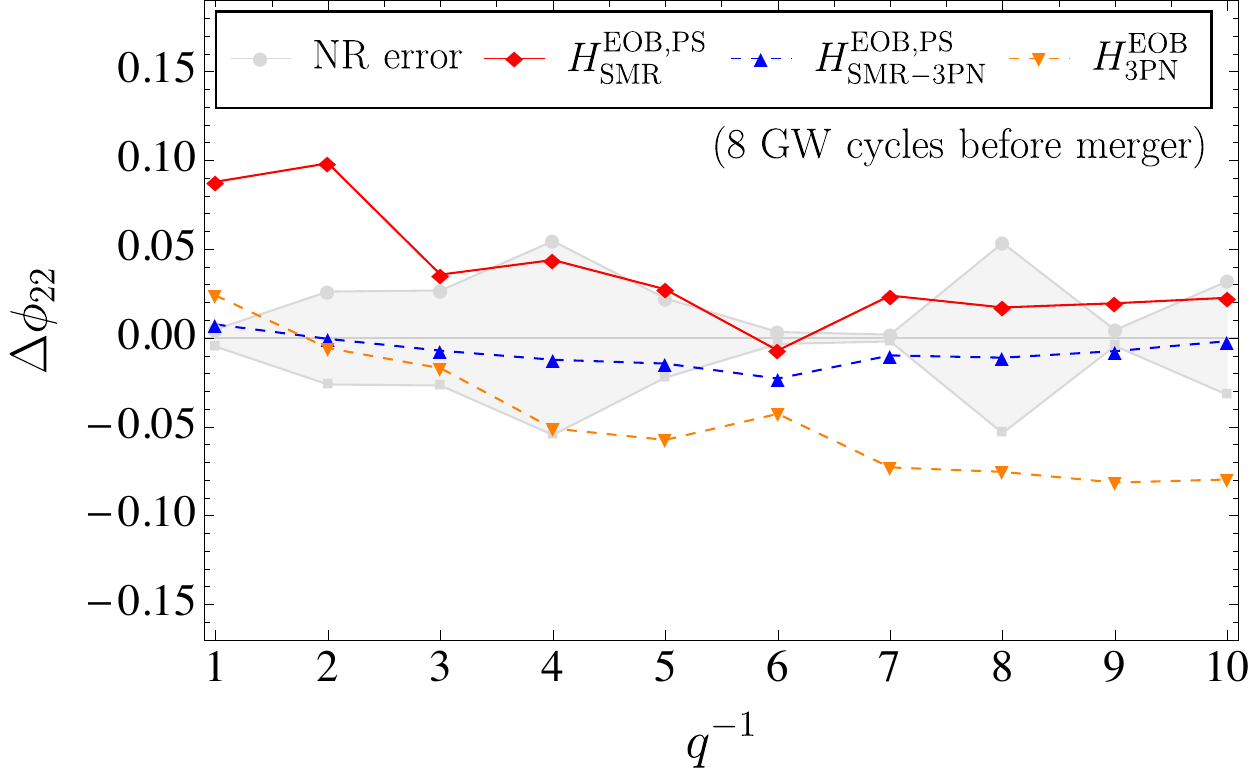}
	\includegraphics[width=.5\columnwidth]{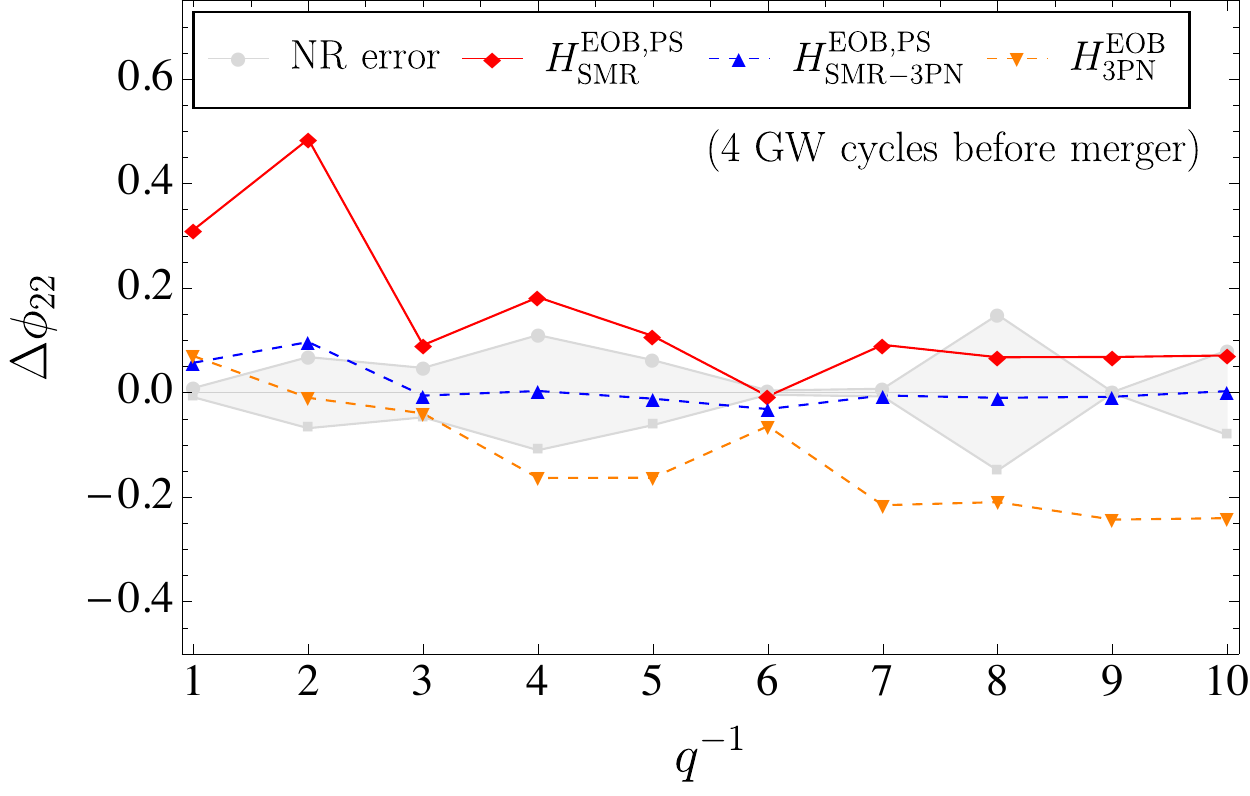}
	\caption{We compare the dephasing of $H_{\text{SMR}}^{\text{EOB,PS}}$, $H_{\text{SMR-PN}}^{\text{EOB,PS}}$ and $H_{\text{3PN}}^{\text{EOB}}$ after they have been aligned with the NR simulations from Table~\ref{table:simulations}. For each $q$, we snapshot the dephasing of the EOB models and the NR simulation at a time corresponding to 4 and 2 orbits before the merger of the binary system in the NR simulation. From Ref.~\cite{Antonelli:2019fmq}. }
	\label{fig:dphivsqnew}
\end{figure*}
\begin{align}\label{eq:wavs}
&h^\text{NR}_{22}=A_\text{NR}(t)e^{i\phi_\text{NR}(t)}\,,\\
&h^\text{EOB}_{22}=A_\text{EOB}(t+\Delta t)e^{i[\phi_\text{EOB}(t+\Delta t)+\Delta\phi]}\,.
\end{align} 
In Fig.~\ref{fig:wavephasenew}, we show results for the comparison between waveforms (upper panel) and phases (lower panel) of NR simulations and EOB models (with and without SMR information) for comparable mass ratios (see Fig.~\ref{fig:wavephase} for the $q=1/10$ case).
For clarity, the upper panel only includes the $H_{\text{SMR}}^{\text{EOB,PS}}$ and $H_{\text{SMR-3PN}}^{\text{EOB,PS}}$ models that contain SMR information, and the NR simulation. We show here the real parts of Eq.~\eqref{eq:wavs}, from which we infer that the SMR models do not accumulate a significant amount of dephasing in the $\sim$20 orbital cycles in which the comparison is performed, showing very good agreement. In the lower panel, we quantify this agreement by assessing the dephasings of SMR models from NR, and comparing them to those from 3PN models in both the DJS and PS gauges that have been aligned in the same manner. 
Interestingly, $H_{\text{SMR}}^{\text{EOB,PS}}$ and $H_{\text{SMR-3PN}}^{\text{EOB,PS}}$ fare much better than the 3PN model in the same gauge, $H_{\text{3PN}}^{\text{EOB,PS}}$, even if we are considering equal masses. Their dephasings are comparable to those of the EOB 3PN DJS Hamiltonian ($H_{\text{3PN}}^{\text{EOB}}$), which we recall is written in the gauge used in the \texttt{SEOBNR}~\cite{Pan:2011gk,Taracchini:2013rva,Bohe:2016gbl,Cotesta:2018fcv} and \texttt{TEOBResumS}~\cite{Nagar:2018zoe} families of models. In the $q=1/10$ case, they have a smaller dephasing than any other PN model considered in this study, including $H_{\text{3PN}}^{\text{EOB}}$, see Fig.~\ref{fig:wavephase}. 
The above results seem to corroborate the main findings of the binding-energy comparison above and of Refs.~\cite{Barausse:2011dq,Tiec:2013twa}. A more thorough analysis is in order to assess the robustness of this important result. For instance, we can study how the dephasing of the above models varies as a function of the mass ratio, employing 10 NR simulations for quasi-circular nonspinning binaries with mass ratios from $q=1$ to $q=1/10$ from the SXS catalog (the details of which can be found in Table~\ref{table:setofNR} of \chap~\ref{chap:three}). When comparing different NR simulations, one should keep in mind their different lengths. Moreover, employing the \emph{same} time windows for all simulations in the alignment procedures would result in encompassing different numbers of GW cycles $N_\text{GW}$. Both of the above features tend to make comparisons less clear and, ultimately, less trustworthy.
To keep both parameters under control, we align our models with time windows that are dictated by the number of GW cycles to merger $\Delta N_\text{GW}(t) \equiv N_\text{GW}(t)-N_\text{GW}^\text{merg}
$ of the NR simulations. That is, for each mass ratio we fix a different time-window [$t^\text{alig}_1,t^\text{alig}_2$], corresponding to the \textit{same} interval of cycles to merger [$\Delta N_\text{GW}(t^\text{alig}_1)$, $\Delta N_\text{GW}(t^\text{alig}_2)$]. 
A first advantage of choosing alignment windows like this is that they depend on the position of the NR merger (peak of $h_{22}^{\text{NR}}$), which is a quantifiable feature of every NR simulation.
Moreover, this choice allows us to assess trends across the mass ratios fairly, since the waveforms thus aligned are compared in the same range of GW cycles.\footnote{The only trade-off is that the same number of cycles at smaller mass ratios imply regions of stronger gravity. We would expect PN models to degrade in accuracy there, with SMR models improving instead. This is not a problem for our comparison, and in fact it highlights the point that the PN models are fundamentally limited to weak fields, while SMR ones are not.}
In Fig.~\ref{fig:dphivsqnew}, we plot the dephasing for the three models that perform best in Fig.~\ref{fig:wavephasenew}-- $H_{\text{3PN}}^{\text{EOB}}$, $H_{\text{SMR}}^{\text{ EOB,PS}}$ and $H_{\text{SMR-3PN}}^{\text{EOB,PS}}$-- which we align between 34 and 24 GW cycles before merger and compare in the last 24.
For every simulation, we calculate the dephasing 8 and 4 GW cycles before merger to show the robustness of the trends.
We notice that the 3PN EOB waveform in the DJS gauge starts degrading in accuracy as the mass ratio is decreased, while the SMR and SMR-3PN ones improve, as expected. Remarkably, for most $q$'s, the SMR-3PN model only dephases by a few hundredths of a radian
up to a 4 GW cycles before merger, including in the equal-mass ratio case.  
We therefore conclude that the good agreement between EOBSMR models and NR simulations in Fig.~\ref{fig:wavephasenew} was not accidental, and is most pronounced when SMR and PN information are synergetically included in the same waveform (especially for comparable masses). 
The analysis suggests that EOBSMR models could potentially form a new basis for EOB models for arbitrary mass ratios, from the equal-mass cases detected by LIGO to the extreme-mass-ratio cases with LISA.

\section{Assessing biases from waveform mismodelling and confusion noise}
\label{sec:PE}

The ultimate goal of approximations to the two-body problem and the synergies we have discussed thus far is to reach accurate waveform models. These are needed to extract information out of GW detections in a robust manner, as free as possible from misinterpretations. For this reason, there is a strong connection between waveform modelling and data analyses. The present section is dedicated to reviewing the basics of the detection and parameter estimation (PE) techniques used in GW astrophysics, with a particular focus on the role of (and need for) accurate waveform models. We discuss the problem of mismodelling biases, especially in the context of future detectors, in which the problem will be exacerbated by the presence of confusion noise from multiple overlapping signals.

\subsection{Detectors}
\label{sec:detectors}

Gravitational-wave detectors can be idealized as a set of test masses. In the \emph{proper detector frame}, in which an origin is fixed and coordinates are specified through an appropriate rigid ``ruler'' measuring the displacement of these test masses due to the passage of GWs, it can be found (see Sec.~1.3 of Maggiore's book~\cite{Maggiore:1900zz}) that the deviation $\zeta^\alpha$ between nearby geodesics $x^\alpha$ and $x^\alpha +\zeta^\alpha$ is
	\begin{equation}\label{eq:geo_dev}
	\ddot\zeta^\alpha = \frac{1}{2}\ddot h^{\text{TT}}\zeta^\alpha\,.
	\end{equation}
This can be further shown to amount to a perturbation of the masses through a Newtonian force $F= m 	\ddot\zeta^\alpha$, whose effect can be measured.
Several concepts for GW detectors have been developed, including resonant bar detectors~\cite{Pizzella:2016ooh}, atomic interferometers~\cite{Howl:2021giy} and pulsar-timing arrays~\cite{Detweiler:1979wn} (which stack millipulsar measurements in what is effectively a galactic-scale interferometer). The LIGO-Virgo network is based on laser interferometry~\cite{TheLIGOScientific:2014jea,Buikema:2020dlj,Davis:2021ecd}. 
Taking the LIGO configuration as  an example, a laser beam is shot towards a beam splitter, which forms two new beams that move along orthogonal arms cavities of $L\sim4$km in length. Test-mass mirrors are hung by wire at the end of each cavity. 
After several interactions of the beam with the mirrors, the two beams meet again at the beam splitter. If GWs are passing by, test masses are perturbed and light leaks out of the beam splitter depending on the relative phase between the beams, reaching a photodector that measures its intensity and, along with it, the signal.

In a realistic scenario, measuring the variation in intensity from a signal is made more difficult by the presence of noise~\cite{Weiss_detector,TheLIGOScientific:2014jea}. 
The advanced LIGO and Virgo detectors are mainly sensitive to frequencies $20\text{Hz}\lesssim
f \lesssim 1\text{kHz}$. Below $f\sim 20$ Hz, they are limited by seismic noise, i.e., noise in the gravitational field due to seismic waves propagating on Earth's crust. Above $f\sim 100$ Hz, detectors are limited by shot noise generated by fluctuations in the number of photons detected. 
These sources of noise will still be the main limitations for updates of current networks and future third-generation ground-based detectors (alongside quantum noise from radiation pressure). Still, the Cosmic Explorer~\cite{Reitze:2019iox} and Einstein Telescope~\cite{Punturo:2010zz} will improve the sensitivity of LIGO and Virgo by roughly an order of magnitude in the sensitivity bucket $1\lesssim f/\text{Hz}\lesssim 1000$, granting detections in the thousands per year~\cite{Samajdar:2021egv,Pizzati:2021gzd,Himemoto:2021ukb,Relton:2021cax}. They will also carve out a currently inaccessible region of the frequency band $\mathcal{O}(\text{few})\lesssim f/\text{Hz}\lesssim 10$. The spaceborne LISA detector~\cite{Audley:2017drz}, to be launched in the 2030s, also relies on laser interferometry, and the use of three spacecrafts creating a configuration of 6 ``arms'' , 
each carrying test masses free-falling along their geodesics.
LISA operates at a so-far completely unexplored frequency range $10^{-4}\text{Hz}\lesssim f\lesssim 1\text{Hz}$ (corresponding to binaries with much higher masses than currently detected with the LIGO-Virgo network), with a sensitivity bucket around $f\sim 10^{-3}\text{Hz}$, and must confront different sources of noise~\cite{Armano:2009zz}. 

\subsection{Detection techniques}
\label{sec:detection}

The data stream $\vec d(t)$ coming out of GW detectors is a time series $\vec d(t)$ composed of a detector noise component $n(t)$ and a potential signal $\vec h(t;\boldsymbol{\theta})$ with parameters $\boldsymbol{\theta}$, 
\begin{equation}\label{data_stream}
\vec d(t)= \vec n(t)+\vec h(t;\boldsymbol{\theta})\,.
\end{equation}
Strains are below the noise level for current GW searches, $|h|\ll |n|$. The noise component in~\eqref{data_stream} is assumed to be a stationary random process following a Gaussian distribution, and additive with respect to the potential signal in the data. Under these assumptions, the probability of a particular noise realization is
\begin{equation}\label{noise_prob}
p(\vec n)\propto \exp\left[-\frac{1}{2}(\vec n|\vec n)\right],
\end{equation}
where $(\cdot|\cdot)$ defines the inner product in Fourier space, with $\hat a(f)$ the Fourier transform of $\vec a(t)$~\cite{Finn:1992wt},
\begin{equation}\label{innprod}
(\vec a|\vec b)=4\, \text{Re}\int_0^\infty df\, \frac{\hat a^*(f)\hat b(f)}{S_n(f)}\,.
\end{equation}
The quantity $S_n(f)$ is the power spectral density (PSD) of the detector, corresponding to the auto-correlation function of noise~\cite{Finn:1992wt,Flanagan:1997kp},
\begin{equation}\label{PSD}
S_n(f) = 2\int_{-\infty}^\infty \langle n(t)n(t+\tau)\rangle e^{-2i\pi f\tau}\,.
\end{equation}
In a Bayesian setting, one may reduce the problem of detecting signals from the data to comparing a hypothesis that the data stream $\vec d(t)$ only contains noise $\vec n(t)$ (hypothesis $\mathcal{H}_0$) to the hypothesis that it also contains a signal (hypothesis $\mathcal{H}_1$)~\cite{Finn:1992wt}. We use Bayes theorem
\begin{equation}
\label{Bayes}
p(\boldsymbol{\theta}|\vec d,\mathcal{A}) =\frac{ p(\vec d|\boldsymbol{\theta},\mathcal{A}) p(\boldsymbol{\theta},\mathcal{A})}{p(\vec d,\mathcal{A})}\,,
\end{equation}
where $p(a|b,\mathcal{A})$ stands for the conditional probability that ``a'' depends on ``b'' and all the assumptions $\mathcal{A}$ that underlie the statistical analysis (the conditional quantities are separated by commas). Bayes theorem relates the likelihood $p(\vec d|\boldsymbol{\theta},\mathcal{A})$, parameter priors $p(\boldsymbol{\theta},\mathcal{A})$, and model evidence $p(\vec d,\mathcal{A})$ to give the posterior probability $p(\boldsymbol{\theta}|\vec d,\mathcal{A})$ that the parameters $\boldsymbol{\theta}$ depend on the given realization of the data.
According to the Neyman-Pearson lemma, the optimal statistic for testing a simple hypothesis $\mathcal{H}_0$ (say, no detection) versus $\mathcal{H}_1$ (say, positive detection) is the ratio of likelihoods~\cite{Neyman:1933wgr}, $p(\vec d|\mathcal{H}_1,\mathcal{A})/p(\vec d|\mathcal{H}_0,\mathcal{A})$,
which can further be expanded using Eq.~\eqref{noise_prob} for the (null) hypothesis $p(\vec d|\mathcal{H}_0,\mathcal{A})$, and the Whittle likelihood~\cite{whittle:1957}
	\begin{equation}\label{likelihood}
	p(\vec d|\boldsymbol\theta)\propto\, \exp\bigg[-\frac{( \vec d-\vec h(t;\boldsymbol\theta)| \vec d- \vec h(t;\boldsymbol\theta))}{2}\bigg]\,,
	\end{equation}
for the (signal) hypothesis $p(\vec d|\mathcal{H}_1,\mathcal{A})$. The ratio of likelihoods then becomes
	\begin{equation}\label{step2}
	\frac{p(\vec d|\mathcal{H}_1,\mathcal{A})}{p(\vec d|\mathcal{H}_0,\mathcal{A})} = \exp(\vec d|\vec h(t;\boldsymbol\theta))\exp\left[-\frac{1}{2}(\vec h(t;\boldsymbol\theta)|\vec h(t;\boldsymbol\theta))\right]\,.
	\end{equation}
	As a result of the above, we have that the probability of detection $p(\mathcal{H}_1|\vec d,\mathcal{A})$ increases when $(\vec d(t) | \vec h(t;\boldsymbol{\theta}))$ increases, implying that 
	the latter can be used as a proxy for the probability of detection (including in the choice of thresholds to claim any such occurrences). 
	Because of this, the inner product $(\vec d(t) | \vec h(t;\boldsymbol{\theta}))$ is central to the technique of matched filtering, which we now describe.
	We first normalize $(\vec d(t) | \vec h(t;\boldsymbol{\theta}))$ defining
	the signal-to-noise ratio (SNR) $\rho$,
	\begin{equation}\label{eq:SNR}
	\rho \equiv \frac{(\vec d(t) | \vec h(t;\boldsymbol{\theta}))}{\sqrt{(\vec h(t;\boldsymbol{\theta}) | \vec h(t;\boldsymbol{\theta}))}}\,.
	\end{equation}
	The numerator of Eq.~\eqref{eq:SNR} is such that the detector output $\hat d^*(f)$ is filtered by the \emph{Wiener filter}, $\hat W(f)\equiv \hat h(f;\boldsymbol{\theta})/S_n(f)$,
	\begin{equation}
	(\vec d(t) | \vec h(t;\boldsymbol{\theta})) = 4\, \text{Re}\int_0^\infty df\, \frac{\hat d^*(f) \hat h(f;\boldsymbol{\theta})}{S_n(f)} \equiv 4\, \text{Re}\int_0^\infty \hat W(f) \hat d^*(f) df\,.
	\end{equation}
	The choice of the Weiner filter implies that the SNR is maximized when $\hat d(f) = \hat h(f)$. The maximized SNR is also known as the \emph{optimal} SNR,\footnote{Notice that the optimal SNR can also be defined as the expected value of Eq.~\eqref{eq:SNR}, as it corresponds to the (most likely, expected) case in which the noise realization in the data vanishes. The optimal SNR is the one used and quoted in the rest of the present \chap and in the last \chap of the thesis.} $\rho^2_\text{opt} \equiv (\vec h(t;\boldsymbol{\theta}) | \vec h(t;\boldsymbol{\theta}))$.
	The signal $\hat h(f;\boldsymbol{\theta})$ that filters the detector output is a \emph{template} depending on parameters $\boldsymbol{\theta}$ (that remain unknown at the stage of GW searches), which is drawn from a discrete set of pre-computed values that span the parameter space (the \emph{template bank})~\cite{Usman:2015kfa}. During the GW search process, the highest SNR in the template bank is compared to the SNR one would expect without noise. If the former is higher than a pre-established threshold, above which it is unlikely that the SNR is due to noise only, a ``trigger'' (candidate GW event) is recorded. In realistic searches, given the discreteness of the template bank, one can only hope to get ``close enough'' to a signal with a template. In practice, one requires that the \emph{faithfulness} between the maximum SNR attainable for a signal $\rho_\text{max} = \rho(\boldsymbol{\theta}_\text{max})$ and the maximum SNR found in the template bank $\rho_\text{max}^{\text{bank}}= \rho(\boldsymbol{\theta}_\text{max}^{\text{bank}})$ is above a certain value. The faithfulness $F$ is defined by $\rho_\text{max}^{\text{bank}} \approx \rho_\text{max} [1-F(\vec h(t;\boldsymbol{\theta}_\text{max}^\text{bank}),\vec h(t;\boldsymbol{\theta}_\text{max})]$, where $F(\vec h(t;\boldsymbol{\theta}_\text{max}^\text{bank}),\vec h(t;\boldsymbol{\theta}_\text{max}))$ is an overlap maximised over time of coalescence $t_c$ and phase $\phi_0$,
	\begin{equation}
	F
	= \text{max}_{t_c,\phi_0}\frac{(\vec h(t;\boldsymbol{\theta}_\text{max}^\text{bank})|\vec h(t;\boldsymbol{\theta}_\text{max}))}{\sqrt{(\vec h(t;\boldsymbol{\theta}_\text{max}^\text{bank})|\vec h(t;\boldsymbol{\theta}_\text{max}^\text{bank}))(\vec h(t;\boldsymbol{\theta}_\text{max})|\vec h(t;\boldsymbol{\theta}_\text{max}))}}\,.
	\end{equation}
	The faithfulness attains its maximum value at $F(\vec h(t;\boldsymbol{\theta}_\text{max}^\text{bank}),\vec h(t;\boldsymbol{\theta}_\text{max}))=1$, though this is not reachable in practice. In LIGO-Virgo searches, a value $0.97$ is considered sufficient to establish that a template is ``close enough'' to the underlying signal in the data~\cite{Sathyaprakash:1991mt,Owen:1995tm,DalCanton:2017ala}. Searches must also deal with other issues, for instance, noise glitches that behave as triggers and need more sophisticated analysis strategies to be dealt with~\cite{Allen:2004gu,Babak:2012zx}. However, the basic description of matched filtering presented above already clearly highlights the importance of having accurate and faithful templates to avoid missing signals in the data.

\subsection{Source characterization}
\label{sec:sourcechar}

Once a trigger has been established to be a GW event, one may begin characterizing the potential signal through PE studies aimed at inferring its parameters $\boldsymbol{\theta}$. Under the gaussian and stationary noise approximation, the probability~\eqref{noise_prob} can be rewritten using \eqref{data_stream} as the Whittle likelihood~\eqref{likelihood}.
The goal of PE studies is to maximize this, finding the parameters $\boldsymbol\theta_{\text{bf}}$ that best fit (``bf'') the signal in the data. Maximizing the Whittle likelihood implies minimizing $( \vec d-\vec h(t;\boldsymbol\theta_{\text{bf}})| \vec d- \vec h(t;\boldsymbol\theta_{\text{bf}}))$, which can be done analytically within the linear signal approximation.
In this approximation, one expands the parameters around the true values, $\boldsymbol\theta_\text{bf}=\boldsymbol\theta_{\text{true}}+\Delta \boldsymbol\theta$, assuming high SNRs such that the deviation is small, $|\Delta \boldsymbol\theta/\boldsymbol\theta_\text{bf}|\ll 1$. We then have that
\begin{equation}\label{hoftheta}
\vec h(t;\boldsymbol\theta_\text{bf}) = \vec h(t;\boldsymbol\theta_{\text{true}}) + \partial_i h \Delta\boldsymbol\theta^i+\mathcal{O}(\Delta\boldsymbol\theta^2)\,,
\end{equation}
with $\partial_i h \equiv \partial h/\partial\theta^i$.
If we further include a notion of ``exact'' waveform $h_e$ as the one perfectly fitting the signal in the data and ``model'' waveform $h_m$ used in actual searches, then the likelihood is minimised when
\begin{align}\label{eq:CV_derivation}
0=&( \vec d(t)-\vec h_m(t;\boldsymbol\theta_{\text{bf}})| \vec d(t)- \vec h_m(t;\boldsymbol\theta_{\text{bf}}))\\
=&( \vec n(t) + \vec h_e(t; \boldsymbol\theta_{\text{tr}})-\vec h_m(t;\boldsymbol\theta_{\text{bf}})|  \vec n(t) + \vec h_e(t; \boldsymbol\theta_{\text{tr}})- \vec h_m(t;\boldsymbol\theta_{\text{bf}}))\nonumber\\
\approx& ( \vec n(t) + \delta \vec h(\boldsymbol\theta_{\text{tr}}) -\Delta\theta^i\partial_i\vec h_m(t;\boldsymbol\theta_{\text{bf}})|  \vec n(t) + \delta \vec h(\boldsymbol\theta_{\text{tr}})-\Delta\theta^i\partial_i\vec h_m(t;\boldsymbol\theta_{\text{bf}}))\nonumber,
\end{align}
where in the last line we have expanded $ \vec h_m$ as in Eq.~\eqref{hoftheta}, defined the residual $\delta \vec h(\boldsymbol\theta_{\text{tr}})=\vec h_e(t; \boldsymbol\theta_{\text{tr}})-\vec h_m(t;\boldsymbol\theta_{\text{tr}})$, and assumed that $\vec h_m (t;\boldsymbol\theta_{\text{tr}})\approx \vec h_m (t;\boldsymbol\theta_{\text{bf}})$~\cite{Cutler:2007mi}. Plugging the above expression in~\eqref{likelihood}, one can rearrange it to obtain
\begin{equation}
p(\vec d|\boldsymbol\theta)=\sqrt{\frac{|\Gamma|}{2\pi}}\exp\left(-\frac{1}{2} \sum_{i,j} b^i\, \Gamma_{ij}(\boldsymbol\theta_\text{bf})\, b^j\right)\label{lik}\,.
\end{equation}
Here we have defined the Fisher matrix~\cite{Finn:1992wt}
\begin{equation}
\Gamma_{ij}(\boldsymbol\theta_\text{bf}) =\left(\frac{\partial  \vec h_m}{\partial \theta^i}\bigg|\frac{\partial  \vec h_m}{\partial \theta^j}\right)\bigg|_{\boldsymbol{\theta}= \boldsymbol{\theta}_\text{bf}} \label{FM}\,,
\end{equation}
which plays the role of the variance-covariance matrix and allows one to estimate parameter errors with
\begin{equation}\label{stime}
\Delta\theta^i = \sqrt{(\Gamma^{-1})^{ii}} \qquad\text{(no summation implied)}\,,
\end{equation}
and the ``bias vector''
\begin{equation}
b^i = \Delta \theta^{i} - [\Gamma^{-1}(\boldsymbol\theta_\text{bf})]^{ik}(\partial_{k}\vec h_m (\boldsymbol\theta_\text{bf})\, |\vec n+\delta\vec h(\boldsymbol\theta_\text{tr}))\label{bvec}\,.
\end{equation}
The role of this vector is to shift the peak of the parameters away from their true values. The overall shift can be split in two contributions~\cite{Flanagan:1997kp,PhysRevD.71.104016,Cutler:2007mi}
\begin{align}
&\Delta \theta^{i}_\text{noise} = (\Gamma^{-1})^{ik}(\partial_{k}\vec h_m|\vec n)\label{eq:delta_theta_noise}\,,\\
&\Delta \theta^{i}_\text{sys}  = (\Gamma^{-1})^{ik}(\partial_{k}\vec h_m|\delta \vec h)\label{eq:delta_theta_syst}\,.
\end{align}
The first contribution is purely due to noise, and must be mitigated with sophisticated statistical techniques. One example are Markov-Chain Monte Carlo (MCMC) methods.
The most used inference pipelines in the context of GW astrophysics are \texttt{LALinference} package~\cite{Veitch:2014wba} and \texttt{bilby}~\cite{Ashton:2018jfp}. In \chap~\ref{chap:six}, we use the publicly-available \texttt{emcee} package to perform MCMC analyses~\cite{ForemanMackey:2012ig}. 
The second contribution in Eq.~\eqref{eq:delta_theta_syst} is the systematic (purely mismodelling) contribution mentioned above, which instead must be mitigated with better waveform predictions.

Under the assumption of zero-mean gaussian noise, and within the linear signal domain, $\Delta \theta^{i}_\text{noise}$ can be thought of in a statistical sense as a random variable $\Delta\widehat{\theta}=\Delta\theta_\text{noise}$ with covariance matrix given by the inverse of the Fisher matrix, $\text{Cov}(\Delta\widehat{\theta},\Delta\widehat{\theta})=\Gamma^{-1}$. As a result, $\Delta\theta_\text{noise}$ contains the same information as the statistical errors estimated from the Fisher matrix, $\Delta\theta_\text{noise}\sim\Delta\theta^i_\text{stat}$, which are easier to obtain. One could use this fact and Eq.~\eqref{eq:delta_theta_syst} to devise a criterion to estabish whether theoretical biases are ``significant''. If the parameter error from waveform mismodelling $\Delta\theta_\text{sys}$ is such that $\Delta\theta_\text{sys}\geq \Delta\theta_\text{stat}$, therefore exceeding the $1\sigma$ width predicted by the covariance matrix, then the parameter is said to be significantly biased.  Alternatively, one may rewrite the criterion defining a function of (generic) parameter shifts $\Delta\theta^i$
\begin{equation}\label{eq:Rdef}
	\mathcal{R}(\Delta\theta^i)\equiv\left|\frac{\Delta\theta^i}{\Delta\theta^i_\text{stat}}\right|\,,
\end{equation}
and requiring that $\mathcal{R}(\Delta\theta^i)<1$ for unbiased results. 
In this thesis we choose to denote the procedure that leads to Eqs.~\eqref{eq:delta_theta_noise} and~\eqref{eq:delta_theta_syst} as the Cutler-Vallisneri (CV) formalism after the authors of Ref.\cite{Cutler:2007mi}, and $\mathcal{R}(\Delta\theta^i_{\text{sys}})<1$ as the CV criterion. We acknowledge that Eqs. \eqref{eq:delta_theta_noise} and \eqref{eq:delta_theta_syst} were known before the work of Cutler and Vallisneri from Refs.~\cite{Flanagan:1997kp} and~\cite{PhysRevD.71.104016}, though the implications were not studied there. In what follows, we extend the CV formalism below to provide biases in the presence of unmodelled and modelled foregrounds of GW signals, as required to analyze data series from future detectors.

\subsection{Predicting biases from multiple GW signals}
\label{sec:5paper}
 
The CV criterion is well suited to data analyses with current detectors, since it applies for signals that are well separated from each other in which the data stream is comprised of a signal $\vec h(t;\boldsymbol{\theta})$ buried in noise $\vec n(t)$. 
In this situation, one does not expect any other biases other than the waveform mismodelling for the source of interest.\footnote{Mismodelling biases can be further split into biases from using incorrect waveform models, from noise mismodelling (glitches), and calibration errors. We group these into a generic mismodelling error. }
However, the criterion fails for the data analyses with future spaceborne and ground-based detectors such as LISA, the Cosmic Explorer, and the Einstein Telescope. The reason is that, while one needs to dig signals out of noise with current detectors, future detectors will require one to dig signals out of a \emph{foreground} of other signals. Clearly, in the latter case one may not assume that signals are well separated from each other, implying that other potential sources of biases may arise from the presence of \emph{multiple overlapping} signals in the data stream. We address two main questions relevant for future detectors.
\begin{enumerate}
	\item How can we predict the biases on the parameters of a source of interest in the presence of \emph{unmodelled foregrounds} of other sources?
	\item How can we predict the biases on the parameters of a source of interest due to the \emph{inaccurate fitting} of other sources?
\end{enumerate}
Regarding the first question, the data stream~\eqref{data_stream} from which the CV criterion is deduced
only assumes that the noise $\vec n(t)$ is stationary and Gaussian. In sections~\ref{sec:detection} and~\ref{sec:sourcechar}, this is taken to be the instrumental noise that is folded in the PSD, e.g., $S_{\text{n}}(f)$ in Eq.~\eqref{PSD}. 
With future detectors, however, the overall noise will have to include the confusion noise from unmodelled foreground of other GW signals~\cite{Crowder_2007,B_aut_2010,Robson:2017ayy,Roebber:2020hso,Korol:2020hay,Samajdar:2021egv,Pizzati:2021gzd}. 
The correct way to treat such confusion noise would be to combine the instrumental and a random confusion noise $\Delta  \vec H_{\text{conf}}$ into a single noise term, $\vec n + \Delta  \vec H_{\text{conf}}$. Then the standard parameter estimation formalism of Sec.~\ref{sec:sourcechar} can be used, with the substitution $S_{\text{n}} (f) \rightarrow S_{\text{n}}(f) + S_{\text{conf}}(f)$ in the inner product, where $ S_{\text{conf}}(f)$ is an appropriate quantity that 
folds the confusion noise estimator $\Delta H_{\text{conf}}$ into the PSD. 
The inference uncertainties would then be given by the inverse of the Fisher matrix, $\Gamma^{-1}_{\text{n}+\text{conf}}=\Sigma_{\text{n}+\text{conf}}$, where
\begin{equation}
\Gamma_{\text{n}+\text{conf}}^{ij} =4\text{Re}\int_{0}^\infty \frac{(\partial_k \hat{h}_m(f) \partial_l \hat{h}^{\star}_m(f))}{S_{\text{n}}(f)+S_{\text{conf}}(f)}.
\label{eq:confvarnew}
\end{equation}
However, obtaining the PSD correction $S_{\text{conf}}(f)$ is hard in practice, which prompts us to look for cheaper alternatives that can still be used for exploratory future-detector studies.
One approach is to consider a random confusion-noise realization, $\Delta \vec H_{\text{conf}}$, and add it the data stream~\eqref{data_stream}
\begin{equation}\label{noisetotnew}
 \vec d(t) =  \vec h_{e}(t;\boldsymbol{\theta}_{\text{tr}}) + \vec  n(t) + \Delta  \vec H_{\text{conf}}(t;\boldsymbol\theta^{(i)})\,.
\end{equation}
We have used ``exact'' waveforms $  \vec h_{e}(t;\boldsymbol{\theta}_{\text{tr}})$ that encapsulate the true waveform.
By adding a particular noise realization, one loses the random nature of the confusion noise, which becomes a \emph{deterministic} superposition of $N$ signals
\footnote{
Treating confusion noise deterministically is somewhat unnatural, since such contribution is in principle random. However, it can be shown that it leads to conservative estimates. That is, the predictions for the biases  obtained with Eq.~\eqref{eq:bias_confnew} do not underestimate the actual biases one may have with future detectors' data.  The argument can be found in Sec.~\ref{sec:Source_Confusion_Bias}.
}
\begin{equation}\label{confres}
\Delta  \vec  H_\text{conf} (t;\boldsymbol\theta^{(i)})= \sum_{i=1}^{N}  \vec h^{(i)}_e (t;\boldsymbol\theta^{(i)})\,.
\end{equation} 
Notice that the subscript here refers to the $i^\text{th}$ source in the $N$ unresolved signals, each depending on its own set of parameters.  The size of the catalogue and the mass probability density function are chosen arbitrarily at this stage. We stress in fact that our goal is to test the formalism to predict biases. More realistic studies, which require an understanding of as-of-yet poorly constrained rates for massive BHs, are to be performed in future work. One can then follow the same routine presented in~\eqref{eq:CV_derivation}, but starting from Eq.~\eqref{noisetotnew} instead of~\eqref{data_stream}, and obtain the parameter errors from noise realizations~\eqref{eq:delta_theta_noise} and theoretical mismodelling~\eqref{eq:delta_theta_syst}, as well as an extra contribution from confusion noise (see also Sec.~\ref{sec:Source_Confusion_Bias}),
\begin{equation}\label{eq:bias_confnew}
\Delta\theta^{i}_{\text{conf}} = (\Gamma^{-1})^{ij}(\partial_j   \vec h_e| \Delta  \vec H_\text{conf}).
\end{equation}
In analogy to the statement of the CV criterion, confusion noise biases the parameter $\theta^i$ whenever $\mathcal{R}(\Delta\theta^i_\text{conf})>1$, with $\mathcal{R}$ defined in Eq.~\eqref{eq:Rdef}. 

We now generalise the data stream~\eqref{noisetotnew} to include multiple \emph{incorrectly} modelled signals, as relevant to address the second question posed above. The goal is to provide a metric to assess how much the incorrect removal of signals impacts parameter estimates of a source of interest.
We consider a data stream of the form
\begin{equation}
\vec d(t) = \vec h(t;\Theta) + \vec n(t) + \Delta \vec H_{\text{conf}},
\label{eq:general_data}
\end{equation}
with $\vec h(t;\boldsymbol\Theta)$ now collecting all the signals that are fitted out of the data,\footnote{For ease of presentation, all signals here are assumed to be of the same type, and modelled with the same template. See Sec.~\ref{sec:Source_Confusion_Bias} for a more general argument that does not rely on these assumptions. }
\begin{equation}
\vec h(t;\boldsymbol\Theta) =\sum_{i=1}^{N} \vec h^{(i)}_{e}(t;\boldsymbol{\theta}_i)\quad \text{for $\boldsymbol{\Theta} = \{\boldsymbol{\theta}_{1},\ldots,\boldsymbol{\theta}_{N}\}$}\,.
\end{equation}
The indices $i=[1,N]$ run from 1 to $N$ fitted sources, each depending on a different set of parameters $\boldsymbol{\theta}_i$.
In the data stream, we model signals with the (``exact'') waveform $\vec h^{(i)}_{e}(t;\boldsymbol{\theta}_i)$, though we assume we fit signals out with an approximate ``model'' waveform $\vec h^{(i)}_{m}(t;\boldsymbol{\theta}_i)$: each time a signal is fitted out, a ``residual'' (waveform error) $\delta\vec  h^{(i)}= \vec h^{(i)}_{e}(t;\boldsymbol{\theta}_i)-\vec h^{(i)}_{m}(t;\boldsymbol{\theta}_i)$ is left into the data and added to the confusion noise.
The vector of parameters $\boldsymbol{\Theta} = \{\boldsymbol{\theta}_{1},\ldots,\boldsymbol{\theta}_{N}\}$ runs through all the sources that are recognised to be present in the data. For any given parameter in $\boldsymbol{\Theta}$, there is exactly one waveform $\vec h^{(i)}_{e}(t;\boldsymbol{\theta}_i)$ in the above sum that depends on that parameter. Thus the derivatives of the signal reduce to derivatives of the specific waveform template. The combined Fisher matrix has a block structure, with the on-diagonal blocks being the Fisher matrices for the individual sources, and the off-diagonal blocks being overlaps of waveform derivatives of one source with waveform derivatives of another source. Through calculating the Fisher matrix on parameters $\boldsymbol{\Theta}$, one is able to extract precision measurements on individual parameters taking into account \emph{all} parameter correlations. The derivation of the biases is analogous to the derivation of the CV expressions~\eqref{eq:delta_theta_noise} and~\eqref{eq:delta_theta_syst}. The parameter errors can then be found to be 
\begin{equation}
\Delta\Theta^{i}=(\Gamma^{-1})^{ij}\left(\frac{\partial\vec  h^{(i)}_m}{\partial \Theta^{j}}\bigg| \vec n(t) + \delta \vec h + \Delta\vec  H_{\text{conf}} \right)\label{multiparnew}\,,
\end{equation}
with the waveform systematic errors of Eq.~\eqref{eq:delta_theta_syst} being replaced by the combination of detector and confusion noises and the residuals of \emph{all} the waveforms that have been incorrectly fitted out of the data.
The theoretical errors $\delta \vec h$ and $\Gamma_{ij}$ are given by the sum of differences between ``exact'' and ``model'' waveforms and the \emph{joint} Fisher matrix respectively,
\begin{equation*}
\delta \vec h= \sum_{i = 1}^{N}(\vec h_{e}^{(i)}(t;\boldsymbol{\theta}^{(i)}) - \vec h_{m}^{(i)}(t;\boldsymbol{\theta}^{(i)})),
\quad
\Gamma_{ij} = \left(\frac{\partial\vec  h_{m}^{(i)}}{\partial \Theta^{i}}\bigg\rvert \frac{\partial\vec  h_{m}^{(i)}}{\partial \Theta^{j}}\right).
\end{equation*}
Equation~\eqref{multiparnew} can be split into a noise induced error $\Delta\Theta^{i}_\text{noise}$, a confusion noise bias $\Delta\Theta^{i}_\text{conf}$, and a theoretical mismodelling bias $\Delta\Theta^{i}_\text{sys}$,
\begin{align}\label{eq:bias_gen}
&\Delta\Theta^{i}_\text{noise}=(\Gamma^{-1})^{ij}\left(\frac{\partial \vec h_{m}^{(i)}}{\partial \Theta^{j}}\bigg| \vec n(t)\right)\\
&\Delta\Theta^{i}_\text{conf}=(\Gamma^{-1})^{ij}\left(\frac{\partial\vec  h_{m}^{(i)}}{\partial \Theta^{j}}\bigg| \Delta \vec H_{\text{conf}} \right)\\
&\Delta\Theta^{i}_\text{sys}=(\Gamma^{-1})^{ij}\left(\frac{\partial\vec  h_{m}^{(i)}}{\partial \Theta^{j}}\bigg|\delta\vec  h \right)\,.
\end{align}
We extend the CV criterion to give significant biases whenever the combination of confusion and mismodelling biases satisfies $\mathcal{R}(\Delta\Theta^{i}_\text{conf} + \Delta\Theta^{i}_\text{sys})>1$. We conclude this discussion by briefly showing how the formalism above could be used to (qualitatively) estimate the biases one may face with future detectors. We present two examples, covering both the confusion noise and source overlap cases.  The codes used for these examples are publicly available at  \text{\url{https://github.com/aantonelli94/GWOP}}.

\subsubsection*{First example: confusion noise}

In the first example we consider, a PE study on a signal of interest is carried out in the presence of a confusion noise term $\Delta H_\text{conf}$ from unmodelled sources that are drawn from a realistic mass distribution. The (frequency-domain) data stream in this case is
\begin{equation}\label{signew}
	\hat d(f) = \hat h(f;\boldsymbol\theta_\text{tr})+ \Delta H_\text{conf}(f;\boldsymbol\theta)+ \hat n(f).
\end{equation}
Noise is modelled as a normal distribution with zero mean and variance $\sigma^{2}(f_{i}) = N_{t}S_{n}(f_{i})/4\Delta t$ (with $N_{t}$ the length of the padded signal in the time domain), obtained discretizing Eq. \eqref{PSD}. We employ the analytical fit for LISA's PSD $S_n$ from~\cite{Robson_2019}. The signal $\hat h(f;\boldsymbol\theta_\text{tr})$ is modelled directly in the fourier domain with a \texttt{TaylorF2} template, which reads 
\begin{align}
	\hat h(f) &= \mathcal{A}\left(\pi G M f\right)^{-7/6}e^{-i\psi(f)}\,,  \quad
	\mathcal{A} = -\sqrt{\frac{5}{24}}\frac{\left(G\mathcal{M}_c\right)^{5/6}}{D_\text{eff}\pi^{2/3}}\,.\nnm
\end{align}
Here, $\mathcal M_c\equiv M \nu^{3/5}$ is the chirp mass in terms of total mass and symmetric mass ratio, and $D_\text{eff}$ the effective distance.
The phase $\psi(f)$ is the 3.5PN one in Ref.~\cite{Cutler:2007mi}, with the inclusion of a spin-orbit parameter $0\leq\beta\leq 9.4$ at 1.5PN order (as defined in Ref.~\cite{Berti:2004bd}).  The ``true'' (injected) parameters for this example are $\boldsymbol{\theta}_\text{tr} = \{\log \mathcal{M}_c = 83.34, \nu = 0.210, \beta = 5.00\}$. Notice that ``$\log$'' is the natural logarithm. Together with the chosen mass-ratio, the mass $\log \mathcal{M}_c = 83.34$ corresponds to a total mass $M \sim 10^6 M_\odot$. The other parameters in the template chosen (but not sampled through) are the effective distance $D_\text{eff} = 1$Gpc, phase at coalescence $\phi_c = 0$, and time at coalescence $t_c$ given by the chirp time at 3.5PN~\cite{Allen:2005fk}. The binary is observed from $f_0=0.25\, m$Hz until the (Schwarzschild) ISCO frequency for this mass, $f_\text{max} = 2.2\, m$Hz (corresponding to 4.4 days of observation). Under these parameters, the signal has a high enough SNR with $\rho\sim 4200$ for the Fisher matrix predictions to be valid.  The confusion term $\Delta H_\text{conf}(f; \boldsymbol\theta)$ is built from a mock catalogue of $N=1000$ sources that samples from uniform distributions $D_\text{eff}\sim U[1,5]$, $\beta\sim U[0.001,9.4]$Gpc, $\nu\sim U[0.001,0.25]$, and $\phi_c\sim U[0, 2\pi]$, with $t_c$ given by the individual chirp times of the individual signals of the foreground.
The total masses of the binaries in this catalog are taken to follow a standard probability density function for massive BHs~\cite{Gair:2010bx,Gair:2010yu,Sesana:2010wy},
\begin{equation}
\frac{dN}{dM} = \frac{\alpha\, M^{\alpha - 1}}{M^{\alpha}_\text{max}-M^{\alpha}_{\text{min}}},
\end{equation}
where the masses' range is $M_\text{min}=10^4M_\odot < M< 10^7 M_\odot=M_\text{max}$ and $\alpha=0.03$ is the fit in Ref.~\cite{Gair:2010yu} to the inactive massive BHs of~\cite{Greene:2007xw}. 
The signals in the catalogue that have an SNR $\rho_\text{obs}  <  15$ are added to $\Delta H_\text{conf}$ (that is, they are considered ``missed'' by search pipelines and included in the confusion noise). For $N=1000$ events in the mock catalogue, about $N_\text{U}=\mathcal{O}(160)$ have SNRs below that threshold. 

\begin{figure*}
	\centering
	\includegraphics[height = 7.5cm, width =1.1\linewidth]{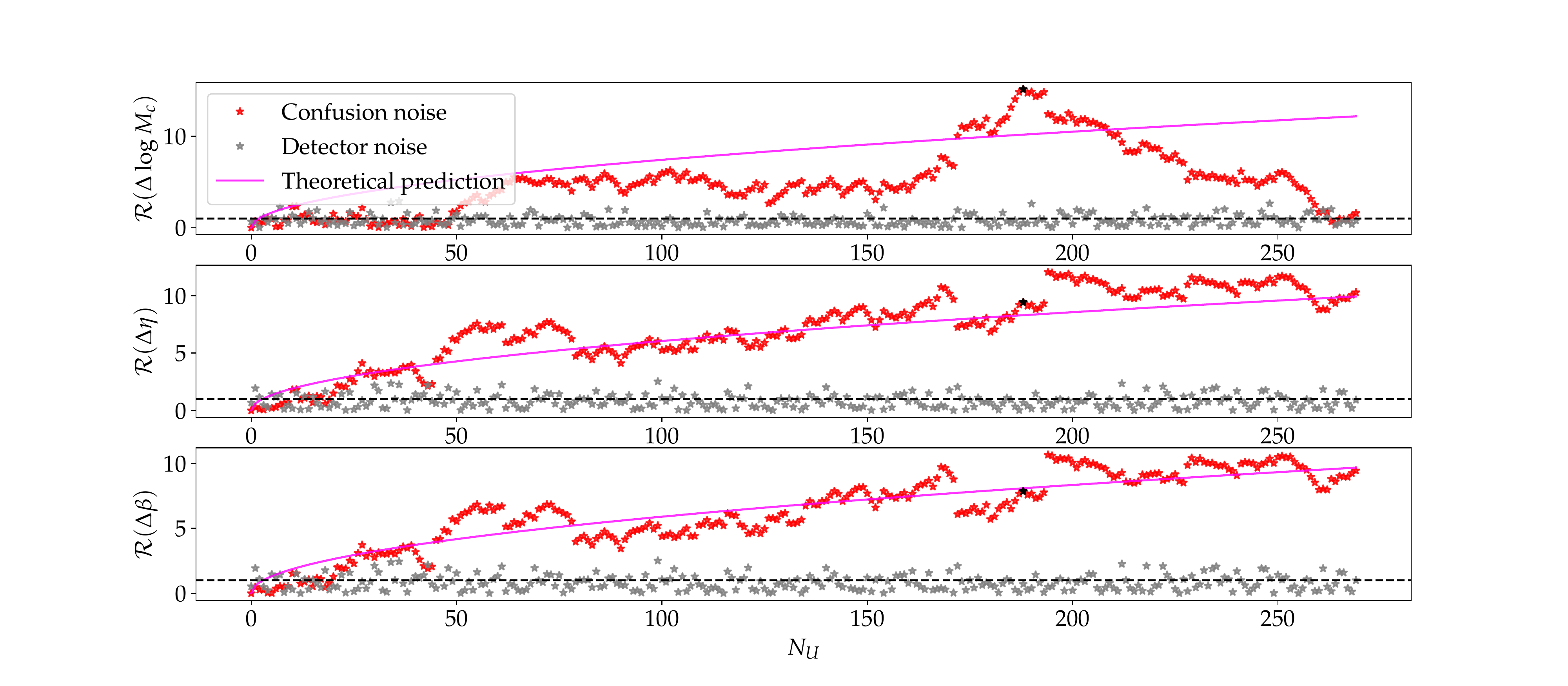}
	\caption{In red, the accumulation of bias on the parameters of the reference signal from massive BH binaries that have not been resolved, as a function of the number of unresolved sources $N_U$ (see main text). 
	In gray, the statistical errors arising from instrumental noise fluctuations. The noise is independently generated for each data set and so we expect the ${\cal R}$ values to follow a $N(0,1)$ distribution, which is consistent with what is seen in the figure. In black, the data point with the largest bias in ${\cal M}_c$, for which the results were verified using an MCMC simulation, see Fig.~\ref{fig:conf_noise_biasnew}. 
	}
	\label{fig:conf_noise_LISAnew}
\end{figure*}

\begin{figure}
	\includegraphics[width=.7\linewidth]{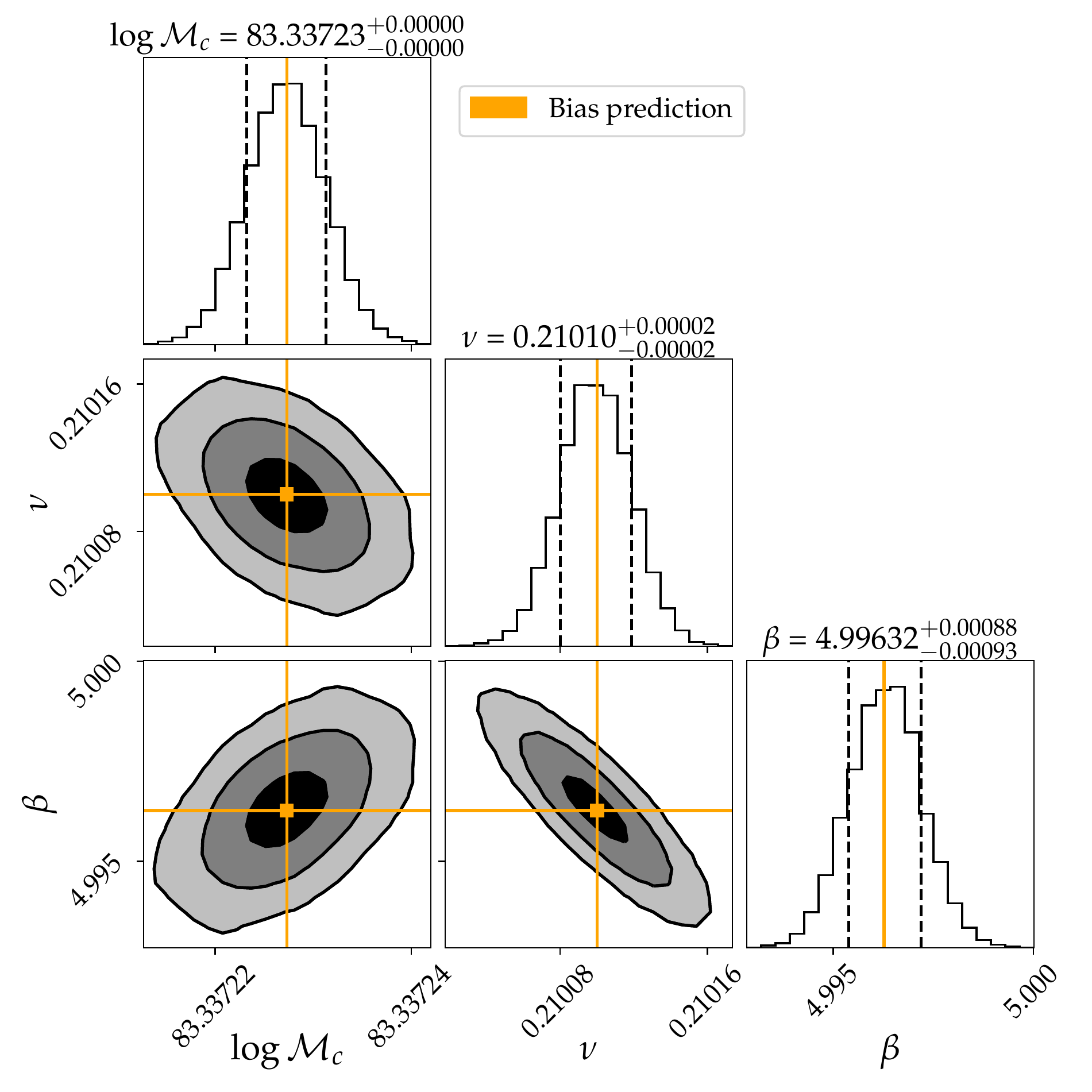}
	\centering
	\caption{
	MCMC posteriors and predictions from the Fisher formalism for the largest-bias case in Fig.~\ref{fig:conf_noise_LISAnew}.
	The panels in the diagonal are the 1D posteriors of the parameters (with 1$\sigma$ confidence interval represented by dashed lines), while the other panels are the 2D marginalized posteriors. The bias predictions (orange) are obtained adding $\Delta\theta^{i}_{\text{conf}}$ -- Eq.\eqref{eq:bias_confnew} -- to the true values, and show very good agreement with the peak of the MCMC posteriors. The black, dark gray and light gray regions in the 2D posteriors represents the 1$\sigma$, 2$\sigma$ and 3$\sigma$ intervals, respectively.
}
	\label{fig:conf_noise_biasnew}
\end{figure}

From the $\Delta H_\text{conf}$ and signal $\hat h(f,\boldsymbol{\theta}_\text{tr})$ specified above, a prediction for the inference biases from confusion noise $\Delta\theta^i_\text{conf}$ can be calculated with Eq.~\eqref{eq:bias_confnew}. In this case, the template to take derivatives of in the overlap and Fisher matrix is the signal template $\hat h(f,\boldsymbol{\theta}_\text{tr})$ itself, while the Fisher matrix is the $3\times3$ Fisher matrix from the single source of interest (since this is the only source that is modelled in the data stream~\ref{signew}). From the predictions $\Delta\boldsymbol{\theta}_\text{conf} = \{\Delta\log\mathcal{M}_c,\Delta\nu,\Delta\beta\}$, the bias ratios  $R(\Delta\theta^i_\text{conf})$ are calculated via Eq.~\eqref{eq:Rdef}.
Figure~\ref{fig:conf_noise_LISAnew} shows the accumulation of bias from foreground noise by plotting (in red) the ratio $\mathcal{R}$ as a function of $N_U$, the number of unresolved sources in the mock catalogue.\footnote{Here the data set with $N+1$ confusion sources consistently includes the same sources as the $N$ confusion sources data set, plus one additional source.} The formalism predicts significant biases ($\mathcal{R}>1$) after a handful of sources have been added. 

While more thorough analyses must be carried out with realistic situations before any conclusions are drawn from this analysis, we can assess the bounty of the Fisher predictions with independent calculations. For instance, we can confirm the analysis with MCMC techniques.  These numerical calculations are more expensive to obtain, but they present the most stringent test for the Fisher-based predictions of parameter precision and biases. In the high-SNR limit, the covariances from both Fisher and Bayesian approaches must agree. The MCMC runs we carry out sample the posterior $p(\boldsymbol{\theta}|\vec d,\mathcal{A})$ from Eq. \eqref{Bayes} using uniform priors and the Whittle likelihood \eqref{eq:whittle_likelihood}. The sampling is carried out with the publicly available \texttt{emcee} package \cite{ForemanMackey:2012ig}.
The check between the predictions of Fig. \ref{fig:conf_noise_LISAnew} and MCMC analysis can be performed for the data set that gives the largest  bias (${\cal R} \sim 8$) in chirp mass (the black data point in Fig.~\ref{fig:conf_noise_LISAnew}), as this is the case for which it is most likely that our Fisher estimates break down. The MCMC posteriors are obtained injecting the data stream~\eqref{signew}, but recovering the parameters of the source of interest assuming no confusion terms in the Whittle likelihood. The omission of the confusion term is cause for the large bias that can be checked with the analytical predictions from this section. The result of the MCMC runs and the predictions for the shift in the peak of the likelihood due to the confusion sources and noise, computed adding $\Delta\theta^i_\text{noise}$ and $\Delta\theta^i_\text{conf}$ to the true parameter values, can be found in Fig.~\ref{fig:conf_noise_biasnew}. The results show an excellent match between the position of the parameter's peaks inferred from the MCMC analysis and the bias prediction from the Fisher-based formalism. A similar agreement holds picking other random data realizations, suggesting that the Fisher-based bias predictions we obtain always match the expensive MCMC runs, at a computational cost that is negligible in comparison.

\subsubsection*{Second example: incorrect fitting of overlapping sources}

\begin{figure}
	\includegraphics[width=.7\linewidth]{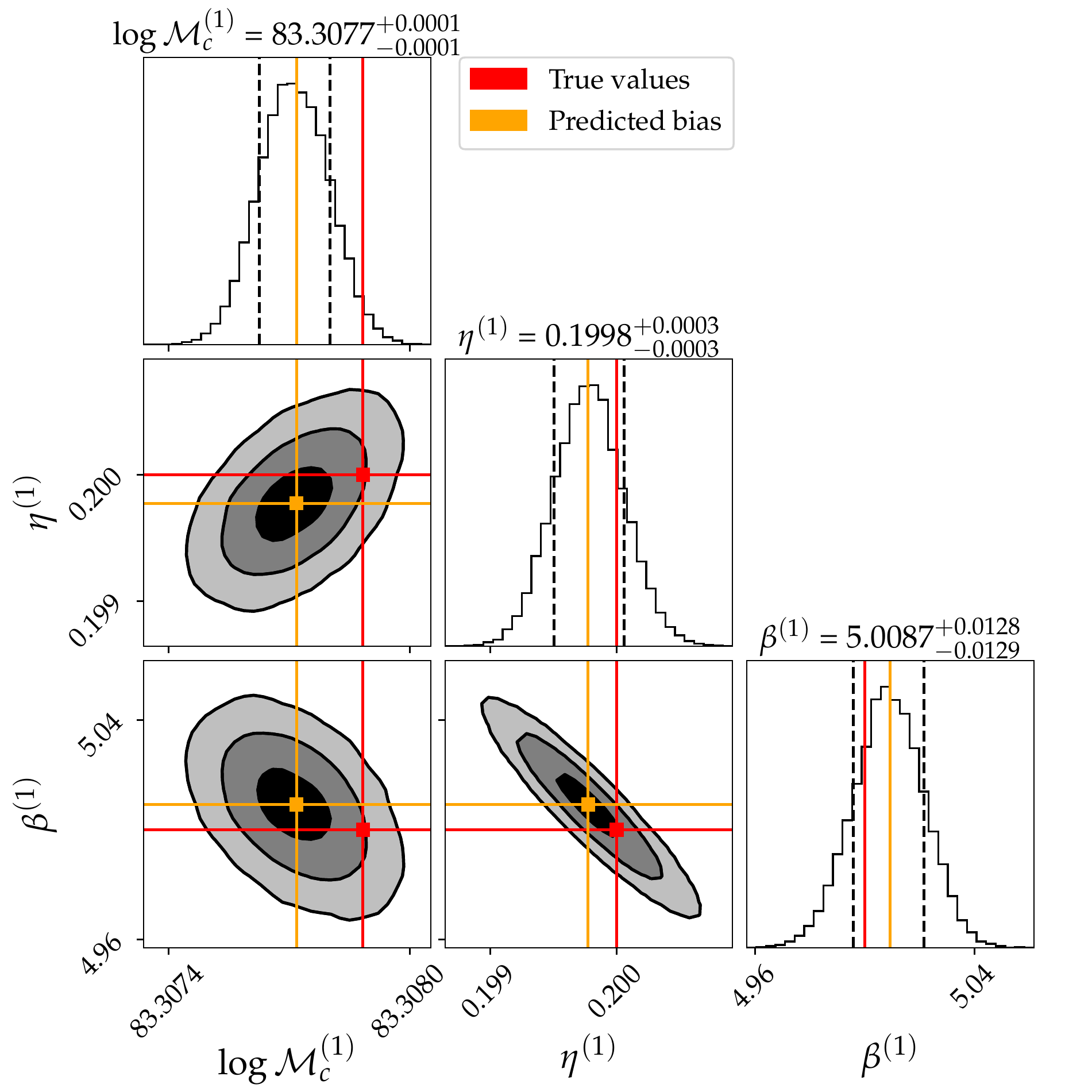}
	\centering
	\caption{ Posterior distributions for the parameters of a reference signal, computed using MCMC, when 2 mismodelled overlapping signals are removed from the data (with parameters given in Table~\ref{tab:example_2new}). The true values are reported in red, while the prediction (obtained adding the predicted parameter shift from Eq.\eqref{eq:bias_gen} to the true values) is reported in orange. The Fisher-matrix based and MCMC analyses show excellent agreement as to the location of the biased parameters. Adapted from Ref.~\cite{Antonelli:2021vwg}.}
	\label{fig:inacc_rem_biasnew}
\end{figure}

In the second and final example, we study the reliability of the method in predicting  biases on the parameters of a source of interest in the case in which several incorrectly modelled overlapping are fitted out of the data. Consider the following data stream in LISA
\begin{equation}\label{data_inc_remnew}
\hat d(f) = \hat h^{(1)}_e(f; \boldsymbol\theta^{(1)}) + \hat h^{(2)}_e(f; \boldsymbol\theta^{(2)})
+ \hat h^{(3)}_e(f; \boldsymbol\theta^{(3)}).
\end{equation}
No detector or confusion noise is included, but three sources are simultaneously fitted. We assume that the three sources are observed at the same time and overlap until each coalesce (which happens for the parameters chosen at the coalescence time $t_c$ reported in Table \ref{tab:example_2new}).
The first source $\hat h^{(1)}_e(f; \boldsymbol\theta^{(1)})$ is the signal of interest, and it depends on parameters $\boldsymbol\theta^{(1)}$ whose inference biases we want to inspect. All three sources are injected with ``exact'' waveforms $\hat h^{(i)}_e(f; \boldsymbol\theta^{(i)})$ (modelled with \texttt{TaylorF2} templates, exactly as in the previous example). However, the PE study of the latter two sources is carried out with approximate templates $h_m^{(2,3)}(f; \boldsymbol\theta^{(2,3)},\epsilon)$ modelled inserting a
auxiliary parameter $\epsilon \in [0,1]$ in the 3.5PN phase contribution as follows
	\begin{equation}\label{epsdef}
		\psi_{\text{3.5PN}}^{(\epsilon)}:=\pi  \left(\frac{77096675}{254016}+\frac{378515}{1512}\eta -(1-\epsilon)\frac{74045}{756}\eta ^2\right)\,.
	\end{equation}
The other coefficients up to 3.5PN are given, e.g., in Sec.IIIB of \cite{Cutler:2007mi}. The true PN  waveform  has $\epsilon=0$  and we will take a value $\epsilon=0.3$ in our approximate templates below.
The waveforms in the Fisher matrix and the Whittle likelihood for the MCMC analyses are the approximate templates. This situation mimicks the inaccurate removal of the two mismodelled sources, whose residuals,
\begin{equation}\label{deltah}
\delta h = \sum_{i=2}^{3} \hat h^{(i)}_e(f;\boldsymbol\theta^{(i)})-\hat h^{(i)}_m(f;\boldsymbol\theta^{(i)}, \epsilon=0.3)\,
\end{equation}
leave a potential bias on the parameters $\boldsymbol\theta^{(1)}$ of the signal of interest. The parameters for all three sources can be found in Table~\ref{tab:example_2new}. The bias predictions for the parameters (accounting for correlations) are obtained from $\Delta\Theta^i_\text{sys}$ in Eq.~\eqref{eq:bias_gen}, where the relevant parameter space for the ($9\times9$) joint Fisher matrix is $\boldsymbol\Theta =\{\boldsymbol\theta^{(1)},\boldsymbol\theta^{(2)},\boldsymbol\theta^{(3)}\}$ (with each subset $\boldsymbol\theta^{(i)}=\{\log \mathcal{M}_c^{(i)},\nu^{(i)},\beta^{(i)}\}$) and $\delta h$ is given by~\eqref{deltah}. Predictions for the biases on the first source's parameters $\boldsymbol\theta^{(1)}$ are tested in Fig.~\ref{fig:inacc_rem_biasnew} against MCMC results obtained again using \texttt{emcee}, with injected data \eqref{data_inc_remnew}, and approximate templates in the Whittle likelihood for the second and third source . Adding the parameter shifts $\Delta\boldsymbol{\theta}^{(1)}$ to the true parameters, the predictions for the biased parameters' location are obtained. In Fig.~\ref{fig:inacc_rem_biasnew}, the predicted biases shows excellent agreement with the shifted posteriors of the MCMC analyses. 

\begin{table}
	\centering
	\caption{Parameter configurations for the second example.  $\rho_h \equiv\sqrt{(h|h)}$  and $\rho_{\delta h}\equiv\sqrt{(\delta h|\delta h)}$ refer to the SNR of signal of interest and residual, respectively.}
	\label{tab:example_2new}
	\begin{tabular}{lccc|ccc|cr} 
		\hline
		i & $M/M_\odot$ & $\nu$ & $\beta$ & $D_\text{eff}$ & $t_c$ & $\phi_c$ & $\rho_h$ & $\rho_{\delta h}$\\
		\hline
		1 & $2\cdot 10^6$  & 0.20 & 5.0 & 10 $\text{Gpc}$& 6 h & 0 & 83 & - \\
		2 & $1\cdot 10^6$  & 0.23 & 1.0 & 3 $\text{Gpc}$& 48 h & $\pi$ & 790 & 31 \\
		3 & $4\cdot 10^6$  & 0.08 & 2.4 & 2 $\text{Gpc}$& 6 h & 0.9 & 2216 & 76 \\
		\hline
	\end{tabular}
\end{table}

The excellent agreement shown with MCMC analyses in both examples also suggest the possibility of using the formalism for exploratory studies of both next-generation ground-based and spaceborne detectors, for which sophisticated MCMC analyses are hard to carry out. One could for instance study how much the confusion noise from BH and neutron-star binaries that is expected with future ground-based detectors~\cite{Samajdar:2021egv,Pizzati:2021gzd,Himemoto:2021ukb} affects the parameter estimation of sources that might be used to test GR, or whether the removal of high-SNR sources (such as supermassive BH binaries in LISA) can affect the PE of lower-SNR sources (such as EMRIs). These investigations are left for future work. 

\newcommand{\AEI}{\affiliation{Max Planck Institute for Gravitational Physics (Albert Einstein Institute), Am M\"uhlenberg 1, Potsdam 14476, Germany}}
\newcommand{\Maryland}{\affiliation{Department of Physics, University of Maryland, College Park, MD 20742, USA}}

\chapter{Energetics of two-body Hamiltonians in post-Minkowskian gravity}
\label{chap:two}

\textbf{Authors}\footnote{Originally published in \emph{ Phys.Rev.D} 99 (2019) 10, 104004.}: Andrea Antonelli, Alessandra Buonanno, Jan Steinhoff, Maarten van de Meent, Justin Vines.
\newline

\textbf{Abstract:} Advanced methods for computing perturbative, quantum-gravitational scattering amplitudes show great promise for improving our knowledge of classical gravitational dynamics.  This is especially true in the weak-field and arbitrary-speed (post-Minkowskian, PM) regime, where the conservative dynamics at 3PM order has been recently determined for the first time, via an amplitude calculation.  
Such PM results are most relevantly applicable to relativistic scattering (unbound orbits), while bound/inspiraling binary systems, 
the most frequent sources of gravitational waves for the LIGO and Virgo detectors, are most suitably modeled by the weak-field and slow-motion (post-Newtonian, PN) approximation.  Nonetheless, it has been suggested that PM results can independently lead to improved modeling of bound binary dynamics, especially when taken as inputs for effective-one-body (EOB) models of inspiraling binaries. Here, we initiate a quantitative study of this possibility, by comparing PM, EOB and PN predictions for
the binding energy of a two-body system on a quasi-circular
inspiraling orbit against results of numerical relativity (NR) simulations. 
The binding energy is one of the two central ingredients (the other being the gravitational-wave energy flux) that enters the computation of gravitational waveforms employed by LIGO and Virgo detectors, and for (quasi-)circular orbits it provides an accurate diagnostic of the conservative sector of a model. 
We find that, whereas 
3PM results do improve the agreement with NR with respect to 2PM (especially when used in the EOB framework), it is crucial to push 
PM calculations at higher orders if one wants to achieve better performances than current waveform models used for LIGO/Virgo data analysis.

\section{Introduction}
\label{sec:intro}

Gravitational waves (GWs) from binary black holes (BHs) and neutron
stars (NSs)~\cite{Abbott:2016blz,TheLIGOScientific:2016pea,TheLIGOScientific:2017qsa,LIGOScientific:2018mvr}
encode information about the structure of compact objects and their
interaction via (strong, dynamical) gravitational fields.  The
continuously improving network of GW detectors~\cite{TheLIGOScientific:2014jea,TheVirgo:2014hva,Aso:2013eba,LIGOIndia}
offers unprecedented insights into astrophysics and fundamental physics. Likewise, a continuous improvement in the accuracy of existing GW predictions, using both numerical and analytical methods, is necessary in order to continue the successful story of GW astronomy.

Regarding GW predictions for compact binary coalescence, numerical and
analytical methods complement each other well, since the analytic
post-Newtonian (PN, weak-field and slow-motion) approximation (e.g., see Refs~\cite{Blanchet:2013haa,Schafer:2018kuf,Futamase:2007zz,Poisson:2014,Goldberger:2007hy,Rothstein:2014sra}) 
relies on the separation between the orbit's and the body's scales
being large, while current numerical-relativity (NR) simulations (e.g., see Refs.~\cite{Mroue:2013xna,Chu:2015kft,Husa:2015iqa}) become inefficient in this regime.  Since
the orbital separation shrinks over time due to energy and angular
momentum loss via GW emission, a synergistic approach between both
methods is needed to predict the complete inspiral-merger-ringdown
(IMR) sequence for the compact binaries now routinely detected by
ground-based GW observatories~\cite{LIGOScientific:2018mvr}.

The effective-one-body (EOB) formalism~\cite{Buonanno:1998gg,Buonanno:2000ef} improves 
the accuracy of the (perturbative) PN two-body dynamics  (see, e.g., Refs.\cite{LeTiec:2011bk,Damour:2011fu,LeTiec:2011dp,Tiec:2013twa,Ossokine:2017dge}) by resumming PN results in such a 
way as to include the exact test-particle limit. EOB waveforms~\cite{Taracchini:2013rva,Pan:2013rra,Bohe:2016gbl,Nagar:2018zoe} are an important 
class of IMR waveform models employed in LIGO/Virgo searches and inference studies~\cite{
	Abbott:2016blz,Abbott:2016nmj,TheLIGOScientific:2016pea,
	TheLIGOScientific:2016wfe,TheLIGOScientific:2016src,Abbott:2016izl,TheLIGOScientific:2017qsa,LIGOScientific:2018mvr}. Because of 
the more accurate description of the dynamics toward merger (with respect to PN), 
EOB waveforms are also employed to build another class of IMR waveforms, the phenomenological waveform models 
(e.g., see Ref.~\cite{Khan:2015jqa}). In order to improve EOB waveform models in the entire binary's parameter space (i.e., large-mass ratios and large 
spins), better understand the uniqueness and robustness of the EOB resummation, and gain confidence in its range of
applicability, it is important to extend the EOB formalism to highly relativistic 
bound and unbound orbits. The large mass-ratio case, which is relevant for space-based and third-generation 
ground-based detectors and requires a very accurate modeling of fast-motion effects, is one important example~\cite{Yunes:2009ef,Damour:2009sm,Yunes:2010zj,Barausse:2011dq,Akcay:2012ea}, which we will follow up elsewhere~\cite{Antonelliea:2019}. Here, we focus on the post-Minkowskian (PM)
approximation (i.e., weak field and fast motion)~\cite{Blanchet:2013haa,Poisson:2014,Bertotti:1956,Bertotti:1960,Rosenblum:1978zr,Bel:1981be,Damour:1981bh,Portilla:1979xx,Portilla:1980uz,Westpfahl:1979gu,Westpfahl:1985,Schafer:1986,Westpfahl:1987}
applicable to scattering binaries (see also
Refs.~\cite{Ledvinka:2008tk,Foffa:2013gja,Damour:2016gwp,Bini:2017xzy,Vines:2017hyw,Damour:2017zjx,Bini:2018ywr,Blanchet:2018yvb,Bjerrum-Bohr:2013bxa,Holstein:2004dn,Bjerrum-Bohr:2018xdl,Neill:2013wsa,Vaidya:2014kza,Cheung:2018wkq,Kosower:2018adc}
for recent applications).

A crucial ingredient of the EOB formalism is the energy map \cite{Buonanno:1998gg} 
between the two-body and the effective one-body description.  While the
energy map was established as a natural choice up to 4PN order
\cite{Buonanno:1998gg,Damour:2000we,Damour:2015isa}, its properties become more 
apparent and unique (at least at 1PM) when extending the conservative EOB Hamiltonian to 1PM and 2PM 
orders \cite{Damour:2016gwp,Damour:2017zjx}.  One can also gain insight into EOB spin
maps at 1PM and 2PM orders~\cite{Vines:2017hyw,Vines:2018gqi,Bini:2018ywr}. These results, together with the prospect of
creating a waveform model for scattering binaries, certainly provide a
good motivation to push the PM knowledge to higher orders. Quite interestingly, 
profiting from recent advances in the area of scattering amplitudes
\cite{Parke:1986gb,Bern:1994zx,Bern:1994cg,Britto:2004nc,Britto:2004ap,Britto:2005fq,Bern:2008qj,Bern:2010yg,Bern:2010ue,
	Guevara:2017csg,Arkani-Hamed:2017jhn,Ochirov:2018uyq}, the conservative Hamiltonian 
for a two-body system has been recently derived at 2PM order~\cite{Damour:2017zjx,Cheung:2018wkq} and 
3PM order~\cite{Bern:2019nnu}.  

However, the most frequent sources of GWs for LIGO/Virgo experiments
are bound/inspiraling binaries, instead of unbound/scattering ones.
It is not clear \textit{a priori} whether the source modeling 
of coalescing binaries for GW detectors will take real advantage of PM results, thus motivating 
to push PM calculations at even higher orders and extend them to the dissipative 
sector. It is also unclear whether 
insights on (and explicit resummations for) the EOB
Hamiltonian from PM results are already useful to improve the accuracy
of quasi-circular, inspiral waveforms for LIGO/Virgo analyses. Here, 
we start to shed light on these important inquiries by comparing PM, EOB and PN predictions for
the binding energy of a two-body system on a quasi-circular
inspiraling orbit against results of NR simulations. Indeed, the binding energy is 
one of the two central ingredients (the other being the GW energy flux) 
that enters the computation of gravitational waveforms (e.g., see Refs.~\cite{Buonanno:2009zt}). 
Thus, assessing the accuracy of PM predictions against the (``exact'') NR results, and quantifying the differences 
with respect to the EOB/PN results currently used in building waveform models for LIGO/Virgo analyses, is 
very relevant and timely, given also the strong interest that PM calculations have recently generated in the theoretical high-energy physics community.

This paper is organized as follows. In Sec.~\ref{sec:EOB3PM} we take
full advantage of the most recent PM results appeared in the
literature~\cite{Cheung:2018wkq, Bern:2019nnu} and extend to 3PM order the
PM EOB Hamiltonian originally derived by Damour~\cite{Damour:2017zjx}
at 2PM order. In Sec.~\ref{sec:ener} we compare various binding-energy
curves obtained from PM, EOB and mixed PM-PN against each other and
NR, and discuss the implications of PM calculations for LIGO/Virgo
source modeling. Section~\ref{sec:concl} contains our final remarks 
and discusses future work.  In Appendix ~\ref{appendixA}, we first briefly discuss the
special role of the nonlocal-in-time (tail) part of the two-body
Hamiltonian at 4PN order. Then, we derive an extension of the 3PM EOB Hamiltonian computed in this paper 
to 4PN order, including such tail terms, using the 4PN EOB Hamiltonian in 
Ref.~\cite{Damour:2015isa}. In Appendix ~\ref{appendixB}, we start to explore 
how to improve the use of PM results in the EOB framework by presenting an 
alternative EOB Hamiltonian at 3PM order for circular orbits.

Henceforth, we work in units in which the speed of light $c=1$.

\section{An effective-one-body Hamiltonian at third post-Minkowskian order}
\label{sec:EOB3PM}

Damour \cite{Damour:2017zjx} and Cheung, Rothstein, and Solon (henceforth, CRS)
\cite{Cheung:2018wkq} have each given results for the
Hamiltonian governing the conservative dynamics of a
two-body system at 2PM order.  Damour's EOB Hamiltonian was deduced by
matching an ansatz to the gauge-invariant scattering angle function, 
first derived at 2PM order by Westpfahl \cite{Westpfahl:1985}, noting that
the complete local-in-time, gauge-invariant information content of the (conservative) 
Hamiltonian is encoded in the scattering angle computed from the Hamiltonian.
Westpfahl's 2PM result for the scattering angle has since been
rederived by Bjerrum-Bohr et al.\ \cite{Bjerrum-Bohr:2018xdl} by
applying the eikonal approximation to scattering amplitudes
for massive scalars exchanging gravitons at one-loop order.  The CRS
2PM Hamiltonian was deduced by directly matching between those same amplitudes and amplitudes computed
from an effective (classical) field theory.  As was noted later in Ref.~\cite{Bern:2019nnu}, and as we show in this section, the CRS 2PM Hamiltonian also leads to (and is determined by) Westpfahl's 2PM scattering angle.

Recently, Bern et al.\ \cite{Bern:2019nnu} (henceforth, BCRSSZ) have 
extended the computation of the classical-limit amplitudes to two-loop order, accomplished the matching to a 3PM Hamiltonian, and given the 3PM scattering angle.  Here we provide an independent derivation of the 3PM scattering angle from the 3PM 
Hamiltonian of Ref.~\cite{Bern:2019nnu}, and we use the scattering angle to extend the EOB Hamiltonian 
of Ref.~\cite{Damour:2017zjx} to 3PM order.  

We consider a two-body system composed of non-spinning black holes 
with rest masses $m_1$ and $m_2$, total mass $M = m_1+m_2$, 
reduced mass $\mu = m_1\,m_2/M$, and symmetric mass ratio $\nu = \mu/M$. 
The 3PM Hamiltonian of Ref.~\cite{Bern:2019nnu}, given
in the binary's center-of-mass frame and in an isotropic 
gauge, reads
\begin{equation}
H(\bs r,\bs p)=H_0(\bs p^2)
+\sum_{n=1}^3\frac{G^n}{r^n}c_n(\bs p^2)
+\mc O(G^4),
\label{3PMH}
\end{equation}
\begin{equation}
H_0(\bs p^2)=\sqrt{m_1^2+\bs p^2}+\sqrt{m_2^2+\bs p^2},
\end{equation}
where $\bs r$ and $\bs p$ are the radial separation and its conjugate momentum, respectively.  
The functions $c_1$, $c_2$ and $c_3$ are explicitly given in Eqs.~(10) of 
BCRSSZ.  These functions determine (and are determined by) the coefficients in the 3PM scattering-angle function, 
as follows. (Henceforth, we refer to the Hamiltonian above as $H_{\text{3PM}}$.)

Since we neglect black-hole's spins, the binary's orbital plane is fixed. We introduce polar coordinates $(r,\phi)$ in the orbital plane,
with conjugate momenta $(p_r,p_\phi\equiv L)$ satisfying the standard relation
\begin{equation}\label{psq}
\bs p^2=p_r^2+\frac{L^2}{r^2}.
\end{equation}
Note that $L=p_\phi$ is a constant of motion due to axial symmetry.  We denote with $E = H(\bs r,\bs p)=H(r,p_r,L)$ the total conserved energy of the binary system.  Using the Hamilton-Jacobi formalism, it can be shown 
(e.g., see Ref.~\cite{Damour:2017zjx}) that the total change in the angle coordinate $\phi$ for a scattering orbit is
given by
\begin{equation}
\Delta\phi=\pi+\chi(E,L)=-2\int_{r_\mr{min}}^\infty dr\, \frac{\doe}{\doe L}p_r(r,E,L)\,,
\label{scattangle}
\end{equation}
where $\chi$ is generally called the scattering angle, and vanishes for free motions. The radial momentum $p_r(r,E,L)$ is obtained by 
solving $H(r,p_r,L)=E$ for $p_r$ (taking the branch $p_r>0$ in Eq.~(\ref{scattangle})), while $r_\mr{min}$ is the appropriate 
root of $p_r=0$. 

The solution for $p_r$ resulting from the 3PM Hamiltonian (\ref{3PMH}) can be obtained from Eq.~(\ref{psq}) after we solve for $\bs p^2$, working perturbatively in $G$.  To conveniently express the dependence on the energy $E$, we define the quantities\footnote{ We notice that the true relative velocity at infinity for a scattering orbit is the $v$ in $\gamma=(1-v^2)^{-1/2}$, with $\gamma$ given in terms of the energy and masses by Eq.~\eqref{gG}.  The same quantity is called $\hat{\mc E}_\mr{eff}$ in Ref.~\cite{Damour:2016gwp}; at zeroth order (or, at infinity, to all orders), it is the quantity called $\sigma$ in BCRSSZ.  The $\Gamma$ in the right-hand side of Eq.~\eqref{gG} is called $h$ in Ref.~\cite{Damour:2017zjx}; at infinity, it is the variable $\gamma$ in BCRSSZ.}
\be
\gamma=\frac{E^2-m_1^2-m_2^2}{2m_1m_2},
\qquad
\Gamma\equiv\frac{E}{M}=\sqrt{1+2\nu(\gamma-1)}\,.
\label{gG}
\ee
From a straightforward calculation, using the results for $\{c_n(\bs p^2)\}_{n=1}^3$ from Eqs.~(10) of BCRSSZ, we find
\be
\label{psquared}
\bs p^2(r,E)=p_0^2(E)
+\sum_{n=1}^3\frac{G^n}{r^n}f_n(E)
+\mc O(G^4),
\ee
with
\begin{subequations}
	\begin{align}
		p_0^2&=\mu^2\frac{\gamma^2-1}{\Gamma^2},\\
		f_1&=2\mu^2 M\frac{2\gamma^2-1}{\Gamma},\\
		f_2&=\frac{3}{2}\mu^2M^2\frac{5\gamma^2-1}{\Gamma},\\
		f_3&=\mu^2M^3\Bigg[\Gamma\frac{18\gamma^2-1}{2}
		-4\nu\gamma\frac{14\gamma^2+25}{3\Gamma}
		+\frac{3}{2}\frac{\Gamma-1}{\gamma^2-1}(2\gamma^2-1)(5\gamma^2-1)
		\nnm\\
		&\quad-8\nu\frac{4\gamma^4-12\gamma^2-3}{\Gamma\sqrt{\gamma^2-1}}\sinh^{-1}\sqrt{\frac{\gamma-1}{2}}
		\,\Bigg].
	\end{align}
\end{subequations}
Combining Eqs.~\eqref{psq}, \eqref{scattangle}, and \eqref{psquared} and evaluating the integral, we find
\begin{alignat}{3}
	\frac{\chi}{2}&=-\int_{r_\mr{min}}^\infty dr\frac{\doe}{\doe L}\sqrt{p_0^2-\frac{L^2}{r^2}+\sum_n\frac{G^n}{r^n}f_n}-\frac{\pi}{2}
	\\\nnm
	&=\frac{G}{L}\frac{f_1}{2p_0}+\frac{G^2}{L^2}\frac{\pi f_2}{4}
	\quad+\frac{G^3}{L^3}\bigg(p_0f_3+\frac{f_1f_2}{2p_0}-\frac{f_1^3}{24p_0^3}\bigg)+\mc O(G^4).
\end{alignat}
Thus, the 3PM (half) scattering angle is given by
\be
\frac{1}{2}\chi(E,L)=\sum_{n=1}^3\bigg(\frac{GM\mu}{L}\bigg)^n\chi_n(E)+\mc O(G^4),
\ee
with coefficients
\begin{subequations}\label{chi_n}
	\begin{align}
		\chi_1&=\frac{2\gamma^2-1}{\sqrt{\gamma^2-1}},\\
		\chi_2&=\frac{3\pi}{8}\frac{5\gamma^2-1}{\Gamma},\\
		\chi_3&=\frac{64\gamma^6-120\gamma^4+60\gamma^2-5}{3(\gamma^2-1)^{3/2}}-\frac{4}{3}\frac{\nu}{\Gamma^2}\gamma\sqrt{\gamma^2-1}(14\gamma^2+25)
		\nnm\\
		&\quad-8\frac{\nu}{\Gamma^2}(4\gamma^4-12\gamma^2-3)\sinh^{-1}\sqrt{\frac{\gamma-1}{2}},
	\end{align}
\end{subequations}
which agrees with Eq.~(12) of BCRSSZ.
When we take the limit $\nu\to0$ at fixed $\gamma$, implying $\Gamma\to1$, the scattering angle reduces to the one for a test particle with energy-per-mass $\gamma$ and angular-momentum-per-mass $L/\mu$, following a geodesic in a Schwarzschild spacetime with mass $M$.
Note that $\chi_1$ is the same as the Schwarzschild value; $\chi_2$ is the Schwarzschild value over $\Gamma$. 
The first line of $\chi_3$ coincides with its Schwarzschild value.

Damour has shown in Ref.~\cite{Damour:2017zjx} that an EOB Hamiltonian valid at 3PM order (for the conservative dynamics) can be obtained directly from the scattering-angle coefficients. The real EOB Hamiltonian $H^\mr{EOB}(\bs r,\bs p)$ is given in terms of the effective Hamiltonian $H^\mr{eff}(\bs r,\bs p)$ via the EOB energy map~\cite{Buonanno:1998gg},
\begin{equation}
H^{\text{EOB}} = M\,\sqrt{ 1 + 2 \nu\,\left (\frac{H^{\text{eff}}}{\mu} - 1 \right ) }\,,
\label{3PMEOBH}
\end{equation}
and $H^\mr{eff}$ reduces to the Hamiltonian for Schwarzschild geodesics $H_\mr S$ as $\nu\to0$.
The Schwarzschild-geodesic Hamiltonian (for a test particle of mass $\mu$) is given in Schwarzschild coordinates in the equatorial plane, with $H_\mr S(\bs r,\bs p) \equiv H_\mr S(r,p_r,L)$, by
\be
H_\mr S^2 =
\left(1-\frac{2GM}{r}\right)\,\left [\mu^2+\frac{L^2}{r^2}+\left(1-\frac{2GM}{r}\right)p_r^2\right ]\,.
\label{Hs}
\ee
Defining the reduced (dimensionless) quantities
\begin{gather}\label{reducedHu}
	\hat H^\mr{eff}=\frac{H^\mr{eff}}{\mu},\quad \hat H_\mr S=\frac{H_\mr S}{\mu},\quad u=\frac{GM}{r},\\
	\hat{p}_r = \frac{p_r}{\mu},\quad l \equiv \hat{p}_\phi = \frac{L}{GM\mu},
\end{gather}
the effective Hamiltonian of Ref.~\cite{Damour:2017zjx}---which we will refer to as the post-Schwarzschild (PS) effective Hamiltonian---is given through 3PM order by Eq.~(5.13) of
Ref.~\cite{Damour:2017zjx} as
\begin{align}
	(\hat H^\mr{eff,PS})^2&=
	\hat H_\mr S^2
	+(1-2u)\Big[u^2q_\mr{2PM}+u^3q_\mr{3PM}
	+\mc O(G^4)\Big],
	\label{3PMeffH}\\
	\hat H_\mr S^2 &= (1-2u) \left[1 + l^2 u^2 + (1-2u) \hat{p}_r^2 \right], \label{HShat}
\end{align}
where the functions $q_\mr{2PM}(\hat H_\mr S,\nu)$ and $q_\mr{3PM}(\hat H_\mr S,\nu)$ are determined by the scattering-angle coefficients via Eqs.~(5.6) and (5.8) of Ref.~\cite{Damour:2017zjx}. (Notice the absence of a $q_\mr{1PM}(\hat H_\mr S,\nu)$ term, which vanishes identically in the EOB formulation. Indeed, the energy map \eqref{3PMEOBH} applied to the unmodified Schwarzschild-geodesic Hamiltonian \eqref{Hs} precisely reproduces the two-body dynamics at 1PM order \cite{Damour:2016gwp,Damour:2017zjx}.)

Inserting our coefficients (\ref{chi_n}) into those equations yields
\begin{subequations}\label{q2q3}
	\begin{align}
		q_\mr{2PM}&=\frac{3}{2}(5\hat H_\mr S^2-1)\Bigg(1-\frac{1}{\sqrt{1+2\nu(\hat H_\mr S-1)}}\Bigg),\label{qtwo}
		\\
		q_\mr{3PM}&=-\frac{2\hat H_\mr S^2-1}{\hat H_\mr S^2-1}q_\mr{2PM}
		+\frac{4}{3}\nu\hat H_\mr S\frac{14\hat H_\mr S^2+25}{1+2\nu(\hat H_\mr S-1)}\nnm\\
		&\quad+\frac{8\nu}{\sqrt{\hat H_\mr S^2-1}}\frac{4\hat H_\mr S^4-12\hat H_\mr S^2-3}{1+2\nu(\hat H_\mr S-1)}\sinh^{-1}\sqrt{\frac{\hat H_\mr S-1}{2}}. \label{qthree}
	\end{align}
\end{subequations}
The resultant 3PM EOB Hamiltonian, $H^{\text{EOB}}_{\text{3PM}}$, for the two-body description, is obtained by plugging (\eqref{3PMeffH}) into Eq.~(\eqref{3PMEOBH}). The $H^{\text{EOB}}_{\text{3PM}}$ and BCRSSZ Hamiltonians are
equivalent in the sense that they lead to the same scattering angle when
expanded at 3PM order.  In the PN expansion, they are both complete up to 2PN order.  We discuss in Appendix \ref{appendixA} how to augment the above 3PM EOB Hamiltonian with additional PN information at 3PN and 4PN orders.

\begin{figure*}
	\includegraphics[width=\linewidth]{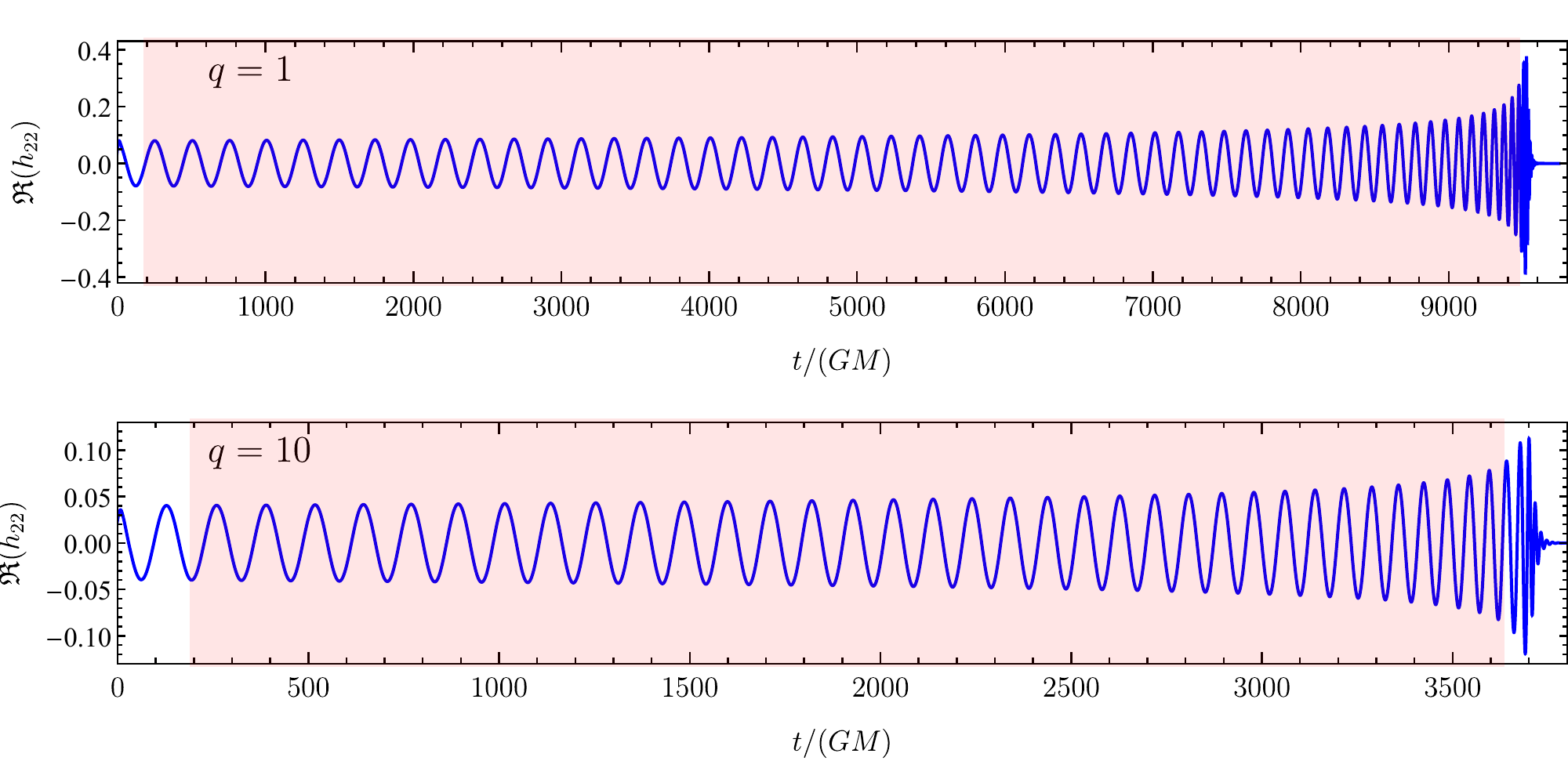}
	\caption{{\textbf{ NR simulations.}} In this paper the energetics of various approximants are compared against two NR simulations of non-spinning binary black holes produced by the Simulating eXtreme Spacetimes (SXS) collaboration~\cite{Mroue:2013xna,Chu:2015kft}. The top (bottom) panel shows the waveform (more specifically the real part of the $l=m=2$ mode of the strain, $\mathfrak{R}(h_{22})$) of the simulation with mass-ratio $q=1$ ($q=10$), identified in the SXS catalog as SXS ID: 0180 (SXS ID: 0303). In both panels, the red shading shows the segment of the simulation used for the binding energy's comparison in all figures of this paper.
		\label{fig:waveforms}
}
\end{figure*}

\section{Energetics of binary systems with post-Minkowskian Hamiltonians}
\label{sec:ener}

Gravitational waveforms emitted by inspiraling binaries are constructed from the binary's binding energy and GW energy flux (e.g., see Refs.~\cite{Buonanno:2009zt}). To assess the relevance for LIGO/Virgo analysis of the recently derived conservative two-body dynamics in PM theory, we compute one of these building blocks, the binding energy, for a variety of PM, PN and EOB approximants. We then compare these with results from NR simulations.

\begin{figure*}[t]
	\includegraphics[width=.47\columnwidth]{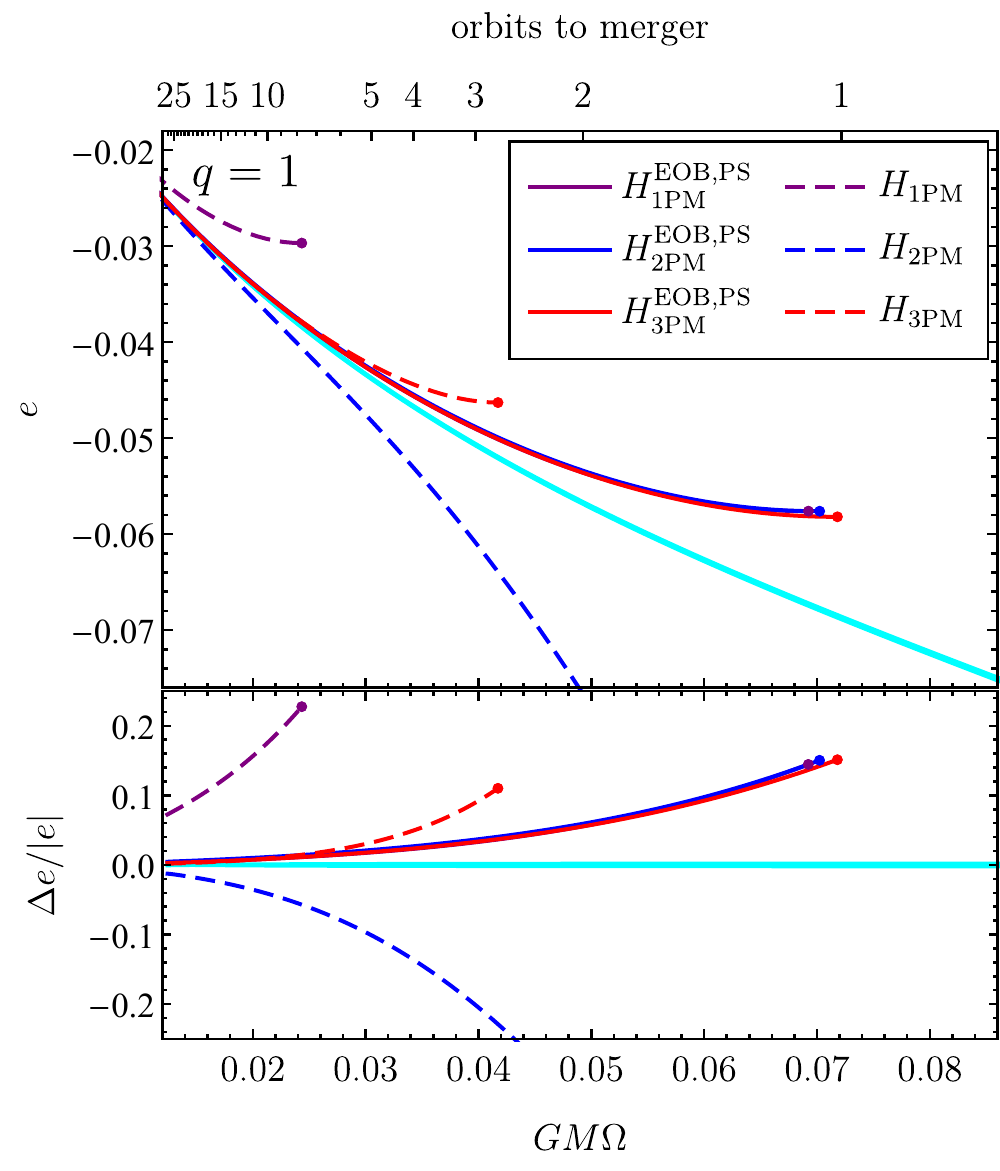}
	\hspace{12pt}
	\includegraphics[width=.47\columnwidth]{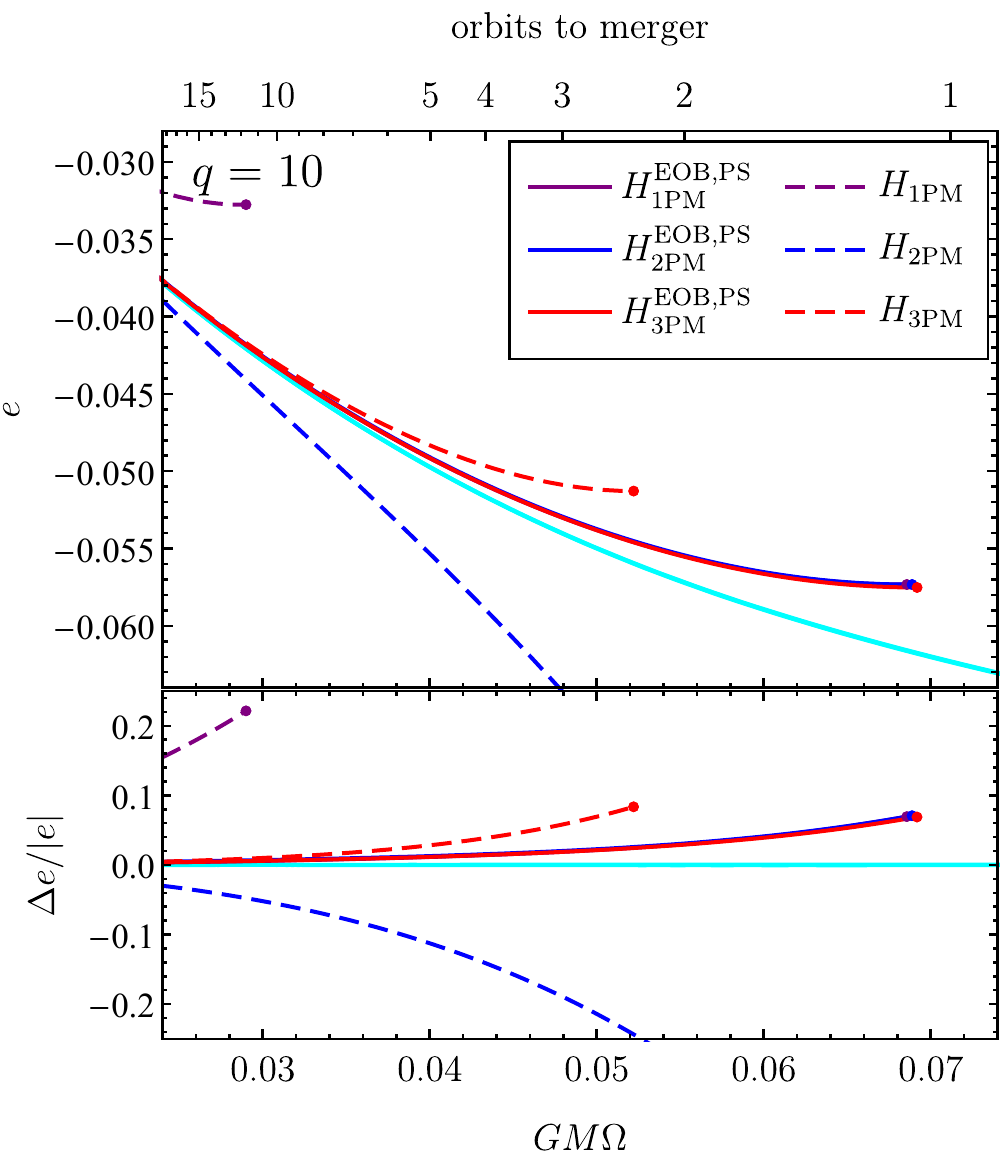}
	\caption{{\textbf {Energetics of PM Hamiltonians}.} We compare to NR the binding energy as a function of orbital frequency $GM\Omega$ from both PM and PM-EOB Hamiltonians for a nonspinning binary black hole with mass ratio $q=1$ (left panel) and $q=10$ (right panel). The dots at the end of the curves mark the ISCOs, when present in the corresponding two-body dynamics. The NR binding energy and its error are in cyan. The top $x$-axis shows the number of orbits until merger. In the lower panel we show the fractional difference between the approximants and the NR result. \label{fig:energywPM}}
\end{figure*}

\begin{figure*}[t]
	\includegraphics[width=.47\columnwidth]{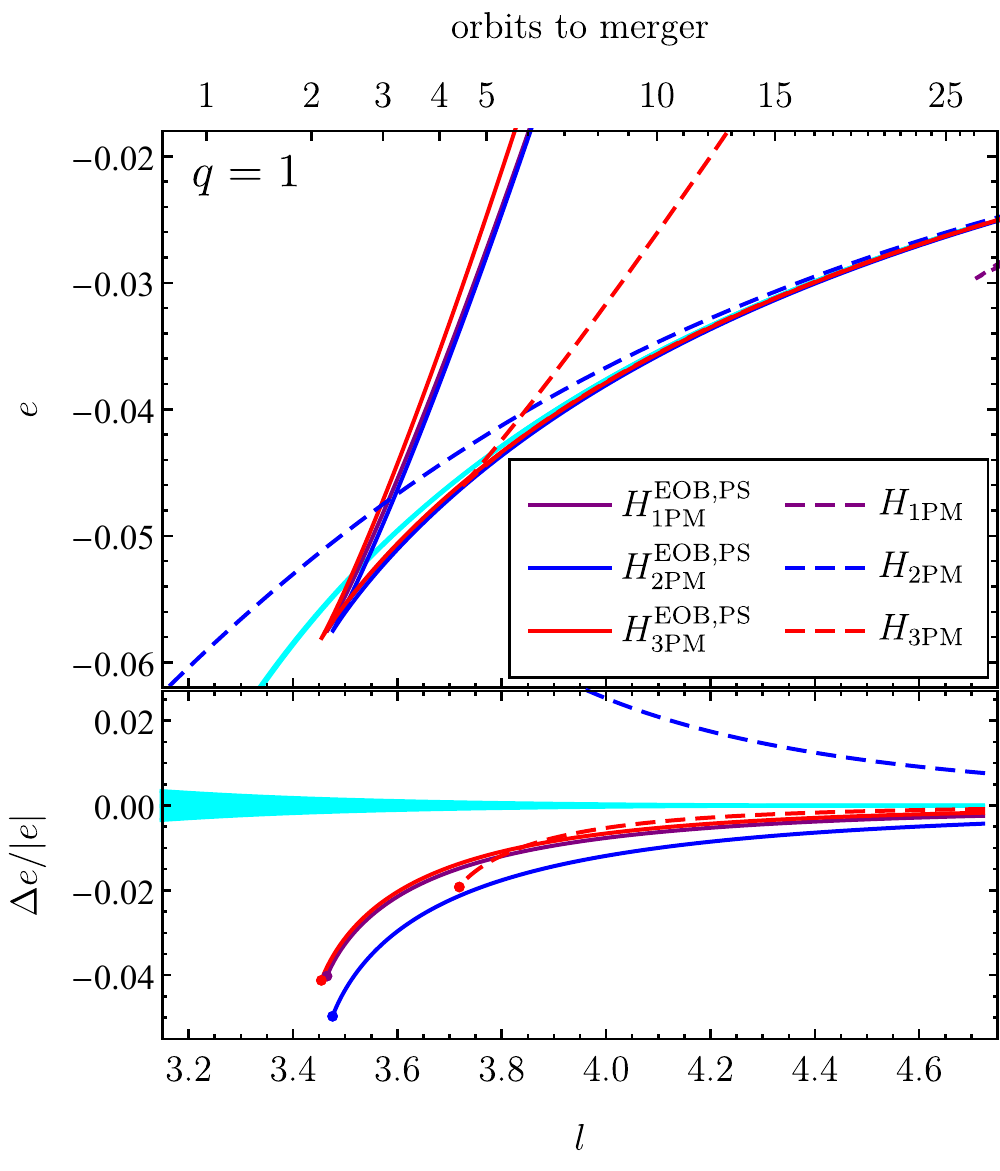}
	\hspace{12pt}
	\includegraphics[width=.47\columnwidth]{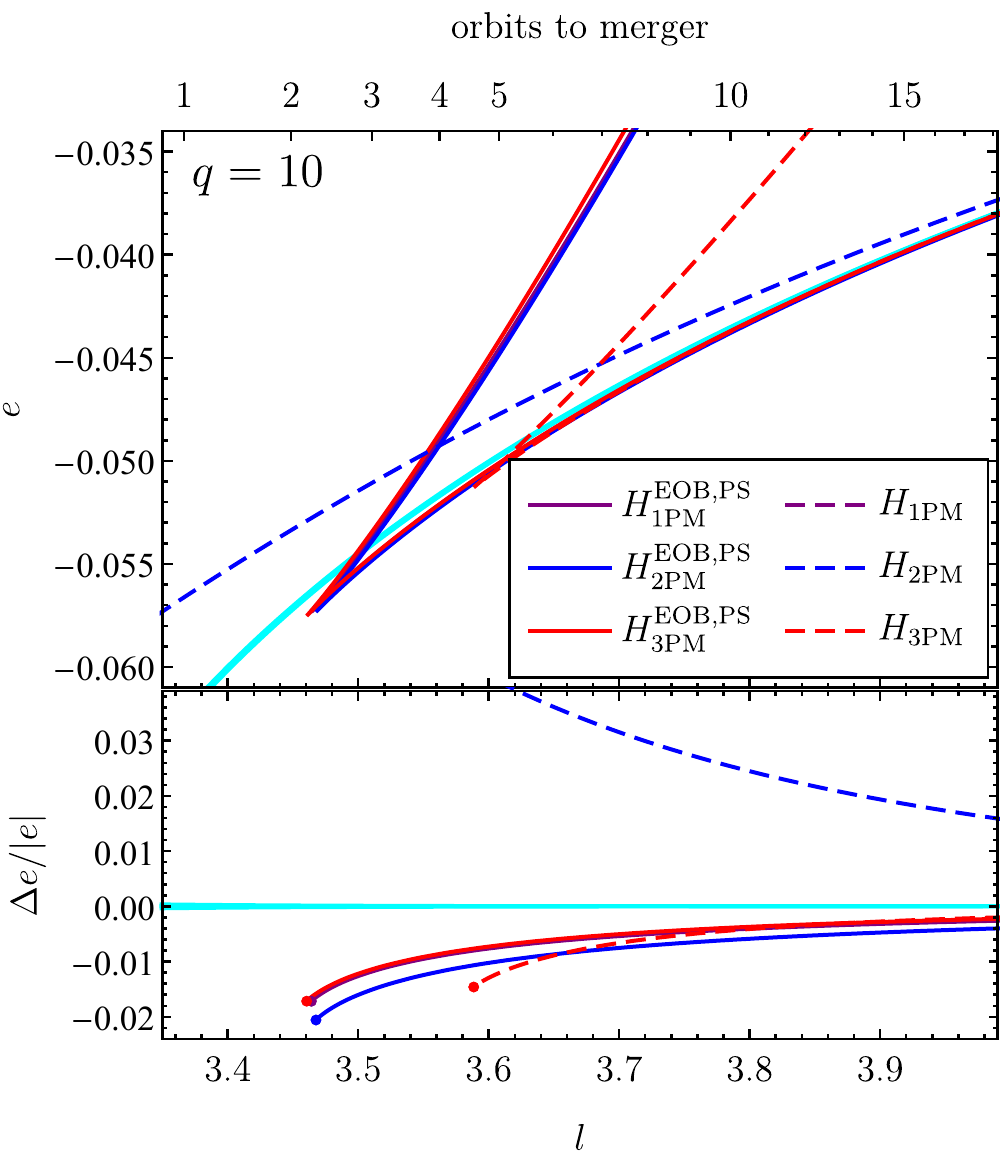}
	\caption{{\textbf{ Energetics of PM Hamiltonians.}} Same as in Fig.~\ref{fig:energywPM} but versus the dimensionless angular momentum $l=L/(G \mu M)$. The cusps signal the presence of the ISCO, where the branches of stable and unstable circular-orbit solutions meet. Note that the orbital-frequency range in the plots ends about $1.4$ and $1.8$ GW cycles, for $q=1$ and $q=10$, respectively, before the two black holes merge. \label{fig:energyJPM}}
\end{figure*}

We recall that the total energy $E$, linear momentum $\vec{P}$, and angular momentum $\vec{L}$ of a gravitating two-body system in an asymptotically flat spacetime are nearly\footnote{The quantities $E$, $\vec{P}$, and $\vec{L}$ are only invariant up to the fixing of a frame at infinity.} gauge-invariant quantities. It is convenient to introduce the dimensionless binding energy $e \equiv (E - M) / \mu$ and angular momentum $l \equiv |\vec{L}| / (G M \mu)$.  For an inspiraling binary of non-spinning black holes, the energy and angular momentum monotonically decrease over time and trace out a curve $e(l)$ for each set of binary parameters. This $e(l)$-curve is a gauge-invariant relation that can be used to compare analytic predictions in PN, PM and EOB frameworks against numerical-relativity (NR) results~\cite{Damour:2011fu,LeTiec:2011dp,Nagar:2015xqa,Ossokine:2017dge}. In absence of radiation reaction and for circular orbits, the relation $e(l)$ encodes the conservative dynamics.  For quasi-circular inspirals, the $e(l)$-relation still depends most sensitively on the conservative dynamics.  Hence, it is a good indicator for the accuracy of the binding energy derived from the PN, PM and EOB Hamiltonians for circular orbits.

Now, $E$ and $\vec{L}$ are not directly extracted from NR simulations as a function of time, but instead it is the gravitational radiation leaving the binary system (more precisely, the ``news function'') that is extracted~\cite{Damour:2011fu,Ossokine:2017dge}. The radiated energy and angular momentum fluxes as functions of time can then be integrated to yield the energy $E$ and angular momentum $\vec{L}$ of the binary at a given (retarded) time, which can then be combined into the relation $e(l)$~\cite{Damour:2011fu,Ossokine:2017dge}. The integration constants can be adjusted to match $E$ and $\vec{L}$ at the initial time of the simulation~\cite{Damour:2011fu,Nagar:2015xqa}, which has the disadvantage that one has to accurately track the fluxes during the initial junk-radiation phase.  A better approach is to fix the integration constants to match the energy and angular momentum of the final (merged) black hole; but even in this case, a further tuning of the integration constants is needed to achieve agreement with analytical models of the early inspiral (e.g., see Ref.~\cite{Ossokine:2017dge}). In the following, we use the binding energy from NR simulations as extracted and tuned in Ref.~\cite{Ossokine:2017dge}.

Similarly, the relation $e(l)$ can be obtained by solving the Hamilton equations with radiation-reaction effects for a given PN, PM and EOB Hamiltonian (e.g., as done in Refs.~\cite{Damour:2011fu,LeTiec:2011dp}).  However, for most of the analysis in this section, we neglect radiation-reaction effects (which have been shown to make EOB Hamiltonians accurate past the last-stable orbit and all the way down to merger~\cite{Damour:2013tla}), and construct the $e(l)$ curves using an adiabatic sequence of circular orbits instead. For this reason, we should not expect exact agreement with the NR results, which do include radiation-reaction effects. Our motivation for this choice is that $e(l)$ only depends on the Hamiltonian model (and not the radiation-reaction model), so it is easier to interpret our results and put them into context for future improvements of the Hamiltonian (e.g., when higher-order PN and PM results become available). More importantly, previous investigations~\cite{Damour:2011fu,LeTiec:2011dp} have shown that at least until the innermost-stable circular orbit (ISCO, where we terminate the comparison with NR results), the difference between the binding energy computed from a sequence of circular orbits and from a quasi-circular inspiral is not very large (typically no more than 5-10\%, as we discuss below and in Fig.~\ref{fig:evolq1}).

\begin{table}
	\caption{{\textbf{Two-body Hamiltonians}.} A summary of the Hamiltonians used in this paper to compute the binding energy and compare it with NR predictions. \label{table:models}}
	\begin{tabular}{lp{7cm}p{5cm}}
		$H_{m\text{PM}}$
		&	PM Hamiltonian	& \cite{Cheung:2018wkq,Bern:2019nnu} \\
		$H_{m\text{PM}}^{\text{ EOB,PS}}$					
		&	PM EOB Hamiltonian & \cite{Damour:2017zjx} and this paper \\
		$H^{\text{EOB,PS}}_{m{\text{PM}} + n{\text{PN}}}$
		&       PM EOB Hamiltonian with PN information when $n\geq m$ & \cite{Damour:2017zjx} and this paper	\\
		$H_{n\text{PN}}^{\text{EOB}}$	
		&	PN EOB Hamiltonian used in LIGO/Virgo data-analysis & \cite{Buonanno:1998gg,Damour:2000we,Damour:2015isa} \\
		$H_{3\text{PM}}^{\text{ EOB},\widetilde{PS}}$ & alternative 3PM EOB Hamiltonian & this paper\\
		$H_{n\text{PN}}$	& PN Hamiltonian & \cite{Jaranowski:1997ky}	\\
	\end{tabular}
\end{table}

In the absence of radiation reaction, the Hamilton equations for a generic Hamiltonian $H(r, p_r, L)$ describing a two-body system of nonspinning black holes read,
\begin{subequations}
	\begin{align}
		\dot{r} &= \frac{\partial H}{\partial p_r}\,, &
		\dot{L} &= - \frac{\partial H}{\partial \phi}=0\,, \\
		\Omega \equiv \dot{\phi} &= \frac{\partial H}{\partial L}\,, &
		\dot{p}_r &= - \frac{\partial H}{\partial r} \,.
	\end{align}
\end{subequations}
Note that $L \equiv G M \mu l = \text{const}$. For circular orbits, $p_r = 0$, $r = r_\text{circ} = \text{const}$. Furthermore, $\dot{p}_r=0$ and consequently it follows from the Hamilton equations that $({\partial H}/{\partial r})_{r = r_\text{circ}}=0$, which determines $r_\text{circ}(l)$ and hence the circular-orbit relation $e(l) \equiv  [H(r_\text{circ}(l), 0,  l) - M]/\mu$. The relation $\Omega(l) \equiv (\partial H/\partial L)_{r = r_\text{circ}} = e'(l)$ determines a second gauge-invariant relation. Inverting this relation gives $l(\Omega)$, which can be combined with $e(l)$ to give $e(\Omega)$.

\begin{figure*}[t]
	\includegraphics[width=.47\linewidth]{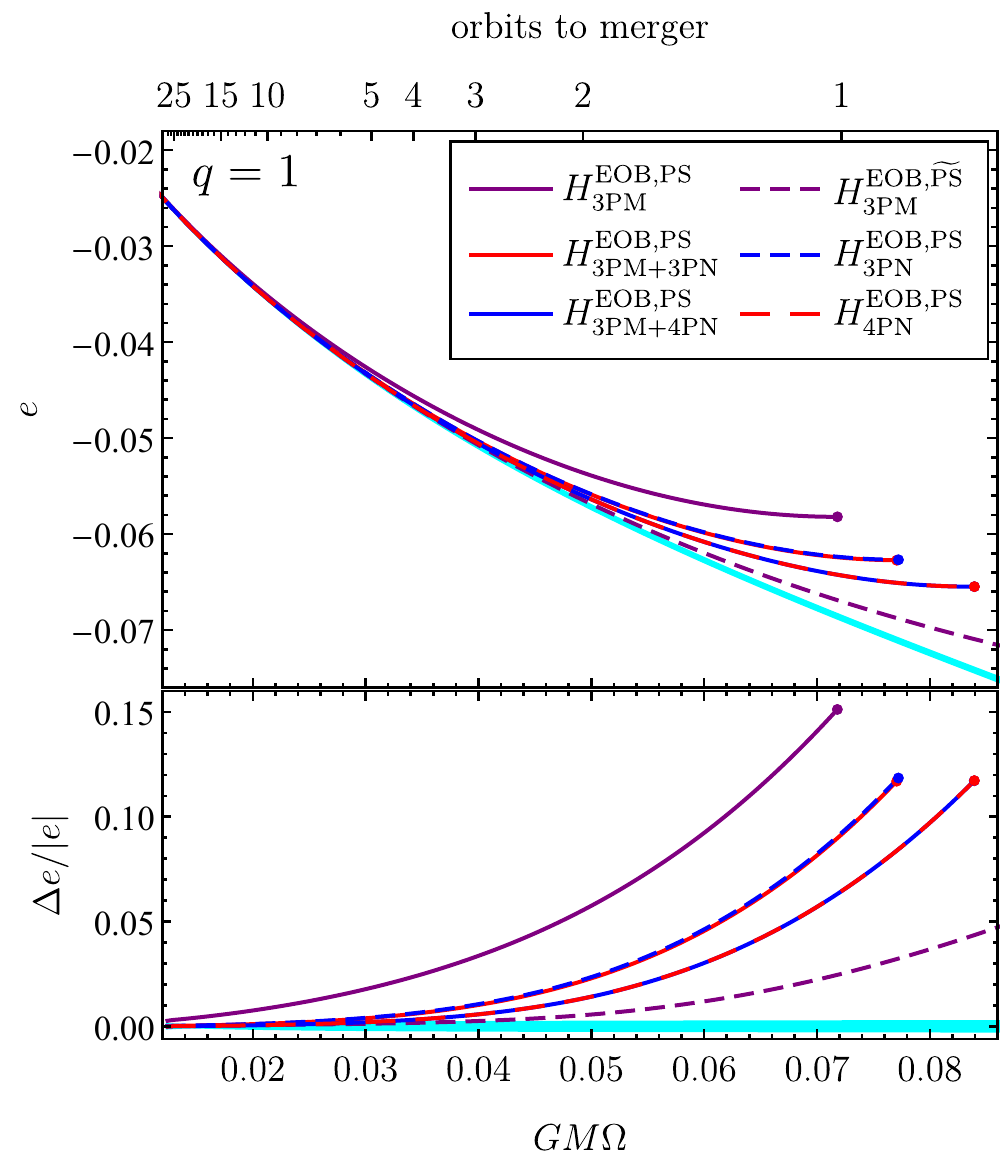}
	\hspace{12pt}
	\includegraphics[width=.47\linewidth]{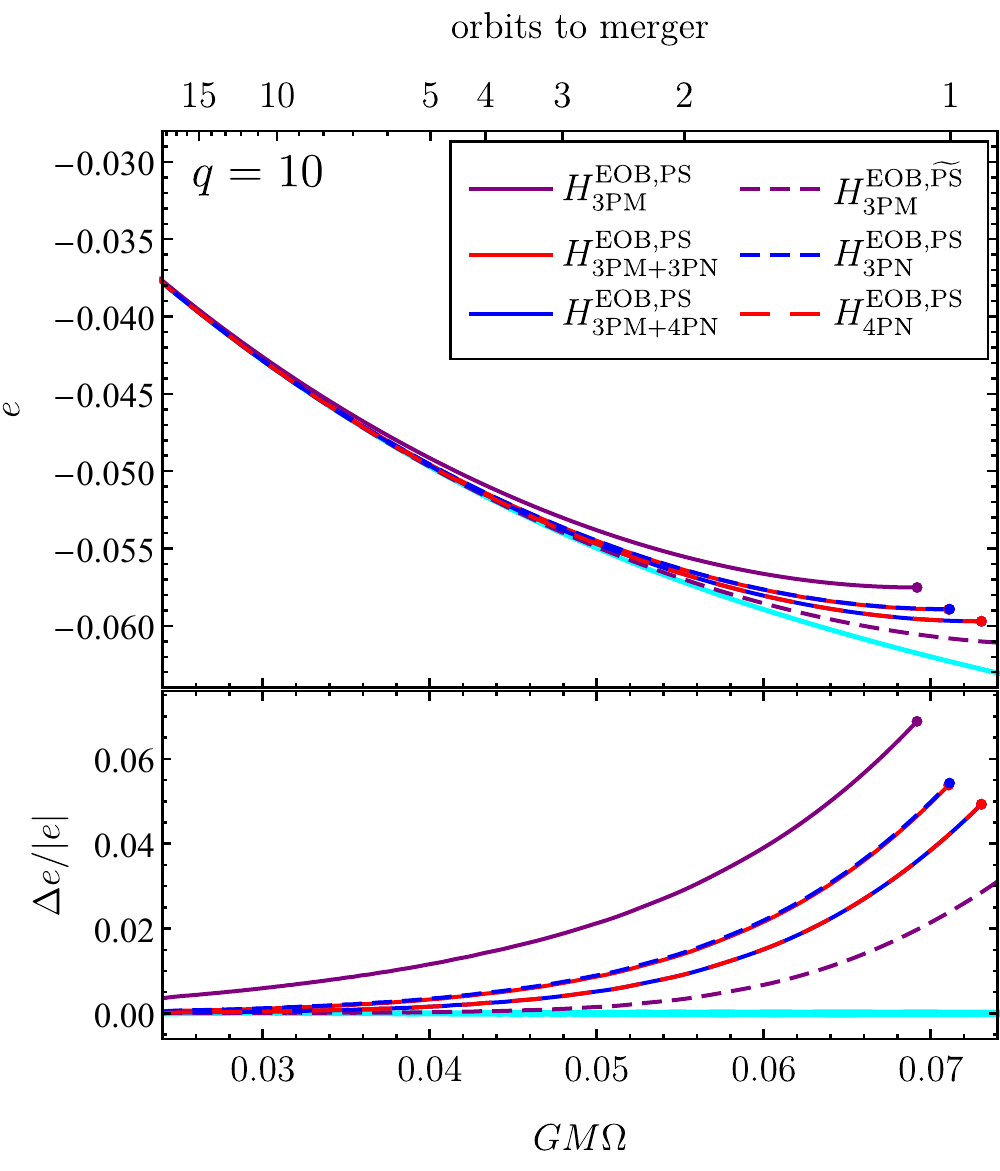}
	\caption{{\textbf{ Energetics of PM Hamiltonians augmented by PN information.}} Same as in Fig.~\ref{fig:energywPM} but now we compare to NR the binding energy of PM EOB Hamiltonians augmented by PN information. Notice that adding 3PM information at 3PN or above does not lead to a visible difference from plain PN EOB Hamiltonians (the 3PM-3PN and 3PN curves, as well as the 3PM-4PN and 4PN ones, are essentially on top of each other). Also included is a curve for an alternative 3PM EOB Hamiltonian,  $H^{\text{EOB},\widetilde{PS}}_{3 \text{PM}}$, derived in Appendix \ref{appendixB}.\label{fig:energymixed}}
\end{figure*}

Given a Hamiltonian there are different ways to determine the $e(l)$
and $e(\Omega)$ relationships. For example, we can solve for them
order-by-order in a systematic PN expansion yielding $e$ and $l$ as a
power series in $(GM\Omega)^{2/3}$ (e.g., see Eq.~(232) in Ref.~\cite{Blanchet:2013haa}). 
However, we note that if we were just expanding the binding energy computed from the 
PM Hamiltonian in powers of $(GM\Omega)^{2/3}$ and then truncating it, all extra information obtained
through the PM approximation would be lost. Nevertheless, performing 
such PN expansion of the binding energy provides an important consistency
check between different Hamiltonians --- for example one readily verifies that
starting from the 3PM Hamiltonian of Ref.~\cite{Bern:2019nnu} one
obtains the well known PN circular orbit relations $e(\Omega)$ and
$l(\Omega)$ to order  $(GM\Omega)^{2}$. To gauge the
additional information present in the PM Hamiltonians we opt for a
different approach, where we treat the various PN, EOB and PM 
approximants as exact Hamiltonians and determine the relations $e(l)$
and $e(\Omega)$ numerically (i.e. without any further expansions). 
We discuss in Fig.~\ref{fig:vanPMvsPNq1} below, differences in the PN 
binding energy when we build it from the exact (unexpanded)  PN Hamiltonian and the 
systematically PN expanded one.  

We consider only stable (and marginally stable) circular orbits, for which the Hamiltonian is minimal, $ 0 \leq ({\partial^2 H}/{\partial r^2})_{r = r_\text{circ}}$. Here equality corresponds to a saddle point of the Hamiltonian, which indeed exists for most --- but not all --- of the Hamiltonians under investigation. This is the well-known ISCO and corresponds to an angular momentum $l = l_\text{ISCO}$. 

For simplicity, we restrict the discussion to NR simulations of nonspinning binary black holes with 
mass ratios $q = 1$ and $10$~\cite{Ossokine:2017dge}. In Fig.~\ref{fig:waveforms} we display the 
NR waveforms. Those simulations span about $56$ and $36$ GW cycles (corresponding to $\sim 28$ and $\sim 18$ orbital cycles), for $q=1$ and $q=10$, respectively, before merger. 
We highlight in Fig.~\ref{fig:waveforms} the portion of the waveform that we use to compare with 
the binding-energy approximants. As can be seen, the comparisons with NR extend up to 
about $1.4$ and $1.8$ GW cycles, for $q=1$ and $q=10$, respectively, before the two black holes merge. 
Thus, our comparisons of analytic models to NR predictions extend to the late inspiral of a binary evolution, a stage characterized by high velocity and strong gravity.

We compare NR predictions against analytic results obtained with PM, EOB and PN Hamiltonians, summarized in Table~\ref{table:models}. Notably, we compute results with the Hamiltonian at $m {\text{PM}}$ orders with $m = 1, 2, 3$~\cite{Cheung:2018wkq, Bern:2019nnu} (labeled $H_{m\text{PM}}$),  and with the EOB Hamiltonian of Refs.~\cite{Damour:2016gwp,Damour:2017zjx} and this paper at $m {\text{PM}}$ orders with $m = 1, 2,3$ (labeled $H^{\text{EOB, PS}}_{m\text{PM}}$). We also compare results with the PM EOB Hamiltonian augmented with PN results up to 4PN order (labeled $H^{\text{EOB,PS}}_{m{\text{PM}} + n{\text{ PN}}}$), as derived in Appendix \ref{appendixA}. Furthermore, the (original) EOB Hamiltonian employed in LIGO/Virgo data 
analysis~\cite{Taracchini:2013rva,Bohe:2016gbl} is built from the EOB Hamiltonian of Refs.~\cite{Buonanno:1998gg,Damour:2000we,Damour:2015isa}, and it resums perturbative PN results differently from the PM EOB Hamiltonian. To understand the impact of the different 
resummation, and also highlight the accuracy that PM results would need to achieve in order to motivate their use in waveform modeling, we also show results with such an EOB Hamiltonian (labeled $H^{\text{EOB}}_{n{\text{PN}}}$). Finally, we also employ the PN Hamiltonian 
from Ref.~\cite{Jaranowski:1997ky} (labeled $H_{n\text{PN}}$), and an alternative 3PM EOB Hamiltonian presented for circular orbits 
in Appendix \ref{appendixB} (labeled $H^{\text{ EOB},\widetilde{PS}}_{\text{3PM}}$).

\begin{figure}
	\includegraphics[width=\linewidth]{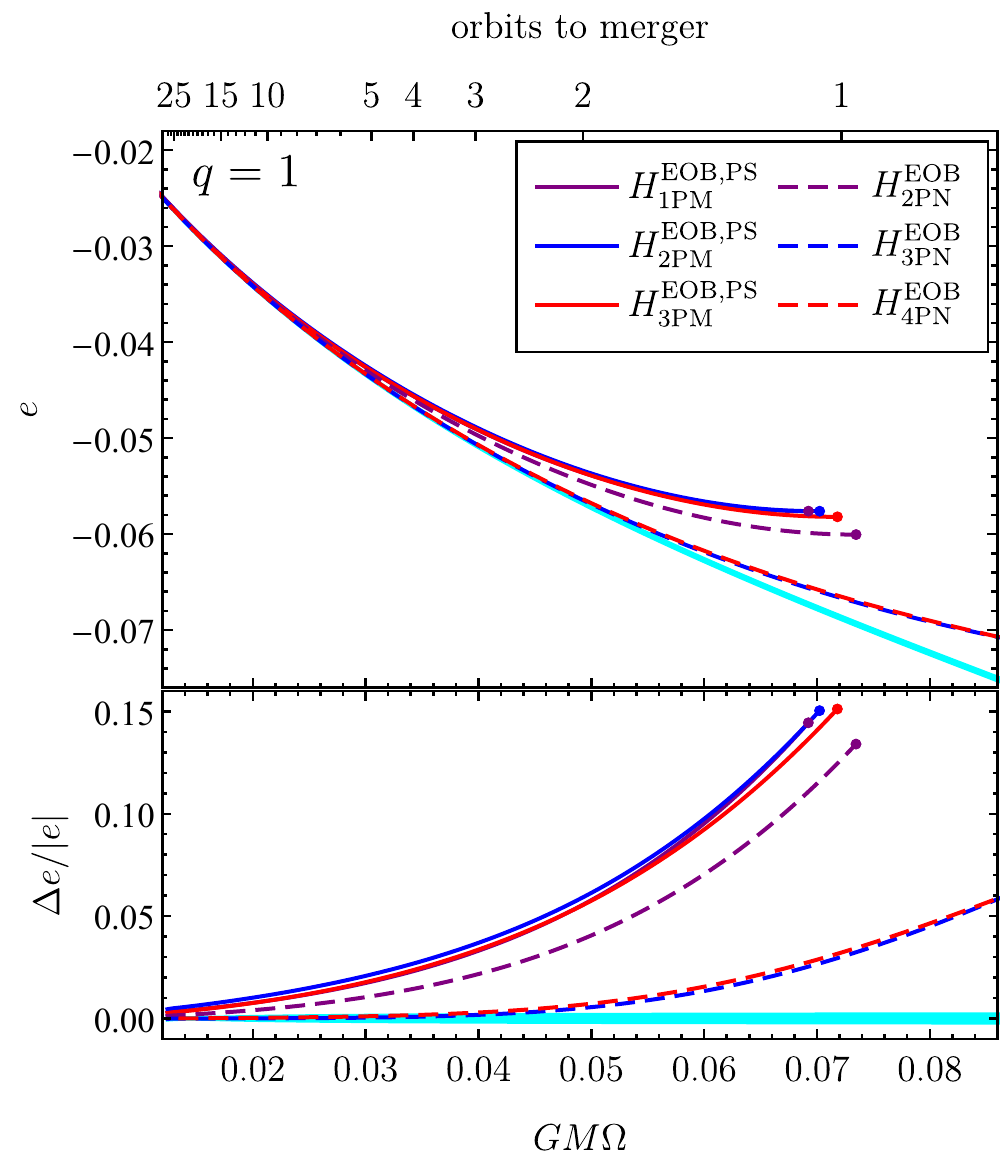}
	\caption{{\textbf{ Energetics of PM EOB Hamiltonian and the EOB Hamiltonian used in LIGO/Virgo data-analysis.}} Same as in Fig.~\ref{fig:energywPM} and Fig.~\ref{fig:energymixed}, but now we show how the $H^{\text{EOB,PS}}_{m\text{PM}}$ Hamiltonian compares with the (original)  $H^{\text{EOB}}_{n\text{PN}}$ Hamiltonian currently employed at 4PN order to build waveform models for LIGO/Virgo data-analysis. We observe that $H^{\text{EOB}}_{n\text{PN}}$ Hamiltonians still produce $e(\Omega)$-curves substantially closer to NR result than the 3PM approximant.\label{fig:PMvsclassicEOB}}
\end{figure}

\begin{figure}
	\includegraphics[width=\linewidth]{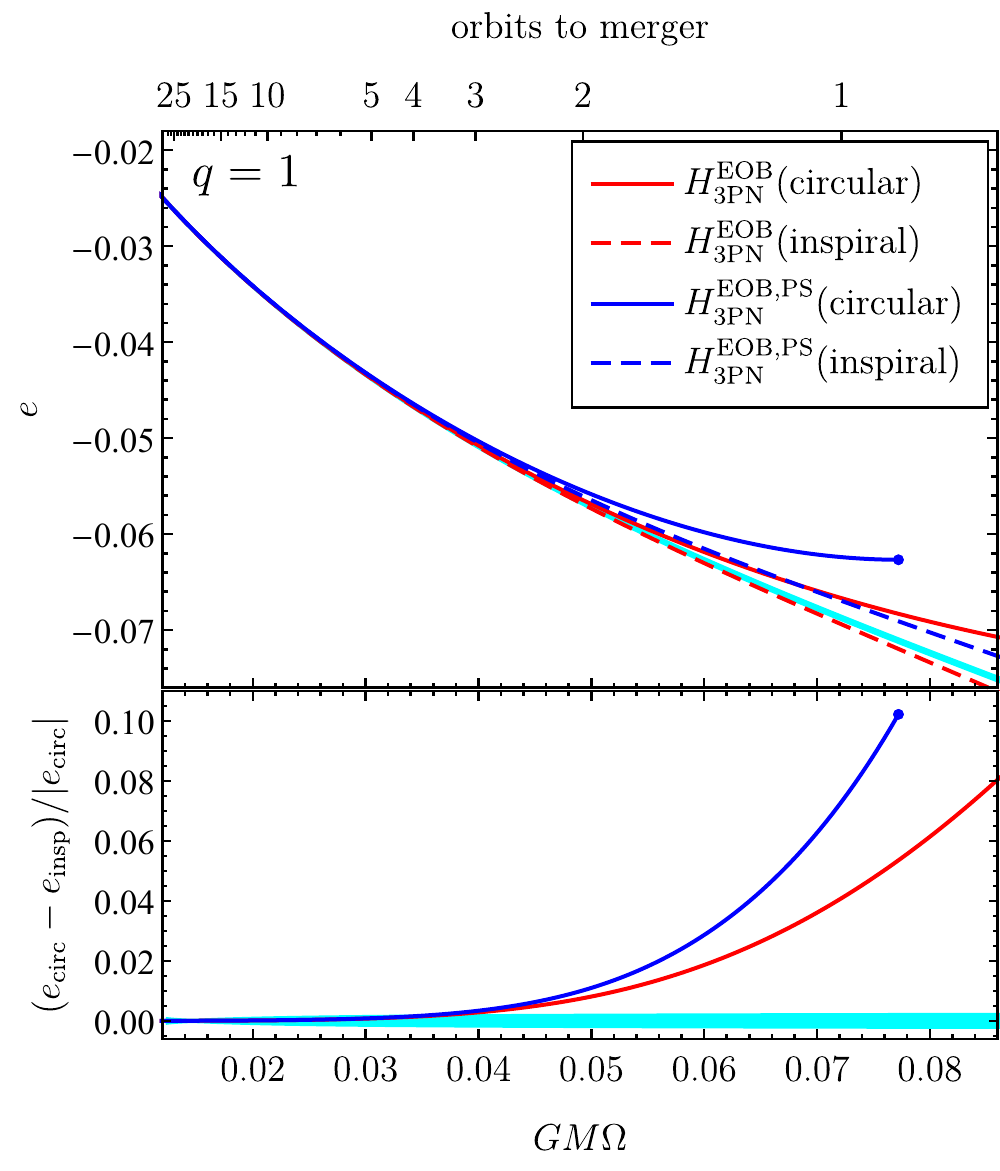}
	\caption{{\textbf{ Energetics of circular versus inspiral  PN approximants.}}		We show the binding energy obtained from the $H^{\text{EOB}}_{3\text{PN}}$ and $H^{\text{EOB,PS}}_{n\text{PN}}$ either through an adiabatic sequence of circular orbits or numerically evolving the Hamilton equations with a suitable radiation-reaction 		force for a quasi-circular inspiral.  The bottom panel shows the relative difference between circular and inspiral curves. This gives an indication of the magnitude of the impact which should be kept in mind when interpreting the other figures. By comparison the size of the NR error --- indicated in gray --- is very small.\label{fig:evolq1}}
\end{figure}

In Figs.~\ref{fig:energywPM} and \ref{fig:energyJPM} we compare the
binding energy computed in NR with the ones from PM and PM EOB
Hamiltonians versus either the binary's orbital frequency (Fig.~\ref{fig:energywPM}) or angular momentum
(Fig.~\ref{fig:energyJPM}), for mass ratios $q=1$ and $q=10$. We
clearly see the improvement of the PM binding energy from 1PM to 3PM,
especially at low frequency. The PM-EOB binding energies
generally show better agreement with NR, but they have a much
smaller range of variation from 1PM to 3PM. The 3PM result does
slightly better than 1PM, while 2PM is worse than the other two.
Overall those results demonstrate the value and relevance of pushing
PM calculations at higher order, and of further exploring how to
use PM results to improve EOB models.

To understand the impact of PM calculations for LIGO/Virgo analyses,
it is important to compare the PM binding energy with current
approximants used to build waveform models. 
Let us emphasize again
that perturbative PM calculations (weak-field/fast-motion), suitable
for unbound/scattering orbits, are not necessarily expected to
improve, when available at low PM orders, the predictions obtained
in perturbative PN calculations (weak-field/slow-motion), suitable
for bound/inspiraling orbits, which are the LIGO/Virgo GW
sources. It is
instructive to understand how the PM binding
energy compares with the PN binding energy, which at $n$PN order we
expect to be more accurate than the one at $n$PM order. For this
study we restrict to the 3PM EOB Hamiltonian and augment it with 3PN
and 4PN information, as derived explicitly in
Appendix~\ref{appendixA}. We display results in
Fig.~\ref{fig:energymixed}. Interestingly, the figure shows that the
mixed PM-PN Hamiltonian does not improve much over a PN Hamiltonian.
This means that currently the known PM Hamiltonian does not improve
in accuracy compared to PN ones (as usual, regarding NR as the
``true'' result). However, it is important to note that so far the PM
information has been incorporated into the EOB Hamiltonian in one
particular way, as proposed in Ref.~\cite{Damour:2017zjx}, in the
$H^{\text{EOB,PS}}_{m \text{PM}}$ curves.  We note in
Fig.~\ref{fig:energymixed} that one alternatively resummed EOB-3PM
Hamiltonian, $H^{\text{ EOB},\widetilde{PS}}_{3 \text{PM}}$, defined in
Appendix~\ref{appendixB}, shows better agreement with NR.  In the
PM expansion, this Hamiltonian is perturbatively equivalent to
$H^{\text{EOB,{PS}}}_{3 \text{PM}}$ up to 3PM order, i.e., they differ
only by 4PM-order terms.  The variation between those two curves
thus gives some indication of the variability expected from 4PM
order, and motivates calculations at higher PM order. 

\begin{figure}
	\includegraphics[width=\linewidth]{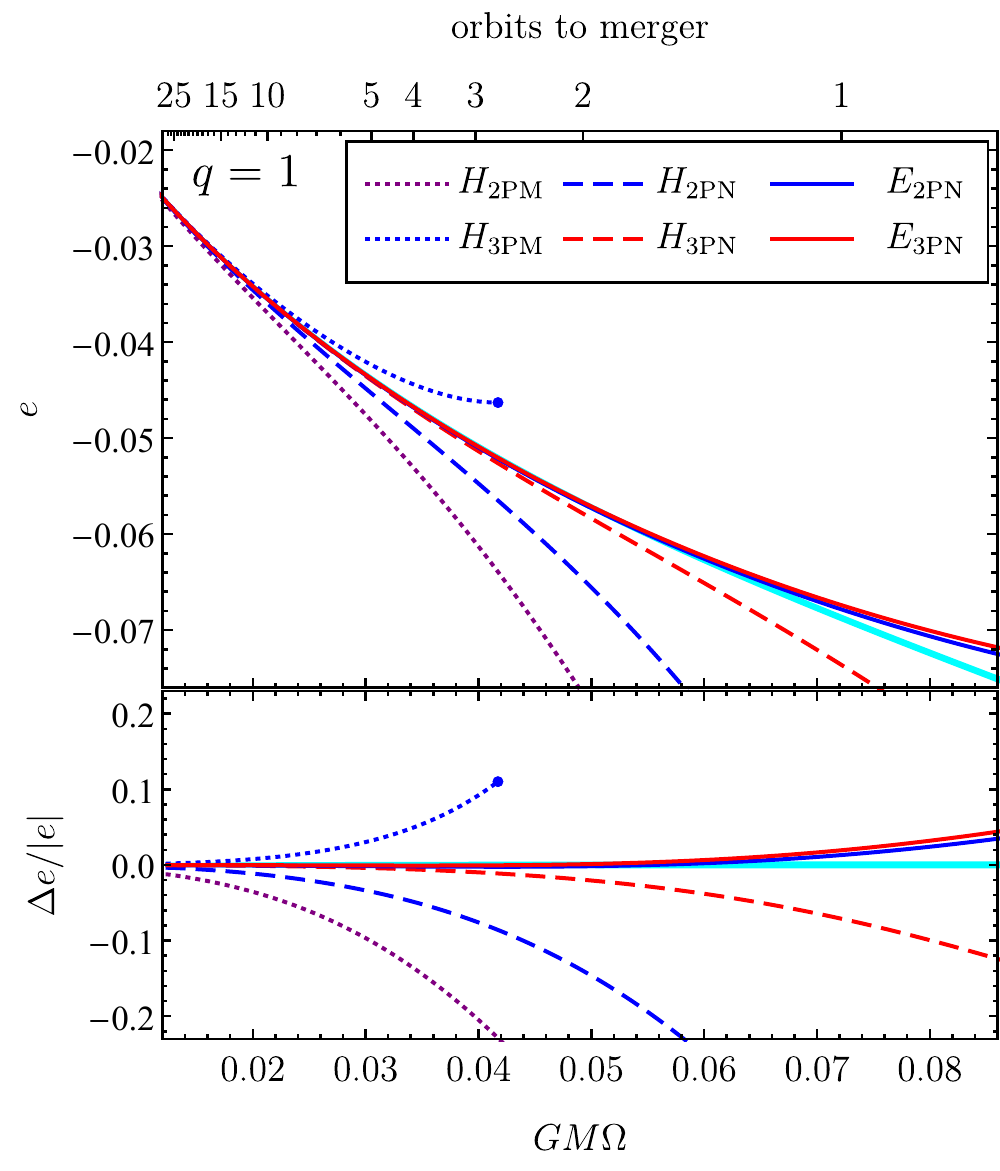}
	\caption{{\textbf{ Energetics of PM and PN approximants versus analytic PN calculations.}} As in previous figures, the dotted and dashed curves show the binding energy obtained numerically from the $H^{\text{EOB}}_{m \text{PM}}$ and $H^{\text{EOB}}_{n\text{PN}}$ Hamiltonians. The solid curves labeled ``$n$PN'' show the binding energy computed order-by-order in $GM\Omega$~\cite{Blanchet:2013haa}.\label{fig:vanPMvsPNq1}}
\end{figure}

In Fig.~\ref{fig:PMvsclassicEOB}, for mass ratio $q=1$, we
show how the $H^{\text{EOB,PS}}_{m \text{PM}}$ Hamiltonian compares with the 
(original) $H^{\text{EOB}}_{n \text{PN}}$ Hamiltonian, currently employed (after further improvements from NR simulations) 
in LIGO/Virgo searches and data analysis~\footnote{We note that the upper right panel of Fig. 4 in Ref.~\cite{Ossokine:2017dge} also shows a comparison between the binding energy from the EOB Hamiltonian and NR predictions. However, the agreement to NR differs from ours in Fig.~\ref{fig:PMvsclassicEOB}, because Ref.~\cite{Ossokine:2017dge} employs the EOB Hamiltonian where the potential for circular orbits has been resummed 
	as suggested in Ref.~\cite{Barausse:2009xi}, and it computes the binding energy through a quasi-circular inspiral.}. We find that $H^{\text{EOB}}_{n \text{PN}}$
always leads to a binding energy that is closer to the NR one. Thus,
we find that insights on (and explicit resummations for) the EOB
Hamiltonian from current PM results are not yet sufficient to improve
the accuracy of quasi-circular inspiral waveforms for LIGO/Virgo data
analysis. This is not entirely surprising, because the currently known
3PM level only covers completely the 2PN level of the PN
approximation; there is much room (hope) for improvement coming from
4PM. The conclusion is that it will be very useful to
extend the knowledge of PM calculations to higher orders --- for
example at least 4PM, but even 5PM order. 

Before ending this section we remark that the comparison
results that we have illustrated  depend on several choices. 
First of all, we have decided to compare the binding energy
extracted from NR simulations to results obtained from an adiabatic
sequence of circular orbits, instead of the ones from the Hamilton equations 
with radiation-reaction force. To illustrate the impact of this choice we compare in 
Fig.~\ref{fig:evolq1} the binding energies of $H^{\text{EOB}}_{3 \text{PN}}$
and $H^{\text{EOB,PS}}_{n \text{PN}}$ obtained by evolving the Hamilton equations with 
a suitable radiation-reaction force 
(labeled ``inspiral'') and using an adiabatic sequence of circular orbits (labeled ``circular''). 
The difference is small early in the evolution and grows as the inspiral approaches the ISCO,
where we observe a typical difference in the binding energy of 5\% to
10\% (for $q=1$).

Lastly, Fig.~\ref{fig:vanPMvsPNq1} demonstrates the difference of calculating
$e(\Omega)$ numerically, treating the various approximants of the
Hamiltonian as exact, and analytically as an expansion in
$(GM\Omega)$. The plots show the results of calculating $e(\Omega)$
numerically from $m$PM and $n$PN Hamiltonians treated as ``exact'', and also
the curves from the analytically computed binding-energy $E_{n\text{PN}}(\Omega)$ truncated at 2PN
(i.e., $(GM\Omega)^{6/3}$ with respect to leading term) and 3PN (i.e., $(GM\Omega)^{8/3}$) order (see Eq.~(232) in 
Ref.~\cite{Blanchet:2013haa}) (labeled $E_{n\text{PN}}$). As already noticed 
in Ref.~\cite{Buonanno:2005xu}, the differences can be quite substantial. However, it is
worth re-emphasizing that if one calculates $e(\Omega)$ analytically
starting from either $H_{3\text{PM}}$ or $H_{2\text{PN}}$ one recovers the
2PN result exactly.

\section{Conclusions}
\label{sec:concl}

The study of the energetics conducted in this work, using currently available PM Hamiltonians 
up to third order, highlights two main points.  Firstly, the binding energy 
for circular orbits computed with the 3PM Hamiltonian of Ref.~\cite{Bern:2019nnu}
and the 3PM EOB Hamiltonian of Sec.~\ref{sec:EOB3PM} are closer to NR predictions than 
the ones computed at lower PM orders, especially for small
frequencies (or high angular momenta) (see Figs.~\ref{fig:energywPM} and \ref{fig:energyJPM}). 
This suggests that similar improvements can be made by pushing PM calculations to higher orders, leading to
a more accurate modeling of the inspiral phase.

Secondly, we find that higher-order PM calculations of the conservative 
two-body dynamics would be needed to improve the agreement to NR and compete with (the
conservative part of) currently available waveform models used in LIGO/Virgo 
data analysis (see Figs.~\ref{fig:energymixed} and \ref{fig:PMvsclassicEOB}).
This is not surprising, since the 3PM order contains complete PN information only up to 2PN order, but also not obvious \textit{a priori}, since the 3PM approximation contains information not available in any of the PN expansions.

Furthermore, we have found that the PM EOB Hamiltonian of Sec.~\ref{sec:EOB3PM} (originally 
derived at 2PM order in Ref.~\cite{Damour:2017zjx}) gives 
good agreement against NR (and better agreement than the 3PM Hamiltonian of 
Ref.~\cite{Bern:2019nnu}), albeit not at the level of the PN 
EOB Hamiltonian~\cite{Buonanno:1998gg,Damour:2000we,Damour:2015isa} used to 
build waveform models for LIGO/Virgo data analysis (see Figs.~\ref{fig:energymixed} and ~\ref{fig:PMvsclassicEOB}). 
Relatedly, in Fig.~\ref{fig:energymixed} we have also shown the binding energy computed with a 3PM EOB Hamiltonian 
that we have derived in Appendix \ref{appendixB} and that differs from the one of Sec.~\ref{sec:EOB3PM} at 4PM order. 
Interestingly, we have found that such an alternative EOB Hamiltonian has much better agreement with NR than 
the one of Sec.~\ref{sec:EOB3PM} (e.g., confront the lower panels of Fig.~\ref{fig:energymixed} 
and Fig.~\ref{fig:PMvsclassicEOB}), reaching agreement similar to the EOB Hamiltonian 
employed to construct waveform models for LIGO/Virgo detectors (the latter would still do much better in the low frequency early inspiral not covered by the NR simulation). This rather 
encouraging result  motivates a more comprehensive study of EOB resummations of PM results. 

We recall that there are several caveats that underlie our investigation. 
To begin with, we have chosen to work in the circular-orbit approximation, rather than 
incorporating radiation-reaction effects and evolving the two-body system along an 
inspiraling orbit. This choice was dictated by the desire of clearly singling out 
the contribution coming from the conservative dynamics in the PM, PN and EOB descriptions. 
We have also decided to treat perturbative PM and PN Hamiltonians as exact when computing 
the binding energy. The effect of each of these choices has been illustrated 
in Figs.~\ref{fig:evolq1} and \ref{fig:vanPMvsPNq1}.

It is relevant to extend the above comparisons to quasi-circular inspiraling orbits (i.e., 
including  radiation-reaction effects), and we plan to do so in the near future.  
It would also be very interesting to perform the comparisons for scattering/unbound orbits, 
i.e. a setting closer to the natural domain of the PM approximation. It would be desirable,
for instance, to compare observables like the scattering angle against
NR simulations, as initiated in Ref.~\cite{Damour:2014afa}. Here, in a regime of high 
impact velocity and large impact parameter, the PM Hamiltonians are expected to behave better
than pure PN ones.

\section*{Acknowledgments}
We thank the authors of Ref.~\cite{Bern:2019nnu} for discussions and for sharing with us the results of their 3PM Hamiltonian 
while finalizing their paper. It is also a pleasure to thank Sergei Ossokine for making available to us the NR data for the binding energy and Chris Kavanagh for useful discussions.  

MvdM was supported by European Union's Horizon 2020 research and innovation program under grant agreement No.~705229.

  

\newcommand{\Htot}{$\Ham_{\text{SMR-3PN}}$}
\newcommand{\p}{\hat{p}}
\newcommand{\aaa}{a}
\newcommand{\ddd}{d}
\newcommand{\zzz}{\Delta z}
\newcommand{\EEE}{\hat{E}_{\text{SMR}}}
\newcommand{\X}{f}
\newcommand{\Ebindsmr}{\hat{E}^{\text{SF}}_{\text{bind}}}
\newcommand{\Ebindeob}{\hat{E}^{\text{EOB}}_{\text{bind}}}
\newcommand{\lnE}{\ln E_\text{S}^{-2}}

\chapter[Quasicircular inspirals and plunges from nonspinning effective-one-body$\dots$]{Quasicircular inspirals and plunges from nonspinning effective-one-body Hamiltonians with gravitational self-force information}
\label{chap:three}

\textbf{Authors}\footnote{Originally published in \emph{Phys.Rev.D} 101 (2020) 2, 024024.}: Andrea Antonelli, Maarten van de Meent, Alessandra Buonanno, Jan Steinhoff, Justin Vines.
\newline

\textbf{Abstract:}  the self-force program aims at accurately modeling relativistic
two-body systems with a small mass ratio (SMR). In the context of
the effective-one-body (EOB) framework, current results from this program
can be used to determine the effective metric
components at linear order in the mass ratio, resumming
post-Newtonian (PN) dynamics around the test-particle limit in the
process. It was shown in [Akcay et al., Phys. Rev. D \textbf{86}
(2012)] that, in the original (standard) EOB gauge,
the SMR contribution to the metric component  $g^\text{eff}_{tt}$ exhibits a coordinate singularity at the light-ring (LR) radius. In this paper, we adopt a different gauge for the
EOB dynamics and obtain a Hamiltonian that is free of poles at the
LR, with complete circular-orbit information at linear order in the
mass ratio and non-circular-orbit and higher-order-in-mass-ratio
terms up to 3PN order.  We confirm the absence of the LR-divergence in
such an EOB Hamiltonian via plunging trajectories through the LR radius.  Moreover, we compare the
binding energies and inspiral waveforms of EOB models with SMR, PN and mixed
SMR-3PN information on a quasi-circular inspiral against
numerical-relativity predictions. We find good agreement
between NR simulations and EOB models with SMR-3PN information for both equal and unequal
mass ratios. In particular, when compared to EOB inspiral waveforms with only 3PN information, EOB Hamiltonians with
SMR-3PN information improves the modeling of binary systems
with small mass ratios $q \lesssim 1/3$, with a dephasing accumulated in $\sim$30 gravitational-wave (GW) cycles being of the order of few hundredths of a radian up to 4 GW cycles before merger.

\section{Introduction}

Solving the two-body problem in General Relativity (GR) remains a challenge of both theoretical interest and astrophysical relevance. 
Albeit an analytical solution is lacking, advances in numerical relativity
(NR) in the past decades provided the first numerical evolutions of
merging compact objects \cite{Pretorius:2005gq,Campanelli:2005dd,Baker:2005vv}, as 
well as catalogs of waveforms \cite{Mroue:2013xna,Jani:2016wkt,Healy:2017psd,Dietrich:2018phi, Boyle:2019kee}. On the
analytical side of the problem, approximations to the binary motion and gravitational 
radiation, via expansions in one or more small parameters, have been applied to different 
domains of validity~\cite{Blanchet:2013haa,Barack:2018yvs,Tiec:2014lba}, providing us 
with a variety of waveform models.

The effective-one-body (EOB) framework is a synergistic approach that allows us to resum information from several analytical approximations.
NR-calibrated inspiral-merger-ringdown models based on EOB theory \cite{Pan:2011gk,Taracchini:2012ig,Taracchini:2013rva,Bohe:2016gbl,Cotesta:2018fcv} were employed by LIGO-Virgo experiments to detect gravitational waves (GWs) and infer astrophysical and cosmological information from them~\cite{Abbott:2016blz,Abbott:2016nmj,Abbott:2017vtc,Abbott:2017oio,Abbott:2017gyy,TheLIGOScientific:2017qsa,LIGOScientific:2018mvr,TheLIGOScientific:2016wfe,LIGOScientific:2018mvr,TheLIGOScientific:2016src}. In view of the expected increase in the signal-to-noise ratio of signals detected with upcoming LIGO-Virgo runs, and next generation detectors in space (LISA~\cite{Audley:2017drz}) and on Earth (Einstein Telescope~\cite{Punturo:2010zz} and Cosmic Explorer~\cite{Evans:2016mbw}),
it is important and timely to include more physics and build more accurate waveforms in the EOB approach. 

Historically, physical information from the two-body problem has
mostly entered EOB theory via the post-Newtonian (PN) expansion
\cite{Buonanno:1998gg, Buonanno:2000ef, Damour:2000we}, valid for
bound orbits at large distances and for velocities smaller than the
speed of light $v^2/c^2\sim\text{G}M/\R c^2 \ll 1$ (here $M=m_1+m_2$
is the total mass, with $m_1$ the mass of the primary and $m_2$ the
mass of the secondary body).  PN conservative-dynamics information has
so far been calculated up to fourth order, in the nonspinning
case, using the Arnowitt-Deser-Misner (ADM)
\cite{Jaranowski:2015lha,Damour:2014jta,Damour:2016abl}, Fokker
\cite{Bernard:2015njp,Bernard:2016wrg,Bernard:2017bvn} and
effective-field-theory approaches \cite{Foffa:2019rdf,Foffa:2019yfl} (which were also employed to determine the 5PN gravitational interaction in the static limit~\cite{Foffa:2019hrb,Blumlein:2019zku}). In the quasi-circular-orbit limit,
4PN information has been successfully included in the EOB dynamics in
the form of an expansion in the inverse radius $u \equiv \text{G}M/\R
c^2 \ll 1$ and in the momenta $\boldsymbol{p}^2$, with exact
dependence on the symmetric mass ratio $\nu=m_{1}m_{2}/M^{2}$
\cite{Damour:2015isa}. 
Further resummations of this PN expansion form the core of the EOB waveform models \cite{Damour:2008gu,Damour:2009kr,Barausse:2009xi,Pan:2011gk,Hinderer:2016eia}.
Post-Minkowskian (PM) information, valid in the weak field $\text{G}M/\R c^2 \ll 1$, but for all velocities $v^2/c^2\leq 1$, has also provided valuable insight in the structure of EOB Hamiltonians, for both spinning and non-spinning bound systems \cite{Damour:2017zjx, Damour:2016pm,Vines:2017pm,Vines:2018gqi,Antonelli:2019ytb}.

The self-force (SF) program, initiated in Refs.~\cite{Mino:1996nk,Quinn:1996am}  and based on an expansion of Einstein's equations in the small mass ratio (SMR) $q=m_2/m_1$,
has been successful in the calculation of the gravitational SF of a small body around Schwarzschild \cite{Barack:2007tm,Barack:2010tm},  and recently Kerr black-holes \cite{Shah:2012gu,vandeMeent:2015lxa,vandeMeent:2016pee,vandeMeent:2017bcc}, to first order in the mass ratio and for generic bound orbits. The results, corroborated by the use of several gauges and numerical techniques (see, e.g., Ref.~\cite{Barack:2018yvs} and references therein), have been already used to evolve extreme-mass-ratio-inspirals (EMRIs) around a Schwarzschild black-hole \cite{Warburton:2011fk,Osburn:2015duj} and they represent a key input for EMRI waveform modeling schemes recently developed \cite{vandeMeent:2018rms} and under development \cite{Hinderer:2008dm}.

As the SF program employs different gauge-dependent schemes to obtain
its results \cite{Barack:2018yvs}, it is paramount to be able to check
results via gauge-invariant quantities, such as the innermost-stable--circular-orbit (ISCO)-shift
\cite{Barack:2009ey}, periastron advance 
\cite{Barack:2010ny,LeTiec:2011bk,vandeMeent:2016hel}, spin-precession
\cite{Dolan:2013roa,Akcay:2016dku,Akcay:2017azq, Kavanagh:2017wot,Bini:2014ica,Bini:2018ylh},
tidal invariants \cite{Dolan:2014pja,Bini:2014zxa} and the Detweiler redshift
\cite{Detweiler:2008ft,Barack:2011ed,Akcay:2015pza,Kavanagh:2015lva,Shah:2013uya,Johnson-McDaniel:2015vva}.
For a particle with four-velocity $\tilde{u}^\alpha$ normalized in an
effective metric $\tilde{g}_{\alpha\beta}=g^\text{(0)}_{\alpha\beta}+h^{\text{R}}_{\alpha\beta}$
[i.e., moving around a Schwarzschild background
$g^\text{(0)}_{\alpha\beta}$ perturbed by a regularized metric
$h_{\alpha\beta}^{\text{R}}$ and such that
$\tilde{g}_{\alpha\beta}\tilde{u}^\alpha\tilde{u}^\beta=-1+\mathcal{O}(\nu)$],
the Detweiler redshift is defined as the ratio between proper time
measured in an orbit around the effective metric
$\tilde{g}_{\alpha\beta}$, $d\tilde{\tau}$, and coordinate time,
$d\tau$ \footnote{As pointed out in Ref.~\cite{Barack:2018yvs}, $z$
	does not correspond to the gravitational redshift due to the use of
	the regularized perturbation $h^{\text{R}}_{\alpha\beta}$ in its
	definition. It does only in the full geometry, e.g., including a
	singular metric $h^{\text{S}}_{\alpha\beta}$ at the location of the
	particle such that the body perturbation is $h_{\alpha\beta}\equiv
	h^{\text{R}}_{\alpha\beta}+h^{\text{S}}_{\alpha\beta}$. A sounder
	physical description can be obtained if the small companion is a
	black hole, since the Detweiler redshift can then be linked to the
	surface gravity $\kappa$ of the small
	body~\cite{Zimmerman:2016ajr}. }:
$z \equiv (\tilde{u}^t)^{-1}=d\tilde{\tau}/d\tau$
\cite{Detweiler:2008ft,Barack:2018yvs}.  Recently, the
Detweiler redshift has been used for cross-cultural studies between
approximations to the two-body problem in GR
\cite{Detweiler:2008ft,LeTiec:2011ab,Tiec:2014lba,Akcay:2015pza, Zimmerman:2016ajr}, and it has provided an important
benchmark to check PN and SMR results in the small-mass-ratio and
weak-field domain, in which both PN and SMR frameworks are expected to
be valid. This synergistic program has been extended 
to NR simulations of equal--mass-ratio binaries with the computation of 
the Detweiler redshift in Ref.~\cite{Zimmerman:2016ajr}.

As pointed out in Ref.~\cite{Damour:2009sf}, gauge-invariant SMR
quantities such as the Detweiler redshift can be also used to inform
the conservative sector of EOB Hamiltonians
\cite{LeTiec:2011dp,Damour:2009sf,Barausse:2011dq,LeTiec:2011bk}.
There are two ways in which this valuable information
could be incorporated into the EOB approach: it can be either used to partially determine high-order PN coefficients of EOB Hamiltonians \cite{kavanaghetal:2015,Hopper:2015icj,Kavanagh:2016idg,Bini:2013rfa,Bini:2014nfa,Bini:2015bla, Bini:2015bfb,Bini:2015xua,Bini:2016dvs,Bini:2016qtx} or it can be used to resum PN dynamics around the test-body limit \cite{LeTiec:2011dp,Barausse:2011dq,Akcay:2012ea}. Here, we focus on the latter approach. 

Currently available EOB Hamiltonians informed with the Detweiler redshift cannot be reliably evolved near the Schwarzschild light-ring (LR) radius, i.e., $\R=3\text{G}M/c^2$. Such an issue, hereafter called the \textit{LR-divergence} problem,  appears as a coordinate singularity of the effective Hamiltonian at the Schwarzschild LR \cite{LeTiec:2011dp,Akcay:2012ea}. In this paper we address the problem and, adopting a different EOB gauge, we obtain a Hamiltonian with SMR information that exhibits no divergence at the LR radius. This result allows us to use the precious near-LR, strong-field information from SF calculations in the evolutions of EOB Hamiltonians.

The organization of the paper is as follows.
In Sec.~\ref{sec:prelim} we review the LR-divergence arising from informing the conservative sector of standard EOB Hamiltonians with the Detweiler redshift and we discuss how a different EOB gauge (introduced in Ref.~\cite{Damour:2017zjx} in the context of PM calculations) helps to solve the issue.
In Sec.~\ref{sec:EOBSMR}, we inform the conservative sector of EOB Hamiltonians in the alternative gauge with circular-orbit information from the Detweiler redshift, and with both non-circular-orbit and higher-order-in-mass-ratio information from the PN approximation.
In Sec.~\ref{sec:evolutions}, we evolve quasi-circular inspirals from this LR-divergence-free Hamiltonian and show that the evolution of the orbital separation crosses the LR radius without encountering singularities. Moreover, we perform systematic comparisons against NR predictions of phase and binding energy for non-spinning systems with mass ratios $1/10 \leq q \leq 1$. We conclude in Sec.~\ref{sec:conclusions}. In Appendix~\ref{sec:appred} we present high-precision fits to the Detweiler redshift with improved data in the strong field. 
We use geometric units G=$c$=1 throughout the paper.

\section{On gauges and the light-ring divergence}
\label{sec:prelim}

We begin by noting some conventions to be used in the following sections. In the present paper, we do not consider spinning systems; we denote the reduced mass by $\mu=(m_1 m_2)/M$ and the total mass by $M=m_1+m_2$.  We work with generalized (polar) coordinates ${q}_a\equiv (\R,\phi)$ in the orbital plane, with canonically conjugate momenta $p_a\equiv(p_r,p_\phi)$, and we often employ the mass-reduced inverse orbital separation $u \equiv M/\R$ and the mass-reduced momenta $\p_r \equiv \Pp_r/\mu$ and $\p_\phi\equiv \Pp_\phi/(M\mu)$.

\subsection{The light-ring divergence}

In the EOB approach, the real two-body motion is mapped to the
effective motion of a test body in an effective \textit{deformed}
Schwarzschild spacetime with coordinates $(t,r,\theta,\phi)$, with the deformation parameter being the symmetric mass ratio $\nu$. 
The mapping can be obtained via a dictionary between the action integrals $I_a=(2\pi)^{-1}\oint
{\Pp}_a\, d{q}_a$ of a two-body system in the
center-of-mass frame and those of a test-body moving in the effective
metric $g_{\mu\nu}^{\text{eff}}$. Considering orbits in the equatorial plane $\theta=\pi/2$, identifying the radial and angular
action integrals of real and effective systems, i.e., setting
$I_\text{\R}^{\text{real}}=I_\text{\R}^{\text{eff}}$ and
$I_\phi^{\text{real}}=I_\phi^{\text{eff}}$, the EOB approach allows a
simple relation between the real
$H\subreal(r,p_r,p_\phi,\nu)$ and effective
$H_{\text{eff}}(r,p_r,p_\phi,\nu)$ Hamiltonians
\cite{Buonanno:1998gg}:
\begin{equation}\label{enmap}
H\subreal \equiv M \hat H\subreal =M\sqrt{1+2\nu\bigg(\frac{H_{\text{eff}}}{\mu}-1\bigg)}\,.
\end{equation}
$H_{\text{eff}}$ describes the motion of a test body with mass $\mu$ and is determined by a mass-shell constraint of the form~\cite{Damour:2000we}
\begin{equation}\label{HamJac}
g^{\mu\nu}_\text{eff}\Pp_\mu\Pp_\nu+\mu^2+Q(r,p_r,p_\phi,\nu)
=0,
\end{equation}
where the effective metric is given by
\begin{align}\label{effmetric}
ds^{2}=-A(\R,\nu)dt^{2}+[A(\R,\nu)\bar D(\R,\nu)]^{-1}d\R^{2}+\R^{2} d\Omega^{2}\,,
\end{align}
with the potentials $A(\R,\nu)$ and $\bar D(\R,\nu)$ depending on the orbital separation $r$ and the symmetric mass ratio $\nu$. In terms of the inverse radius $u=M/r$, they reduce to $A_0(u)=1-2u$ and $\bar D_0=1$ in the test particle limit ($\nu\rightarrow0$). Inserting the inverse of the metric~\eqref{effmetric} into Eq.~\eqref{HamJac}, and using $\Pp_\mu=(-H_\text{eff},\Pp_r,\Pp_\theta=0,\Pp_\phi)$, the mass-reduced effective Hamiltonian $\Ham_{\text{eff}}\equiv H_{\text{eff}}/\mu$ is found to be \cite{Damour:2000we}
\begin{align}\label{HeffDJS}
\Ham_{\text{eff}}^2&=A(u,\nu)\big[1+\p_{\phi}^{2}u^{2}+A(u,\nu)\bar D(u,\nu)\p_{r}^{2}+\hat{Q}(u,\p_r,\p_\phi,\nu)\big],
\end{align}
with $\hat Q\equiv Q/\mu^2$.  The non-geodesic function $ Q$ in
Eq.~\eqref{HamJac} has been introduced to extend the EOB Hamiltonian 
through 3PN order without changing the mapping (\ref{enmap}) (for a geodesic 
one-body motion at 3PN order with an energy map different from 
(\ref{enmap}) see Appendix A in Ref.~\cite{Damour:2000we}). Its
mass-reduced form $\hat{Q}(u,\p_r,\p_\phi,\nu)$ in Eq.~\eqref{HeffDJS} generically depends on both the mass-reduced
radial momentum $\p_{r}$ and the mass-recuded angular momentum $\p_{\phi}$. Reference \cite{Damour:2000we} 
showed that at 3PN order $\hat{Q}$ must be
fourth order in the momenta, and that the non-geodesic term 
is not uniquely fixed. By setting some of the free parameters to zero, it is possible 
to make the function  $\hat{Q}(u,\p_r,\p_\phi,\nu)$ depend only on the radial momentum  
[i.e., $\hat Q(u,\p_r,\p_\phi,\nu)\rightarrow\hat
Q(u,\p_r,\nu)$]. Since 2000, this choice of $\hat{Q}$ 
has been adopted in several EOB papers [although see Refs.~\cite{Damour:2002vi,Buonanno:2007pf} for alternative choices 
of $\hat{Q}$]. Henceforth, we shall 
denote the $\hat{Q}$ function that only depends on the radial momentum 
as $\hat{Q}^{\text{DJS}}({u},\p_r,\nu)$, after the 
initials of the three authors of
Ref.~\cite{Damour:2000we}. We refer to the DJS EOB Hamiltonian as 
the Hamiltonian that uses the $\hat{Q}^{\text{DJS}}({u},\p_r,\nu)$ function. 
Note that in this gauge, the angular momentum $\p_{\phi}$ only appears in the second term in brackets in
Eq.~\eqref{HeffDJS}. Moreover, in the circular orbit limit ($\p_{r}=0$) the
conservative dynamics information is fully described by the $A(u,\nu)$
potential in this gauge, as found at 2PN order~\cite{Buonanno:1998gg}.  
The 4PN expressions for $A(u,\nu)$, $\bar D(u,\nu)$
and $\hat Q^{\text{DJS}}({u},\p_{r},\nu)$ in the DJS gauge, for quasi-circular orbits,  
are obtained mapping Eq.~\eqref{enmap} to the 4PN-expanded Hamiltonian and can be
found in Ref.~\cite{Damour:2015isa}.

The first efforts to incorporate SMR quantities in EOB Hamiltonians sought to do so 
using the gauge of Eq.~\eqref{HeffDJS} with $\hat Q(u,\hat p_{r},\hat p_{\phi},\nu)\rightarrow \hat Q^{\text{DJS}}(u,\hat p_{r},\nu)$
\cite{Barausse:2011dq,Akcay:2015pjz,Barack:2010ny,Akcay:2012ea}. In this gauge, the function $A(u, \nu)$, having the complete dynamical information for circular orbits, allows a linear-in-$\nu$ expansion about the Schwarzschild limit:
\begin{align}\label{ADnuexp}
&A(u, \nu)=1-2u+\nu \aaa (u)+\mathcal{O}(\nu^2)\,.
\end{align}
The $\aaa(u)$ function resums the complete circular-orbit PN dynamics in linear order in $\nu$.
References~\cite{LeTiec:2011dp,Barausse:2011dq} obtained an expression for $\aaa(u)$ employing the linear-in-$\nu$ correction to the Detweiler redshift. Notably, the Detweiler redshift is expanded around the Schwarzschild background, $z(x)=\sqrt{1-3x}+\nu \zzz(x)+\mathcal{O}(\nu^2)$ [where $x\equiv (M\Omega)^{2/3}$ is the gauge-independent inverse radius], and the $\zzz$ correction is linked to $\aaa(u)$ via the first law of binary black-hole mechanics \cite{LeTiec:2011ab}.
The resulting expression reads: 

\begin{equation}\label{Asf}
\aaa(u)= \zzz(u)\sqrt{1-3u}-u\bigg(1+\frac{1-4u}{\sqrt{1-3u}}\bigg)\,.
\end{equation}
In Eq.~\eqref{Asf}, $\zzz$ depends on the gauge-dependent inverse radius $u$, rather than its gauge-independent counterpart $x$. This is only correct if we restrict to first order in $\nu$, since $x=u+\mathcal{O}(\nu)$.
The quantity $\zzz(x)$, has been fitted with data extending to the LR \cite{Akcay:2012ea}, allowing precious strong-field information to enter the EOB dynamics. 

The form of $\aaa(u)$ is suggestive of trouble arising at the Schwarzschild light ring, i.e., at $u_{\text{LR}}=1/3$, where the second term in Eq.~\eqref{Asf} diverges. In principle, this divergence might be tamed by the behaviour of the redshift $\zzz(u)$ appearing in the first term in brackets, but data for the redshift up to the LR show that this is not the case and that $\aaa(u)$ indeed diverges there~\cite{Akcay:2012ea}. This is worrisome, as $\aaa(u)$ directly enters the effective Hamiltonian and, via the energy map, the EOB-resummed dynamics. The EOB dynamics thus contains a divergence \textit{for generic orbits} (e.g., for any value of $\p_{\phi}$ and $\p_{r}$).	
It was pointed out in Ref.~\cite{Akcay:2012ea} that the LR-divergence is a phase-space coordinate singularity that arises due to the use of the DJS gauge, and that can be solved adopting a different gauge in which the function $\hat Q$ grows as $\hat Q\propto \p_\phi^3$ when $\p_\phi\rightarrow\infty$ and $\p_r\rightarrow0$. 

It is worth mentioning that the argument in Ref.~\cite{Akcay:2012ea} stems from a similar LR divergence that has appeared when including tidal effects 
in the EOB approach \cite{Bini:2012gu}. Tidal effects enter the potential $A(u)$ via a correction in a tidal expansion akin to Eq.~\eqref{ADnuexp}: $A(u)=A_{\text{2pp}}+\mu_\text{T} a_\text{T}(u,\nu)+\mathcal{O}(\mu_\text{T}^2)$, where $A_{\text{2pp}}$ is the two point-particle (pp) EOB potential \cite{Bini:2012gu} and $\mu_\text{T}$ the small tidal parameter. It has been found in Ref.~\cite{Bini:2012gu} that, in the extreme-mass-ratio limit and for circular orbits, the first-order correction scales as $a_\text{T}(u,\nu)\propto (1-3u)^{-1}$ when $u\rightarrow u_\text{LR}$. An alternative EOB Hamiltonian that includes dynamical tides without introducing poles at the LR has been introduced in Ref.~\cite{Steinhoff:2016rfi}; this has been achieved by abandoning the DJS gauge (see, e.g., their Appendix D).

\subsection{The post-Schwarzschild effective-one-body gauge}

Reference~\cite{Damour:2017zjx} has shown that it is possible to obtain a different EOB gauge, hereafter the \textit{post-Schwarzschild} (PS) gauge, solving Eq.~\eqref{HamJac} with the Schwarzschild limit of the metric~\eqref{effmetric}.
The mass-reduced effective Hamiltonian thus obtained has the following form:
\begin{equation}\label{HeffEG}
\Ham_{\text{eff}}^{\text{PS}}=\sqrt{\Ham_{\text{S}}^{2}+(1-2u)\hat{Q}^{\text{PS}}(u,\nu,\Ham_{\text{S}})}\,,
\end{equation}
where $\Ham_{\text{S}}$ is the Schwarzschild Hamiltonian:
\begin{equation}\label{Hs}
\Ham_{\text{S}}(u,\p_{r},\p_{\phi})=\sqrt{(1-2u)\left [1+\p_{\phi}^{2}u^{2}+(1-2u)\p_{r}^{2}\right ]}\,.
\end{equation} 

In Ref.~\cite{Damour:2017zjx}, the PS function $\hat{Q}^{\text{PS}}$ has been derived to 2PM order via a scattering-angle calculation and to 3PN order via a canonical transformation from the DJS Hamiltonian at 3PN. In Ref.~\cite{Antonelli:2019ytb}, these calculations have been extended to 3PM and 4PN orders, respectively (the latter only in the near-circular orbit limit).

It is noticed that, in PS EOB Hamiltonians, all the information on the two-body problem with $\nu \neq 0$ is contained in $\hat{Q}^{\text{PS}}(u,\nu,\Ham_{\text{S}})$. This feature and the fact that circular-orbit dynamics is contained also in the $\hat{Q}$ function, significantly differentiate PS Hamiltonians from DJS ones. The PS gauge is uniquely fixed resumming the angular and radial momenta into the Schwarzschild Hamiltonian \eqref{Hs}. The powers of such momenta are furthermore not bound in any way, due to the generic functional dependence of $\hat{Q}^{\text{PS}}(u,\nu,\Ham_{\text{S}})$ on $\Ham_{\text{S}}$. In principle, then, arbitrary powers of $\p_{\phi}$ are contained in $\hat{Q}^{\text{PS}}(u,\nu,\Ham_{\text{S}})$ via $\Ham_{\text{S}}$. In particular, differently from $\hat{Q}^{\text{DJS}}(u,\nu,\p_r)$,  powers of momentum enter at second order 
in $\hat{Q}^{\text{PS}}(u,\nu,\Ham_{\text{S}})$ instead of fourth order.

The unconstrained dependence of $\hat{Q}^{\text{PS}}$ on $\Ham_{\text{S}}$ makes the use of PS Hamiltonians very appealing in the context of our work. It was shown in Ref.~\cite{Damour:2017zjx} that, in the high energy limit for which $\p_{\phi}\rightarrow\infty$, 
the LR-divergence can be captured by the coefficient of a term proportional to $\Ham_{\text{S}}^3$. This result is in agreement with a point made in the conclusions of Ref.~\cite{Akcay:2012ea}. As it approaches the LR radius, the effective mass moving in a deformed-Schwarzschild background described by Eqs.~\eqref{ADnuexp} and~\eqref{Asf} has a divergent-energy behaviour that must be removed with an appropriate energy-corrected mass-ratio parameter $\tilde\nu=\nu \Ham_{\text{S}}$.
In the next section, building from this knowledge and making use of a simple ansatz for $\hat{Q}^{\text{PS}}(u,\nu,\Ham_{\text{S}})$, we construct a Hamiltonian in the PS gauge that contains information from $\zzz$, while remaining analytic at the LR.

\section{Conservative dynamics of post-Schwarzschild Hamiltonians}
\label{sec:EOBSMR}

\subsection{Information from circular orbits}\label{sec:preamble}
In this section, we link the conservative sector of the PS EOB Hamiltonian to the SMR contribution to $\zzz$. 
Following Ref.~\cite{Barausse:2011dq}, we do so matching, at fixed frequency, the circular orbit binding energy at linear order in $\nu$ from the EOB Hamiltonian with the binding energy in the same limit from SF results. The latter is obtained in Ref.~\cite{LeTiec:2011dp} and is a consequence of the first law of binary-black-hole mechanics. As a function of 
$\zzz$ and the gauge-invariant inverse radius $x$, it reads \cite{LeTiec:2011dp}:

\begin{equation}\label{EsEsf}
\Ebindsmr=\frac{1-2x}{\sqrt{1-3x}}-1+\nu \EEE(x,\zzz,\zzz')+\mathcal{O}(\nu^2)\,,
\end{equation}
\begin{align}\label{Ezold}
\EEE(x,\zzz,\zzz')=&-1+\sqrt{1-3 x}-\frac{x}{3}
\zzz'(x)+\frac{\zzz(x)}{2}+\frac{(7-24 x) x}{6
	(1-3 x)^{3/2}}\,.
\end{align}
The prime denotes differentiation with respect to $x$.
We find it useful to rewrite the redshift as:

\begin{equation}\label{zsmr}
\zzz(x)=\frac{\zzero(x)}{1-3x}+\frac{\zone(x)}{\sqrt{1-3x}}+\frac{
	\ztwo(x)}{1-3x}\lnE(x)
\,,
\end{equation}
In the above expression, we have defined $E_\text{S}(x)\equiv (1-2x)/\sqrt{1-3x}$.
In Appendix~\ref{sec:appred}, $\zzero(x)$, $\zone(x)$ and $\ztwo(x)$ are fitted to high-precision SF data and such to be analytic at the LR. 
Equation~\eqref{Ezold} then reads:
	\begin{align}\label{Esf}
	\EEE=&\sqrt{1-3 x}-1+\frac{(7-24 x) x}{6 (1-3 x)^{3/2}}
	\nonumber\\
	&
	+\frac{1}{2(1-3x)}
	\bigg[\zzero(x)+\zone(x)\sqrt{1-3 x}
	+\ztwo(x)\lnE(x)\bigg]
	\nonumber\\
	&
	-\frac{x}{3(1-3x)}\bigg\{\frac{3\zzero(x)}{1-3 x}+\frac{3 \zone(x)}{2 \sqrt{1-3 x}}
	\nonumber\\
	&
	+\bigg[\frac{1-6 x}{(1-2x)(1-3x)}+\frac{3\lnE(x)}{(1-3x)}\bigg]\ztwo(x)
	\nonumber\\
	&
	+(\zzero)'(x)+ \sqrt{1-3 x}\,(\zone)'(x)+ (\ztwo)'(x)\lnE(x)\bigg\}\,.
	\end{align}
For the remainder of this section, we consider the PS EOB Hamiltonian $H\subreal$, i.e. Eq.~\eqref{enmap} with $H_{\text{eff}}/\mu$ replaced by $\hat H^{\text{PS}}_{\text{eff}}$ of Eq.~\eqref{HeffEG}. We propose an ansatz for $\hat Q^{\text{PS}}$ of the following form:

\begin{equation}\label{HSMR}
\hat{Q}^{\text{PS}}_{\text{SMR}}(u,\nu,\Ham_{\text{S}})=\nu\big[\X_{0}(u)\Ham_{\text{S}}^{5}+\X_{1}(u)\Ham_{\text{S}}^{2}+\X_{2}(u)\Ham_{\text{S}}^{3}\ln\Ham^{-2}_{\text{S}}\big]\,.
\end{equation}
In the rest of this section, when matching to the SMR results, we limit to circular orbits; thus we use $\Ham_{\text{S}}(u,\p_r=0,\p_\phi)$ in Eq.~(\ref{HSMR}). The role of the $\Ham_{\text{S}}^5$ term is to capture the global divergence $(1-3x)^{-2}$ of Eq.~\eqref{Esf}\footnote{In principle, a $\Ham_{\text{S}}^3$ term will suffice to capture the divergence. However, we find that this minimal choice leads to evolutions that are not well behaved for systems with comparable masses.}, while the second term $\Ham_{\text{S}}^2$ is devised to incorporate the $\sqrt{1-3x}$ terms appearing in the numerator of the same equation, which would make the Hamiltonian imaginary after the light ring. The term proportional to $\ln\Ham_{\text{S}}^{-2}$ incorporates the logs in the fit that would make the Hamiltonian non-smooth at the light ring.
Setting $\Pp_r=0$ and using:
\begin{equation}
\dot{\Pp}_r=-\frac{\partial  H\subreal}{dr}(r,\Pp_r=0,\Pp_\phi^{\text{circ}},\nu)=0\,,
\end{equation}
the (mass-reduced) circular-orbit momentum $\p_\phi^{\text{circ}}$ as a function of the inverse radius $u$ is determined at linear order in $\nu$ (with $\X_i'(u)=d\X_i/du$):
	\begin{align}\label{pu}
	\p_{\phi}^{\text{circ}}(u,\nu)=&\frac{1}{\sqrt{u(1-3u)}}+\nu\, 
	\frac{(1-2 u)^2 }{4 (1-3
		u)^3 \sqrt{u}}\bigg[2(1-2u)^3\X_{0}(u)\nonumber\\
	&+2(1-3u)^{3/2}\X_1 (u)+2(1-2u)(1-3u)\X_2(u)\lnE (u)\nonumber\\
	&-(1-2u)^4\X'_{0}(u)-(1-2u)(1-3u)^{3/2}\X'_{1}(u)\nonumber\\
	&-(1-2u)^2(1-3u)\lnE(u)\X'_2(u)\bigg]+\mathcal{O}(\nu^2)\,.
	\end{align}
We further use the relation:

\begin{equation}\label{omega}
\Omega=\frac{\partial H\subreal}{d\Pp_\phi}(r,\Pp_r=0,\Pp_{\phi}^{\text{circ}},\nu)\,,
\end{equation}
and exploit its link to the gauge-independent inverse radius $x$ given by $x=(M\Omega)^{2/3}$.
Inserting Eq.~\eqref{pu} in Eq.~\eqref{omega} and inverting the obtained expression at linear order in $\nu$,
we establish a link between the gauge-dependent $u$ and the gauge-independent $x$ inverse radii:
	\begin{align}\label{ux}
	u^{\text{circ}}&(x,\nu) = x+
	\frac{x\,\nu}{6 (1-3
		x)^{3/2}}
	\bigg\{
	4-20x+24 x^2-(4-12x) \sqrt{1-3 x}\nonumber\\
	&-10 (1-2 x)^4 \X_0(x)-4 \sqrt{1-3 x} \left(1-5x+6 x^2\right) \X_1(x)\nonumber\\
	&+\big[4 - 28 x + 64 x^2 - 48 x^3 -(6-42x+96x^2-72x^3)\lnE(x)\big] \X_2(x) \nonumber\\
	&+\left(1-10x+40x^2-80x^3+80x^4-32x^5\right) \X_0'(x)\nonumber\\
	&+\sqrt{1-3 x}\left(1-7x+16x^2-12x^3\right)
	\X_1'(x)\nonumber\\
	&+\left(1-9x+30x^2-44x^3+24x^4\right)\lnE(x) \X_2'(x)
	\bigg\}+\mathcal{O}(\nu^2)\,.
	\end{align}
To calculate the (mass-reduced) gauge-invariant, circular-orbit binding energy at linear order in $\nu$ from $H\subreal$, we employ the definition :
\begin{equation}\label{Ebind}
\Ebindeob \equiv  (H\subreal-M)/\mu\,,
\end{equation}
Inserting Eqs.~\eqref{pu} and \eqref{ux} in $H\subreal$ and retaining only terms up to first order in the mass ratio, we get:
	\begin{align}\label{Eeob}
	\Ebindeob&(x,\nu)=\frac{1-2x}{\sqrt{1-3 x}}-1-\frac{\nu}{6(1-3x)^3}
	\bigg\{
	(1-3x)(6-37x+59 x^2-12 x^3)
	\nonumber\\
	&-2(1-3 x)^{3/2} (3-14x+12x^2)
-\left(3-7x-18 x^2\right) (1-2 x)^4 \X_0(x)
    \nonumber\\
    &-(1-3x)^{3/2} \left(3-16 x+20 x^2\right)
	\X_1(x)\nonumber\\
	&+(1-3x)(1-2 x)^2  \big[2x (1-6 x)-\left(3-9x-6 x^2\right) \lnE(x)\big]\X_2(x)
	\nonumber\\
	&+2 x (1-2 x)^5 (1-3 x) \X_0'(x)+2x (1-3 x)^{5/2} (1-2 x)^2 \X_1'(x)\nonumber\\
	&+2x (1-3 x)^2 (1-2 x)^3 \lnE(x) \X_2'(x)
	\bigg\}+\mathcal{O}(\nu^2)\,.
	\end{align}
Matching Eq.~\eqref{EsEsf} [with correction given by Eq.~\eqref{Esf}] and Eq.~\eqref{Eeob}, we obtain differential equations to be solved for $\X_0(x)$, $\X_1(x)$ and $\X_2(x)$.
Further splitting the $\X_i$ coefficients as follows:
\begin{align}
&\X_0(x)=\tilde{\X}_0(x)+\sum_{i=0}^{i=2}\X_0^{(i)}(x)\Delta z^{(i)}(x)\label{X0ans1}
\\
&\X_1(x)=\tilde{\X}_1(x)+\sum_{i=0}^{i=2}\X_1^{(i)}(x)\Delta z^{(i)}(x)\label{X1ans1}\,,\\
&\X_2(x)=\tilde{\X}_2(x)+\sum_{i=0}^{i=2}\X_2^{(i)}(x)\Delta z^{(i)}(x)\label{X2ans1}\,,
\end{align}
and imposing that the Hamiltonian coefficients be analytic at the LR radius (i.e., that they do not contain  $\sqrt{1-3x}$ or $\lnE(x)$ terms),
we obtain the following non-zero solutions\footnote{Similarly to what is done in Ref.~\cite{Barausse:2011dq}, we impose that the PN expansion cannot admit half-integer powers of $x$. This allows us to set all constants of integration to zero.}:
\begin{subequations}
	\begin{align}
	&\tilde{\X}_0(x)=-\frac{x (1-3x)\left(1-4x
		\right)}{(1-2x)^5}\,,\\
	&\tilde{\X}_1(x)=-\frac{x 
	}{(1-2 x)^2}\,,\\
	&\X_0^{(0)}(x)=\frac{1-3x}{(1-2x)^5}\,,\\
	&\X_1^{(1)}(x)=\frac{1}{(1-2x)^2}\,,\\
	&\X_2^{(2)}(x)=\frac{1}{(1-2x)^3}\,.
	\end{align}
\end{subequations}
The $\X_{i}(x)$ coefficients are readily found via Eqs.~\eqref{X0ans1}, ~\eqref{X1ans1} and ~\eqref{X2ans1} and then inserted in the non-geodesic term in the effective Hamiltonian \eqref{HSMR} to obtain:
\begin{align}\label{QSMR}
\frac{\hat{Q}^{\text{PS}}_{\text{SMR}}}{\nu}(u,\nu,\Ham_{\text{S}})=&(1-3u)\bigg[\frac{\zzero(u)}{(1-2u)^5}-\frac{(1-4 u)\text{ }u}{(1-2 u)^5}\bigg]\Ham_{\text{S}}^5\\
&+\bigg[\frac{\zone(u)}{(1-2 u)^2}-\frac{u}{(1-2
	u)^2}\bigg]\Ham_{\text{S}}^2+\frac{\ztwo(u)}{(1-2u)^3}\Ham_{\text{S}}^{3}\ln\Ham_{\text{S}}^{-2}\,.\nonumber
\end{align}
We see that the resulting Hamiltonian concisely resums the complete circular-orbit PN dynamics at linear order in $\nu$. The non-geodesic function $\hat{Q}^{\text{PS}}_{\text{SMR}}$ does not contain any term divergent at the LR, as $\zzero(u)$, $\zone(u)$ and $\ztwo(u)$ are constructed to be analytic there.

\subsection{Information from non-circular orbits and from higher orders in the mass ratio}\label{sec:noncirc}

The calculation in Sec.~\ref{sec:preamble} is carried out in the circular-orbit limit at linear order in the mass ratio. However, it is  possible to include more physical information to the Hamiltonian, coming both from non-circular-orbit terms and from terms at higher orders in the mass ratio. For instance, self-force information for mildly eccentric orbits can be obtained via the SMR correction to the periastron advance $\rho_\text{SF}$ \cite{Barack:2010ny}, which can then be linked to the EOB potentials. This was the strategy used in Refs.~\cite{Damour:2009sf,Barausse:2011dq} to obtain an expression for the potential $\bar D(r)$ in terms of $\zzz(u)$ and $\rho_\text{SF}(u)$ and introduce non-circular SF data into the EOB Hamilonian up to the Schwarzschild ISCO (i.e., $u_\text{ISCO}=1/6$). Alternatively, one can exploit the generalized redshift~\cite{Barack:2011ed} and link it to $\bar D(r)$, as done in Refs.~\cite{Tiec:2015cxa,Akcay:2015pjz}. 
Here, we insert generic-orbit PN information in our Hamiltonian and leave the inclusion of non-circular SMR information in $\hat Q^{\text{PS}}$  to future work.

Post-Schwarzschild EOB Hamiltonians with PN information from generic-orbits have been already considered in the literature. For example, the PS Hamiltonians at 3PN order has been investigated in Ref.~\cite{Damour:2017zjx}.
Using the PN parameters $Y \equiv (\hat H^{2}_{\text{S}}-1)\sim \mathcal{O}(1/c^2)$ and $u$, its expression is given by:
\begin{align}
\label{QPScoeff}
\hat Q^{\text{PS}}_{\text{3PN}}=&3\nu u^{2}Y+5\nu u^{3}+\bigg(3\nu-\frac{9}{4}\nu^{2}\bigg)u^{2}Y^{2}
\\
&+\bigg(27\nu-\frac{23}{4}\nu^{2}\bigg)u^{3}Y+\bigg(\frac{175}{3}\nu-\frac{41\pi^{2}}{32}\nu-\frac{7}{2}\nu^{2}\bigg)u^{4}\,.\nonumber
\end{align}
As discussed, the above Hamiltonian contains two-body information that is not captured by the calculation leading to $\hat{Q}^{\text{PS}}_{\text{SMR}}$ and that we wish to add to it.

To this end, we consider a mixed SMR-3PN non-geodesic function of the following form:

\begin{equation}\label{Qtot}
\hat Q^{\text{PS}}_{\text{SMR-3PN}}=\hat Q^{\text{PS}}_{\text{SMR}}+\Delta\hat Q^{\text{PS}}\,,
\end{equation}
where $\hat Q^{\text{PS}}_{\text{SMR}}$ is given by Eq.~\eqref{QSMR} and contains all the circular-orbit terms at linear order in $\nu$, while $\Delta\hat Q^{\text{PS}}$ is fixed demanding that it contains all the additional PN information from Eq.~\eqref{QPScoeff}, in such a way not to contribute to the linear-in-$\nu$ binding energy in the circular-orbit limit. 

We opt to further split $\Delta\hat Q^{\text{PS}}$ into two contributions: $\Delta\hat Q^{\text{PS}}_{\text{extra}}$ collects the extra terms up to 3PN order (including both non-circular 3PN terms at linear order in $\nu$ and $\nu^2$ terms), while $\Delta\hat Q^{\text{PS}}_{\text{count.}}$ is a counterterm whose functionality is explained below. We then have:
\begin{equation}\label{Q4pnsplit}
\Delta\hat Q^{\text{PS}} \equiv \Delta\hat Q^{\text{PS}}_{\text{extra}}-\Delta\hat Q^{\text{PS}}_{\text{count.}}\,.
\end{equation}
The former contribution is readily obtained calculating the difference between Eq.~\eqref{QPScoeff} and the 3PN expansion of Eq.~\eqref{QSMR}\footnote{That is, Eq.~\eqref{QSMR} is expanded in the PN parameters $u$ and $Y=\hat H_{\text{S}}^2-1$. The redshift functions $\zzero(u)$, $\zone(u)$ and $\ztwo(u)$ also need to be PN expanded: their expressions are obtained matching the 3PN expansion of the redshift from Ref.~\cite{Kavanagh:2016idg} and Eq.~\eqref{zsmr}.}. The result reads:
\begin{align}\label{Qpn1}
\Delta\hat Q^{\text{PS}}_{\text{extra}}=& 3\nu u^2 Y+\bigg(3\nu-\frac{9}{4} \nu^2 \bigg) u^2Y^2+3 \nu u^3 \nonumber\\
&+\bigg(22\nu-\frac{23}{4} \nu^2 \bigg) u^3Y+\bigg(16 \nu-\frac{7}{2} \nu^2 \bigg)  u^4\,.
\end{align}
In the PS gauge $\hat Q^{\text{PS}}$ depends on momenta via $\hat H_\text{S}(u,\p_{\R},\p_{\phi})$, which cannot be separated into circular and non-circular orbit contributions.
Because of that, the linear-in-$\nu$ portion of Eq.~\eqref{Qpn1} contributes to the linear-in-$\nu$ binding energy for circular orbits.
Therefore, the addition of $\Delta\hat Q^{\text{PS}}_{\text{extra}}$ to $\Delta\hat Q^{\text{PS}}_{\text{SMR}}$ spoils the matching between EOB and SF binding energies for circular orbits at linear order in the mass ratio guaranteed by the sole presence of $\Delta\hat Q^{\text{PS}}_{\text{SMR}}$.

The matching between the two binding energies can be maintained with a particular choice of the second contribution to Eq.~\eqref{Q4pnsplit}, i.e., $\Delta\hat Q^{\text{PS}}_{\text{count.}}$. We choose a counterterm that starts at 4PN, in order not to spoil the agreement at 3PN for generic orbits guaranteed by Eq.~\eqref{Qpn1}:
\begin{equation}\label{Qpn2}
\Delta\hat Q^{\text{PS}}_{\text{count.}}=\nu \big[q_{(3,2)}u^3Y^2+q_{(4,1)}u^4Y+q_{(5,0)}u^5\big]\,.
\end{equation}
We impose that the linear-in-$\nu$ binding energy from $\Delta\hat Q^{\text{PS}}$ from Eq.~\eqref{Q4pnsplit} [calculated as done for Eq.\eqref{Eeob} in Sec.~\ref{sec:preamble}] vanishes and we obtain: 
\begin{equation}\label{qcoeff}
q_{(3,2)}=9\,;\, q_{(4,1)}=96\,;\, q_{(5,0)}=112\,.
\end{equation}

The final PN correction $\Delta\hat Q^{\text{PS}}$ thus contains all the extra information from generic orbits at 3PN that is not captured by $\hat Q^{\text{PS}}_{\text{SMR}}$, without contributing to the linear in mass ratio binding energy for circular orbits.
The exercise above can be repeated at one PN order higher to obtain $\Delta\hat Q^{\text{PS}}$ at 4PN starting from the 4PN EOB Hamiltonian in the PS gauge.~\cite{Antonelli:2019ytb}. Such a computation does not present major differences from the calculation above: the only feature changing is the counterterm, which needs to start at 5PN and include logarithmic terms. We have decided not to include $\Delta\hat Q^{\text{PS}}$ at 4PN in this paper, as the 4PN Hamiltonian from which it is constructed is only valid for near-circular orbits. The $\Delta\hat Q^{\text{PS}}$ at 3PN that we obtain here is instead valid for generic orbits.

\begin{table}
	\caption{{\textbf{Two-body EOB Hamiltonians.}}
		\label{table:models}}
	\centering
		\begin{tabular}{lp{7cm}p{2.5cm}}
			\hline
			$H_{\text{SMR}}^{\text{EOB,PS}}$
			& SMR Hamiltonian in PS gauge	& This paper \\
		
			$H_{\text{SMR-3PN}}^{\text{EOB,PS}}$					
			& SMR-3PN Hamiltonian in PS gauge & This paper \\
		
			$H_{\text{SMR}}^{\text{EOB}}$
			& SMR Hamiltonian in the DJS gauge (with LR divergence)	& \cite{Barausse:2011dq} \\
		
			$H_{n{\text{PN}}}^{\text{EOB,PS}}$
			&   $n$PN  Hamiltonian in PS gauge & \cite{Damour:2017zjx}	\\
			
			$H_{n\text{PN}}^{\text{EOB}}$	
			&	$n$PN  Hamiltonian in DJS gauge& \cite{Buonanno:1998gg,Damour:2000we} 	\\
		
		\end{tabular}
\end{table}

\section{Inspirals in effective-one-body theory}
\label{sec:evolutions}

\subsection{Plunging through the light ring with small mass-ratio Hamiltonians}

\begin{figure}[t]
	\includegraphics[width=\columnwidth]{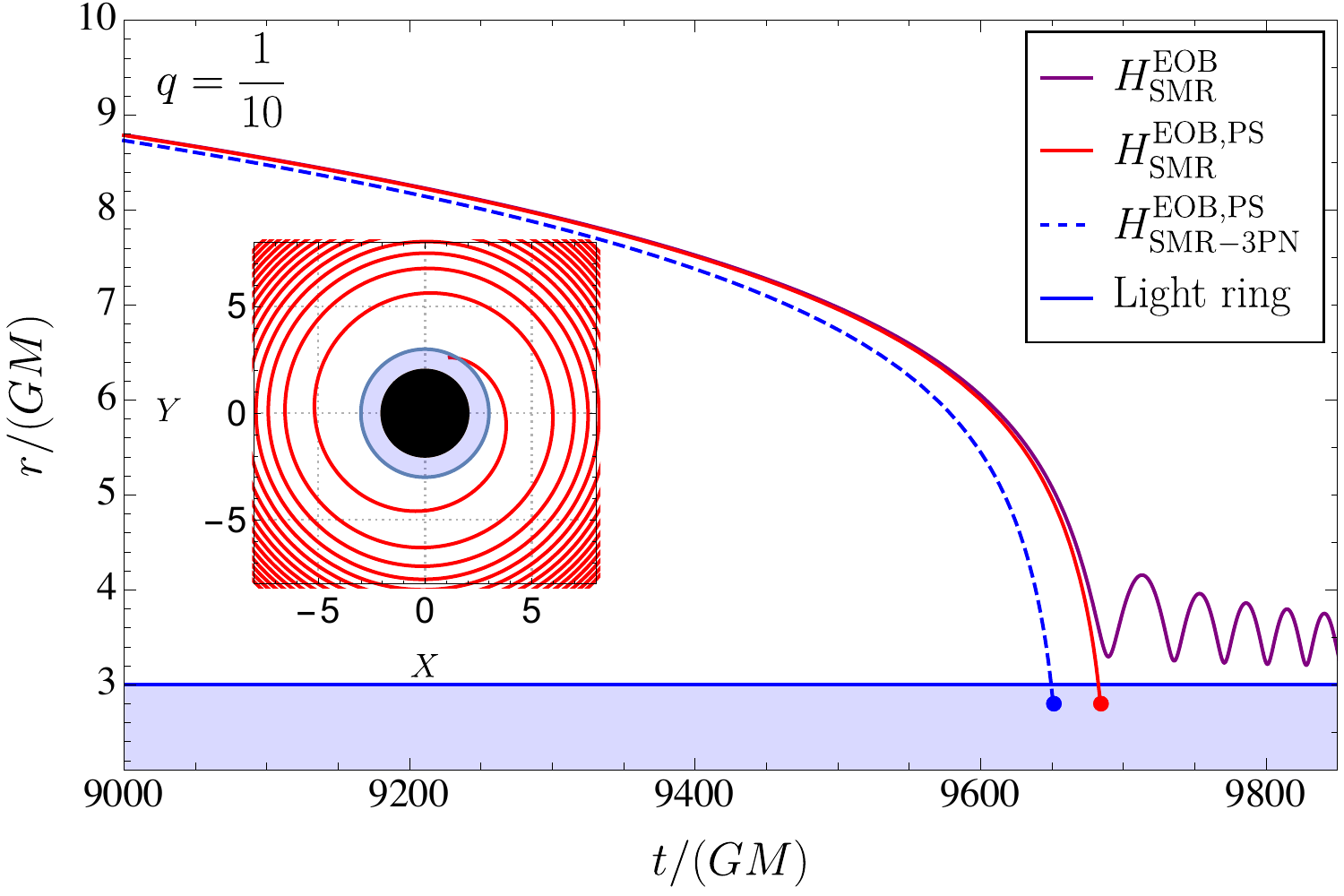}
	\caption{\textbf{Plunges through the light-ring radius:} the evolved orbital separation for the SMR Hamiltonians is presented. The effective masses of models $H_{\text{SMR}}^{\text{EOB,PS}}$ and $H_{\text{SMR-3PN}}^{\text{EOB,PS}}$
		plunge through the LR radius $r_S=3M$. Conversely, the plunge of the effective mass of $H_{\text{SMR}}^{\text{EOB}}$ presents unphysical features associated to the LR-divergence.
	}
	\label{fig:orbitsep}
\end{figure}

In this section, we evolve the EOB Hamiltonians constructed in Secs.~\ref{sec:preamble} and ~\ref{sec:noncirc} [i.e., Eq.~\eqref{HeffEG} with non-geodesic functions~\eqref{QSMR} and~\eqref{Qtot}], and the EOB Hamiltonian with SMR information in the DJS gauge. We refer to them as $H_{\text{SMR}}^{\text{EOB,PS}}$, $H_{\text{SMR-3PN}}^{\text{EOB,PS}}$ and $H_{\text{SMR}}^{\text{EOB}}$, see Table~\ref{table:models} (which also includes our notation for the PN Hamiltonians in both DJS and PS gauges). 

The EOB approach comprises of a conservative sector, discussed in detail in Sec.~\ref{sec:prelim}, and a dissipative sector, responsible for the slow GW-driven inspiral of the compact bodies towards merger. The basic set of equations for inspiraling orbits in the EOB framework are the Hamilton equations augmented with a radiation-reaction force $\mathcal{F}_{\text{RR}}$.
In terms of a generic mass-reduced EOB Hamiltonian $\Ham^{\text{EOB}}(\hat r,\p_{r_{*}},\p_\phi)$, the equations read \cite{Buonanno:2000ef,Buonanno:2007pf,Damour:2007xr,Pan:2011gk}:
\begin{subequations}
	\begin{align}
	& \frac{d\hat r}{d\hat t}=\frac{A(\hat r)}{\sqrt{D(\hat r)}}\frac{\partial \Ham^{\text{EOB}} }{\partial \p_{r_{*}}}\,, \label{HH1}\\
	&\frac{d\phi}{d\hat t}=\frac{\partial \Ham^{\text{EOB}}}{\partial \p_{\phi}}\,, \label{HH2}\\
	&\frac{d\p_{ r_{*}}}{d\hat t}=-\frac{A(\hat r)}{\sqrt{D(\hat r)}}\frac{\partial \Ham^{\text{EOB}}}{\partial \hat r}+\mathcal{F}_{\text{RR}}\frac{\p_{r_{*}}}{\p_{\phi}}\,, \label{HH3}\\
	&\frac{d\p_{\phi}}{d\hat t}=\mathcal{F}_{\text{RR}}\,, \label{HH4}
	\end{align}
\end{subequations}
where we have introduced the mass-reduced radius $\hat r \equiv r/M$ and coordinate time $\hat t \equiv t/M$ and used the mass-reduced radial momentum $\p_{r_{*}}$ conjugate to the radius $r_{*}$ in tortoise coordinates,
defined for generic potentials $A(\hat r)$ and $D(\hat r)$\footnote{Here $D(\hat r)$ is the inverse of $\bar D(\hat r)$ mentioned in Sec.~\ref{sec:prelim}.} by:
\begin{equation}\label{prprstar}
\frac{d\hat r_{*}}{d\hat r} \equiv \frac{\sqrt{D(\hat r)}}{A(\hat r)}
=\frac{\p_{r}}{\p_{r_{*}}}
\,.
\end{equation}
In the evolution of the EOB Hamiltonian in the DJS gauge we use the PN-expanded expressions for $A(\hat r)$, $D(\hat r)$ and $\hat Q^{\text{DJS}}$ at the required PN order~\cite{Buonanno:1998gg,Damour:2000we,Damour:2015isa} (i.e., we use their 2PN and 3PN expressions in the evolutions of $H_{2\text{PN}}^{\text{EOB}}$ and $H_{3\text{ PN}}^{\text{EOB}}$, respectively), whereas we use their test-body limits in the evolutions of Hamiltonians in the PS gauge\footnote{The effective Hamiltonian in the PS gauge \eqref{HeffEG} is obtained solving the Hamilton-Jacobi equations with the Schwarzschild metric. The $A(\hat r)$ and $D(\hat r)$ are therefore fixed by their Schwarzschild limits.}. The Hamiltonians in both gauges depend on $\p_{ r_{*}}$, rather than $\p_{ r}$. 

The radiation reaction force $\mathcal{F}_{\text{RR}}$ drives the inspiral of the system and it contains semi-analytical two-body information \cite{Damour:2007xr,Damour:2008gu,Pan:2010hz}. In this paper, we employ its non-Keplerian form (with $\hat \Omega \equiv d\phi/d\hat t=M\Omega$):
\begin{equation}
^{\text{nK}}\mathcal{F}_{\text{RR}}=-\frac{1}{\nu\hat \Omega}\frac{dE}{dt}\,,
\end{equation}
where $dE/dt$ is the GW flux for quasi-circular orbits~\cite{Damour:2008gu}:
\begin{equation}
\frac{dE}{dt}=\frac{\hat\Omega^2}{8\pi}\sum_{l=2}^{l_\text{max}=8}\sum_{m=l-2}^{l} m^2 \big|\hat rh_{lm}\big|^2\,.
\end{equation}
The modes $h_{lm}$ are built from PN theory, but resummed multiplicatively (see e.g., Ref.~\cite{Damour:2008gu}). Here, we use the resummation of the (non-spinning) modes and flux presented in Ref.~\cite{Pan:2011gk} (which coincides with the state-of-the-art modes and flux used in the EOB waveform model for 
LIGO/Virgo data-analsyis~\cite{Bohe:2016gbl}, when spins are set to zero). We do not include 
the ``next-to-quasi-circular'' (NQC) coefficients~\cite{Bohe:2016gbl}, or any calibration parameter obtained 
imposing better agreement with numerical-relativity waveforms. Our main motivation here is to compare how well the conservative 
EOB-dynamics of SMR models compare to PN ones and with NR.

\begin{table}
	\caption{{\textbf{Set of non-spinning NR simulations and alignment time-windows.}}  We list the SXS IDs, the mass ratios $q$ and the number of orbital cycles $N_{\text{orb}}^{\text{merg}}$ from the beginning of the simulation up to the binary black-hole merger (peak of $h^{\text{NR}}_{22}$), as reported in the SXS catalog. We further include the time $t^\text{alig}_\text{in}$ at which the alignment procedure starts, the time $t^\text{alig}_\text{fin}$ at which it ends (in units of $M$) and the estimated NR error at merger $\Delta \phi^\text{merg}_\text{NR}$ (in radians).
		\label{table:setofNR}}
	\centering
		\begin{tabular}{ccc  ccc}
			SXS ID:						&	q$^{-1}$	& $N_{\text{orb}}^{\text{merg}}$	&    $t^\text{alig}_\text{in}$					&	$t^\text{alig}_\text{fin}$		&$\Delta \phi^\text{merg}_\text{NR}$ 	\\
			\hline
			0180	&	1	& 28.18	& 820	&	2250   & $\pm$0.25 \\
			1222	&	2	& 28.76	& 1000	&	2555  & $\pm$1.26 \\
			1221	&	3	& 27.18	& 1800	&	3000  & $\pm$0.21 \\
			1220	&	4	& 26.26	& 1800	&	3000   & $\pm$1.82 \\
			0056	&	5	& 28.81	& 1500	&	3000   & $\pm$0.39 \\
			0181	&	6	& 26.47	& 1000	&	2500   & $\pm$0.01 \\
			0298	&	7	& 19.68	& 780	&	2180   & $\pm$0.10 \\
			0063	&	8	& 25.83	& 1140	&	2540   & $\pm$0.85 \\
			0301	&	9	& 18.93	& 780	&	2180   & $\pm$0.13 \\
			0303	&	10	& 19.27	& 700	&	1900   & $\pm$0.49 \\
			
		\end{tabular}
\end{table}

The result of the evolved orbital separations $\hat r$ of both DJS and PS  Hamiltonians  for $q=1/10$ are reported in Fig.~\ref{fig:orbitsep}.
Focusing on the evolution in the DJS case, it is seen that the pole in the conservative part of the DJS Hamiltonian affects the motion of the effective body close to the LR radius. That is, $H_{\text{SMR}}^{\text{EOB}}$ diverges at $\hat r_{\text{S}}^{\text{LR}}=3$, at which point it acts as an infinite potential barrier that the effective mass cannot cross. Conversely, the effective mass plunges through the Schwarzschild LR radius in the cases of $H_{\text{SMR}}^{\text{EOB,PS}}$ and $H_{\text{SMR-3PN}}^{\text{EOB,PS}}$. This finding confirms that there is no unphysical behaviour at the LR radius for SMR Hamiltonians in the PS gauge. 
To conclude, we also notice that the evolutions of the $H_{\text{SMR}}^{\text{EOB,PS}}$ and $H_{\text{SMR-3PN}}^{\text{EOB,PS}}$ models (red and blue dots) stop soon after the LR radius. In principle we would expect them to stop at the Schwarzschild horizon ($u=1/2$). This is not the case in Fig.~\ref{fig:orbitsep}. In the PS gauge, the orbital frequency scales as:
\begin{equation}
\Omega=\frac{\partial H\subreal}{d\Pp_\phi}\propto
\frac{\partial  H_\text{eff}^{\text{PS}}}{\partial p_\phi}\propto \frac{\partial  H_\text{eff}^{\text{PS}}}{\partial  H_\text{S}} \frac{\partial  H_\text{S}}{\partial p_\phi}\,.
\end{equation}
where in the first proportionality relation we have used the energy map~\eqref{enmap} and in the second we have exploited the fact that the PS effective Hamiltonian only depends on the angular momentum $p_\phi$ via $ H_\text{S}$. The factor $\partial  H_\text{S}/\partial p_\phi$ vanishes at $u=1/2$ (corresponding to the usual Schwarzschild horizon). However, we also find that, with our Hamiltonian ansatz, the $\partial \hat H_\text{eff}^{\text{PS}}/\partial  H_\text{S}$ factor develops a zero just below the LR radius. Consequently, $\Omega$ vanishes at this point and we stop the evolution. We note that having little model dynamics after the peak of the frequency,  while not presenting an issue by itself, could pose problems in the modelling of EOB waveforms and frequencies during the transition between plunge and merger-ringdown phases.

\subsection{Comparisons against numerical relativity}
\label{sec:energetics}

\begin{figure}
	\includegraphics[width=\columnwidth]{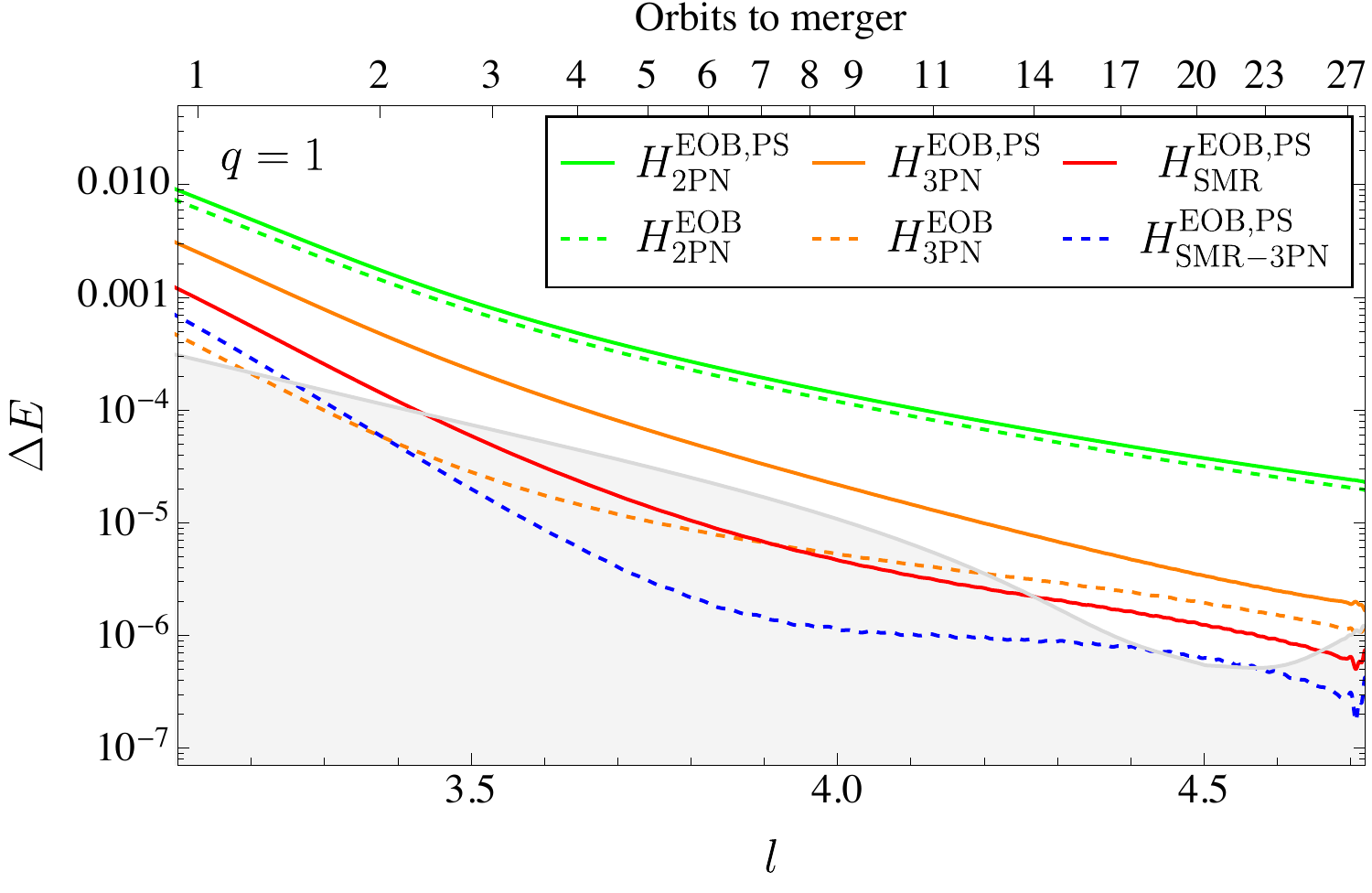}
	\includegraphics[width=\columnwidth]{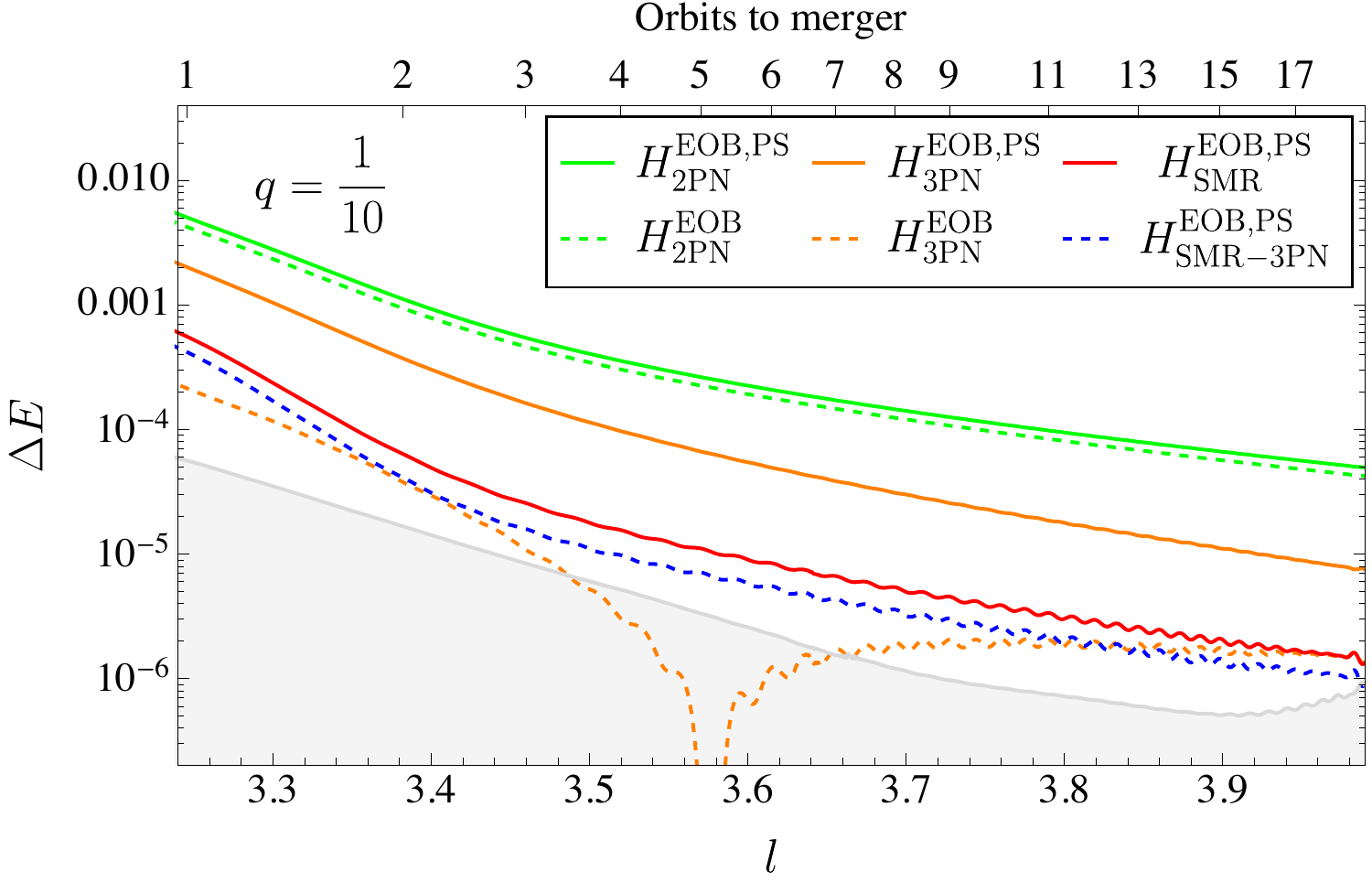}
	\caption{\textbf{SMR vs PN binding energies:} we compare the difference $\Delta E$ in binding energy from NR for our SMR Hamiltonians versus angular momentum $l$. We compare it to similar results for PN models up to third order, in both PS and DJS gauges. The \textit{estimated} NR error is shown in grey.}
	\label{fig:compEJ}
\end{figure}

\begin{figure}
	\includegraphics[width=\columnwidth]{figs/eoq1.pdf}
	\includegraphics[width=\columnwidth]{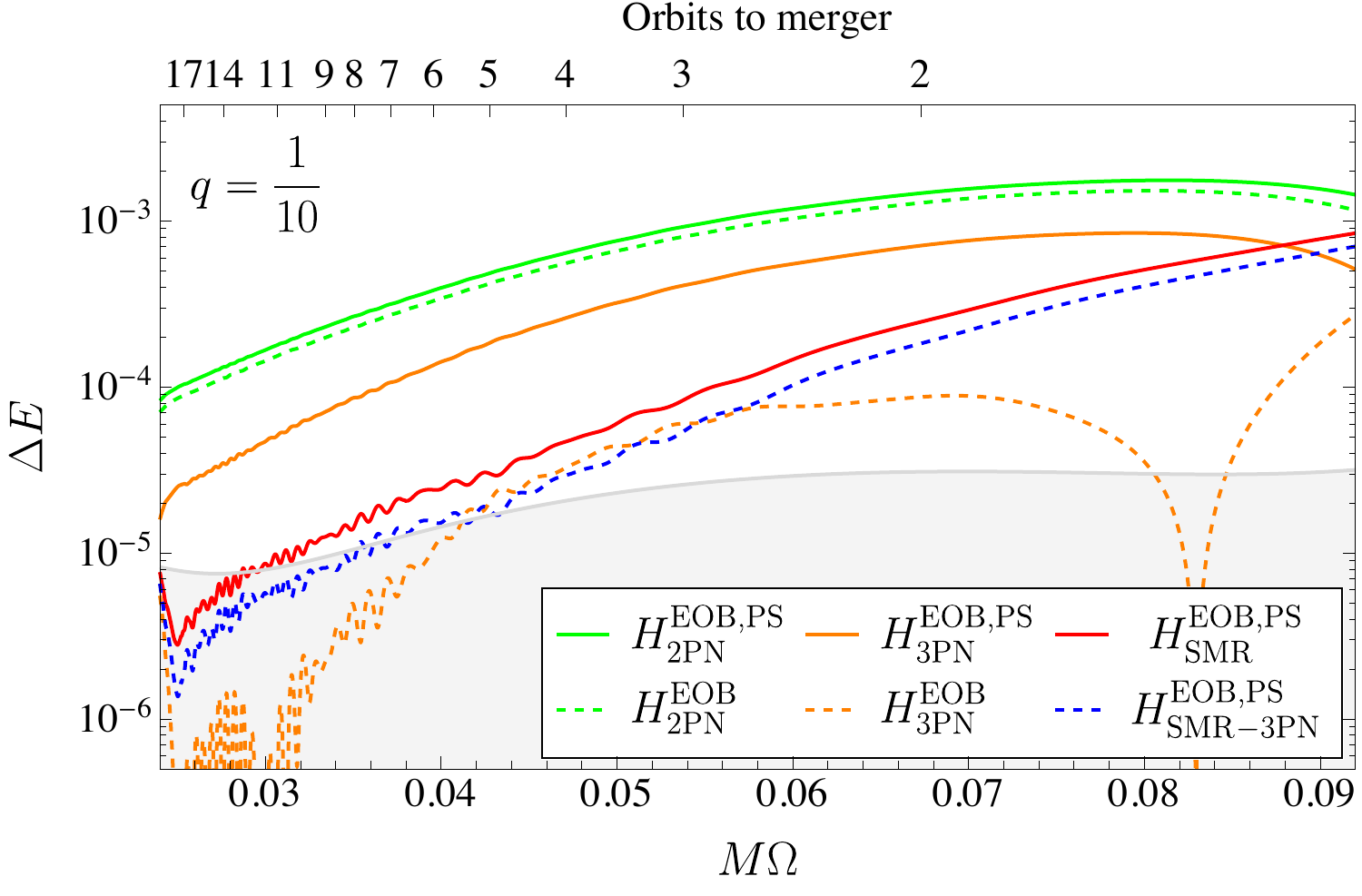}
	\caption{\textbf{SMR vs PN binding energies:} we compare the difference $\Delta E$ in binding energy from NR for our SMR Hamiltonians versus frequency $(M\Omega)$. We compare it to similar results for PN models up to third order, in both PS and DJS gauges. The \textit{estimated} NR error is shown in grey.}
	\label{fig:compEO}
\end{figure}

Here we study the energetics of the $H_{\text{SMR}}^{\text{EOB,PS}}$ and $H_{\text{SMR-3PN}}^{\text{EOB,PS}}$ models and the PN EOB models in both gauges via comparisons of their binding energies against NR predictions. The main reason why we choose to compare SMR models to PN ones is to assess how useful SMR information could be in improving the EOB models currently in use, which are  based on PN information.

The (quasi) gauge-invariant relations between the dimensionless circular orbit binding energy $E \equiv (H-M)/\mu$ and angular momentum $ l \equiv \p_\phi=\Pp_\phi/(M\mu)$ (and orbital frequency $\hat\Omega$) are used to draw comparisons against NR.
This type of comparisons is useful to understand how information of the real two-body motion is resummed into the conservative dynamics~\cite{Antonelli:2019ytb}.
In contrast to Ref.~\cite{Antonelli:2019ytb} and Sec.~\ref{sec:EOBSMR} of this paper, where the binding energy is calculated in the circular-orbit limit, the binding energies appearing in this section are obtained evolving the EOB Hamiltonians along quasi-circular orbits. This more closely matches the procedure used to extract the binding energy from NR simulations of quasi-circular inspirals, providing clearer comparisons \cite{Ossokine:2017dge}.
Finally, we calculate the dephasing $\Delta\phi_{22} \equiv  \phi_\text{NR}-\phi_\text{EOB}$ of the ($\ell,m$)=(2,2) modes of the $H_{\text{SMR}}^{\text{EOB,PS}}$ and $H_{\text{SMR-3PN}}^{\text{EOB,PS}}$ models against NR results. While more thorough comparisons aimed at using the models for LIGO inference studies would need a systematic calculation of the unfaithfulness (see e.g., Refs.~\cite{Pan:2011gk,Taracchini:2013rva,Bohe:2016gbl,Cotesta:2018fcv}), we find these comparisons illustrative to contextualize the $H_{\text{SMR}}^{\text{EOB,PS}}$ and $H_{\text{SMR-3PN}}^{\text{EOB,PS}}$ models in this paper.

\begin{table*}
	\caption{{\textbf{Details of the dephasing comparison.}} We report the dephasing (in radians) of the SMR and 3PN models in both gauges at 8 and 4 GW cycles before NR merger, as found using the time-windows of Table~\ref{table:setofNR}. We also report the corresponding estimated NR error, which we denote by $\Delta \phi_\text{NR}$. The error for each NR simulation is estimated taking the phase differences between the highest two resolutions of the NR simulation (at fixed extrapolation order) and between two successive extrapolation orders (at fixed resolution), and adding them in quadrature.
		\label{table:simulations0}}
	\centering
		\begin{tabular}{cc | cccc c | cccc c  cccc c| ccc ccc }
			&\multirow{2}{*}{$q^{-1}$}
			&\multicolumn{4}{c}{8 GW cycles before merger}
			&\multirow{3}{*}{$\Delta\phi_{\text{NR}}$ }
			\\
			
			& 
			
			& $\Delta \phi_{\text{SMR}}^{\text{EOB,PS}}$
			& $\Delta \phi_{\text{SMR-3PN}}^{\text{EOB,PS}}$
			& $\Delta \phi_{\text{3PN}}^{\text{EOB,PS}}$
			& $\Delta \phi_{\text{3PN}}^{\text{EOB}}$
			&
			
			\\
			\hline
			&	1	& ~0.111	& -0.033	&	-0.971   & ~0.032 & $\pm$0.032\\
			&	2	& ~0.112	& -0.061	&	-1.342   & -0.023 & $\pm$0.105\\
			& 	3	& ~0.050	& -0.021	&	-0.617   & -0.023 & $\pm$0.093\\
			&	4	& ~0.046	& -0.038	&	-0.859   & -0.078 & $\pm$0.203\\
			&	5	& ~0.037	& -0.034	&	-0.846   & -0.086 & $\pm$0.023\\
			&	6	& -0.035	& -0.064	&	-0.433   & -0.093 & $\pm$0.006\\
			&	7	& ~0.024	& -0.009	&	-0.462   & -0.070 & $\pm$0.001\\
			&	8	& ~0.021	& -0.021	&	-0.676   & -0.107 & $\pm$0.057\\
			&	9	& ~0.017	& -0.005	&	-0.368   & -0.068 & $\pm$0.002\\
			&	10	& ~0.022	& -0.001	&	-0.413   & -0.076 & $\pm$0.033\\
			
		\end{tabular}
	
	\begin{tabular}{cc | cccc c | cccc c  cccc c| ccc ccc }
		&\multirow{3}{*}{$\Delta\phi_{\text{NR}}$ }
		&\multicolumn{4}{c}{4 GW cycles before merger }  
		& \multirow{3}{*}{$\Delta\phi_{\text{NR}}$ }	
		\\
		
%
		&
		
		& $\Delta \phi_{\text{SMR}}^{\text{EOB,PS}}$
		& $\Delta \phi_{\text{SMR-3PN}}^{\text{EOB,PS}}$
		& $\Delta \phi_{\text{3PN}}^{\text{EOB,PS}}$
		& $\Delta \phi_{\text{3PN}}^{\text{EOB}}$
		&
		\\
		\hline
		&	1
		& ~0.352 & -0.012 & -2.630 & ~0.084 & $\pm$0.056\\ 
		&	2
		& ~0.512 & -0.021 & -5.586 & -0.043 & $\pm$0.224\\ 
		&	3
		& ~0.111 & -0.026 & -1.209 & -0.048 & $\pm$0.144\\ 
		&	4
		& ~0.187 & -0.041 & -2.540 & -0.212 & $\pm$0.372\\ 
		&	5
		& ~0.125 & -0.044 & -2.077 & -0.211 & $\pm$0.064\\ 
		&	6
		& -0.041 & -0.082 & -0.599 & -0.126 & $\pm$0.007\\ 
		&	7
		& ~0.092 & -0.003 & -1.403 & -0.211 & $\pm$0.009\\ 
		&	8
		& ~0.076 & -0.025 & -1.660 & -0.260 & $\pm$0.155\\ 
		&	9
		& ~0.063 & -0.005 & -1.185 & -0.220 & $\pm$0.012\\ 
		&	10
		& ~0.070 & -0.004 & -1.245 & -0.233 & $\pm$0.083\\
		
	\end{tabular}
\end{table*}

\begin{table}
	\caption{{\textbf{Alternative alignment time-windows.}}Time-windows (in units of $M$) employed for Fig.~\ref{fig:dphivsq}: here, $t^\text{alig}_\text{in}$ is the time corresponding to 34 GW cycles before merger for each NR simulation, whereas $t^\text{alig}_\text{fin}$ is chosen to encompass 10 GW cycles. The time at merger is given by $t_\text{merg}$.
		\label{table:simulations}}
	\centering
		\begin{tabular}{cccc | cccc }
			q$^{-1}$			&   $t^\text{alig}_\text{in}$					&	$t^\text{alig}_\text{fin}$ &	$t_\text{merg}$			& q$^{-1}$			&   $t^\text{alig}_\text{in}$					&	$t^\text{alig}_\text{fin}$ &	$t_\text{merg}$	\\
			\hline
			1	& 5107	& 6911 & 9517	&	6   & 2971 & 4254 & 6000 \\
			2	& 5406	& 7078 & 9384   &   7   & 776 & 2083 & 4142\\
			3   & 3940	& 5532 & 7858   & 8 & 2652 & 3918& 5956\\
			4   & 3479	& 4975 & 7200	& 9 & 513 & 1732 & 3692\\
			5   & 4206	& 5641 & 7864   & 10  &  587 & 1771 & 3691	\\	
			\end{tabular}		
\end{table}

We employ a set of ten non-spinning NR simulations from the Simulating eXtreme Spacetimes (SXS) collaboration \cite{Mroue:2013xna,Chu:2015kft}, with mass ratios $1/10\leq q\leq 1$. We summarize the details of these simulations in Table~\ref{table:setofNR}.
A description of how the $E(l)$ and $E(\hat\Omega)$ curves were calculated for a subset of these simulations can be found in Ref.~\cite{Ossokine:2017dge}.

We evolve EOB Hamiltonians with PN information up to third order, since 3PN is the order at which PS-gauge Hamiltonians can be uniquely derived for generic orbits (see the Appendix of Ref.~\cite{Antonelli:2019ytb} for more details). It is worthwhile to mention that the $H_{\text{3PN}}^{\text{EOB}}$ Hamiltonian has better energetics and phases performances against NR than both $H_{\text{4PN}}^{\text{EOB}}$ and the SEOBNR Hamiltonian used as a baseline for the current generation of EOB waveform models (defined, e.g., in the Appendix of Ref.~\cite{Steinhoff:2016rfi}), when calibration and NQC parameters are turned off. Restricting ourselves to comparisons with $H_{\text{3PN}}^{\text{EOB}}$ only, we are therefore not running the risk to overestimate the performance of SMR models when comparing them to PN results.

Let us begin comparing the $E(l)$ and $E(\hat\Omega)$ curves. In Figs.~\ref{fig:compEJ} and \ref{fig:compEO}, the difference $\Delta E \equiv |E_\text{NR}-E_\text{EOB}|$ is plotted for a variety of EOB models and for mass ratios $q=1$ and $q=1/10$. We choose to present results for these mass ratios only as we find them to be representative of the behaviour of the models across the parameter range considered in this study.
Considering the $E(l)$ relations first and focusing on the SMR models, it is seen that for $q=1/10$ both $H^{\text{EOB,PS}}_{\text{SMR}}$ and $H^{\text{EOB,PS}}_{\text{SMR-3PN}}$ perform better against NR than the 3PN model in the same gauge, e.g., $H^{\text{EOB,PS}}_{{\text{3PN}}}$.
The $H^{\text{EOB,PS}}_{\text{SMR-3PN}}$ model also performs better than both in the comparable-mass case.
A similar finding is obtained investigating the $E(\hat\Omega)$ curves, see Fig.~\ref{fig:compEO}. 
Taken together, these results highlight the importance of SMR results to improve the modeling of both equal- and unequal-mass systems within the EOB approach.
It is also seen that, for both mass ratios considered and for both $E(l)$ and $E(\hat\Omega)$ curves, $H^{\text{EOB,PS}}_{\text{SMR-3PN}}$ improves the predictions of $H^{\text{EOB,PS}}_{\text{SMR}}$, suggesting that generic orbit terms are important when considering quasi-circular orbit binding energies (especially in the equal-mass-ratio case).

PN Hamiltonians in the PS gauge generically perform worse in binding energy comparisons than Hamiltonians in the DJS gauge, as found out in the adiabatic approximation already in Ref.~\cite{Antonelli:2019ytb}.
This finding suggests that, notwithstanding the already good agreement between SMR models and NR simulations for both mass ratios, a better description for the EOB dynamics than the one provided by the PS gauge could be pursued in order to maximize the performance of evolutions from both PN and SMR EOB models.
\begin{figure*}[t]
	\includegraphics[width=.47\columnwidth]{figs/wavephaseq1.pdf}
	\hspace{12pt}
	\includegraphics[width=.47\columnwidth]{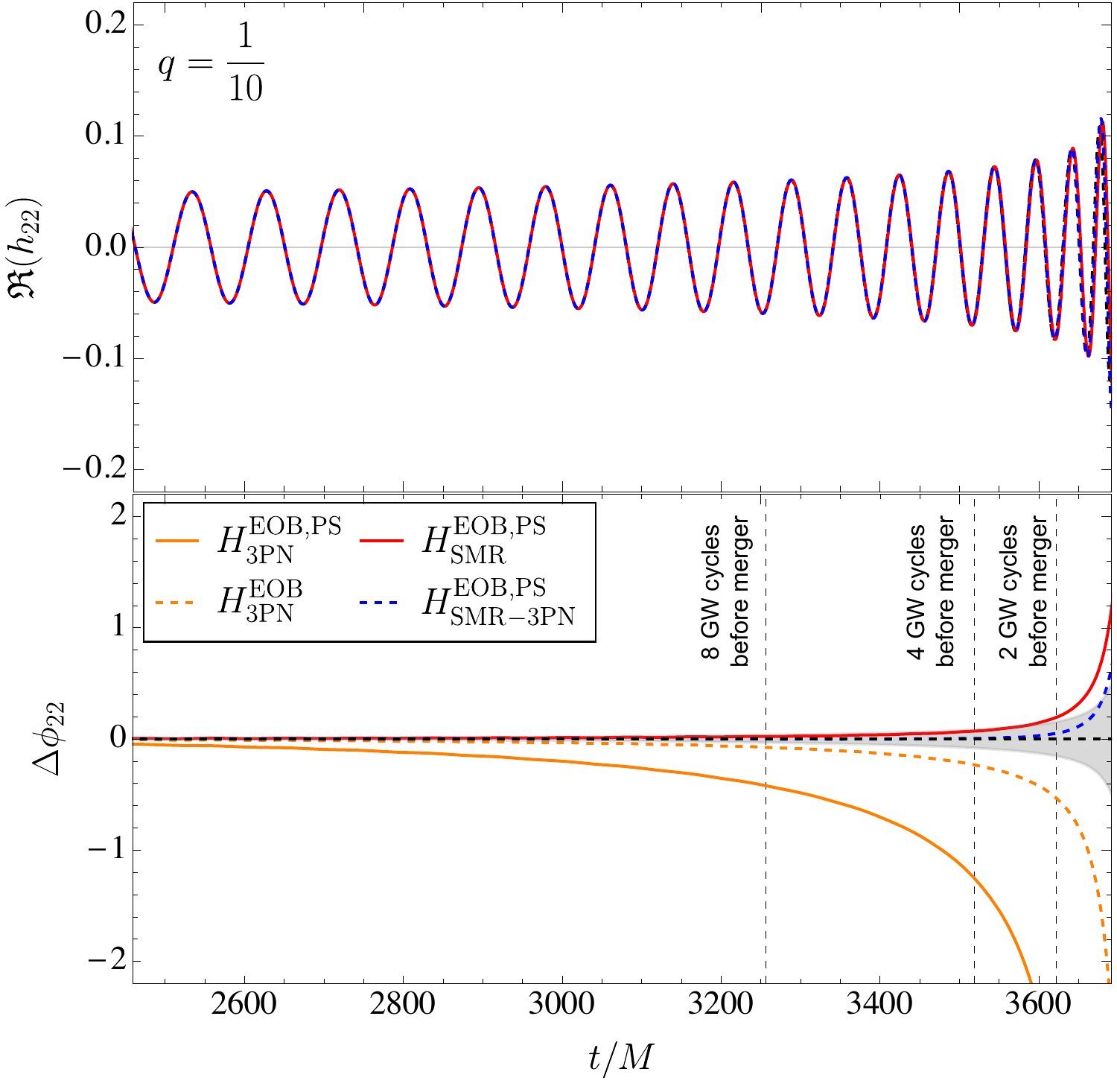}
	\caption{\textbf{Dephasing of EOB models:} in the top panels, the real parts $\mathfrak{R}(h_{22})$ of the ($\ell,m$)=(2,2) mode EOB waveform for the SMR, SMR-3PN models are shown and compared to the NR waveforms (in dashed-black, overlapping with the EOB waveforms up to few GW cycles to merger).
		In the lower panels, the dephasing of SMR and PN EOB models from the NR simulations is calculated.
		Also shown are the times corresponding to 8, 4 and 2 GW cycles before NR merger.  }
	\label{fig:wavephase}
\end{figure*}
\begin{figure*}[t]
	\includegraphics[width=\columnwidth]{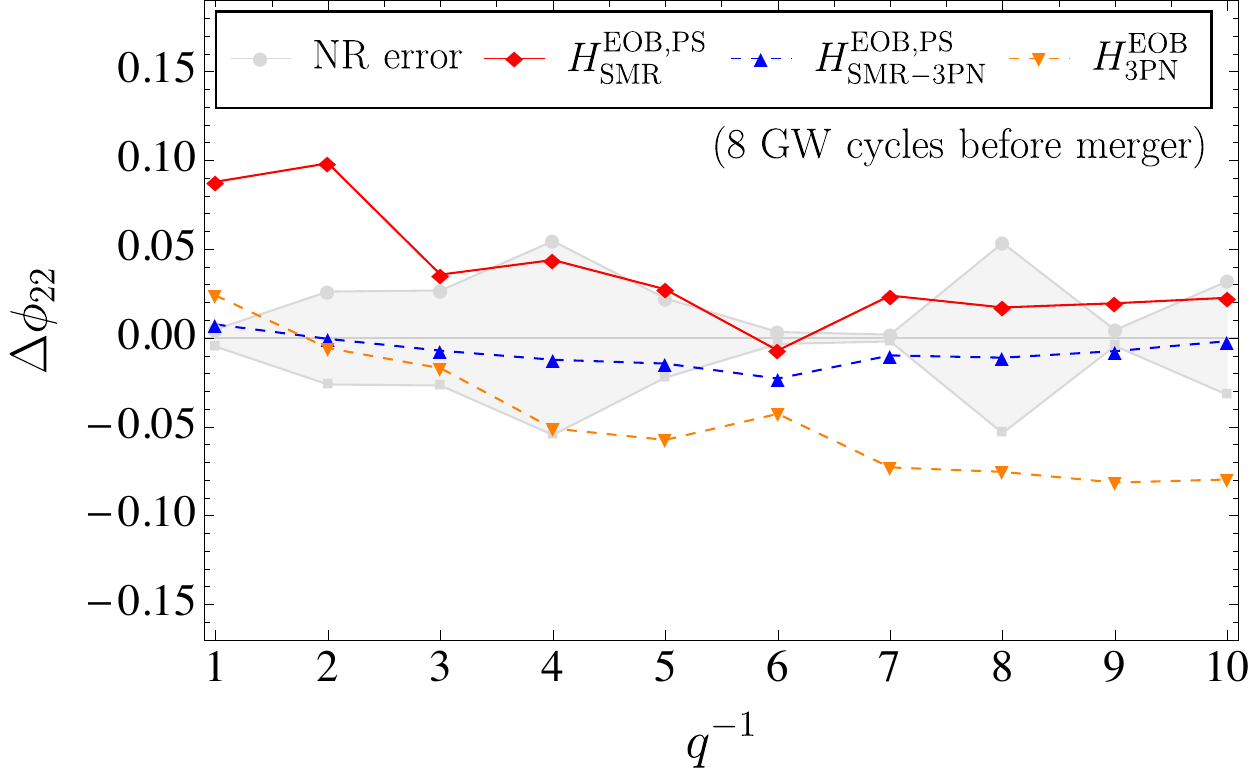}
	\hspace{12pt}
	\includegraphics[width=\columnwidth]{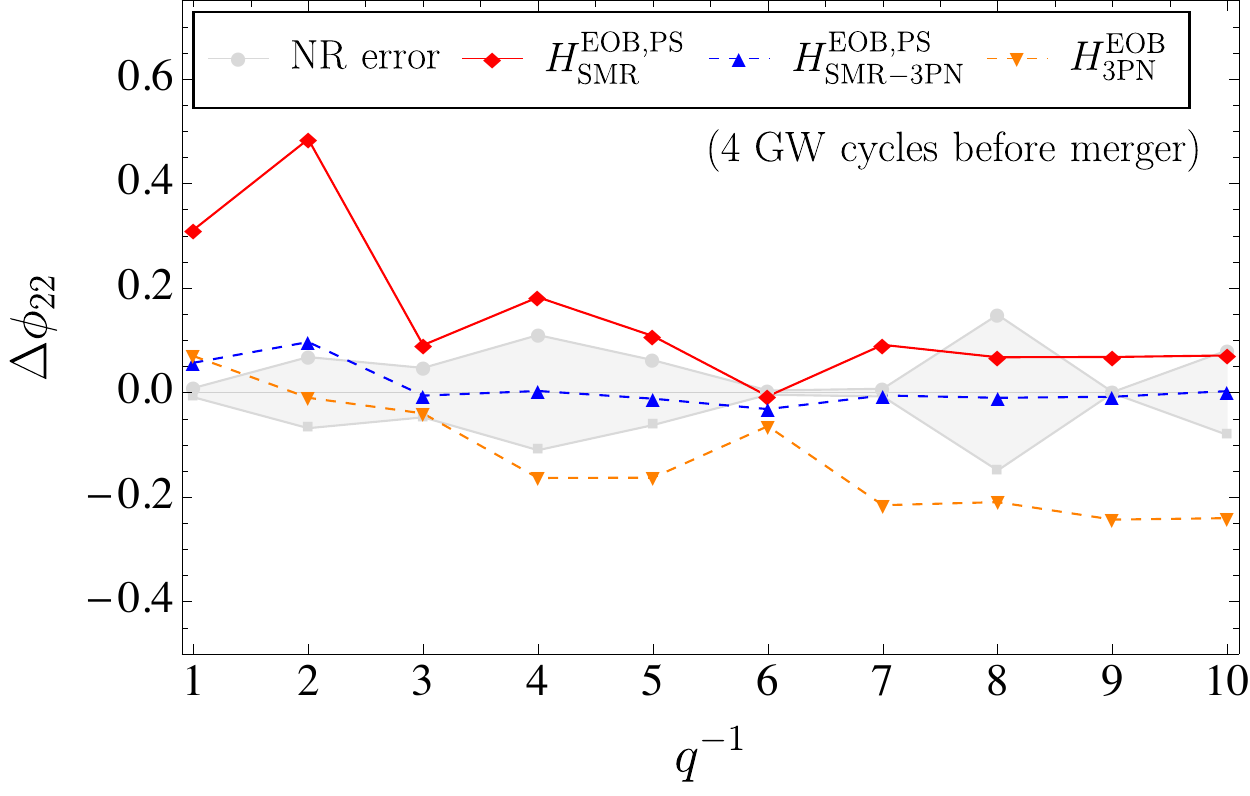}
	\caption{\textbf{Dephasing vs mass ratio:} we compare the dephasing of $H_{\text{SMR}}^{\text{EOB,PS}}$, $H_{\text{SMR-PN}}^{\text{EOB,PS}}$ and $H_{\text{3PN}}^{\text{EOB}}$ after they have been aligned with the NR simulations from Table~\ref{table:simulations}. For each $q$, we snapshot the dephasing of the EOB models and the NR simulation at a time corresponding to 4 and 2 orbits before the merger of the binary system in the NR simulation. }
	\label{fig:dphivsq}
\end{figure*}
We complete our comparison study with the dephasing $\Delta\phi_{22}$ of the ($\ell,m$)=(2,2) modes from the EOB models and the NR simulations.
For a proper comparison, the EOB and NR waveforms must be aligned for each $q$. Here we use the alignment procedure outlined in Ref.~\cite{Pan:2011gk}, which amounts to minimizing the function:
\begin{equation}
\Xi(\Delta t,\Delta \phi)=\int_{t^\text{alig}_1}^{t^\text{alig}_2}[\phi_\text{NR}(t)-\phi_\text{EOB}(t+\Delta t)-\Delta\phi]^2 dt\\,
\end{equation} 
over the time and phase shifts, $\Delta t$ and $\Delta\phi$. 
The integrating interval [$t^\text{alig}_1,t^\text{alig}_2$] defines the time-domain window in which the alignment is performed: conservatively, it must be chosen in the inspiral of the NR simulation, large enough to average out the numerical noise and such as to avoid junk radiation at the beginning of the NR simulation \cite{Pan:2011gk}. From the alignment procedure described above, one can obtain the phase and amplitude time-shift to be applied to the EOB model to align it with the NR waveforms, i.e., the aligned waveforms are:
\begin{align}
&h^\text{NR}_{22}=A_\text{NR}(t)e^{i\phi_\text{NR}(t)}\,,\label{h22nr}\\
&h^\text{EOB}_{22}=A_\text{EOB}(t+\Delta t)e^{i[\phi_\text{EOB}(t+\Delta t)+\Delta\phi]}\label{h22eob}\,.
\end{align} 

Our choices for the time-windows are reported in Table~\ref{table:setofNR}.
In Fig.~\ref{fig:wavephase}, we show the results of our phase comparisons for $q=1$ and $q=1/10$ up to merger. For clarity, the upper panels only include the $H_{\text{SMR}}^{\text{EOB,PS}}$ and $H_{\text{SMR-3PN}}^{\text{EOB,PS}}$ models and the NR simulations. They show the real parts of Eqs.~\eqref{h22nr} and ~\eqref{h22eob}, from which we infer that the SMR models do not accumulate a significant amount of dephasing. Overall, they are in very good agreement with NR for both $q=1$ and $q=1/10$.
It is important to place the above results in context. In the lower panel, the dephasing of SMR models from NR is compared to that of 3PN models\footnote{In this comparison we do not include 2PN models, which we find to have much larger dephasing than the 3PN models shown.}.
Interestingly, even in the equal-mass-ratio case $H_{\text{SMR}}^{\text{EOB,PS}}$ and $H_{\text{SMR-3PN}}^{\text{EOB,PS}}$ compare much better than the 3PN model in the same gauge, e.g., $H_{\text{3PN}}^{\text{EOB,PS}}$. Their dephasing is comparable to  $H_{\text{3PN}}^{\text{EOB}}$. In the $q=1/10$ case,
they have a smaller dephasing than any other PN model considered in this study. 
In Table~\ref{table:simulations0}, we report the dephasing that the $H_{\text{SMR}}^{\text{EOB,PS}}$, $H_{\text{SMR-3PN}}^{\text{EOB,PS}}$, $H_{\text{3PN}}^{\text{EOB}}$ and $H_{\text{3PN}}^{\text{EOB,PS}}$ models accumulate up to 8 and 4 GW cycles before merger for all mass ratios (with the corresponding estimated NR error)\footnote{We have checked that shifting the time-windows by $\Delta t=\pm 100\,M$,  our $\Delta \phi$'s only change by a few hundredths of a radian.}.

Next, we want to study how the dephasing of the above models varies as a function of $q$. It would be tempting to compare the $\Delta \phi$'s reported in Table~\ref{table:simulations0} at a fixed number of cycles before merger. While this remains a valid possibility, such a comparison would neither take into account the different lengths of the NR simulations used in this set, nor the different number of GW cycles encompassed by the time-windows of Table~\ref{table:setofNR}.
To keep both parameters under control, we realign our models with alternative time-windows that are dictated by the number of GW cycles to merger $\Delta N_\text{GW}(t) \equiv N_\text{GW}(t)-N_\text{GW}^\text{merg}
$ of the NR simulations. That is, for each mass ratio we fix a different time-window [$t^\text{alig}_1,t^\text{alig}_2$], corresponding to the \textit{same} interval of cycles to merger [$\Delta N_\text{GW}(t^\text{alig}_1)$, $\Delta N_\text{GW}(t^\text{alig}_2)$]. 
The benefits of this choice are two-fold. To begin with, the alignment windows thus calculated depends on the position of the NR merger (peak of $h_{22}^{\text{NR}}$), which is a quantifiable feature of every NR simulation.
Moreover, this choice allows us to assess trends across the mass ratios fairly, since the waveforms thus aligned are compared in the same range of GW cycles. A caveat for this alignment method is that the GW cycles of evolutions with smaller $q$ lie in a regime of stronger gravity.

We choose to align the EOB models to NR in an interval of $N_\text{GW}$ such that 
$[\Delta N_\text{GW}(t^\text{alig}_1), \Delta N_\text{GW}(t^\text{alig}_2)]=[-34, -24]$, corresponding to the time-windows reported in Table~\ref{table:simulations}.
This choice stems from the length of the shortest NR simulation, e.g., $q=1/9$, which counts $N_\text{GW}^\text{merg}=37.86$ GW cycles at merger (the first $\sim$3GW cycles of this simulation are neglected in order to avoid junk radiation). In Fig.~\ref{fig:dphivsq}, we plot the dephasing for the three models that perform best in Fig.~\ref{fig:wavephase}: that is, $H_{\text{3PN}}^{\text{EOB}}$, $H_{\text{SMR}}^{\text{ EOB,PS}}$ and $H_{\text{SMR-3PN}}^{\text{EOB,PS}}$ and study the trends across $q$. For every simulation, we calculate the dephasing 8 and 4 GW cycles before merger to show the robustness of the trends\footnote{We have also checked that the trends are unaffected by variations in the number of orbital cycles in the alignment window.}.
Noticeably, the 3PN EOB waveform in the DJS gauge starts degrading in accuracy as the mass ratio is decreased, while the SMR and SMR-3PN ones improve: remarkably, for most $q$'s, the SMR-3PN model only dephases by a few hundredths of a radian
up to a 4 GW cycles before merger.
Moreover, we notice that SMR models start performing better than $H_{\text{3PN}}^{\text{ EOB}}$ for $q \lesssim 1/3$, hinting again to the 
fact that SMR information, when reorganized in the EOB framework, could be used to model systems that are very close to the equal-mass-ratio regime \cite{LeTiec:2011bk,Barausse:2011dq}.

The picture emerging from Fig.~\ref{fig:dphivsq} is that the SMR-3PN model is the most consistent of the two models with SMR information, corroborating the findings for $q=1$ and $q=1/10$ in the binding energy comparisons. The small dephasing of the SMR-3PN model suggests that the Hamiltonian upon which it is based is a possible starting point to develop a new generation of EOB waveform models able to tackle the currently challenging intermediate-mass-ratio regime.

\section{Discussion and Conclusions}
\label{sec:conclusions}

The complete EOB Hamiltonian at linear order in SMR from Ref.~\cite{LeTiec:2011dp} suffers from a coordinate singularity at the LR radius in the deformed Schwarzschild background. Building on Refs.~\cite{Akcay:2012ea,Damour:2017zjx}, we have constructed two Hamiltonians in the post-Schwarzschild (PS) reformulation of the EOB approach \cite{Damour:2017zjx,Antonelli:2019ytb} (both with the SMR correction to the Detweiler redshift and with mixed SMR-3PN information), and checked 
that they are not affected by poles  at the LR radius (and related unphysical features) by studying plunging trajectories.

We have then explored the merits of the SMR and mixed SMR-3PN Hamiltonians via comparisons of their waveforms and binding energies, and those of PN Hamiltonians in different gauges, against NR predictions. Ultimately, we find that:
\begin{enumerate} 
	\item For both $q=1$ and $q=1/10$, the binding energies of SMR and SMR-3PN EOB models (see Figs.~\ref{fig:compEJ} and~\ref{fig:compEO}) generally compare better against NR than the binding energy of the PS Hamiltonian with 3PN information.
	\item The generic orbit 3PN information in the SMR-3PN EOB Hamiltonian improves the binding energy and phase comparisons of SMR EOB models.
	\item PN Hamiltonians in the EOB-PS gauge have binding energies that compare worse than those from PN Hamiltonians in the standard EOB gauge, confirming the findings of Ref.~\cite{Antonelli:2019ytb} and extending their validity to non-adiabatic evolutions.
	\item The SMR-3PN EOB model agrees remarkably well against NR simulations, see Fig.~\ref{fig:dphivsq}. The dephasing up to 4 GW cycles before merger is a few hundredths of a radian for $q \lesssim 1/3$ and a tenth of a radian for $q>1/3$. The only EOB PN model with comparable dephasing is the 3PN EOB Hamiltonian in the DJS gauge for $q \gtrsim 1/3$.
\end{enumerate}

The construction of the SMR EOB Hamiltonian in this paper depends on a number of choices. First of all, we chose to fix the coordinate freedom in the effective Hamiltonian using the PS gauge. This was chosen because of its relative simplicity, while allowing a natural path towards avoiding singularities at the light ring. However, there may exist different choices that are equally (or more) effective. Second, while the EOB Hamiltonian in principle applies to generic orbits, we fix the linear-in-$\nu$ part only by comparing to the circular-orbit binding energy. Consequently, there is considerable freedom in the ``non-circular-orbit'' part of the Hamiltonian. In practice, we fix this freedom by choosing the specific functional dependence of the effective Hamiltonian on $\Ham_{\text{S}}$ given by Eq.~\eqref{HSMR}. This choice is in part restricted by the requirement that the Hamiltonian be analytic, but other options are available. Third and finally, SMR data for the binding energy extends only to the light ring. The Hamiltonian in the region $u>\tfrac{1}{3}$ therefore depends only on the analytic extension of the redshift data. Given that this data is known only to finite numerical precision, there is some freedom in the choice of the exact analytical form of its fit. This choice can also affect the relative size of the different coefficient functions in Eq.~\eqref{HSMR}.

Our investigation opens up further avenues of research. To begin with, one can study whether it is possible to uniquely fix other EOB gauges that could accommodate the Detweiler redshift (without introducing a LR-divergence) and study their merits via comparisons against NR. As discussed already in Ref.~\cite{Akcay:2012ea}, to solve the LR-divergence arising in this context the non-geodesic function $\hat{Q}$ needs a term proportional to $p_{\phi}^3$, possibly resummed in another quantity (as done in the PS gauge using $\Ham_\text{S}$). It would be quite interesting to see whether other gauges that allow solving the LR-divergence also improve the comparisons against NR predictions. One concrete example of different resummation that was shown to improve the comparisons of the conservative sector of post-Minkowskian Hamiltonians in PS form has been given in the Appendix of Ref.~\cite{Antonelli:2019ytb}. It is worthwhile to study whether a similar choice  could work for the SMR and SMR-3PN models herein presented.
The hope is that using different resummations, and including information from the second order in the SMR, one could obtain a considerably improved EOB Hamiltonian that, after further calibration to NR, would be very useful for LIGO/Virgo analyses in the near-future.

Further research endeavours could be directed towards informing the EOB with different SMR quantities than the circular orbit Detweiler redshift. An example of a quantity that still needs to be fully exploited is the generalized redshift \cite{Barack:2011ed,Akcay:2015pza}, which includes information for arbitrarily eccentric orbits.
We envision using EOB Hamiltonians at linear and higher orders in the
mass ratio for inference studies in the future detectors' era, when
precise models will be needed to properly characterize high
signal-to-noise systems, possibly having rather small mass ratios. In
order for this program to be achieved, not only should the
conservative sector be optimized with both results at second order in
$q$ and (potentially) a better resummation, but information from other
crucial physical quantities should also be incorporated: notably missing
features in our analysis are the spin and eccentricity. Furthermore, a
more comprehensive study of the dissipative sector must be pursued. It
would be desirable, for instance, to include more self-force
information in the flux. Lastly, we would also need to build the full 
inspiral, merger and ringdown waveforms, and calibrate them to NR simulations.  
We leave these important investigations to future work.

\section{Acknowledgments}
	It is a pleasure to thank Sergei Ossokine for providing us with the NR data used for the comparisons, and Tanja Hinderer for the original version of the EOB evolution code in Mathematica used in this paper. A.A. would like to further thank Sergei Ossokine and Roberto Cotesta for many fruitful discussions over the preparation of this paper. MvdM was supported by European Union's Horizon 2020 research and innovation programme under grant agreement No.~705229.
	This work makes use of the Black-Hole Perturbation Toolkit \cite{BHPToolkit}.

  

\newcommand{\bpm}{\begin{pmatrix}}
\newcommand{\epm}{\end{pmatrix}}
\newcommand{\vinf}{v}

\chapter[Gravitational spin-orbit coupling through third-subleading PN order]{Gravitational spin-orbit coupling through third-subleading post-Newtonian order: from first-order self-force to arbitrary mass ratios}
\label{chap:four}

\textbf{Authors}\footnote{Originally published in \emph{Phys.Rev.Lett.} 125 (2020) 1, 011103 .}: Andrea Antonelli, Chris Khavanagh, Mohammed Khalil, Jan Steinhoff, Justin Vines.
\newline

\textbf{Abstract:} exploiting simple yet remarkable properties of relativistic gravitational scattering, we use first-order self-force (linear-in-mass-ratio) results to obtain arbitrary-mass-ratio results for the complete third-subleading post-Newtonian (4.5PN) corrections to the spin-orbit sector of spinning-binary conservative dynamics, for generic (bound or unbound) orbits and spin orientations.
We thereby improve important ingredients of models of gravitational waves from spinning binaries, and we demonstrate the improvement in accuracy by comparing against aligned-spin numerical simulations of binary black holes.

\section{Introduction}

The success of gravitational-wave (GW) astronomy in the next decades relies on significantly improved theoretical predictions of GW signals from coalescing binaries of \emph{spinning} compact objects such as black holes (BHs).
A network of GW detectors~\cite{TheLIGOScientific:2014jea,TheVirgo:2014hva} has now observed dozens of signals from binary BHs,
measuring distributions of the BHs' masses and spins and extrinsic properties, 
enabling diverse applications in astro- and fundamental physics~\cite{LIGOScientific:2018mvr,LIGOScientific:2018jsj,LIGOScientific:2019fpa,Abbott:2019yzh}: e.g., discerning binary BH formation channels~\cite{LIGOScientific:2018jsj}, measurement of the Hubble constant~\cite{Abbott:2019yzh}, and tests of general relativity (GR)~\cite{LIGOScientific:2019fpa}.
The search for and parameter estimation of GW signals require accurate predictions, from the inspiral (treated by analytic approximations) to the last orbits and merger of the binary (treated by numerical relativity, NR).
The current accuracy of theoretical predictions, from combined analytic and numerical methods, will likely become insufficient when current detectors reach design sensitivity around 2022~\cite{Purrer:2019jcp}.
More accurate predictions for gravitational waves are thus key to enable the 
physics applications mentioned above.

The primary relevant analytic approximation is the post-Newtonian (PN, weak-field and slow-motion) approximation.
The conservative orbital dynamics is known for nonspinning binaries to the fourth-subleading PN order~\cite{Damour:2014jta,Bernard:2016wrg,Foffa:2019rdf,Foffa:2019yfl,Blumlein:2020pog} (with partial results at the fifth~\cite{Ledvinka:2008tk,Blanchet:2018yvb,Foffa:2019hrb,Blumlein:2019zku,Bini:2019nra} and sixth~\cite{Blumlein:2020znm,Cheung:2020gyp,Bini:2020nsb}), but only to second-subleading order (or next-to-next-to-leading order, N$^2$LO) in the spin-orbit sector~\cite{Hartung:2011te,Levi:2015uxa,Bohe:2012mr}.  The gravitational spin-orbit couplings, linear in the component bodies' spins, are analogous to those in atomic physics.
Recently, the three-loop Feynman integrals at N$^3$LO in the spin-orbit case were calculated~\cite{Levi:2020kvb}, leaving however plenty of tensorial lower-loop integrals as a comparably large computational task.
Innovations that complement these massive algebraic manipulations are thus of great potential value.

In this Letter, we follow a line of reasoning which leads to a complete result for the sought-after N$^3$LO-PN spin-orbit dynamics (at 4.5PN order for rapidly spinning binaries), requiring relatively little computational effort by building on a diverse array of previous results.
We extend to the spinning case a novel approach based on special properties of the gauge-invariant scattering-angle function~\cite{Bini:2019nra,Damour:2019lcq,Vines:2018gqi}, which encodes the complete binary dynamics (both bound and unbound).
The weak-field approximation of the scattering angle is strongly constrained by results in the small-mass-ratio approximation\footnote{We define the small-mass-ratio limit as $q=\frac{m_1}{m_2}\ll1$, where $m_{1,2}$ are the masses of the compact objects.}, as treated in the gravitational \emph{self-force} paradigm~\cite{Barack:2018yvs}.
The scattering-angle constraints imply that known first-order (linear-in-mass-ratio) self-force results with spin~\cite{Bini:2016dvs,Kavanagh:2016idg,Kavanagh:2017wot} uniquely fix the full N$^3$LO-PN spin-orbit dynamics for arbitrary mass ratios.
This result completes the 4.5PN conservative dynamics of (rapidly) spinning binaries, together with the NLO cubic-in-spin couplings~\cite{Levi:2019kgk} (see also \cite{Siemonsen:2019dsu}).

As applications, we compute quantities which can be employed to improve waveform models for GW astronomy: the circular-orbit aligned-spin binding energy and the effective gyro-gravitomagnetic ratios.
The former is a crucial ingredient in the construction of faithful models (together with the GW energy flux), for which we quantify the accuracy gain due to the present results by comparing to NR simulations. 
The latter parametrize spin effects in the \texttt{SEOBNR} waveform codes~\cite{Bohe:2016gbl,Babak:2016tgq,Cotesta:2018fcv,Ossokine:2020kjp} used in LIGO-Virgo searches and inference analyses~\cite{LIGOScientific:2018mvr} and in the upcoming \texttt{TEOBResumS} waveform models~\cite{Nagar:2018plt,Nagar:2018zoe}.
The gyro-gravitomagnetic ratios are analogous to the famous ``g-factor'' describing the anomalous magnetic dipole moment of the electron, where contributions at the fifth subleading order were obtained~\cite{Aoyama:2012wj} and lead to spectacular agreement with experiment~\cite{Odom:2006zz}.
Regarding the gravitational analog, experimental constraints on the gyro-gravitomagnetic ratios are so far seemingly out of reach. In fact, only two GW events were observed to contain nonvanishing spin effects with 90\% confidence~\cite{LIGOScientific:2018mvr} (see also Refs.~\cite{Zackay:2019tzo,Huang:2020ysn}).
However, this will change, e.g., when systems with precessing spins are observed in the future, since the precession of the orbital plane leads to a characteristic modulation of the emitted GWs.
This may allow improved tests of GR and inference of spins.  Measuring BH spins and their orientations is also important for discriminating binary formation channels~\cite{LIGOScientific:2018jsj}.

We begin by extending the link between weak-field scattering and the self-force approximation~\cite{Vines:2018gqi,Bini:2019nra,Damour:2019lcq} to the spin-orbit sector.
Using existing self-force results, we are then able to uniquely determine the N$^3$LO-PN spin-orbit dynamics, as encoded in the gauge-invariant scattering angle.
We continue by calculating the gyro-gravitomagnetic ratios and circular-orbit aligned-spin binding energy.
We compare to NR simulations to quantify the accuracy improvement and present our conclusions.
$G$ denotes Newton's constant, and $c$ the speed of light.

\section{The mass dependence of the scattering angle}

\begin{figure}
	\centering
	\includegraphics[width=0.9\linewidth]{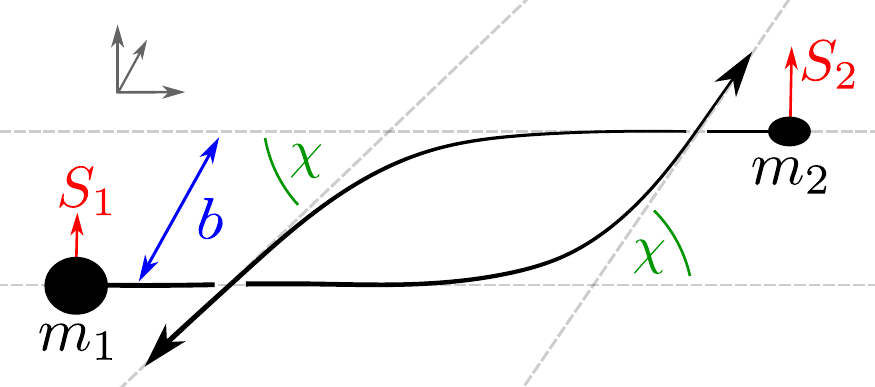}
	\caption{Illustration of aligned-spin scattering BHs.}
	\label{fig:scattering}
\end{figure}

The local-in-time conservative dynamics of a two-massive-body system (without spin or higher multipoles) is fully encoded in the system's gauge-invariant scattering-angle function $\chi(m_1,m_2,v,b)$ ~\cite{Damour:2016gwp,Damour:2017zjx}.  This gives the angle $\chi$ by which both bodies are deflected in the center-of-mass frame, as a function of the masses $m_\ms a$ ($\ms a=1,2$), the asymptotic relative velocity $v$, and the impact parameter $b$.  Based on the structure of iterative solutions in the weak-field (post-Minkowskian) approximation, it has been argued in Sec.~II of Ref.~\cite{Damour:2019lcq} that this function exhibits the following simple dependence on the masses (at fixed $v$ and $b$), through the total mass $M=m_1+m_2$ and the symmetric mass ratio $\nu=m_1m_2/M^2$,
\bse\label{mastereq}
\begin{alignat}{3}\label{chiform}
\frac{\chi}{\Gamma} &= \frac{GM}{b}X_{G^1}^{\nu^0}(v)+\Big(\frac{GM}{b}\Big)^2X_{G^2}^{\nu^0}(v)
\\\nnm
&\quad +\Big(\frac{GM}{b}\Big)^3\Big[X_{G^3}^{\nu^0}(v)+\nu X_{G^3}^{\nu^1}(v)\Big]
\\\nnm
&\quad +\Big(\frac{GM}{b}\Big)^4\Big[X_{G^4}^{\nu^0}(v)+\nu X_{G^4}^{\nu^1}(v)\Big]+\mc O\Big(\frac{GM}{b}\Big)^5,
\end{alignat}
where $\Gamma=E/Mc^2$, with $E^2=(m_1^2+m_2^2+2m_1m_2\gamma)c^4$ being the squared total energy, and $\gamma=(1-v^2/c^2)^{-1/2}$ the asymptotic relative Lorentz factor.  The remarkable fact to be noted here is that the $\mc O(\frac{GM}{b})^{1,2}$ terms are independent of $\nu$, while the $\mc O(\frac{GM}{b})^{3,4}$ terms depend linearly on $\nu$.

As will be argued in detail in future work,${}^{\ref{foot:deriv}}$ this result generalizes straightforwardly to the case of spinning bodies in the aligned-spin configuration, i.e., spins pointing in the direction of the orbital angular momentum (as shown in Fig.~\ref{fig:scattering}).  The aligned-spin dynamics is fully described by the aligned-spin scattering-angle function $\chi(m_\ms a,S_\ms a,v,b)$~\cite{Vines:2018gqi}. Here, $S_\ms a=m_\ms a ca_\ms a$ are the signed spin magnitudes, positive if aligned as in Fig.~\ref{fig:scattering}, negative if anti-aligned.  At the spin-orbit (linear-in-spin) level, the form of Eq.~(\ref{chiform}) holds, with the $X$ functions acquiring additional (linear) dependence on the spins \emph{only} through the dimensionless ratios $a_\ms a/b={S_\ms a}/{m_\ms acb}$, as follows:\footnote{\label{foot:deriv}Note that our Eq.~(\ref{mastereq}) is equivalent to Eqs.~(2.14) and (2.15) of Ref.~\cite{Damour:2019lcq}, but with all the functions $\ms Q^{n\mr{PM}}_{\cdots}(\gamma)$ on the right-hand side of (2.15) replaced by functions $\ms Q^{n\mr{PM}}_{\cdots}(\gamma,a_1/b,a_2/b)$ which are linear in $a_1/b$ and $a_2/b$, and with the additional constraints imposed by symmetry under $(m_1,a_1)\leftrightarrow(m_2,a_2)$.  The arguments leading to this result are very much analogous to those for the spinless case as given in Ref.~\cite{Damour:2019lcq} --- using the structure of the PM expansion, Poincar\'e symmetry, dimensional analysis, etc.\ --- with the given mass dependence holding at fixed ``geometric quantities,'' except that these are now $v,b,a_1,a_2$ instead of just $v$ (or $\gamma$) and $b$.  The rescaled spins $a_\ms a=S_\ms a/m_\ms a c$ and the ``covariant'' (Tulczyjew-Dixon) worldlines (separated by the ``covariant'' impact parameter $b$) are identified as the appropriate geometrical (mass-independent) quantities, because it is in terms of these variables that the first-order metric perturbation is linear in the masses.}
\begin{alignat}{3}\label{replaceX}
X_{G^{n}}^{\nu^m} &\to X_{G^{n}}^{\nu^m}(v)+\frac{a_+}{b}X_{G^na_+}^{\nu^m}(v)+\delta\frac{a_-}{b}X_{G^na_-}^{\nu^m}(v),
\end{alignat}
\ese
where $a_\pm = a_2\pm a_1$ and $\delta=(m_2-m_1)/M$, with the special constraints $X_{G^1a_-}^{\nu^0}=0=X_{G^3a_-}^{\nu^1}$; cf.\ Eq.~(4.32) of Ref.~\cite{Vines:2018gqi}, where this is seen to hold through N$^2$LO in the PN expansion.  It is crucial to note that the impact parameter $b$ in Eq.~\eqref{mastereq}, is the (``covariant'') one orthogonally separating the asymptotic worldlines defined by the Tulczyjew-Dixon condition~\cite{Dixon:1979,Tulczyjew:1959} for each spinning body~\cite{Vines:2018gqi,Vines:2017hyw}.

Now, the fourth order in $G M / b$ encodes the complete spin-orbit dynamics at N$^3$LO in the PN expansion, and according to Eq.~\eqref{mastereq} only terms up to linear order in the mass ratio $\nu$ appear on the right-hand side (noting $\delta\to\pm1$ as $\nu\to0$)---that is, first-order self-force (linear-in-$\nu$) results can be employed to fix the functions $X_{G^n\cdots}^{\nu^m}(v)$ for $n \leq 4$.

\section{Scattering angle, Hamiltonian, and binding energy}

We now connect the scattering angle to an ansatz for a local-in-time binary Hamiltonian including spin-orbit interactions.
If nonlocal-in-time (tail) effects are present, this step requires extra care~\cite{Bini:2019nra}, but this is not the case at the N$^3$LO-PN spin-orbit level.
Crucially, the Hamiltonian describes the dynamics for both unbound (scattering) and bound orbits.
The latter are not only most relevant for GW astronomy, but are also where the vast majority of self-force results are available.
Hence, a gauge-dependent Hamiltonian allows us to connect the scattering angle \eqref{mastereq} with known self-force results.

Let us parametrize our binary Hamiltonian $H(\vec r,\vec p,\vec S_1,\vec S_2)$ in the effective-one-body (EOB)~\cite{Buonanno:1998gg} form,
\begin{equation}\label{HEOB}
H = M c^2 \sqrt{1 + 2 \nu \left( \frac{H_\text{eff}}{\mu c^2} - 1 \right)} ,
\end{equation}
where $H_\text{eff}(\vec r,\vec p,\vec S_1,\vec S_2)$ is the effective Hamiltonian and $\mu = M \nu$ is the reduced mass, with canonical Poisson brackets $\{r^i,p_i\}=\delta^i_j$, $\{S_\ms a^i,S_\ms a^j\}=\epsilon^{ij}{}_kS_\ms a^k$, and all others vanishing.  At the spin-orbit level, to linear order in the spins, parity invariance implies that $H$ can depend on the spins only through the scalars $\vec L\cdot \vec S_\ms a$, where $\vec L=\vec r\times\vec p$ is the canonical orbital angular momentum.  Thus, a generic Hamiltonian ansatz is of the form
\begin{equation}
H_\text{eff} = H_\text{eff}^\text{ns} + \frac{1}{c^2 r^3} \vec{L} \cdot (g_S \vec{S} + g_{S^*} \vec{S}^*), \label{Heff}
\end{equation}
where $H_\text{eff}^\text{ns}(\vec r,\vec p)$ is the nonspinning Hamiltonian. We use the conventional spin combinations $\vec{S} = \vec{S}_1 + \vec{S}_2$, $\vec{S}^* = \frac{m_2}{m_1} \vec{S}_1 + \frac{m_1}{m_2} \vec{S}_2$, while $g_S(\vec r,\vec p)$ and $g_{S^*}(\vec r,\vec p)$ are the effective gyro-gravitomagnetic ratios.  In specializing to the case of aligned spins, in which $\vec S_\ms a=S_\ms a\hat{\vec L}$ are (anti)parallel to $\vec L=L\hat{\vec L}$ ($L=|\vec L|$), the motion is confined to the plane orthogonal to the angular momenta, and Eq.~\eqref{Heff} simplifies to
\begin{equation}
H_\text{eff} =  H_\text{eff}^\text{ns} + \frac{1}{c^2 r^3} L  (g_S S + g_{S^*} S^*), \quad \text{(aligned)} \label{Heffaligned}
\end{equation}
where, crucially, $g_S$ and $g_{S^*}$ are unmodified by this specialization (as they are independent of the spins).  The aligned-spin Hamiltonian is therefore sufficient to reconstruct the generic-spin Hamiltonian, up to the spin-orbit level.  We can adopt polar coordinates $(r,\varphi)$ in the orbital plane, with canonically conjugate momenta $(p_r,L)$, and the Hamiltonian is independent of $\varphi$ due to rotation invariance.  Then $H^\text{ns}_\text{eff}$, $g_S$, and $g_{S^*}$ are each functions of $(r,p_r,L)$.   We take $H_\text{eff}^\text{ns}$ to be given to 4PN order by Eqs.~(5.1) and (8.1) in Ref.~\cite{Damour:2015isa}.  Considering the freedom under canonical transformations, it can be shown that there exists a gauge in which $g_S$ and $g_{S^*}$ are independent of $L$ \cite{Damour:2008qf,Nagar:2011fx,Barausse:2011ys}; we adopt this choice and parametrize our spin-orbit Hamiltonian with the undetermined gyro-gravitomagnetic ratios $g_S(r, p_r)$ and $g_{S^*}(r, p_r)$.
Each term in a PN-expanded ansatz for $g_S$ and $g_{S^*}$ carries a certain power in $c$, from which the PN order can be read off; we include terms up to $c^{-6}$ here.
($c^{-2}$ corresponds to one PN order and $c \rightarrow \infty$ to the Newtonian limit.)

To ascribe physical significance to the spin-orbit Hamiltonian, we point to
the striking similarity with the electromagnetic spin-orbit interactions in atomic physics, which makes $g_S$ and $g_{S^*}$ analogous to the ``g-factor'' of the electron (except that $g_S$ and $g_{S^*}$ depend on dynamical variables). 
This is no accident, since the gravito-magnetic field generated, e.g., by a rotating mass, can be interpreted to exert a Lorentz-like force.
The relativistically preferred geometrical interpretation is that gravito-magnetic fields are dragging inertial/free-falling reference frames, as impressively demonstrated by the Gravity Probe B satellite experiment~\cite{Everitt:2011hp}.

We constrain the ansatz for the Hamiltonian by requiring that it reproduces (i) the mass dependence of the scattering angle \eqref{mastereq}, (ii) the $\nu\to0$ limit of the scattering angle, for a spinning test particle in a Kerr background, as obtained, e.g., by integrating Eq.~(65) of Ref.~\cite{Bini:2017pee}, and (iii) certain gauge-invariant self-force observables, namely, the Detweiler-Barack-Sago redshift~\cite{Detweiler:2008ft,Barack:2011ed,Bini:2016dvs,Hopper:2015icj,Bini:2016qtx,Kavanagh:2015lva,Kavanagh:2016idg,Bini:2015mza} and the spin-precession frequency~\cite{Dolan:2013roa,Akcay:2016dku,Akcay:2017azq,Bini:2014ica,Bini:2015mza,Kavanagh:2017wot,Bini:2018ylh,Bini:2018aps} for bound eccentric aligned-spin orbits, to linear order in the mass ratio.
The scattering angle $\chi$ is obtained from the Hamiltonian \eqref{HEOB} via Eq.~(4.10) of Ref.~\cite{Vines:2018gqi}, with the translation from the total energy $E=H$ and canonical orbital angular momentum $L$ to the asymptotic relative velocity $v$ and ``covariant'' impact parameter $b$ accomplished by Eqs.~(4.13) and (4.17) of Ref.~\cite{Vines:2018gqi}.
The redshifts $z_\ms a$ and spin-precession frequencies $\Omega_\ms a$  ($\ms a=1,2$) are given by
\begin{equation}\label{1law}
z_\ms a =\bigg\langle \frac{\partial H}{\partial m_\ms a} \bigg\rangle, \qquad
\Omega_\ms a =\bigg\langle \frac{\partial H}{\partial S_\ms a}\bigg\rangle,
\end{equation}
where $\langle\cdots\rangle$ denotes an average over one period of the radial motion, following from a first law of binary mechanics for eccentric aligned-spin orbits~\cite{LeTiec:2011ab,Blanchet:2012at,Tiec:2015cxa,Fujita:2016igj}.  The procedure for expressing these quantities, in the small-mass-ratio limit, in terms of variables used in self-force calculations is detailed in Ref.~\cite{Bini:2019lcd}.  In this process, to reach the N$^3$LO-PN accuracy in the spin-orbit sector, it is necessary to include the nonspinning 4PN part of the Hamiltonian, including the nonlocal tail part~\cite{Damour:2014jta}, given as an expansion in the orbital eccentricity as in Ref.~\cite{Damour:2015isa}.  After lengthy calculation, working consistently in the small-mass-ratio and PN approximations, we obtain, from our Hamiltonian ansatz, expressions for the redshift $z_1$ and precession frequency $\Omega_1$ of the smaller body, which can be directly compared with the self-force results in Eq.~(4.1) of Ref.~\cite{Kavanagh:2016idg}, Eq.~(23) of Ref.~\cite{Bini:2016dvs} and Eq.~(20) of Ref.~\cite{Bini:2019lcd} for the redshift, and Eq.~(3.33) of Ref.~\cite{Kavanagh:2017wot} for the precession frequency.
The resultant constraints uniquely fix $g_S(r, p_r)$ and $g_{S^*}(r, p_r)$ at N$^3$LO, via an overdetermined system of equations. 

From the Hamiltonian, we can finally calculate the \emph{aligned-spin} circular-orbit binding energy $E_b = H-Mc^2$ as a function of the circular-orbit frequency $\omega=d\varphi/dt=\doe H/\doe L$.
This is a gauge-invariant relation that can be compared to NR.
We decompose $E_b$ into nonspinning and spin-orbit (SO) parts, and further into PN orders, as in
\begin{equation}\label{bindSO}
E_b^\text{SO} = E_{b,\text{LO}}^\text{SO} + E_{b,\text{NLO}}^\text{SO} + E_{b,\text{N$^2$LO}}^\text{SO} + E_{b,\text{N$^3$LO}}^\text{SO} + \dots . \\
\end{equation}
We can decompose the $g_S$, $g_{S^*}$, and $\chi_\text{SO}$ results from the previous discussion in the same way.
The N$^3$LO pieces of all these quantities are the main results of this Letter:
	\begin{align}
	\label{chiNNNLOSO}
	\frac{\chi_\text{SO}^\text{N$^3$LO}}{\Gamma}&=
	\frac{\vinf}{c\, b} \Big(a_+ \quad \delta a_-\Big)\Bigg\{ \bigg[\frac{1}{4}\bpm 177 \nu \\ 0 \epm \frac{\vinf^6}{c^6}\bigg] \left(\frac{GM}{\vinf^2b}\right)^3
	 \nonumber\\
	&+\pi \Bigg[ \frac{3}{4} \bpm -91 + 13\nu \\ -21 + \nu \epm \frac{\vinf^2}{c^2}
	-\frac{1}{8}\bpm 1365 - 777\nu \\ 315 - 45\nu \epm\frac{\vinf^4}{c^4} \nonumber\\
	&-\frac{1}{32} \bpm 1365 - \left(\frac{23717}{3} - \frac{733\pi^2}{8}\right)\nu \\ 315 - \left(\frac{257}{3} + \frac{251\pi^2}{8}\right)\nu \epm\frac{\vinf^6}{c^6} \Bigg] \left(\frac{GM}{\vinf^2b}\right)^4 \Bigg\},\\
	c^6 g_S^\text{N$^3$LO} &= \frac{\nu}{1152} \left(-80399+1446\pi^2+13644\nu-63\nu^2\right)\frac{(GM)^3}{r^3} \nonumber\\
	&+ \frac{3\nu}{64} \left(-1761+2076\nu+23\nu^2\right) \frac{p_r^2}{\mu^2} \frac{(GM)^2}{r^2}
	\nonumber\\
	&+ \frac{\nu}{128} \left(781+3324\nu-771\nu^2\right) \frac{p_r^4}{\mu^4} \frac{GM}{r} + \frac{7\nu}{128} \left(1-36\nu-95\nu^2\right) \frac{p_r^6}{\mu^6} ,\\
	c^6 g_{S^*}^\text{N$^3$LO} &= - \frac{1}{384} \left[1215+2(7627-246\pi^2)\nu-4266\nu^2+36\nu^3 \right] \frac{(GM)^3}{r^3} \nonumber\\
	&- \frac{3}{64} \left(15+558\nu-1574\nu^2-36\nu^3\right) \frac{p_r^2}{\mu^2} \frac{(GM)^2}{r^2}
	\nonumber\\
	&+ \frac{1}{128} \left(-1105-106\nu+702\nu^2-972\nu^3\right) \frac{p_r^4}{\mu^4} \frac{GM}{r} 
	\nonumber\\
	&- \frac{7}{128} \left(45+50\nu+66\nu^2+60\nu^3\right) \frac{p_r^6}{\mu^6} ,\\
	E_{b,\text{N$^3$LO}}^\text{SO} &= - \frac{\nu c^3}{G M} \frac{v_\omega^{11}}{c^{11}} \bigg[ S \bigg( 45 - \frac{19679+174\pi^2}{144} \nu + \frac{1979}{36} \nu^2 + \frac{265}{3888} \nu^3 \bigg)
	\nonumber\\
	&+ \frac{S^*}{8} \bigg( \frac{135}{2} - 565 \nu + \frac{1109}{3} \nu^2 + \frac{50}{81} \nu^3 \bigg) \bigg] , \label{EbNNNLO}
	\end{align}
where $v_\omega = (GM\omega)^{1/3}=x^{1/2}c$. 
One needs to add our Eq.~\eqref{chiNNNLOSO} to Eq.~(4.32b) in Ref.~\cite{Vines:2018gqi} to obtain the complete spin-orbit scattering-angle contribution through N$^3$LO-PN and through $\mc O(\frac{GM}{b})^4$.
The lower-order corrections to $E_b^\text{SO}$ can be found in Eq.~(5.4) of Ref.~\cite{Levi:2015uxa}, and the lower-order gyro-gravitomagnetic ratios in Eqs.~(55) and (56) of Ref.~\cite{Nagar:2011fx} (see also Ref.~\cite{Damour:2008qf,Barausse:2011ys}). Through the results for $g_S$ and $g_{S^*}$ presented above, one can straightforwardly improve the \texttt{SEOBNR} waveform models~\cite{Bohe:2016gbl,Babak:2016tgq,Cotesta:2018fcv,Ossokine:2020kjp} used in contemporary gravitational-wave data analysis~\cite{LIGOScientific:2018mvr}.
Likewise, one can use them to improve the upcoming \texttt{TEOBResumS} waveform models~\cite{Nagar:2018plt,Nagar:2018zoe}.
The other main waveform model used by LIGO-Virgo data analysis~\cite{LIGOScientific:2018mvr} is the \texttt{IMRPhenom} family~\cite{Hannam:2013oca,Husa:2015iqa,Khan:2015jqa,Khan:2019kot,Garcia-Quiros:2020qpx,Pratten:2020fqn,Pratten:2020ceb}, which can also be improved using our results, though less directly.

\section{Comparison to NR}
\begin{figure}
	\centering
	\includegraphics[width=\linewidth]{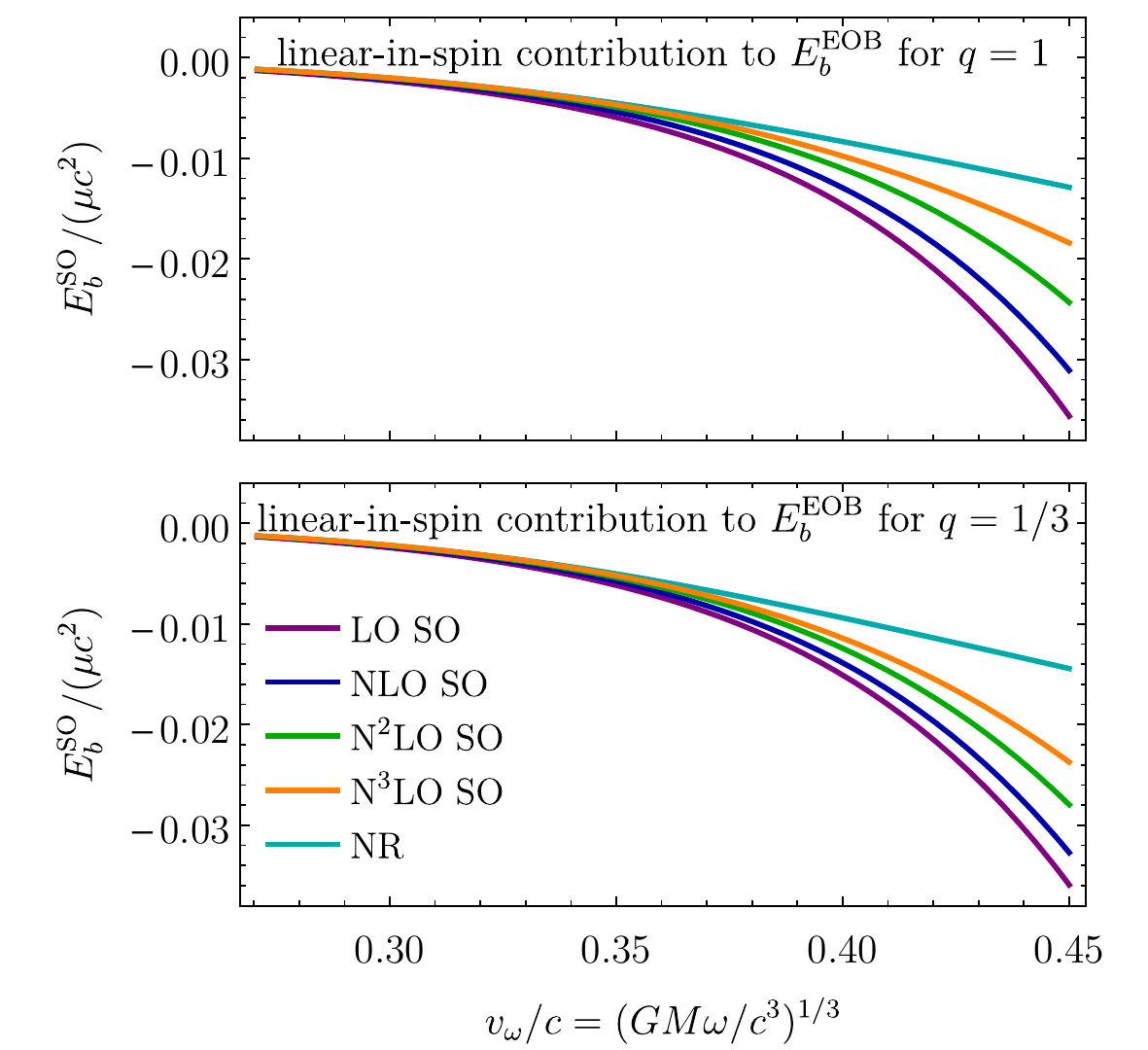}
	\caption{
		Comparison of the gauge-invariant relation between the circular-orbit aligned-spin spin-orbit binding energy $E_b$ and $v_\omega$.
		The figure shows results obtained numerically from the (PN-resummed) EOB Hamiltonian \eqref{HEOB} and NR results from Refs.~\cite{Ossokine:2017dge,SXS}.
		The linear-in-spin contribution is isolated using Eq.~\eqref{EbSO} with spin magnitudes $\hat{a} = 0.6$ and for mass ratios $q=1$ and $1/3$.}
	\label{fig:bindingenergy}
\end{figure}

We now quantify the improvement in accuracy from the new N$^3$LO spin-orbit correction. The circular-orbit aligned-spin binding energy is a particularly good diagnostic for this, since it encapsulates the conservative dynamics of analytical models, and can be obtained from accurate NR simulations~\cite{Damour:2011fu,Nagar:2015xqa}.
Of particular interest for us is the possibility to (approximately) isolate the linear-in-spin (spin-orbit) contribution by combining the binding energy for two configurations with spins parallel and anti-parallel to the direction of the angular momentum  as follows \cite{Dietrich:2016lyp,Ossokine:2017dge}
\begin{equation}
\label{EbSO}
E_b^\text{SO}(\nu,\hat{a},\hat{a}) = \frac{1}{2} \left[ E_b(\nu,\hat{a},\hat{a}) - E_b(\nu,-\hat{a},-\hat{a})\right],
\end{equation}
with dimensionless spin $\hat{a} = \hat{a}_{\ms a} \equiv c S_{\ms a} / (G m_{\ms a}^2)$.
The result, based on recent NR simulations~\cite{SXS,Ossokine:2017dge}, is shown in Fig.~\ref{fig:bindingenergy}.
The figure also shows the spin-orbit binding energy extracted numerically from the EOB Hamiltonian~\eqref{HEOB}, combining two binding energies for different spin directions in the same way as in the NR case.
The N$^3$LO spin-orbit result shows a clear advantage over the N$^2$LO one, and that improvement is more pronounced for equal masses than for slightly unequal masses.
[The N$^3$LO PN binding energy (\ref{EbNNNLO}) is very similar to the EOB one for the shown mass ratios.]
This indicates that an inclusion of the N$^3$LO into existing waveform models may lead to improvements even in the strong-field regime, 
otherwise only accessible by computationally-expensive NR simulations.
Recall that gravitational waves are observed from low frequencies (where approximation methods are applicable) to high frequencies (where PN theory is expected to break down).

\section{Conclusions}
Currently-operating (second-generation) gravitational-wave detectors require accuracy improvements for GW predictions by the time they reach design sensitivity around 2022, which become even more stringent for future upgrades and the upcoming third generation of detectors~\cite{Purrer:2019jcp}.
The detector upgrades~\cite{LIGOaplus} in the coming years and a concurrent growing network of observatories~\cite{Aso:2013eba,LIGOIndia} also imply an increased number of detections~\cite{Aasi:2013wya}, making it overall more likely to observe binaries oriented ``edge on'' instead of ``face on,'' which allows measuring precession and extracting spin values with higher accuracy.
The accurate modeling of GW modulations caused by precession, and also the phase accuracy in the aligned-spin case and the contingent improvement in the estimation of spin parameters, motivate us to push predictions for gravitational spin effects to higher orders.

For this purpose, we extended to spin-orbit couplings a link between the weak-field and small-mass-ratio approximations, via the scattering-angle function, as proposed in the nonspinning case in Ref.~\cite{Bini:2019nra,Damour:2019lcq} (see also Ref.~\cite{Vines:2018gqi}).  We employed existing self-force results~\cite{Bini:2016dvs,Kavanagh:2016idg,Kavanagh:2017wot} to uniquely determine a N$^3$LO PN spin-orbit binary Hamiltonian.
We calculated the effective gyro-gravitomagnetic ratios as they would enter the \texttt{SEOBNR}~\cite{Bohe:2016gbl,Babak:2016tgq,Cotesta:2018fcv,Ossokine:2020kjp} and \texttt{TEOBResumS}~\cite{Nagar:2018plt,Nagar:2018zoe} waveform models, and we obtained the gauge-invariant scattering angle and circular-orbit binding energy for aligned spins.
Since the spin-orbit interaction is universal, our results are applicable to generic spinning binaries, e.g., binaries containing neutron stars.

In Fig.~\ref{fig:bindingenergy} we compared the EOB-resummed binding energy against NR results.
The EOB resummation shows a nice convergent behavior towards NR (for aligned spins) even in the strong field regime, which is usually not expected for asymptotic series expansions like the PN one.
More importantly, the new contribution obtained in this Letter roughly halves the gap to NR in the high-frequency regime compared to earlier N$^2$LO results for $q=1$.
This indicates that improved (resummed) analytical predictions based on our result can be trusted to higher frequencies, which may alleviate the need for longer and computationally very expensive NR waveforms.
Hence, it is of particular value and urgency to improve the accuracy of the PN-approximate analytic part of GW models.

A clear avenue for future work is to consider higher orders in spin (and higher multipoles).
In particular, in a forthcoming publication, we fix the $S_1S_2$ couplings at N$^3$LO (5PN order) for aligned spins using known self-force results.
It seems reasonable to expect that complete quadratic-in-spin contributions at N$^3$LO, for BHs, for aligned and perhaps even generic spins, should be within reach of first-order self-force computations.  These would require both further self-force observables and new conceptual developments, in particular, generalizations of first-law relations to include higher orders in spin and higher multipoles, and to the case of generic spin orientations (the precessing case).
Future first-order self-force results for unbound orbits may also enable obtaining spin effects to fourth order in the weak-field (post-Minkowskian) approximation---for generic masses and velocities---for BH scattering events (only the second order is currently known~\cite{Bini:2018ywr}).
While this scenario is unlikely to be of astrophysical relevance, it is still very interesting to consider from a conceptual point of view: after all, scattering encounters are the most elementary form of interaction.

\section{Acknowledgments}

We are grateful to Maarten van de Meent for helpful discussions, and to Alessandra Buonanno for comments on an earlier version of this manuscript.
We also thank Sergei Ossokine and Tim Dietrich for providing NR data for the binding energy and for related useful suggestions.

  

\newcommand{\NNNLOSO}{N$^3$LO-PN SO}
\newcommand{\NNNLOPN}{N$^3$LO-PN}
\newcommand{\spinonespintwo}{spin$_1$-spin$_2$}
\newcommand{\sonestwo}{S$_1$S$_2$}

\newcommand{\vk}{\varkappa}
\newcommand{\vs}{\varsigma}
\newcommand{\vt}{\vartheta}
\newcommand{\up}{\upsilon}
\newcommand{\W}{W}

\chapter[Gravitational spin-orbit and aligned spin$_1$-spin$_2$ couplings\dots]{Gravitational spin-orbit and aligned spin$_1$-spin$_2$ couplings through third-subleading post-Newtonian orders}
\label{chap:five}

\textbf{Authors}\footnote{Originally published in \emph{Phys.Rev.D}  102 (2020) 124024.}: Andrea Antonelli, Chris Khavanagh, Mohammed Khalil, Jan Steinhoff, Justin Vines.
\newline

\textbf{Abstract:} The study of scattering encounters continues to provide new insights into the general relativistic two-body problem. The local-in-time conservative dynamics of an aligned-spin binary, for both unbound and bound orbits, is fully encoded in the gauge-invariant scattering-angle function, which is most naturally expressed in a post-Minkowskian (PM) expansion, and which exhibits a remarkably simple dependence on the masses of the two bodies (in terms of appropriate geometric variables). This dependence links the PM and small-mass-ratio approximations, allowing gravitational self-force results to determine new post-Newtonian (PN) information to all orders in the mass ratio. In this paper, we exploit this interplay between relativistic scattering and self-force theory to obtain the third-subleading (4.5PN) spin-orbit dynamics for generic spins, and the third-subleading (5PN) spin$_1$-spin$_2$ dynamics for aligned spins. We further implement these novel PN results in an effective-one-body framework, and demonstrate the improvement in accuracy by comparing against numerical-relativity simulations.

\section{Introduction}

The burgeoning field of gravitational-wave (GW) astronomy has already shown its potential to revolutionize our understanding of our universe \cite{Abbott:2019yzh}, gravity \cite{LIGOScientific:2019fpa}, and the nature of compact objects \cite{LIGOScientific:2018jsj,LIGOScientific:2018mvr}, such as black holes (BHs) and neutron stars. The detection of compact-binary GW sources and the accurate inference of their parameters is contingent on having accurate theoretical predictions for their coalescence. As a result of this, a variety of techniques, both analytical and numerical, have been developed to understand the coalescence of binary compact objects, with the final goal of providing faithful waveform models that can be used in GW data analysis.

\emph{Post-Newtonian} (PN) theory, the best known of the analytical techniques, has provided the foundation for the analytical studies of the two-body problem in general relativity which are most directly useful for gravitational-wave astronomy \cite{Blanchet:2013haa,Schafer:2018kuf,Rothstein:2014sra,Goldberger:2007hy,Futamase:2007zz,Pati:2000vt,Porto:2016pyg,Levi:2018nxp}. In this approximation, most applicable to bound systems, one simultaneously assumes weak gravitational potential and small velocities, i.e., $GM/rc^2 \sim v^2/c^2 \ll 1$. The PN expansion is thus a powerful tool for describing the early inspiral of the binaries observed by LIGO and Virgo~\cite{Abbott:2016blz,LIGOScientific:2018mvr}.
PN studies have been carried out at high orders both in the nonspinning~\cite{Damour:2014jta,Damour:2015isa,Bernard:2016wrg,Bernard:2017ktp,Foffa:2019hrb,Foffa:2019rdf,Foffa:2019yfl,Blumlein:2019zku,Blumlein:2020pog,Bini:2019nra,Bini:2020wpo, Bini:2020hmy,Bini:2020nsb,Bini:2020uiq} and in the spinning sectors, including spin-orbit (SO)~\cite{Hartung:2011te,Hartung:2013dza,Marsat:2012fn, Bohe:2012mr,Levi:2015uxa,Levi:2020kvb}, bilinear-in-spin ({\spinonespintwo}, {\sonestwo})~\cite{Hartung:2011ea,Levi:2011eq,Levi:2014sba,Levi:2020uwu} and spin-squared (S$^2$)~\cite{Levi:2015ixa,Levi:2015msa,Levi:2016ofk,Levi:2020uwu} couplings, as well as cubic and higher-in-spin corrections~\cite{Levi:2019kgk,Levi:2014gsa,Levi:2020lfn,Vines:2016qwa,Siemonsen:2019dsu}.
PN information on the spin dynamics has also been included in effective-one-body (EOB) waveform models~\cite{Nagar:2011fx,Barausse:2011ys,Bohe:2016gbl,Babak:2016tgq,Cotesta:2018fcv,Ossokine:2020kjp,Nagar:2018plt,Nagar:2018zoe,Khalil:2020mmr}.

In parallel to PN formalisms, the small-mass-ratio approximation, based on gravitational \emph{self-force} (GSF) theory, has also seen rapid development (see Ref.~\cite{Barack:2018yvs} and references therein for a review). As suggested by the name, the expansion parameter in this limit is the mass ratio of the two bodies $q=m_1/m_2 \ll 1$. The leading order in this approximation is given by the geodesic motion of a test body in a Schwarzschild or Kerr background. Successive corrections,  which can be interpreted as a force moving the body away from geodesic motion, are due to the perturbation of the background sourced by the small body's nonzero stress-energy tensor.
This self-force effect on the motion of a nonspinning body has currently been numerically calculated to first order in $q$ for generic orbits in Kerr spacetime~\cite{vandeMeent:2017bcc}. In a recent breakthrough~\cite{Pound:2019lzj}, the second-order-in-$q$ binding energy in a Schwarzschild background has been calculated and compared to predictions from the first law of binary black-hole mechanics~\cite{LeTiec:2011ab}.
Meanwhile, much activity has led to the analytic calculation at very high PN orders (but at first order in $q$) of gauge-invariant quantities, such as the Detweiler redshift~\cite{Detweiler:2008ft,Bini:2013rfa,Kavanagh:2015lva,Johnson-McDaniel:2015vva,Hopper:2015icj,Bini:2015bfb,Kavanagh:2016idg,Bini:2018zde,Bini:2019lcd,Bini:2020zqy} and the precession frequency~\cite{Dolan:2013roa,Bini:2014ica,Akcay:2016dku,Kavanagh:2017wot,Akcay:2017azq,Bini:2018ylh,Bini:2019lcd}, including effects of the smaller body's spin. This has quite naturally led to related activity in confronting and validating the PN and GSF approximations~\cite{LeTiec:2011ab,Blanchet:2011aha,Akcay:2015pza} in the domain which both are valid, i.e., for large orbital separations and small mass ratios, as well as in constructing EOB models based on both approximations~\cite{Barausse:2011dq,Akcay:2012ea,Akcay:2015pjz,Antonelli:2019fmq}.

Recently, there has also been rapid advance in understanding and employing \emph{post-Minkowskian} (PM) techniques, using a weak-field approximation $GM/rc^2\ll 1$ in a background Minkowski spacetime, with no restriction on the relative velocity of the two bodies \cite{Damour:2016gwp,Damour:2017zjx,Bjerrum-Bohr:2018xdl,Cheung:2018wkq,Bern:2019nnu,Bern:2019crd}. This approximation most naturally applies to the weak-field scattering of compact objects, in which possibly relativistic velocities can be reached. Recent advances in PM gravity and in our understanding of the scattering of compact objects have been spearheaded by modern on-shell scattering-amplitude techniques, developed originally in the context of quantum particle physics (see, e.g., Ref.~\cite{Bern:2019crd} and references therein).

Scattering amplitudes were used in Ref.~\cite{Bjerrum-Bohr:2018xdl} to calculate the nonspinning 2PM ($\mathcal{O}(G^2)$, one-loop) scattering angle, reproducing with astonishing efficiency the decades-old results of Westpfhal~\cite{Westpfahl:1980mk,Westpfahl:1979gu} obtained by classical methods; an equivalent canonical Hamiltonian at 2PM order was derived from amplitudes in Ref.~\cite{Cheung:2018wkq}. The scattering angle plays a key role in PM gravity: it encodes the complete local-in-time conservative dynamics of the system (at least in a perturbative sense) and it can be used to specify a Hamiltonian in a given unique gauge~\cite{Damour:2017zjx}, which can in turn be used for unbound as well as \emph{bound} systems (with potential relevance for improving waveform models~\cite{Antonelli:2019ytb}); see in particular Refs.~\cite{Kalin:2019rwq,Kalin:2019inp}.
In Refs.~\cite{Bern:2019nnu,Bern:2019crd}, the scattering angle and a corresponding Hamiltonian have been obtained at 3PM (two-loop) order for nonspinning systems, and the results have been confirmed and expounded upon in  Refs.~\cite{Cheung:2020gyp,Blumlein:2020znm,Bini:2020wpo,Kalin:2020fhe}.

The PM approximation for two-\emph{spinning}-body systems was first tackled only very recently, with the SO dynamics at the 1PM and 2PM levels first derived by classical means in Refs.~\cite{Bini:2017xzy,Bini:2018ywr}.  These results have since been confirmed by amplitudes methods in Ref.~\cite{Bern:2020buy}, which also gave the 1PM and 2PM dynamics for the \sonestwo~sector, rounding out the current state of the art for generic-spin PM results beyond tree level.  Several other works have also considered amplitudes methods in relation to spinning two-body systems, also beyond the SO and \sonestwo~sectors (beyond the dipole level in the bodies' multipole expansions), in particular for special cases such as bodies with black-hole-like spin-induced multipole structure and/or for the aligned-spin configuration (in which the bodies' spins are [anti-]parallel to the orbital angular momentum); see, e.g., \cite{Siemonsen:2019dsu,Aoude:2020onz,Chung:2020rrz} and references reviewed therein.

These works demonstrate that the study of gravitational scattering continues to provide novel results and useful insights on the relativistic two-body problem, with implications for precision gravitational-wave astronomy yet to be explored.  A particularly powerful example of such an insight concerns the nontrivially simple dependence of the scattering-angle function on the masses \cite{Damour:2019lcq} (see also \cite{Vines:2018gqi,Bern:2019crd,Kalin:2019rwq}). This was exploited in Refs.~\cite{Bini:2019nra,Bini:2020wpo} to obtain almost all the 5PN dynamics (with the exception of 2 out of 36 coefficients in the EOB Hamiltonian; see also Refs.~\cite{Foffa:2019hrb,Blumlein:2019zku}) from first-order self-force calculations (while appropriately dealing with nonlocal-in-time tail terms). This approach has also been used in Refs.~\cite{Bini:2020nsb, Bini:2020uiq} to obtain most of the 6PN dynamics.  An extension of this approach to spinning systems was used by the current authors in Ref.~\cite{Antonelli:2020aeb} to obtain the next-to-next-to-next-to-leading order (N$^3$LO) SO PN dynamics.

In this paper, we provide details for the calculation of the {\NNNLOSO} dynamics presented in Ref.~\cite{Antonelli:2020aeb}, which completes the PN knowledge at 4.5PN order together with the NLO S$^3$ dynamics from Ref.~\cite{Levi:2019kgk} (see also \cite{Siemonsen:2019dsu}).
Furthermore, we extend our analysis to include a derivation of the N$^3$LO {\sonestwo} effects, contributing at 5PN order, for the case of spins aligned with the orbital angular momentum. We note that partial results of the {\NNNLOSO} and N$^3$LO {\sonestwo} dynamics have previously been presented in Refs.~\cite{Levi:2020kvb,Levi:2020uwu}, where all terms at $G^4$ were calculated within the powerful effective field theory framework using Feynman integral calculus. The latter of these references gives further results for \emph{all} quadratic-in-spin terms at  N$^3$LO.

Our derivations are organized in the following procedures:
\begin{enumerate}
	\item
	We argue that the scattering angle for an aligned-spin binary has a simple dependence on the masses (when expressed in terms of appropriate geometrical variables), which extends the result of Ref.~\cite{Damour:2019lcq} for nonspinning binaries. 
	This mass dependence implies that the 4PM part of the scattering angle, which encodes the N$^3$LO PN dynamics, is determined by terms up to linear order in the mass ratio.
	We use analytic results for the test-spin scattering angle to fix all terms at zeroth order in the mass ratio, leaving the linear terms to be fixed by first-order GSF results.
	\item
	Assuming the existence of a PN Hamiltonian at the desired 4.5PN SO and 5PN {\sonestwo} orders, and making use of its associated mass-shell constraint with undetermined coefficients, we calculate the scattering angle and match it to the constrained form from step 1. This procedure fixes its lower orders in velocity at 3PM and 4PM orders, leaving but half of the linear-in-mass-ratio coefficients to be determined by GSF calculations. We construct the bound-orbit radial action from the scattering angle (via the Hamiltonian dynamics), noting its simple dependence on the bodies' masses.
	\item
	From the radial action, we calculate the redshift and spin-precession invariants and compare them with GSF results available in the literature to determine the remaining coefficients of the scattering angle. Vital to this step is the first law of spinning binary mechanics~\cite{LeTiec:2011ab,Blanchet:2012at,Tiec:2015cxa}, which is used to relate the radial action to the redshift and precession frequency, and for which we herein discuss an extension to arbitrary-mass-ratio aligned-spin eccentric orbits.
\end{enumerate}
(Although we work with aligned spins throughout, we note that the aligned SO result actually fixes the SO Hamiltonian also for precessing spins \cite{Antonelli:2020aeb}.)

The paper is organized as follows. Sections~\ref{sec:massdep},~\ref{Irsec} and~\ref{sec:N3LOSO} discuss points 1, 2 and 3, respectively.
In Sec.~\ref{sec:NRcomp}, we implement the new PN results in the scattering angle in an EOB model, and use it to compare our results against NR simulations.
We conclude in Sec.~\ref{sec:conc} with a discussion of results and potential future work.
Finally, Appendix A contains expressions for tail terms in the radial action, while Appendix B contains explicit expressions for a certain mapping between variables used to connect redshift and precession-invariant results from the radial action to GSF results in the literature, which have been previously erroneously (yet innocuously) reported in the literature.

\subsection*{Notation}

We use the metric signature $(-,+,+,+)$, and use units in which the speed of light is $c = 1$.
For a binary of compact objects with masses $m_1$ and $m_2$, we use the following combinations of the masses
\begin{gather}
M= m_1 + m_2, \quad \mu = \frac{m_1m_2}{M}, \quad \nu = \frac{\mu}{M}, \nonumber\\
q = \frac{m_1}{m_2}, \quad \delta =\frac{m_2 - m_1}{M}, \label{massmap}
\end{gather}
with $m_1<m_2$.
We often make use of the rescaled versions of the canonical spins $\bm{S}_1$ and $\bm{S}_2$, i.e.,
\begin{gather}
\bm{a}_1 = \frac{\bm{S}_1}{m_1}, \qquad 
\bm{a}_2 = \frac{\bm{S}_2}{m_2},
\end{gather}
and define the following combinations of spins
\begin{gather}
\bm{S}=\bm{S}_1+\bm{S}_2, \quad \bm{S}_* = \frac{m_2}{m_1}\bm{S}_1+\frac{m_1}{m_2}\bm{S}_2, \nonumber\\
\bm{a}_\mr b =\frac{\bm{S}}{M}, \quad \bm{a}_\mr t =\frac{\bm{S}_*}{M}.
\end{gather}
The relative position and momentum 3-vectors are denoted by $\vec{r}$ and $\vec{p}$, respectively.
Using an implicit Euclidean background, it holds that
\begin{equation}
\vec p^2 = p_r^2 + \frac{L^2}{r^2}, \quad
p_r= \bm{n}\cdot\bm{p}, \quad
\bm{L}=\bm{r}\times\bm{p},
\end{equation}
where $\bm{n}=\bm{r}/r$ with $r=|\vec r|$, and $\vec L$ is the orbital angular momentum with magnitude $L$.

\section{The mass dependence of the scattering angle}
\label{sec:massdep}
Here we argue that the structure of the PM expansion, applied to the conservative orbital dynamics of a two-massive-body system, leads to simple constraints on the dependence of the scattering-angle function on the bodies' masses, at fixed geometric quantities characterizing the incoming state.  We closely follow the arguments given in Sec.~II of Ref.~\cite{Damour:2019lcq} for the nonspinning case, considering only the local-in-time, conservative part of the dynamics, while generalizing to the case of spinning bodies, finally, in the aligned-spin configuration.  

The motion of a two-point-mass system (the nonspinning case) is effectively governed by the coupled system of (i) geodesic equations for the worldlines of the two point masses, using the full two-body spacetime metric (with a suitable regularization or renormalization procedure), and (ii) Einstein's equations for the metric, sourced by effective point-mass energy-momentum tensors.  In the case of spinning bodies, to dipolar order in the bodies' multipole expansions, the geodesic equations are replaced by the pole-dipole Mathisson-Papapetrou-Dixon (MPD) equations~\cite{Mathisson:1937zz,Papapetrou:1951pa,Dixon:1970zza},
\bse\label{MPD}
\begin{alignat}{3}\label{MPDp}
\frac{\mr Dp_{\mr i\mu}}{\mr d\tau_\mr i}&=-\frac{1}{2}R_{\mu\nu\rho\sigma}\dot x_\mr i^\nu S_\mr i^{\rho\sigma},
\\\label{MPDS}
\frac{\mr DS_\mr i^{\mu\nu}}{\mr d\tau_\mr i}&=2p_\mr i^{[\mu}\dot x_\mr i^{\nu]},
\\\label{TDSSC}
0&=p_{\mr i\mu}S_\mr i^{\mu\nu},
\end{alignat}
\ese
where, for the $\mr i$th body ($\mr i=1,2$), $p_\mr i^\mu(\tau_\mr i)$ is the linear momentum vector, $S_\mr i^{\mu\nu}(\tau_\mr i)$ is the antisymmetric spin (intrinsic angular momentum) tensor, and $\dot x_\mr i^\mu(\tau_\mr i)$ is the tangent to the body's worldline $x_\mr i(\tau_\mr i)$.   The constraint (\ref{TDSSC}), the ``covariant'' or Tulczyjew-Dixon  spin supplementary condition~\cite{Dixon:1979,Steinhoff:2014kwa,Tulczyjew:1959,fokker1929relativiteitstheorie}, combined with (\ref{MPDp}) and (\ref{MPDS}), uniquely determines a first-order equation of motion for the worldline, $\dot x_\mr i^\mu=\dot x_\mr i^\mu(x_\mr i,p_\mr i,S_\mr i)[g]$.  The corresponding effective energy-momentum tensor,
\begin{alignat}{3}\label{Tmunu}
\begin{aligned}
T^{\mu\nu}(x)&=\sum_\mr i\int d\tau_\mr i\bigg[ p_\mr i^{(\mu}\dot x_\mr i^{\nu)}\frac{\delta^4(x-x_\mr i)}{\sqrt{-g}}+\nabla_\lambda\bigg(S_\mr i^{\lambda(\mu}\dot x_\mr i^{\nu)}\frac{\delta^4(x-x_\mr i)}{\sqrt{-g}}\bigg)\bigg],
\end{aligned}
\end{alignat}
sources Einstein's equations,
\be\label{Einstein}
R_{\mu\nu}-\frac{1}{2}Rg_{\mu\nu}=8\pi GT_{\mu\nu}.
\ee
In the PM scheme, an iterative solution to these equations is obtained as an expansion in $G$ of the worldlines, momenta and spins,
\begin{alignat}{3}\label{expand_xpS}
x_\mr i^\mu(\tau_\mr i)&=x_{\mr i0}^\mu(\tau_\mr i)+G x_{\mr i1}^\mu(\tau_\mr i)+G^2 x_{\mr i2}^\mu(\tau_\mr i)+\cdots,
\nnm\\
p_\mr i^\mu(\tau_\mr i)&=p_{\mr i0}^\mu(\tau_\mr i)+G p_{\mr i1}^\mu(\tau_\mr i)+G^2 p_{\mr i2}^\mu(\tau_\mr i)+\cdots,
\\\nnm
S_\mr i^{\mu\nu}(\tau_\mr i)&=S_{\mr i0}^{\mu\nu}(\tau_\mr i)+G S_{\mr i1}^{\mu\nu}(\tau_\mr i)+G^2 S_{\mr i2}^{\mu\nu}(\tau_\mr i)+\cdots,
\end{alignat}
and of the metric,
\be\label{expand_g}
g_{\mu\nu}(x)=\eta_{\mu\nu}+G h_1{}_{\mu\nu}(x)+G^2 h_2{}_{\mu\nu}(x)+\cdots,
\ee
where $\eta_{\mu\nu}$ is the Minkowski metric, which we henceforth use instead of the full metric $g_{\mu\nu}$ for all 4-vector manipulations (index raising and lowering, dot products and squares of vectors, etc.).

At the leading orders in (\ref{expand_xpS}), given by the solutions to (\ref{MPD}) with $g=\eta$, each body moves inertially in flat spacetime,
\begin{alignat}{3}\label{xpS0}
\begin{aligned}
x_{\mr i0}^\mu(\tau_\mr i)&=y_{\mr i}^\mu+u_{\mr i}^\mu\tau_\mr i,
\\
p_{\mr i0}^\mu(\tau_\mr i)&=m_\mr i u_{\mr i}^\mu,
\\
S_{\mr i0}^{\mu\nu}(\tau_\mr i)&=m_\mr i\epsilon^{\mu\nu}{}_{\rho\sigma}u_{\mr i}^\rho a_{\mr i}^\sigma.
\end{aligned}
\end{alignat}
Here, $y_\mr i^\mu$ are constant displacements from the origin at $\tau_\mr i=0$, and $u_\mr i^\mu$ are constant 4-velocities, with $u_\mr i^2=-1$, so that $\tau_\mr i$ are the (Minkowski) proper times, and $p_\mr i^2=-m_\mr i^2$ where $m_\mr i$ are the constant rest masses.  The zeroth-order spin tensors $S_{\mr i0}^{\mu\nu}$ are also constant, and, being orthogonal to $u_{\mr i\mu}$, have been parametrized in terms of a constant mass-rescaled (Pauli-Lubanski, ``covariant'') spin vector,
\be
a_\mr i^\mu=-\frac{1}{2m_\mr i}\epsilon^\mu{}_{\nu\rho\sigma}u_{\mr i}^\nu S_{\mr i0}^{\rho\sigma},
\ee
with dimensions of length, the magnitude of which would measure the radius of the ring singularity of a corresponding (linearized) Kerr black hole.  We identify the zeroth-order geometric (mass-independent) quantities, $y_\mr i^\mu$, $u_\mr i^\mu$ and $a_\mr i^\mu$, with those characterizing the asymptotic incoming state, along with the masses $m_1$ and $m_2$.

Inserting (\ref{xpS0}) into (\ref{Tmunu}) (with $g=\eta$) yields the zeroth-order stress-energy tensor, which serves as a source for the first-order metric perturbation $h_{1\mu\nu}$ in the linearization of (\ref{Einstein}).  The solution for the trace-reversed $\bar h_1^{\mu\nu}=h_1^{\mu\nu}-\frac{1}{2}\eta^{\mu\nu}h_{1\rho}{}^\rho$, in harmonic gauge ($\doe_\mu \bar h_1^{\mu\nu}=0$), reads
\begin{alignat}{3}\label{h1PM}
\bar h_1^{\mu\nu}(x)&=4 \sum_{\mr i}m_\mr i \Big(u_\mr i^\mu u_\mr i^\nu+u_\mr i^{(\mu}\epsilon^{\nu)}{}_{\rho\sigma\lambda}u_\mr i^\rho a_{\mr i}^\sigma\doe^\lambda\Big)\frac{1}{r_\mr i},
\end{alignat}
where $r_\mr i=\{(x-y_\mr i)^2+[u_\mr i\cdot(x-y_\mr i)]^2\}^{1/2}$ is the (Minkowski) distance of the field point $x$ from the (zeroth-order, flat geodesic) worldline $x_{\mr i0}=y_\mr i+u_\mr i \tau_\mr i$ in its rest frame, and $\doe_\mu$ is the flat covariant derivative. (Note that the result for the first-order field (\ref{h1PM}) would be the same whether we used the physical retarded Green's function or the time-symmetric Green's function, given the nature of the zeroth-order source, constant momentum and spin along a flat-spacetime geodesic.)
A key property to be noted here is that $h_1$ is linear in the masses $m_\mr i$, while having a more intricate dependence on the geometric quantities $y_\mr i^\mu$, $u_\mr i^\mu$ and $a_\mr i^\mu$.  (It is linear in the spins $a_\mr i^\mu$ here only because we are working to linear order in the spins, to dipolar order in the multipole expansions.)

In the next step of the iterative scheme, one uses $g=\eta+h_1$ in the bodies' equations of motion (\ref{MPD}) to solve for the first-order perturbations in (\ref{expand_xpS}) [for which it is sufficient to integrate the RHSs of (\ref{MPDp}) and (\ref{MPDS}) along the zeroth-order motion (\ref{xpS0}), and to regularize by simply dropping the divergent self-field contribution].
Importantly, one finds that $x_{\mr i1}^\mu$, $p_{\mr i1}^\mu/m_\mr i$ and $S_{\mr i1}^{\mu\nu}/m_\mr i$ are each linear functionals of $h_{1\mu\nu}(x)$, and are thus linear in the masses.  From Poincar\'e symmetry, it follows that these results can depend on the positions $y_\mr i$ only through the vectorial impact parameter $b^\mu=y_1^\mu-y_2^\mu$, where the $y_\mr i^\mu$ here are chosen along the two zeroth-order worldlines by the conditions $u_1\cdot b=u_2\cdot b=0$ (at mutual closest approach).  For example, the impulse (net change in momentum) for body 1, $\Delta p_1^\mu=Gp_{11}^\mu(\tau_1\to\infty)+\mc O(G^2)$, is given by\footnote{Results equivalent to the first two lines of Eq.~(\ref{Deltap1PM}) were first derived in Ref.~\cite{Bini:2017xzy}, and the last line results from an expansion in spins of the all-orders-in-spin results for black holes from Ref.~\cite{Vines:2017hyw}, both references having worked from purely classical considerations; see also \cite{Maybee:2019jus,Guevara:2019fsj} for derivations from quantum scattering amplitudes.}
\begin{alignat}{3}\label{Deltap1PM}
\Delta p_1^\mu&=\frac{2Gm_1m_2}{\sqrt{\gamma^2-1}}\bigg[{-}(2\gamma^2-1)\frac{b^\mu}{b^2}
\\\nnm
&\quad+\frac{2\gamma}{b^4}(2b^\mu b^\nu-b^2\eta^{\mu\nu})\epsilon_{\nu\rho\sigma\lambda}u_1^\rho u_2^\sigma (a_1^\lambda+a_2^\lambda)
\\\nnm
&+2\frac{2\gamma^2-1}{b^6}(4b^\mu b^\nu b^\rho-3b^2 b^{(\mu}\Pi^{\nu\rho)})a_{1\nu}a_{2\rho}\bigg]
+\mc O(G^2),
\end{alignat}
where
\be\label{gamma}
\gamma=-u_1\cdot u_2
\ee
is the asymptotic relative Lorentz factor, and $\Pi^\mu{}_\nu=\epsilon^{\mu\rho\alpha\beta}\epsilon_{\nu\rho\gamma\delta}u_{1\alpha}u_{2\beta}u_1^\gamma u_2^\delta/(\gamma^2-1)$ is the projector into the plane orthogonal to both $u_1$ and $u_2$.  Here, as below, we work to linear order in each spin, $a_1$ and $a_2$, keeping the cross term.  We note again in (\ref{Deltap1PM}) the simple dependence on the masses, with an overall factor of $m_1m_2$, at fixed geometric quantities $b^\mu$, $u_\mr i^\mu$ and $a_\mr i^\mu$.  
In continuing the iterative PM solution, the $\mc O(G^n)$ terms in the bodies' degrees of freedom (\ref{expand_xpS}) correct the source (\ref{Tmunu}) for the field equation (\ref{Einstein}), determining the $\mc O(G^{n+1})$ metric perturbation in (\ref{expand_g}); the latter, via the bodies' equations of motion (\ref{MPD}), determines the $\mc O(G^{n+1})$ corrections in (\ref{expand_xpS}).  As in Ref.~\cite{Damour:2019lcq} we are assuming here a systematic use of the time-symmetric Green's function, to pick out the conservative sector of the dynamics.  It becomes evident from the structure of these expansions that the $\mc O(G^n)$ metric perturbation $h_n^{\mu\nu}$ in (\ref{expand_g}) can be expressed as a homogeneous polynomial of degree $n$ in the masses,
\begin{alignat}{3}\label{h1h2}
h_1^{\mu\nu}(x)&=m_1h_{m_1}^{\mu\nu}(x)+m_2h_{m_2}^{\mu\nu}(x),
\nnm\\
h_2^{\mu\nu}(x)&=m_1^2h_{m_1^2}^{\mu\nu}(x)+m_2^2h_{m_2^2}^{\mu\nu}(x)+m_1m_2h_{m_1m_2}^{\mu\nu}(x),
\nnm\\
&\cdots
\end{alignat}
where the $h_{\cdots}^{\mu\nu}$ on the RHSs are functions only of the (asymptotic incoming) geometric quantities $(y_\mr i^\mu,u_\mr i^\mu,a_\mr i^\mu)$ and the field point $x$.  The first line of (\ref{h1h2}) matches (\ref{h1PM}). Similarly, the $\mc O(G^n)$ corrections $x_{\mr in}^\mu$, $p_{\mr in}^\mu/m_\mr i$, $S_{\mr in}^{\mu\nu}/m_\mr i$ for the body degrees of freedom (\ref{expand_xpS}) will be homogeneous polynomials of degree $n$ in the masses; this is the crucial point for the following analysis (and for its conceivable extensions beyond the aligned-spin case).  The zeroth-order quantities $x_{\mr i0}^\mu=y_\mr i^\mu+u_\mr i^\mu\tau_\mr i$, $p_{\mr i0}^\mu/m_\mr i=u_\mr i^\mu$ and $S_{\mr i0}^{\mu\nu}/m_\mr i=\epsilon^{\mu\nu}{}_{\rho\sigma}u_\mr i^\rho a_\mr i^\sigma$ from (\ref{xpS0}) are (taken to be) independent of the masses, as is the zeroth-order metric $h_0=\eta$; they, along with the masses, both (i) fully parametrize the asymptotic incoming state and (ii) can be used to parametrize all the higher-order corrections.

Let us now specialize to the case of aligned spins, in which both spin vectors $a_\mr i^\mu$ are (anti-)parallel to the orbital angular momentum, all of which remain constant throughout the scattering, while the orbital motion is confined to the fixed plane orthogonal to the angular momenta (just as for the nonspinning case).  This entails $u_1\cdot a_\mr i=u_2\cdot a_\mr i=0$ and $b\cdot a_\mr i=0$.  Choosing $\hat z^\mu$ (with $\hat z^2=1$) to be the direction of the orbital angular momentum ($\propto-\epsilon_{\mu\nu\rho\sigma}u_1^\nu u_2^\rho b^\sigma$), let us write $a_\mr i^\mu=a_\mr i\hat z^\mu$ for the constant rescaled spin vectors (equal to their incoming values), where the scalars $a_\mr i$ are positive for spins aligned with $\hat z^\mu$ and negative for anti-aligned.  Crucially, in this case, the only nontrivial independent Lorentz-invariant scalars that can be constructed from the vectors $u_\mr i^\mu$, $a_\mr i^\mu$ and $b^\mu$ are the magnitude $b=(b^2)^{1/2}$ of the impact parameter and the two spin lengths $a_1$ and $a_2$, all three with dimensions of length, and the dimensionless Lorentz factor $\gamma=-u_1\cdot u_2$.

Now consider the extension to higher orders in $G$ of the impulse $\Delta p_1^\mu$ (\ref{Deltap1PM}), which equals $-\Delta p_2^\mu$ (under the conservative dynamics) as the total momentum $p_1^\mu+p_2^\mu$ is conserved.  Its magnitude $\ms Q:=(\Delta p_{1\mu}\Delta p_1^\mu)^{1/2}$ must be a Lorentz-invariant scalar.  In the aligned-spin case, given the previous discussion, and due to Poincar\'e symmetry and dimensional analysis, it must be a function only of the dimensionless scalar $\gamma$ and the dimension-length scalars $b$, $a_1$, $a_2$, $Gm_1$ and $Gm_2$.  Given also the conclusion from above that, in (\ref{expand_xpS}) with $\mr i=1$, $p_{1n}^\mu/m_1$ is a homogeneous polynomial of degree $n$ in the masses, with the leading $n=1$ result seen in (\ref{Deltap1PM}), it follows that the magnitude $\ms Q$ of the impulse must take the following form through fourth order in $G$ (through 4PM order),
\bse\label{QQ}
\begin{alignat}{3}\label{masterQ}
\ms Q&=\frac{2Gm_1m_2}{b}\bigg[\ms Q^\mr{1PM}
\\\nnm
&\quad+\frac{G}{b}\bigg(m_1\ms Q^\mr{2PM}_{m_1}+m_2\ms Q^\mr{2PM}_{m_2}\bigg)
\\\nnm
&\quad+\frac{G^2}{b^2}\bigg(m_1^2\ms Q^\mr{3PM}_{m_1^2}+m_2^2\ms Q^\mr{3PM}_{m_2^2}+m_1m_2 \ms Q_{m_1m_2}^\mr{3PM}\bigg)
\\\nnm
&\quad+\frac{G^3}{b^3}\bigg(m_1^3\ms Q^\mr{4PM}_{m_1^3}+m_2^3\ms Q^\mr{4PM}_{m_2^3}
\\\nnm
&\qquad\qquad+m_1^2m_2 \ms Q_{m_1^2m_2}^\mr{4PM}+m_1m_2^2 \ms Q_{m_1m_2^2}^\mr{4PM}\bigg)\bigg]+\mc O(G^5),
\end{alignat}
where the $\ms Q$'s on the RHS are functions of the dimensionless scalars $\gamma$, $a_1/b$ and $a_2/b$,
\begin{alignat}{3}\label{Qcoeffs}
\ms Q^{n\mr{PM}}_{m_1^im_2^j}&=\ms Q^{n\mr{PM}}_{m_1^im_2^j}(\gamma,\frac{a_1}{b},\frac{a_2}{b})
\\\nnm
&=\ms Q^{n\mr{PM}}_{m_1^im_2^ja^0}(\gamma)
\\\nnm
&\quad+\frac{a_1}{b}\ms Q^{n\mr{PM}}_{m_1^im_2^ja_1}(\gamma)
+\frac{a_2}{b}\ms Q^{n\mr{PM}}_{m_1^im_2^ja_2}(\gamma)
\\\nnm
&\quad+\frac{a_1a_2}{b^2}\ms Q^{n\mr{PM}}_{m_1^im_2^ja_1a_2}(\gamma)
\end{alignat}
\ese
(with $i+j=n-1$).  In the second equality, we have expanded to linear order in each spin (assuming regular limits as the spins go to zero), and we are finally left with a set of undetermined functions depending only on the Lorentz factor $\gamma$.

Furthermore, $\ms Q$ must be invariant under an exchange of the two bodies' identities, $(m_1,a_1)\leftrightarrow(m_2,a_2)$.  At 1PM order, this tells us that $\ms Q^\mr{1PM}(\gamma,a_1/b,a_2/b)$ is symmetric under $a_1\leftrightarrow a_2$, and thus $\ms Q^{\mr{1PM}}_{a_1}=\ms Q^{\mr{1PM}}_{a_2}$, so that the third line of (\ref{Qcoeffs}) in this case is proportional to $a_1+a_2$.  Indeed, the explicit expression for $\ms Q^\mr{1PM}$ is given by the magnitude of the aligned-spin specialization of (\ref{Deltap1PM}) (divided by $2Gm_1m_2/b$),\footnote{Note that this is the expansion to linear order in the spins of the result (80) from~\cite{Vines:2017hyw} for a two-black-hole system,
	\be
	\ms Q^\mr{1PM}=\bigg(\frac{2\gamma^2-1}{\sqrt{\gamma^2-1}}-2\gamma\frac{a_1+a_2}{b}\bigg)\bigg(1-\frac{(a_1+a_2)^2}{b^2}\bigg)^{-1},
	\ee
	to all orders in the spin-multipole expansion at 1PM order.}
\be\label{Q1PM}
\ms Q^\mr{1PM}=\frac{2\gamma^2-1}{\sqrt{\gamma^2-1}}\bigg(1+2\frac{a_1a_2}{b^2}\bigg)-2\gamma\frac{a_1+a_2}{b}.
\ee
At 2PM order, the $1\leftrightarrow2$ symmetry tells us that each of the two functions in the second line of (\ref{masterQ}) determines the other,
\be\label{Q2PMsymm}
\ms Q^\mr{2PM}_{m_1}(\gamma,\frac{a_1}{b},\frac{a_2}{b})
=
\ms Q^\mr{2PM}_{m_2}(\gamma,\frac{a_2}{b},\frac{a_1}{b}).
\ee
This function, like $\ms Q^\mr{1PM}$, is in fact fully determined by the (extended) test-body limit of $\ms Q/(m_1m_2)$ --- the limit where one of the masses, say, $m_1$, goes to zero, while keeping fixed $m_2$, $a_2$ and $a_1$ (and $\gamma$ and $b$).  The result for $\ms Q/m_1$ in this limit can be consistently determined by solving the pole-dipole MPD equations (\ref{MPD}) for a spinning test body in a stationary Kerr background;
we will present explicit results from this procedure below in terms of the scattering-angle function.  This test-body limit, with $m_1\to0$, determines all of the functions $\ms Q^{n\mr{PM}}_{m_2^{n-1}}$ with no powers of $m_1$, for all $n$, and the $1\leftrightarrow2$ symmetry also tells us that
\be\label{Qmnm1}
\ms Q^{n\mr{PM}}_{m_1^{n-1}}(\gamma,\frac{a_1}{b},\frac{a_2}{b})
=
\ms Q^{n\mr{PM}}_{m_2^{n-1}}(\gamma,\frac{a_2}{b},\frac{a_1}{b}).
\ee
The only remaining functions in (\ref{masterQ}), those not determined by the test-body limit and exchange symmetry, are $\ms Q^\mr{3PM}_{m_1m_2}$, $\ms Q^\mr{4PM}_{m_1^2m_2}$ and $\ms Q^\mr{4PM}_{m_1m_2^2}$.  They are however still constrained by the exchange symmetry as follows. Firstly,
\be\label{Q3PMsymm}
\ms Q^\mr{3PM}_{m_1m_2}(\gamma,\frac{a_1}{b},\frac{a_2}{b})
=
\ms Q^\mr{3PM}_{m_1m_2}(\gamma,\frac{a_2}{b},\frac{a_1}{b}),
\ee
which implies that the third line of (\ref{Qcoeffs}) for $\ms Q^\mr{3PM}_{m_1m_2}$ (like for $\ms Q^\mr{1PM}$ above) is proportional to $a_1+a_2$.  Secondly,
\be
\ms Q^\mr{4PM}_{m_1^2m_2}(\gamma,\frac{a_1}{b},\frac{a_2}{b})
=
\ms Q^\mr{4PM}_{m_1m_2^2}(\gamma,\frac{a_2}{b},\frac{a_1}{b}),
\ee
so that one of these two functions determines the other.

Taking all of these constraints from exchange symmetry, we can eliminate all of the $\ms Q$'s with more $m_1$'s in the subscript for those with more $m_2$'s, while those with the same number of $m_1$'s and $m_2$'s must be symmetric under $a_1\leftrightarrow a_2$.  First focusing on the nonspinning ($a^0$) part of (\ref{masterQ}), this becomes
\begin{alignat}{3}
&\ms Q_{a^0}=\frac{2Gm_1m_2}{b}\bigg[\ms Q^\mr{1PM}_{a^0}
+\frac{G}{b}(m_1+m_2)\ms Q^\mr{2PM}_{m_2a^0}
\nnm\\
&\quad+\frac{G^2}{b^2}\bigg((m_1^2+m_2^2)\ms Q^\mr{3PM}_{m_2^2a^0}+m_1m_2 \ms Q_{m_1m_2a^0}^\mr{3PM}\bigg)
\\\nnm
&+\frac{G^3}{b^3}\bigg((m_1^3+m_2^3)\ms Q^\mr{4PM}_{m_2^3a^0}
+m_1m_2(m_1+m_2) \ms Q_{m_1m_2^2a^0}^\mr{4PM}\bigg)\bigg],
\end{alignat}
recalling that all the $\ms Q$'s on the right-hand side are functions only of $\gamma$ [henceforth dropping $+\mc O(G^5)$].
Introducing the total rest mass $M=m_1+m_2$ and the symmetric mass ratio $\nu=m_1m_2/M^2=\mu/M$ as in (\ref{massmap}), and noting
\begin{alignat}{3}
m_1+m_2&=M,
\nnm\\
m_1^2+m_2^2&=M^2(1-2\nu),
\\\nnm
m_1^3+m_2^3&=M^3(1-3\nu),
\end{alignat}
this becomes 
\begin{alignat}{3}\label{Qa0}
\ms Q_{a^0}&=\frac{2Gm_1m_2}{b}\bigg[\ms Q^\mr{1PM}_{a^0}
+\frac{GM}{b}\ms Q^\mr{2PM}_{m_2a^0}
\nnm\\
&\quad+\Big(\frac{GM}{b}\Big)^2\bigg(\ms Q^\mr{3PM}_{m_2^2a^0}+\nu \tilde{\ms Q}_{m_1m_2a^0}^\mr{3PM}\bigg)
\\\nnm
&\quad+\Big(\frac{GM}{b}\Big)^3\bigg(\ms Q^\mr{4PM}_{m_2^3a^0}+\nu \tilde{\ms Q}_{m_1m_2^2a^0}^\mr{4PM}\bigg)\bigg],
\end{alignat}
where we defined $\tilde{\ms Q}_{m_1m_2a^0}^\mr{3PM}:={\ms Q}_{m_1m_2a^0}^\mr{3PM}-2\ms Q^\mr{3PM}_{m_2^2a^0}$ and $\tilde{\ms Q}_{m_1m_2^2a^0}^\mr{4PM}:={\ms Q}_{m_1m_2^2a^0}^\mr{4PM}-3\ms Q^\mr{4PM}_{m_2^3a^0}$, still functions only of $\gamma$.  Remarkably, through 4PM order, this is just linear in the mass ratio $\nu$ at fixed $M$.  Precisely the same manipulations go through for the $a_1a_2$ terms, replacing $a^0$ with $a_1a_2$ in all the subscripts and with an overall factor of $a_1a_2/b^2$ on the right-hand side.

Next consider just the 1PM and 2PM terms of the SO ($a^1$) part of (\ref{masterQ}), after accounting for the exchange symmetry in the same way as in the previous paragraph (with $\ms Q^\mr{1PM}_{a_1}=\ms Q^\mr{1PM}_{a_2}$, $\ms Q^\mr{2PM}_{m_1a_1}=\ms Q^\mr{2PM}_{m_2a_2}$ and $\ms Q^\mr{2PM}_{m_1a_2}=\ms Q^\mr{2PM}_{m_2a_1}$); we find
\begin{alignat}{3}\label{Qa12}
&\ms Q_{a^1}+\mc O(G^3)=\frac{2Gm_1m_2}{b}\bigg[\frac{a_1+a_2}{b}\ms Q^\mr{1PM}_{a_2}
\\\nnm
&+\frac{G}{b}\bigg(
\frac{m_1a_1+m_2a_2}{b}\ms Q^\mr{2PM}_{m_2a_2}+\frac{m_2a_1+m_1a_2}{b}\ms Q^\mr{2PM}_{m_2a_1}
\bigg)\bigg].
\end{alignat}
We recognize in the second line the following spin combinations often used in the PN and EOB literature,
\begin{alignat}{3}
\begin{aligned}
S&:=m_1a_1+m_2a_2=S_1+S_2,
\\
S_*&:=m_2a_1+m_1a_2=\frac{m_2}{m_1}S_1+\frac{m_1}{m_2}S_2.
\end{aligned}
\end{alignat}
We will find it convenient to rescale each of these by the total rest mass $M$, defining
\begin{alignat}{3}\label{abat}
\begin{aligned}
a_\mr b&:=\frac{S}{M}=\frac{m_1a_1+m_2a_2}{m_1+m_2},
\\
a_\mr t&:=\frac{S_*}{M}=\frac{m_2a_1+m_1a_2}{m_1+m_2},
\end{aligned}
\end{alignat}
where b stands for background (or big) and t stands for test (or tiny).  The (first) reason for these labels is that, in the extended test-body limit [$m_1\to0$ at fixed $m_2$ (or $M$) and fixed $a_1$ and $a_2$], we see that $a_\mr b\to a_2$ becomes the spin-per-mass of the big background object with mass $M=m_2$, and $a_\mr t\to a_1$ becomes the spin-per-mass of the tiny spinning test body with negligible mass (with a further reason explained below).  Note that $a_\mr b+a_\mr t=a_1+a_2$.  Now extending (\ref{Qa12}) to 4PM order, from (\ref{masterQ}) accounting for exchange symmetry, using our new notation, we find
\begin{alignat}{3}\label{Qa1234}
&\ms Q_{a^1}=\frac{2Gm_1m_2}{b^2}\bigg[\ms Q^\mr{1PM}_{a_2}(a_\mr b+a_\mr t)
\\\nnm
&\quad+\frac{GM}{b}\bigg(
\ms Q^\mr{2PM}_{m_2a_2}a_\mr b+\ms Q^\mr{2PM}_{m_2a_1}a_\mr t
\bigg)
\\\nnm
&+\Big(\frac{GM}{b}\Big)^2\bigg(
\ms Q^\mr{3PM}_{m_2^2a_2}a_\mr b+\ms Q^\mr{3PM}_{m_2^2a_1}a_\mr t+\nu \tilde{\ms Q}^\mr{3PM}_{m_1m_2a_2}(a_\mr b+a_\mr t)
\bigg)
\\\nnm
&+\Big(\frac{GM}{b}\Big)^3\bigg(
\ms Q^\mr{4PM}_{m_2^3a_2}a_\mr b+\ms Q^\mr{4PM}_{m_2^3a_1}a_\mr t
+\nu\Big[ \tilde{\ms Q}^\mr{4PM}_{m_1m_2^2a_2}a_\mr b+\tilde{\ms Q}^\mr{4PM}_{m_1m_2^2a_2}a_\mr t\Big]
\bigg)\bigg],
\end{alignat}
where we defined $\tilde{\ms Q}^\mr{3PM}_{m_1m_2a_2}={\ms Q}^\mr{3PM}_{m_1m_2a_2}-{\ms Q}^\mr{3PM}_{m_2^2a_2}-{\ms Q}^\mr{3PM}_{m_2^2a_1}$, $\tilde{\ms Q}^\mr{4PM}_{m_1m_2^2a_2}:={\ms Q}^\mr{4PM}_{m_1m_2^2a_2}-2\ms Q^\mr{4PM}_{m_2^3a_2}-\ms Q^\mr{4PM}_{m_2^3a_1}$ and $\tilde{\ms Q}^\mr{4PM}_{m_1m_2^2a_2}:={\ms Q}^\mr{4PM}_{m_1m_2^2a_2}-\ms Q^\mr{4PM}_{m_2^3a_2}-2\ms Q^\mr{4PM}_{m_2^3a_1}$, all still functions only of $\gamma$.  We see that (\ref{Qa1234}), like (\ref{Qa0}), is linear in the symmetric mass ratio $\nu$ (at fixed $M$, $a_\mr b$ and $a_\mr t$).

Now, just as in Eq.~(2.14) of \cite{Damour:2019lcq} --- following from conservation of the total momentum $p_1^\mu+p_2^\mu$ and simple geometry and kinematics (which is identical for the nonspinning and aligned-spin cases) --- the scattering angle $\chi$, by which both bodies are deflected in the system's center-of-mass (cm) frame, is related to the magnitude $\ms Q$ of the impulse by
\be\label{chiQ}
\sin\frac{\chi}{2}=\frac{\ms Q}{2p_\infty},
\ee
where $p_\infty$ (called ``$P_\mr{c.m.}$'' by Damour) is the magnitude of the bodies' equal and opposite spatial momenta in the cm frame, at infinity,
\be\label{pcm}
p_\infty=\frac{m_1m_2}{E}\sqrt{\gamma^2-1}.
\ee
Here, $E$ is the total energy in the cm frame,
\begin{alignat}{3}\label{Etotal}
E^2&=m_1^2+m_2^2+2m_1m_2\gamma
\nnm\\
&=M^2(1+2\nu(\gamma-1)),
\end{alignat}
determined by the asymptotic Lorentz factor $\gamma$ and the rest masses.  Note also the definition of the asymptotic relative velocity $v$ as used e.g.\ in \cite{Vines:2018gqi,Siemonsen:2019dsu,Antonelli:2020aeb},
\be\label{gammav}
v=\frac{\sqrt{\gamma^2-1}}{\gamma}\quad\Leftrightarrow\quad\gamma=\frac{1}{\sqrt{1-v^2}}.
\ee
We will find it convenient to define yet another variable equivalent to $\gamma$ or $v$, namely
\be
\ve:=\gamma^2-1=\gamma^2v^2=\Big(\frac{p_\infty E}{m_1m_2}\Big)^2,
\ee
which, like $v^2$, can serve as a PN expansion parameter, and unlike $v$, is real for both unbound and bound orbits,
\begin{alignat}{3}
\begin{aligned}
\tr{unbound:}\quad E>M\quad\Leftrightarrow\quad\ve&>0, 
\\
\tr{bound:}\quad E<M\quad\Leftrightarrow\quad\ve&<0, 
\end{aligned}
\end{alignat}
noting that $v=i\sqrt{1-\gamma^2}/\gamma$ and $p_\infty$ are imaginary for bound orbits.  (Note that our $\ve=\gamma^2v^2$ is Damour's ``$p_\infty^2=p_\mr{eob}^2$'' [the squared momentum per mass of the effective test body], while our $p_\infty$ is Damour's ``$P_\mr{c.m.}$''.)
We will also find it convenient to define a notation for the dimensionless ratio $\Gamma$ (Damour's ``$h$'') between the total energy and the total rest mass,
\be\label{defGamma}
\Gamma:=\frac{E}{M}=\sqrt{1+2\nu(\gamma-1)},
\ee
with $\Gamma>1$ ($\gamma>1$) for unbound orbits, and $\Gamma<1$ ($\gamma<1$) for bound orbits.  Then $p_\infty=\mu\gamma v/\Gamma=\mu\sqrt{\ve}/\Gamma$.

With this notation in order, we can take our simplified result for the impulse magnitude $\ms Q$ (\ref{masterQ}) [namely the sum of (\ref{Qa0}), its analogous $a_1a_2$ version, and the SO part (\ref{Qa1234})], insert it into (\ref{chiQ}), and solve for the aligned-spin scattering angle $\chi$.  After this process, $\chi/\Gamma$ turns out to be linear in $\nu$ in the same way that $\ms Q$ is, thanks to the facts that the sine function is odd in its argument and that $\Gamma^2$ is linear in $\nu$.  The result can be expressed as follows,
\bse\label{masterchi}
\begin{alignat}{3}\label{chiform}
\frac{\chi}{\Gamma} &= \frac{GM}{b\sqrt{\ve}}\ms X_{G^1}^{\nu^0}
\\\nnm
&\quad+\Big(\frac{GM}{b\sqrt{\ve}}\Big)^2\ms X_{G^2}^{\nu^0}
\\\nnm
&\quad +\Big(\frac{GM}{b\sqrt{\ve}}\Big)^3\Big[\ms X_{G^3}^{\nu^0}+\nu \ms X_{G^3}^{\nu^1}\Big]
\\\nnm
&\quad +\Big(\frac{GM}{b\sqrt{\ve}}\Big)^4\Big[\ms X_{G^4}^{\nu^0}+\nu \ms X_{G^4}^{\nu^1}\Big]+\mc O\Big(\frac{GM}{b}\Big)^5,
\end{alignat}
where each $\ms X_{G^k}^{\nu^m}$ takes the form
\begin{alignat}{3}\label{Xkm}
\ms X_{G^k}^{\nu^m} 
&= \ms X_{k}^{m}(\ve)
\\\nnm
&\quad+\frac{a_\mr b}{b\sqrt{\ve}}\ms X_{k}^{m\mr b}(\ve)+\frac{a_\mr t}{b\sqrt{\ve}}\ms X_{k}^{m\mr t}(\ve)
\\\nnm
&\quad+\frac{a_1a_2}{b^2\ve}\ms X_{k}^{m\times}(\ve),
\end{alignat}
\ese
with $\times$ standing for the ``cross term'' $a_1a_2$, and with the special constraints
\be\label{speccons}
\ms X^{0\mr b}_{1}=\ms X^{0\mr t}_{1},\qquad \ms X^{1\mr b}_{3}=\ms X^{1\mr t}_{3},
\ee
recalling from (\ref{abat}) that $Ma_\mr b=m_1a_1+m_2a_2$ and $M a_\mr t=m_2a_1+m_1a_2$.\footnote{In Ref.~\cite{Antonelli:2020aeb}, the expression of the result (\ref{masterchi}) for the mass dependence of the scattering angle differed in that (i) we did not pull a factor of $1/\sqrt{\ve}$ out of the $X$'s for every factor of $1/b$, (ii) we used $v$ instead of $\ve$, and (iii) we used $a_+$ and $\delta\,a_-$ in place of $a_\mr b$ and $a_\mr t$, with $a_\pm:=a_2\pm a_1$ and $\delta:=(m_2-m_1)/M$; the equivalence of the two expressions is apparent since
	\begin{alignat}{3}
	\begin{aligned}
	a_++\delta\,a_-&=2a_\mr b,
	\\
	a_+-\delta\,a_-&=2a_\mr t.
	\end{aligned}
	\end{alignat}}
All the $\ms X$'s on the right-hand side of (\ref{Xkm}) are dimensionless and are functions \emph{only} of the dimensionless $\ve=\gamma^2-1$; they can be expressed in terms of the above $\ms Q(\gamma)$'s alone.

We see that the 1PM and 2PM terms in (\ref{masterchi}) are independent of the symmetric mass ratio $\nu$ and are thus fully preserved in the (\emph{extended}) test-body limit $\nu\to0$ (at fixed $M$, or equivalently $m_1\to 0$ at fixed $M$, \emph{and} at fixed $a_1$, $a_2$, $b$ and $\gamma$), while the 3PM and 4PM terms are linear in $\nu$.  This allows us to deduce the complete 1PM and 2PM results for $\chi/\Gamma$ from its test-body limit, and the complete 3PM and 4PM results from first-order self-force (linear-in-mass-ratio) calculations.

The special constraints (\ref{speccons}) are consequences of the $1\leftrightarrow2$ symmetry, as seen in the $G^1\nu^0$ and $G^3\nu^1$ SO terms in (\ref{Qa1234}).  This is a prediction of the above arguments which our considerations below will be able to test, rather than to rely on.  For the case of the $G^3\nu^1$ SO terms, which we will determine (in a PN expansion) below from matching to first-order self-force calculations, we will allow $\ms X^1_{3\mr b}$ and $\ms X^1_{3\mr t}$ to be independent --- in fact, $\ms X^1_{3\mr b}$ will be determined by the redshift invariant in a Kerr background and $\ms X^1_{3\mr t}$ by the spin-precession invariant in a Schwarzschild background --- and we will find from the matching procedure that they are indeed equal through the considered PN orders.
The fact that the complete content of Eqs.~(\ref{masterchi}) holds through N$^2$LO in the PN expansion can be seen in Eqs.~(4.32) of Ref.~\cite{Vines:2018gqi}.

The $\nu^0$ terms in (\ref{masterchi}) can be determined by solving the MPD equations of motion (\ref{MPD}) for a spinning (pole-dipole) test body in a stationary background Kerr spacetime.  An integrand for the test-spin-in-Kerr aligned-spin scattering angle function, to all PM orders, was derived in Ref.~\cite{Bini:2017pee}; see, e.g., their Eq.~(66) (which also includes pole-dipole-quadrupole terms for a test black hole).  The results of the integration are as follows, to all orders in $\ve$ (to all PN orders at each PM order), extending Eq.~(5.5) of Ref.~\cite{Vines:2018gqi} to 4PM order in the spin-orbit and bilinear-in-spin terms.  The nonspinning parts are
\bse\label{X0s}
\begin{alignat}{3}\label{X0a0}
\ms X_{1}^{0}&=2\frac{1+2\ve}{\sqrt{\ve}}=2\frac{2\gamma^2-1}{\sqrt{\gamma^2-1}}=2\frac{1+v^2}{v\sqrt{1-v^2}},
\\\nnm
\ms X_{2}^{0}&=\frac{3\pi}{4}(4+5\ve)=\frac{3\pi}{4}(5\gamma^2-1),
\\\nnm
\ms X_{3}^{0}&=2\frac{-1+12\ve+72\ve^2+64\ve^3}{3\ve^{3/2}},
\\\nnm
\ms X_{4}^{0}&=\frac{105\pi}{64}(16+48\ve+33\ve^2),
\end{alignat}
the SO parts are
\begin{alignat}{3}\label{X0a1}
\ms X_{1}^{0\mr b}a_\mr b+\ms X_{1}^{0\mr t}a_\mr t&=-4\gamma\sqrt{\ve}(a_\mr b+a_\mr t),
\\\nnm
\ms X_{2}^{0\mr b}a_\mr b+\ms X_{2}^{0\mr t}a_\mr t&=-\frac{\pi}{2}\gamma(2+5\ve)(4a_\mr b+3a_\mr t),
\\\nnm
\ms X_{3}^{0\mr b}a_\mr b+\ms X_{3}^{0\mr t}a_\mr t&=-4\gamma\frac{1+12\ve+16\ve^2}{\sqrt{\ve}}(3a_\mr b+2a_\mr t),
\\\nnm
\ms X_{4}^{0\mr b}a_\mr b+\ms X_{4}^{0\mr t}a_\mr t&=-\frac{21\pi}{16}\gamma(8+36\ve+33\ve^2)(8a_\mr b+5\mr a_t),
\end{alignat}
and the bilinear-in-spin parts are
\begin{alignat}{3}\label{X0a1a2}
\ms X_{1}^{0\times}&=4\sqrt{\ve}(1+2\ve),
\\\nnm
\ms X_{2}^{0\times}&=\frac{3\pi}{2}(2+19\ve+20\ve^2),
\\\nnm
\ms X_{3}^{0\times}&=8\frac{1+38\ve+128\ve^2+96\ve^3}{\sqrt{\ve}},
\\\nnm
\ms X_{4}^{0\times}&=\frac{105\pi}{16} (24+212\ve+447\ve^2+264\ve^3)
\end{alignat}
\ese
with $\gamma=\sqrt{1+\ve}$.\footnote{Note that, through 2PM order and up through the SO terms, the first two lines of the right-hand side of (\ref{chiform}), with (\ref{X0a0}) and (\ref{X0a1}) plugged into the first two lines of (\ref{Xkm}), correctly give either (i) the aligned-spin scattering angle for a spinning test body with rescaled spin $a_\mr t$ in a Kerr background with mass $M$ and rescaled spin $a_\mr b$, or (ii) the rescaled aligned-spin scattering angle $\chi/\Gamma$ for the arbitrary-mass two-spinning-body system, using the ``spin maps'' (\ref{abat}); this is a further reason for the labels $a_\mr t$ and $a_\mr b$.  This gives a different ``EOB scattering-angle mapping,'' an alternative to Eq.~(3.16) of \cite{Vines:2018gqi}, which produces the 1PM and 2PM SO terms in the two-body scattering angle from its extended test-body limit.  [Note however that this different mapping fails at quadratic order in the spins, while Eq.~(3.16) of \cite{Vines:2018gqi} still holds, according to all known results.]}

The $\nu^1$ terms  in (\ref{masterchi}), at 3PM and 4PM orders, can be determined in a PN expansion (here, an expansion in $\ve$) from first-order self-force results (as well as from consistency with lower orders), as we will explicitly demonstrate below for the spin parts.  We will use the known nonspinning coefficients through 4PM-3PN order \cite{Vines:2018gqi},
\bse\label{X1s}
\begin{alignat}{3}\label{X1a0}
\ms X_{3}^{1}&=-\frac{8+94\ve+313\ve^2+\mc O(\ve^3)}{12\sqrt{\ve}},
\\\nnm
\ms X_{4}^{1}&=\pi \bigg[{-}\frac{15}{2}+\bigg(\frac{123}{128}\pi^2-\frac{557}{8}\bigg)\ve+\mc O(\ve^2)\bigg],
\end{alignat}
noting the transcendental $\zeta(2)$ contribution in the last term (the 4PM-3PN term).  We will parametrize 
the SO coefficients as
\begin{alignat}{3}\label{SOchipar}
\ms X_{3}^{1\mr i}&=\frac{\gamma}{\sqrt{\ve}}\Big(\ms X^{1\mr i}_{30}+\ms X^{1\mr i}_{31}\ve+\ms X^{1\mr i}_{32}\ve^2+\ms X^{1\mr i}_{33}\ve^3+\mc O(\ve^4)\Big),
\nnm\\
\ms X_{4}^{1\mr i}&=\pi\gamma\Big(\ms X^{1\mr i}_{41}+\ms X^{1\mr i}_{42}\ve+\ms X^{1\mr i}_{43}\ve^2+\mc O(\ve^3)\Big),
\end{alignat}
with $\mr i=\mr b, \mr t$, and the bilinear-in-spin coefficients as
\begin{alignat}{3}\label{S1S2chipar}
\ms X_{3}^{1\times}&=\frac{1}{\sqrt{\ve}}\Big(\ms X^{1\times}_{30}+\ms X^{1\times}_{31}\ve+\ms X^{1\times}_{32}\ve^2+\ms X^{1\times}_{33}\ve^3+\mc O(\ve^4)\Big),
\nnm\\
\ms X_{4}^{1\times}&=\pi\Big(\ms X^{1\times}_{41}+\ms X^{1\times}_{42}\ve+\ms X^{1\times}_{43}\ve^2+\mc O(\ve^3)\Big).
\end{alignat}
\ese
We have included all the same powers of $\ve$ present in the $\nu^0$ coefficients (\ref{X0s}), up to the orders in $\ve$ which will contribute at the N$^3$LO PN level.  (We have also factored out $\gamma=\sqrt{1+\ve}$ in the SO terms and $\pi$ in the 4PM terms, following the patterns at $\nu^0$.) For these $\ms X^{1\cdots}_{kn}$, which are all pure numbers, $k$ gives the PM order, and $n$ gives the maximum PN order (N$^n$LO) which determines that coefficient.  This labeling and the consistency and sufficiency of this ansatz for the scattering angle will become evident in the matching between the scattering angle and a canonical Hamiltonian described in the following section.

Finally, it is important to note that the impact parameter $b$ appearing everywhere in this section is the distance orthogonally separating the two spinning bodies' asymptotic incoming worldlines as defined by the ``covariant'' or Tulczyjew-Dixon condition~\cite{Dixon:1979,Steinhoff:2014kwa,Tulczyjew:1959,fokker1929relativiteitstheorie}, Eq.~(\ref{TDSSC}) above, for each body---the so-called ``proper'' or ``covariant'' impact parameter $b\equiv b_\mr{cov}$ \cite{Vines:2018gqi,Guevara:2018wpp,Siemonsen:2019dsu}.  This is crucial to the above argument because only with the covariant condition (\ref{TDSSC}) (or something equivalent to it at 0PM order) does it hold that the first-order field (\ref{h1PM}) is linear in the masses.  Below, we will also work with the canonical orbital angular momentum $L\equiv L_\mr{can}=p_\infty b_\mr{can}$, where $b_\mr{can}$ is the impact parameter orthogonally separating the asymptotic incoming worldlines defined by cm-frame Newton-Wigner conditions~\cite{pryce1948mass,newton1949localized} for each body.  This coincides with the conserved canonical orbital angular momentum $L$ appearing in a canonical Hamiltonian formulation of aligned-spin two-body dynamics~\cite{Barausse:2009aa,Vines:2016unv}.  [Note that, for the aligned-spin case, the covariant/Pauli-Lubanski spin vectors $m_\mr ia_\mr i^\mu$ used above coincide with the canonical spin vectors $S_\mr i^\mu$ (spatial vectors in the cm frame) which would be associated with the cm-frame Newton-Wigner conditions, and thus so do the aligned-spin (signed) magnitudes, $S_\mr i=m_\mr ia_\mr i$.]  As shown in \cite{Vines:2017hyw,Vines:2018gqi}, the canonical $L=:L_\mr{can}$ is related to the covariant $b$ by
\begin{alignat}{3}\label{DeltaL}
L&=L_\mr{cov}+\Delta L,
\\\nnm
L_\mr{cov}&=p_\infty b=\frac{\mu}{\Gamma}\gamma v b=\frac{\mu}{\Gamma}\sqrt{\ve} b,
\\\nnm
\Delta L&=\Big(\sqrt{m_1^2+p_\infty^2}-m_1\Big)a_1+\Big(\sqrt{m_2^2+p_\infty^2}-m_2\Big)a_2
\\\nnm
&=M\frac{\Gamma-1}{2}\bigg(a_\mr b+a_\mr t-\frac{a_\mr b-a_\mr t}{\Gamma}\bigg).
\end{alignat}
Solving this for $b$, inserting the result into (\ref{masterchi}) [or (\ref{masterchi2})], and re-expanding to bilinear order in the (mass-rescaled) spins $a_1$ and $a_2$, one obtains the final parametrized form for the aligned-spin scattering angle function $\chi(E,L;m_\mr i,a_\mr i)$ used in the following matching calculations.

Let us finally rewrite the scattering angle to include both the $\nu^0$ and $\nu^1$ terms in single coefficients (or which could allow mass dependence differing from that deduced above), and which would accommodate general quadratic-in-spin terms, with sums over i and j implied,
\be\label{masterchi2}
\frac{\chi}{\Gamma}=\sum_{k\ge1}\Big(\frac{GM}{b\sqrt{\ve}}\Big)^k\bigg[\ms X_k(\ve,\nu)+\frac{a_\mr i}{b\sqrt{\ve}}\ms X_k{}^{\mr i}(\ve,\nu)+\frac{a_\mr i a_\mr j}{b^2{\ve}}\ms X_k{}^{\mr i\mr j}(\ve,\nu)\bigg]
\ee
$+\mc O(a^3)$, with 
\begin{alignat}{3}
a_\mr i \ms X_k{}^{\mr i}&=a_\mr b \ms X_k{}^{\mr b}+a_\mr t \ms X_k{}^{\mr t},
\\\nnm
a_\mr i a_\mr j\ms X_k{}^{\mr i\mr j}&=a_1 a_2 \ms X_k{}^{\!\times}+\mc O(a_1^2,a_2^2).
\end{alignat} 
Our prediction for the mass-ratio dependence of the $k$PM coefficients ${\ms X}_k{}^{\mr A}=\{{\ms X}_k,{\ms X}_k{}^{\mr b},{\ms X}_k{}^{\mr t},{\ms X}_k{}^{\!\times}\}$ is that 
\be\label{joinXs}
{\ms X}_k{}^{\mr A}(\ve,\nu)=\left\{
\begin{array}{cc}
	{\ms X}_k^{0\mr A}(\ve),\quad &k=1,2 
	\\
	{\ms X}_k^{0\mr A}(\ve)+\nu {\ms X}_k^{1\mr A}(\ve),\quad &k=3,4
\end{array}\right..
\ee
The $\nu^0$ coefficients $\ms X^{0\mr A}_k(\ve)$ from the extended test-body limit are given explicitly in (\ref{X0s}), and the $\nu^1$ coefficients $\ms X^{1\mr A}_k(\ve)$ which we will determine from self-force results are parametrized in a PN expansion in (\ref{X1s}).  Note that we will also be able to use the self-force results to test the fact that there are no $\nu^1$ terms  at 1PM and 2PM orders in this parametrization of the scattering angle.  The fact that there are no $\nu^2$ or higher terms through 4PM order cannot be probed with first-order self-force results, but has already been confirmed by arbitrary-mass PN results through N$^2$LO.  Our prediction for the mass dependence will yield new arbitrary-mass results at the N$^3$LO PN level once we have fixed the PN expansions of the coefficients $\ms X^{1\mr A}_k$ from first-order self-force calculations.

\section[From the unbound scattering angle to the bound radial action via\dots]{From the unbound scattering angle to the bound radial action via canonical Hamiltonian dynamics}
\label{Irsec}

Besides the mass dependence of the scattering angle function established in the previous section, and the inputs of test-body results (discussed above) and first-order self-force results (discussed below), the other central ingredient in our derivation is the assumption of the existence of a (local-in-time) canonical Hamiltonian governing the aligned-spin conservative dynamics in the cm frame, for generic (\emph{both} bound and unbound) orbits, with the Hamiltonian having well-defined (regular, polynomial) PN and PM expansions. Through the desired 4.5PN order in the SO sector and 5PN {\sonestwo} one, we can safely ignore nonlocal-in-time (tail) contributions in the final dynamics/scattering angle. While these do appear at the 4PN level in the nonspinning sector~\cite{Damour:2014jta} (see e.g., Ref.~\cite{Bini:2017wfr} for a translation into a nonlocal-in-time scattering angle), they only start appearing at 5.5PN order in the spinning one. This can most easily be seen in the first line of Eq.(68a) in Ref.~\cite{Siemonsen:2017yux}, where the linear-in-spin tails are a relative 1.5PN order from the leading quadrupolar contributions to the tail. [As mentioned at the very end of this section, we find it necessary to include tail terms at 4PN order in the nonspinning sector to make contact with available results in the GSF literature.]

Our ultimate goal in this section is to take the gauge-invariant scattering-angle function $\chi$ for unbound orbits, parametrized in the previous section, and derive from it a parametrized expression for the gauge-invariant radial-action function $I_r$ which characterizes bound orbits, from which we can derive all the bound-orbit gauge invariants to be compared with self-force results in Sec.~\ref{sec:smallq} below.  

We do this by passing through the gauge-dependent canonical Hamiltonian dynamics.  It is to some extent true that this process (as we implement it here) can be bypassed by using relationships between gauge invariants for unbound and bound orbits found in \cite{Kalin:2019inp}, but not entirely.  Those relationships yield $I_r$ through $\mc O(G^4)$ from $\chi$ through $\mc O(G^4)$, but the complete PN expansion of $I_r$ through N$^3$LO extends to $\mc O(G^8)$ (for the spin terms).  The extra terms in $I_r$ are obtained here via the canonical Hamiltonian dynamics, which determines them from (the PN re-expansion of) $\chi$ through $\mc O(G^4)$.  Note that $\chi$ through $\mc O(G^4)$ does not contain the complete PN expansion of $\chi$ through N$^3$LO, nor through LO, since even the Newtonian scattering angle has contributions at all orders in $G$.  But the PN expansion of the 4PM scattering angle, $\chi$ through $\mc O(G^4)$, does contain the complete information of the N$^3$LO PN Hamiltonian (contained in its $O(G^4)$ truncation), which determines the N$^3$LO PN radial action $I_r$ (contained in its $O(G^8)$ truncation).

We begin in Sec.~\ref{sec:massshell} by discussing canonical Hamiltonians for aligned-spin binaries, the resultant equations of motion, and their gauge freedom under canonical transformations, in a PM-PN expansion.  We fix a unique gauge by imposing simplifying conditions not on the Hamiltonian function $H$ itself, but on its corresponding ``mass-shell constraint'' (or ``impetus formula''\cite{Kalin:2019rwq}), which is simply a rearrangement of the expression of the Hamiltonian, in which the squared momentum is given as a function of the Hamiltonian $H$ (of the energy $E=H$).  In Sec.~\ref{sec:angle}, we describe how the scattering-angle function can be derived from the canonical mass-shell constraint, or vice versa (with our gauge-fixing for the mass shell), and derive the explicit relationships between the scattering-angle coefficients and the mass-shell coefficients.  Finally, in Sec.~\ref{sec:radialaction}, we compute the radial action $I_r$, and point out a hidden simplicity in its dependence on the mass ratio, when expressed in terms of appropriate (covariant rather than canonical) variables, which is a simple consequence of the mass dependence of the scattering angle $\chi$ and the relationship between $\chi$ and $I_r$ discovered in \cite{Kalin:2019inp}.

\subsection{The canonical Hamiltonian and/or the mass-shell constraint}\label{sec:massshell}

For an aligned-spin binary canonical Hamiltonian,
\begin{alignat}{3}
&H(r,\phi,p_r,L;m_\mr i,a_\mr i)=H(r,p_r,L;m_\mr i,a_\mr i)
\end{alignat}
the dynamical variables (depending on a time parameter $t$) are polar coordinates $(r,\phi)$ in the orbital plane, with $r$ being the orbital separation, and their conjugate momenta $(p_r,p_\phi\equiv L)$.  The Hamiltonian does not depend on the angular coordinate $\phi$ due to the system's axial symmetry, and it otherwise depends only on the constant masses and spins $(m_\mr i,a_\mr i)=(m_1,m_2,a_1,a_2)$.  The Hamiltonian equations of motions read
\begin{alignat}{3}\label{Hameqs}
\dot r&=\frac{\doe H}{\doe p_r},\qquad & \dot p_r&=-\frac{\doe H}{\doe r},
\\\nnm
\dot \phi&=\frac{\doe H}{\doe L},\qquad & \dot L&=-\frac{\doe H}{\doe\phi}=0,
\end{alignat}
where we note that the canonical orbital angular momentum $L$ is a constant of motion.

Such a Hamiltonian is not unique, but is subject to a type of gauge freedom, namely under canonical transformations: diffeomorphisms of the phase space which preserve the canonical form (\ref{Hameqs}) of the equations of motion.  In a quite general gauge (one which encompasses all gauges encountered in previous PN or PM aligned-spin Hamiltonians), the Hamiltonian takes the following form through quadratic order in the spins, through 4PM order,
\begin{alignat}{3}\label{Hgeneral}
&H=H_0(\bs p^2;m_\mr i)+\sum_{k=1}^4\frac{G^k}{r^k}\bigg[c_k(\bs p^2,\frac{L^2}{r^2};m_\mr i)
\\\nnm
&\quad+\frac{La_\mr i}{r^2}c_k^\mr i(\bs p^2,\frac{L^2}{r^2};m_\mr j)+\frac{a_\mr ia_\mr j}{r^2}c_k^{\mr i\mr j}(\bs p^2,\frac{L^2}{r^2};m_\mr k)\bigg]+\mc O(G^5),
\end{alignat}
where
\be\label{psqprsq}
\bs p^2=p_r^2+\frac{L^2}{r^2},
\ee
is the total squared canonical linear momentum.  Here, $H_0$ is the 0PM (free) Hamiltonian, and the functions $c_k$, $c_k^\mr i$ and $c_k^{\mr i\mr j}$ encode respectively the nonspinning, spin-orbit, and quadratic-in-spin gravitational couplings at the $k$PM orders.
The $c$'s are assumed to have regular Taylor series around $L^2=0$ and $\bs p^2=0$.  We will work here with the standard (gauge) choice for the free Hamiltonian in the cm frame,
\be\label{H0PM}
H_0=\sqrt{m_1^2+\bs p^2}+\sqrt{m_2^2+\bs p^2},
\ee
such that, as $r\to\infty$, the magnitude $\sqrt{\bs p^2}$ of the canonical linear momentum corresponds to the two bodies' physical equal and opposite spatial momenta in the cm frame.

The expression (\ref{Hgeneral}) of the Hamiltonian can be solved, working perturbatively in $G$, for $\bs p^2(r,E,L; m_\mr i,a_\mr i)$, where $E\equiv H(r,p_r,L; m_\mr i,a_\mr i)$ is the total energy; one finds
\begin{alignat}{3}\label{psqgeneral}
\bs p^2&=p_\infty^2(E;m_\mr i)+\sum_{k\ge1}\frac{G^k}{r^k}\bigg[f_k(E,\frac{L^2}{r^2};m_\mr i)
\\\nnm
&\qquad+\frac{La_\mr i}{r^2}f_k^\mr i(E,\frac{L^2}{r^2};m_\mr j)+\frac{a_\mr ia_\mr j}{r^2}f_k^{\mr i\mr j}(E,\frac{L^2}{r^2};m_\mr k)\bigg],
\end{alignat}
where the 0PM part $p_\infty^2$ is found by (exactly) inverting (\ref{H0PM}), $H_0(\bs p^2)=E$ $\;\Leftrightarrow\;$ $p_\infty^2(E)=\bs p^2$,
\be
p_\infty^2=\frac{(E^2-m_1^2-m_2^2)^2-4m_1^2m_2^2}{4E^2}=\mu^2\frac{\gamma^2-1}{\Gamma^2},
\ee
which we recognize as the same $p_\infty$ from (\ref{pcm}). The functions $f_k$, $f^\mr i_k$ and $f^{\mr i\mr j}_k$ are determined by (and carry all of the information of) the $c_k^{\cdots}$ coefficients in the Hamiltonian (\ref{Hgeneral}).  Importantly, the $f^{\cdots}_k$ functions will have regular limits as $\gamma^2-1=\ve\to0$ (as $p_\infty\to0$) and as $L^2\to0$, given our assumption that the $c^{\cdots}_k$ functions were regular as $\bs p^2\to 0$ and $L^2\to 0$.  The quantities $\gamma$, $\ve$ and $\Gamma$ are all defined in terms of the energy $E$ and the rest masses just as in the previous section.

As discussed in Ref.~\cite{Vines:2018gqi} (through N$^2$LO in the PN expansion, and as we have explicitly verified through N$^3$LO), it is possible to find a perturbative canonical transformation which brings the Hamiltonian (\ref{Hgeneral}) into a ``quasi-isotropic'' form, i.e., a form in which the $c$'s depend only $\bs p^2$ and not on $L^2/r^2$.
Furthermore, the freedom in canonical transformations [among Hamiltonians of the form (\ref{Hgeneral})] is completely fixed once one imposes this quasi-isotropic-Hamiltonian condition \emph{and} uniquely specifies a 0PM Hamiltonian $H_0$, as we have done in (\ref{H0PM}).  For such a quasi-isotropic Hamiltonian, one finds that the corresponding ``mass shell constraint,'' the expression for $\bs p^2$ (\ref{psqgeneral}), has nonspinning and SO coefficients $f_k$  $f_k^\mr i$ which are independent of $L^2/r^2$, but its quadratic-in-spin coefficients $f_k^{\mr i\mr j}$ have terms at zeroth and first orders in $L^2/r^2$.  However, there also exists a different (non-quasi-isotropic) gauge for the Hamiltonian (\ref{Hgeneral}) (one with $L^2/r^2$ terms in $c_k^{\mr i\mr j}$) such that its mass shell constraint (\ref{psqgeneral}) is quasi-isotropic, with the $f_k$, $f_k^\mr i$ and $f^{\mr i\mr j}_k$ \emph{all} depending only on $E$ (and the masses) and not on $L^2/r^2$.
Because both the scattering angle and the radial action are more directly related to the $f$ coefficients in the mass shell, we will find it convenient to adopt this quasi-isotropic-mass-shell gauge (which is also unique with a given choice for $H_0$), specializing (\ref{psqgeneral}) to the form
\begin{alignat}{3}\label{psqiso}
\bs p^2&=p_\infty^2(E;m_\mr i)+\sum_{k\ge1}\frac{G^k}{r^k}\bigg[f_k(E;m_\mr i)+\frac{La_\mr i}{r^2}f_k^\mr i(E;m_\mr j)+\frac{{a_\mr ia_\mr j}}{r^2}f_k^{\mr i\mr j}(E;m_\mr k)\bigg].
\end{alignat}
Regrouping in terms of powers of $r$ instead of powers of $G$, we have
\be\label{massshellsimple}
p_r^2+\frac{L^2}{r^2}=\bs p^2=p_\infty^2+\sum_{k\ge1}\frac{G^k}{r^k}\tilde f_k,
\ee
where we define
\be\label{fktotal}
\tilde f_{k}=f_k+\frac{La_\mr i}{G^2}f_{k-2}^\mr i+\frac{{a_\mr ia_\mr j}}{G^2}f_{k-2}^{\mr i\mr j},
\ee
with $f^{\cdots}_{-1}=f^{\cdots}_{0}=0$, and we need to extend the sum to $k=6$ (while dropping the nonspinning $f_5$ and $f_6$).  Our starting point for the following calculations will be this ansatz for the mass shell constraint, which is fully equivalent to an ansatz for a Hamiltonian of the form (\ref{Hgeneral}) modulo gauge freedom.  Our fundamental assumption is the existence of such a canonical Hamiltonian.  We will find that the coefficients $f_k^{\cdots}(E;m_\mr i)$ are uniquely determined by the expansion of the scattering-angle function to $k$PM order.

\subsection{The scattering angle}\label{sec:angle}

As shown in \cite{Damour:2017zjx}, the scattering angle $\chi(E,L;m_\mr i,a_\mr i)$ for an unbound orbit can be found directly from the canonical mass-shell constraint as follows.  The constraint (\ref{massshellsimple}) can be solved for the radial momentum $p_r(r,E,L;m_\mr i,a_\mr i)$, and then the scattering angle is given by the integral
\begin{alignat}{3}\label{chidef}
\pi+\chi(E,L)&=-\int_\infty^\infty\mr d r\, \frac{\doe}{\doe L}p_r(r,E,L)
\\\nnm
&=-2\int_{r_\mr{min}}^\infty\mr d r\, \frac{\doe}{\doe L}\sqrt{p_\infty^2-\frac{L^2}{r^2}+\sum_{k\ge1}\frac{G^k}{r^k}\tilde f_k},
\end{alignat}
where $r_\mr{min}$ is the largest real root of $p_r=0$.
In the direct evaluation of this integral, it would matter that the $\tilde f_k$ in (\ref{fktotal}) depend on $L$ (in the SO terms).  But let us define an antiderivative of $\pi+\chi$ with respect to $L$ to be ``the unbound radial action,'' 
\bse\label{defW}
\be\label{WdinvL}
\W=-\frac{1}{2\pi}\Big(\frac{\doe}{\doe L}\Big)^{-1}(\pi+\chi),
\ee
which is essentially a \emph{partie finie} of the radial action integral for unbound orbits,
\be
\W(E,L)=\frac{1}{2\pi}\mr{Pf}\int_\infty^\infty \mr d r\, p_r(r,E,L).
\ee
\ese
The eikonal phase~\cite{Bjerrum-Bohr:2018xdl,Kabat:1992tb,Akhoury:2013yua,Bern:2020buy} is $\W/\hbar$ (up to a constant).
For the expression of $\W$ in terms of the $\tilde f_k$, it does not matter that the $\tilde f_k$ depend on $L$.  That expression will be identical to the $L$-antiderivative of the nonspinning scattering angle expressed in terms of the nonspinning $f_k$, with $f_k\to\tilde f_k$, so this reduces the evaluation of the integral for the spinning case to the nonspinning problem, using the coefficient mapping (\ref{fktotal}).  The results of the nonspinning integral (for $\chi$, from which constructing $\W$ is trivial) have been tabulated at high orders, e.g., in \cite{Bjerrum-Bohr:2019kec}.  One finds
\be\label{WPM}
2\pi \W=-\pi L-\frac{G\ln L}{p_\infty}\tilde \chi_1+\sum_{k\ge 2}\frac{G^k}{p_\infty^k L^{k-1}}\frac{\tilde \chi_k}{k-1},
\ee
where $\tilde \chi_k$ are the entries of Table 1 in \cite{Bjerrum-Bohr:2019kec} with $f_k\to\tilde f_k$; the first few read
\begin{alignat}{3}\label{tildechis}
\tilde \chi_1&=\tilde f_1,\phantom{\bigg|}
\\\nnm
\tilde \chi_2&=\frac{\pi}{2}p_\infty^2\tilde f_2,
\\\nnm
\tilde \chi_3&=2p_\infty^4\tilde f_3+p_\infty^2\tilde f_1\tilde f_2-\frac{\tilde f_1^3}{12},
\\\nnm
\tilde \chi_4&=\frac{3\pi}{8}p_\infty^4(2p_\infty^2 \tilde f_4+\tilde f_2^2+2\tilde f_1\tilde f_3),
\\\nnm
& \cdots
\end{alignat}
The scattering angle $\chi$ is then given by
\be
\pi+\chi=-2\pi\frac{\doe\W}{\doe L},
\ee
with the $L$-derivative acting also inside the $\tilde f_k$ in (\ref{fktotal}).  
To obtain $\W$ or $\chi$ through quadratic order in spins and through 4PM order, $\mc O(G^4)$, counting both the $G^k$ in (\ref{WPM}) and the $1/G^2$ in (\ref{fktotal}), we must include parts of the contributions up to $\tilde f_6$ and up to $\tilde\chi_8$.  The resultant explicit expression of the scattering angle $\chi$ in terms of the $f_k^{\cdots}$ coefficients up to 4PM order and quadratic order in spins is 
	\begin{alignat}{3}\label{chifs}
	\chi&=\frac{G}{p_\infty L}f_1+\frac{\pi G^2}{2L^2} f_2
	+\frac{G^3}{ p_\infty^3L^3}\bigg[{-}\frac{1}{12}f_1^3+p_\infty^2 f_1f_2+2p_\infty^4 f_3\bigg]\nonumber\\
	&+\frac{3\pi G^4}{8L^4}\bigg[f_2^2+2f_1f_3+2p_\infty^2f_4\bigg]
+a_\mr i\Bigg\{\frac{Gp_\infty}{L^2}f_1^\mr i+\frac{\pi G^2}{2L^3}\bigg[f_1f_1^\mr i+p_\infty^2 f_2^\mr i\bigg]
	\nonumber\\
	&+\frac{G^3}{p_\infty L^4}\bigg[\frac{3}{4}f_1^2 f_1^\mr i+3 p_\infty^2\Big(f_2 f_1^\mr i+f_1 f_2^\mr i\Big)+2p_\infty^4 f_3^\mr i\bigg]
\nonumber\\
&+\frac{3\pi G^4}{4 L^5}\bigg[2 f_1 f_2 f_1^\mr i+f_1^2f_2^\mr i+2p_\infty^2\Big(f_3f_1^\mr i+f_2 f_2^\mr i+f_1 f_3^\mr i\Big)+p_\infty^4f_4^\mr i\bigg]\Bigg\}
\nonumber\\
&+a_\mr ia_\mr j\Bigg\{\frac{2Gp_\infty}{L^3}f_1^{\mr i\mr j}+\frac{3\pi G^2}{16L^4}\bigg[4f_1f_1^{\mr i\mr j}+p_\infty^2\Big(3 f_1^\mr if_1^\mr j+4f_2^{\mr i\mr j}\Big)\bigg]
\nonumber\\
&	+\frac{G^3}{p_\infty L^5}\bigg[f_1^2f_1^{\mr i\mr j}+4p_\infty^2\Big(f_2f_1^{\mr i\mr j}+f_1 f_1^\mr if_1^\mr j+f_1f_2^{\mr i\mr j}\Big)
	+\frac{8}{3}p_\infty^4\Big(2f_1^\mr if_2^\mr j+f_3^{\mr i\mr j}\Big)\bigg]
\nonumber\\
&	+\frac{15\pi G^4}{64 L^6}\bigg[
	8f_1f_2f_1^{\mr i\mr j}+5f_1^2f_1^\mr if_1^\mr j+4f_1^2f_2^{\mr i\mr j}
\nonumber\\
&	+2p_\infty^2\Big(4f_3 f_1^{\mr i\mr j}+5 f_2f_1^\mr if_1^\mr j+4 f_2f_2^{\mr i\mr j}+10f_1f_1^\mr if_2^\mr j+4f_1 f_3^{\mr i\mr j}\Big)
\nonumber\\
&	+p_\infty^4\Big(5f_2^\mr if_2^\mr j+10 f_1^\mr if_3^\mr j+4f_4^{\mr i\mr j}\Big)\bigg]\Bigg\}+\mc O(a^3)+\mc O(G^5).
	\end{alignat}
We see that the $k$PM coefficients $f_k^{\cdots}$ first enter in the $G^k$ terms; however, they do not enter those terms at the leading orders in $p_\infty$ (in the PN expansion of each PM coefficient).  Recalling that all of the $f$'s are finite as $p_\infty\to0$ ($\ve\to 0$), we see that, within each set of square brackets multiplying $G^k$, the lowest orders in $p_\infty$ do not depend on $f_k^{\cdots}$, rather only on the lower-PM-order $f$'s (with some exceptions at $G^1$ and $G^2$).  Similarly, for the scattering-angle coefficients at even higher orders in $G$ (some of which will be relevant below), the lower orders in their PN expansions will be determined by coefficients from lower orders in $G$ already appearing here.

This gives the scattering angle $\chi$ in terms of the mass-shell coefficients $f_k$, $f_k^\mr i$, $f_k^{\mr i\mr j}$, as an expansion in the canonical orbital angular momentum $L$.  Equating that expression to a parametrization of $\chi$ of the form (\ref{masterchi2}) in terms of the covariant impact parameter $b$, using the translation (\ref{DeltaL}) while re-expanding in spins, one can solve for the $f$ coefficients in the mass shell in terms of the $\ms X$ coefficients in the scattering angle (or vice versa), order by order in the PM expansion.  Recall $p_\infty=\mu\sqrt{\ve}/\Gamma$.  Rewriting $\Delta L=L-p_\infty b$ from (\ref{DeltaL}) as a sum over (effective) spins,
\bse\label{dLxi}
\be
\Delta L=\frac{\mu}{\Gamma}\xi^\mr ia_\mr i=\frac{\mu}{\Gamma}\Big(\xi^\mr ba_\mr b+\xi^\mr ta_\mr t\Big),
\ee
with
\begin{alignat}{7}
\xi^\mr b&=\frac{(\Gamma-1)^2}{2\nu}&&=2\nu\Big(\frac{\gamma-1}{\Gamma+1}\Big)^2&&=\frac{\nu\ve^2}{8}+\mc O(\ve^3),
\nnm\\
\xi^\mr t&=\frac{\Gamma^2-1}{2\nu}&&=\gamma-1&&=\frac{\ve}{2}+\mc O(\ve^2),
\end{alignat}
\ese
the results for the $f$'s through 2PM order are as follows: nonspinning,
\bse\label{f12}
\begin{alignat}{3}\label{f012}
f_1&=\mu^2 M \frac{\sqrt{\ve}}{\Gamma}\ms X_1,
\\\nnm
f_2&=\frac{2\mu^2 M^2}{\pi\Gamma}\ms X_2,
\end{alignat}
spin-orbit,
\begin{alignat}{3}
f_1^\mr i&= \frac{\mu M}{\sqrt{\ve}}\Big(\ms X_1{}^{\mr i} + \ms X_1\xi^\mr i\Big),
\\\nnm
f_2^\mr i&=\frac{\mu M^2}{\ve}\bigg[\frac{2}{\pi}\ms X_2{}^{\mr i}-\Gamma \ms X_1 \ms X_1{}^{\mr i}+\Big(\frac{4}{\pi}\ms X_2-\Gamma (\ms X_1)^2\Big)\xi^\mr i\bigg]
\end{alignat}
and quadratic in spin,
\begin{alignat}{3}
f_1^{\mr i\mr j}&= \frac{\mu^2 M}{2\Gamma\sqrt{\ve}}\Big( \ms X_1{}^{\mr i\mr j} +2 \ms X_1\ms X_1{}^{\mr i}\xi{}^{{\mr j}}+ \ms X_1\xi^\mr i\xi^\mr j\Big),
\\\nnm
f_2^{\mr i\mr j}&=\frac{\mu^2 M^2}{\Gamma\ve}\bigg[\frac{4}{3\pi}\ms X_2{}^{\mr i\mr j}-\frac{1}{2}\Gamma\ms X_1\ms X_1{}^{\mr i\mr j}-\frac{3}{4}\Gamma\ms X_1{}^\mr i\ms X_1{}^\mr j
\\\nnm
&\quad+\Big(\frac{4}{\pi}\ms X_2{}^\mr i-\frac{5}{2}\Gamma\ms X_1\ms X_1{}^\mr i\Big)\xi^\mr j+\Big(\frac{4}{\pi}\ms X_2-\frac{5}{2}\Gamma (\ms X_1)^2\Big)\xi^\mr i\xi^\mr j\bigg],
\end{alignat}
\ese
with symmetrization over i and j understood.  These 1PM and 2PM results are exact (to all orders in $\ve$).  With our predicted mass-ratio dependence from the previous section, we have, for $k=1,2$, $\ms X_k(\ve,\nu)=\ms X_k^0(\ve)$, $a_\mr i\ms X_k{}^\mr i(\ve,\nu) =a_\mr b\ms X_k^{0\mr b}(\ve)+a_\mr t\ms X_k^{0\mr t}(\ve)$, and $a_\mr ia_\mr j\ms X_k{}^{\mr i\mr j}(\ve,\nu)=a_1a_2\ms X_k^{0{\times}}(\ve)+\mc O(a_1^2,a_2^2)$, all independent of $\nu$, and the $\ms X_k^{0{\cdots}}(\ve)$ from the extended test-body limit are given explicitly by (\ref{X0s}).  Though it is not immediately obvious here, each of these $f$'s has a finite limit as $\ve\to0$, as is required by our Hamiltonian ansatz.  We will need the expansions of the $f_1^{\cdots}$ up to $\mc O(\ve^3)$, and of the $f_2^{\cdots}$ up to $\mc O(\ve^2)$.  Along with $f_3^{\cdots}$ up to $\mc O(\ve^1)$ and $f_4^{\cdots}$ at $\mc O(\ve^0)$, we will then have a complete mass-shell constraint (\ref{psqiso}) up to N$^3$LO in the PN expansion, which could be solved for the corresponding canonical Hamiltonian (\ref{Hgeneral}).

At 3PM and 4PM orders, one can also solve for the $f$'s in terms of the $\ms X$'s, obtaining exact expressions analogous to the above.  But we will now work in a PN expansion, an expansion in $\ve$, while enforcing our predicted mass-ratio dependence [which (\ref{f12}) did not].  For the nonspinning coefficients, using the known results (\ref{X0a0}) and (\ref{X1a0}) for the $\ms X$'s, we find
\begin{alignat}{3}
\frac{f_3}{\mu^2 M^3}&=\frac{17-10\nu}{2}+\frac{36-91\nu+13\nu^2}{4}\ve+\mc O(\ve^2),
\nnm\\
\frac{f_4}{\mu^2M^4}&=8+\Big(\frac{41}{32}\pi^2-\frac{160}{3}\Big)\nu+\frac{7}{2}\nu^2+\mc O(\ve),
\end{alignat}
through the orders that contribute to the N$^3$LO PN level.  Here again we note the finite limits as $\ve\to0$.  For the spinning contributions, we must enforce that all the $f$'s have finite limits as $\ve\to0$, which will fix some of the unknown coefficients in our parametrization (\ref{X1s}) of the $\nu^1$ parts of the scattering angle, or relationships between them, from consistency with the lower-order $f$'s and $\ms X$'s [recall the discussion following (\ref{chifs})].  At the SO level, this determines or constrains the lower-PN-order scattering-angle coefficients,
\begin{alignat}{3}\label{lowPNchiSO}
\ms X_{30}^{1\mr i}a_\mr i&=0,\phantom{\bigg|}
\\\nnm
\ms X_{31}^{1\mr i}a_\mr i&=10(a_\mr b+a_\mr t),
\\\nnm
\ms X_{41}^{1\mr i}a_\mr i&=\frac{21}{2}a_\mr b+9a_\mr t,
\\\nnm
\ms X_{42}^{1\mr i}a_\mr i&=\frac{3}{4}\Big(68a_\mr b+49 a_\mr t+2 \ms X^{1\mr i}_{32}a_\mr i\Big),
\end{alignat}
and expressions for $f_3^\mr i$ and $f_4^\mr i$ which are explicitly regular as $\ve\to0$ and depend on the remaining unknowns $\ms X^{1\mr i}_{32}$, $\ms X^{1\mr i}_{33}$, and $\ms X^{1\mr i}_{43}$, with $\mr i=\mr b,\mr t$,
\bse\label{f34SO}
\begin{alignat}{3}
&\frac{f_3^\mr ia_\mr i}{\mu M^3}=\frac{-6+4\nu-5\nu^2}{2}a_\mr b+\frac{-3-31\nu-9\nu^2}{4}a_\mr t+\frac{\nu}{2}\ms X_{32}^{1\mr i}a_\mr i
\nnm\\
&\qquad\quad+\bigg[\frac{-24+172\nu-276\nu^2+21\nu^3}{16}a_\mr b
\\\nnm
&+\frac{-166\nu-90\nu^2+9\nu^3}{8}a_\mr t+\frac{\nu}{4} \Big(\ms X_{32}^{1\mr i}+2\ms X_{33}^{1\mr i}\Big)a_\mr i\bigg]\ve+\mc O(\ve^2),
\end{alignat}
and
\begin{alignat}{3}
\frac{f_4^\mr ia_\mr i}{\mu M^4}&=\bigg({-}2-\frac{811}{8}\nu-4\nu^2+\frac{13}{8}\nu^3\bigg)a_\mr b
\\\nnm
&\quad+\bigg(\frac{1}{8}-\frac{1577}{12}\nu+\frac{41}{16}\pi^2\nu+\frac{35}{4}\nu^2+\frac{3}{2}\nu^3\bigg)a_\mr t
\\\nnm
&\quad+\nu\Big[(4+\nu)\ms X_{32}^{1\mr i}-2\ms X^{1\mr i}_{33}+\frac{4}{3}\ms X^{1\mr i}_{43}\Big]a_\mr i+\mc O(\ve).
\end{alignat}
\ese
Similarly, for the bilinear-in-spin coefficients, we find
\begin{alignat}{3}\label{lowPNchiS1S2}
\ms X^{1{\times}}_{30}&=0,
\qquad 
\ms X^{1{\times}}_{31}=8,
\qquad
\ms X^{1{\times}}_{41}=\frac{15}{2},
\nnm\\
\ms X^{1{\times}}_{42}&=\frac{45}{32}\Big({-}22+\ms X^{1{\times}}_{32}\Big),
\end{alignat}
while $f_{k}^{\mr i\mr j}a_\mr ia_\mr j=f_k^\times a_1a_2+\mc O(a_1^2,a_2^2)$ with $f_3^\times$ and $f_4^\times$ given in terms of the remaining unknowns $\ms X^{1{\times}}_{32}$, $\ms X^{1{\times}}_{33}$, and $\ms X^{1{\times}}_{43}$ (and remaining unknowns from the SO level) by
\bse\label{f34S1S2}
\begin{alignat}{3}
\frac{f_3^\times}{\mu^2M^3}&=\frac{5}{2}+\Big(\frac{9}{2}+\frac{3}{8}\ms X_{32}^{1{\times}}\Big)\nu-\nu^2
\\\nnm
&\quad+\frac{3}{8}\bigg[4-15\nu-16\nu^2+4\nu^3+2\nu(1-2\nu)\ms X_{32}^{1\mr b}
\\\nnm
&\qquad\quad+4\nu^2\ms X_{32}^{1\mr t}-\frac{\nu^2}{2}\ms X_{32}^{1{\times}}+\nu \ms X_{33}^{1{\times}}\bigg]\ve+\mc O(\ve^2),
\end{alignat}
and
\begin{alignat}{3}
\frac{f_4^\times}{\mu^2M^4}&=2+\frac{187}{4}\nu-21\nu^2+\frac{13}{8}\nu^3
\\\nnm
&\quad+\frac{\nu}{4}(19+2\nu)\ms X_{32}^{1\mr b}+\frac{\nu}{2}(10-\nu)\ms X_{32}^{1\mr t}
\\\nnm
&\quad-\frac{3}{4}\nu(4+\nu)\ms X_{32}^{1{\times}}+\frac{3}{2}\nu\ms X_{33}^{1{\times}}+\frac{16}{15}\nu \ms X_{43}^{1{\times}}+\mc O(\ve).
\end{alignat}
\ese
We now have a complete expression of the mass-shell constraint (\ref{psqiso}) through N$^3$LO in the PN expansion and through bilinear order in spins, which could be solved for the corresponding canonical Hamiltonian. It depends on the remaining unknown (dimensionless, numerical) coefficients $\ms X^{1{\mr A}}_{32}$, $\ms X^{1{\mr A}}_{33}$, and $\ms X^{1{\mr A}}_{43}$ with $\mr A=\{\mr b,\mr t,{\times}\}$, from (\ref{X1s}).  Recall, for $\ms X^{1{\mr A}}_{kn}$, $k$ is the PM order, and $n$ is the relative PN order.

\subsection{The radial action}\label{sec:radialaction}

For a bound orbit ($\gamma^2-1=\gamma^2v^2=\ve<0$), the same canonical mass-shell constraint (\ref{psqiso}) governs the motion.  The (gauge-dependent) radial momentum function $p_r(r,E,L;m_\mr i,a_\mr i)$ is still given by
\be
p_r=\pm\sqrt{p_{\infty}^2-\frac{L^2}{r^2}+\sum_k\frac{G^k}{r^k}\bigg[f_k+\frac{L a_\mr i}{r^2}f_k^\mr i+\frac{a_\mr ia_\mr j}{r^2}f_k^{\mr i\mr j}\bigg]},
\ee
but now $p_\infty^2=(\mu/\Gamma)^2\ve$ is negative.  As a result, $p_r^2(r)$ has two positive real roots $r=r_\pm$ between which $p_r^2$ is positive, with $r_+$ being the largest real root, and the trajectory oscillates between these radial turning points $r_\pm$.  The \emph{canonical radial action} function $I_r(E,L,m_\mr i,a_\mr i)$ is defined as the integral of $p_r\mr dr$ over one period of the radial motion,
\begin{equation}\label{defIr}
2\pi I_r:=\oint \mr dr\,p_r=\int_{r_-}^{r_+}\mr dr \Big({+}\sqrt{p_r^2}\Big)+\int_{r_+}^{r_-}\mr dr \Big({-}\sqrt{p_r^2}\Big)=2\int_{r_-}^{r_+}\mr dr \sqrt{p_r^2},
\end{equation}
and it is a gauge-invariant function, from which one can derive several other gauge-invariant functions physically characterizing bound orbits~\cite{Kalin:2019inp,Kalin:2019rwq}.  Like the ``unbound radial action'' $\W$ (the $L$-antiderivative of the scattering-angle $\chi$) (\ref{defW}), the bound radial action $I_r(E,L,m_\mr i,a_\mr i)$ encodes the complete gauge-invariant information content of the canonical Hamiltonian (governing both unbound and bound orbits) (at least up to the N$^3$LO PN level) --- though in a subtly different way, concerning orders in the PM-PN expansion of $I_r$ versus that of $\W$.

It was shown in \cite{Kalin:2019inp} that the periastron-advance angle, 
$
\Phi=2\pi+\Delta\Phi=-2\pi\doe I_r/\doe L
$,
the angle swept out by a bound orbit during one period of the radial motion, is related to the scattering angle, $\pi+\chi=-2\pi \doe \W/\doe L$, by
\begin{equation}
\Phi(E,L,m_\mr i,a_\mr i)=2\pi+ \chi(E,L,m_\mr i,a_\mr i)+ \chi(E,-L,m_\mr i,-a_\mr i),
\end{equation}
where the right-hand side requires an analytic continuation from $E>M$ (unbound, for which $\chi$ is real) to $E<M$ (bound, for which $\chi$ is complex), as detailed below.
It follows from a straightforward extension of their argument that a particular $L$-antiderivative of this relation holds, giving the bound radial action $I_r$ in terms of the unbound radial action $\W$,
\begin{equation}\label{WtoIr}
I_r(E,L,m_\mr i,a_\mr i)=\W(E,L,m_\mr i,a_\mr i)-\W(E,-L,m_\mr i,-a_\mr i),
\end{equation}
as can also be verified by explicit calculation.

Consider the unbound radial action in the form (\ref{WPM}), after replacing $\tilde\chi_1$ using (\ref{tildechis}) and (\ref{f012}),
\be\label{Wto}
\W=-\frac{ L}{2}-GM\mu\frac{1+2\ve}{\sqrt{\ve}}\frac{\ln L}{\pi }+\frac{1}{2\pi }\sum_{k\ge 2}\frac{G^k}{p_\infty^k L^{k-1}}\frac{\tilde \chi_k}{k-1}.
\ee
In continuing this from the unbound case, $\ve>0$, $p_\infty^2>0$, to the bound case, $\ve<0$, $p_\infty^2<0$, the second term with $1/\sqrt{\ve}$ becomes imaginary, as do all of the terms in the sum with $k$ odd, having odd powers of $p_\infty=(\mu/\Gamma)\sqrt{\ve}$.  Note, from (\ref{tildechis}) and (\ref{fktotal}), and from the fact that all of the $f$'s have regular Taylor series in $\ve$ about $\ve=0$, that all of the $\tilde\chi_k$ are still real for the bound case, and that the $\tilde\chi_k$ are unchanged by $(L,a_\mr i)\to(-L,-a_\mr i)$.  Thus, plugging the continuation of (\ref{Wto}), with $\sqrt{\ve}=i\sqrt{-\ve}$, into (\ref{WtoIr}), we see that all of the odd-$k$ terms are canceled; after using $\ln L-\ln(-L)=\ln(-1)=-i\pi$ (choosing the branch which yields the physically sensible result), we are left with
\be\label{Irtildechi}
I_r=-L+GM\mu\frac{1+2\ve}{\sqrt{-\ve}}+\frac{1}{\pi }\sum_{l\ge 1}\frac{G^{2l}}{p_\infty^{2l} L^{2l-1}}\frac{\tilde \chi_{2l}}{2l-1},
\ee
which is real for bound orbits.  Only the $\tilde \chi_k$ with $k$ even ($k=2l$) remain, and those with $k$ odd are gone (except for $\tilde\chi_1$).  This may make it seem as though we have lost information in passing from $\W$ to $I_r$, but in fact we have not, as long as we are sure to keep all terms in the consistent PN expansion of $I_r$ (at least up to the N$^3$LO PN level); this is due to relationships between the $\tilde \chi_k$ as discussed below (\ref{chifs}).  

As we will make clearer below, the complete PN expansion of $I_r$ up to N$^3$LO is contained in its PM expansion up to $\mc O(G^6)$ for the nonspinning terms and up to $\mc O(G^8)$ for the spin-orbit and quadratic-in-spin terms.   This can be computed directly from (\ref{Irtildechi}), recalling that the $\tilde\chi_k$ are the entries of Table 1 of \cite{Bjerrum-Bohr:2019kec} with $f_k\to\tilde f_k$, as in (\ref{chifs}) above, with the $\tilde f_k$ given by (\ref{fktotal}).  We need again the contributions from $f_k$, $f_k^\mr i$, $f_k^{\mr i\mr j}$ up to $k=4$, contained in the $\tilde f_k=f_k+f_{k-2}^\mr iLa_\mr i/G^2+f_{k-2}^{\mr i\mr j}a_\mr ia_\mr j/G^2$ up to $k=6$.  To reach the all the $G^8$ quadratic-in-spin terms, we must take the sum in (\ref{Irtildechi}) up to $l=6$, involving parts of $\tilde\chi_{12}$.

This process yields the radial action $I_r$ through the N$^3$LO PN level as an expansion in the inverse canonical orbital angular momentum $L\equiv L_\mr{can}$.  To express the results of that process, it will be advantageous to use the covariant orbital angular momentum $L_\mr{cov}$, which we define for the bound-orbit case by 
\be\label{LcovL}
L_\mr{cov}:=L-\Delta L,
\ee
with $\Delta L(E,a_\mr i)$ still given by the last two lines of (\ref{DeltaL}) or by (\ref{dLxi}), in which we note that everything is still real for bound orbits [unlike in the second line of (\ref{DeltaL}), where we would need to continue to imaginary $b$ to keep $L_\mr{cov}=p_\infty b$ real].  

In fact, the expression of the radial action (mostly) in terms of $L_\mr{cov}$ is simply related to the expression of the scattering angle in terms of $L_\mr{cov}$, as follows.  Taking the form (\ref{masterchi2}) for the scattering angle and eliminating $b$ in favor of $L_\mr{cov}=(\mu/\Gamma)\sqrt{\ve}b$,
\begin{alignat}{3}\label{chiLcov}
\chi&=\Gamma\sum_{k\ge1}\Big(\frac{GM}{b\sqrt{\ve}}\Big)^k\bigg[\ms X_k+\frac{a_\mr i}{b\sqrt{\ve}}\ms X_k{}^{\mr i}+\frac{a_\mr i a_\mr j}{b^2{\ve}}\ms X_k{}^{\mr i\mr j}\bigg]
\\\nnm
&=\Gamma\sum_{k\ge1}\Big(\frac{GM\mu}{\Gamma L_\mr{cov}}\Big)^k\bigg[\ms X_k+\frac{\mu a_\mr i}{\Gamma L_\mr{cov}}\ms X_k{}^{\mr i}+\frac{\mu^2a_\mr i a_\mr j}{\Gamma^2L_\mr{cov}^2}\ms X_k{}^{\mr i\mr j}\bigg],
\end{alignat}
and then using (\ref{WdinvL}), being sure to match up the constant of integration with (\ref{Wto}), we find
\begin{alignat}{3}
\W&=-\frac{ L}{2}-GM\mu\,\ms X_1\frac{\ln L_\mr{cov}}{2\pi }+\frac{1}{2\pi}\sum_{k\ge2}\frac{(GM\mu)^k}{(\Gamma L_\mr{cov})^{k-1}}\frac{\ms X_k}{k-1}
\nnm\\
&\quad+\frac{1}{2\pi}\sum_{k\ge1}\Big(\frac{GM\mu}{\Gamma L_\mr{cov}}\Big)^k\bigg[\mu a_\mr i\frac{\ms X_k{}^{\mr i}}{k}+\frac{\mu^2a_\mr i a_\mr j}{\Gamma L_\mr{cov}}\frac{\ms X_k{}^{\mr i\mr j}}{k+1}\bigg].
\end{alignat}
Then applying (\ref{WtoIr}), as we did between (\ref{Wto}) and (\ref{Irtildechi}), noting $L_\mr{cov}\to-L_\mr{cov}$ under $(L,a_\mr i)\to(-L,-a_\mr i)$, we are left with
\begin{alignat}{3}\label{Ir}
I_r&=-L+GM\mu\frac{1+2\ve}{\sqrt{-\ve}}+\frac{1}{\pi}\sum_{l\ge1}\frac{(GM\mu)^{2l}}{(\Gamma{L_\mr{cov}})^{2l-1}}\bigg[\frac{\ms X_{2l}}{2l-1}
\nnm\\
&\qquad+\frac{\mu a_\mr i}{\Gamma L_\mr{cov}}\frac{\ms X_{2l}{}^\mr i}{2l}+\frac{\mu^2 a_\mr ia_\mr j}{(\Gamma L_\mr{cov})^2}\frac{\ms X_{2l}{}^{\mr i\mr j}}{2l+1}\bigg],
\end{alignat}
where these $\ms X_k{}^{\cdots}(\ve,\nu)$ are precisely the same coefficients from the scattering angle in (\ref{chiLcov}).  These coefficients up through $k=2l=4$ are those we gave or parametrized above in (\ref{X0s}) and (\ref{X1s}), with (\ref{joinXs}).  Recollecting them here, while using the constraints (\ref{lowPNchiSO}) and (\ref{lowPNchiS1S2}) obtained in matching between the scattering angle and the canonical mass shell, we have the $G^2$ coefficients which are independent of $\nu$ and are known exactly,
\bse\label{X_2l}
\begin{alignat}{3}
\ms X_{2}&=\frac{3\pi}{4}(4+5\ve),
\\\nnm
\ms X_{2}{}^{\mr i}a_\mr i&=-\frac{\pi}{2}\gamma(2+5\ve)(4a_\mr b+3a_\mr t),
\\\nnm
\ms X_{2}{}^{\times}&=\frac{3\pi}{2}(2+19\ve+20\ve^2),
\end{alignat}
and the $G^4$ coefficients which are linear in $\nu$,
\begin{alignat}{3}
\ms X_{4}&=\frac{105\pi}{64}(16+48\ve+33\ve^2)
\\\nnm
&\quad+\pi \bigg[{-}\frac{15}{2}+\bigg(\frac{123}{128}\pi^2-\frac{557}{8}\bigg)\ve+\mc O(\ve^2)\bigg]\nu,
\\\nnm
\ms X_{4}{}^{\mr i}a_\mr i&=-\frac{21\pi}{16}\gamma(8+36\ve+33\ve^2)(8a_\mr b+5\mr a_t)
\\\nnm
&\quad+\pi\gamma\bigg[\frac{21}{2}a_\mr b+9a_\mr t+\frac{3}{4}\Big(68a_\mr b+49 a_\mr t+2 \ms X^{1\mr i}_{32}a_\mr i\Big)\ve
\\\nnm
&\qquad\qquad+\ms X^{1\mr i}_{43}a_\mr i\ve^2+\mc O(\ve^3)\bigg]\nu,
\\\nnm
\ms X_{4}{}^{\times}&=\frac{105\pi}{16} (24+212\ve+447\ve^2+264\ve^3)
\\\nnm
&\quad+\pi\bigg[\frac{15}{2}+\frac{45}{32}\Big({-}22+\ms X^{1{\times}}_{32}\Big)\ve+\ms X^{1\times}_{43}\ve^2+\mc O(\ve^3)\bigg]\nu.
\end{alignat}
As mentioned above, for the complete expression of the radial action at the N$^3$LO PN level, we need the low orders in the PN expansions of $\ms X_6{}^{\cdots}$ and (for the spin terms) $\ms X_8{}^{\cdots}$.  We have obtained these from the procedure to compute the radial action described in the paragraph containing (\ref{Irtildechi}) and the following paragraph, in which the inputs are the $f_k^{\cdots}$ up to $k=4$ found in the previous subsection, finally changing variables using (\ref{LcovL}) to bring the result into the form (\ref{chiLcov}).  At $G^6$, we find the nonspinning
\begin{alignat}{3}\label{X6a0}
\frac{\ms X_6}{5\pi}&=\frac{231}{4}+\Big(\frac{123}{128}\pi^2-\frac{125}{2}\Big)\nu+\frac{21}{8}\nu^2+\mc O(\ve),
\end{alignat}
spin-orbit,
\begin{alignat}{3}\label{X6a1}
\frac{\ms X_6{}^\mr ia_\mr i}{{15\pi}}&=\Big({-}99+\frac{127}{4}\nu-\frac{5}{4}\nu^2\Big)a_\mr b
\\\nnm
&\qquad+\Big({-}\frac{231}{4}+\frac{167}{8}\nu-\frac{9}{8}\nu^2\Big)a_\mr t
+\frac{1}{4} \nu\ms X_{32}^{1\mr i}a_\mr i
\\\nnm
&\quad+\bigg[\Big({-}693+\frac{4989}{16}\nu-\frac{123}{32}\pi^2\nu-\frac{225}{16}\nu^2\Big)a_\mr b
\\\nnm
&\qquad+\Big({-}\frac{1617}{4}+\frac{733}{4}\nu-\frac{123}{64}\pi^2\nu-\frac{182}{16}\nu^2\Big)a_\mr t
\\\nnm
&\qquad+\nu\Big(\frac{7-3\nu}{8}\ms X_{32}^{1\mr i}-\frac{5}{4}\ms X_{33}^{1\mr i}+\ms X_{43}^{1\mr i}\Big)a_\mr i\bigg]\ve+\mc O(\ve^2),
\end{alignat}
and bilinear-in-spin, 
\begin{alignat}{3}\label{X6a2}
\frac{\ms X_6{}^\times}{35\pi}&=\frac{495}{4}-\frac{123\nu}{16}-\frac{9}{8}\nu^2+\frac{3}{32}\nu\ms X_{32}^{1{\times}}
\\\nnm
&\quad+\bigg[\frac{10197}{8}-\frac{4835}{32}\nu+\frac{123}{128}\pi^2\nu-\frac{399}{32}\nu^2
\\\nnm
&\qquad-\frac{3}{8}\nu(1+2\nu)\ms X_{32}^{1\mr b}-\frac{3}{4}\nu(1-\nu)\ms X_{32}^{1\mr t}
\\\nnm
&\quad+\frac{9}{64}\nu(2-\nu)\ms X_{32}^{1{\times}}-\frac{15}{32}\nu\ms X_{33}^{1{\times}}+\frac{2}{5}\nu\ms X_{43}^{1{\times}}\bigg]\ve+\mc O(\ve^2).
\end{alignat}
At $G^8$, spin-orbit,
\begin{alignat}{3}
\frac{\ms X_8{}^{\mr i}a_\mr i}{35\pi}&=\Big({-}715+\frac{23947}{48}\nu-\frac{41}{8}\pi^2\nu-\frac{97}{2}\nu^2+\frac{13}{16}\nu^3\Big)a_\mr b
\nnm\\\nnm
&\quad+\Big({-}\frac{6435}{16}+\frac{6883}{24}\nu-\frac{41}{16}\pi^2\nu-\frac{277}{8}\nu^2+\frac{3}{4}\nu^3\Big)a_\mr t
\\
&\quad+\nu\Big(\frac{2-\nu}{2}\ms X_{32}^{1\mr i}- \ms X_{33}^{1\mr i}+\frac{2}{3}\ms X_{43}^{1\mr i}\Big)a_\mr i+\mc O(\ve),
\end{alignat}
and bilinear-in-spin
\begin{alignat}{3}
\frac{\ms X_8{}^\times}{315\pi}&=\frac{5005}{16}-\frac{6599}{96}\nu+\frac{41}{128}\pi^2\nu-\frac{199}{32}\nu^2+\frac{5}{16}\nu^3
\nnm\\\nnm
&\quad-\frac{1}{8}\nu(1+2\nu)\ms X_{32}^{1\mr b}-\frac{1}{4}\nu(1-\nu)\ms X_{32}^{1\mr t}
\\
&\quad+\frac{3}{64}\nu(2-\nu)\ms X_{32}^{1{\times}}-\frac{3}{32}\nu\ms X_{33}^{1{\times}}+\frac{1}{15}\nu\ms X_{43}^{1{\times}}+\mc O(\ve).
\end{alignat}
\ese

Note that the $\ms X_6{}^{\cdots}$ coefficients in (\ref{X6a1}) and (\ref{X6a2}) are exactly quadratic in $\nu$, in spite of the fact that the $f$'s from which they are constructed, in (\ref{f34SO}) and (\ref{f34S1S2}), are cubic in $\nu$.  Less surprisingly, the $\ms X_6{}^{\cdots}$ are cubic in $\nu$, and more surprisingly the $\ms X_4{}^{\cdots}$ are linear in $\nu$ and the $\ms X_2{}^{\cdots}$ are independent of $\nu$.  This is all in fact a simple consequence of (i) the link (\ref{Ir}) between the scattering-angle coefficients $\ms X_k{}^{\cdots}$ and the radial-action coefficients, and (ii) the (straightforward) extension of the predicted mass-ratio dependence (\ref{joinXs}) to $k$PM order:  $\ms X_k{}^{\cdots}$ is a polynomial of degree $\lfloor{\frac{k-1}{2}}\rfloor$ in $\nu$.  This is the spinning analog of the ``hidden simplicity'' of the mass dependence of (the local-in-time part of) the radial action (which is the complete radial action through the N$^3$LO PN level) emphasized in Ref.~\cite{Bini:2020wpo}; here in the spin terms, this is crucially dependent on expressing $I_r$ in (\ref{Ir}) in terms the covariant $L_\mr{cov}$ rather than the canonical $L$.

Finally, we can make the PN order counting explicit by restoring factors of $1/c$.  Through N$^3$LO, (\ref{Ir}) reads
\begin{alignat}{3}\label{Irexp}
I_r&=\Bigg[{-}L+GM\mu\frac{1+2\ve}{c\sqrt{-\ve}}
+\frac{1}{c^2}\frac{(GM\mu)^2}{\pi\Gamma{L_\mr{cov}}}\ms X_2
\\\nnm
&\quad
+\frac{1}{c^4}\frac{(GM\mu)^4}{3\pi(\Gamma L_\mr{cov})^3}\ms X_4+\frac{1}{c^6}\frac{(GM\mu)^6}{5\pi(\Gamma L_\mr{cov})^5}\ms X_6+\mc O(\frac{1}{c^8})\Bigg]
\\\nnm
&+\frac{\mu}{c}  a_\mr i\Bigg[\frac{(GM\mu)^2}{2\pi(\Gamma L_\mr{cov})^2}\ms X_{2}{}^\mr i
+\frac{1}{c^2}\frac{(GM\mu)^4}{4\pi(\Gamma L_\mr{cov})^4}\ms X_{4}{}^\mr i
\\\nnm
&\quad
+\frac{1}{c^4}\frac{(GM\mu)^6}{6\pi(\Gamma L_\mr{cov})^6}\ms X_{6}{}^\mr i
+\frac{1}{c^6}\frac{(GM\mu)^8}{8\pi(\Gamma L_\mr{cov})^8}\ms X_{8}{}^\mr i+\mc O(\frac{1}{c^8})\Bigg]
\\\nnm
&+\mu^2  {a_\mr ia_\mr j}\Bigg[
\frac{(GM\mu)^2}{3\pi(\Gamma L_\mr{cov})^3}\ms X_2{}^{\mr i \mr j}
+\frac{1}{c^2}\frac{(GM\mu)^4}{5\pi(\Gamma L_\mr{cov})^5}\ms X_{4}{}^{\mr i \mr j}
\\\nnm
&\quad
+\frac{1}{c^4}\frac{(GM\mu)^6}{7\pi(\Gamma L_\mr{cov})^7}\ms X_{6}{}^{\mr i \mr j}
+\frac{1}{c^6}\frac{(GM\mu)^8}{9\pi(\Gamma L_\mr{cov})^9}\ms X_{8}{}^{\mr i \mr j}+\mc O(\frac{1}{c^8})\Bigg],
\end{alignat}
with all the coefficients, to the orders in $\ve=\gamma^2-1=\mc O(c^{-2})$ contributing here at N$^3$LO, relative $\mc O(c^{-6})$, given explicitly by (\ref{X_2l}). These depend on the remaining unknowns $\ms X_{kn}^{1\mr A}$ from the parametrization of the scattering angle, at $k$PM order and relative $n$PN order. 

In all the above manipulations, it was consistent to keep the nonspinning, spin-orbit, and bilinear-in-spin terms all through the same relative PN orders, here relative 3PN order, N$^3$LO.  However, in matching to self-force results, due to certain changes of variables discussed below, the treatment of the N$^3$LO spin-orbit and bilinear-in-spin terms will require the inclusion of the 4PN nonspinning terms.  We thus need to add to (\ref{Irexp}) the 4PN nonspinning part of the radial action for bound orbits, which includes contributions from the nonlocal-in-time tail integrals.  We present in Appendix \ref{app:4PN} the additional terms at 4PN order, which have been computed from (\ref{defIr}) applied to the 4PN EOB Hamiltonian derived in \cite{Damour:2015isa}, valid in an expansion in eccentricity (about the circular orbit limit) to sixth order.  Replacing the first two lines of (\ref{Irexp}) with (\ref{Ir4PN}) yields the final form of the radial-action function which we will use to compute the gauge-invariant quantities to be compared with self-force calculations.

\section{Third-subleading post-Newtonian spin-orbit and spin$_1$-spin$_2$ couplings}
\label{sec:N3LOSO}

The remaining unknowns in the parametrization of the scattering-angle function~(\ref{X1s}) can be fixed with available self-force results. The key feature here is the existence of a Hamiltonian/radial action allowing us to connect the scattering-angle to the redshift and spin-precession invariants that, in the small-mass-ratio limit, can be matched to expressions independently calculated in GSF literature. A vital step in this calculation is the first law of BBH mechanics, which we extend to aligned-spins and eccentric orbits.

\subsection{The first law of BBH mechanics}\label{firstlaw}

The first law of BBH mechanics~\cite{LeTiec:2011ab} was first derived for nonspinning point particles in circular orbits in Ref.~\cite{LeTiec:2011ab}, then generalized to spinning particles on circular orbits in Ref.~\cite{Blanchet:2012at}, to nonspinning particles in eccentric orbits in Refs.~\cite{Tiec:2015cxa,Blanchet:2017rcn}, and to precessing eccentric orbits of a point mass in the small mass-ratio approximation~\cite{Fujita:2016igj}.
In the following, we briefly review the arguments leading to these incarnations of the first law for binaries, making explicit how they apply to generic mass-ratio aligned-spin systems on eccentric orbits.

Let us follow Ref.~\cite{Blanchet:2012at} and start out with an action $\mathcal{S}$ for the binary,
\begin{equation}
\label{fullaction}
\mathcal{S} = \mathcal{S}_\text{grav} + \mathcal{S}_1 + \mathcal{S}_2 \,,
\end{equation}
where the compact objects are approximated by effective point-particles moving along worldlines $x_{\mr i}^\mu(\tau_{\mr i})$,
\begin{equation}
\label{ppaction}
\mathcal{S}_{\mr i} = \int \mr d \tau_{\mr i} \! \left[ - m_{\mr i} + \frac{1}{2} S_{\mr i\mu\nu} \Lambda_{\mr i c}{}^\mu \frac{D \Lambda_{\mr i}^{c\nu}}{d \tau} + \lambda_{\mr i}^\mu S_{\mr i\mu\nu} \dot{x}_{\mr i}^{\nu} + \dots \right] \!,
\end{equation}
and the gravitational action $\mathcal{S}_\text{grav}$ is given by the Einstein-Hilbert one with appropriate gauge-fixing and boundary terms.
Here $\Lambda_{\mr i}^{c\mu}$ are frame transformations between the coordinate frame and a body-fixed frame (labeled by $c=0,1,2,3$) that is Lorentz-orthonormal ($\Lambda_{\mr i c}{}^{\mu} \Lambda_{\mr i d \mu} = \eta_{cd}$).
We take $\tau_{\mr i}$ to be the (full-metric) proper times from now on.
The equations of motion are obtained by varying the action with respect to the dynamical variables $X_A = \{ x_{\mr i}, S_{\mu\nu}, \Lambda_{\mr i}^{c \mu}, \lambda_{\mr i}^\mu, g_{\mu\nu} \}$, leading to Eqs.~\eqref{MPD}--\eqref{Einstein}, see, e.g., Refs.~\cite{Steinhoff:2014kwa,Vines:2016unv}.
The dots in Eq.~\eqref{ppaction} represent nonminimal (curvature) couplings to the worldline that may carry undetermined coefficients.
These terms also include couplings of quadratic and higher orders in spin related to spin-induced multipole moments of the body~\cite{Steinhoff:2014kwa}.

Let us write the action as an integral of a Lagrangian $L$ over coordinate time $t$ as
\begin{equation}
\label{defL}
\mathcal{S} = \int \mr dt \, L \,.
\end{equation}
We can vary the Lagrangian $L$ not only with respect to the dynamical variables $X_A$, but also vary certain constants appearing in the action, e.g., the masses $C_B = \{ m_1, m_2 \}$.
Furthermore, taking the dynamical variables $X_A$ on-shell (fulfilling their equations of motion) after variation, we arrive at (using summation convention for $A$, $B$)
\begin{equation}
\delta L = \frac{\partial L}{\partial C_B} \delta C_B + \underbrace{\frac{\delta L}{\delta X_A}}_{=0 \text{ (on-shell)}} \delta X_A+ \text{(td)} \,,
\end{equation}
with a total time derivative (td).
Now, if one performs a transformation of the dynamical variables $X_A \rightarrow X'_{A'}$, which may depend on the $C_B$, then on-shell it holds
\begin{equation}
\delta L = \frac{\partial L}{\partial C_B} \delta C_B + \bigg[ \underbrace{\frac{\delta L}{\delta X_A}}_{0} \frac{\delta X_A}{\delta X'_{A'}} \frac{\partial X'_{A'}}{\partial C_B} + \text{(td)} \bigg] \delta C_B + \underbrace{\frac{\delta L}{\delta X_A}}_{0} \frac{\delta X_A}{\delta X'_{A'}} \delta X'_{A'} + \text{(td)} \,.
\end{equation}
Also allowing for changes of the Lagrangian of the form $L = L' + \text{(td)}$, we arrive at
\begin{equation}
\label{firstlawmaster}
\left \langle \left( \frac{\partial L'}{\partial C_B} \right)_{X'_{A'}} \right \rangle = \left \langle \left( \frac{\partial L}{\partial C_B} \right)_{X_{A}} \right \rangle \quad \text{(on-shell)}\,,
\end{equation}
where the subscripts indicate quantities that are kept fixed during differentiation and with $\langle \dots \rangle$ an appropriate on-shell averaging that removes the total time derivatives.

For generic bound orbits, one can average the conservative motion in Eq.~\eqref{firstlawmaster} over an infinite time in order to remove total time derivatives, which can be traded for a phase-space average in regions where the motion is ergodic; see, e.g., Refs.~\cite{Hinderer:2008dm,Fujita:2016igj}.
For the aligned-spin case where the motion is confined to a plane, all oscillatory behavior can be removed by an average over a single orbit~\cite{Tiec:2015cxa} (defined as an oscillation cycle of the radial distance $r$); this is the averaging used in the present paper.
Further specializing to circular orbits, the radial distance is constant and hence the average becomes trivial~\cite{Blanchet:2012at}.
Finally, note that another benefit of the averaging in Eq.~\eqref{firstlawmaster} is that it helps to make expressions manifestly gauge invariant~\cite{Fujita:2016igj}, which is important when matching PN Hamiltonians to (eccentric-orbit) self-force results.

It is straightforward to generalize the discussion from Lagrangians $L'$ to Hamiltonians $H'$.
Hamilton's dynamical equations for some pairs of canonical variables $(q^{\ms c}, p_{\ms c})$ are equivalently encoded by Hamilton's action principle,
\begin{equation}
\label{HamP}
0=\delta \mathcal{S} = \delta \int \mr dt \bigg[ \underbrace{\sum_{\ms c} p_{\ms c} \frac{d q^{\ms c}}{dt} - H'}_{L'} \bigg] \,.
\end{equation}
Noting that the dynamical variables are now $X'_{A'} = \{q^{\ms c}, p_{\ms c}\} $, and that the kinematic $p \dot{q}$-terms in $L'$ are independent of the $C_B$, we see that either Lagrangian in Eq.~\eqref{firstlawmaster} can be replaced by \emph{minus} a Hamiltonian (i.e., it can be applied also to canonical transformations between two Hamiltonians).
The rather general on-shell relation~\eqref{firstlawmaster} is interesting on its own, aside from facilitating the derivation of the first law of binary dynamics as demonstrated below.

We are now in a position to elaborate on the redshift variables $z_{\mr i}$~\cite{LeTiec:2011ab, Blanchet:2012at, Tiec:2015cxa},
\begin{equation}
\label{redshiftdef}
z_{\mr i} \equiv \left \langle \frac{\mr d \tau_{\mr i}}{\mr d t} \right \rangle = - \left \langle \frac{\partial L}{\partial m_{\mr i}} \right \rangle \,,
\end{equation}
where the first equality is the definition of $z_{\mr i}$ adopted by us and the second equality is a consequence of the definition of $L$~\eqref{defL} together with the original point-particle action~\eqref{ppaction}, $\int \mr dt \, L \sim - m_{\mr i} \int \mr dt \, \mr d\tau_{\mr i} / \mr d t$.
We note that this relation holds to all orders in spin if the coefficients in the nonminimal couplings (the dots) in Eq.~\eqref{ppaction} are normalized such that no further explicit dependence on the masses $m_{\mr i}$ arises~\cite{Bini:2020zqy}.
Now, several nontrivial transformations of the original action~\eqref{fullaction} are performed to arrive at a PN Hamiltonian (see, e.g., Refs.~\cite{Levi:2015msa,Blanchet:2012at,Blanchet:2017rcn}):
a transformation to SO(3)-canonical (Newton-Wigner) variables for the spin degrees of freedom, integrating out the orbital/near-zone metric or tetrad field (calculating the ``Fokker action''), reduction of higher-order time derivatives via further variable transformations, a Legendre transform to the Hamiltonian $H$, specialization to the cm system, and eventually reducing nonlocal-in-time tail contributions to local ones.
However, all of these transformations fall into the class of transformations $(X_A, L) \rightarrow (X'_{A'}, L')$ discussed above, so we may apply Eq.~\eqref{firstlawmaster} (with $L' \rightarrow - H$) to Eq.~\eqref{redshiftdef} and conclude that the redshift variables $z_{\mr i}$ can be obtained from a PN Hamiltonian $H$ via
\begin{equation}
\label{zham}
z_{\mr i} = \left \langle \frac{\partial H}{\partial m_{\mr i}} \right \rangle \,.
\end{equation}

Beside the redshift, let us introduce the (averaged) spin precession frequency $\Omega_{\mr i}$ as another important observable~\cite{Blanchet:2012at},
\begin{equation}
\label{OmegaSdef}
\Omega_{\mr i} \equiv \left \langle \left| \vec{\Omega}_{\mr i}^\text{inst} \right| \right \rangle \,.
\end{equation}
The (instantaneous, directed) precession frequency $\vec{\Omega}_{S_{\mr i}}^\text{inst}$ can be read off from the equations of motion for the SO(3)-canonical spin vectors $S_{\mr i}^i$ generated by the Hamiltonian $H$,
\begin{equation}
\label{OmegaSinsetdef}
\frac{\mr d \vec{S}_{\mr i}}{\mr d t} = \vec{\Omega}_{\mr i}^\text{inst} \times \vec{S}_{\mr i} \,, \qquad
\vec{\Omega}_{\mr i}^\text{inst} \equiv \frac{\partial H}{\partial \vec{S}_{\mr i}} \,.
\end{equation}
Indeed, this describes a precession of the spin vector;
it is straightforward to see that the spin length $S_{\mr i} \equiv (\vec{S}_{\mr i} \cdot \vec{S}_{\mr i})^{1/2}$ is constant,
\begin{equation}
\frac{\mr d ( \vec{S}_{\mr i} \cdot \vec{S}_{\mr i} )}{\mr d t} = 2 \vec{S}_{\mr i} \cdot \vec{\Omega}_{\mr i}^\text{inst} \times \vec{S}_{\mr i} = 0 \,.
\end{equation}

From now on, as in previous sections, we simplify the discussion to nonprecessing (aligned or anti-aligned) spins, so that $\vec{\Omega}_{\mr i}^\text{inst} \parallel \vec{S}_{\mr i}$ and $d \vec{S}_{\mr i} / dt = 0$.
That is, the spin degrees of freedom become nondynamical and can be dropped from the set of dynamical variables.\footnote{More precisely, their contribution to the kinematic terms in Hamilton's principle~\eqref{HamP} (have to) vanish or turn into total time derivatives.}
We can now include the spin lengths into our set of constants, $C_B = \{ m_{\mr i}, S_{\mr i} \}$.
Furthermore, the spin-direction component of the defining relation for $\vec{\Omega}_{\mr i}^\text{inst}$~\eqref{OmegaSinsetdef} reads $| \vec{\Omega}_{\mr i}^\text{inst} | = \partial H / \partial S_{\mr i}$.
Hence Eq.~\eqref{OmegaSdef} becomes
\begin{equation}
\label{precessham}
\Omega_{S_{\mr i}} = \left \langle \frac{\partial H}{\partial S_{\mr i}} \right \rangle \qquad \text{(nonprecessing)} .
\end{equation}
We have now arrived at the important Eqs.~\eqref{zham} and~\eqref{precessham} for the (gauge-invariant) observables $z_{\mr i}$ and $\Omega_{\mr i}$, that could be used to relate a PN Hamiltonian $H$ to self-force results~\cite{Bini:2019lcd,Bini:2019lkm}.
But here, for the purpose of matching to self-force, we perform a canonical transformation to different phase-space variables that simplify explicit calculations and connects to the radial action introduced above.

As a first step in that direction, we choose the (nonprecessing) motion to be in the equatorial plane $\theta = \pi /2$, removing the polar angle $\theta$ and its canonical conjugate momentum $p_\theta$ from the phase space;
the Hamiltonian is now of the form discussed in Sec.~\ref{sec:massshell}.
Furthermore, since we consider a system where the Hamilton-Jacobi equation is separable, one can construct a special canonical transformation (for bound orbits) where the \emph{constant} action variables
\begin{align}
I_r &= \frac{1}{2\pi}\oint \mr dr \, p_r , &
I_\phi &= \frac{1}{2\pi}\oint \mr d\phi \, p_\phi = L \,,
\end{align}
are the new momenta~\cite{Goldstein:2000}, with the cm orbital angular momentum of the binary $p_\phi \equiv L = \text{const}$ conjugate to the azimuthal angle $\phi$.
The advantage of these variables for our purpose is that the averaging $\langle \dots \rangle$ over one radial period becomes trivial due to the integral over one radial period $\oint$ in their definition.
The canonical conjugates to $I_r$, $I_\phi$ are the so-called angle variables $q_r$, $q_\phi$ and evolve linear in time, i.e., their angular frequencies $\Omega_r = \dot{q}_r$, $\Omega_\phi = \dot{q}_\phi$ are constant~\cite{Goldstein:2000}; overall Hamilton's equations of motion for the new, canonically transformed, Hamiltonian $H'(I_r, I_\phi = L; C_B)$ read
\begin{align}
\label{hamactionangle}
\Omega_r &= \frac{\partial H'}{\partial I_r} = \text{const} \,, &
\Omega_\phi &= \frac{\partial H'}{\partial L} = \text{const} \,, \\
\dot{I}_r &= -\frac{\partial H'}{\partial q_r} = 0 \,, &
\dot{L} &= -\frac{\partial H'}{\partial q_\phi} = 0 \,.
\end{align}
Recalling that $C_B = \{ m_{\mr i}, S_{\mr i} \}$, we can apply Eq.~\eqref{firstlawmaster} (with both Lagrangians replaced by Hamiltonians) for the canonical transformation to action-angle variables as well.
Equations~\eqref{zham} and~\eqref{precessham} then turn into
\begin{equation}
\label{observables}
z_{\mr i} = \frac{\partial H'}{\partial m_{\mr i}} \,, \qquad
\Omega_{\mr i} = \frac{\partial H'}{\partial S_{\mr i}} \,,
\end{equation}
where the averaging over one radial period is inconsequential and can be dropped.
Collecting Eqs.~\eqref{hamactionangle} and~\eqref{observables}, we see that the differential of the cm energy $E \equiv H'$ can be written as
\begin{equation}\label{1law}
\mr d E = \Omega_r \mr d I_r + \Omega_\phi \mr d L + \sum_{\mr i} ( z_{\mr i} \mr d m_{\mr i} + \Omega_{\mr i} \mr d S_{\mr i} ) .
\end{equation}
In analogy to the first law of thermodynamics for the differential of the internal energy, this can be called the first law of conservative spinning binary dynamics for nonprecessing bound orbits (covering eccentric orbits and generic mass ratios).
It also resembles the first law of BH thermodynamics, which provides a relation for the differential of the Arnowitt-Deser-Misner (ADM) energy $\mr d m_{\mr i}$ of an isolated BH and can be generalized to other compact objects as well~\cite{Carter:2010}.
Recall that Eq.~\eqref{1law} is valid to all orders in spin, if the coefficients of possible nonminimal coupling terms denoted by dots in Eq.~\eqref{ppaction} are normalized such that no additional dependence on $m_{\mr i}$ arises.
It would be interesting to consider these coefficients as part of the constants $C_B$ in future work.

Since the fundamental function introduced in the last section that generates observables for bound orbits is the radial action $I_r(E, L; m_{\mr i}, S_{\mr i})$, we consider the first law~\eqref{1law} in the form
\begin{equation}
2\pi \, \mr dI_r = T_r \mr dE - \Phi \mr dL - \sum_{\mr i} ( \mathcal{T}_{\mr i} \mr d m_{\mr i} + \Phi_{\mr i} \mr d S_{\mr i} ) \,,
\end{equation}
where we have introduced
\begin{align}
T_r&= \frac{2\pi}{\Omega_r} =\oint \mr dt\,, &
\Phi&=\Omega_\phi T_r=\oint \mr d \phi \,, \\
\mathcal{T}_{\mr i}&=z_{\mr i} T_r=\oint \mr d \tau_{\mr i} \,, &
\Phi_{\mr i} &= \Omega_{\mr i} T_r \,.
\end{align}
As a consequence of the first law, we hence obtain
\begin{align}\label{periods}
&\frac{T_r}{2\pi}=\bigg(\frac{\partial I_r}{\partial E}\bigg)_{L,m_{\mr i},S_{\mr i}},\\
&\frac{\Phi}{2\pi}=-\bigg(\frac{\partial I_r}{\partial L}\bigg)_{E,m_{\mr i},S_{\mr i}},\\
&\frac{\mathcal{T}_{\mr i}}{2\pi}=-\bigg(\frac{\partial I_r}{\partial m_{\mr i}}\bigg)_{E,L,m_{\mr j},S_{\mr i}},\\
&\frac{\Phi_{\mr i}}{2\pi}=-\bigg(\frac{\partial I_r}{\partial S_{\mr i}}\bigg)_{E,L,m_{\mr i},S_{\mr j}}.
\end{align}
Now the redshift variables can be calculated, from a given radial action $I_r$, as the ratio of proper and coordinate times,
\begin{equation}\label{redgen}
z_{\mr i}= \frac{\mathcal{T}_{\mr i}}{T_r}\,,
\end{equation}
which manifestly agrees with the (inverse of the) Detweiler-Barack-Sago redshift invariant calculated in GSF literature~\cite{Detweiler:2008ft,Barack:2011ed}.
The spin-precession frequency $\Omega_{\mr i}$ is given by $\Omega_{\mr i}= \Phi_{\mr i} / T_r$ from which we obtain the spin-precession invariant~\cite{Dolan:2013roa}
\begin{equation}\label{spininv}
\psi_{\mr i} = \frac{\Omega_{\mr i}}{\Omega_\phi} = \frac{\Phi_{\mr i}}{\Phi} \,.
\end{equation}

\subsection{Comparison with self-force results}
\label{sec:smallq}

Starting from the radial action~\eqref{Ir}, we calculate the redshift $z_1$ and spin-precession invariants $\psi_1$ of the small body  using Eqs.~\eqref{redgen} and~\eqref{spininv}. To compare with results available in the literature, we express them in terms of the gauge-invariant variables
\footnote{
	Note that the denominator for $\iota$ in Eq.~\eqref{xiota} is of 1PN order, which effectively scales down the PN ordering in such a way that manifestly nonlocal-in-time (4PN nonspinning) terms appear in the N$^3$LO correction to the spin-precession invariant. For this reason, we have included the 4PN nonspinning tail terms in the radial action as discussed at the end of the previous section.
}
\begin{equation}\label{xiota}
x = (G M \Omega_\phi)^{2/3},   \quad  \iota = \frac{3 x}{\Phi/(2\pi)-1}\,.
\end{equation}
which are linked to $(\ve,L)$ via Eqs.~\eqref{hamactionangle} and~\eqref{periods}. 
The expressions we obtain for $z_1(x,\iota)$ and $\psi (x,\iota)$ agree up to N$^2$LO with those in Eq.~(50) of Ref.~\cite{Bini:2019lcd} and Eq.~(83) of Ref.~\cite{Bini:2019lkm}. The full expressions up to N$^3$LO are lengthy, which is why we provide them as a \texttt{Mathematica} file in the Supplemental Material.

Next, we expand $U_1\equiv z_1^{-1}$ and $\psi_1$ to first order in the mass ratio $q$, first order in the massive body's  spin $ a_2$, and zeroth order in the spin of the smaller companion $a_1$,
\bse
\begin{align}
&U_1=  U^\text{(0)}_{1a^0} +\hat{a}\, U^\text{(0)}_{1a}+q\left(\delta U^\text{GSF}_{1a^0} +\hat{a}\, \delta U^\text{GSF}_{1a}\right)+\mathcal{O}(q^2,\hat{a}^2)\,,\\
&\nonumber\\
&\psi_1=  \psi^\text{(0)}_{1a^0} +\hat{a}\, \psi^\text{(0)}_{1a}+q\left(\delta \psi^\text{GSF}_{1a^0} +\hat{a}\, \delta \psi^\text{GSF}_{1a}\right)+\mathcal{O}(q^2,\hat{a}^2)\,,
\end{align}
\ese
with $\hat a= a_2/m_2$. 
In performing that expansion, we make use of the gauge-independent variables $y$ and $\lambda$, which are related to $x$ and $\iota$ via
\bse
\begin{align}\label{ylambda}
y&= (Gm_2 \Omega_\phi)^{2/3} = \frac{x}{(1 + q)^{2/3}} \,,\\
\lambda &=\frac{3y}{\Phi/(2\pi)-1} = \frac{\iota}{(1 + q)^{2/3}} \,. 
\end{align}
\ese

To compare the 1SF corrections $\delta U^{\text{GSF}}_{1\cdots}$ and $\delta \psi^{\text{GSF}}_{1\cdots}$ with those derived in the literature, we express the redshift and spin-precession invariants in terms of the Kerr-geodesic variables $(u_p,e)$, where $e$ is the eccentricity and $u_p$ is the inverse of the dimensionless semilatus rectum (see Appendix~\ref{Kerrvar} for details.)
The terms needed to solve for the N$^3$LO SO unknowns are $\delta U^\text{GSF}_{1a}$ and  $\delta\psi_{1\, a^0}^\text{GSF}$, for which we obtain
\bse
\begin{align}\label{ua0}
&\delta U^\text{GSF}_{1a}=\left(3-\frac{7 e^2}{2}-\frac{e^4}{8}\right) u_p^{5/2}+\left(18-4
e^2-\frac{117 e^4}{4}\right) u_p^{7/2} \nonumber\\
&+\bigg[\frac{251}{4}+\frac{1}{2}\ms X^{1\mr b}_{32}+
\frac{287 e^2}{2}-e^4 \left(\frac{11099}{32}+\frac{15 }{16}\ms X^{1\mr b}_{32}\right)\bigg]u_p^{9/2}\nonumber\\
&+\bigg[\frac{239}{2}-\frac{5 }{4}\ms X^{1\mr b}_{32}-\frac{5
}{2}\ms X^{1\mr b}_{33}+\frac{4 }{3}\ms X^{1\mr b}_{43}\nonumber\\
&\qquad+e^2 \left(\frac{35441}{24}-\frac{41 \pi ^2}{8}-\frac{11
}{4}\ms X^{1\mr b}_{32}-\frac{5 }{2}\ms X^{1\mr b}_{33}+2 \ms X^{1\mr b}_{43}\right)\nonumber\\
&\qquad +e^4 \left(-\frac{230497}{96}+\frac{205 \pi ^2}{32}+\frac{195
}{32}\ms X^{1\mr b}_{32}+\frac{135 }{16}\ms X^{1\mr b}_{33}-5
\ms X^{1\mr b}_{43}\right)\bigg]u_p^{11/2}\,,
\end{align}
\begin{align}
\label{psia0}
\delta\psi_{1\, a^0}^\text{GSF} &=-u_p+\left(\frac{9}{4}+e^2\right) u_p^2\nonumber\\
&+
\bigg[\frac{933}{16}-\frac{123 \pi ^2}{64}-\frac{1}{4}\ms X^{1\mr t}_{32}+e^2 \left(\frac{79}{2}-\frac{123
	\pi ^2}{256}-\frac{3 }{8}\ms X^{1\mr t}_{32}\right)\bigg]u_p^3\nonumber\\
&\quad+\bigg[-\frac{277031}{2880}+\frac{1256 \gamma_E }{15}+\frac{15953 \pi
	^2}{6144}+\frac{11 }{8}\ms X^{1\mr t}_{32}+\frac{5
}{4}\ms X^{1\mr t}_{33}-\frac{2 }{3}\ms X^{1\mr t}_{43}\nonumber\\
&+\frac{296 }{15}\ln 2+\frac{729}{5}\ln 3+\frac{628}{15}\ln u_p+e^2
\bigg(\frac{20557}{480}+\frac{536 \gamma_E }{5}-\frac{55217 \pi
	^2}{4096}\nonumber\\
&+\frac{55 }{16}\ms X^{1\mr t}_{32}+\frac{25
}{8}\ms X^{1\mr t}_{33}-2 \ms X^{1\mr t}_{43}+\frac{11720 }{3}\ln 2-\frac{10206 }{5}\ln 3+\frac{268 }{5}\ln u_p\bigg)\bigg]u_p^4\,.
\end{align}
\ese
These results can be directly compared with the GSF results in Eq.~(4.1) of Ref.~\cite{Kavanagh:2016idg}, Eq.~(23) of Ref.~\cite{Bini:2016dvs} and Eq.~(20) of Ref.~\cite{Bini:2019lcd} for the redshift, and Eq.~(3.33) of Ref.~\cite{Kavanagh:2017wot} for the precession frequency.
At N$^2$LO, as expected, our expressions depend on the scattering-angle coefficients. Upon matching these with the above-mentioned equations in the literature, we get the following four constraints (at each order in eccentricity):
\bse
\begin{align}
&u_p^{9/2} \bigg[\frac{1}{2}\ms X^{1\mr b}_{32}-\frac{97}{4}+e^4\left(\frac{1455}{32}-\frac{15}{16}\ms X^{1\mr b}_{32}\right)\bigg]=0\,, \\
&u_p^3 \bigg[\frac{97}{8}-\frac{1}{4}\ms X^{1\mr t}_{32}
+e^2 \left(\frac{291}{16}-\frac{3}{8}\ms X^{1\mr t}_{32}\right)\bigg]=0\,,
\end{align}
\ese
which can be consistently solved for the two unknowns
\begin{equation}
\label{NNLOsol}
\ms X^{1\mr b}_{32}=\ms X^{1\mr t}_{32}=\frac{97}{2}\,.
\end{equation}
Note that the special constraint~\eqref{speccons}, due to symmetry under interchanging the two bodies' labels $1\leftrightarrow 2$, is thus satisfied. Similarly, at N$^3$LO order, after substituting in the N$^2$LO coefficients, it holds that
\bse
\begin{align}
&u_p^{11/2} \bigg[-\frac{26881}{72}+\frac{241 \pi ^2}{96}-\frac{5
}{2}X^{1\mr b}_{33}+\frac{4 }{3}X^{1\mr b}_{43}\\
&\qquad +e^2 \bigg(-\frac{1846}{3}+\frac{241 \pi ^2}{64}-\frac{5
}{2}X^{1\mr b}_{33}+2 X^{1\mr b}_{43}\bigg)\nonumber\\
&\qquad +e^4 \bigg(\frac{276775}{192}-\frac{1205 \pi ^2}{128}+\frac{135
}{16}X^{1\mr b}_{33}-5 X^{1\mr b}_{43}\bigg)\bigg]=0, \nonumber\\
&u_p^4 \bigg[\frac{8381}{48}-\frac{41 \pi ^2}{16}+\frac{5
}{4}\ms X^{1\mr t}_{33}-\frac{2}{3}\ms X^{1\mr t}_{43} \\
&\qquad +e^2 \bigg(\frac{17647}{32}-\frac{123 \pi ^2}{16}+\frac{25}{8}\ms X^{1\mr t}_{33}-2 \ms X^{1\mr t}_{43}\bigg)\bigg]=0\nonumber\,.
\end{align}
\ese
These five equations can be consistently solved for the remaining four unknowns in the N$^3$LO SO scattering angle,
\begin{gather}\label{NNNLOsol}
\ms X^{1\mr b}_{33}=\ms X^{1\mr t}_{33}=\frac{177}{4},  \\ \ms X^{1\mr b}_{43}=\frac{17423}{48}-\frac{241 \pi ^2}{128}, \quad \ms X^{1\mr t}_{43}=\frac{2759}{8}-\frac{123 }{32}\pi ^2 .\nonumber
\end{gather}
Again, the special constraint~\eqref{speccons} is satisfied by $\ms X^{1\mr b}_{33}$ and $\ms X^{1\mr t}_{33}$. 
Considering the {\sonestwo} dynamics, the relevant constraints can be obtained from the linear-in-spin correction to the spin-precession invariant, which in terms of the remaining unknown coefficients $\ms X^{1\times}_{ij}$ reads
\begin{align}\label{psia1}
\delta\psi_{1\, a^1}^\text{GSF}& = -\frac{u_p^{3/2}}{2}-\left(\frac{41}{8}+\frac{e^2}{8}\right)
u_p^{5/2}\nonumber\\
&- \left[\frac{63}{32}+\frac{123 \pi ^2}{64}+\frac{3 }{16}\ms X^{1\times}_{32}+e^2 \left(\frac{71}{4}+\frac{123 \pi ^2}{256}+\frac{9 }{32}\ms X^{1\times}_{32}\right)\right]u_p^{7/2}\nonumber\\
&+\bigg[\frac{75841 \pi ^2}{6144}-\frac{4496717}{5760}+\frac{1256\gamma_E }{15}+\frac{39 }{32}\ms X^{1\times}_{32}+\frac{15}{16}\ms X^{1\times}_{33}-\frac{8}{15}\ms X^{1\times}_{43}\nonumber\\
&\qquad+\frac{296}{15}\ln 2+\frac{729}{5}\ln 3+\frac{628 }{15}\ln u_p\nonumber\\
&+e^2 \bigg(\frac{7703 \pi
	^2}{4096} -\frac{1016249}{640}+\frac{536 \gamma_E}{5}+\frac{195 }{64}\ms X^{1\times}_{32}+\frac{75
}{32}\ms X^{1\times}_{33}-\frac{8 }{5}\ms X^{1\times}_{43}\nonumber\\
&\qquad+\frac{11720 }{3}\ln 2-\frac{10206 }{5}\ln 3+\frac{268}{5}\ln u_p\bigg)\bigg]u_p^{9/2}\,.
\end{align}	
At N$^2$LO, this can be matched to Eqs.~(52) and (56) of Ref.~\cite{Bini:2019lkm} to get the two constraints (at each order in $e$)
\begin{equation}
u_p^{7/2} \bigg[\frac{75}{8}+\frac{3 }{16}\ms X^{1\times}_{32}+e^2\left(\frac{225}{16}+\frac{9}{32}\ms X^{1\times}_{32}\right)\bigg]=0\,,
\end{equation}
which can be solved  for
\begin{equation}\label{NNLOS1S2}
\ms X^{1\times}_{32} = -50\,.
\end{equation}
Similarly, at N$^3$LO it holds that
\begin{align}
u_p^{9/2} &\bigg[-\frac{6299}{16}+\frac{123 \pi ^2}{32}+\frac{15
}{16}\ms X^{1\times}_{33}-\frac{8 }{15}\ms X^{1\times}_{43}+\\
\qquad &e^2\bigg(-\frac{41943}{32}+\frac{369 \pi ^2}{32}+\frac{75
}{32}\ms X^{1\times}_{33}-\frac{8 }{5}\ms X^{1\times}_{43}\bigg)\bigg]=0\,. \nonumber
\end{align}
Each order in eccentricity is solved for the remaining {\sonestwo} unknown coefficients
\begin{equation}\label{NNNLOS1S2}
\ms X^{1\times}_{33} =-\frac{1383}{5}, \qquad \ms X^{1\times}_{43}=-\frac{9795}{8}+\frac{1845 \pi ^2}{256}\,.
\end{equation}

Combining the solutions obtained in this section with the results of Sec.~\ref{sec:massdep} yields the scattering angle containing the complete local-in-time conservative SO and {\sonestwo} dynamics through the third-subleading PN order
\begin{align}
\frac{\chi}{\Gamma}=\quad &\left(\frac{GM}{b\sqrt{\ve}}\right)2\frac{1+2\ve}{\sqrt{\ve}}+\left(\frac{GM}{b\sqrt{\ve}}\right)^2\frac{3\pi}{4}(4+5\ve)\nonumber\\
-&\left(\frac{GM}{b\sqrt{\ve}}\right)^3\frac{1}{\sqrt{\ve}}\bigg[2\frac{1-12\ve-72\ve^2-64\ve^3}{3\ve}\nonumber\\
&+\nu\left(\frac{8+94\ve+313\ve^2}{12}+\mc O(\ve^3)\right)\bigg]\nonumber\\
+&\left(\frac{GM}{b\sqrt{\ve}}\right)^4\pi\bigg[\frac{105}{64}(16+48\ve+33\ve^2)\nonumber\\
&+\nu \bigg({-}\frac{15}{2}+\bigg(\frac{123}{128}\pi^2-\frac{557}{8}\bigg)\ve+\mc O(\ve^2)\bigg)\bigg]\nonumber\\
-\left(\frac{a_b}{b\sqrt{\ve}}\right)\bigg\{& \left(\frac{GM}{b\sqrt{\ve}}\right)4\gamma \sqrt{\ve}+\left(\frac{GM}{b\sqrt{\ve}}\right)^22\pi \gamma\, (2+5\ve)\, \nonumber\\
+&\left(\frac{GM}{b\sqrt{\ve}}\right)^3\frac{\gamma}{\sqrt{\ve}}\bigg[12(1+12\ve+16\ve^2)\nonumber\\
&-\nu\left(10\ve+\frac{97}{2}\ve^2+\frac{177}{4}\ve^3+\mc O(\ve^4)\right)\bigg]\nonumber\\
+&\left(\frac{GM}{b\sqrt{\ve}}\right)^4\pi\gamma\bigg[\frac{21}{2}(8+36\ve+33\ve^2)\nonumber\\
&-\nu\left(\frac{21}{2}+\frac{495}{4}\ve+\left(\frac{17423}{48}-\frac{241\pi^2}{128}\right)\ve^2+\mc O(\ve^3)\right)\bigg]\bigg\}\nonumber\\
-\left(\frac{a_t}{b\sqrt{\ve}}\right)\bigg\{&\left(\frac{GM}{b\sqrt{\ve}}\right)4\gamma \sqrt{\ve}+\left(\frac{GM}{b\sqrt{\ve}}\right)^2\frac{3\pi}{2} \gamma\, (2+5\ve)\, \nonumber\\
+&\left(\frac{GM}{b\sqrt{\ve}}\right)^3\frac{\gamma}{\sqrt{\ve}}\bigg[8(1+12\ve+16\ve^2)\nonumber\\
&-\nu\left(10\ve+\frac{97}{2}\ve^2+\frac{177}{4}\ve^3+\mc O(\ve^4)\right)\bigg]\nonumber\\
+&\left(\frac{GM}{b\sqrt{\ve}}\right)^4\pi\gamma\bigg[\frac{105}{16}(8+36\ve+33\ve^2)\nonumber\\
&-\nu\left(9+\frac{219}{2}\ve+\left(\frac{2759}{8}-\frac{123}{32}\pi^2\right)\ve^2+\mc O(\ve^3)\right)\bigg]\bigg\}\nonumber\end{align}
\begin{align}\label{chifin}
+\left(\frac{a_1a_2}{b^2\ve}\right)\bigg\{&\left(\frac{GM}{b\sqrt{\ve}}\right)\frac{4}{\sqrt{\ve}}(\ve+2\ve^2)+\left(\frac{GM}{b\sqrt{\ve}}\right)^2\frac{3\pi}{2}(2+19\ve+20\ve^2)\nonumber\\
+&\left(\frac{GM}{b\sqrt{\ve}}\right)^3\frac{1}{\sqrt{\ve}}\bigg[8(1+38\ve+128\ve^2+96\ve^3)\nonumber\\
&+\nu\left(8\ve-50\ve^2-\frac{1383}{5}\ve^3+\mc O(\ve^4)\right)\bigg]\nonumber\\
+&\left(\frac{GM}{b\sqrt{\ve}}\right)^4\pi\bigg[
\frac{105}{16} (24+212\ve+447\ve^2+264\ve^3)
\nonumber\\
&+\nu\left(\frac{15}{2}-\frac{405}{4}\ve+\left(-\frac{9795}{8}+\frac{1845\pi^2}{256}\right)\ve^2+\mc O(\ve^3)\right)\bigg]\bigg\}\,.
\end{align}	
Importantly, we have checked that all the above results can be reproduced by starting from a Hamiltonian ansatz (rather than a radial action), constraining it via the mass-ratio dependence of the scattering angle (calculated via~{(\ref{chidef})}), and obtaining the redshift and spin-precession invariants through Eqs.~\eqref{zham} and~\eqref{precessham}.

\section{Effective-one-body Hamiltonian and comparison with numerical relativity}
\label{sec:NRcomp}
In this section, we quantify the improvement in accuracy from the new N$^3$LO SO and S$_1$S$_2$ corrections using numerical relativity (NR) simulations as means of comparison. We do this using an EOB Hamiltonian, calculated using the scattering angle obtained above, since the resummation of PN results it grants is expected to improve the agreement with NR in the high-frequency regime.

The EOB Hamiltonian is calculated from an effective Hamiltonian $H^\text{eff}$ via the energy map
\begin{equation}
\label{EOBHam}
H^\text{EOB} = M \sqrt{1 + 2\nu \left(\frac{H^\text{eff}}{\mu} - 1\right)},
\end{equation}
where we use for the effective Hamiltonian an aligned-spin version of the Hamiltonian for a nonspinning test mass in a Kerr background (denoted SEOB$_\text{TM}$ in Ref.~\cite{Khalil:2020mmr}) with SO and S$_1$S$_2$ PN corrections. The effective Hamiltonian is given by
\begin{align}
\label{Heff}
{H}^\text{eff} &= \bigg[
A
\left( \mu^2 + p^2  + B_{p_r} p_r^2  + B_{L} \frac{L^2a^2}{r^2} + \mu^2Q  
\Big)\right]^{1/2}  \nonumber\\
&\quad  
+ \frac{GMr}{\Lambda} L \left(g_S S + g_{S^*} S^*\right),
\end{align}
where $\Lambda = (r^2+a^2)^2 - \Delta a^2$ with $\Delta = r^2 - 2GMr + a^2$. The Kerr spin $a$ is mapped to the binary's spins via $a = a_1 + a_2$, and the potentials are taken to be
\begin{subequations}
	\begin{align}
	A &= \frac{\Delta r^2}{\Lambda} \left(A^0 +  A^\text{SS}\right), \\
	B_{p_r} &= \left(1 - \frac{2GM}{r} + \frac{a^2}{r^2}\right) \left(A^0 D^0 +  B_{p_r}^\text{SS}\right)  - 1, \\
	B_L &= - \frac{r^2 + 2GM r}{\Lambda}, \\
	Q &= Q^0 + Q^\text{SS},
	\end{align}
\end{subequations}
i.e., we factorize the PN corrections to the Kerr potentials. 
The zero-spin corrections $A^0(r),~D^0(r)$ and $Q^0(r)$ are given by Eq.~(28) of Ref.~\cite{Khalil:2020mmr} and are based on the 4PN nonspinning Hamiltonian derived in Ref.~\cite{Damour:2015isa}.
The SO corrections are encoded in the gyro-gravitomagnetic factors $g_S$, and $g_{S^*}$, while the S$_1$S$_2$ corrections are added through $A^\text{SS}, \, B_{p_r}^\text{SS}$, and $Q^\text{SS}$.

For those PN corrections, we choose a gauge such that $g_S$, and $g_{S^*}$ are independent of $L$~\cite{Damour:2008qf,Nagar:2011fx,Barausse:2011ys}; we write an ansatz such that, up to N$^3$LO,
\begin{align}
g_S(r, p_r) &= 2 \sum_{i=0}^{3}\sum_{j=0}^{i} \alpha_{ij}\frac{p_r^{2(i-j)}}{c^{2i} r^j}, \nonumber\\
g_{S^*}(r, p_r) &= \frac{3}{2} \sum_{i=0}^{3}\sum_{j=0}^{i} \alpha_{ij}^*\frac{p_r^{2(i-j)}}{c^{2i} r^j},
\end{align}
for some unknown coefficients $\alpha_{ij}$ and $\alpha_{ij}^*$.
For the S$_1$S$_2$ corrections, $A^\text{SS}$ and $B_{p_r}^\text{SS}$ start at NLO and are independent of $p_r$, while $Q^\text{SS}$ starts at NNLO and depends on $p_r^4$ or higher powers of $p_r$, i.e., we use an ansatz of the form
\begin{align}
A^\text{SS} &= S_1S_2 \left(\frac{\alpha_4^A}{c^6r^4} + \frac{\alpha_5^A}{c^8r^5} + \frac{\alpha_6^A}{c^{10}r^6}\right), \nonumber\\
B^\text{SS} &= S_1S_2 \left(\frac{\alpha_3^B}{c^4r^3} + \frac{\alpha_4^B}{c^6r^4} + \frac{\alpha_5^B}{c^8r^5}\right), \nonumber\\
Q^\text{SS} &= S_1S_2 \left(\alpha_{34}^Q \frac{p_r^4}{c^6r^3} + \alpha_{44}^Q \frac{p_r^4}{c^8r^4} + \alpha_{36}^Q \frac{p_r^6}{c^8r^3}\right).
\end{align}
To determine those unknowns, we calculate the scattering angle from such an ansatz using Eq.~\eqref{chidef} (which entails inverting the EOB Hamiltonian for $p_r$ in a PN expansion, differentiating with respect to $L$, and integrating with respect to $r$). We then match the result of that calculation to the scattering angle calculated in the previous section and solve for the unknown coefficients in the Hamiltonian ansatz. This uniquely determines all the coefficients of the spinning part of the Hamiltonian since our choice for the ansatz fixes the gauge dependence of the Hamiltonian. (See Sec.~\ref{sec:massshell} for a discussion of the gauge freedom in the Hamiltonian.)

We obtain the gyro-gravitomagnetic factors
	\bse
	\label{gyros}
	\begin{align}
	g_S &= 2 \bigg\lbrace
	1 + \frac{\nu}{c^2}\left[-\frac{27}{16} \frac{p_r^2}{\mu^2} - \frac{5 }{16}\frac{GM}{r}\right]
	\nonumber\\
	&+\frac{\nu}{c^4}\bigg[
	\left(\frac{5}{16}+\frac{35 \nu}{16}\right) \frac{p_r^4}{\mu^4}
	-\left(\frac{21}{4}-\frac{23 \nu}{16}\right)\frac{p_r^2}{\mu^2}\frac{GM}{r}
	-\left(\frac{51}{8}+\frac{\nu}{16}\right)\frac{(GM)^2}{r^2}
	\bigg] \nonumber\\
	&+\frac{\nu}{c^6} \bigg[
	\bigg(-\frac{80399}{2304}+\frac{241 \pi ^2}{384}+\frac{379 \nu }{64}-\frac{7 \nu ^2}{256}\bigg)\frac{(GM)^3}{r^3} 
	\nonumber\\
	&\quad\qquad+ \left(-\frac{5283}{128}+\frac{1557 \nu }{32}+\frac{69 \nu ^2}{128}\right) \frac{p_r^2}{\mu^2} \frac{(GM)^2}{r^2} \nonumber\\
	&\quad\qquad+ \left(\frac{781}{256}+\frac{831 \nu }{64}-\frac{771 \nu ^2}{256}\right) \frac{p_r^4}{\mu^4} \frac{GM}{r} 
	+  \left(\frac{7}{256}-\frac{63 \nu }{64}-\frac{665 \nu ^2}{256}\right) \frac{p_r^6}{\mu^6}
	\bigg]
	\bigg\rbrace, \\
	g_{S^*} &= \frac{3}{2} \bigg\lbrace
	1 + \frac{1}{c^2} \left[-\left(\frac{3 \nu }{2}+\frac{5}{4}\right) \frac{p_r^2}{\mu^2}
	-\left(\frac{3}{4}+\frac{\nu }{2}\right)\frac{GM}{r}\right]
	\nonumber\\
	&\quad
	+\frac{1}{c^4} \bigg[
	\left(\frac{15 \nu ^2}{8}+\frac{5 \nu }{3}+\frac{35}{24}\right) \frac{p_r^4}{\mu^4}
	+ \left(\frac{19 \nu ^2}{8}-\frac{3 \nu }{2}+\frac{23}{8}\right)\frac{p_r^2}{\mu^2}\frac{GM}{r}
	\nonumber\\
	&\quad\qquad\qquad+ \left(-\frac{\nu ^2}{8}-\frac{13 \nu }{2}-\frac{9}{8}\right) \frac{(GM)^2}{r^2}
	\bigg] \nonumber\\
	&\quad 
	+\frac{1}{c^6}
	\bigg[
	\bigg(-\frac{135}{64}+\frac{41 \pi ^2 \nu }{48}-\frac{7627 \nu }{288} +\frac{237 \nu ^2}{32} -\frac{\nu ^3}{16}\bigg) \frac{(GM)^3}{r^3} 
	\nonumber\\
	&\quad\qquad+ \left(-\frac{15}{32}-\frac{279 \nu }{16}+\frac{787 \nu ^2}{16} + \frac{9 \nu ^3}{8}\right) \frac{p_r^2}{\mu^2} \frac{(GM)^2}{r^2} \nonumber\\
	&\quad\qquad 
	+ \left(-\frac{1105}{192}-\frac{53 \nu }{96}+\frac{117 \nu ^2}{32}-\frac{81 \nu ^3}{16}\right) \frac{p_r^4}{\mu^4} \frac{GM}{r} 
	\nonumber\\
	&\quad\qquad+ \left(-\frac{105}{64}-\frac{175 \nu }{96}-\frac{77 \nu ^2}{32}-\frac{35 \nu ^3}{16}\right) \frac{p_r^6}{\mu^6}
	\bigg]
	\bigg\rbrace,
	\end{align}
	\ese
	and the S$_1$S$_2$ corrections
	\bse
	\label{SSpots}
	\begin{align}
	A^\text{SS} &= \frac{S_1S_2}{G^2M^2\mu^2} \bigg\lbrace
	\frac{(GM)^4}{c^6r^4}\left(2\nu - \nu^2\right)
	+ \frac{(GM)^5}{c^8r^5}\left(\frac{17 \nu }{2}+\frac{113 \nu ^2}{8}+\frac{3 \nu^3}{4}\right) \nonumber\\
	&\quad\qquad\qquad
	+ \frac{(GM)^6}{c^{10}r^6}\left(\frac{61 \nu}{2}-\frac{41 \pi^2 \nu^2}{16}+\frac{3791 \nu^2}{48}+\frac{25 \nu^3}{4}+\frac{21 \nu^4}{32}\right) 
	\bigg\rbrace, \\
	B_{p_r}^\text{SS} &=\frac{S_1S_2}{G^2M^2\mu^2} \bigg\lbrace
	\frac{(GM)^3}{c^4r^3}\left(6\nu + \frac{9}{2} \nu^2\right)
	+ \frac{(GM)^4}{c^6r^4}\left(20 \nu+26 \nu ^2+10 \nu ^3\right)
	\nonumber\\
	&+ \frac{(GM)^5}{c^8r^5}\left(67 \nu+328 \nu ^2-\frac{375 \nu ^3}{4} -\frac{37 \nu ^4}{16}\right)\bigg\rbrace, \\
	Q^\text{SS} &= \frac{S_1S_2}{G^2M^2\mu^2} \bigg\lbrace
	\frac{1}{c^6}\frac{p_r^4}{\mu^4}\frac{(GM)^3}{r^3} \left(+\frac{5 \nu }{2}+\frac{45 \nu ^2}{8}-\frac{25 \nu^3}{4}\right) 
	\nonumber\\
	&+ \frac{1}{c^8} \bigg[
	\frac{p_r^6}{\mu^6}\frac{(GM)^3}{r^3} \left(-\frac{7 \nu }{4}-\frac{63 \nu^2}{16}-\frac{35 \nu^3}{8}+\frac{245 \nu^4}{32}\right) \nonumber\\
	&\quad\qquad\qquad
	+ \frac{p_r^4}{\mu^4}\frac{(GM)^4}{r^4} \left(-13 \nu+\frac{183 \nu^2}{4} -63 \nu^3 -\frac{437 \nu^4}{16}\right)
	\bigg]\bigg\rbrace.
	\end{align}
	\ese
Importantly, the factors $g_S$ and $g_{S^*}$, obtained here for the aligned-spin case, also fix the generic-spin case by simply writing the odd-in-spin part of the effective Hamiltonian as
\begin{equation}
H_\text{eff}^\text{odd} = \frac{GMr}{\Lambda} \bm{L} \cdot \left(g_S \bm{S} + g_{S^*} \bm{S}^*\right),
\end{equation}
with $g_S$ and $g_{S^*}$ unmodified since they are independent of the spins (see Ref.~\cite{Antonelli:2020aeb} for more details.)
However, the spin$_1$-spin$_2$ corrections in Eq.~\eqref{SSpots} are only for aligned spins since the generic-spins case has additional contributions proportional to $(\bm{n}\cdot \bm{S}_1)(\bm{n}\cdot \bm{S}_2)$, where $\bm{n}=\bm{r}/r$.
Such terms vanish for aligned spins and cannot be fixed from aligned-spin self-force results or be removed by canonical transformations.

\begin{figure*}
	\centering
	\includegraphics[width=0.49\linewidth]{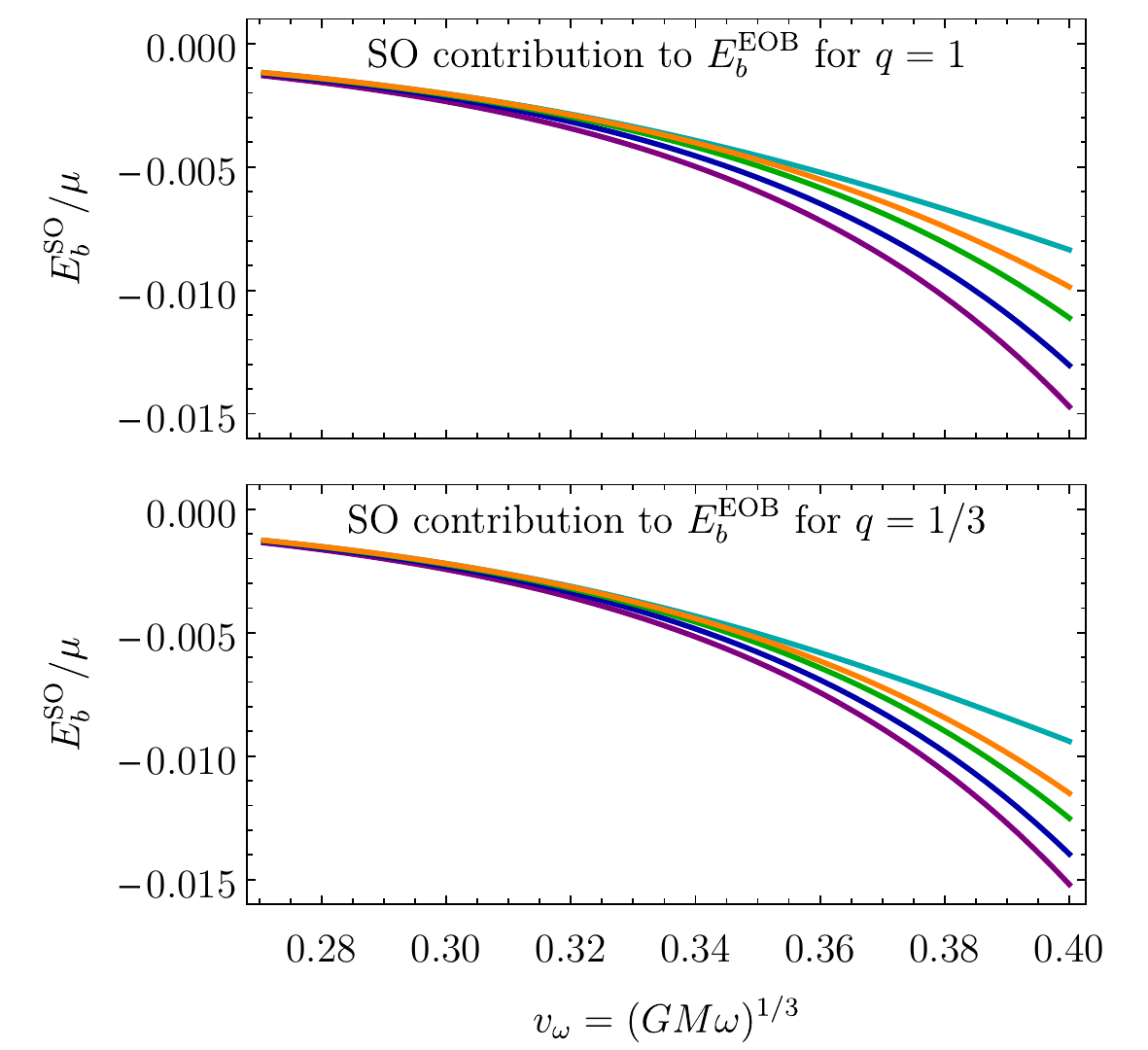}
	\includegraphics[width=0.49\linewidth]{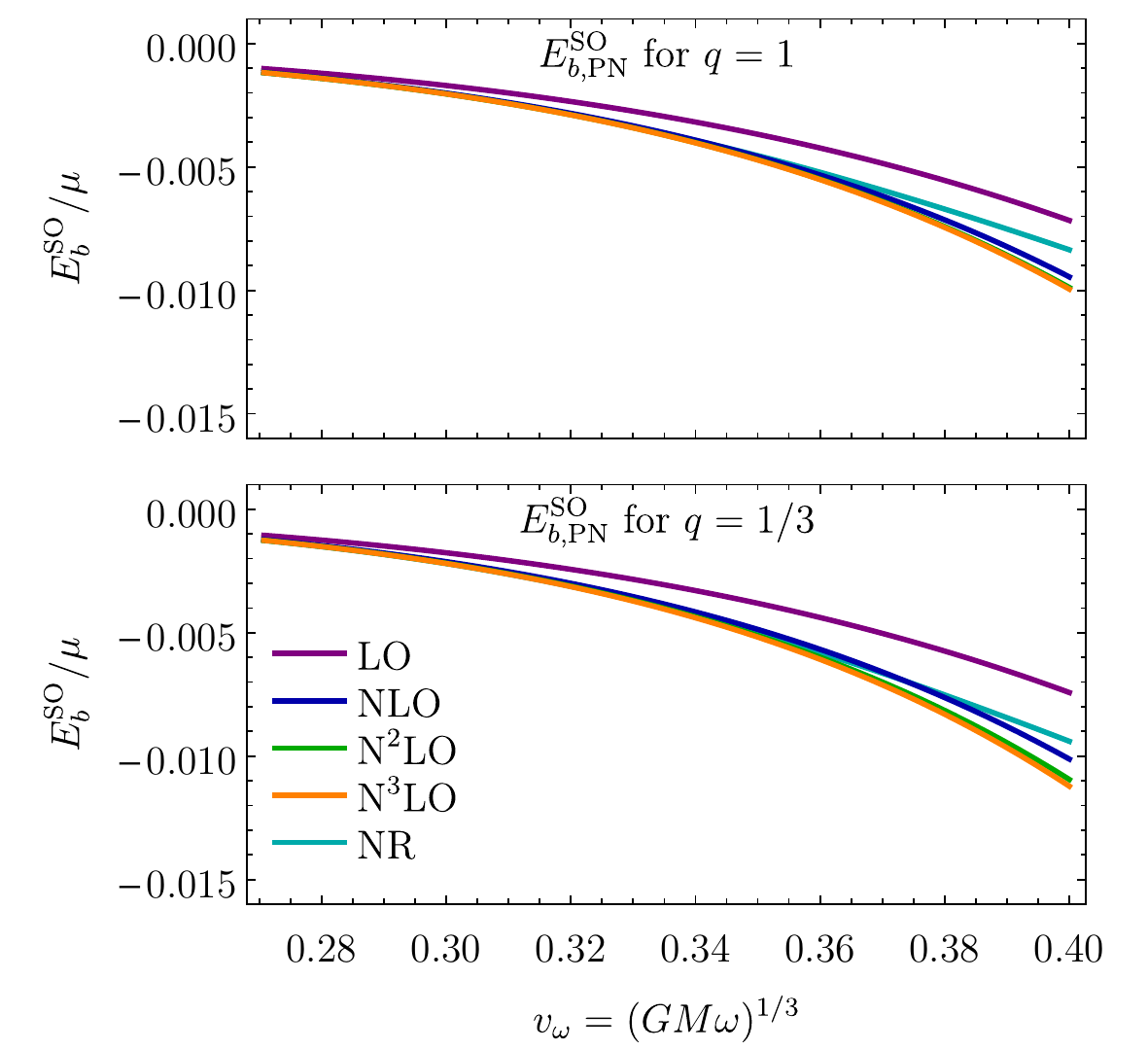}
	\caption{Binding energy versus the velocity parameter $v_\omega$ for the SO contribution to the EOB (left panels) and PN-expanded (right panels) binding energies for mass ratios $q=1$ (top panels) and $q=1/3$ (bottom panels).}
	\label{fig:bindEnSO} 
\end{figure*}

\begin{figure*}
	\centering
	\includegraphics[width=0.49\linewidth]{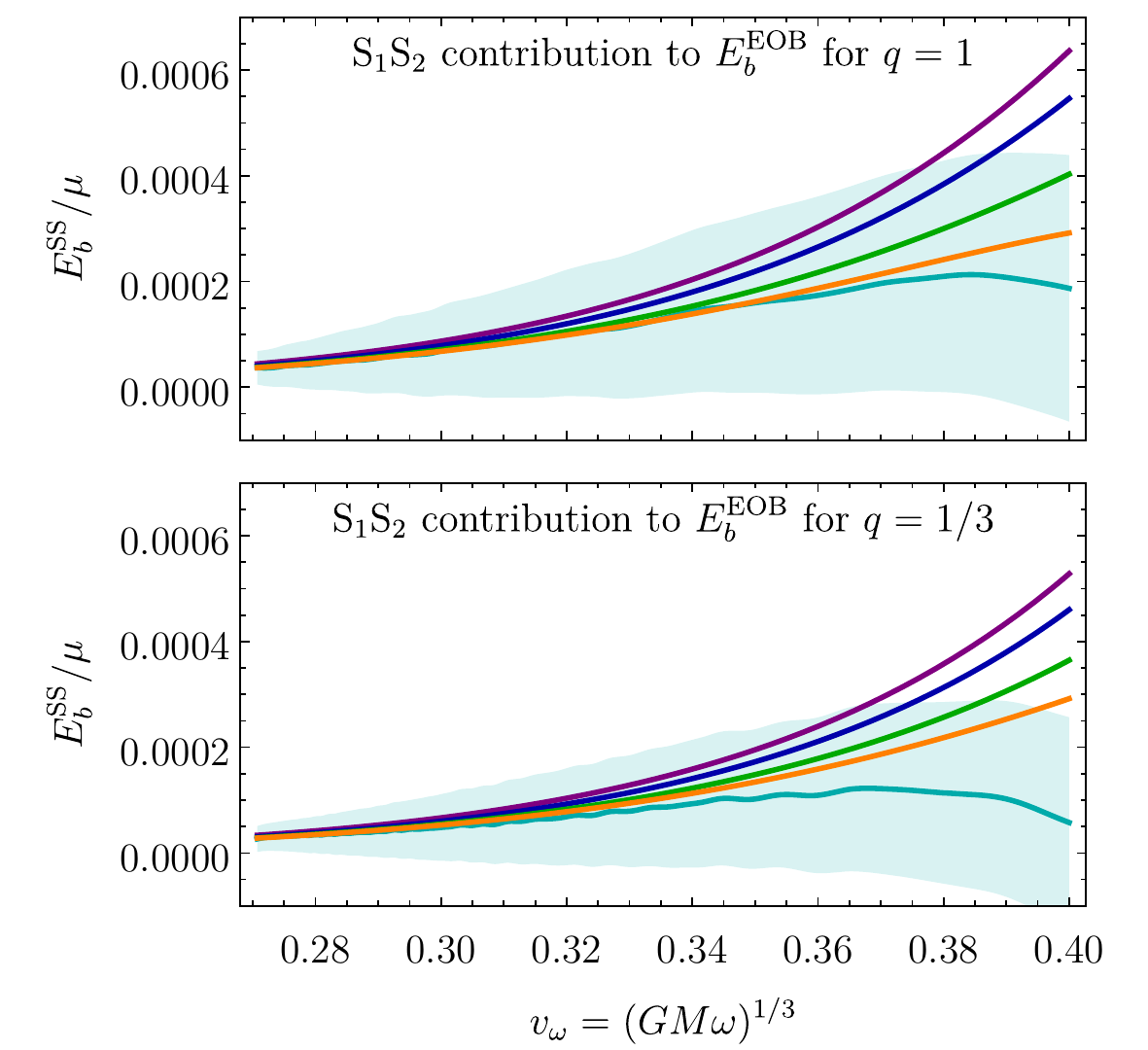}
	\includegraphics[width=0.49\linewidth]{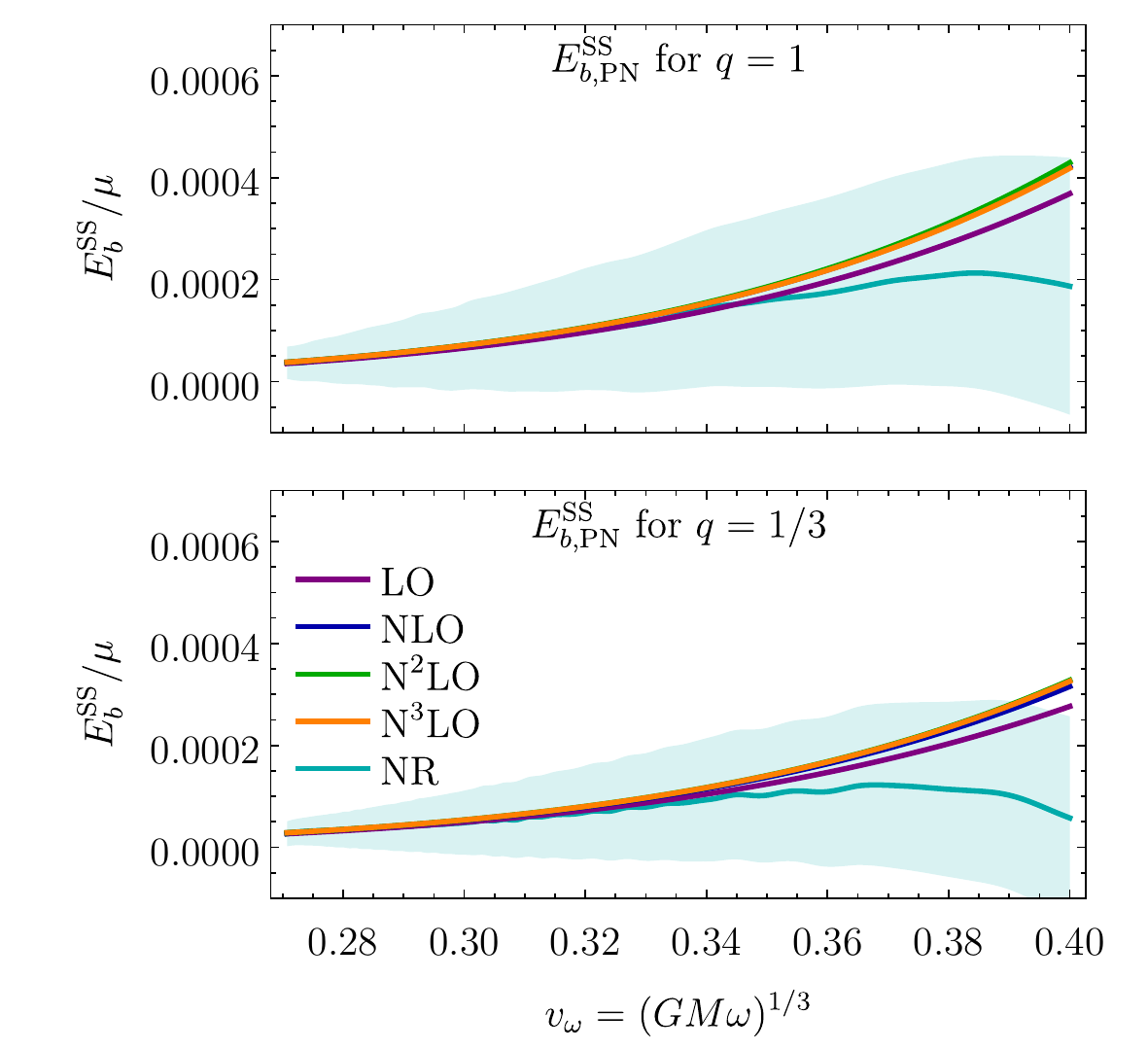}
	\caption{Binding energy versus the velocity parameter $v_\omega$ for the S$_1$S$_2$  contribution to the EOB (left panels) and PN-expanded (right panels) binding energies for mass ratios $q=1$ (top panels) and $q=1/3$ (bottom panels). The NR error is indicated by the shaded regions.}
	\label{fig:bindEnSS} 
\end{figure*}

For comparison with NR, a particularly good quantity to consider is the binding energy, since it encapsulates the conservative dynamics of analytical models, and can be obtained from accurate NR simulations~\cite{Damour:2011fu,Nagar:2015xqa}. The NR data for binding energy that we use were extracted in Ref.~\cite{Ossokine:2017dge} from the Simulating eXtreme Spacetimes (SXS) catalog \cite{SXS}.
The binding energy calculated from NR simulations is defined by
\begin{equation}
E_b^\text{NR} = E_\text{ADM} - E_\text{rad} - Mc^2,
\end{equation}
where $E_\text{rad}$ is the radiated energy, and $E_\text{ADM}$ is the ADM energy at the beginning of the simulation. 
We then calculate the binding energy from the EOB conservative Hamiltonian using $E_b = H^\text{EOB} - Mc^2$ for exact circular orbits at different orbital separations, i.e., we neglect the radiation reaction due to the emitted GWs. As a result of this assumption, the circular-orbit binding energy we calculate is not expected to agree  with NR in the last few orbits.

To obtain the binding energy from a Hamiltonian in an analytical PN expansion, we set $p_r=0$ for circular orbits and perturbatively solve $\dot{p}_r=0=-\partial H/\partial r$ for the angular momentum $L$. The orbital frequency $\omega$ is given by $\omega = \partial H / \partial L$ from which we define the velocity parameter
\begin{equation}
v_\omega = (GM\omega)^{1/3}.
\end{equation}
Expressing the PN-expanded Hamiltonian in terms of $v_\omega$ yields, for the SO part,
	\begin{align}
	\label{EbPNSO}
	E_{b,\text{PN}}^\text{SO} &= \frac{\nu}{G M} \bigg\lbrace
	v_\omega^5 \left[-\frac{4}{3} S - S^* \right]
	+ v_\omega^7 \left[ S \left(\frac{31 \nu }{18}-4\right) + S^* \left(\frac{5 \nu }{3}-\frac{3}{2} \right)\right] \nonumber\\
	&\quad
	+ v_\omega^9 \left[\frac{S}{24}  \left(- 324+ 633 \nu  - 14 \nu ^2\right) + \frac{S^*}{8} \left( - 27 + 156 \nu -5 \nu ^2\right) \right] \nonumber\\
	&\quad + v_\omega^{11} \bigg[ S \left(- 45 + \frac{19679+174\pi^2}{144} \nu - \frac{1979}{36} \nu^2 - \frac{265}{3888} \nu^3 \right)
	\nonumber\\
	&\qquad- \frac{S^*}{8} \left(\frac{135}{2} - 565 \nu + \frac{1109}{3} \nu^2 + \frac{50}{81} \nu^3 \right) \bigg] \bigg\rbrace,
	\end{align}
	while for the S$_1$S$_2$ part,
	\begin{align}
	\label{EbindS1S2}
	E_{b,\text{PN}}^\text{SS} &= \frac{S_1 S_2}{G^2 M^3} \bigg[v_\omega^6 
	+ v_\omega^8 \left(\frac{5}{6} + \frac{5}{18}\nu \! \right)
	+ v_\omega^{10} \left(\frac{35}{8}-\frac{1001}{72}\nu -\frac{371 }{216}\nu ^2 \! \right)  \nonumber\\
	&+ v_\omega^{12} \left( \frac{243}{16} + \frac{123 \pi^2-4214}{32}\nu + \frac{147}{8}\nu ^2 + \frac{13}{16}\nu^3 \! \right) \! \bigg].
	\end{align}
The same steps can be performed numerically to obtain the EOB binding energy without a PN expansion.

To examine the effect of the new N$^3$LO terms on the binding energy, we isolate the SO and the S$_1$S$_2$ contributions to the binding energy by combining configurations with different spin orientations (parallel or anti-parallel to the orbital angular momentum), as explained in Refs.~\cite{Dietrich:2016lyp,Ossokine:2017dge}. For the SO contribution, we use
\begin{equation}
\label{EbSO}
E_b^\text{SO}(\nu,\hat{a},\hat{a}) = \frac{1}{2} \left[ E_b(\nu,\hat{a},\hat{a}) - E_b(\nu,-\hat{a},-\hat{a})\right] + \mathcal{O}(\hat{a}^3),
\end{equation}
while for the S$_1$S$_2$ contribution, we use
\begin{align}
\label{EbSS}
E_b^\text{SS}(\nu,\hat{a},\hat{a}) &= E_b(\nu,\hat{a},0) + E_b(\nu,0,-\hat{a}) - E_b(\nu,\hat{a},-\hat{a}) \nonumber\\
&\quad - E_b(\nu,0,0)  + \mathcal{O}(\hat{a}^3).
\end{align}

In Fig.~\ref{fig:bindEnSO}, we plot the SO contribution to the EOB and PN-expanded binding energies versus the velocity parameter $v_\omega$ for spin magnitudes $\hat{a} = 0.6$. We also plot the NR results by combining the binding energies of configurations with different spins using results from Refs.~\cite{SXS,Ossokine:2017dge}.
From the figure, we see that, adding each PN order improves agreement of the EOB binding energy with NR, especially in the high-frequency regime, with better improvement for equal masses than for unequal masses.  In contrast, the PN binding energy, plotted using Eq.~\eqref{EbPNSO}, seems \emph{not} to converge towards NR in the high-frequency regime, with little difference between the N$^2$LO and N$^3$LO SO orders. 
Figure~\ref{fig:bindEnSS} shows the S$_1$S$_2$ contribution to the EOB and PN binding energies. As in the SO case, adding the new N$^3$LO significantly improves agreement of the EOB binding energy to NR, especially for equal masses, but there is little difference between PN orders for the PN binding energy.

Note that Figs.~\ref{fig:bindEnSO} and \ref{fig:bindEnSS} should not be interpreted as a direct comparison between PN and EOB dynamics since our results were obtained for simplicity using exact circular-orbits, which leads to a very different behavior than for an \emph{inspiraling} binary;  Refs.~\cite{Ossokine:2017dge,Antonelli:2019fmq,Nagar:2015xqa}, for example, show that EOB results are significantly better than PN when taking into account the binary evolution.
Let us also stress that while the EOB and PN curves are based on the same PN information, the EOB Hamiltonian represents a particular resummation of the PN results.
We leave the exploration of other resummations and a calibration to NR for future work.

\section{Conclusions}
\label{sec:conc}

GW astronomy allows a multitude of applications in fundamental and astrophysics \cite{Abbott:2019yzh,LIGOScientific:2019fpa,LIGOScientific:2018jsj,LIGOScientific:2018mvr} that rely on accurate waveform models for inferring the source parameters.  In this paper, we improved the PN description of spinning compact binaries using information from relativistic scattering and self-force theory, which is an extension of the approach introduced and used in Refs.~\cite{Bini:2019nra,Damour:2019lcq,Bini:2020wpo} for the nonspinning case.
We started by extending the arguments from Ref.~\cite{Damour:2019lcq} to show that the scattering angle for an aligned-spin binary has a simple dependence on the masses. This allowed us to determine the SO and aligned {\sonestwo} couplings through N$^3$LO in a PN expansion using GSF results for the redshift and precession frequency of a small body on an eccentric orbit in a Kerr background.
This result is neatly encapsulated in the gauge-invariant aligned-spin scattering-angle function, given explicitly in Eq.~\eqref{chifin}.  The derivation presented here provides the full details for the recently reported result at SO level in Ref.~\cite{Antonelli:2020aeb}, while extending the analysis to aligned {\sonestwo} couplings. 

Using these new PN results, we calculated the circular-orbit binding energy, the EOB gyro-gravitomagnetic factors, and implemented these results in an EOB Hamiltonian. 
To illustrate the effect of the new N$^3$LO terms, we compared the binding energy with NR simulations (see Figs.~\ref{fig:bindEnSO} and \ref{fig:bindEnSS},) showing an improvement over the N$^2$LO.
These results could be implemented in state-of-the-art \texttt{SEOBNR}~\cite{Bohe:2016gbl,Babak:2016tgq,Cotesta:2018fcv,Ossokine:2020kjp} and \texttt{TEOBResumS}~\cite{Nagar:2018plt,Nagar:2018zoe} waveform models used in LIGO-Virgo searches and inference analyses~\cite{LIGOScientific:2018mvr}.

While it is arguable whether PM results already provide a useful resummation of the PN ones~\cite{Antonelli:2019ytb}, the present work shows that, with the crucial contribution of GSF theory, advances in PM theory already allow one to advance the PN knowledge in the spin sector. We thus beseech further research to explore synergies between GSF, PM, and PN theory, along the lines of Refs.~\cite{Bini:2019nra,Siemonsen:2019dsu,Antonelli:2020aeb,Bini:2020wpo,Bini:2020nsb,Bini:2020uiq} and the present paper. 
One could, for instance, extend the results in this paper to N$^3$LO S$^2$ couplings, i.e.\ at quadratic order in each spin.  This is an important step to complete the aligned-spin 5PN dynamics for BBHs. However, we leave such a calculation for future work, since it would require currently unavailable GSF results.

One can envision further important work at the interface between the PM and GSF approximations. With knowledge of first-order GSF theory, one can in principle determine the full 3PM and 4PM scattering angle in a completely independent way from techniques employed, e.g., in Ref.~\cite{Bern:2019nnu}. To this end, one could calculate the PM expansion of GSF gauge-invariant quantities for bound orbits directly (e.g., expansions in $u_p$ valid at all orders in the eccentricity $e$). This enterprise would have to take great care in the inclusion of tail terms in the dynamics, as well as in the analytical continuation of such results to scattering systems. Should these quantities be calculated, one could exploit the method herein presented to fix the 3PM and 4PM scattering angles without further PN re-expansions.
Even better would be a direct GSF treatment of scattering orbits and the scattering angle. This is likely to first come in the form of numerical calculations at first-order in the mass ratio. It will however be worth exploring whether ``experimental mathematics'' techniques can be used to obtain analytic expressions for the 4PM scattering angle by pushing such numerical calculations to extreme precision (see, e.g., Ref.~\cite{Johnson-McDaniel:2015vva} for an example along these lines in the GSF literature).

Finally, we stress that it is paramount to check our results with more established PN calculations (e.g., with the EFT approach, as was done partially at N$^3$LO in Refs.~\cite{Levi:2020kvb,Levi:2020uwu}), as they have been obtained with a so-far completely unexplored method in the spinning sector that is begging to be further scrutinized.

\section*{Acknowledgments}
We are grateful to Alessandra Buonanno and Maarten van de Meent for helpful discussions.
We also thank Sergei Ossokine and Tim Dietrich for providing NR data for the binding energy and for related useful suggestions.
  

\newcommand{\bk}{{\bm k}}
\newcommand{\bfe}{{\tbm e}}
\newcommand{\pg}[1]{{\color{red}#1}}

\newcommand{\DP}{\Delta_4}
\newcommand{\bq}{{\bm{q}}}
\newcommand{\bp}{{\bm{p}}}
\newcommand{\bJ}{{\bm{J}}}
\newcommand{\bS}{{\mathbf S}}
\newcommand{\bGamma}{{\bf{\Gamma}}}
\newcommand{\ff}{f(\eta, \bx, \bp)}
\newcommand{\G}{\mathcal{G}}
\renewcommand{\AA}{\mathcal{A}}
\newcommand{\OO}{\mathcal{O}}
\newcommand{\HH}{\mathcal{H}}
\newcommand{\ep}{\epsilon}
\newcommand{\de}{\delta}
\newcommand{\De}{\Delta}
\newcommand{\si}{\sigma}
\newcommand{\we}{\wedge}
\newcommand{\opt}{{\hat{ \cal T}}}
\newcommand{\al}{\alpha}
\renewcommand{\b}{\beta}
\newcommand{\Ga}{\Gamma}
\newcommand{\Om}{\Omega}
\newcommand{\La}{\Lambda}
\newcommand*{\tin}{_{\text{in}}}
\newcommand*{\ti}{\tilde}
\newcommand{\trr}{\text{Tr}}
\newcommand{\dd}{\text{d}}
\newcommand{\pd}{\partial}
\newcommand{\nn}{\nonumber}
\newcommand{\bss}{\boldsymbol{\mathcal{S}}}
\newcommand{\bom}{\boldsymbol{\omega}}
\newcommand{\obs}{_{\rm O}}
\newcommand{\Gal}{_{\rm G}}
\newcommand{\Sou}{_{\rm S}}
\newcommand{\GW}{{_{\rm GW}}}
\voffset=-0.3in
\newcommand{\ob}[1]{\textcolor{blue}{\bf [Ollie: #1]}}
\newcommand{\andrea}[1]{\textcolor{orange}{\bf [AA: #1]}}
\newcommand{\jg}[1]{\textcolor{red}{\bf [JG: #1]}}
\newcommand{\ooo}[1]{\textcolor{brown}{#1}}
\newcommand{\vecorbf}[1]{\bf #1}

\chapter[Noisy neighbours]{Noisy neighbours: inference biases from overlapping gravitational-wave signals}
\label{chap:six}

\textbf{Authors\footnote{Manuscript~\cite{Antonelli:2021vwg} accepted for publication in the \emph{Monthly Notices of the Royal Astronomical Society}.} }:
	Andrea Antonelli, Ollie Burke, Jonathan R. Gair.
\newline

\textbf{Abstract:} Understanding and dealing with inference biases in gravitational-wave (GW) parameter estimation when a plethora of signals are present in the data is one of the key challenges for the analysis of data from future GW detectors. Working within the linear signal approximation, we describe generic metrics to predict inference biases on GW source parameters in the presence of confusion noise from unfitted foregrounds, from overlapping signals that coalesce close in time to one another, and from residuals of other signals that have been incorrectly fitted out. We illustrate the formalism with simplified, yet realistic, scenarios appropriate to third-generation ground-based (Einstein Telescope) and space-based (LISA) detectors, and demonstrate its validity against Monte-Carlo simulations. We find it to be a reliable tool to cheaply predict the extent and direction of the biases. 
Finally, we show how this formalism can be used to correct for biases that arise in the sequential characterisation of multiple sources in a single data set, which could be a valuable tool to use within a global-fit analysis pipeline.

\section{Introduction} 

In the analysis of data from future gravitational-wave (GW) detectors, we will be confronted with the prospect of detecting and performing parameter inference on sources that overlap with other resolved or unresolved signals. The presence of such additional signals in the data or their incomplete removal through inaccurate waveform templates, might lead to biases in the parameter estimates for the source of interest, if they are not properly accounted for. While this possibility is relevant for imminent upgrades of the LIGO-Virgo-KAGRA detectors' network~\cite{Aasi:2013wya}, the odds of this happening are higher with future ground-based and space-based detectors such as the Einstein Telescope (ET)\cite{Punturo:2010zz}, Cosmic Explorer (CE)~\cite{Reitze:2019iox} and the Laser Interferometer Space Antenna (LISA)\cite{Audley:2017drz}. 
The former is expected to detect thousands of GW signals from low-mass black holes and neutron stars~\cite{Punturo:2010zz}, the latter is guaranteed to detect tens of thousands of white dwarf binaries in the Milky Way, and is also expected to detect signals from mergers involving supermassive black holes ~\cite{Audley:2017drz}. For these future detectors, one will \emph{have to} take into account the possible presence of signals or high-SNR residuals lurking in the data. As this problem is only of peripheral relevance to analyses for the current LIGO-Virgo detector network, it has attracted relatively limited attention in the literature. In the context of ground-based detector networks, the detectability of confusion noise from a population of unresolved signals has been considered~\cite{Regimbau:2009rk}, but not the impact of the presence of that confusion foreground on parameter estimation for resolved sources. There have also been some recent Bayesian parameter estimation studies for second and third generation detectors, which computed the bias that arises in parameter estimation for a source due to the presence of another source with an overlapping merger~\cite{Samajdar:2021egv,Pizzati:2021gzd,Himemoto:2021ukb,Relton:2021cax}, and the impact of simultaneous fitting of two sources on the individual parameter precisions~\cite{1853026}. These studies were limited to just two sources and did not consider the impact of waveform modelling uncertainties. 
In the LISA context, there have been studies of the detectability of confusion foregrounds from unresolved extreme mass-ratio inspirals~\cite{Bonetti_2020}, and extensive exploration of the simultaneous global-fit of the thousands of galactic binary sources expected to be present in LISA data~\cite{Robson:2017ayy,Littenberg:2020bxy,Karnesis:2021tsh}. The latter global-fit analyses tackle the problem head-on by considering the simultaneous inference on parameters of an unknown number of sources in the data stream. Clearly, this is a formidable task due to the exceptionally large parameter space and complexity of the likelihood surface. 
It is thus important to have independent procedures to 
aid global-fit search pipelines (and potentially confirm the results).

We use semi-analytic methods based on the Fisher formalism 
to cheaply assess when confusion from other sources, and/or imperfect subtraction of those sources due to waveform errors, is likely to be problematic, in the sense of leading to significant biases in parameter estimation for a source of interest. We leverage existing metrics for the ``goodness'' of individual waveform models based on the linear signal (Fisher matrix) formalism to derive generic metrics to assess the inference biases on source parameter characterisations. We describe how to apply this approach to several cases of relevance: i) parameter estimation in the presence of ``confusion noise'' from unfitted signals in the data; ii) parameter estimation for two overlapping signals with approximately coincident coalescence times; iii) parameter estimation for a population of sources using inaccurate waveform models; and iv) the case in which both confusion noise and mismodelling errors contribute to the final biases. Finally, we will show how these results can be used to mitigate biases in a sequential-fitting pipeline for LISA.
Our analysis is related to previous work by~\cite{Flanagan:1997kp}, \cite{PhysRevD.71.104016} and~\cite{Cutler:2007mi}, in which expressions are provided for the error on parameters due to the presence of noise and due to waveform errors. While their work has been mainly considered in the context of individual signals in the data, two observations make it relevant and easily extendible to the above applications. Firstly, no assumptions are made on the source of the noise
appearing in their expressions, meaning that the observed noise can be made into a linear combination of detector noise and confusion noise [with applications to points i) and ii)]. Secondly, no assumptions are made about the dimensions of the parameter space, meaning that expressions relevant to points iii) and iv) can be derived from them. 

We illustrate these metrics for several cases of relevance to future ground-based and space-based detectors. We take the ET and LISA instruments as our examples and use simplified, but realistic models for the gravitational waveforms. We consider the following, increasingly more complex, situations:
\begin{itemize}
	\item The parameter estimation of a single LISA massive black hole source in the presence of other unfitted massive black holes forming a foreground [Sec.~(\ref{sec:source_conf})].
	\item The parameter estimation of a single ET source in the presence of an overlapping signal with time of coalescence a fraction of a second from the former [Sec.~(\ref{subsec:overlap_merger})].
	\item The parameter estimation of a single LISA source in the presence of two overlapping sources which have been incorrectly fitted out of the data [Sec.~(\ref{subsec:biases_innacurate_removal})].
	\item The simultaneous inference in LISA of a few overlapping sources, subject to waveform errors, detector noise and unresolved signals present in the data stream [Sec.~(\ref{sec:results})].
\end{itemize}
We find that unfitted foregrounds or incorrectly removed sources may lead to significant biases [as discussed in sections~Sec.~(\ref{sec:source_conf}) and~Sec.~(\ref{subsec:biases_innacurate_removal})], but that biases from confusion noise and waveform inaccuracies could deconstructively interfere [as discussed in Sec.~(\ref{sec:results})]. We qualitatively confirm one of the main results of~\cite{Samajdar:2021egv,Pizzati:2021gzd,Himemoto:2021ukb,Relton:2021cax} in Sec.~(\ref{subsec:overlap_merger}), showing that biases arise when the difference between the coalescence times of two overlapping signals is smaller than a few tens of waveform periods, corresponding to a fraction of a second for the ground-based detector examples considered in those studies. We find that the formalism herein developed is capable of predicting the biases very well (as confirmed with MCMC analyses), which makes it a useful tool for exploratory studies of future detectors. 

Finally, in Section~\ref{sec:GF}, we introduce the \emph{local-fit strategy} as a possible approach to the global-fit in LISA data analysis.
This method separately fits the parameters of individual sources, and then uses the 
Fisher-based formalism presented in (\ref{sec:Source_Confusion_Bias}) to correct the biases that result in these estimates from ignoring the other sources in the data.
Although a well-designed algorithm could, in principle, deliver a simultaneous fit to all sources for comparable computational cost, the local-fit algorithm is likely to be much easier to implement and to optimise.
We believe that this algorithm could therefore be used to aid global-fit strategies, for example by providing a quick estimate of the parameters of all sources, that could be used as a starting point for a simultaneous-fitting algorithm that then delivers the final joint posterior distribution.

The paper is organised as follows: Sec.~(\ref{sec:DA_concepts}) contains a review of basic data-analysis concepts needed throughout the paper; Sec.~(\ref{sec:Source_Confusion_Bias}) contains the description of the Fisher formalism herein developed; Sec.~(\ref{sec:model&noise}) contains a brief review of our choices of waveform models; Sec.~(\ref{sec:illustrations}) discusses the illustrations of the formalism described above; Sec~(\ref{sec:GF}) describes the local-fit strategy; and Sec.~(\ref{conclusions}) summarises our findings and describes some possible future avenues of investigation. In appendix~(\ref{app:geometry}), we discuss a geometrical interpretation for the errors from noise and the biases from mismodelling~\cite{Cutler:2007mi}; in appendix~(\ref{app:num_routines}) we describe the numerical methods used to  obtain the results reported in the previous sections; in appendix~(\ref{app:Fisher_Matrix}) we describe how we computed the Fisher matrices and how these were verified using MCMC analyses; finally, in appendix~(\ref{app:ET_gf}) we complement the LISA results of Sec.~\ref{sec:results} with results for ET. 

\section{Data Analysis Concepts} \label{sec:DA_concepts}
The data stream observed by a gravitational wave detector is a superposition of noise $n(t)$ intrinsic to the detector and a gravitational wave signal $h_{e}$ with ``true'' parameters $\boldsymbol{\theta}_{\text{tr}}$
\begin{equation}\label{eq:data}
	d(t) = h_{e}(t;\boldsymbol{\theta}_{\text{tr}}) + n(t).
\end{equation}
In general, the gravitational wave component is a combination of the signals from a number of individual sources. The consequences of this will be made explicit in Section~\ref{sec:Source_Confusion_Bias}. 
In this analysis, we make the usual assumption that the noise $n(t)$ is both stationary and Gaussian with zero mean. As a consequence of stationarity, the covariance of the noise in the frequency domain can be expressed by~\cite{wiener1930generalized,khintchine1934korrelationstheorie}
\begin{equation}\label{eq:Wiener-Khinchin-Theorem_freq}
	\langle \hat{n}(f)\hat{n}^{\star}(f')\rangle = \frac{1}{2}\delta(f - f')S_{n}(f).
\end{equation}
Here and throughout this paper, hatted quantities will denote the continuous time Fourier transform. In the above, $\delta$ denotes the Dirac delta function and $\langle \cdot \rangle$ denotes an ensemble averaging process. The quantity $S_{n}(f)$ denotes the (one-sided) power spectral density, which describes the distribution of power of the noise in the frequency domain. 

The ``loudness'' of a signal can be represented by the optimal matched filtering signal to noise ratio (SNR), the square of which is given by
\begin{equation}\label{eq:cont_SNR}
	\rho^{2} = (h|h) = 4\int_{0}^{\infty}\frac{|\hat{h}(f)|^{2}}{S_{n}(f)}df
\end{equation}
where we have defined the inner product for real valued time-series,
\begin{equation}\label{eq:inn_prod}
	(a|b) = 4\text{Re}\int_{0}^{\infty}\frac{\hat{a}(f)\hat{b}^{\star}(f)}{S_{n}(f)}df.
\end{equation}
To make inference on parameters, one requires a probabilistic model on the data stream for given unknown parameters $\boldsymbol{\theta}$. As the noise $n(t)$ is stationary and Gaussian, the Whittle (log) likelihood can be used~\cite{whittle:1957} 
\begin{equation}\label{eq:whittle_likelihood}
	\log p(d|\boldsymbol{\theta}) \propto -\frac{1}{2}(d - h_{m}|d - h_{m}).
\end{equation}
We note that the gravitational wave component of the data stream in Eq.~\eqref{eq:data},  $h_{e}(t;\boldsymbol{\theta}_{\text{tr}})$, is the true signal which depends on parameters $\boldsymbol{\theta}_{\text{tr}}$ that we wish to infer. In \eqref{eq:whittle_likelihood}, we are denoting the signal by $h_{m}$, to allow for the possibility that there is a difference between the approximate waveform templates used to analyse the data, and the true signal, $h_{e}$, present in the data stream.

Finally, to quantify the precision of measurements on parameters, we will make use of the the linear signal approximation (LSA)~\cite{Finn:1992wt}. By considering a small perturbation $\boldsymbol{\theta} = \boldsymbol{\theta}_{\text{tr}} + \Delta \boldsymbol{\theta}$, one can expand the waveform model in the vicinity of the best-fit parameters as
\begin{equation}\label{eq:LSA}
	h_{m}(t;\boldsymbol{\theta}) \approx h_{m}(t;\boldsymbol{\theta}_{\text{tr}}) + \partial_{i}h_{m}(t;\boldsymbol{\theta}_{\text{tr}})\Delta \theta^{i},
\end{equation}
which is valid for $|\Delta \theta_{\text{bf}}^{i}|\ll 1$.
We are using the standard notation $\partial_{i} = \partial/\partial \theta^i$.
Substituting \eqref{eq:LSA} into \eqref{eq:whittle_likelihood} and restricting to the case that the model and true waveform agree, $h_{m} = h_{e}$ for all $\boldsymbol{\theta}$, one obtains
\begin{align}\label{eq:fishy_likelihood}
	-2\log p(d|\boldsymbol{\theta}) &= (\Delta \theta^{i} - \Delta \theta_{\text{noise}}^{i}) \Gamma_{ij}(\Delta \theta^{j} - \Delta \theta_{\text{noise}}^{j}). \\
	\Delta \theta_{\text{noise}}^{i} &= (\Gamma^{-1})^{ij}(\partial_{j}h|n)\,,
\end{align}
where $\Gamma_{ij}$ is the Fisher matrix, with components
\begin{equation}\label{eq:Fish_Matrix}
	\Gamma_{ij} = (\partial_{i}h|\partial_{j}h).
\end{equation}
In the derivation of Eq.\eqref{eq:fishy_likelihood}, we neglected higher order terms which scale like $\mathcal{O}(\rho^{-1})$. Thus this representation of the likelihood is only valid for high SNR. Notice that \eqref{eq:fishy_likelihood} is Gaussian and centered on $\theta^{i}_{\text{bf}} = \theta^{i}_{\text{tr}} + \Delta\theta^{i}_{\text{noise}}$. 
Defining the statistic $\widehat{\Delta\theta^{i}} = \Delta\theta^{i}_{\text{noise}}$, one observes 
\begin{equation}
	\mathbb{E}[\widehat{\Delta\theta^{i}}] = 0, \quad \text{Cov}(\widehat{\Delta\theta^{i}},\widehat{\Delta\theta^{j}}) = (\Gamma^{-1})^{ij} + \mathcal{O}(\rho^{-1}).
\end{equation}
This implies that the statistic $\widehat{\Delta\theta^{i}}$ is unbiased with co-variance equal to the inverse of the Fisher matrix. In other words, the shift in the peak of the likelihood due to noise fluctuations is consistent with its width. 

In the derivation of \eqref{eq:fishy_likelihood}, we assumed the model template was consistent with the true gravitational waveform in the data set. We can relax this assumption and now consider $h_{e}\neq h_{m}$, 
which leads to a mismodelling error $\delta h(\boldsymbol{\theta})= h_{e}(t;\boldsymbol{\theta}) - h_{m}(t;\boldsymbol{\theta})$. 
The maximum of the likelihood function is at a set of parameter values $\boldsymbol{\theta}_{\text{bf}}$ that are a solution to
\begin{equation}
	(\partial_{i}h_{m}(t;\boldsymbol{\theta}_{\text{bf}})|d - h_{m}(t;\boldsymbol{\theta}_{\text{bf}})) = 0.
\end{equation}
Using the LSA \eqref{eq:LSA} and considering a perturbation $\boldsymbol{\theta}_{\text{tr}} = \boldsymbol{\theta}_{\text{bf}} + \Delta \boldsymbol{\theta}$ and a data stream $d(t) = h_e(t;\boldsymbol{\theta}_\text{tr})+n(t)$ including the true gravitational waveform, one obtains
\begin{align}\label{eq:CV_1step}
	d- h_m &= n +\delta h (\boldsymbol\theta_{\text{tr}}) + h_m(\boldsymbol\theta_{\text{tr}})-  h_m(\boldsymbol\theta_{\text{bf}})\nonumber\\
	&\approx  n + \delta h (\boldsymbol\theta_{\text{bf}}) - \Delta\theta^i \partial_i  h_m (\boldsymbol\theta_{\text{bf}})\,,
\end{align}
where in the last line we take $\delta\vec h (\boldsymbol\theta_{\text{tr}})\approx \delta\vec h (\boldsymbol\theta_{\text{bf}})$. With all waveform models evaluated at the best-fit parameters, we deduce that~\cite{Cutler:2007mi}
\begin{align}
	(\partial_i h_m|  d-h_m)
	&\approx(\partial_i h_m|  n)+(\partial_i h_m| \delta h) - \Delta\theta^j\Gamma_{ij}=0\,,\nonumber\\
	\Longleftrightarrow \Delta\theta^i &= (\Gamma^{-1})^{ij}\left[(\partial_j h_m| n)+(\partial_j h_m| \delta h)\right] \label{eq:CV_altorithm_multiple}\,.
\end{align}
where we now separate $\Delta\theta$ into an error from instrumental noise, $\Delta\theta_{\text{noise}}^i$ and a theoretical bias $\Delta\theta_{\text{sys}}^i$,
\begin{align}
	\Delta\theta_{\text{noise}}^i &= (\Gamma^{-1})^{ij}(\partial_{j}h_{m}|n), \label{eq:width_noise}\\
	\Delta\theta_{\text{sys}}^i& = (\Gamma^{-1})^{ij}(\partial_j h_{m}| \delta h). \label{eq:width_sys}
\end{align}
This expression for systematic errors first appeared in~\cite{Flanagan:1997kp}, see their Eq.~(6.11), although the implications were not studied in that paper. A much more through analysis was given in~\cite{Cutler:2007mi}. A geometrical intuition for the origin of Eqs.~(\ref{eq:width_noise},\ref{eq:width_sys}) is given in appendix~\ref{app:geometry}. A \texttt{python} tutorial on how to use equations \eqref{eq:width_noise} and \eqref{eq:width_sys} can be found \href{https://github.com/OllieBurke/Noisy_Neighbours/tree/main/Basic_CV_Formalism}{here}.

Generally speaking, a waveform model $h_{m}$ is ``good enough'' for parameter estimation if and only if $\Delta\theta^i_\text{sys}\lesssim \Delta\theta^i_\text{noise}$. The quantity $\Delta\theta^i_\text{noise}$ is a zero-mean random variable, so this inequality should hold in an average sense. 
The $1\sigma$ deviation of $\Delta \theta_{\text{noise}}^{i}$ is $\Delta\theta^{i}_{\text{stat}} = \sqrt{\Gamma^{-1})^{ii}}$, so we define the function
\begin{equation}\label{eq:R_func}
	\mathcal{R}(\Delta\theta) := \rvert \Delta\theta^{i}/\Delta\theta_{\text{stat}}^{i}\rvert,
\end{equation}
and consider biases on the parameter $\theta$ arising from systematic effects to be significant 
whenever $\mathcal{R}(\Delta\theta)>1$.
To conclude this section, we note that the statistical error $\Delta\boldsymbol\theta_\text{stat} \sim
\rho^{-1}$, while the systematic error $\Delta\boldsymbol\theta_\text{sys} \sim
\rho^{0}$. This implies that biases from modelling errors are independent of the SNR, while statistical errors become smaller as the SNR increases. Therefore, we expect systematics to become more important for loud sources.

\section{Generalisations} \label{sec:Source_Confusion_Bias}

We now generalise the formalism represented by Eqs.~\eqref{eq:width_noise} and~\eqref{eq:width_sys} to two new cases, the first being the presence of \emph{confusion noise} from signals that have not been fitted for in parameter estimation, and the second being the inclusion of multiple signals in the data stream that are incorrectly modelled with approximate waveforms. 

\subsection{Source Confusion Bias}


The likelihood \eqref{eq:whittle_likelihood} only assumes that the noise $n(t)$ is both stationary and Gaussian (with zero mean). The noise $n(t)$ is usually assumed to be instrumental and modelled through the PSD via~\eqref{eq:Wiener-Khinchin-Theorem_freq}. However, in third-generation or space-based detectors there may be additional astrophysical contributions to the data stream from unresolved 
foregrounds of other GW signals~\cite{Crowder_2007,B_aut_2010,Robson:2017ayy,Roebber:2020hso,Korol:2020hay,Samajdar:2021egv,Pizzati:2021gzd,Karnesis:2021tsh}. 
This confusion noise $\Delta H_{\text{conf}}$ can be represented as part of the signal component of the data stream~\eqref{eq:data},
\begin{equation}\label{noisetot}
	d(t) = h_{e}(t;\boldsymbol{\theta}_{\text{tr}}) + n(t) + \Delta H_{\text{conf}}(t;\boldsymbol\theta^{(i)})\,.
\end{equation}
To understand when such confusion foregrounds can lead to biases, one may consider it to be  
a (deterministic) superposition of $N$ signals,
\begin{equation}\label{confres}
	\Delta  H_\text{conf} (t;\boldsymbol\theta^{(i)})= \sum_{i=1}^{N} h^{(i)}_e (t;\boldsymbol\theta^{(i)})\,.
\end{equation}
Equation~\eqref{eq:CV_1step} now becomes
\begin{equation}
	d- h_m = n + \Delta  H_\text{conf}+ \delta h (\boldsymbol\theta_{\text{tr}}) + h_m(\boldsymbol\theta_{\text{tr}})-  h_m(\boldsymbol\theta_{\text{bf}})\,,
\end{equation} 
from which we deduce the extra contribution to the biases~\eqref{eq:width_noise} and~\eqref{eq:width_sys} that originates from the source confusion term is
\begin{equation}\label{eq:bias_conf}
	\Delta\theta^{i}_{\text{conf}} = (\Gamma^{-1})^{ij}(\partial_j  h_m| \Delta H_\text{conf}).
\end{equation}
By analogy with~\eqref{eq:R_func}, source confusion from unfitted signals can be said to bias parameter estimates when its size exceeds the 1$\sigma$ deviations arising from instrumental noise fluctuations, which is true if $\mathcal{R}(\Delta\theta_\text{conf})>1$.
To summarise, when inferring the parameters of a single source, the total error is given by the sum of statistical error from noise fluctuations and the biases from source confusion and waveform errors through
\begin{equation}\label{eq:total_bias_one_source}
	\Delta\theta^{i} = \Delta\theta_{\text{noise}}^{i} + \Delta\theta_{\text{sys}}^{i} + \Delta\theta_{\text{conf}}^{i},
\end{equation}
with the above terms from left to right given by Eqs.(\ref{eq:width_noise},\ref{eq:width_sys},\ref{eq:bias_conf}) respectively.  

In general, the confusion noise  contribution to \eqref{eq:total_bias_one_source}
depends on the particular sources from the unresolved population that are present in the data and so it is a random quantity. The correct way to handle this is to marginalise the likelihood of the corrected data stream, $d(t) - \Delta  H_\text{conf} (t;\boldsymbol\theta^{(i)})$, over the distribution of possible confusion backgrounds, $p(\Delta  H_\text{conf})$. This is a computationally expensive procedure and it is therefore difficult to obtain insights in that way. An alternative avenue to understanding when confusion is important, is to use the formalism described here to work with the bias induced by the confusion noise, $\Delta\theta_{\text{conf}}^{i}$, which is also a random quantity. 
We can characterise it at the order of the linear signal approximation through its mean and variance. Since the total confusion noise contribution is a superposition of contributions from $N$ independent sources, the mean and variance of the total contribution is $N$ times the mean and variance of the contribution from a single source, $h_{e}(\boldsymbol{\theta}_{\text{conf}})$, which are
\begin{align}
	\mu_{\text{conf}}^i 
	&= \int (\Gamma^{-1})^{ij}(\partial_j  h_m| h_e (\boldsymbol\theta_{\text{conf}})) \,p_{\text{pop}}(\boldsymbol\theta_{\text{conf}}) \, {\text{ d}} \boldsymbol\theta_{\text{conf}}\,, \label{eq:mu_conf} \\
	\Sigma_{\text{conf}}^{ij} &= \int (\Gamma^{-1})^{ik}(\partial_k  h_m| h_e (\boldsymbol\theta_{\text{conf}})) \times \nonumber \\
	&(\Gamma^{-1})^{jl}(\partial_l  h_m| h_e (\boldsymbol\theta_{\text{conf}})) \,p_{\text{pop}}(\boldsymbol\theta_{\text{conf}}) \,  {\text{ d}} \boldsymbol\theta_{\text{conf}} \nonumber \\
	&  - \mu_{\text{conf}}^i \mu_{\text{conf}}^j. \label{eq:sigma_conf}
\end{align}
Here, $p_{\text{pop}}(\boldsymbol\theta_{\text{conf}})$ is the probability density function of the population of confusion sources.
We would normally expect the mean to be close to zero, since for some sources in the population the bias would be positive and others negative and so it averages to zero (though this is not guaranteed to be the case). Regardless, the variance does not vanish, driving the total error to grow like a random walk as the total number of sources contributing to the confusion background increases. 

For large $N$, we can find a scaling relationship for the total bias using the central limit theorem 
\begin{align}
	&\frac{\sqrt{N}}{(N\Sigma_{\text{conf}})^{1/2}}(\Delta\theta_{\text{conf}} - N\mu_{\text{conf}}) \rightarrow \mathcal{N}(0,1) \\ &\Longrightarrow \Delta\theta^{i} - N \mu_{\text{conf}}^i \sim \sqrt{N} (\Sigma_{\text{conf}}^{ii})^{1/2}\,X,\label{eq:confnoiseth}
\end{align}
where $X$ is a standard Normal random variable. 
This behaviour will be investigated further in Sec.(\ref{sec:source_conf}).

In appendix \ref{App:confusion_noise}, we give a treatment of the confusion noise under the assumption $\Delta H_{\text{conf}}$ is a stationary time-series. In this prescription, making reference to the discussion above Eq.\eqref{eq:confvar}, the power of the confusion noise is folded into the PSD to form a combined noise PSD $S_{n}(f) \mapsto S_{n}(f) + S_{\text{conf}}(f)$. In realistic scenarios, due to the relative orientation of the galactic center with respect to the detector plane, the confusion noise will exhibit time-dependent amplitude modulations --- a non-stationary effect. In this work we will not treat $\Delta H_{\text{conf}}$ as a stationary time series and instead include it as an arbitrary superposition of sinusoids present in the data stream. We will treat both $n(t)$ and $\Delta H(t)$ as independent sources of noise and do not combine them into a single noise component $N(t)$.

\subsection{Biases due to waveform modelling errors}
We now generalise Equations~\eqref{noisetot} and \eqref{eq:total_bias_one_source} to the case of inference on multiple sources within the data stream. Similar ideas can be found in~\cite{Robson:2017ayy} for the case of massive black holes and galactic binaries in LISA. Here, we extend their discussion and include a prescription for the effect of waveform errors and confusion noise, generalising their results to multiple source types with an arbitrary number of sources.
We suppose there are $J$ different types of source in the data. We suppose that there are $N_j$ sources of type $j$ in the data stream, indexed by $i$, which each depend on a set of $m_j$ parameters, denoted by $\boldsymbol{\theta}^{(j)}_i$, which determine the corresponding gravitational waveform, $h^{j}(t;\boldsymbol{\theta}^{(j)}_i)$. The complete data stream can be written as
\begin{align}
	d(t) &= h(t;\boldsymbol{\Theta}) + n(t) + \Delta H_{\text{conf}}\nonumber\\
	&= \sum_{j=1}^J \sum_{i=1}^{N_j} h_{e}^{(j)}(t;\boldsymbol{\theta}^{(j)}_i) + n(t) + \Delta H_{\text{conf}}.
	\label{eq:general_data}
\end{align}
Here we have introduced a composite vector of parameters, $\boldsymbol{\Theta} = \{\boldsymbol{\theta}^{(j)}_i\}^{j = 1,\dots,J}_{i = 1,\dots,N_{j}}$, such that $\Theta_{N_{<j}+(i-1)m_j+k} = (\boldsymbol{\theta}^{(j)}_i)_k$, where $N_{<j} = \sum_{l=1}^{j-1} N_l m_l$. 
For any given parameter in $\boldsymbol{\Theta}$, there is exactly one waveform in the above sum that depends on that parameter. Thus the derivatives of the signal reduce to derivatives of the specific waveform template. The combined Fisher matrix has a block structure, with the on-diagonal blocks being the Fisher matrices for the individual sources, and the off-diagonal blocks being formed from overlaps of waveform derivatives of one source with waveform derivatives of another source. Through calculating the Fisher matrix on parameters $\boldsymbol{\Theta}$, one is able to estimate the expected precision of  measurements on individual parameters, taking into account \emph{all} parameter correlations. This is (an estimate for) the precision that would be achieved in a simultaneous coherent fit to all sources in the data.

Without loss of generality, we illustrate this considering two classes of sources, with one source in the first class ($j=1$, $N_{1} = 1$) and an arbitrary number $N_{2}$ of sources in the second ($j=2$). This split is only made for ease of exposition, and is quite arbitrary as the sources could always be relabelled so that the first source is the source of interest. We want to estimate the impact of confusion due to the presence of the population of (fitted) sources of type 2, on the precision of parameter estimation for source 1. We define the following quantities
\begin{align}
	&\Gamma^{(1)}_{jk}= \left(\partial_j  h^{(1)}(\boldsymbol\theta^{(1)})\big|\partial_k  h^{(1)}(\boldsymbol\theta^{(1)})\right)\label{eq:gamma_source_1_jk}\\
	& \left(\Gamma^{(2)}_i\right)_{jk} = \left(\partial_j  h^{(2)} (\boldsymbol\theta^{(2)}_i) \big| \partial_k h^{(2)} (\boldsymbol\theta^{(2)}_i)\right)\\
	&\left(\Gamma^{ {\text{ mix}}}_i\right)_{jk}  = \left(\partial_j h^{(1)}(\boldsymbol\theta^{(1)})\big|\partial_k  h^{(2)}(\boldsymbol\theta^{(2)}_i)\right)\label{eq:gamma_source_mix}.
\end{align}
Here $\Gamma^{(1)}$ is the Fisher matrix for the source of type 1, $\Gamma^{(2)}_i$ is the Fisher matrix for the $i$'th source of type 2 ($i = 1,\dots, N_2$) and $\Gamma^ {\text{ mix}}_i$ is the mixed Fisher matrix for the source of type 1 and the $i$'th source of type 2.
In what follows, we find it useful to combine the Fisher matrix contributions of the entire population of sources in a more compact form. One can write Eqs.(\ref{eq:gamma_source_1_jk}-\ref{eq:gamma_source_mix}) as
\begin{align}
	&\left(\Gamma^{(2)}\right)_{m_2(i-1)+j,m_2(l-1)+k}=\left(\partial_j  h^{(2)}(\boldsymbol{\theta}^{(2)}_{i}) \big| \partial_k  h^{(2)}(\boldsymbol{\theta}^{(2)}_{l})\right) \label{eq:convoluted_fish_matrix}\\
	&\Gamma^ {\text{ mix}}_{j,m_2(i-1)+k} =(\partial_j  h^{(1)}(\boldsymbol\theta^{(1)})|\partial_k  h^{(2)}(\boldsymbol{\theta}^{(2)}_{i})).\label{gammamix}
\end{align}
The Fisher matrix for the full analysis and its inverse are therefore
\begin{equation}\label{eq:inv_joint_matrix}
	\Gamma = \left( \begin{array}{cc} \Gamma^{(1)}&\Gamma^{\text{mix}} \\ (\Gamma^ {\text{mix}})^T&\Gamma^{(2)}\end{array}\right); \quad 
	\Gamma^{-1} = \left( \begin{array}{cc} \Gamma^{-1}_{11}&\Gamma^{-1}_{12} \\ (\Gamma^{-1}_{12})^T&\Gamma^{-1}_{22}\end{array}\right)
\end{equation}
with the components of the inverse\footnote{We note also that
	\begin{align*}
		\Gamma^{-1}_{11} &= (\Gamma^{(1)})^{-1} +  (\Gamma^{(1)})^{-1}  \Gamma^{\text{mix}}\Gamma^{-1}_{22}\,  (\Gamma^ {\text{mix}})^{T} (\Gamma^{(1)})^{-1} \\
		\Gamma^{-1}_{22} &= (\Gamma^{(2)})^{-1} +  (\Gamma^{(2)})^{-1}  \Gamma^{\text{mix}}\Gamma^{-1}_{11}\,  (\Gamma^ {\text{mix}})^{T} (\Gamma^{(2)})^{-1} 
	\end{align*}
	which can sometimes be cheaper to compute than Eq.~\eqref{eq:fish_matrix_22}.}
\begin{align}
	\Gamma^{-1}_{11} &= \left( \Gamma^{(1)} - \Gamma^{\text{mix}} (\Gamma^{(2)})^{-1} (\Gamma^{\text{mix}})^T\right)^{-1}\,, \label{eq:fish_matrix_11}\\
	\Gamma^{-1}_{22} &= \left( \Gamma^{(2)} - (\Gamma^{\text{mix}})^T (\Gamma^{(1)})^{-1} \Gamma^{\text{mix}}\right)^{-1}\,, \label{eq:fish_matrix_22}\\
	\Gamma^{-1}_{12} &= -\Gamma^{-1}_{11}\, \Gamma^{\text{mix}} (\Gamma^{(2)})^{-1}\,. \label{eq:fish_matrix_12}
\end{align}
The components $\Gamma^{-1}_{11} $ encode the measurement precisions for source 1. If the degree of correlation between the source types is small, i.e., $|\Gamma^{\text{mix}}| \ll 1$, we can approximate this as
\begin{equation}
	\Gamma^{-1}_{11} \approx (\Gamma^{(1)})^{-1} + (\Gamma^{(1)})^{-1} \Gamma^{\text{mix}} (\Gamma^{(2)})^{-1} (\Gamma^{\text{mix}})^T (\Gamma^{(1)})^{-1}. \label{eq:coverrorapprox}
\end{equation}
The first term is the measurement precision when there are no sources in the data, while the second represents the degradation in the precision due to confusion with the other sources. We can understand the form of the second term as follows. If the other sources were ignored when fitting for source 1, the parameter bias would be given by Eq.~\eqref{eq:width_sys}
\begin{equation}
	\Delta \theta^{(1),i}_{\text{sys}}= (\Gamma^{(1)})^{-1}_{ij} (\partial_j h_m^{(1)} | \boldsymbol{h}^{(2)})
	\label{eq:explainbias}
\end{equation}
where we are combining all of the sources of type 2 into the single term $\boldsymbol{h}^{(2)}$. This bias is dominated by the contribution from the true waveform. When we simultaneously fit for the sources of type 2, we imperfectly remove these signals, leaving a residual in the data of the form $\partial_j \boldsymbol{h}^{(2)} \Delta \theta_2^j$, where again we are combining the parameters of all of the sources of type 2 into a single parameter vector, $\boldsymbol{\theta}_2$. The parameter error, $\Delta \boldsymbol{\theta}_2$, is a random variable with covariance matrix $\langle \Delta \theta_2^j \Delta \theta_2^k \rangle = (\Gamma^{(2)})^{-1}_{jk}$. The bias on source 1 parameters can be approximated by $\Delta \theta^{(1),i}_{\text{sys}} \approx (\Gamma^{(1)})^{-1}_{ik}(\partial_{k}h^{(1)}|\partial_{l}\boldsymbol{h}^{(2)}\Delta\theta_{2}^{l})$. The covariance of the induced systematic error in the parameters of source 1 is then 
\begin{align*}
	\langle \Delta \theta^{(1),i}_{\text{sys}} \Delta \theta^{(1),j}_{\text{sys}} \rangle &= (\Gamma^{(1)})^{-1}_{ik} (\partial_k h_m^{(1)} |\partial_l \boldsymbol{h}^{(2)}) \langle \Delta \theta_2^l \Delta \theta_2^m \rangle \\
	& \hspace{1cm} (\partial_n h_m^{(1)} |\partial_m \boldsymbol{h}^{(2)}) (\Gamma^{(1)})^{-1}_{jm} \\
	&=\left[(\Gamma^{(1)})^{-1} \Gamma^{\text{mix}} (\Gamma^{(2)})^{-1} (\Gamma^{\text{mix}})^T (\Gamma^{(1)})^{-1}\right]_{ij},
\end{align*}
which is the second term from Eq.~\eqref{eq:coverrorapprox}. 
There is nothing that can be done to mitigate uncertainties of this type, which arise from an over-abundance of sources in the data. However, as described above, additional uncertainties can arise from their inaccurate modelling.
Previous studies have focused on biases from inaccurate modelling of the target source, but it is also important to ask if the inaccurate modelling of a large number of other sources can leave a sufficient residual in the data to cause problems.

To estimate this, we define $\delta h^{(1)} =  h^{(1)}_e -  h^{(1)}_m$ as the difference between the exact $h_e$ and template $h_m$ waveforms for the source of type 1, and similarly $\delta h^{(2)}_i = h^{(2)}_e(\boldsymbol\theta^{(2)}_i) -  h^{(2)}_m(\boldsymbol\theta^{(2)}_i)$ for the $i$'th source of type 2. We also define $\delta h = \delta h^{(1)} + \sum_{i=1}^{N_2} \delta h^{(2)}_i$ as the combination of all waveform residuals. 
Let us define the bias vector $\boldsymbol{b}$ 
\begin{multline*}
	\boldsymbol{b} = (b^{(1)}_{1},\ldots,b^{(1)}_{m_{1}},(b^{(2)}_{1})_{1},\ldots, \\ (b_{1}^{(2)})_{m_{2}},\ldots,(b^{(2)}_{N_{2}})_{1}, \ldots, (b^{(2)}_{N_{2}})_{m_{2}})^{T},
\end{multline*}
such that $\boldsymbol{b} = [\boldsymbol{b}^{(1)},\boldsymbol{b}^{(2)}]\in\mathbb{R}^{(m_{1} + N_{2}m_{2})\times 1}$ with individual components given by
\begin{align}
	b^{(1)}_j &= (\partial_j  h^{(1)} (\boldsymbol\theta^{(1)})| \delta  h ),  \\
	(b^{(2)}_i)_j &= (\partial_j h^{(2)}(\boldsymbol\theta^{(2)}_i) | \delta  h ). \nonumber 
\end{align}
Note that the bias defined here is only the contribution from modelling errors. The full shift in the peak of the likelihood may be found from a similar expression, with $n(t)$ and $\Delta H_{\text{conf}}$ added to $\delta h$ in the inner products. The quantity $b^{(1)}_{j}$ for $j = 1,\ldots,m_{1}$ are the components $\boldsymbol{b}$ for the first source of type 1. The quantity $(b_{i}^{(2)})_{j}$ are the $j$th components of $\boldsymbol{b}$ with respect to the $i$th source of type 2. The vector $\boldsymbol{b}$ can be written more concisely as
\begin{align}
	& b_j =  b^{(1)}_j \quad \mbox{for } j=1,\ldots,m_1, \label{eq:b_vec_sys_1}\\ 
	& b_{m_1+m_2(i-1)+j} = (b^{(2)}_i)_j \mbox{ for } i=1,\ldots,N_2; j=1,\ldots,m_2,\label{eq:b_vec_sys_2}
\end{align}
The biases computed from Eq.~\eqref{eq:width_sys} are given by $\Delta\boldsymbol\Theta = \Gamma^{-1}\boldsymbol{b}$ and are thus
\begin{equation}\label{eq:master}
	\Delta \boldsymbol\Theta := \begin{pmatrix}
		\Delta\boldsymbol\theta^{(1)}  \\[6pt]\Delta\boldsymbol\theta^{(2)}
	\end{pmatrix}=\Gamma^{-1} \begin{pmatrix}\boldsymbol{b}^{(1)}  \\[6pt]\boldsymbol{b}^{(2)}
	\end{pmatrix} 
	\,,
\end{equation}
Using Eqs.(\ref{eq:convoluted_fish_matrix}-\ref{eq:inv_joint_matrix}) and Eqs.(\ref{eq:b_vec_sys_1},\ref{eq:b_vec_sys_2}), the bias in the source parameters of the signal of type 1 is
\begin{equation}\label{eq:source_1_bias}
	\Delta \theta^{(1)}_i = (\Gamma^{-1}_{11})^{ij}  b_{j} + (\Gamma^{-1}_{12})^{im} b_{m_1+m}\,,
\end{equation}
with components of $(\Gamma^{-1}_{11})$ and $(\Gamma^{-1}_{12})$ defined in Eqs.(\ref{eq:fish_matrix_11},\ref{eq:fish_matrix_12}). Using the approximation that led to Eq.~\eqref{eq:coverrorapprox}, that the elements of $\Gamma^{\text{mix}} (\Gamma^{(2)})^{-1} (\Gamma^{\text{mix}})^T$ are much smaller than those of $\Gamma^{(1)}$, we can approximate Eq.~\eqref{eq:source_1_bias} as
\begin{multline}\label{multibias}
	\Delta \theta^{(1)}_i \approx [(\Gamma^{(1)})^{-1}]^{ij} b_{j} - \\ [(\Gamma^{(1)})^{-1}]^{ij} (\Gamma^{\text{mix}})_{jl} [(\Gamma^{(2)})^{-1}]^{lm} b_{m_1+m} 
\end{multline}
We see that there are two contributions to the parameter bias on the single source of type 1: the standard CV bias \eqref{eq:width_sys} arising from mismodelling of that source; and an extra correction due to mismodelling of overlapping sources. If the sources from each source type are orthogonal, $\Gamma^{\text{mix}}\rightarrow 0$, then the presence of other sources does not contribute a parameter bias.

In testing the formalism below, we drop the source type indices for simplicity. 
The waveform and shift in the peak of the likelihood will be denoted
\begin{align}
	h(t;\boldsymbol\Theta) &=\sum_{i=1}^{N} h_{e}(t;\boldsymbol{\theta}_i)\quad \text{for $\boldsymbol{\Theta} = \{\boldsymbol{\theta}_{1},\ldots,\boldsymbol{\theta}_{N}\}$}\nonumber\,,\\
	\Delta\Theta^{i}&=(\Gamma^{-1})^{ij}\left(\frac{\partial h}{\partial \Theta^{j}}\bigg| n(t) + \delta h + \Delta H_{\text{conf}} \right)\nonumber \\
	&= \Delta\Theta^{i}_\text{noise} + \Delta\Theta^{i}_\text{sys} + \Delta\Theta^{i}_\text{conf}\label{multipar}\,,
\end{align} 
with total theoretical error $\delta h$ and Fisher matrix denoted
\begin{equation*}
	\Gamma_{ij} = \left(\frac{\partial h}{\partial \Theta^{i}}\bigg\rvert \frac{\partial h}{\partial \Theta^{j}}\right) \label{eq:FM_one_source_type},\quad  \delta h= \sum_{i = 1}^{N}(h_{e}^{(1)}(t;\boldsymbol{\theta}_{i}^{(1)}) - h_{m}^{(1)}(t;\boldsymbol{\theta}_{i}^{(1)})).
\end{equation*}
The $\Gamma$ appearing in~\eqref{multipar} is the joint Fisher matrix $\Gamma \in \mathbb{R}^{(N\times m)\times(N\times m)}$, with $m$ the dimension of each parameter space $\boldsymbol{\theta}_{1},\ldots, \boldsymbol{\theta}_{N}$.
Equation~\eqref{multipar} is separated into a noise induced error, $\Delta\Theta^{i}_\text{noise}$, and biases split into a confusion noise contribution, $\Delta\Theta^{i}_\text{conf}$, and a contribution from theoretical errors, $\Delta\Theta^{i}_\text{sys}$. From Eq.\eqref{eq:R_func}, biases are then significant whenever $\mathcal{R}(\Delta\Theta^{i}_\text{conf} + \Delta\Theta^{i}_\text{sys})>1$.

\section{Modelling signals and noise}\label{sec:model&noise}

To illustrate the above formalism, we will consider a number of simplified scenarios. For all of these we will model the signals using the \texttt{TaylorF2} waveform model
\begin{align}\label{eq:signal_model_SPA}
	\hat h(f) &= \mathcal{A}\left(\frac{\pi G M f}{c^3}\right)^{-7/6}e^{-i\psi(f)}\,,  \\
	\mathcal{A} &= -\sqrt{\frac{5}{24}}\frac{c}{D_\text{eff}\pi^{2/3}}\left(\frac{G\mathcal{M}_c}{c^3}\right)^{5/6}\,.
\end{align}
Here, $M_c:=M \eta^{3/5}$ is the chirp mass and $D_\text{eff}$ the effective distance.
For this reason $D_\text{eff}$ should be treated effectively as an overall scaling factor, and not as a physical distance parameter. 
We retain only the leading-order amplitude $\mathcal{A}$~\cite{Allen:2005fk} in the waveform. The phase is PN-expanded in the velocity $v:=(\pi M G f/c^3)^{1/3}$ and reads
\begin{equation}
	\psi(f) = 2 \pi  f t_c - \phi_c+\frac{3v^{-5}}{128 \eta} \left(1 +\sum_{n=2}^{n=7} v^n \psi_{\frac{n}{2}\text{PN}}\right)\,,
\end{equation}
with coefficients up to 3.5PN as given in Sec.IIIB of \cite{Cutler:2007mi}.
The constant portion of the phase depends on the time and phase at coalescence, $t_c$ and $\phi_c$. We have only included spin-orbit interactions in the 1.5PN phase through the spin parameter $\beta$, defined in~\cite{Berti:2004bd}. We remark that $\beta$ satisfies the inequality $|\beta|\lesssim 9.4$.
We take the above \texttt{TaylorF2} model to be the exact waveform $\hat h_e(f;\boldsymbol\theta)$.
In these examples, for simplicity we will treat the phase, $\phi_c$, time of coalescence, $t_{c}$, and distance, $D_\text{eff}$, as perfectly-known parameters. 
Notice that we also ignore the effect of the detector response function.
Ignoring the detector response is a restrictive simplification, since over the observation time in either ground-based or spaceborne detectors we would expect the phase and amplitudes of the signal to be modulated by detector motion. Moreover, the angular dependence introduced by the detector response leads to a multi-modal and generally non-gaussian likelihood \cite{Cornish:2020vtw,Marsat:2020rtl}, which our Fisher matrix cannot reproduce. 
As
the purpose of our examples is to illustrate the formalism of Sec.(\ref{sec:Source_Confusion_Bias}) the simplifications we make here are not a serious restriction, though the impact must be assessed in future studies.

To evaluate the modelling error we need an estimate for the waveform uncertainty, which is necessarily not known exactly. If this is completely unconstrained, then modelling errors lead to non-estimable ``stealth biases'' in waveform parameters \cite{Vallisneri:2013rc}. However, in practice we generally have an idea of how large modelling uncertainties are, by comparing two different waveform models, or two different orders of expansion of the same waveform model. Given an estimated waveform difference, we can use the formalism described here to assess if that model is good enough to avoid significant systematic errors in parameter estimation.
To represent modelling inaccuracies, we represent the approximate waveform by modifying the smallest contribution in the 3.5PN phase contribution
\begin{equation}
	\psi_{\text{3.5PN}}^{(\epsilon)}:=\pi  \left(\frac{77096675}{254016}+\frac{378515}{1512}\eta -(1-\epsilon)\frac{74045}{756}\eta ^2\right)\,,
\end{equation}
(for $\epsilon \in [0,1]$). The true PN  waveform  has $\epsilon=0$  and we will take a (fixed) value of $\epsilon \neq 0$ to represent the approximate model. 
Finally, we model confusion noise as a superposition of \texttt{TaylorF2} models, unless otherwise specified (see Sec.~\ref{sec:results}).

We generate detector noise in both ET and LISA using Eq. \eqref{eq:Wiener-Khinchin-Theorem_freq} and the PSDs found in~\cite{Robson_2019} (LISA) and~\cite{regimbau2012mock} (ET). 
More details on how we generate our signals and noise realisations are found in Appendix~\ref{app:num_routines}. In Appendix~\ref{app:Fisher_Matrix} we describe how the waveform derivatives (and Fisher matrices) are calculated, and outline the MCMC techniques used to verify them. 

\section{Results} \label{sec:illustrations}

In this section, we present four illustrations for the formalism described in Sec.~\ref{sec:Source_Confusion_Bias}. The first one concerns confusion and detector noise only. The second concerns the overlap of two signals with coincident coalescence. The third concerns theoretical errors from incorrectly removed waveforms only. The fourth considers all of the above combined.

\subsection{Biases from detector and confusion noise}\label{sec:source_conf}

\begin{figure*}
	\centering
	\includegraphics[height = 7.5cm, width = 15cm]{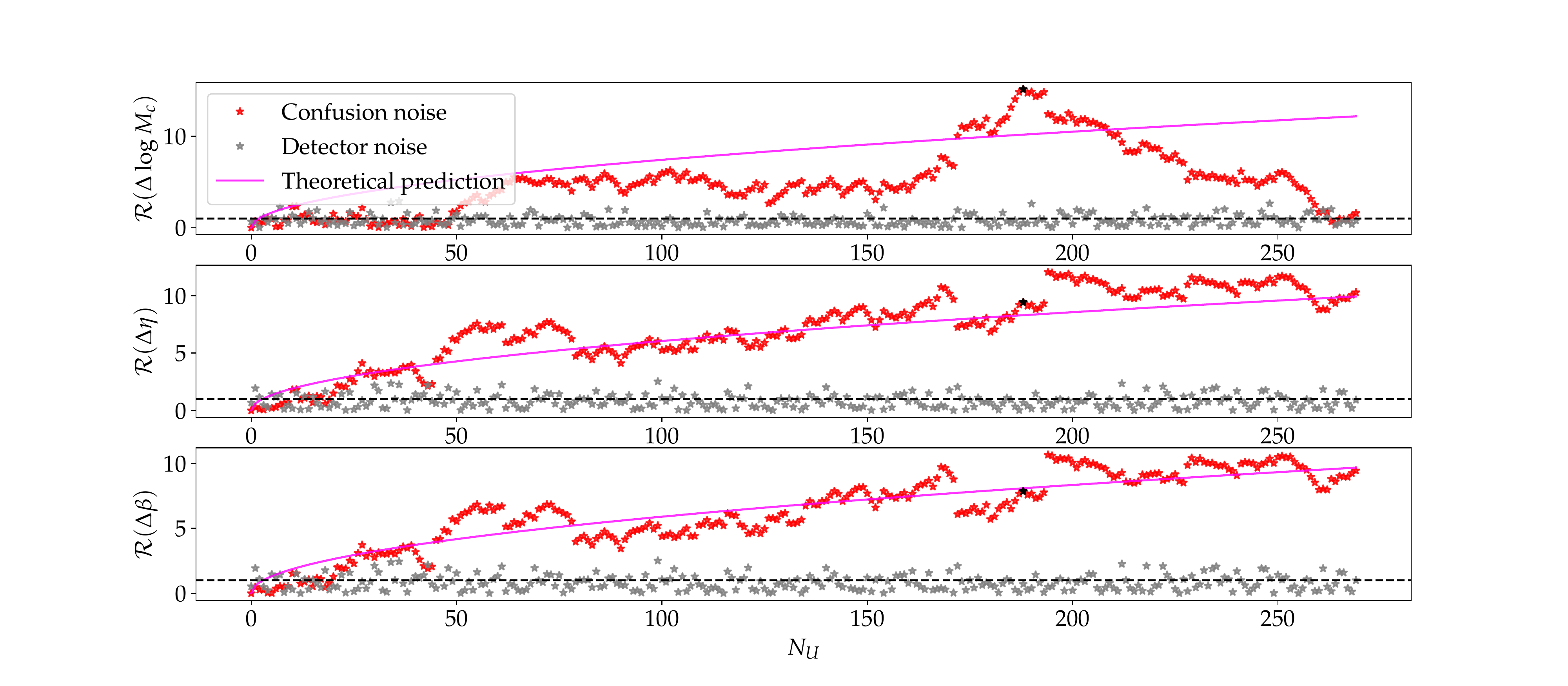}
	\caption{\emph{Accumulation of bias from population of overlapping signals}. In red, the accumulation of bias on the parameters of the reference signal from massive black hole binaries that have not been resolved. In gray, the statistical errors arising from instrumental noise fluctuations. The noise is independently generated for each data set and so we expect the ${\cal R}$ values to follow a $N(0,1)$ distribution, which is consistent with what is seen in the figure. In purple, the theoretical prediction, which follows a $\sqrt{N}$ behaviour according to Eq.\eqref{eq:confnoiseth}. In black, the data point with the largest bias in ${\cal M}_c$, for which the results were verified using an MCMC simulation, giving the posterior shown in Figure~\ref{fig:conf_noise_bias}. We note that these panels are not independent, as they represent one-dimensional marginals of a three dimensional distribution that has large correlations. 
	}
	\label{fig:conf_noise_LISA}
\end{figure*}

In this exploration, we consider a single reference signal in the LISA band and a confusion noise $\Delta H_\text{conf}$ of binaries that follow a realistic mass distribution. Our aim is to understand how much the combined effects of the confusion signals affect recovery of the parameters of the reference signal, and whether we can predict the biases using the formalism described above. The data stream we consider is
\begin{equation}\label{sig}
	\hat d(f) = \hat h_e(f;\boldsymbol\theta_\text{tr})+ \Delta H_\text{conf}(f;\boldsymbol\theta)+ \hat n(f).
\end{equation}
We recover the reference signal perfectly by modelling it with the exact waveform $\hat h_e(f; \boldsymbol\theta_\text{tr})$ of~\eqref{eq:signal_model_SPA} in both the Fisher matrix and the MCMC sampling algorithms. We therefore expect no biases from modelling errors. 
We use the following configuration of true (injected) parameters, 
\begin{equation}
	\boldsymbol{\theta}_\text{tr} = \{\log \mathcal{M}_c = 83.34, \eta = 0.210, \beta = 5.00\}\,,
\end{equation}
which correspond to a spinning binary of total mass $M = 2\times 10^6 M_\odot$. We complete the full set of parameters by choosing an effective distance $D_\text{eff} = 1$Gpc and phase at coalescence $\phi_c = 0$, with time at coalescence given by the chirp time (see appendix~\ref{app:num_routines}). We begin observing the binary at $f_0=0.25\, m$Hz and stop at $f_\text{max} = 2.2\, m$Hz, corresponding to the ISCO frequency in a Schwarzschild spacetime for the chosen total mass. That is, we observe the binary until it chirps $\sim$4.4 days after we have started observing it. These choices lead to an SNR of $\rho\sim 4200$ for this signal, for which we expect the Fisher formalism to be a very good approximation.

To construct $\Delta H_\text{conf}(f; \boldsymbol\theta)$, we first build a mock catalogue of $N=800$ sources, which are sampled from uniform distributions
\begin{align}
	\beta&\sim U[0.001,9.4] \nonumber\\
	\eta&\sim U[0.001,0.25] \nonumber\\
	\phi_c&\sim U[0, 2\pi]
\end{align}
and $t_c$ given by the individual chirp times. We distribute sources uniformly in volume by sampling distances $D_\text{eff}^3\sim U[1,125]\text{ Gpc}^3$.
We let the total masses of the binaries in this catalog follow a standard probability density function for massive black holes~\cite{Gair:2010bx,Gair:2010yu,Sesana:2010wy},
\begin{equation}
	\frac{dN}{dM} = \frac{\alpha\, M^{\alpha - 1}}{M^{\alpha}_\text{max}-M^{\alpha}_{\text{min}}},
\end{equation}
where the masses' range is $M_\text{min}=10^4M_\odot < M< 10^7 M_\odot=M_\text{max}$ and $\alpha=0.03$ is the fit in Ref.~\cite{Gair:2010yu} to the inactive massive black holes of~\cite{Greene:2007xw}. 
We can directly sample the total masses using
\begin{equation}
	\log M =\alpha^{-1} \log\big[(M^{\alpha}_\text{max}-M^{\alpha}_{\text{min}})\, u +M^{\alpha}_{\text{min}}\big],
\end{equation}
with $u \sim U[0,1]$. 
For each element of the catalogue, we compute the waveform of the binary using the exact model $h_e$. For those mass draws for which the frequency array of the binary is longer than that of the reference signal, we cut the former to be of the same length as the latter. Otherwise, we stop the evolution of the binary at its ISCO to avoid introducing an artificial portion of the waveform into the analysis. If the waveform has an observed SNR $\rho_\text{obs} = \rho + N(0,1)$~\cite{Sathyaprakash:2009xs}, where $N(0,1)$ is a standard normal distribution, such that
$\rho_\text{obs}  < \rho_\text{threshold} = 15$
then we consider the binary as ``missed'', retain the waveform and add it to $\Delta H_\text{conf}$ in a cumulative fashion. In our example, for $N=800$ events in the mock catalogue, $N_\text{U}=\mathcal{O}(270)$ have SNRs below the threshold and are thus unresolved. The final SNR of $\Delta H_\text{conf}$ is $\sim$170 in this case. 

Once $\Delta H_\text{conf}$ is obtained and the data stream~\eqref{sig} is thus fully specified, we predict the biases from confusion noise $\Delta\boldsymbol\theta_{\text{conf}}$ [namely, using~\eqref{multipar} retaining only $\Delta H_\text{conf}$ in the bias vector], which we can compare to the statistical error $\Delta\boldsymbol\theta_{\text{noise}}$ [found from~\eqref{multipar} with $n$ only]. We show the accumulation of the biases from confusion noise in Fig.~\ref{fig:conf_noise_LISA} by plotting the ratio $\mathcal{R}(\Delta\boldsymbol\theta_{\text{conf}})$. In this plot, calculations with different numbers of sources use different noise realisations, but consistent source catalogues, i.e., the data set with $N+1$ confusion sources includes the same sources as the $N$ confusion sources data set, plus one additional source. The ratio, $\mathcal{R}(\Delta\boldsymbol\theta_{\text{noise}})$, of the noise-induced shift in the peak of the likelihood to the expected standard deviation of this quantity, is  also  shown  and can be seen to hover around the value of $1$, as expected. Conversely, we find that the formalism predicts significant biases ($\mathcal{R}>1$) from the accumulation of missed signals drawn from a simple, but realistic distribution of the masses. We plot the theoretical prediction from Eq.~\eqref{eq:confnoiseth} of Sec.~\ref{sec:Source_Confusion_Bias} on top of the found ratios, showing that they (qualitatively) follow the expected $\sqrt{N}$ behaviour. We note that we do not expect the bias to precisely track the theoretical prediction. As sources are added the bias follows a random walk, and Eq.~\eqref{eq:confnoiseth} gives an approximate 1-$\sigma$ boundary to that random walk. We have tried many confusion noise realisations with $N_{U} \gg 1000$ and in all cases the accumulation of the bias follows a similar pattern. The realisation used in this figure happens to track the theoretical prediction quite well, but is reasonably typical.

\begin{figure}
	\includegraphics[width=\linewidth]{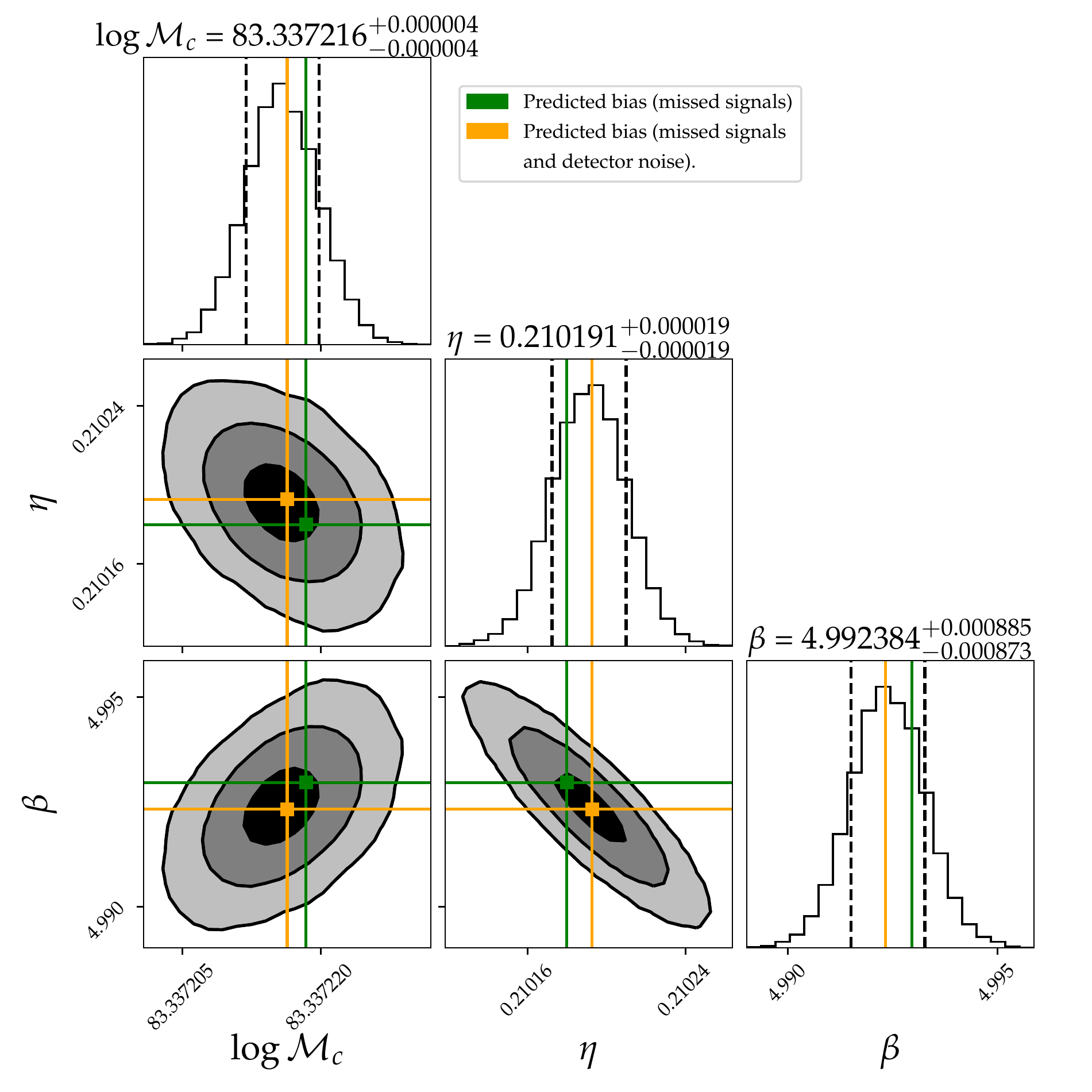}
	\caption{\emph{Biases from source confusion and detector noise}. MCMC posteriors and predictions from the Fisher formalism for the largest-bias case in Fig.~\ref{fig:conf_noise_LISA}. The values in green are predictions considering source confusion only. Those in orange combine biases from source confusion and detector noise (which we cannot access in a realistic situation). The true values are well beyond the range of the plot at $\sim 15 \sigma$ for each parameter (see Fig.~\ref{fig:conf_noise_LISA}). }
	\label{fig:conf_noise_bias}
\end{figure}

To assess whether these predictions are sound, we confirm them with an MCMC analysis for the data set that gives the largest bias (${\cal R} \sim 15$) in chirp mass, indicated by the black data point in Fig.~\ref{fig:conf_noise_LISA}). The result of the MCMC run and the predictions for the shift in the peak of the likelihood due to the confusion sources and noise, computed with  Eq.~\eqref{multipar}, are shown in Fig~\ref{fig:conf_noise_bias}. Even in this most extreme case, we can clearly see that the predictions for the bias match the MCMC posterior very well, demonstrating that the formalism works well in estimating source confusion from missed signals.
We remark that in this example the SNR of the residuals is lower than the SNR of the signal we are inferring from the data stream. This is a regime in which we would expect that the linear signal approximation is valid. In scenarios in which the SNR of the ``missed'' signals is larger than that of the target source, the linear signal approximation might cease to be valid, but this formalism should at least provide an indication that systematic biases are ``large''. 

\subsection{Biases from overlapping signals with coincident coalescence}
\label{subsec:overlap_merger}

\begin{figure*}
	\includegraphics[width=\linewidth]{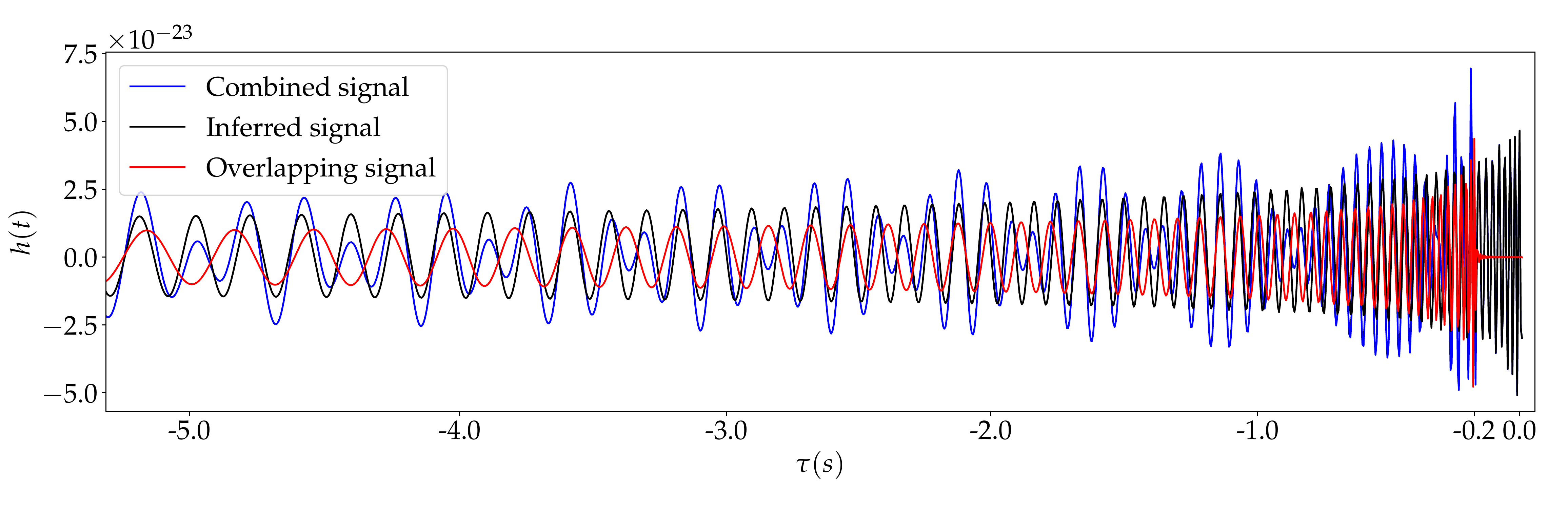}
	\caption{\emph{Waveforms for overlapping signals}. We plot the waveforms for signal $h_e^{(1)}(t)$ in black; this represents the ``inferred source'' for which we are attempting to recover the parameters. We plot the waveform of the overlapping signal $h_e^{(2)}(t)$ in red; the signal has a coalescence time at $\tau = -0.2s$ relative to the one of the inferred source. The sum of the two signals is shown in blue.}
	\label{fig:TD_waveforms}
\end{figure*}
A particularly interesting class of overlapping signals that has attracted attention in the recent literature are those where the coalescence times $t_c$ are nearly simultaneous. Such a scenario could be relevant to mergers of massive black holes observed by LISA or to stellar-origin binary black-holes (BBH) observed by ET and Cosmic Explorer (CE), but this will depend on the rate of such mergers and, therefore, the probability that mergers happen within the same time period. Quantitative studies of the rate of overlapping mergers have been carried out for advanced LIGO and CE. In~\cite{Samajdar:2021egv,Pizzati:2021gzd,Himemoto:2021ukb,Relton:2021cax}, the authors conclude that coincident (meaning merger times within 2 seconds) mergers of BBH binaries will occur tens of times per year for CE, and binary neutron star (BNS) mergers could occur coincidentally with other BNS or BBH mergers hundreds or even thousands of times per year.

The same papers, as well as~\cite{1853026}, also present the first Bayesian inference analyses with overlapping signals, with some critical differences. \cite{1853026} studies the simultaneous inference of overlapping neutron star binaries, in such a way that no biases on the parameters are expected from confusion noise. \cite{Relton:2021cax} performs a similar analysis for the second-generation LIGO-Voyager detector, \cite{Samajdar:2021egv} for pairs of BBH-BBH, BBH-BNS and BNS-BNS systems using  \texttt{LAL-inference}~\cite{Veitch:2014wba}, and \cite{Pizzati:2021gzd} for BBH pairs with \texttt{bilby}~\cite{Ashton:2018jfp}. However, in these last two papers, inference is performed for one binary only, treating the second as confusion noise. They find that biases occur when the difference between the coalescence times $\tau=t_c^{(2)}-t_c^{(1)}$ of signals ``(1)'' and ``(2)'' is sufficiently small, roughly $\tau\lesssim 0.5s$. Here we analyse a similar scenario to that of \cite{Pizzati:2021gzd}, interpreting the bias as arising from a single confusion source, to see whether the analytic formalism presented here can reproduce that result without the need for expensive Bayesian posterior computation. Notice that a (joint) Fisher-matrix analysis is presented in \cite{Himemoto:2021ukb} for a similar scenario, though the similarities end there.

\begin{figure}
	\includegraphics[width=\linewidth]{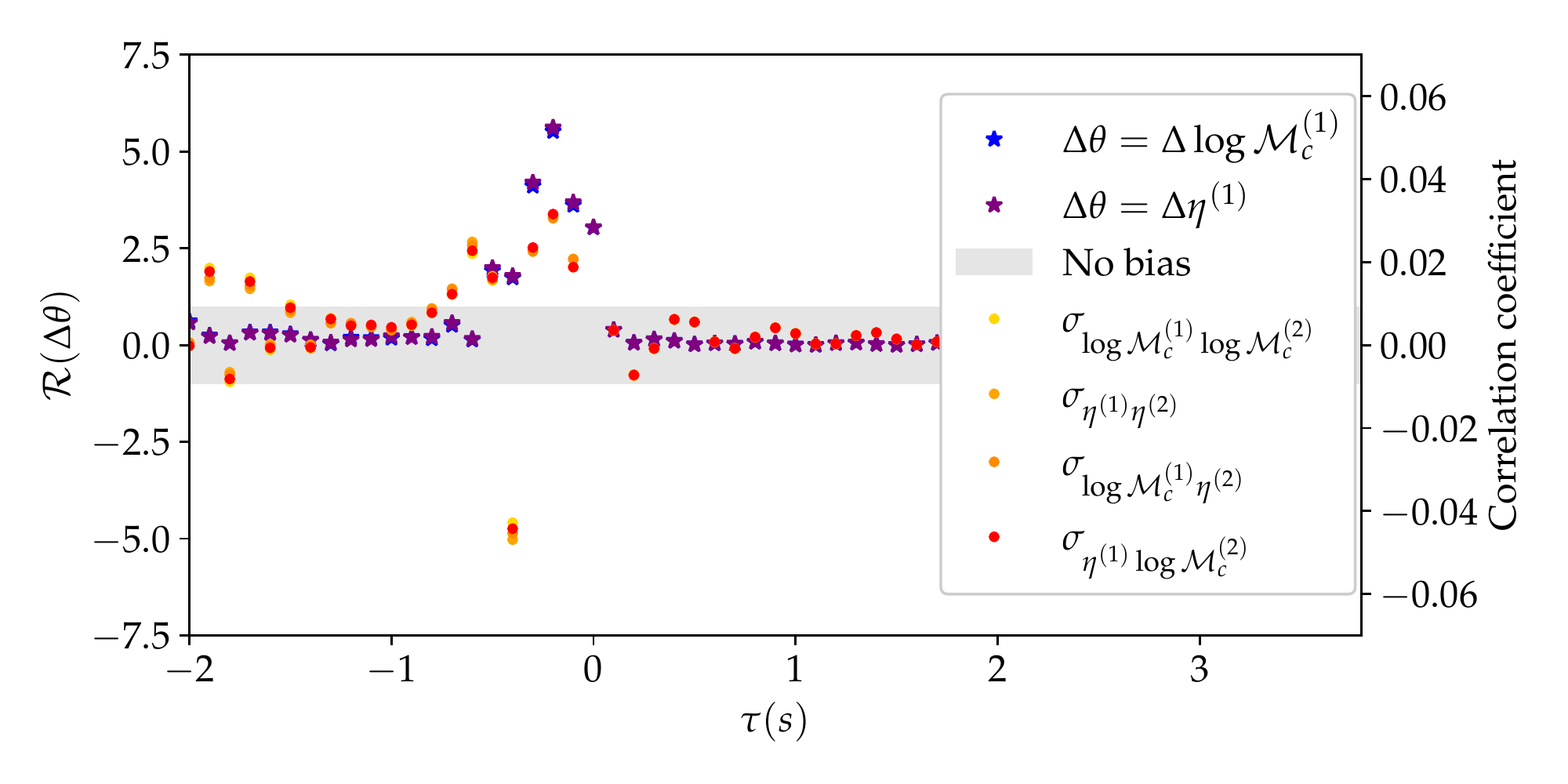}
	\caption{\emph{Biases from an overlapping signal as a function of the difference in coalescence time}. In cold (blue, purple) colors, we plot the bias ratios for the parameters of signal ``(1)'' due to the unaccounted-for presence of signal ``(2)'', as a function of the coalescence time difference $\tau$ between the two signals. The relevant scale is the y-axis on the left, where we see that biases $\mathcal{R}>1$ can arise. In gray, we indicate the region where we regard biases as not significant ($\mathcal{R}<1$). In warm colors, we plot the correlation coefficients (with relevant y-axis on the right), defined in Eq.~\eqref{eq:corrcoeff}. We see that the largest correlations $\sigma\gtrsim 0.05$ correspond to the largest biases ($\sim 6\sigma$).}
	\label{fig:R_theta_correlations}
\end{figure}

We consider an ET data stream composed of a signal $h^{(1)}$ to be inferred and a missed signal $h^{(2)}$ that creates confusion noise \begin{equation}
	\hat d(f) =\hat h_e^{(1)} (f; \boldsymbol{\theta}^{(1)}) + \hat h_e^{(2)} (f; \boldsymbol{\theta}^{(2)})\,.
\end{equation}
For this example we ignore waveform errors and detector noise. The biases arise solely due to the confusion noise $\hat h_e^{(2)}$, and can be predicted from~\eqref{multipar} setting $n=\delta h=0$. 
The parameter space of the Fisher matrix is $\boldsymbol\theta^{(1)}=\{\log \mathcal{M}_c^{(1)},\eta^{(1)}\}$, with true parameters $\boldsymbol{\theta}^{(1)}_\text{tr}=\{15.4 M_\odot, 0.243\}$ (corresponding to a binary with component masses $m_1= 21M_\odot$ and $m_2 = 15 M_\odot$). We take the signal to be nonspinning ($\beta^{(1)}=0$) with an effective distance $D_\text{eff}^{(1)}=5$Gpc, and phase and times at coalescence $\phi^{(1)}_c = \pi/3$ and $t_c^{(1)}=0s$. The SNR for this source is $\rho(h^{(1)})\sim 75$. For the overlapping signal, we pick component masses $m_1 = 25M_\odot$ and $m_2 = 20M_\odot$, a nonspinning configuration $\beta^{(2)}=0$, an effective distance $D_\text{eff}^{(2)}=10$Gpc, and phase at coalescence $\phi^{(0)}_c = \pi/3$. We let $t_c^{(2)}$ vary as a free parameter. For a nominal value of $t_c^{(2)}=-0.2s$, the SNR for the overlapping source is $\rho(h^{(2)})\sim 46$. In Fig.~\ref{fig:TD_waveforms}, we plot time-domain waveforms for this particular configuration.

We now turn to the problem of predicting the biases on $ \boldsymbol{\theta}^{(1)}$. From Eqs.~\eqref{multipar} and~\eqref{eq:R_func}, we compute the bias ratio $\mathcal{R}(\Delta\boldsymbol\theta_\text{conf}^{(1)})$ due to the presence of confusion noise, varying $\tau :=t^{(2)}_c-t^{(1)}_c$ between $\tau = -2.0$ and $\tau = 2.0$. The results are shown in Fig.~\ref{fig:R_theta_correlations}. In this Figure, we plot both the ratios $\mathcal{R}$ and the Pearson correlation coefficients\footnote{These correlations are calculated using the joint Fisher matrix, which fundamentally assumes that we have resolved \emph{both} signals. In this case, we would expect no biases from the overlapping signal. In the bias ratios calculation, we treat the second signal as unfitted, which leads us to the shown biases from confusion noise. 
	Regardless of this difference in treating the Fisher matrix, we conclude that Pearson correlations can be a guide to understand where biases would occur if the overlapping signal were not inferred, as suggested in~\cite{Pizzati:2021gzd}.
}, defined as 
\begin{equation}
	\sigma_{\theta_1 \theta_2} = \frac{(\Gamma^{-1})_{\theta_1 \theta_2}}{\sqrt{(\Gamma^{-1})_{\theta_1 \theta_1}(\Gamma^{-1})_{\theta_2 \theta_2}}}\,.\label{eq:corrcoeff}
\end{equation}
We notice that non-trivial biases start appearing when $|\tau|\lesssim 0.5$, which correspond to the largest correlation coefficients ($\sigma_{\theta_1 \theta_2}\sim 0.05$). We therefore (qualitatively) confirm the main result of~\cite{Pizzati:2021gzd} [and of \cite{Samajdar:2021egv,Himemoto:2021ukb,Relton:2021cax} indirectly]. Notice that because of our choice of data input and parameters, our comparisons with the results of~\cite{Pizzati:2021gzd} can only be qualitative. They consider noise in Advanced LIGO, while we consider ET (picking a noiseless realization in the data stream).
Furthermore, they model their signals with a different approximant (\texttt{IMRPhenomv2}), include detector response functions, sample through masses with different true values and include additional parameters in the analysis, specifically the phase, $\phi_c$, and time, $t_c$, at coalescence, and luminosity distance, $d_L$. 

To check the reliability of our bias predictions, we have also compared them against posteriors from an MCMC run for a configuration with the $\tau$ leading to the largest biases ($\sim 6\sigma$, for the $\tau=-0.2s$ configuration shown in Fig.~\ref{fig:TD_waveforms}): we obtain excellent agreement, at the level of the accuracy shown by the (orange) prediction in Fig.~\ref{fig:conf_noise_bias}. This example illustrates the advantage of our formalism, namely that the biases can be cheaply and reliably predicted. Our formalism will be a valuable tool for extending previous Bayesian analyses into regions of parameter space that are difficult to sample with fully Bayesian techniques.

\subsection{Biases from the inaccurate removal of signals}\label{subsec:biases_innacurate_removal}

\begin{figure}
	\includegraphics[width=\linewidth]{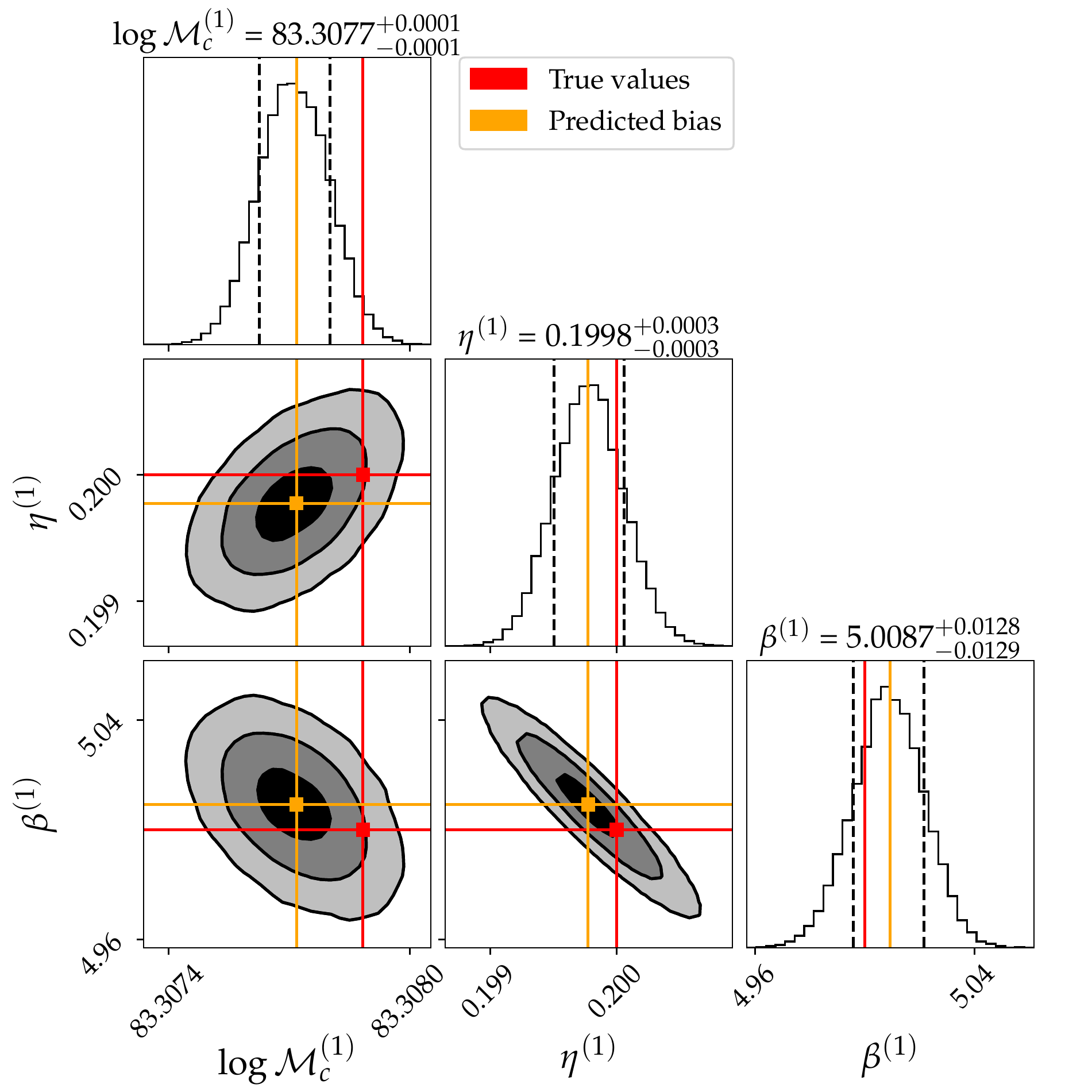}
	\caption{ \emph{Biases from the inaccurate removal of loud sources}. Posterior distributions for the parameters of a reference signal, computed using MCMC, when 2 mismodelled overlapping signals are removed from the data (with parameters given in Table~\ref{tab:example_2}). We also show the biases predicted using our formalism.}
	\label{fig:inacc_rem_bias}
\end{figure}

We now consider the situation in which the confusion sources are not ``missed'', but incorrectly fitted out. To simulate this, we consider a LISA data stream,
\begin{table}
	\centering
	\caption{Parameter configurations for the signals in Sec.~\ref{subsec:biases_innacurate_removal}. We also report the SNR of the source, $\rho_h$ and of the residual $\rho_{\delta h}$. Notice that we do not consider waveform errors for the first (reference) source here, implying its residual is zero. We sample all the sources from $f=0.5$mHz and stop at 2mHz ($T_\text{obs}=0.3$ days), the earliest chirp time for these masses. }
	\label{tab:example_2}
	\begin{tabular}{lccc|ccc|cr} 
		\hline
		i & $M/M_\odot$ & $\eta$ & $\beta$ & $D_\text{eff}$ & $t_c$ & $\phi_c$ & $\rho_h$ & $\rho_{\delta h}$\\
		\hline
		1 & $2\cdot 10^6$  & 0.20 & 5.0 & 10 $\text{Gpc}$& 6 h & 0 & 83 & - \\
		2 & $1\cdot 10^6$  & 0.23 & 1.0 & 3 $\text{Gpc}$& 48 h & $\pi$ & 790 & 31 \\
		3 & $4\cdot 10^6$  & 0.08 & 2.4 & 2 $\text{Gpc}$& 6 h & 0.9 & 2216 & 76 \\
		\hline
	\end{tabular}
\end{table}
\begin{equation}\label{data_inc_rem}
	\hat d(f) = \hat h^{(1)}_e(f; \boldsymbol\theta^{(1)}) + \hat h^{(2)}_e(f; \boldsymbol\theta^{(2)})
	+ \hat h^{(3)}_e(f; \boldsymbol\theta^{(3)})
\end{equation}
where the signal ``(1)'' is our reference signal, which we assume is modelled perfectly, and the other sources are incorrectly subtracted using approximate templates $\hat h_m(f; \boldsymbol\theta^{(2,3)},\epsilon=0.3)$. In such a procedure, we expect biases to arise only from the residual that the incorrectly modelled signals leave in the data stream~\eqref{data_inc_rem},
\begin{equation}
	\delta h = \sum_{i=2}^{3} \hat h^{(i)}_e(f;\boldsymbol\theta^{(i)})-\hat h^{(i)}_m(f;\boldsymbol\theta^{(i)}, \epsilon=0.3)\,.
\end{equation}
In this case, the  relevant parameter space is $\boldsymbol\Theta =\{\boldsymbol\theta^{(1)},\boldsymbol\theta^{(2)},\boldsymbol\theta^{(3)}\}$, where we pick each subset to be $\boldsymbol\theta^{(i)}=\{\log \mathcal{M}_c^{(i)},\eta^{(i)},\beta^{(i)}\}$. The joint Fisher matrix $\Gamma$ is therefore a 9$\times$9 matrix (calculated using $\hat h_m$). We report the true source parameters in Table~\ref{tab:example_2}. We calculate the biases $\Delta\boldsymbol\theta^{(1)}$ on the reference signal's parameters using~\eqref{eq:source_1_bias}(or equivalently~\eqref{multipar}), which leads us to
\begin{align}\label{R_res_inacc}
	&\mathcal{R}(\Delta\log \mathcal{M}_c^{(1)})=1.98 >1\nonumber\\
	&\mathcal{R}(\Delta\eta^{(1)})=0.84\nonumber\\
	&\mathcal{R}(\Delta\beta^{(1)})=0.74\,.
\end{align}
Biases are then significant for the chirp mass in this case. These predictions can be checked with an MCMC analysis, see Fig.~\ref{fig:inacc_rem_bias}. We find that the formalism can accurately predict the biases from the inaccurate removal of signals. 

The fact that each contribution to $\delta h$ in Eqs.~(\ref{eq:source_1_bias},\ref{multipar}) affects the parameters of each source equally suggests that residuals effectively behave as missed sources and confusion noise. In fact, we can rewrite the data stream analysed in Fig.~\ref{fig:inacc_rem_bias} in the form
\begin{align}\label{data_inc_rem_check}
	\hat d(f) =& \hat h^{(1)}_e(f; \boldsymbol\theta^{(1)}) +\nonumber\\
	&\hat h^{(2)}_m(f; \boldsymbol\theta^{(2)},\epsilon=0.3) + \hat h^{(3)}_m(f; \boldsymbol\theta^{(3)},\epsilon=0.3)+\delta h,
\end{align}
which explicitly separates out the modelled part using the models employed by the MCMC analysis and the calculation of the joint Fisher matrix. Doing so leaves an extra term, $\delta h$, which plays the role of the confusion noise caused by the residuals.
One can check that the biases predicted from the data stream~\eqref{data_inc_rem_check} (and obtained using the joint Fisher matrix with $\hat h_m$) match the predictions reported in Fig.~\ref{fig:inacc_rem_bias}.
An important implication of this equivalence of results is that significant biases may arise from the incorrect removal of a very large number of signals drawn from the same population, in direct analogy with the findings of the previous section. We have not checked this directly, since adding a considerable number of fitted sources dramatically increases the
dimensionality of $\boldsymbol\Theta$, making the implementation of the joint Fisher matrix difficult.

\subsection{Waveform errors \& confusion noise} \label{sec:results}

\begin{figure*}
	\centering
	\includegraphics[height = 15cm, width = 15cm]{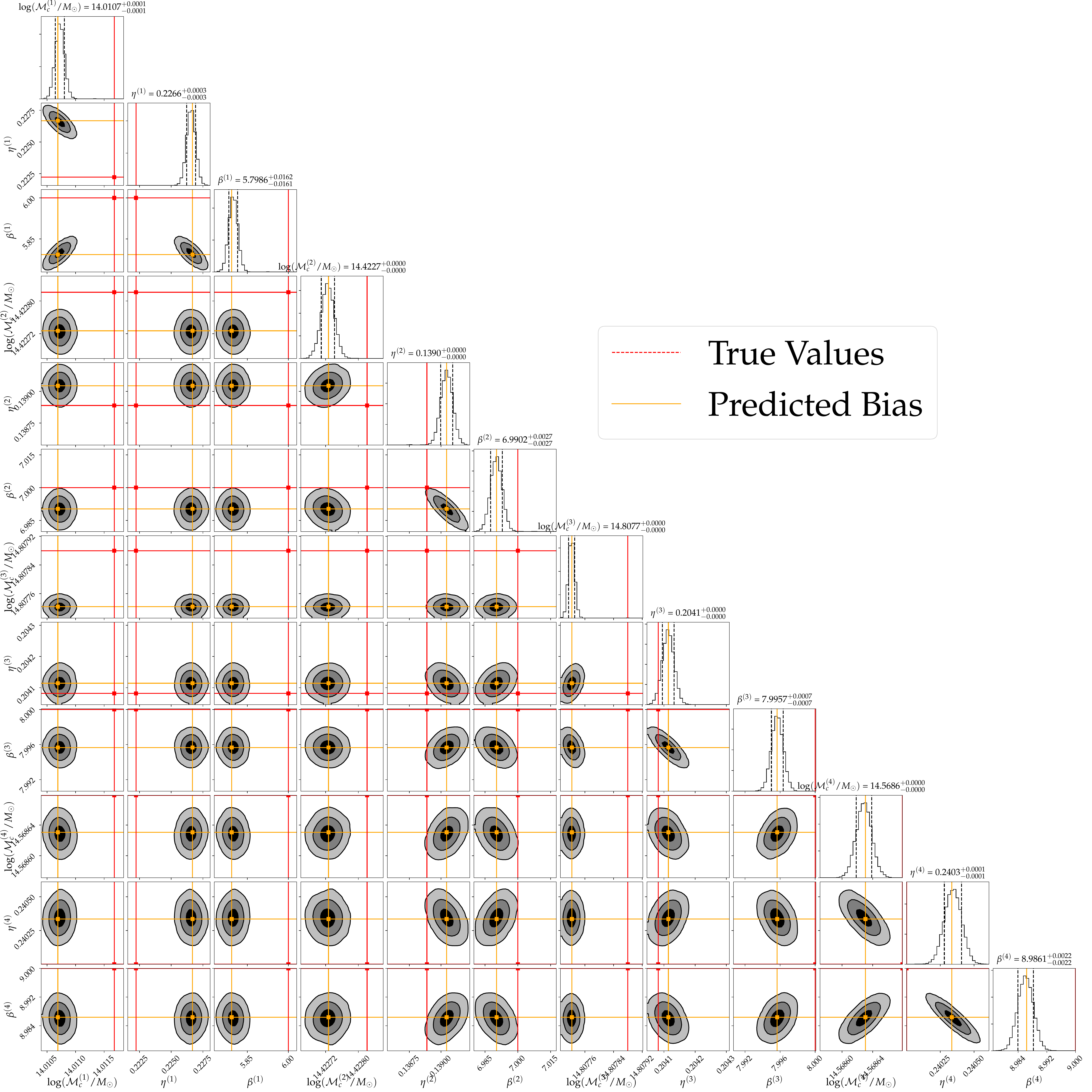}
	\caption{Triangle plot of the one-dimensional (on the diagonal) and two-dimensional marginalised posterior distributions for the inferred parameters in the LISA scenario considered in Sec.~\ref{sec:results}. The red lines indicate the true parameters and orange lines indicate the biases predicted from \eqref{multipar}.}
	\label{fig:LISA_Corner}
\end{figure*}

\begin{figure*}
	\centering
	\includegraphics[height = 5.5cm, width = 14cm]{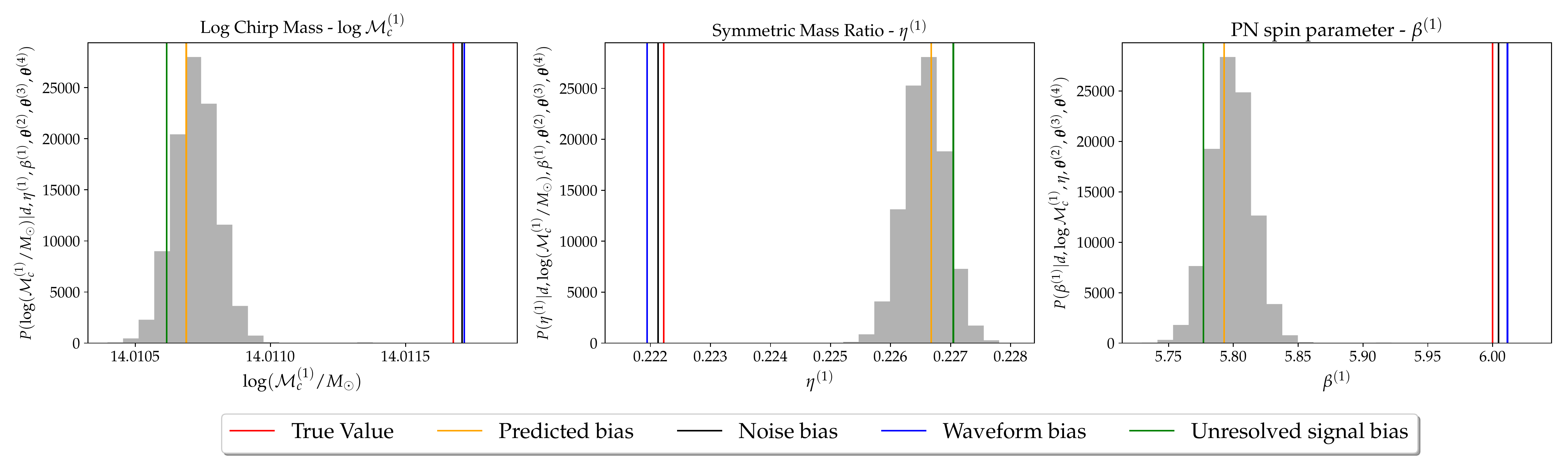}
	\caption{(top/bottom left to right) The grey histograms are the posterior samples for $\log M_{c}^{(1)},\eta^{(1)}$ and $\beta^{(1)}$ for the LISA scenario considered in Sec.~\ref{sec:results}. The red lines indicate the true parameters, blue lines the biases arising from the use of inaccurate waveforms as templates, the black ones the noise induced shift in the peak of the likelihood, the green lines the biases due to unresolved signals and the orange lines show the approximation to the total bias computed from \eqref{multipar}.}
	\label{fig:posteriors_LISA}
\end{figure*}

We now bring together the ideas described in sections (\ref{sec:source_conf}) and (\ref{subsec:biases_innacurate_removal}), and show that the formalism developed in Sec.~(\ref{sec:Source_Confusion_Bias}) can accurately predict biases on parameter estimates when we simultaneously fit $N_{\text{fit}}$ sources with inaccurate waveforms, while confusion and detector noise are also present in the data stream. We show this here for LISA, but an ET example may be found in Appendix~\ref{app:ET_gf}. The data stream in this case is
\begin{align}
	\hat{d}(f) &=  \sum_{i=1}^{N_{\text{fit}}}\hat{h}^{(i)}_{e}(f;\boldsymbol{\theta}^{(i)}_{\text{tr}}) + \Delta H_\text{conf} + \hat{n}(f).\label{eq:global_eq}
\end{align}
We assume $\Delta H_{\text{conf}}$ arises from the galactic foreground of white-dwarf binaries (WDB). LISA is guaranteed to detect WDBs in their thousands or even tens of thousands~\cite{Crowder_2007,B_aut_2010} (depending on the imposed SNR threshold), but there will also be millions of unresolved WDBs radiatig in the LISA band. Here we assume that WDBs with $\rho < 8$ have been folded into the PSD~\cite{B_aut_2010}. 
We additionally assume that only WDBs with $\rho > 15$ have been detected by dedicated pipelines, which leaves us with missed WDBs with SNRs in the range $8< \rho< 15$. To simulate these sources, we construct a superposition of signals, see Eq.~\eqref{confres}, with frequencies chosen from $f_{i}\in(10^{-4},10^{-3})$Hz. For simplicity, we only retain the leading PN term in the waveform, computed for random masses drawn from $(m_{1},m_{2}) \sim 10^{2}\cdot U^{2}[0.3,1]M_{\odot}$.
We finally draw effective distances $D_{\text{eff}} \sim 10^{4}\cdot U[1,3]\text{pc}$. We discard binaries not in the specified range of SNRs, until $N_U=1000$ are found.
To complete the input data stream, we add $N_{\text{fit}} =4$ fitted signals with waveform errors $\epsilon = 0.04$ and source parameters $\boldsymbol{\theta}^{(i)}$ given in Tab.\ref{tab:sec_5.3_table}. We choose initial frequencies $f_{0} = 10^{-4}$Hz and sample the sources simultaneously with a maximum frequency given by the highest ISCO frequency among the fitted sources. For simplicity, we set $(\phi_{c},t_{c}) = (0,T_{\text{min}})$ for all sources, where $T_{\text{min}}$ is the minimum chirping time allowed over all parameter configurations. The SNRs are of order $\mathcal{O}(10^{3})$ for all fitted sources. 

\begin{table}
	\centering
	\caption{Parameters for the simultaneously-fitted LISA signals in Sec.~\ref{sec:results}.}
	\label{tab:sec_5.3_table}
	\begin{tabular}{c|cccc}
		$i$                   & $M/M_{\odot}$ & $\eta$ & $\beta$ & $D_{\text{eff}}/\text{Gpc}$   \\ \hline
		1                     & $3 \times 10^{6}$                &  0.222      &  6       & 2    \\
		2                     & $6 \times 10^{6}$               &   0.139     &  7     &  3      \\
		3                     &    $7 \times 10^{6}$            &   0.204     &    8      & 1   \\
		4                     &    $5 \times 10^{6}$            &   0.240     &    9     &  1 
	\end{tabular}
\end{table}

Corner plots displaying \emph{all} parameter biases can be found in Fig.~\ref{fig:LISA_Corner}. We see that the predicted biases from~\eqref{multipar} are in remarkable agreement with the posteriors from the MCMC algorithm.  
Additionally, in Fig.\ref{fig:posteriors_LISA} we show how the total shift in the peak of the posterior of the parameters $\boldsymbol{\theta}^{(1)}$ of the first source, computed from Eq.~\eqref{eq:source_1_bias}, breaks down into its constituent contributions. 
Firstly, we see that biases from confusion noise, unresolved sources or waveform residuals can deconstructively interfere, i.e., the combined contribution can be smaller than the worst of the individual contributions. 
Secondly, we notice that there are large biases from confusion noise, which implies that if global-fit analyses miss $\mathcal{O}(1000)$ WDBs, this will lead to a significant bias in parameter estimates for other GW sources. We have further explored how biases change when the threshold is taken to be any value $\rho_\text{th} \in [8,15]$. We have tested that, when this threshold is increased towards $\rho_\text{th}=15$, biases tend to increase as the SNR of $\Delta H_\text{conf}$ increases. While the model used here is approximate, it suggests that the completeness of LISA data analysis algorithms needs to be sufficiently high down to sufficiently low threshold SNRs for biases on other parameters to be minimized.

\section{global-fit schemes}\label{sec:GF}

So far, we have defined the global-fit as the simultaneous search for and parameter estimation of \emph{all} gravitational wave signals in the LISA data stream. In Sec.~\ref{sec:results}, this was achieved by assuming the number of signals (and the associated parameter space) present in the data stream was known precisely. However, in a realistic scenario, we will not know how many signals are present in the data. Furthermore, 
the number of signals present at any given time may be large, leading to a prohibitively large parameter space. Consider, for example, the simultaneous inference of an extreme mass-ratio inspiral (a small compact object inspiraling into a super massive black hole) and a massive black-hole binary. Both systems will have parameter spaces $\gtrsim$ 14 dimensions, requiring parameter estimation algorithms to sample from a $\gtrsim 28$ dimensional posterior. This could stretch the capabilities of current inference techniques (especially when correlations between parameters of different sources are high). The problem is likely to worsen as more signals are included in the model. One solution is to use state-of-the-art parameter estimation techniques that are able to efficiently sample such complicated, high-dimensional posterior distributions. In principle, such methods would be no more computationally expensive than the method we describe here. However, it is likely to be difficult to design an algorithm that can robustly and efficiently sample from the full global-fit posterior, and so it is valuable to consider alternative approaches that are easier to implement, and more robust. We will describe one such alternative idea in this section. 
We begin by proposing an (expensive) iterative approach to sample reduced portions of the parameter space. Then, using the formalism developed above, we illustrate how to cheaply correct for the biases arising within the first few parameter estimation simulations. The final posterior estimates will not be as accurate as those from a simultaneous global-fit, and so this algorithm cannot fully replace a general global-fit analysis. However, the approach is worth exploring as it could provide a quicker and easier way to obtain an accurate initial estimate of the source parameters and their uncertainties. This could then be used to assist the global-fit, for example by providing a starting point for further sampling and refinement, or by providing a proposal distribution to use within the global-fit sampler, or by just providing a cross-check of the results~\footnote{We note here that cross checks are likely to be useful only in the domain in which the Fisher matrix is a good approximation for all the considered parameters. The range of applicability of the Fisher matrix, whose extent is to be substantiated with future analyses, may be further restricted with the addition of realistic features such as the detector response functions.}, to ensure that the global-fit sampler has converged. 

\subsection{Parameter Estimation through local-fits}
Let $h^{(\mathcal{A})} \in \mathcal{A}$ and $h^{(\mathcal{B})} \in \mathcal{B}$ denote a set of distinct signals with parameters $\boldsymbol{\theta}^{\mathcal{A}}_{\text{tr}}$ and $\boldsymbol{\theta}^{\mathcal{B}}_{\text{tr}}$ we wish to infer. The joint data stream is given by 
\begin{equation}\label{eq:gf_joint_data_stream}
	d(\mathcal{A},\mathcal{B}) = \sum_{\mathcal{A}} h^{(\mathcal{A})}_{e}(\boldsymbol{\theta^{\mathcal{A}}_{\text{tr}}}) + \sum_{\mathcal{B}} h^{(\mathcal{B})}_{e}(\boldsymbol{\theta^{\mathcal{B}}_{\text{tr}}}) + n(t).
\end{equation} 
For simplicity, we ignore effects coming from unresolved signals. Global-fit pipelines are concerned with the data stream \eqref{eq:gf_joint_data_stream} with the goal to simultaneously infer both signal sets $h^{(\mathcal{A})} \in \mathcal{A}$ and $h^{(\mathcal{B})} \in \mathcal{B}$.

In a local-fit procedure, we consider performing parameter estimation \emph{only} on signal set $\mathcal{A}$ and treat signals from the set $\mathcal{B}$ as missed signals. 
We write this data stream as
\begin{align}
	d(\mathcal{A}|\mathcal{B}) &= \sum_{\mathcal{A}} h^{(\mathcal{A})}_{e}(\boldsymbol{\theta^{\mathcal{A}}_{\text{tr}}}) + \Delta H_{\text{conf}} + n(t),  \label{eq:gf_A_g_B}\\
	\Delta H_{\text{conf}} &= \sum_{\mathcal{B}} h^{(\mathcal{B})}_{e}(\boldsymbol{\theta^{\mathcal{B}}_{\text{tr}}}).
\end{align}
The best fit parameters for $\mathcal{A}$ obtained in this stage can be denoted $\boldsymbol{\theta}^{\mathcal{A}|\mathcal{B}}_{\text{bf}}$, the conditioning on ${\cal B}$ indicating that the estimate was obtained with ${\cal B}$ present in the data. In the second step, we use the recovered parameters $\boldsymbol{\theta}^{\mathcal{A}|\mathcal{B}}_{\text{bf}}$ to subtract out an estimate of $h^\mathcal{A}$ from the joint data stream using our approximate model
\begin{equation}\label{eq:gf_PE_B_A_step_1}
	d(\mathcal{B}|\mathcal{A}_{\text{res}}) = d(\mathcal{A},\mathcal{B}) - \sum_{\mathcal{A}}h_{m}^{(\mathcal{A})}(\boldsymbol{\theta}^{\mathcal{A}|B}_{\text{bf}}).
\end{equation}
Then one estimates the parameters of signal $\mathcal{B}$ using the data stream \eqref{eq:gf_PE_B_A_step_1} with signal templates representing signals in $\mathcal{B}$. This will yield parameters $\boldsymbol{\theta}^{\mathcal{B}|\mathcal{A}_{\text{res}}}_{\text{bf}}$, where $\mathcal{A}_{\text{res}}$ indicates that this analysis was done on a ``residual data set'' from which an estimate of $h^{\cal A}$ had been subtracted. This estimate can be used to \emph{update} the initial data stream $d(\mathcal{A}|\mathcal{B})$, now denoted $d(\mathcal{A}|\mathcal{B}_{\text{res}}) = d(\mathcal{A},\mathcal{B}) - \sum_{\mathcal{B}}h^{(\mathcal{B})}(\boldsymbol{\theta}^{\mathcal{B}|\mathcal{A}_{\text{res}}}_{\text{bf}})$. Again, we can perform parameter estimation on signals $\mathcal{A}$, now with residuals from $\mathcal{B}$ in the data stream, using this updated data array and recovering $\boldsymbol{\theta}^{\mathcal{A}|\mathcal{B}_{\text{res}}}_{\text{bf}}$. These recovered parameters should be \emph{closer} to the true parameters than $\boldsymbol{\theta}^{\mathcal{A}|\mathcal{B}}_{\text{bf}}$. We can continue this scheme by then searching over \begin{equation}\label{eq:gf_PE_B_A_step_2}
	d(\mathcal{B}|\mathcal{A}_{\text{res}}) = d(\mathcal{A},\mathcal{B}) - \sum_{\mathcal{A}}h_{m}^{(\mathcal{A})}(\boldsymbol{\theta}^{\mathcal{A}|\mathcal{B}_{\text{res}}}_{\text{bf}}),
\end{equation} recovering parameters, then searching over $d(\mathcal{A}|\mathcal{B}_{\text{res}})$, and so on and so forth. What we would find is that the recovered parameters for both $\boldsymbol{\theta}^{\mathcal{A}}_{\text{bf}}$ and $\boldsymbol{\theta}^{\mathcal{B}}_{\text{bf}}$ tend towards the ``true'' parameters, i.e., the parameters that would have been recovered if a global fit procedure was carried out. 
An advantage of this procedure is that it sidesteps issues arising from sampling the joint posterior for $\mathcal{A}$ and $\mathcal{B}$, but a clear disadvantage is that it requires a number of repeated parameter inference calculations. Computationally, this is expensive and time consuming. As an alternative, we propose that one can use the algorithm presented in Sec.(\ref{sec:Source_Confusion_Bias}) to correct the biases found above. In doing so, one may be able to get a reliable estimate of the true parameters $\boldsymbol{\theta}^{(\mathcal{A})}_{\text{tr}}$ and $\boldsymbol{\theta}^{(\mathcal{B})}_{\text{tr}}$ without having to iterate, i.e., using just the first two parameter inference calculations.

\subsection{Correcting biases in the local-fit analysis}

Before we talk about the details of our algorithm, it is instructive to discuss the source of the biases in parameters $\boldsymbol{\theta}^{\mathcal{A}|\mathcal{B}}_{\text{bf}}$ and $\boldsymbol{\theta}^{\mathcal{B}|\mathcal{A}_{\text{res}}}_{\text{bf}}$. 
For the data stream \eqref{eq:gf_A_g_B}, the bias in the recovered parameter $\boldsymbol{\theta}^{\mathcal{A|B}}_{\text{bf}}$ is sourced by
\begin{equation}\label{eq:gf_A_g_B_h}
	\delta h^{\mathcal{A}|\mathcal{B}} = \sum_{\mathcal{B}} \hat{h}^{(\mathcal{B})}_{e}(\boldsymbol{\theta}_{\text{tr}}) +
	\sum_{\mathcal{A}}\left[\hat{h}_{e}^{(\mathcal{A})}(\boldsymbol{\theta}^{(\mathcal{A})}_{\text{tr}}) - 
	\hat{h}_{m}^{(\mathcal{A})}(\boldsymbol{\theta}^{(\mathcal{A})}_{\text{tr}})\right]  + \hat{n}(f),
\end{equation}
and similarly the bias in $\boldsymbol{\theta}^{\mathcal{B}}_{\text{tr}}$ when performing PE on the data stream \eqref{eq:gf_PE_B_A_step_1}
\begin{multline}\label{eq:gf_B_g_A_res_h}
	\delta h^{\mathcal{B}|\mathcal{A}_{\text{res}}} = \sum_{\mathcal{A}} \left[\hat{h}_{e}^{(\mathcal{A})}(\boldsymbol{\theta}^{(\mathcal{A})}_{\text{tr}}) - \hat{h}_{m}^{(\mathcal{A})}(\boldsymbol{\theta}^{\mathcal{A}|\mathcal{B}}_{\text{bf}})\right]
	+ \\
	\sum_{\mathcal{B}} \left[\hat{h}_{e}^{(\mathcal{B})}(\boldsymbol{\theta}^{(\mathcal{B})}_{\text{tr}}) - 
	\hat{h}_{m}^{(\mathcal{B})}(\boldsymbol{\theta}^{(\mathcal{B})}_{\text{tr}})\right]
	+ \hat{n}(f).
\end{multline}
In Eq.\eqref{eq:gf_A_g_B_h}, the first term is the bias due to missed signals $\mathcal{B}$, the second term the residuals due to incorrect subtraction of the true signals and finally the noise. The noise related bias should be consistent with the width of the posterior. Also, the errors due to inaccurate waveforms should decrease as more accurate waveforms are developed. Thus, we believe it is reasonable to assume that the dominant contribution to the bias comes from the first term in Eq.\eqref{eq:gf_A_g_B_h}. A similar story can be told for Eq.\eqref{eq:gf_B_g_A_res_h} where we expect the first term will dominate and the latter two will be subdominant corrections. 
Finally, we do not have access to the true parameters $\boldsymbol{\theta}^{\mathcal{A}}_{\text{tr}}$ and $\boldsymbol{\theta}^{\mathcal{B}}_{\text{tr}}$, nor the exact models for $h_{e}^{(\mathcal{A})}$ or $h_{e}^{(\mathcal{B})}$. We make a further approximation for the $\mathcal{B}$ true parameters $\boldsymbol{\theta}^{\mathcal{B}}_{\text{tr}} \approx  \boldsymbol{\theta}^{\mathcal{B}|\mathcal{A}_{\text{res}}}_{\text{bf}}$ and assume that $h_{e} \approx h_{m}$. We have access to these parameters from our first parameter estimation run on signal set  $\mathcal{B}$ using the data stream $d(\mathcal{B}|\mathcal{A}_{\text{res}})$. From this information, we can approximate both Eqs.\eqref{eq:gf_A_g_B_h} and Eq.\eqref{eq:gf_B_g_A_res_h} by 
\begin{align}
	\delta h^{\mathcal{A}|\mathcal{B}} &\approx \sum_{\mathcal{B}} \hat{h}^{(\mathcal{B})}_{m}(\boldsymbol{\theta}^{\mathcal{B}|\mathcal{A}_{\text{res}}}_{\text{bf}}) \label{eq:gf_A_B_approximate}\\
	\delta h^{\mathcal{B}|\mathcal{A}_{\text{res}}} &\approx \sum_{\mathcal{A}} \left[\hat{h}_{m}^{(\mathcal{A})}(\boldsymbol{\theta}^{(\mathcal{A})}_{\text{tr}}) - \hat{h}_{m}^{(\mathcal{A})}(\boldsymbol{\theta}^{\mathcal{A}|\mathcal{B}}_{\text{bf}})\right]. \label{eq:gf_B_A_res_approximate}
\end{align}
A similar complication arises from our lack of access to $\boldsymbol{\theta}^{\mathcal{A}}_{\text{tr}}$ in Eq.\eqref{eq:gf_B_A_res_approximate}. However, the true parameter $\boldsymbol{\theta}^{\mathcal{A}}_{\text{tr}}$ can be estimated by calculating the CV bias using $\delta h^{\mathcal{A}|\mathcal{B}}$ from Eq.\eqref{eq:gf_A_B_approximate} with the Fisher matrix and numerical derivatives calculated at parameter values $\boldsymbol{\theta}^{\mathcal{A}|\mathcal{B}}_{\text{bf}}$. This will produce an estimate of the bias, $\Delta\boldsymbol{\theta}^{\mathcal{A}|\mathcal{B}}$, which can be  subtracted from $\boldsymbol{\theta}^{\mathcal{A}|\mathcal{B}}_{\text{bf}}$, to give an updated estimate of $\boldsymbol{\theta}^{\mathcal{A}}$ that should lie closer to the true parameters, $\boldsymbol{\theta}^{\mathcal{A}}_{\text{tr}}$. This new parameter $\widehat{\boldsymbol{\theta}^{\mathcal{A}|\mathcal{B}}_{\text{bf}}} =  \boldsymbol{\theta}^{\mathcal{A}|\mathcal{B}}_{\text{bf}} - \Delta \boldsymbol{\theta}^{\mathcal{A}|\mathcal{B}}$ can be used to approximate $\boldsymbol{\theta}^{\mathcal{A}}_{\text{tr}}$ in Eq.\eqref{eq:gf_B_A_res_approximate}. Finally, using parameter values $\boldsymbol{\theta}^{\mathcal{B}|\mathcal{A}}_{\text{res}}$ to evaluate waveform derivatives and Fisher matrices, one can compute a new estimate of the bias in the $\mathcal{B}$ set signal parameters, $\Delta\boldsymbol{\theta}^{\mathcal{B}|\mathcal{A}_{\text{res}}}$ by using Eq.\eqref{eq:gf_B_A_res_approximate} with $\widehat{\boldsymbol{\theta}^{\mathcal{A}|\mathcal{B}}_{\text{bf}}} \approx \boldsymbol{\theta}^{A}_{\text{tr}}.$ This new bias can be used to update our best guess for the true parameters if the set of $\mathcal{B}$ signals, namely $\widehat{\boldsymbol{\theta}^{\mathcal{B}|\mathcal{A}_{\text{res}}}_{\text{bf}}} =  \boldsymbol{\theta}^{\mathcal{B}|\mathcal{A}_{\text{res}}}_{\text{bf}} - \Delta \boldsymbol{\theta}^{\mathcal{B}|\mathcal{A}_{\text{res}}}$. By construction, the parameter values $\widehat{\boldsymbol{\theta}^{\mathcal{A}|\mathcal{B}}_{\text{bf}}}$ and  $\widehat{\boldsymbol{\theta}^{\mathcal{B}|\mathcal{A}_{\text{res}}}_{\text{bf}}}$ should lie closer to the true values $\boldsymbol{\theta}^{\mathcal{A}}_{\text{tr}}$ and $\boldsymbol{\theta}^{\mathcal{B}}_{\text{tr}}$ respectively. 

To summarise, the algorithm is as follows
\begin{enumerate}
	\item Calculate $\boldsymbol{\theta}^{\mathcal{A}|\mathcal{B}}_{\text{bf}}$ and $\boldsymbol{\theta}^{\mathcal{B}|\mathcal{A}_{\text{res}}}_{\text{bf}}$ by performing PE on signals $\mathcal{A}$ and $\mathcal{B}$ using data streams $d(\mathcal{A}|\mathcal{B})$ then $d(\mathcal{B}|\mathcal{A}_{\text{res}})$.
	\item Calculate 
	\begin{equation}
		\delta h^{\mathcal{A}|\mathcal{B}}_{\text{conf}} \approx \sum_{\mathcal{B}} \hat{h}_{m}(\boldsymbol{\theta}^{\mathcal{B}|\mathcal{A}_{\text{res}}}_{\text{bf}})
	\end{equation}
	and then compute an estimate of the bias on the parameters specific to $\mathcal{A}$, 
	denoted $\Delta \boldsymbol{\theta}^{\mathcal{A}|\mathcal{B}}_{\text{bf}}$, evaluating the waveform derivatives at the parameter values $\boldsymbol{\theta}^{\mathcal{A}|\mathcal{B}}_{\text{bf}}$. Set new best fit parameters for $\mathcal{A}$ 
	as   $\widehat{\boldsymbol{\theta}^{\mathcal{A}|\mathcal{B}}_{\text{bf}}} =  \boldsymbol{\theta}^{\mathcal{A}|\mathcal{B}}_{\text{bf}} - \Delta \boldsymbol{\theta}^{\mathcal{A}|\mathcal{B}}_{\text{bf}}.$
	\item Then calculate
	\begin{equation}
		\delta h^{\mathcal{B}|\mathcal{A}_{\text{res}}}_{\text{conf}} = \hat{h}_{m}(\widehat{\boldsymbol{\theta}^{\mathcal{A}|\mathcal{B}_{\text{res}}}_{\text{bf}}}) -  \hat{h}_{m}(\boldsymbol{\theta}^{\mathcal{A}|\mathcal{B}_{\text{res}}}_{\text{bf}}) 
	\end{equation}
	and calculate the CV bias $\Delta \boldsymbol{\theta}^{\mathcal{B}|\mathcal{A}_{\text{res}}}_{\text{bf}}$ on parameters specific to $\mathcal{B}$ 
	using parameter values $\boldsymbol{\theta}^{\mathcal{B}|\mathcal{A}_{\text{res}}}_{\text{bf}}$. Now set new parameters  $\widehat{\boldsymbol{\theta}^{\mathcal{B}|\mathcal{A}_{\text{res}}}_{\text{bf}}} =  \boldsymbol{\theta}^{\mathcal{B}|\mathcal{A}_{\text{res}}}_{\text{bf}} - \Delta \boldsymbol{\theta}^{\mathcal{B}|\mathcal{A}_{\text{res}}}_{\text{bf}}$.
\end{enumerate}

We illustrate the algorithm above by considering a noisy data stream containing two signals, each of which have waveform errors $\epsilon \neq 0$. We lose no generality here since the algorithm presented above is easily generalised to handle a greater number of signals. Thus we consider
\begin{equation}\label{eq:GF_full_data}
	\hat{d}(f) = \underbrace{\hat{h}^{(1)}_{e}(f;\boldsymbol{\theta}^{(1)},\epsilon = 10^{-3})}_{\mathcal{A}} + \underbrace{\hat{h}^{(2)}_{e}(f;\boldsymbol{\theta}^{(2)},\epsilon = 10^{-3})}_{\mathcal{B}} + \hat{n}(f).
\end{equation}
With parameters for the $\mathcal{A}$ and $\mathcal{B}$ sources given in table \ref{tab:GF_params}. The results of applying the local-fit procedure are presented in the next section.
\begin{table}
	\centering
	\caption{This table presents the true parameter values for source 1 $(\mathcal{A})$ and source 2 $(\mathcal{B})$ for the example of the local-fit procedure presented in section~\ref{sec:LFres}. 
		The SNR of each signal within the data stream $\rho_{h}^2 = (h_{\text{e}}|h_{\text{e}})$ is given in the final column.}
	\label{tab:GF_params}
	\begin{tabular}{c|cccc|c} 
		\hline
		$i$ & $M/M_{\odot}$ & $\eta$ & $\beta$ & $D_\text{eff}/\text{Gpc}$ & $\rho_{h}$ \\
		\hline
		1 $(\mathcal{A})$ & $1.2 \times 10^{7}$  & 0.222 & 8 & 2 & $\sim1850$   \\
		2 $(\mathcal{B})$ & $5 \times 10^{6}$ & 0.160 & 7 & 4 & $\sim379$\\
		\hline
	\end{tabular}
\end{table}
\begin{figure*}
	\centering
	\includegraphics[height = 10cm, width = 14cm]{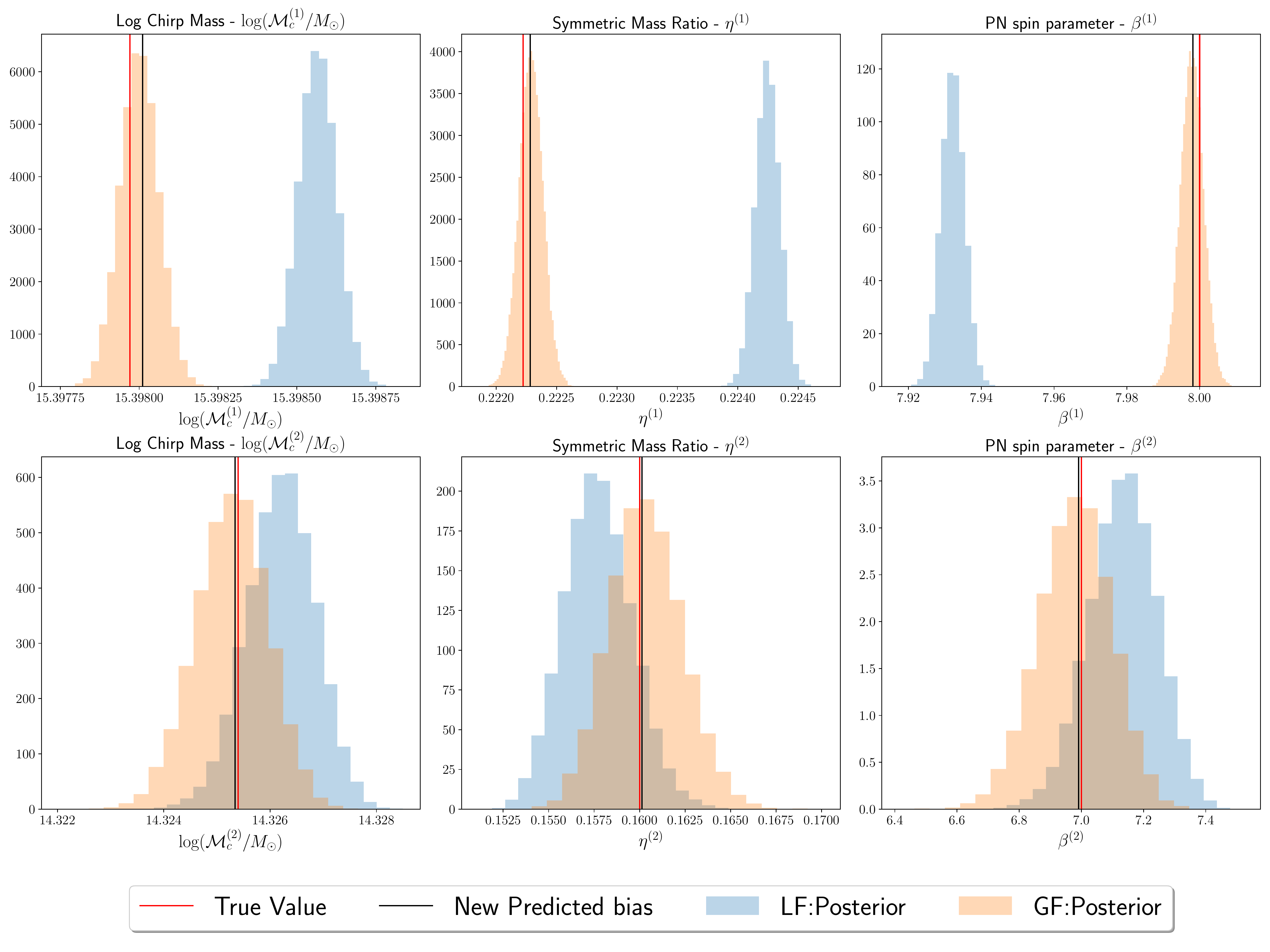}
	\caption{The orange histograms are the global-fit (GF) posteriors from searching the joint data stream $d(\mathcal{A},\mathcal{B})$ for both $\mathcal{A}$ and $\mathcal{B}$ simultaneously. The red lines are true values and black lines the (corrected) new predicted bias using the generalised CV formalism in section \ref{sec:Source_Confusion_Bias}. The blue histograms in the top row are posterior samples from the local-fit (LF) $p(\mathcal{A}|d(\mathcal{A}|\mathcal{B}),\mathcal{B})$ for missed signals $\mathcal{B}$. Similarly, the blue histograms in the bottom row are samples from $p(\mathcal{B}|d(\mathcal{B}|\mathcal{A}_{\text{res}}),\mathcal{A}_{\text{res}})$ for signal $\mathcal{B}$.}
	\label{fig:GF_analysis}
\end{figure*}

\subsection{Results}
\label{sec:LFres}
Following the algorithm above, we present results for the marginalised posteriors in Fig.~\ref{fig:GF_analysis}. In the top row, the blue histogram is the posterior $p(\mathcal{A}|d(\mathcal{A}|\mathcal{B}),\mathcal{B})$ obtained fitting for source $\mathcal{A}$ with source $\mathcal{B}$ in the data, the orange histogram is the posterior for the global-fit solution $p(\mathcal{A},\mathcal{B}|d(\mathcal{A},\mathcal{B}))$, the red lines mark the true parameters and the black line the predicted bias using the formalism. The bottom row of figure \ref{fig:GF_analysis} show corresponding results for the inference of source $\mathcal{B}$, with, for example, the orange histograms representing posterior samples from $p(\mathcal{B}|d(\mathcal{B}|\mathcal{A}_{\text{res}}),\mathcal{A}_{\text{res}})$. In each case, the algorithm is able to correct the bias from the poorly subtracted other signal in the data.
In all cases, after subtracting the predicted bias, the true parameters lie within the $1\sigma$ width of the 
posteriors.

In fig.(\ref{fig:GF_analysis}), the local fit posterior for source $\mathcal{B}$ appears to provide a more conservative estimate on how well we can constrain each parameter in comparison to the global fit analysis. Shifting the posterior by the amount predicted by the preceding algorithm will therefore yield a posterior that is broader, and hence more conservative than that which would be obtained from a full analysis. We are yet to develop a strategy to correct parameter uncertainties from the prior local fit analysis. This implies that one must retain precision measurement statements on parameters from the first two parameter estimation runs on $d(\mathcal{A}|\mathcal{B})$ and $d(\mathcal{B}|\mathcal{A}_{\text{res}}).$ Correcting the widths of the local fit posteriors are beyond the scope of this paper and we leave this for future work.

To conclude this section, we make a few important remarks about the algorithm given above. First of all, the algorithm is likely to be less effective if the recovered best fit parameters are far from the true value. This would cause a breakdown of the linear-signal approximation, which is a key assumption in the generalised CV algorithm presented in \ref{sec:Source_Confusion_Bias}. We also assume that, through many local-fits, we have found all the signals present in the data stream we are studying. Further, the two signals present here are near orthogonal with relatively little correlation between the two signals. If there were significant overlap, then the posteriors for the global-fit procedure would be wider since extra uncertainty would be introduced into the parameters in question. 
This would mean that the procedure presented here, in which we shift a posterior computed with a single source model into the correct location, but do not modify the posterior width, would yield overly optimistic estimates of the source parameters. There are two approaches to address this shortcoming. Firstly, the correlation between sources identified in the data can be evaluated, and any pairs of source with sufficiently high correlation can be reanalysed jointly. Alternatively, it is possible to generate an updated posterior for the parameters of each source by marginalising over the biases due to the other source. The procedure is similar to the algorithm described here, but rather than shift each sample in the $\mathcal{A}$ source posterior by the same amount, given by the best-fit parameters of source $\mathcal{B}$, we instead shift them by an amount given by Eq.~\eqref{eq:explainbias} evaluated for the $\mathbf{h}^{(2)}$ waveform computed as a random sample drawn from the $\mathcal{B}$ source distribution. This approach is beyond the scope of the analysis presented here, but we leave it for future work.

\section{Conclusions}\label{conclusions}


In this paper, we have generalized the approach in \cite{Cutler:2007mi} to provide metrics for the parameter estimation biases on individually resolved sources from the presence of confusion noise from missed signals or incorrectly fitted waveforms. We have illustrated these generalisations with simple (yet realistic) scenarios relevant to the LISA and ET detectors, and we can collect several generic findings:
\begin{itemize}
	\item We find that the presence of altogether missed signals drawn from the same population could lead to significant biases on the parameter estimation of other signals which are instead fitted out of the data. 
	\item We qualitatively confirm one of the main results of~\cite{Samajdar:2021egv,Pizzati:2021gzd,Himemoto:2021ukb,Relton:2021cax}. The coincident arrival of two signals in a ground-based detector, with nearly overlapping mergers, may lead to biases when the difference between coalescence times of the signals is less than a fraction of a second.
	\item We find that residuals in the data arising from the incorrect removal of sources effectively behave like missed signals, and may lead to significant biases. 
	\item We find that biases from confusion noise and waveform inaccuracies may deconstructively interfere with one another.
	\item Our results suggest that galactic binaries which are missed by dedicated searches~\cite{Littenberg:2020bxy}, and not accounted for in confusion noise estimates, may lead to significant biases on the parameter estimation of other typical LISA sources.
	\item We proposed a proof-of-concept global-fit scheme in which, starting from local-fits of LISA sources, guesses for the true parameters are obtained through bias predictions from previous parameter estimation simulations. We find these guesses lie within the $1\sigma$ interval of global-fit posteriors across all sources. This has potential applications to confirm global-fit search algorithms, and as a standalone novel local-fit parameter estimation algorithm.
\end{itemize}
In all the cases outlined above, the formalism we have developed plays an important role in providing a theoretical ground for the described biases and a solid tool to address them. We believe this formalism could be useful in exploratory studies of future GW detectors, to assess under what circumstances we expect the biases described above to appear.
We also believe this formalism is an early but significant step towards an understanding of how to simultaneously infer parameters from multiple signals of different nature with future detectors, as we highlight with our global-fit algorithm scheme. 

There are several ways in which the application of this formalism could be extended. One could perform systematics studies for realistic populations of missed signals using realistically modelled waveforms. One could check whether inaccurately modelled signals could lead to significant biases when several of them are incorrectly subtracted from the data, which our understanding of residuals as missed signals and the biases they lead to strongly suggests. This is a possibility that we have not explored due to the technical challenge in dealing with very large Fisher matrices and MCMC sampling algorithms to sample over such a large parameter space. 
Finally, one could explore further the applications of this formalism for global-fit algorithms, which could be extended to take into account significant overlaps between the signals in the data stream, and to explore correcting the width as well as the peak location of the parameter posteriors.

As a final note, the formalism itself can be extended to take into account brighter confusion sources and more pronounced waveform errors (as would happen with different families of waveform models or models within the same family containing different physics). To do so, one could derive higher order terms in the equations present in Sec.\ref{sec:Source_Confusion_Bias} to describe biases that are farther from the true parameters than those considered in this work. 
\newline

\noindent 
\textit{Acknowledgements.}
The authors thank E.~Berti, D.~Gerosa, M.~P\"urrer,  N.~Tamanini and M.~van de Meent for enlightening discussions. We especially thank R.~Cotesta, M.~Katz and L.~Speri for a careful reading of the manuscript, and R.~Cotesta for collaborating in the early stage of this project as well. The author O.B expresses his gratitude to Sir E. H. John for the vocal support given throughout this work.\\
\noindent
\textit{Data Availability Statement.}
The data underlying this article will be shared on reasonable request to the corresponding author. Antonelli's and Burke's codes relevant to this project can be found at \url{https://github.com/aantonelli94/GWOP} and 
\url{https://github.com/OllieBurke/Noisy_Neighbours}.

  
\chapter{Conclusions \& future work}
\label{chap:seven}

This thesis has focussed on developing accurate waveform models for the coalescence of compact-object binaries, as needed for (near) future data analyses, and on providing metrics to address inference biases in the case in which inaccurate models are used.

The strategy for the first task has been to improve the accuracy of models relied on the use of (semi-)analytical techniques from the relativistic two-body problem, alone or in synergistic ways. We have made use of two-body information from PM, PN and \GSF frameworks, and of the analytical EOB framework to combine them. In chapter~\ref{chap:two} we have focussed on PM theory, studying whether available information at 3PM order~\cite{Bern:2019nnu} can be used to directly improve waveform models through comparisons of the binding energy (as a function of either frequency or angular momentum) against predictions from NR simulations for nonspinning systems in a quasi-circular orbit~\cite{Ossokine:2017dge}. We have found that the PM dynamics does not (yet) encapsulate two-body information in this setting better than the PN models currently in use in GW data analyses~\cite{Damour:2000we,Barausse:2009xi,Nagar:2018zoe}. We argued this is related to the fact that the 3PM dynamics we have used in the analysis only includes complete information up to 2PN order, while 3PN or 4PN terms would be needed for accurate models. Better resummations of the PM dynamics or higher orders in the PM expansion~\cite{Bern:2021dqo} are therefore needed to reach this goal. 
The analysis of Ref.~\cite{Antonelli:2019ytb} has focussed on the conservative dynamics of the binary, and it notably left out the effects of radiation reaction. In future work, it would be interesting to relax this assumption, and check whether improvements from PM theory can be made in the description of GW emission out of the system. 

In chapter~\ref{chap:three} we have focussed on the SMR approximation. In particular, we have presented an EOB model for quasi-circular nonspinning binaries based on the SMR approximation through the Detweiler redshift. This is a viable ``EOBSMR'' model, while other attempts in the literature were hampered by the presence of an unphysical divergence at the light ring of the effective Hamiltonian~\cite{Barausse:2011dq,LeTiec:2011dp,Akcay:2012ea}. After deriving a Hamiltonian that does not suffer from such a divergence, and showing that plunges through the light ring are possible within the new description, we compared the results against NR simulations. We found that the EOBSMR model encapsulates two-body dynamics very well when comparing analytical predictions with numerical ones using the binding energy as a benchmark.
The agreement using the dephasing was even more remarkable, and within numerical error of the NR simulations. We argued that the EOBSMR Hamiltonian is a better base than PN-based models in the same EOB gauge, and comparable to the EOB Hamiltonian that forms the basis of current EOB waveform models used in LIGO-Virgo studies~\cite{Damour:2000we,Barausse:2009xi,Nagar:2018zoe}, especially if one synergistically includes both SMR and PN information. Moreover, the excellent agreement of the EOBSMR model with NR results holds for \emph{comparable masses}, namely systems that one would not expect to be modelled at all by SMR information. This is consistent with the suggestion~\cite{LeTiec:2011bk,LeTiec:2011dp,Tiec:2014lba} that \GSF information could be used to model comparable-mass binaries as well as EMRIs. Overall, the results of chapter~\ref{chap:three} show that the EOBSMR model we present could be a useful basis for future EOB models. Obvious extensions of this work include an inclusion of \GSF results for spinning binaries: the hope would be to replicate the nice agreement even for spinning systems. This could also be a first step towards an implementation of an SMR-based model to be used in LIGO-Virgo-Kagra analyses. This goal would also need us to revisit the radiation-reaction description of the EOB framework (which is currently based on the PN approximation), and it would require one to move away from the quasi-circularity condition that the model in chapter~\ref{chap:three} has been derived in. It would also be interesting to seek connections with other EOB models for EMRIs that are under development~\cite{Yunes:2010zj,Albanesi:2021rby}.

In chapters~\ref{chap:four} and~\ref{chap:five}, motivated by a study that showed that waveform models of spinning binaries may be inaccurate for near-future analyses~\cite{Purrer:2019jcp} and by a method that had been recently exploited for nonspinning systems~\cite{Bini:2019nra}, we have combined constraints from relativistic (PM) scattering and results in double PN-SMR expansions from SF theory to obtain the previously-unknown third-subleading order in the PN expansions of spin-orbit and (aligned) spin$_1$-spin$_2$ couplings for arbitrary masses. 
We have presented two derivations (one in Sec.~\ref{sec:3&4paper} and one in chapter~\ref{chap:five}), and we have incorporated the new results in EOB Hamiltonians. By comparing these to NR simulations of spin-aligned binaries in a circular orbit~\cite{Ossokine:2017dge}, we were able to assess the accuracy gain from the novel terms. We find that they lead to improvements. The results we have obtained fix the third-subleading PN spin-orbit sector for generic spin orientations, but for spin$_1$-spin$_2$ the method is only valid for spin-aligned configurations. Further applications of the method could include a derivation of the remaining terms of the spin-spin couplings (which require more constraints from second-order expansions in the spin of the Detweiler redshift), or their extension to precessing spins.

Finally, in chapter~\ref{chap:six} we focus on inference biases in the presence of multiple overlapping signals and provide metrics to predict such biases in the presence of unmodelled or incorrectly modelled foregrounds. Our work is based on the analysis by Cutler and Vallisneri~\cite{Cutler:2007mi} (see also Refs.~\cite{Flanagan:1997kp,PhysRevD.71.104016}) and on the Fisher matrix. We extend the formalism of Cutler and Vallisneri, which focusses on the mismodelling error of a single source buried in noise, to the case in which a confusion term of many signals is present, or the case in which signals are simultaneously inferred with inaccurate models. The analysis is relevant for future detectors, in which such a term is expected to be present~\cite{Samajdar:2021egv,Pizzati:2021gzd,Robson:2017ayy,Karnesis:2021tsh,Himemoto:2021ukb}. We provided several examples of relevance to both LISA and future ground-based detectors such as the Einstein Telescope and the Cosmic Explorer. We find the method derived in chapter~\ref{chap:six} a reliable tool to predict biases in both the above cases of interest, when compared to MCMC analyses. The predictions can be cheaply computed, implying that the method could be useful to cover the parameter space where MCMC analyses are hard to come by. In particular, this is useful for exploratory studies of future detectors. As such, future work could include analyses of biases from confusion noise in next-generation ground-based detectors~\cite{Samajdar:2021egv,Himemoto:2021ukb} on the parameters of sources of interest such as loud binary black holes (which could be used to test general relativity), or the analyses of biases on a low-SNR source of interest (such as an EMRI) due to the inaccurate removal of a signal with much higher SNR (such as a supermassive black hole). These examples, among others, are the subject of future work. 

  
\backmatter
\pagenumbering{Roman}
\setcounter{page}{1}
\printglossary[style=altlist,title=Glossar]

\printglossary[type=\acronymtype,style=long]

\appendix

\chapter{Appendix}

\section{Canonical transformation from DJS to isotropic gauge at 4PN order}\label{app:DJStoISO@4PN}

The EOB Hamiltonian is given by Eq.~\eqref{eq:enmap}, with effective Hamiltonian~\eqref{Ham_eff}. At 4PN, in the DJS gauge, the $A(u,\nu)$ potential is given by Eq.~\eqref{eq:A4PN}, while $D(u,\nu)$ and $Q(u,\nu,p_r)$ are given by 
\begin{align}
D(u,\nu)=&1+6\nu u^{2}+(52\nu-6\nu^{2})u^{3}+\nnm\\
&\bigg[\nu\bigg(-\frac{23761\pi^{2}}{1536}-\frac{533}{45}+\frac{1184}{15}\gamma_{E}-\frac{6496}{15}\log 2+\frac{2916}{5}\log 3\bigg)+\nnm\\
&\quad\nu^{2}\bigg(\frac{123\pi^{2}}{16}-260\bigg)+\frac{592}{15}\nu\log u\bigg]u^{4}\,,
\end{align}
and
\begin{align}
Q(u,p_{r},\nu)&=\bigg\{2(4-3\nu)\nu u^{2}\nonumber\\
&+\bigg[\bigg(-\frac{5308}{15}+\frac{496256}{45}\log 2-\frac{33048}{5}\log 3\bigg)\nu-83\nu^{2}+10\nu^{3}\bigg]u^{3}\bigg\}p_{r}^{4}\nonumber\\
&+\bigg[\bigg(-\frac{827}{3}-\frac{2358912}{25}\log 2+\frac{1399437}{50}\log 3+\frac{390625}{18}\log 5\bigg)\nu\nonumber\\
&\qquad -\frac{27}{5}\nu^{2}+6\nu^{3}\bigg]u^{2}p_{r}^{6}\,.
\end{align}

The isotropic-gauge Hamiltonian $H^2_{a^0}$ from Eq.~\eqref{HN3LOSO} we want to specify is split into local and nonlocal terms, the latter appearing at 4PN order,
\begin{align}\label{Hnonloc}
H^2_{a^0}=&H^\text{(loc)}_{a^0} + H^\text{(nonloc)}_{a^0}\,,\nnm\\
H^\text{(loc)}_{a^0} =& \sum_{l=1}^4 H^\text{(loc)}_{l\text{PN}} = \sum_{l=1}^4\sum_{j=0}^{l+1} \alpha_{lj} p^{l+1-j}u^j\,,\nnm\\
H^\text{(nonloc)}_{a^0}=&(\alpha_{44}+\alpha_{44}^\text{ln}\ln u)\, p\, u^{4}+(\alpha_{45}+\alpha_{45}^\text{ln}\ln u)u^{5}\,,
\end{align}
with $p^2 = p_r^2 +p_\phi^2/r^2$ as usual. The generating function we use is
\begin{equation}
\begin{split}
\mc G(u, p_{r},p_{\phi})=p_{r}\bigg(&\frac{1}{c^{2}}g_{0}+\frac{1}{c^{4}}g_{1}\, u +\frac{1}{c^{6}}\bigg[g_{2}u^{2}+g_{3}p_{\phi}^{2}u^{3}+g_{4}p_{r}^{2}u\bigg]\\
+\frac{1}{c^{8}}\bigg[g_{5}u^{3}&+g_{6}u^{4}p_{\phi}^{2}+g_{7}u^{2}p_{r}^{2}+g_{8}u^{5}p_{\phi}^{4}+g_{9}u p_{r}^{4}+g_{10}u^{3}p_{\phi}^{2}p_{r}^{2}\bigg]\bigg)\,.
\end{split}
\end{equation}
Finally, we fix the generating function's and Hamiltonian's coefficients imposing
\begin{align}
H^2_{a^0}=&H_{\text{EOB}}+\left\{\mc{G},H_{\text{EOB}}\right\}+\frac{1}{2!}\left\{\mc{G},\left\{\mc{G},H_{\text{EOB}}\right\}\right\}\nonumber\\
&+\frac{1}{3!}\left\{\mc{G},\left\{\mc{G},\left\{\mc{G},H_{\text{EOB}}\right\}\right\}\right\}+\frac{1}{4!}\left\{\mc{G},\left\{\mc{G},\left\{\mc{G},\left\{\mc{G},H_{\text{EOB}}\right\}\right\}\right\}\right\}\,,
\end{align} 
and solving the equations order-by-order in the PN expansion.
In the real gauge~\eqref{eq:realgauge}, the 0PM coefficients are
\begin{align}
\alpha_{10} &= \frac{1}{8} (3 \nu -1),\qquad \alpha_{20} = \frac{1}{16} \left(1-5 \nu +5 \nu ^2\right)\nnm\\
\alpha_{30} & = \frac{5}{128} \left(-1+7 \nu -14 \nu ^2+7 \nu ^3\right), \nnm\\
\alpha_{40}&= \frac{7}{256} \left(1-9 \nu +27 \nu ^2-30 \nu ^3+9 \nu ^4\right).
\end{align}
Which implies the following 1PN and 2PN coefficients,
\begin{align}
& \alpha_{11}= -\frac{3}{2}-\nu,\qquad \alpha_{12}=\frac{1+\nu }{2}\nnm\\
&\alpha_{21}=\frac{5}{8}-\frac{5 \nu }{2}-\nu ^2,\quad \alpha_{22}=\frac{1}{4} \left(10+27 \nu +3 \nu ^2\right),\nnm\\
&\alpha_{23}=-\frac{1}{4} (1+6 \nu )\,,
\end{align}
the 3PN ones,
\begin{align}
& \alpha_{31}=-\frac{7}{16}+\frac{21 \nu }{8}-3 \nu ^2-\nu ^3, \nnm \\
&\alpha_{32}=\frac{3}{16} \left(-9+49 \nu +74 \nu ^2+5 \nu ^3\right),\nnm\\
&\alpha_{33}=\frac{1}{8} \left(-25-130 \nu -107 \nu ^2\right),\nnm \\
&\alpha_{34}=\frac{1}{192} \left(24+\left(2168-123 \pi ^2\right) \nu +336 \nu ^2-24 \nu ^3\right)\,,\nnm
\end{align}
and the 4PN,
\begin{align}
&
\alpha_{41}=\frac{45}{128}-\frac{45 \nu }{16}+\frac{51 \nu ^2}{8}-3 \nu ^3-\nu ^4, \nonumber\\
&\alpha_{42}=\frac{13}{8}-\frac{5245 \nu }{96}+\frac{453 \nu ^2}{32}+\frac{337 \nu ^3}{16}+\frac{35
	\nu ^4}{32}\nnm\\
&\qquad-\frac{73716}{5} \nu  \ln 2+\frac{1399437}{320} \nu  \ln 3+\frac{1953125}{576} \nu  \ln 5,\nnm\\
&\alpha_{43}=\frac{105}{32}+\frac{239797 \nu }{2400}-\frac{2589 \nu ^2}{32}-\frac{487 \nu
	^3}{16}\nnm\\
&\qquad+\frac{12853564}{225} \nu  \ln 2-\frac{27646839 }{1600}\nu \ln 3-\frac{7421875}{576} \nu  \ln 5,\nnm\\
&\alpha_{44} = \frac{105}{32}+\left(\frac{2957}{48}-\frac{41 \pi ^2}{64}\right) \nu ^2+\frac{27 \nu
	^3}{2}-\frac{5 \nu ^4}{16}+\nu  \bigg(\frac{148 \gamma_\text{E} }{15}-\frac{29665 \pi
	^2}{12288}\nnm\\
&\qquad-\frac{420863}{3600}-\frac{5497708\ln 2}{75}+\frac{4510323 \ln 3}{200}+\frac{390625 \ln 5}{24}\bigg),\nnm\\
&\alpha_{45} = -\frac{1}{16}+\frac{44 \gamma  \nu }{15}+\frac{76645 \pi ^2 \nu }{12288}+\frac{1}{384}
\left(-14792+861 \pi ^2\right) \nu ^2-\frac{5 \nu ^3}{8}\nnm\\
&-\frac{12311 \nu
}{720}+\frac{1392508}{45} \nu  \ln 2-\frac{1543293}{160} \nu \ln 3-\frac{1953125}{288} \nu  \ln 5\,,
\end{align}
as well as the following nonlocal coefficients
\begin{align}
&\alpha_{44}^\text{ln}=\frac{74 \nu }{15};
\, \, \, \, \, 
\alpha_{45}^\text{ln}=\frac{22 \nu }{15}\,.
\end{align}
We note in passing that a unique isotropic-gauge Hamiltonian can only be obtained by assuming that there are only two logarithmic terms in the ansatz, and specifically the $u^4$ and $u^5$ contributions in $H^\text{(nonloc)}_{a^0}$.


\section{Scattering-angle predictions from two-body third-subleading Hamiltonians}
\label{app:chipred}

The unspecified functions in Eq.~\eqref{eq:chipredSO} (with $c\equiv 1$) are
\begin{align*}
\chi^\text{(pred)}_{1-}(v)=&\left(\frac{5}{8}-2 \alpha_{13-}+\frac{5 }{8}\nu\right)v^2\nnm\\&+\left[\frac{3}{16}-2 \alpha_{13-}-2 \alpha_{15-}+\left(\frac{3}{16}+5
\alpha_{13-}\right) \nu -\frac{7 \nu ^2}{8}\right]v^4\nnm\\
&+\bigg[\frac{13}{128}-2 \alpha_{13-}-4 \alpha_{15-}-2 \alpha_{17-}+\left(\frac{13}{128}+\frac{33}{4}\alpha_{13-}+7 \alpha_{15-}\right) \nu\nonumber\\
&\qquad-\left(\frac{67}{64}+\frac{35}{4}\alpha_{13-}\right) \nu
^2+\frac{135 \nu ^3}{128}\bigg] v^6,
\end{align*}
\begin{align*}
\chi^\text{(pred)}_{2-}(v)=&\frac{1}{2} + \left(\frac{3}{4}\nu-4 \alpha_{13-}-\alpha_{23-}\right)v^2\nnm\\
+&\bigg[\frac{3}{4}-12 \alpha_{13-}-6 \alpha_{15-}-\alpha_{25-}\nnm\\
&\quad+\left(\frac{3}{4}+9 \alpha_{13-}+\frac{3 }{2}\alpha_{23-}\right)
\nu -\frac{3 \nu ^2}{4}\bigg]v^4\nonumber\\
+&\bigg\{\frac{9}{32}-12 \alpha_{13-}-22 \alpha_{15-}-8 \alpha_{17-}-\alpha_{25-}-\alpha_{27-}\nonumber\\
&\quad+\frac{1}{16} (3+440 \alpha_{13-}+304 \alpha_{15-}+14 \alpha_{23-}+40 \alpha_{25-}) \nu\nonumber\\
&\quad
-\left[14 \alpha_{13-}+\frac{3}{8} (4+5 \alpha_{23-})\right] \nu
^2+\frac{3 \nu ^3}{4}\bigg\}v^6,
\end{align*}
\begin{align*}
\chi^\text{(pred)}_{3-}(v)=&2-\left(\frac{133}{8}+30 \alpha_{13-}+12 \alpha_{23-}-\frac{27 }{8}\nu\right)v^2\nnm\\
+&\bigg[\frac{17}{16}-230 \alpha_{13-}-70 \alpha_{15-}-28 \alpha_{23-}-20
\alpha_{25-}-4 \alpha_{35-}\nonumber\\
&\quad+\left(\frac{135}{16}+65 \alpha_{13-}+16 \alpha_{23-}\right) \nu -2 \nu ^2\bigg]v^4\nnm\\
+&\bigg[\frac{1155}{128}-385\alpha_{13-}-476 \alpha_{15-}-126 \alpha_{17-}\nnm\\
&\qquad-56 \alpha_{25-}-28\alpha_{27-}-4 \alpha_{37-}\nonumber\\
& +\bigg(-\frac{2049}{128}+\frac{1613 }{4}\alpha_{13-}+203 \alpha_{15-}\nnm\\
&\qquad+24 \alpha_{23-}+44 \alpha_{25-}+6 \alpha_{35-}\bigg) \nu\nonumber\\
& +\left(-\frac{959}{64}-\frac{345}{4}\alpha_{13-}-16 \alpha_{23-}\right) \nu ^2+\frac{201 \nu ^3}{128}\bigg]v^6,
\end{align*}
\begin{align*}
\chi^\text{(pred)}_{4-}(v)=&\left[-3 (5+4 \alpha_{13-}+2 \alpha_{23-})+\frac{3}{2}\nu\right]v^2\nnm\\
+&\bigg[-\frac{261}{8}-48 \alpha_{15-}-54 \alpha_{23-}-18 \alpha_{25-}-6
\alpha_{35-}\nonumber\\
&\quad+9 \nu +9 \alpha_{23-} \nu +6 \alpha_{13-} (-38+5
\nu )\bigg]v^4\nnm\\
+&\bigg\{-\frac{3}{16} (-27+4036 \alpha_{13-}+3200 \alpha_{15-}+640 \alpha_{17-}+244 \alpha_{23-}\nonumber\\
&\quad+624 \alpha_{25-}+192 \alpha_{27-}+64
\alpha_{35-}+48 \alpha_{37-}+8 \alpha_{47-})\nonumber\\
&\quad+\bigg(-\frac{1259}{32}+\frac{123 \pi ^2}{128}+\frac{747}{2}\alpha_{13-}+132 \alpha_{15-}\nonumber\\
&\qquad\qquad+\frac{129 }{4}\alpha_{23-}+36 \alpha_{25-}+\frac{15 }{2}\alpha_{35-}\bigg) \nu \nonumber\\
&\quad-\frac{3}{32} (123+320
\alpha_{13-}+56 \alpha_{23-}) \nu ^2\bigg\}v^6,
\end{align*}
\begin{align*}
\chi^\text{(pred)}_{1+}(v)=&-4+\left(-\frac{5}{8}-2 \alpha_{13+}+\frac{41}{8}\nu\right)v^2\nnm\\
&+\left[-\frac{3}{16}-2 \alpha_{13+}-2 \alpha_{15+}+\left(\frac{45}{16}+5
\alpha_{13+}\right) \nu -\frac{13 \nu ^2}{2}\right]v^4\nnm\\
&+\bigg[-\frac{13}{128}-2 \alpha_{13+}-4 \alpha_{15+}-2 \alpha_{17+}\nnm\\
&\qquad+\left(\frac{277}{128}+\frac{33}{4}\alpha_{13+}+7 \alpha_{15+}\right) \nu \nonumber\\
&\qquad-\frac{5}{64} (107+112 \alpha_{13+}) \nu ^2+\frac{975 \nu
	^3}{128}\bigg] v^6,
\end{align*}
\begin{align*}
\chi^\text{(pred)}_{2+}(v)=&-\frac{7}{2}+\left(-12-4 \alpha_{13+}-\alpha_{23+}+\frac{21 }{4}\nu\right)v^2\nnm\\
&-\bigg[\frac{3}{4}+12 \alpha_{13+}+6 \alpha_{15+}+\alpha_{25+}\nnm\\
&\quad-\frac{3}{2} (5+6 \alpha_{13+}+\alpha_{23+}) \nu +\frac{21 \nu
	^2}{4}\bigg]v^4\nnm\\
&+\bigg[-\frac{9}{32}-12 \alpha_{13+}-22 \alpha_{15+}-8 \alpha_{17+}-\alpha_{25+}-\alpha_{27+}\nnm\\
&\quad+\frac{1}{8} (33+220 \alpha_{13+}+152 \alpha_{15+}+7 \alpha_{23+}+20 \alpha_{25+}) \nu
\nnm\\
&\quad+\left(-14 \alpha_{13+}-\frac{3}{8} (31+5 \alpha_{23+})\right) \nu
^2+\frac{21 \nu ^3}{4}\bigg] v^6,
\end{align*}
\begin{align*}
\chi^\text{(pred)}_{3+}(v)=&-10+\left(-\frac{1403}{8}-30 \alpha_{13+}-12 \alpha_{23+}+\frac{215 }{8}\nu\right)v^2\nnm\\
&+\bigg[-\frac{2993}{16}-230 \alpha_{13+}-70 \alpha_{15+}-28 \alpha_{23+}-20 \alpha_{25+}-4 \alpha_{35+}\nnm\\
&\quad+\left(\frac{1753}{16}+65
\alpha_{13+}+16 \alpha_{23+}\right) \nu -\frac{115 \nu ^2}{8}\bigg]v^4\nnm\\
&+\bigg[-\frac{1155}{128}-385 \alpha_{13+}-476 \alpha_{15+}-126 \alpha_{17+}\nnm\\
&\quad-56 \alpha_{25+}-28 \alpha_{27+}-4 \alpha_{37+}\nnm\\
&\quad+\bigg(\frac{7167}{128}+\frac{1613}{4}\alpha_{13+}+203 \alpha_{15+}\nnm\\
&\qquad\quad+24 \alpha_{23+}+44 \alpha_{25+}+6 \alpha_{35+}\bigg) \nu
\nnm\\
&\quad+\left(-\frac{7547}{64}-\frac{345}{4}\alpha_{13+}-16 \alpha_{23+}\right) \nu ^2+\frac{1425 \nu ^3}{128}\bigg] v^6,
\end{align*}
and
\begin{align*}
\chi^\text{(pred)}_{4+}(v)=&\left[-3 (35+4 \alpha_{13+}+2 \alpha_{23+})+\frac{21 }{2}\nu\right]v^2\nnm\\
&+\bigg[-\frac{3}{8} (1169+608 \alpha_{13+}+128 \alpha_{15+}\nnm\\
&\qquad +144 \alpha_{23+}+48 \alpha_{25+}+16 \alpha_{35+})+(123+30 \alpha_{13+}+9
\alpha_{23+}) \nu\bigg]v^4\nnm\\
&+\bigg[-\frac{3}{16} (1211+4036 \alpha_{13+}+3200 \alpha_{15+}+640 \alpha_{17+}+244 \alpha_{23+}\nnm\\
&\quad+624 \alpha_{25+}+192 \alpha_{27+}+64
\alpha_{35+}+48 \alpha_{37+}+8 \alpha_{47+})\nnm\\
&\quad+\bigg(\frac{6911}{32}-\frac{123 \pi ^2}{128}+\frac{747}{2}\alpha_{13+}+132 \alpha_{15+}\nnm\\
&\qquad \quad+\frac{129}{4}\alpha_{23+}+36 \alpha_{25+}+\frac{15 \alpha_{35+}}{2}\bigg) \nu \nnm\\
&\quad-\frac{3}{32} (981+320
\alpha_{13+}+56\alpha_{23+}) \nu ^2\bigg] v^6.
\end{align*}

The unspecified functions in Eq.~\eqref{eq:S1S2chipred} are
\begin{align}
\chi_{1\times}^\text{(pred)}(v)=& - 4 \alpha_{10\times}+ \left[2 (2+\alpha_{10\times}-2 \alpha_{12\times})+(-1+4\alpha_{10\times}) \nu\right]v^2\nnm\\
&+\bigg[\frac{1}{2} (2+\alpha_{10\times}-4 \alpha_{12\times}-8 \alpha_{14\times})\nnm\\
&\quad +\left(-\frac{3}{4}+8\alpha_{12\times}\right) \nu +\left(\frac{1}{2}-4\alpha_{10\times}\right) \nu ^2\bigg]v^4\nnm\\
&+\bigg[\frac{1}{4} (2+\alpha_{10\times}-6 \alpha_{12\times}-24 \alpha_{14\times}-16
\alpha_{16\times})\nnm\\
&\quad+\left(-\frac{3}{8}+9 \alpha_{12\times}+12 \alpha_{14\times}\right) \nu +\left(\frac{3}{4}-3 \alpha_{10\times}-12 \alpha_{12\times}\right) \nu ^2\nnm\\
&\quad+\left(-\frac{5}{16}+4 \alpha_{10\times}\right) \nu ^3
\bigg]
\end{align}
\begin{align}
\chi_{2\times}^\text{(pred)}(v)=& - 3\pi \alpha_{10\times}\nnm\\
+ \bigg[&-\frac{3}{2} \pi  (-13+5 \alpha_{10\times}+4 \alpha_{12\times}+\alpha_{22\times})+\frac{3}{8} \pi  (-5+8 \alpha_{10\times}) \nu\bigg]v^2\nnm\\
+\bigg[&\frac{3}{16} \pi  (29+26 \alpha_{10\times}-80 \alpha_{12\times}-48 \alpha_{14\times}+4 \alpha_{22\times}-8 \alpha_{24\times})\nnm\\
+&\frac{3}{64} \pi  (245-8
\alpha_{10\times}+224 \alpha_{12\times}+32 \alpha_{22\times}) \nu
+\frac{3}{16} \pi  (7-10 \alpha_{10\times}) \nu ^2\bigg]v^4\nnm\\
+&\bigg[\frac{3}{32} \pi  (11+14 \alpha_{10\times}-88 \alpha_{12\times}-304 \alpha_{14\times}-128 \alpha_{16\times}+2 \alpha_{22\times}\nnm\\
&\quad-8 \alpha_{24\times}-16
\alpha_{26\times})\nnm\\
&\quad+\frac{3}{128} \pi  (-25-56 \alpha_{10\times}+1056
\alpha_{12\times}+1024 \alpha_{14\times}+128 \alpha_{24\times}) \nu
\nnm\\
&\quad-\frac{3}{16} \pi  (-49+13 \alpha_{10\times}+72 \alpha_{12\times}+8
\alpha_{22\times}) \nu ^2\nnm\\
&\quad+\frac{3}{64} \pi  (-11+32 \alpha_{10\times}) \nu ^3\bigg],
\end{align}
\begin{align}
\chi_{3\times}^\text{(pred)}&(v)= - 8 \alpha_{10\times}
\nnm\\
&+ \bigg(216-124 \alpha_{10\times}-40 \alpha_{12\times}-16 \alpha_{22\times}+2 \nu +16
\alpha_{10\times}\nu\bigg)v^2\nnm\\
&+\bigg[-59 \alpha_{10\times}-\frac{2}{3} (-633+430 \alpha_{12\times}+140 \alpha_{14\times}+44 \alpha_{22\times}+40 \alpha_{24\times}+8 \alpha_{34\times})\nnm\\
&\quad+\frac{1}{6} (1029+36 \alpha_{10\times}+400 \alpha_{12\times}+80
\alpha_{22\times}) \nu +\frac{55 \nu ^2}{3}\bigg]v^4\nnm\\
&+\bigg[\frac{157\alpha_{10\times}}{2}+\frac{1}{15} (849-5325 \alpha_{12\times}-8820
\alpha_{14\times}-2520 \alpha_{16\times}+310 \alpha_{22\times}\nnm\\
&\quad-920
\alpha_{24\times}-560 \alpha_{26\times}+40 \alpha_{34\times}-80
\alpha_{36\times})+\bigg(-\frac{677}{4}-85 \alpha_{10\times}\nnm\\
&\quad+336 \alpha_{12\times}+224 \alpha_{14\times}+\frac{2 \alpha_{22\times}}{3}+\frac{136\alpha_{24\times}}{3}+\frac{16\alpha_{34\times}}{3}\bigg) \nu
\nnm\\
&\quad+\left(\frac{2939}{10}-26 \alpha_{10\times}-\frac{200\alpha_{12\times}}{3}-\frac{26 \alpha_{22\times}}{3}\right) \nu ^2+\frac{187 \nu ^3}{120}\bigg],
\end{align}
\begin{align}
&\chi_{4\times}^\text{(pred)}(v)= \bigg[\frac{15}{2} \pi  (15-10 \alpha_{10\times}-2 \alpha_{12\times}-\alpha_{22\times})-\frac{15}{8} \pi  (5+4 \alpha_{10\times}) \nu\bigg]v^2\nnm\\
&+\bigg[\frac{15}{8} \pi  (442-137 \alpha_{10\times}-148 \alpha_{12\times}-32
\alpha_{14\times}-34 \alpha_{22\times}-12 \alpha_{24\times}-4 \alpha_{34\times})\nnm\\
&\quad+\frac{15}{16} \pi  (119+44 \alpha_{10\times}+32 \alpha_{12\times}+8\alpha_{22\times}) \nu +\frac{15}{8} \pi  (9+2 \alpha_{10\times}) \nu ^2\bigg]v^4\nnm\\
&+\bigg[\frac{15}{16} \pi  (661+19 \alpha_{10\times}-855 \alpha_{12\times}-768
\alpha_{14\times}-160 \alpha_{16\times}-24 \alpha_{22\times}\nnm\\
&\quad-144
\alpha_{24\times}-48 \alpha_{26\times}-12 \alpha_{34\times}-12
\alpha_{36\times}-2 \alpha_{46\times})\nnm\\
&\quad-\frac{15}{128} \pi  (4519+684
\alpha_{10\times}-2592 \alpha_{12\times}-1152 \alpha_{14\times}-288
\alpha_{24\times}-48 \alpha_{34\times}) \nu \nnm\\
&\quad-\frac{15}{64} \pi  (-1667+66
\alpha_{10\times}+72 \alpha_{12\times}) \nu ^2-\frac{15}{64} \pi  (-37+4\alpha_{10\times}) \nu ^3\bigg].
\end{align}

\section[Effective-one-body Hamiltonian at 3PM order augmented by 3PN and\dots]{Effective-one-body Hamiltonian at 3PM order augmented by 3PN and 4PN information}\label{PM4PN}
\label{appendixA}

The 3PM EOB Hamiltonian given in Sec.~\ref{sec:EOB3PM}, like the BCRSSZ Hamiltonian from which it was derived, encodes the complete conservative dynamics for generic orbits up to 2PN order, as well as partial information at higher PN orders.  Here we discuss how further information from 3PN and 4PN calculations can be added to the 3PM EOB Hamiltonian, focusing on the case of bound (near-circular) orbits.

We recall that the 4PN Hamiltonian as applicable to generic orbits
~\cite{Damour:2014jta,Damour:2016abl,Bernard:2016wrg} 
is \emph{not} a usual local-in-time Hamiltonian, i.e., not a function of instantaneous position and momentum; rather, it contains a contribution which is a nonlocal-in-time functional of the phase-space trajectory---the so-called ``tail'' term.  In Ref.~\cite{Damour:2015isa}, an EOB transcription of the generic nonlocal-in-time 4PN Hamiltonian is evaluated as a usual local-in-time Hamiltonian by implementing an expansion about the circular-orbit limit, i.e., and expansion in small eccentricity or equivalently in small $\hat{p}_r$.  The result for the 4PN (reduced) effective Hamiltonian takes the form
\be
(\hat H^\mr{eff}_\mr{4PN})^2=A\Big(1+l^2u^2+A\,\bar D\,\hat p_r^2+\hat Q\Big),
\ee  
where we recall that 
\be
l=\frac{L}{GM\mu},
\quad
\hat p_r=\frac{p_r}{\mu},
\quad
u=\frac{GM}{r},
\ee
with the potentials $A(u,\nu)$, $\bar D(u,\nu)$, and $\hat Q(u,p_r,\nu)$ at 4PN order given by
\begin{alignat}{3}
	A&=1-2u+2\nu u^3+a_4u^4+(a_{5,\mr c}+a_{5,\ln}\ln u)u^5,
	\nnm\\\nnm
	\bar D&=1+6\nu u^2+\bar d_3 u^3+(\bar d_{4,\mr c}+\bar d_{4,\ln}\ln u)u^4,
	\\\nnm
	\hat Q&=q_{42} \hat p_r^4u^2+(q_{43,\mr c}+q_{43,\ln}\ln u)\hat p_r^4u^3
	\\
	&\quad+(q_{62,\mr c}+q_{62,\ln}\ln u)\hat p_r^6u^2+\mc {\cal O}(\nu \hat p_r^8 u ).
	\label{DJSpotentials}
\end{alignat}
The coefficients up to 2PN order have been written explicitly here, while the 3PN coefficients ($a_4$, $\bar d_3$, $q_{42}$) and 4PN coefficients are functions only of $\nu$ and are given in Eqs.~(8.1) of Ref.~\cite{Damour:2015isa}.  The $A$ and $\bar D$ potentials are complete up to 4PN order, while the $\hat Q$ potential is given at 4PN order as an expansion in $\hat p_r$ (small-eccentricity expansion) up to ${\cal O}(\hat p_r^6)$, and thus is valid only in the near-circular-orbit regime.

One way to add the 3PN and 4PN information to the 3PM EOB Hamiltonian derived in Sec.~\ref{sec:EOB3PM} is to find a canonical transformation which brings the above 4PN Hamiltonian~\cite{Damour:2015isa} into a form matching (the PN expansion of) the following 3PM+4PN ansatz.  As a natural generalization of the 2PM+3PN ansatz in Ref.~\cite{Damour:2017zjx}, we consider a post-Schwarzschild (reduced) effective Hamiltonian of the form
\be
\Big[\hat H^\mr{eff,PS}(u,\hat p_r,l)\Big]^2=
\hat H_\mr S^2
+(1-2u)\hat Q^\mr{PS}(u,\hat H_\mr S,\nu),
\label{HPSapp}
\ee
where $\hat H_\mr S=\sqrt{1-2u}\sqrt{1+l^2u^2+(1-2u)\hat p_r^2}$ is the reduced Schwarzschild Hamiltonian.  Imposing this form, with a dependence on $\hat p_r$ and $l$ only through $\hat H_S$, is seen to fix a unique phase-space gauge choice.  The resultant potential $\hat Q^\mr{PS}$ can be written at 3PM+4PN order as
\begin{alignat}{3}
	\label{QPS}
	\hat Q^\mr{PS}&=u^2q_\mr{2PM}(\hat H_\mr S,\nu)
	+u^3q_\mr{3PM}(\hat H_\mr S,\nu)
	\\\nnm
	&\quad
	+\Delta_\mr{3PN}(u,\hat H_\mr S,\nu)
	+\Delta_\mr{4PN}(u,\hat H_\mr S,\nu)+\mc O(\mr{5PN}).
\end{alignat}
This differs from Eq.~(\ref{3PMeffH}) by the addition of the $\Delta$ terms, which are given as expansions in the two PN small parameters $u$ and $\hat H_S^2-1$ (each $\mc O(1/c^2)$), at the orders needed to find a unique match to the 4PN EOB Hamiltonian of Ref.~\cite{Damour:2015isa}.  At 3PN order, for generic orbits, we need only a single 4PM term [given by Eq.~(6.3) in Ref.~\cite{Damour:2017zjx}], at zeroth order in $\hat H_\mr S^2-1$,
\be
\Delta_\mr{3PN}=\bigg(\frac{175}{3}\nu-\frac{41\pi^2}{32}\nu-\frac{7}{2}\nu^2\bigg)u^4.
\ee
At 4PN order, to match the near-circular-orbit expansion of the potential $\hat Q$ in Eq.~\eqref{DJSpotentials} up to $\mc O(\hat p_r^6)$, we must have 
\begin{alignat}{3}
	\Delta_\mr{4PN}&=\sum_{n=2}^5\alpha_{4n}u^n(\hat H_\mr S^2-1)^{5-n}
	\nnm\\
	&\quad+\Big(\alpha_{44,\ln}u^4(\hat H_\mr S^2-1)+\alpha_{45,\ln}u^5\Big)\ln u,
\end{alignat}
where the $\alpha$'s are functions only of $\nu$.  (The $n=2,3$ terms here arise solely from the nonlocal tail integral, while the $n=4,5$ and $\ln$ terms include local and tail contributions.)  Implementing the canonical transformation from the 4PN EOB Hamiltonian 
of Ref.~\cite{Damour:2015isa}, we find the coefficients
\begin{alignat}{3}
	\alpha_{42}&=
	\bigg({-}\frac{1027 }{12}-\frac{147432}{5}   \ln2
	\nnm\\
	&\quad
	+\frac{1399437}{160}   \ln3
	+\frac{1953125}{288}  \ln5\bigg)\nu\,,
	\\\nnm
	\alpha_{43}&=
	\bigg({-}\frac{78917 }{300}-\frac{14099512}{225}  \ln 2
	\\
	&\quad
	+\frac{14336271}{800} \ln
	3+\frac{4296875}{288} \ln 5\bigg)\nu\,,
	\\\nnm
	\alpha_{44}&=\bigg({-}\frac{43807}{225}
	+\frac{296 \gamma_{\text{E}} }{15}
	-\frac{33601 \pi ^2}{6144}
	\\\nnm
	&\quad-\frac{9771016 }{225}\ln 2
	+\frac{1182681}{100} \ln 3
	+\frac{390625}{36} \ln 5
	\bigg)\nu 
	\\
	&\quad
	+\bigg({-}\frac{405}{4}+\frac{123}{54}\pi^{2}\bigg)\nu^{2}+\frac{13}{2}\nu^{3}\,,
	\\\nnm
	\alpha_{45}&=\bigg({-}\frac{34499}{1800}+\frac{136 }{3} \gamma_{\text{E}}-\frac{29917 }{6144}\pi ^2
	\\\nnm
	&\quad-\frac{254936}{25}\ln 2+\frac{1061181}{400} \ln 3+\frac{390625 }{144}\ln 5\bigg)\nu 				
	\nonumber \\
	&\quad+\bigg({-}\frac{2387}{24}+\frac{205 }{64}\pi ^2\bigg) \nu ^2+\frac{9 }{4}\nu ^3\,,
\end{alignat}
and
\be
\alpha_{44,\ln}=\frac{148}{15}\nu\,,
\qquad
\alpha_{45,\ln}=\frac{68}{3}\nu\,.
\ee
It is important to note, again, that the form of the effective Hamiltonian in Eq.~(\ref{HPSapp}) and (\ref{QPS}) 
(notably the 4PN term $\Delta_\mr{4PN}$ in Eq.~(\ref{QPS}), is only valid for bound orbits
in the small-eccentricity expansion (around the circular-orbit case)\footnote{If we
	included the next term in the small-eccentricity expansion (i.e.,
	a $\mc {\cal O}(\nu \hat p_r^8 u )$ term in \eqref{DJSpotentials}),
	then all coefficients $\alpha_{4n}$ are modified and we need to
	introduce an additional coefficient for $n=1$. This illustrates that our PS transcription of the local EOB Hamiltonian at 4PN order is not optimal when we have only a finite number of terms in the $\hat p_r$ expansion.  A local 4PN-3PM EOB Hamiltonian 
	for scattering orbits could be obtained by matching $\Delta_\mr{4PN}$ to the 4PN scattering angle calculated 
	in Ref.~\cite{Bini:2017wfr} in the large-eccentricity limit.}.

The two-body Hamiltonian in the EOB framework is then obtained by inserting the effective Hamiltonian (\ref{HPSapp}) in 
Eq.~(\ref{3PMEOBH}), thus obtaining $H^\mr{EOB,PS}_\mr{3PM+4PN}$, or $H^\mr{EOB,PS}_\mr{3PM+3PN}$, if we include $\Delta_\mr{3PM}$, but drop $\Delta_\mr{4PN}$.  The Hamiltonian $H^\mr{EOB,PS}_\mr{4PN}$ is obtained by expanding $q_\mr{2PM}$ to $\mc O(\hat H_\mr S^2-1)^3$ and $q_\mr{3PM}$ to $\mc O(\hat H_\mr S^2-1)^2$, while for $H^\mr{EOB,PS}_\mr{3PN}$ we keep one less order of $\hat H_\mr S^2-1$ for each $q$ and drop $\Delta_\mr{4PN}$.

\section{Alternative effective-one-body Hamiltonian at 3PM order for circular orbits}
\label{appendixB}

One straightforward alternative form for a 3PM EOB Hamiltonian can be obtained simply by fully expanding the right-hand-side of Eq.~\eqref{HPSapp} in $G$, to $\mc O(G^3)$.  Here we explicitly state the result of this expansion evaluated at $p_r=0$, which determines the circular-orbit binding-energy approximants:
\begin{align}
	(\hat H_\mr{3PM}^{\mr{eff},\widetilde{\mr{PS}}})^2|_{p_r=0}&=(1-2u)(1+l^2u^2) \nonumber \\
	&\quad+u^2 \tilde q_\mr{2PM}(\gamma_0,\nu) \nonumber \\
	&\quad+u^3\tilde q_\mr{3PM}(\gamma_0,\nu)
	+\mc O(G^4)\,,
\end{align}
where
$
\gamma_0=\sqrt{1+l^2u^2}
$
is the (circular) effective Hamiltonian at zeroth order in $G$, with 
\begin{subequations}
	\begin{align}
		\tilde q_\mr{2PM}(\gamma_0,\nu)&=q_\mr{2PM}(\gamma_0,\nu),
		\\
		\tilde q_\mr{3PM}(\gamma_0,\nu)&=q_\mr{3PM}(\gamma_0,\nu)
		\nonumber \\
		&\quad
		-\frac{3\nu\gamma_0(5\gamma_0^2-1)}{2\Gamma_0^3}-3(10\gamma_0^2-1)\Big(1-\frac{1}{\Gamma_0}\Big)\,,
	\end{align}
\end{subequations}
where the functions 
$q_\mr{2PM}(\hat H_\mr S,\nu)$ and $q_\mr{3PM}(\hat H_\mr S,\nu)$ are given by Eqs.~\eqref{q2q3}, and $\Gamma_0=\sqrt{1+2\nu(\gamma_0-1)}$.


\section{Detweiler-redshift data and fit}\label{sec:appred}

The  linear-in-$\nu$ Detweiler redshift $\zzz$ at a fixed $x$ is given by Refs.~\cite{Detweiler:2008ft,Akcay:2015pza}:
\begin{equation}\label{eq:zfromhuu}
\zzz = -\tfrac{1}{2}\sqrt{1-3x} h_{uu}^{\text{R}}(x)+ \frac{x}{\sqrt{1-3x}},
\end{equation}
where  $h_{uu}^{\text{R}}$ is the double contraction of the regular part of the metric perturbation generated by a particle on a circular orbit with its 4-velocity. We determine $h_{uu}^{\text{R}}$ in a range $0<x<1/3$ to a high precision using the numerical code developed in Ref.~\cite{vandeMeent:2015lxa}. In this code the regular part of the metric perturbation is extracted using the mode-sum formalism. As noted in Ref.~\cite{Akcay:2012ea}, the convergence of the mode-sum decreases drastically as circular orbits approach the light ring. This limits the accuracy with which $h_{uu}^{\text{R}}$ can be obtained. The code from Ref.~\cite{vandeMeent:2015lxa} allows calculations using arbitrary precision arithmetic, which allows us to calculate $\zzz$ much closer to the light ring and at much higher precision than previously done in Ref.~\cite{Akcay:2012ea}. For this paper, we have generated data for $\zzz$ using up to 120 $\ell$-modes, which allows us to obtain $\zzz$ up to $(1-3x)\approx 4\times 10^{-5}$, with relative accuracy $\lesssim 2.5\times10^{-5}$.

To utilize the $\zzz$ data in our SMR EOB model we need an analytic fit to the data. Two aspects of this fit are important to control for the behaviour of the model. First, the model is sensitive to the precise analytical structure of the fit near the light ring. Second, we need to control the behaviour of the fit beyond the light ring $x>1/3$, where we have no self-force data. In light of these two considerations, we want to fit the data with a model having a relatively low number of parameters. To achieve this, we leverage the analytic knowledge of the PN expansion of $\zzz$, which Ref.~\cite{Kavanagh:2015lva} calculated up to 21.5PN order.
We construct a fit of the overall form:
\begin{equation}\label{eq:fitform}
\zzz = Z_0(x) + \frac{(1-2x)^5}{1-3x} Z_{\text{PN}}(x)\left[1+\alpha(x)Z_{\text{ fit}}(x)\right].
\end{equation}
The leading term:
\begin{equation}
Z_0(x) = x\frac{1-4x}{1-3x}+\frac{x}{\sqrt{1-3x}},
\end{equation}
is constructed such that it will exactly cancel the coefficients $\tilde{\X}_0$ and $\tilde{\X}_1$ when matched to the SMR EOB Hamiltonian.

The number of factors $(1-2x)$ in front of the second term has been chosen such that the resulting contribution to the effective Hamiltonian $\hat{Q}^{\text{PS}}_{\text{SMR}}$ vanishes at the horizon of the effective spacetime, $x=1/2$. 
The coefficient function, $Z_{\text{PN}}(x)$ has the form:
\begin{equation}
Z_{\text{PN}}(x) = 2x^3 \sum_{i,j} a_{i,j} x^{i/2} \log^j x,
\end{equation}
where the coefficients $a_{i,j}$ are obtained by requiring that the series expansion of Eq.~\eqref{eq:fitform} matches the 21.5PN expression from Ref.~\cite{Kavanagh:2015lva}. Since these coefficients are numerous and lengthy, and are easily obtained using computer algebra and the expressions available for the Black Hole Peturbation Toolkit \cite{BHPToolkit}, we do not reproduce them explicitly here.

The actual fit $Z_{\text{fit}}$ is multiplied by an attenuation function:
\begin{equation}
\alpha(x) = \exp\Big(\frac{4-x^{-2}}{6}\Big),
\end{equation}
that suppresses the fit exponentially in the weak field regime, ensuring that the PN behaviour of $\zzz$ is unaffected by the fit. The function $\alpha(x)$ has been chosen such that $\alpha(1/2)=1$ and is at its steepest at $x=1/3$.

The fit $Z_{\text{fit}}$ itself is a polynomial in $\beta \equiv 9x(1-3x)(1-2x)$ and $\log[\frac{1-3x}{(1-2x)^2}]$ with arbitrary coefficients. We perform a large number of linear fits for varying combinations of five terms, and compare various ``goodness of fit'' indicators such as the adjusted $R^2$ value and Bayesian Information Criterion. One model that consistently outperformed the others is:
\begin{equation}\label{eq:zfit}
Z_{\text{fit}} = c_0 + c_1 \beta + c_2 \beta^4 + (c_3 \beta+c_4\beta^4)\log\Big[\frac{1-3x}{(1-2x)^2}\Big],
\end{equation}
with:
\begin{subequations}
	\begin{alignat}{2}
	c_0 &=&    0.555947
	&,\\
	c_1 &=&   -2.589868
	&,\\
	c_2 &=&   31.144986
	&,\\
	c_3 &=&    2.440115
	&,\\
	c_4 &=& -179.175818
	&.
	\end{alignat}
\end{subequations}

With this fit the coefficient functions $\X_i$ in Eq.~\eqref{HSMR} become,
\begin{align}
\X_0(x) &= (1-3x) Z_{\text{PN}}(x)\left[1+\alpha(x)(c_0 + c_1 \beta + c_2 \beta^4)\right],\\
\X_1(x) &= 0,\\
\X_2(x) &= (1-2x)^2 Z_{\text{PN}}(x)\left[1+\alpha(x)(c_3 \beta+c_4\beta^4)\right].
\end{align}


\section[The nonspinning 4PN terms in the bound radial action through sixth\dots]{The nonspinning 4PN terms in the bound radial action through sixth order in eccentricity}\label{app:4PN}

Here we present the additional 4PN-order terms in the radial action for bound orbits, computed via (\ref{defIr}) applied to the 4PN EOB Hamiltonian given in \cite{Damour:2015isa}, valid to sixth order in the orbital eccentricity $e$.  Note that the expansion in eccentricity has occurred only in the 4PN terms, at $\mc O(c^{-8})$, where it is sufficient to use the Newtonian relation $e=\sqrt{1+\ve(L/GM\mu)^2}+\mc O(c^{-2})$.  The complete radial action we employ above, through 4PN order for the nonspinning terms and through NNNLO for the spin terms, is obtained by replacing the first two lines of (\ref{Irexp}) with 
\begin{alignat}{3}\label{Ir4PN}
I_r&={-}L+GM\mu\frac{1+2\ve}{c\sqrt{-\ve}}
+\frac{1}{c^2}\frac{(GM\mu)^2}{\pi\Gamma{L_\mr{cov}}}\ms X_2
\\\nnm
&\quad
+\frac{1}{\pi}\sum_{l=2}^4\frac{1}{c^{2l}}\frac{(GM\mu)^{2l}}{(\Gamma L_\mr{cov})^{2l-1}}\frac{\bar{\ms X}_{2l}}{2l-1}+\frac{1}{c^8}\mc O(e^8)+\mc O(\frac{1}{c^{10}}),
\end{alignat}
where
\begin{alignat}{3}
\frac{\bar{\ms X}_4}{3\pi}&=\frac{5}{4}(7-2\nu)+\bigg[\frac{105}{4}+\Big(\frac{41}{128}\pi^2-\frac{557}{24}\Big)\nu\bigg]\ve
\\\nnm
&\quad+\bigg[\frac{1155}{64}+\Big(\frac{65383}{1440}+\frac{33601}{24576}\pi^2-\frac{74}{15}\gamma_\mr E
\\\nnm
&\qquad-\frac{6122}{3}\ln 2+\frac{24057}{20}\ln 3+\frac{74}{15}\ln\frac{cL_\mr{cov}}{GM\mu}\Big)\nu
\\\nnm
&\qquad-\frac{81}{32}\nu^2+\frac{45}{16}\nu^3\bigg]\ve^2+\mc O(\ve^3),
\end{alignat}
\begin{alignat}{3}
\frac{\bar{\ms X}_6}{5\pi}&=\frac{231}{4}+\Big(\frac{123}{128}\pi^2-\frac{125}{2}\Big)\nu+\frac{21}{8}\nu^2
\\\nnm
&\quad+\bigg[\frac{9009}{32}+\Big({-}\frac{64739}{240}+\frac{51439}{4096}\pi^2-\frac{244}{5}\gamma_\mr E
\\\nnm
&\qquad-\frac{60172}{15}\ln 2+\frac{22599}{10}\ln 3+\frac{244}{5}\ln\frac{cL_\mr{cov}}{GM\mu}\Big)\nu
\\\nnm
&\qquad+\Big(\frac{483}{8}-\frac{369}{256}\pi^2\Big)\nu^2+\frac{45}{16}\nu^3\bigg]\ve+\mc O(\ve^2),
\end{alignat}
and
\begin{alignat}{3}
\frac{\bar{\ms X}_8}{7\pi}&=\frac{32175}{64}
+\Big({-}\frac{534089}{720}+\frac{425105}{24576}\pi^2-\frac{170}{3}\gamma_\mr E
\\\nnm
&\qquad-\frac{9982}{5}\ln 2+\frac{21141}{20}\ln 3+\frac{170}{3}\ln\frac{cL_\mr{cov}}{GM\mu}\Big)\nu
\\\nnm
&\qquad+\Big(\frac{4711}{24}-\frac{1025}{256}\pi^2\Big)\nu^2-\frac{15}{8}\nu^3+\mc O(\ve).
\end{alignat}

\section{Kerr-geodesic variables}
\label{Kerrvar}

We provide here the relevant details to compute the change of variables from $(y,\lambda)$ to $(u_p,e)$ needed for comparison with the 1SF calculations of the perturbed redshift and spin precession invariants. Since we are working with perturbed quantities we need only compute this change of variables at the geodesic level. 

The geodesic equations in Kerr spacetime when specialized to the equator $\theta=\tfrac{\pi}{2}$ are
\begin{align}
\dot{t}&=\frac{1}{\Sigma}\left[E\left(\frac{(r^2+a^2)^2}{\Delta}-a^2\right)+a L\left(1-\frac{r^2+a^2}{\Delta}\right)\right],\\
\dot{r}&=\frac{1}{\Sigma}\sqrt{\left(E(r^2+a^2)-a L\right)^2-\Delta(r^2+(L-aE)^2)}, \\
\dot{\phi}&=\frac{1}{\Sigma}\left[(L-aE)+\frac{a}{\Delta}\left(r^2 E-a (L-aE)\right)\right],
\end{align}
where $\dot{{}}\equiv\frac{d}{d\tau}$. The radial motion is commonly parametrized using the Darwin relativistic anomaly $\chi$ as
\begin{equation}
r=\frac{m_2 p}{(1+e\cos \chi)}\,,
\end{equation} 
where $e$  is the eccentricity and  $p$ the (dimensionless) semilatus rectum. This defines the turning points of the orbit to be at $\chi={0,\pi}$. Note that here we use $p$ instead of $u_p\equiv 1/p$ from the text since it makes the equations below simpler.
To determine the constants of motion $E,L$ as functions of $(p,e)$ we set $\dot{r}=0$ at the turning points. While these simultaneous equations can be solved fully, we give their expansion in $a$, which will be sufficient for this work,
\begin{align}\label{eq:pe_of_EL}
E&=\sqrt{\frac{(p-2)^2-4e^2}{p(p-3-e^2)}}-\frac{(e^2-1)^2}{p(p-3-e^2)^{3/2}}a+\mathcal{O}(a^2), \\
L&=\frac{p}{\sqrt{p-3-e^2}}+(3+e^2)\sqrt{\frac{(p-2)^2-4e^2}{p(p-3-e^2)^3}}a+\mathcal{O}(a^2).
\end{align}

Next, we calculate the radial and azimuthal periods $T_{r0}$ and $\Phi_0$ in the Kerr background geometry
\begin{align}
T_{r0}=&\oint dt=\int_{0}^{2\pi}\frac{dt}{d\chi} d\chi\,,\\
\Phi_0=&\oint d\phi=\int_{0}^{2\pi}\frac{d\phi}{d\chi} d\chi\,,
\end{align} 
where 
\begin{align}
\frac{dt}{d\chi}=\frac{\dot{t}}{\dot{r}}\frac{dr}{d\chi}, \qquad 
\frac{d\phi}{d\chi}=\frac{\dot{\phi}}{\dot{r}}\frac{dr}{d\chi}.
\end{align} 
Further expanding the integrands in eccentricity, and integrating order by order in $a$ and $e$ gives for the periods a result of the form
\begin{align}
T_{r0}(p,e)=&T_{r0}^{0}(p,e)+\, T_{r0}^{1}(p,e)a+\mathcal{O}(a^2)\,,\\
\Phi_0(p,e)=&\Phi_0^{0}(p,e)+ \, \Phi_0^{1}(p,e)a+\mathcal{O}(a^2)\,,
\end{align}
with
	\begin{align}
	T_{r0}^{0}=&\frac{2 \pi p^2}{\sqrt{p-6}}\bigg(1+\frac{3 \left(2 p^3-32 p^2+165 p-266\right)}{4 (p-6)^2 (p-2)}e^2  \nonumber\\
	&+\frac{3}{64 (p-6)^4 (p-2)^3} 
	(40 p^7-1296 p^6+17556 p^5-128448 p^4\nonumber\\
	&+546523 p^3-1350786 p^2+1803396
	p-1016920)
	e^4+\mathcal{O}(e^6)\bigg), \\
	T_{r0}^{1}=&-\frac{6 \pi  \sqrt{p} (p+2)}{(p-6)^{3/2}}\bigg(1+\frac{ \left(2 p^3-32 p^2+139 p+6\right)}{4 (p-6)^2 (p+2)}e^2\nonumber\\
	&+\frac{\left(24 p^5-656 p^4+6844 p^3-32576 p^2+60889 p+210\right)}{64 (p-6)^4
		(p+2)}e^4+\mathcal{O}(e^6) \bigg),
	\end{align}
	and
	\begin{align}
	\Phi_0^{0}=&2 \pi  \sqrt{\frac{p}{p-6}}\bigg(1+\frac{3 }{4 (p-6)^2}e^2+\frac{105 }{64 (p-6)^4}e^4+\mathcal{O}(e^6) \bigg),\\
	\Phi_0^{1}=&-\frac{8 \pi }{(p-6)^{3/2}}\bigg(1+\frac{3 (9 p-34)}{4 (p-6)^2 (p-2)}e^2+\nonumber\\
	&\frac{3 \left(739 p^3-6962 p^2+23332 p-28824\right)}{64 (p-6)^4 (p-2)^3}e^4+\mathcal{O}(e^6) \bigg).
	\end{align}
	
	With these we can use Eq.~\eqref{ylambda} to obtain $(y,\lambda)$ to the desired $4.5$PN accuracy by expanding about small $u_p=1/p$ as
	\begin{align}
	y(u_p,e)=&y^{0}(u_p,e)+ a\, y^{a}(u_p,e)+\mathcal{O}(a^2,u_p^{6})\,,\\
	\lambda(u_p,e)=&\lambda^{0}(u_p,e)+ a\, \lambda^{a}(u_p,e)+\mathcal{O}(a^2,u_p^{5})\,,
	\end{align}
	with
	\begin{align}
	y^0(u_p,e)=&\left(1-e^2\right) u_p-2 e^2 \left(-1+e^2\right)
	u_p^2+\left(6 e^2-\frac{23 e^4}{8}\right) u_p^3+\nonumber\\
	&\left(24
	e^2-\frac{13 e^4}{4}\right) u_p^4-\frac{1}{4} e^2
	\left(-480+e^2\right) u_p^5\,,\\
	y^a(u_p,e)=&\frac{2}{3} \left(-1-2 e^2+3 e^4\right)
	u_p^{5/2}+\frac{1}{3} e^2 \left(-52+37 e^2\right)
	u_p^{7/2}+\nonumber\\
	&\left(-102 e^2+\frac{353 e^4}{12}\right)
	u_p^{9/2}+\left(-704 e^2+\frac{763 e^4}{6}\right)
	u_p^{11/2}\,,\nonumber\\
	\lambda^0(u_p,e)=&1-e^2+\frac{1}{4} \left(-18+25 e^2-7 e^4\right) u_p\nonumber\\
	&+\frac{1}{16}
	\left(-36-36 e^2+115 e^4\right) u_p^2\nonumber\\
	&+\frac{3}{64}
	\left(-144-220 e^2+421 e^4\right) u_p^3\nonumber\\
	&+\frac{1}{16}
	\left(-405-807 e^2+1007 e^4\right) u_p^4\nonumber\\
	&+\frac{3}{256}
	\left(-9072-24772 e^2+22501 e^4\right) u_p^5\,,\\
	\lambda^a(u_p,e)=&\frac{4}{3} \left(1-e^2\right)
	\sqrt{u_p}-\frac{2}{3} \left(1-e^2\right)
	u_p^{3/2}+\left(6-\frac{25 e^2}{6}-\frac{13 e^4}{4}\right)
	u_p^{5/2}\nonumber\\
	&+\left(\frac{39}{2}+32 e^2-\frac{202 e^4}{3}\right)
	u_p^{7/2}\nonumber\\
	&+\left(\frac{423}{4}+\frac{1761 e^2}{8}-\frac{26243
		e^4}{96}\right) u_p^{9/2}
	\,.
	\end{align}

\section{Geometrical interpretation of parameter errors}\label{app:geometry}

\begin{figure}
	\includegraphics[width=\linewidth]{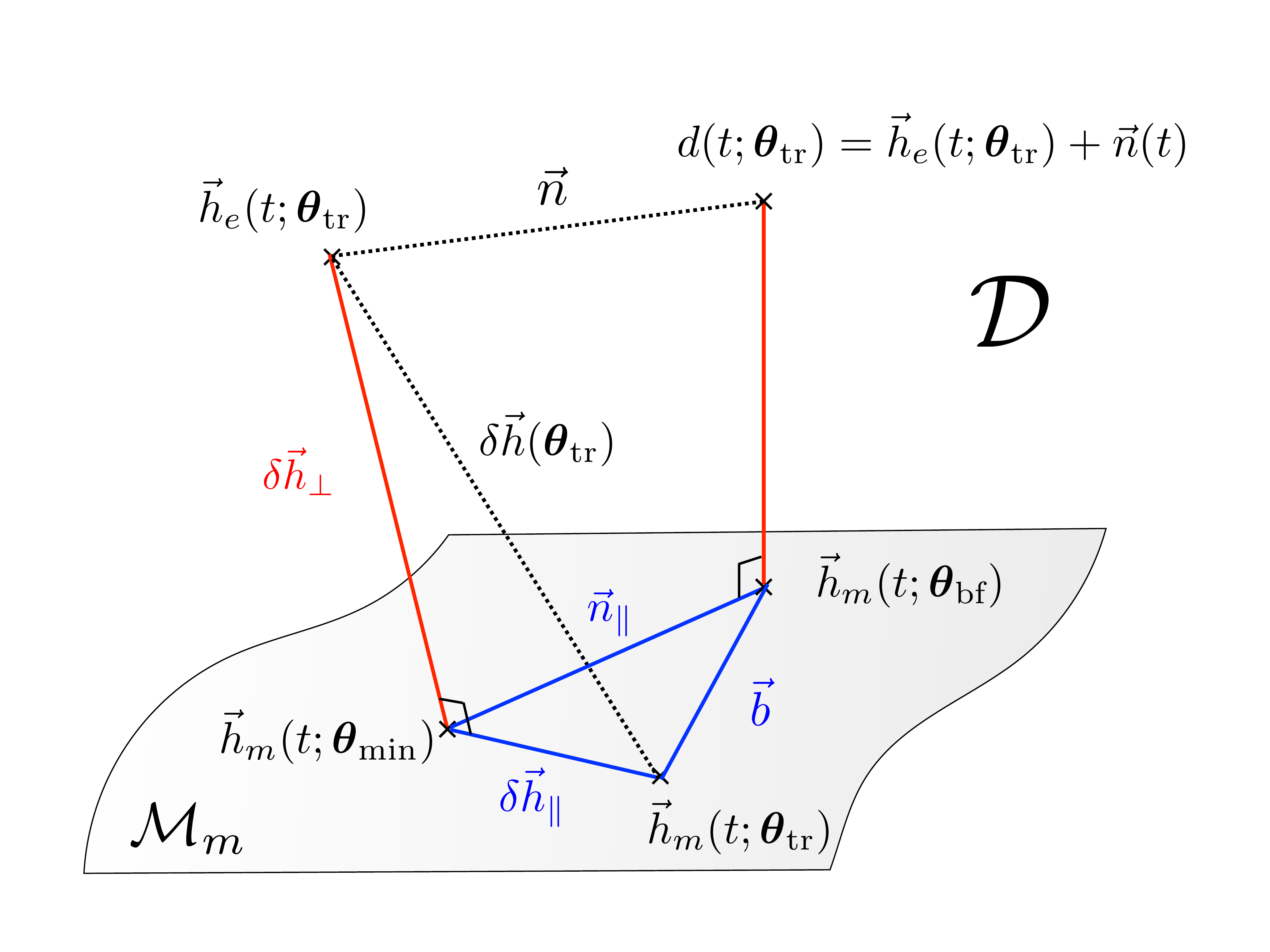}
	\caption{
		Geometrical setup for the CV biases. Represented is the space of signals $\mathcal{D}$ and the various realisations of model and exact templates, with definitions for the parameters as given in the main text. In red, the perpendicular contributions from waveform mismodelling and noise realisations that affect the detectability of the signal. In blue, the contributions to the shifts to the parameters, corresponding to noise-induced errors ($\vec n_\parallel$) and theoretical biases ($\delta\vec h_\parallel$).}
	\label{fig:geometry}
\end{figure}

In this section, we provide a geometrical interpretation for the noise and systematic biases derived in ~\cite{Cutler:2007mi}.
Consider the vector space $\mathcal{D}$ of outputs $\vec d(t;\boldsymbol\theta)$ depending on parameters $\theta^i \in \boldsymbol{\theta}$. Further define two submanifolds $\mathcal{M}_m$ and $\mathcal{M}_e$ of model $\vec h_m(t;\boldsymbol{\theta})$ and fiducial $\vec h_e(t;\boldsymbol{\theta})$ templates, representing both the limiting case of no instrumental noise.
Next, consider the waveform difference $\delta \vec h(\boldsymbol{\theta_\text{tr}}):= \vec h_e(t;\boldsymbol{\theta}_\text{tr})-\vec h_m(t;\boldsymbol{\theta}_\text{tr})$ evaluated at the true parameters.
This can be split into a perpendicular and parallel component. The former is obtained drawing a perpendicular vector $\delta \vec h_\perp$ from $\vec h_e(t;\boldsymbol{\theta}_\text{tr})\in \mathcal{D}$ onto $\mathcal{M}_m$. The projection point is $\vec h_m(t;\boldsymbol\theta_\text{min})$, evaluated at the parameters $\boldsymbol\theta_\text{min}$ that minimise the distance $(\vec h_e - \vec h_m |\vec h_e - \vec h_m)$. Starting from $\vec h_m(t;\boldsymbol\theta_\text{min})$, one can perform a coordinate transformation that maps the model waveform evaluated at $\boldsymbol\theta_\text{min}$ to the same model evaluated at the true parameters $\boldsymbol{\theta}_\text{tr}$. This defines the component $\delta\vec h_\parallel$, see Fig.~(\ref{fig:geometry}). Physically, the $\delta \vec h_\perp$ component corresponds to a ``loss'' of SNR that changes the distance (and therefore affects the likelihood and detectability of the signal only), whereas $\delta h_\parallel \approx (\theta^i_\text{tr} - \theta^i_\text{min}) \partial_i h_m$ corresponds to shifts in the parameters. In what follows, we restrict our attention to vectors in $\mathcal{M}_m$ signalling errors and biases in the parameters, leaving out perpendicular components related to the detectability of the source. 

In a realistic situation, we are confronted with a detector output $\vec d(t; \boldsymbol{\theta}_\text{tr}) = \vec h_e(t; \boldsymbol{\theta}_\text{tr}) +\vec n(t)$ that includes noise. We can project $\vec d$ onto $\mathcal{M}_m$, which defines the model template $\vec h_m(t; \boldsymbol{\theta}_\text{bf})$ evaluated at the best-fit parameters. These are the ones one obtains minimising the argument of the Whittle likelihood, $(\vec d- \vec h_m| \vec d- \vec h_m)$. The new element of $\mathcal{M}_m$, $\vec h_m(t; \boldsymbol{\theta}_\text{bf})$ is connected to  $\vec h_m(t; \boldsymbol{\theta}_\text{min})$ through the parallel component of the noise $\vec n_\parallel$, which can be rewritten as $n_\parallel\approx (\theta^i_\text{bf} - \theta^i_\text{min}) \partial_i h_m$, and to $\vec h_m(t; \boldsymbol{\theta}_\text{tr})$ through a (bias) vector $b\approx (\theta^i_\text{bf} - \theta^i_\text{tr}) \partial_i h_m$, see Fig.~(\ref{fig:geometry}). Then, in this realistic situation the total bias on the PE performed with the model template $\vec h_m$ is given by $\Delta\theta^i:= \theta^i_\text{bf} - \theta^i_\text{tr}$, which itself is formed by two contributions $\Delta\theta^i_\text{noise}:= \theta^i_\text{bf} - \theta^i_\text{min}$ and $\Delta\theta^i_\text{sys}:= \theta^i_\text{tr} - \theta^i_\text{min}$. The former is a statistical error from the noise vector (which averages to zero after many draws of $\vec n$), and we identify it with Eq.~\eqref{eq:width_noise}. The latter is a contribution from waveform mismodelling ($\delta\vec h$) that does not average to zero after many repetitions of the experiment,
and we identify it with the CV bias from theoretical errors~\eqref{eq:width_sys}.

\section{Confusion noise: Stationary treatment}\label{App:confusion_noise}

When the confusion noise is generated by  a very large population of sources, it is common to treat it analogously to the instrumental noise with $f,f'>0$,
\begin{subequations}
	\begin{align}
		\langle \widehat{\Delta H_{\text{conf}}} (f) \rangle &= 0, \\
		\langle \widehat{\Delta H_{\text{conf}}} (f) \widehat{\Delta H^{\star}_{\text{conf}}} (f') \rangle &=  \frac{1}{2} \delta(f-f') S_{\text{conf}}(f), 
		\label{eq:confPSD} \\
		\langle \widehat{\Delta H_{\text{conf}}} (f) \widehat{\Delta H_{\text{conf}}} (f') \rangle &=  0 \label{eq:confPSD_no_conj}
	\end{align}
\end{subequations}
For $S_{\text{conf}}(f)$ the PSD representing the power of the confusion noise at a particular bin of frequency. In this current discussion we are assuming that the confusion noise acts as a \emph{stationary} time-series that is then fully described by an auto-correlation function.

Under these assumptions, the mean bias is zero and the covariance from the confusion background  
takes the alternative form~\footnote{Note that $S_{\text{conf}}$ describes the contribution from the whole astrophysical population, while $\Sigma_{\text{conf}}$ defined in Eq.~\eqref{eq:sigma_conf} was the contribution from a single source in the population. For consistency, we therefore denote the total covariance by $N \Sigma_{\text{conf}}$ in Eq.~\eqref{eq:sigma_conf_PSD}.}

\begin{align}
	&N\Sigma_{\text{conf}}^{ij} = (\Gamma^{-1})^{ik} (\Gamma^{-1})^{jl} \nonumber \\
	&\hspace{0.4cm} \left\langle \int_{-\infty}^\infty \frac{(\partial_k^* h_m (f) \widehat{\Delta H}(f) + \partial_k h_m (f) \widehat{\Delta H}^*(f))}{S_n(f)} \, {text{d}} f \right. \nonumber \\
	& \hspace{0.4cm}\left. \int_{-\infty}^\infty \frac{(\partial_l^* h_m (f') \widehat{\Delta H}(f') + \partial_l h_m (f') \widehat{\Delta H}^*(f'))}{S_n(f')} \,  {text{d}} f'
	\right\rangle \nonumber \\
	&= (\Gamma^{-1})^{ik} (\Gamma^{-1})^{jl} \nonumber \\ 
	&\hspace{0.5cm}2\int_{0}^\infty \frac{(\partial_k h_m^{\star} (f) \partial_l h_m (f) + \partial_l h_m^{\star} (f) \partial_k h_m (f)) S_{\text{conf}}(f)}{S_n^2(f)} \, {text{d}} f
	,\label{eq:sigma_conf_PSD}
\end{align}
Where we have used \eqref{eq:confPSD}-\eqref{eq:confPSD_no_conj} to reach the final equality.
If we use this prescription within the formalism we have here described, we can calculate the total covariance in the parameter estimates arising from instrumental noise and source confusion, which is $\langle (\Delta \theta^i_{\text{noise}} + \Delta \theta^i_{\text{conf}}) (\Delta \theta^j_{\text{noise}} + \Delta \theta^j_{\text{conf}}) \rangle = \Gamma^{-1} + N\Sigma_{\text{conf}}$, with $\Sigma_{text{conf}}$ defined by Eq.~\eqref{eq:sigma_conf}. This results follows because $\langle \Delta \theta^i_{\text{noise}} \Delta \theta^i_{\text{conf}} \rangle=0,$ since the instrumental and astrophysical noises should not depend on one another. 
To calculate the total variance $[\Gamma^{-1} + \Sigma_{\text{conf}}]^{ij}$, we first quote the general result
\begin{equation}\label{CV_paper:step_1}
	\langle (\partial_{i}h_{m}|\widehat{\Delta H_{\text{conf}}})(\partial_{j}h_{m}|\widehat{\Delta H_{\text{conf}}}) \rangle = \Gamma_{ij},
\end{equation}
that is easily proved using \eqref{eq:confPSD}-\eqref{eq:confPSD_no_conj}. We can then re-write $(\Gamma^{-1})^{ij}$ as
\begin{align}\label{CV_paper:step_2}
	(\Gamma^{-1})^{ij} & = \int (\Gamma^{-1})^{ip}\Gamma_{pm}(\Gamma^{-1})^{mj}p_{\text{pop}}(\boldsymbol{\theta}_{\text{conf}})d\boldsymbol{\theta}_{\text{conf}}
\end{align}
since the Fisher matrix is independent of the confusion population and thus population parameters. Integrating over this ensemble of sources is equivalent to taking an ensemble average. Using \eqref{CV_paper:step_1}, \eqref{CV_paper:step_2} and ~\eqref{eq:sigma_conf}, we find 
\begin{align}
	\left[\Gamma^{-1} + \Sigma_{\text{conf}}\right]^{ij} &= (\Gamma^{-1})^{ik}\Sigma_{\text{mix}}^{kl}  (\Gamma^{-1})^{jl},
	\label{eq:biasconfvar}
\end{align}
where
\begin{align}
	\Sigma_{\text{mix}}^{ij} &=4\text{Re}\int_{0}^\infty \frac{(\partial_k \hat{h}_m(f) \partial_l \hat{h}^{\star}_m(f)) (S_{\text{conf}}(f)+S_{\text{n}}(f))}{S_{\text{n}}^2(f)}.
\end{align}

In contrast to this, the standard approach when modelling the confusion background is to combine the instrumental and confusion noises into a single noise term, $N = n + \Delta  H_{\text{conf}}$. Then the standard parameter estimation formalism can be used, with the substitution $S_{\text{n}} (f) \rightarrow S_{\text{n}}(f) + S_{\text{conf}}(f)$ in the inner product~\eqref{eq:inn_prod}. In this case the inference uncertainties are given by the inverse of the Fisher matrix, $\Gamma^{-1}_{\text{n}+\text{conf}}$, where
\begin{equation}
	\Gamma_{\text{n}+\text{conf}}^{ij} =4\text{Re}\int_{0}^\infty \frac{(\partial_k \hat{h}_m(f) \partial_l \hat{h}^{\star}_m(f))}{S_{\text{n}}(f)+S_{\text{conf}}(f)}.
	\label{eq:confvar}
\end{equation}
The variance given by Eq.~\eqref{eq:biasconfvar} is, in general, larger than that predicted by Eq.~\eqref{eq:confvar}. This is because it has been derived by maximizing the standard likelihood as an estimator of the parameters, which is no longer the correct likelihood when random confusion noise is included in the  model. Expression~\eqref{eq:confvar} gives the variance of the true maximum likelihood estimator, which is known to be the minimum variance unbiased estimator and must therefore be smaller than~\eqref{eq:biasconfvar}. Incorporating the confusion noise uncertainty into the PSD is the correct thing to do when Eq.~\eqref{eq:confPSD} is known to be a good approximation, but the formalism described here can be used when that equation is not valid, and to assess when confusion noise is likely to be problematic for parameter estimation. As a final remark, we note that in the limit that there are a large number of sources contributing to the confusion background, the central limit theorem allows us to approximate the probability distribution of the parameter bias correction, $p(\Delta  \boldsymbol\theta_{\text{conf}})$, as a Gaussian with mean $\boldsymbol{\mu}_{\text{conf}}$, given by Eq.~\eqref{eq:mu_conf}, and covariance $\Sigma_{\text{conf}}$. The correct statistical procedure of marginalising the likelihood for $d(t)-\Delta H(t)$ over the confusion noise distribution thus amounts, in the linear signal approximation, to shifting the mean by $\boldsymbol{\mu}_{\text{conf}}$ and adding $\Sigma_{\text{conf}}$ to the covariance. The results described here can therefore be used not only to assess when confusion is important but also to compute leading order corrections to posterior parameter estimates arising from the presence of confusion. 

\section{Numerical Routines}
\label{app:num_routines}
In this appendix, we provide more details on how we sample our signals in the frequency domain. We begin by choosing a starting frequency $f_{0}$ and final frequency determined by the last stable orbit in a Schwarzschild spacetime $f_{\text{max}}= c^{3}/6\sqrt{6}\pi GM$. The calculated time to merger is then predicted through the 3.5PN chirp time (see Eq.(3.5a) of~\cite{Allen:2005fk}). Invoking Shannon's sampling theorem \cite{shannon1949communication}, the spacing between time points $\Delta t$ is chosen to be $\Delta t = 1/(2f_{\text{max}})$. For multiple signals, we choose the minimum sampling interval common to all waveforms for given mass parameters. In doing so, we find the length of the signal $N_{t} = \lfloor t_{obs}/\Delta t \rfloor$ in the time domain. Combining all these elements, one is able to construct a list of sampling frequencies $f = [0,\Delta f, 2\Delta f, \dots, \lfloor (N_{t} - 1)/2\rfloor\Delta f]$ for $\Delta f = 1/N_{t}\Delta t$. Given the discrete Fourier frequencies, it is then possible to construct waveforms using \eqref{eq:signal_model_SPA}.

Noise is generated in the frequency domain with real and imaginary parts drawn separately from Gaussian distributions with equal variance and zero mean. Discretising equation \eqref{eq:Wiener-Khinchin-Theorem_freq}, it's easy to show that the variance of both real and imaginary parts are equivalent to 
\begin{equation}
	\sigma^{2}(f_{i}) = N_{t}S_{n}(f_{i})/4\Delta t.
\end{equation}

Finally, in order to calculate various quantities involving inner products (Fisher matrices, SNRs and likelihoods), we use the discrete analogue of \eqref{eq:inn_prod},

\begin{equation}\label{eq:discrete_inn_prod}
	(a|b) \approx 4\Delta f \ \text{Re} \sum_{i=0}^{\big \lfloor \frac{N_{t}-1}{2} \big 
		\rfloor }\frac{\hat{a}(f_{i})\hat{b}^{\star}(f_{i})}{S_{n}(f_{i})}.
\end{equation}

\section{Fisher Matrices and their validation}\label{app:Fisher_Matrix}

The Fisher Matrix \eqref{eq:Fish_Matrix} can be calculated through inner products of waveform derivatives. We choose to use a second order finite difference method, 
\begin{equation}
	\frac{\partial h_m(f;\Theta^i)}{\partial \Theta^i} \approx \frac{h_m(f;\Theta^i+ \delta\Theta^i)-h_m(f;\Theta^i -\delta\Theta^i)}{2 \delta\Theta^i}
\end{equation}
Fisher matrices in gravitational wave astronomy have high condition numbers, which influence our ability to obtain reliable parameter precision estimates. We invert our Fisher matrices using the high precision arithmetic Python package \texttt{mpmath} \cite{mpmath}. This was done in order to mitigate instabilities arising from computing the inverse of the potentially badly conditioned matrix ~\cite{wen2005detecting,abuse_fisher,porter2009overview,rodriguez2012verifying,gair2013testing,porter2015fisher,amaro2018relativistic,PhysRevD.102.124054}. A criterion to establish the stability of the inverse Fisher matrix based on the (1-norm) absolute value reads $\big|\Gamma^{-1}\Gamma - \texttt{I}\big|_{1} \leq 10^{-3}$, where $\texttt{I}$ is the identity matrix~\cite{Gupta:2020lxa}. We used $\sim$ 500 decimal digits and found  $\Gamma^{-1}\Gamma = \texttt{I} - \epsilon_{ij} \texttt{I}$ with $\max_{i,j}\{|\epsilon_{ij}|\} \approx 10^{-14}$, even with condition numbers $\sim 10^{21}$. This gives us confidence that the numerical inversion of our Fisher matrix is both numerically robust and accurate. 

To validate our results, we carry out a Markov-Chain Monte-Carlo (MCMC) with the goal to match our Fisher matrix results in a high-SNR regime. Our Bayesian analyses are carried out using \texttt{emcee} \cite{ForemanMackey:2012ig} and an appropriate modification of the code developed in~\cite{PhysRevD.102.124054}. The posteriors are sampled with \texttt{emcee} using a Whittle log-likelihood~\eqref{eq:whittle_likelihood} and flat priors. A publicly available implementation of the MCMC illustrations carried out with \texttt{emcee} can be found at \url{https://github.com/aantonelli94/GWOP}.
The latter code is based on a standard Metropolis-Hastings algorithm~\cite{metropolis1953equation}. A publicly available implementation can be found at \url{https://github.com/OllieBurke/Noisy_Neighbours}.  For this algorithm, we chose a proposal distribution equivalent to a multivariate Gaussian with covariance matrix equal to a scaled variate inverse of the Fisher Matrix. By pre-multiplying the inverse Fisher matrix $\Gamma^{-1}$ by $N_{\text{sources}}$, we found better acceptance ratios $\sim 30\%$ [near the optimal acceptance rate for non-single parameter studies~\cite{roberts1997weak}].

\section{Predicting waveform and confusion noise biases with ET}
\label{app:ET_gf}

\begin{figure*}
	\centering
	\includegraphics[height = 11cm,width = 14cm]{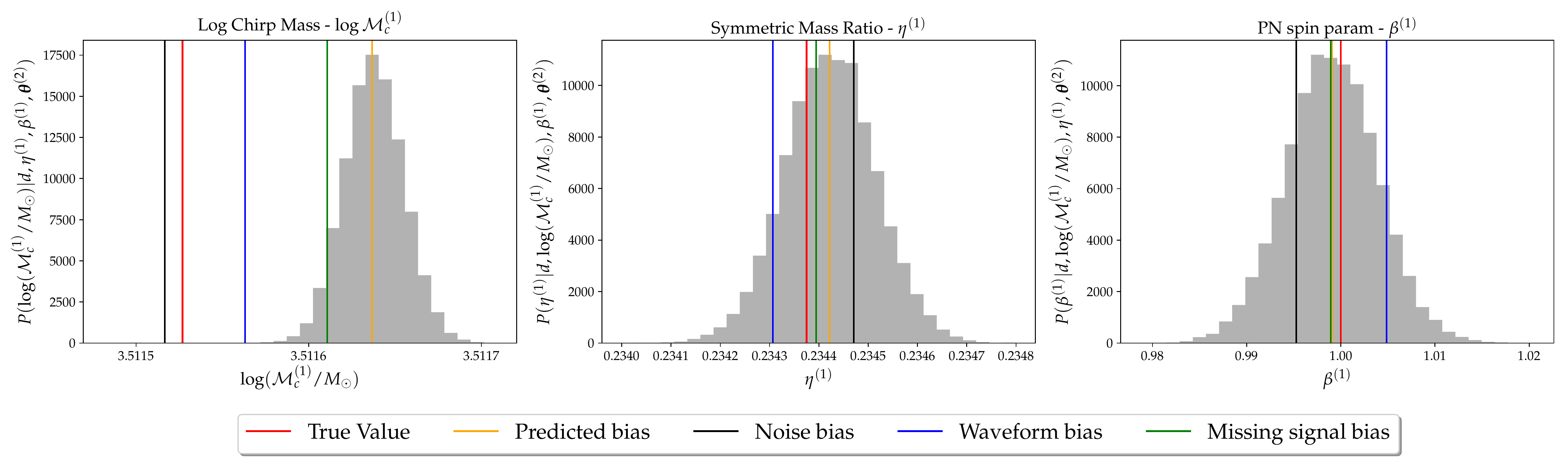}
	\caption{Same as Figure \ref{fig:posteriors_LISA} but for the ET configuration of Appendix~\ref{app:ET_gf}.}
	\label{fig:posteriors_LIGO}
\end{figure*}

In this appendix, we repeat the analysis of Sec.~\ref{sec:results} for a source in ET.
We use the same data stream as~\eqref{eq:global_eq}, modelling $N_\text{fit}=2$ simultaneously-fitted signals in a similar manner. We pick waveform errors $\epsilon = 0.02$ and a starting frequency $f_{0} = 5$Hz. As for confusion noise, we construct it with a series of missed signals which we model without errors. We report the parameters for both fitted and missed sources in Tab.~\eqref{tab:sec_app_table}. The SNRs of the fitted signals are $\mathcal{O}(10^{3})$, those of the missed signals $\lesssim 1000$ (with the lowest $\sim 200$). The SNRs of the missed signals for ET are noticeably high, and would likely be detected in a future analysis. However, for sake of example, treat these signals as missed signals in the parameter estimation scheme. The predictions for the biases of all parameters, Fig.~\ref{app:ET_gf}, show that the formalism can predict the mean of the posterior as remarkably well as in the case of LISA. The individual bias contributions, Fig.~\ref{fig:posteriors_LIGO}, confirm that biases can deconstructively interfere.

\begin{table}
	\centering
	\caption{Parameter configurations for the ET case.}
	\label{tab:sec_app_table}
	\begin{tabular}{c|cccc}
		\multicolumn{1}{l|}{}          & \multicolumn{4}{c|}{Fitted}  \\ \hline
		$i$                    & $M/M_{\odot}$ & $\eta$ & $\beta$ & $D_{\text{eff}}/\text{Mpc}$  \\ \hline
		1                      &   80             & 0.234       &  1    & 400    \\
		2                      &   70             & 0.204       &  5    & 40     \\
		\multicolumn{1}{l|}{}          & \multicolumn{4}{c|}{Missed}   \\ \hline
		$i$                    & $M/M_{\odot}$ & $\eta$ & $\beta$ & $D_{\text{eff}}/\text{Mpc}$  \\ \hline
		1                      & 2.22  & 2.708 & 5.04  & 259.93   \\
		2                      & 2.886 & 0.247 & 3.882 & 253.36   \\
		3                      & 4.395 & 0.2264 & 5.539 & 324.227    \\
		4                      & 6.452 & 0.1991 & 4.404 & 305.828   
	\end{tabular}
\end{table}

\begin{figure*}
	\centering
	\includegraphics[height = 11cm,width = 14cm]{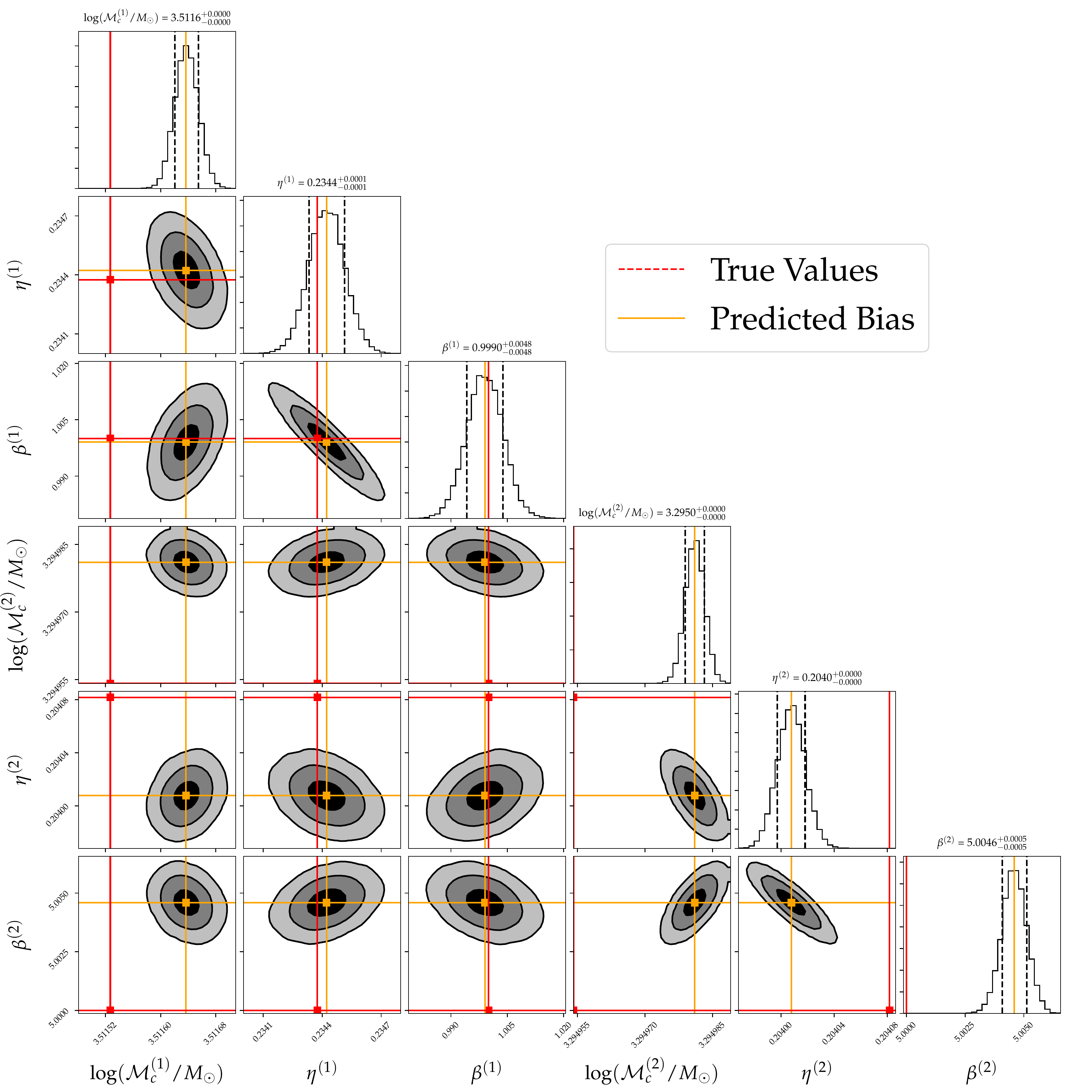}
	\caption{Same as Fig.~\ref{fig:LISA_Corner} but for the ET configuration of Appendix~\ref{app:ET_gf}.}
\end{figure*}

\bibliographystyle{unsrt}
\bibliography{literature}

\begin{thebibliography}{100}

\bibitem{Abbott:2016blz}
B.~P. Abbott et~al.
\newblock {Observation of Gravitational Waves from a Binary Black Hole Merger}.
\newblock {\em Phys. Rev. Lett.}, 116(6):061102, 2016.

\bibitem{Abbott:2017oio}
B.~P. Abbott et~al.
\newblock {GW170814: A Three-Detector Observation of Gravitational Waves from a
  Binary Black Hole Coalescence}.
\newblock {\em Phys. Rev. Lett.}, 119(14):141101, 2017.

\bibitem{TheLIGOScientific:2017qsa}
B.P. Abbott et~al.
\newblock {GW170817: Observation of Gravitational Waves from a Binary Neutron
  Star Inspiral}.
\newblock {\em Phys. Rev. Lett.}, 119(16):161101, 2017.

\bibitem{Abbott:2020niy}
R.~Abbott et~al.
\newblock {GWTC-2: Compact Binary Coalescences Observed by LIGO and Virgo
  During the First Half of the Third Observing Run}.
\newblock 10 2020.

\bibitem{LIGOScientific:2021qlt}
R.~Abbott et~al.
\newblock {Observation of Gravitational Waves from Two Neutron
  Star\textendash{}Black Hole Coalescences}.
\newblock {\em Astrophys. J. Lett.}, 915(1):L5, 2021.

\bibitem{Einstein:1915ca}
Albert Einstein.
\newblock {The Field Equations of Gravitation}.
\newblock {\em Sitzungsber. Preuss. Akad. Wiss. Berlin (Math. Phys. )},
  1915:844--847, 1915.

\bibitem{Einstein:1916cc}
Albert Einstein.
\newblock {Approximative Integration of the Field Equations of Gravitation}.
\newblock {\em Sitzungsber. Preuss. Akad. Wiss. Berlin (Math. Phys. )},
  1916:688--696, 1916.

\bibitem{Kennefick:2007zz}
Daniel Kennefick.
\newblock {\em {Traveling at the speed of thought: Einstein and the quest for
  gravitational waves}}.
\newblock 2007.

\bibitem{Bondi:1957dt}
Hermann Bondi.
\newblock {Plane gravitational waves in general relativity}.
\newblock {\em Nature}, 179:1072--1073, 1957.

\bibitem{PhysRev.117.306}
J.~Weber.
\newblock Detection and generation of gravitational waves.
\newblock {\em Phys. Rev.}, 117:306--313, Jan 1960.

\bibitem{PhysRevLett.22.1320}
J.~Weber.
\newblock Evidence for discovery of gravitational radiation.
\newblock {\em Phys. Rev. Lett.}, 22:1320--1324, Jun 1969.

\bibitem{Hulse:1974eb}
R.~A. Hulse and J.~H. Taylor.
\newblock {Discovery of a pulsar in a binary system}.
\newblock {\em Astrophys. J. Lett.}, 195:L51--L53, 1975.

\bibitem{Weisberg:1981mt}
J.~M. Weisberg, J.~H. Taylor, and L.~A. Fowler.
\newblock {GRAVITATIONAL WAVES FROM AN ORBITING PULSAR}.
\newblock {\em Sci. Am.}, 245:66--74, 1981.

\bibitem{Taylor:1982zz}
J.~H. Taylor and J.~M. Weisberg.
\newblock {A new test of general relativity: Gravitational radiation and the
  binary pulsar PS R 1913+16}.
\newblock {\em Astrophys. J.}, 253:908--920, 1982.

\bibitem{Weiss_detector}
R.~Weiss.
\newblock { Electromagnetically Coupled Broadband Gravitational Antenna}.
\newblock {\em Quarterly Progress Report, Research Laboratory of Electronics,
  MIT}, 1972.

\bibitem{Somiya:2011np}
Kentaro Somiya.
\newblock {Detector configuration of KAGRA: The Japanese cryogenic
  gravitational-wave detector}.
\newblock {\em Class. Quant. Grav.}, 29:124007, 2012.

\bibitem{Cutler:1992tc}
Curt Cutler et~al.
\newblock {The Last three minutes: issues in gravitational wave measurements of
  coalescing compact binaries}.
\newblock {\em Phys. Rev. Lett.}, 70:2984--2987, 1993.

\bibitem{Finn:1992wt}
Lee~S. Finn.
\newblock {Detection, measurement and gravitational radiation}.
\newblock {\em Phys. Rev. D}, 46:5236--5249, 1992.

\bibitem{Finn:1992xs}
Lee~Samuel Finn and David~F. Chernoff.
\newblock {Observing binary inspiral in gravitational radiation: One
  interferometer}.
\newblock {\em Phys. Rev.}, D47:2198--2219, 1993.

\bibitem{Flanagan:1997fn}
{\'E}anna~{\'E}. Flanagan.
\newblock {General relativistic coupling between orbital motion and internal
  degrees of freedom for inspiraling binary neutron stars}.
\newblock {\em Phys. Rev.}, D58:124030, 1998.

\bibitem{Flanagan:1997kp}
Eanna~E. Flanagan and Scott~A. Hughes.
\newblock {Measuring gravitational waves from binary black hole coalescences:
  2. The Waves' information and its extraction, with and without templates}.
\newblock {\em Phys. Rev. D}, 57:4566--4587, 1998.

\bibitem{Sathyaprakash:1991mt}
B.~S. Sathyaprakash and S.~V. Dhurandhar.
\newblock {Choice of filters for the detection of gravitational waves from
  coalescing binaries}.
\newblock {\em Phys. Rev. D}, 44:3819--3834, 1991.

\bibitem{Dhurandhar:1992mw}
S.~V. Dhurandhar and B.~S. Sathyaprakash.
\newblock {Choice of filters for the detection of gravitational waves from
  coalescing binaries. 2. Detection in colored noise}.
\newblock {\em Phys. Rev. D}, 49:1707--1722, 1994.

\bibitem{Abbott:2019yzh}
B.~P. Abbott et~al.
\newblock {A gravitational-wave measurement of the Hubble constant following
  the second observing run of Advanced LIGO and Virgo}.
\newblock 2019.

\bibitem{LIGOScientific:2016lio}
B.~P. Abbott et~al.
\newblock {Tests of general relativity with GW150914}.
\newblock {\em Phys. Rev. Lett.}, 116(22):221101, 2016.
\newblock [Erratum: Phys.Rev.Lett. 121, 129902 (2018)].

\bibitem{LIGOScientific:2019fpa}
B.~P. Abbott et~al.
\newblock {Tests of General Relativity with the Binary Black Hole Signals from
  the LIGO-Virgo Catalog GWTC-1}.
\newblock {\em Phys. Rev. D}, 100(10):104036, 2019.

\bibitem{LIGOScientific:2020tif}
R.~Abbott et~al.
\newblock {Tests of general relativity with binary black holes from the second
  LIGO-Virgo gravitational-wave transient catalog}.
\newblock {\em Phys. Rev. D}, 103(12):122002, 2021.

\bibitem{Unnikrishnan:2013qwa}
C.~S. Unnikrishnan.
\newblock {IndIGO and LIGO-India: Scope and plans for gravitational wave
  research and precision metrology in India}.
\newblock {\em Int. J. Mod. Phys. D}, 22:1341010, 2013.

\bibitem{Punturo:2010zz}
M.~Punturo et~al.
\newblock {The Einstein Telescope: A third-generation gravitational wave
  observatory}.
\newblock {\em Class. Quant. Grav.}, 27:194002, 2010.

\bibitem{Reitze:2019iox}
David Reitze et~al.
\newblock {Cosmic Explorer: The U.S. Contribution to Gravitational-Wave
  Astronomy beyond LIGO}.
\newblock {\em Bull. Am. Astron. Soc.}, 51(7):035, 2019.

\bibitem{Audley:2017drz}
Pau Amaro-Seoane et~al.
\newblock {Laser Interferometer Space Antenna}.
\newblock 2 2017.

\bibitem{Purrer:2019jcp}
Michael P\"urrer and Carl-Johan Haster.
\newblock {Gravitational waveform accuracy requirements for future ground-based
  detectors}.
\newblock {\em Phys. Rev. Res.}, 2(2):023151, 2020.

\bibitem{Relton:2021cax}
Philip Relton and Vivien Raymond.
\newblock {Parameter Estimation Bias From Overlapping Binary Black Hole Events
  In Second Generation Interferometers}.
\newblock 3 2021.

\bibitem{PhysRevD.71.104016}
Mark Miller.
\newblock Accuracy requirements for the calculation of gravitational waveforms
  from coalescing compact binaries in numerical relativity.
\newblock {\em Phys. Rev. D}, 71:104016, May 2005.

\bibitem{Cutler:2007mi}
Curt Cutler and Michele Vallisneri.
\newblock {LISA detections of massive black hole inspirals: Parameter
  extraction errors due to inaccurate template waveforms}.
\newblock {\em Phys. Rev. D}, 76:104018, 2007.

\bibitem{Antonelli:2019fmq}
Andrea Antonelli, Maarten van~de Meent, Alessandra Buonanno, Jan Steinhoff, and
  Justin Vines.
\newblock {Quasicircular inspirals and plunges from nonspinning
  effective-one-body Hamiltonians with gravitational self-force information}.
\newblock {\em Phys. Rev.}, D101(2):024024, 2020.

\bibitem{Antonelli:2019ytb}
Andrea Antonelli, Alessandra Buonanno, Jan Steinhoff, Maarten van~de Meent, and
  Justin Vines.
\newblock {Energetics of two-body Hamiltonians in post-Minkowskian gravity}.
\newblock {\em Phys. Rev.}, D99(10):104004, 2019.

\bibitem{Antonelli:2020aeb}
Andrea Antonelli, Chris Kavanagh, Mohammed Khalil, Jan Steinhoff, and Justin
  Vines.
\newblock {Gravitational spin-orbit coupling through third-subleading
  post-Newtonian order: from first-order self-force to arbitrary mass ratios}.
\newblock {\em Phys. Rev. Lett.}, 125(1):011103, 2020.

\bibitem{Antonelli:2020ybz}
Andrea Antonelli, Chris Kavanagh, Mohammed Khalil, Jan Steinhoff, and Justin
  Vines.
\newblock {Gravitational spin-orbit and aligned spin$_1$-spin$_2$ couplings
  through third-subleading post-Newtonian orders}.
\newblock {\em Phys. Rev. D}, 102:124024, 2020.

\bibitem{Antonelli:2021vwg}
Andrea Antonelli, Ollie Burke, and Jonathan~R. Gair.
\newblock {Noisy neighbours: inference biases from overlapping
  gravitational-wave signals}.
\newblock {\em Mon. Not. Roy. Astron. Soc.}, 507(4):5069--5086, 2021.

\bibitem{Oppenheimer:1939ue}
J.~R. Oppenheimer and H.~Snyder.
\newblock {On Continued gravitational contraction}.
\newblock {\em Phys. Rev.}, 56:455--459, 1939.

\bibitem{May:1966zz}
Michael~M. May and Richard~H. White.
\newblock {Hydrodynamic Calculations of General-Relativistic Collapse}.
\newblock {\em Phys. Rev.}, 141:1232--1241, 1966.

\bibitem{Penrose:1969pc}
R.~Penrose.
\newblock {Gravitational collapse: The role of general relativity}.
\newblock {\em Riv. Nuovo Cim.}, 1:252--276, 1969.

\bibitem{Belczynski:2001uc}
Krzysztof Belczynski, Vassiliki Kalogera, and Tomasz Bulik.
\newblock {A Comprehensive study of binary compact objects as gravitational
  wave sources: Evolutionary channels, rates, and physical properties}.
\newblock {\em Astrophys. J.}, 572:407--431, 2001.

\bibitem{Belczynski:2016obo}
Krzysztof Belczynski, Daniel~E. Holz, Tomasz Bulik, and Richard O'Shaughnessy.
\newblock {The first gravitational-wave source from the isolated evolution of
  two 40-100 Msun stars}.
\newblock {\em Nature}, 534:512, 2016.

\bibitem{Sadowski:2007dz}
Aleksander Sadowski, Krzysztof Belczynski, Tomasz Bulik, Natalia Ivanova,
  Frederic~A. Rasio, and Richard~W. O'Shaughnessy.
\newblock {The Total Merger Rate of Compact Object Binaries In The Local
  Universe}.
\newblock {\em Astrophys. J.}, 676:1162, 2008.

\bibitem{OLeary:2007iqe}
Ryan O'Leary, Richard~W. O'Shaughnessy, and Frederic Rasio.
\newblock {Dynamical Interactions and the Black Hole Merger Rate of the
  Universe}.
\newblock {\em Phys. Rev. D}, 76:061504, 2007.

\bibitem{OLeary:2008myb}
Ryan~M. O'Leary, Bence Kocsis, and Abraham Loeb.
\newblock {Gravitational waves from scattering of stellar-mass black holes in
  galactic nuclei}.
\newblock {\em Mon. Not. Roy. Astron. Soc.}, 395(4):2127--2146, 2009.

\bibitem{Miller:2008yw}
M.~Coleman Miller and Vanessa~M. Lauburg.
\newblock {Mergers of Stellar-Mass Black Holes in Nuclear Star Clusters}.
\newblock {\em Astrophys. J.}, 692:917--923, 2009.

\bibitem{Tsang:2013mca}
David Tsang.
\newblock {Shattering Flares During Close Encounters of Neutron Stars}.
\newblock {\em Astrophys. J.}, 777:103, 2013.

\bibitem{Rodriguez:2015oxa}
Carl~L. Rodriguez, Meagan Morscher, Bharath Pattabiraman, Sourav Chatterjee,
  Carl-Johan Haster, and Frederic~A. Rasio.
\newblock {Binary Black Hole Mergers from Globular Clusters: Implications for
  Advanced LIGO}.
\newblock {\em Phys. Rev. Lett.}, 115(5):051101, 2015.
\newblock [Erratum: Phys.Rev.Lett. 116, 029901 (2016)].

\bibitem{Rodriguez:2016kxx}
Carl~L. Rodriguez, Sourav Chatterjee, and Frederic~A. Rasio.
\newblock {Binary Black Hole Mergers from Globular Clusters: Masses, Merger
  Rates, and the Impact of Stellar Evolution}.
\newblock {\em Phys. Rev. D}, 93(8):084029, 2016.

\bibitem{Campanelli:2006fg}
Manuela Campanelli, C.~O. Lousto, and Yosef Zlochower.
\newblock {Spin-orbit interactions in black-hole binaries}.
\newblock {\em Phys. Rev. D}, 74:084023, 2006.

\bibitem{Rodriguez:2016vmx}
Carl~L. Rodriguez, Michael Zevin, Chris Pankow, Vasilliki Kalogera, and
  Frederic~A. Rasio.
\newblock {Illuminating Black Hole Binary Formation Channels with Spins in
  Advanced LIGO}.
\newblock {\em Astrophys. J. Lett.}, 832(1):L2, 2016.

\bibitem{Vitale:2015tea}
Salvatore Vitale, Ryan Lynch, Riccardo Sturani, and Philip Graff.
\newblock {Use of gravitational waves to probe the formation channels of
  compact binaries}.
\newblock {\em Class. Quant. Grav.}, 34(3):03LT01, 2017.

\bibitem{Farr:2017uvj}
Will~M. Farr, Simon Stevenson, M.~Coleman~Miller, Ilya Mandel, Ben Farr, and
  Alberto Vecchio.
\newblock {Distinguishing Spin-Aligned and Isotropic Black Hole Populations
  With Gravitational Waves}.
\newblock {\em Nature}, 548:426, 2017.

\bibitem{Fishbach:2017dwv}
Maya Fishbach, Daniel~E. Holz, and Ben Farr.
\newblock {Are LIGO's Black Holes Made From Smaller Black Holes?}
\newblock {\em Astrophys. J. Lett.}, 840(2):L24, 2017.

\bibitem{Peters:1964zz}
P.~C. Peters.
\newblock {Gravitational Radiation and the Motion of Two Point Masses}.
\newblock {\em Phys. Rev.}, 136:B1224--B1232, 1964.

\bibitem{Zevin:2017evb}
Michael Zevin, Chris Pankow, Carl~L. Rodriguez, Laura Sampson, Eve Chase,
  Vassiliki Kalogera, and Frederic~A. Rasio.
\newblock {Constraining Formation Models of Binary Black Holes with
  Gravitational-Wave Observations}.
\newblock {\em Astrophys. J.}, 846(1):82, 2017.

\bibitem{Zevin:2018kzq}
Michael Zevin, Johan Samsing, Carl Rodriguez, Carl-Johan Haster, and Enrico
  Ramirez-Ruiz.
\newblock {Eccentric Black Hole Mergers in Dense Star Clusters: The Role of
  Binary\textendash{}Binary Encounters}.
\newblock {\em Astrophys. J.}, 871(1):91, 2019.

\bibitem{Gondan:2018khr}
L\'aszl\'o Gond\'an and Bence Kocsis.
\newblock {Measurement Accuracy of Inspiraling Eccentric Neutron Star and Black
  Hole Binaries Using Gravitational Waves}.
\newblock {\em Astrophys. J.}, 871(2):178, 2019.

\bibitem{Samsing:2018ykz}
Johan Samsing, Daniel~J. D'Orazio, Abbas Askar, and Mirek Giersz.
\newblock {Black Hole Mergers from Globular Clusters Observable by LISA and
  LIGO: Results from post-Newtonian Binary-Single Scatterings}.
\newblock 2 2018.

\bibitem{Rodriguez:2018pss}
Carl~L. Rodriguez, Pau Amaro-Seoane, Sourav Chatterjee, Kyle Kremer,
  Frederic~A. Rasio, Johan Samsing, Claire~S. Ye, and Michael Zevin.
\newblock {Post-Newtonian Dynamics in Dense Star Clusters: Formation, Masses,
  and Merger Rates of Highly-Eccentric Black Hole Binaries}.
\newblock {\em Phys. Rev. D}, 98(12):123005, 2018.

\bibitem{Romero-Shaw:2019itr}
Isobel~M. Romero-Shaw, Paul~D. Lasky, and Eric Thrane.
\newblock {Searching for Eccentricity: Signatures of Dynamical Formation in the
  First Gravitational-Wave Transient Catalogue of LIGO and Virgo}.
\newblock {\em Mon. Not. Roy. Astron. Soc.}, 490(4):5210--5216, 2019.

\bibitem{Ossokine:2020kjp}
Serguei Ossokine et~al.
\newblock {Multipolar Effective-One-Body Waveforms for Precessing Binary Black
  Holes: Construction and Validation}.
\newblock {\em Phys. Rev. D}, 102(4):044055, 2020.

\bibitem{Cotesta:2018fcv}
Roberto Cotesta, Alessandra Buonanno, Alejandro Bohé, Andrea Taracchini, Ian
  Hinder, and Serguei Ossokine.
\newblock {Enriching the Symphony of Gravitational Waves from Binary Black
  Holes by Tuning Higher Harmonics}.
\newblock {\em Phys. Rev.}, D98(8):084028, 2018.

\bibitem{Schutz:1986gp}
Bernard~F. Schutz.
\newblock {Determining the Hubble Constant from Gravitational Wave
  Observations}.
\newblock {\em Nature}, 323:310--311, 1986.

\bibitem{Abbott:2018wiz}
B.~P. Abbott et~al.
\newblock {Properties of the binary neutron star merger GW170817}.
\newblock {\em Phys. Rev. X}, 9(1):011001, 2019.

\bibitem{Maggiore:1900zz}
Michele Maggiore.
\newblock {\em {Gravitational Waves. Vol. 1: Theory and Experiments}}.
\newblock Oxford Master Series in Physics. Oxford University Press, 2007.

\bibitem{Flanagan:2005yc}
E.~E. Flanagan and S.~A. Hughes.
\newblock {The Basics of gravitational wave theory}.
\newblock {\em New J.Phys.}, 7:204, 2005.

\bibitem{Droste}
H.A. Lorentz and J.~Droste.
\newblock {\em Versl.K.Akad.Wet.Amsterdam}, 26:392, 1917.

\bibitem{Einstein:1938yz}
Albert Einstein, L.~Infeld, and B.~Hoffmann.
\newblock {The Gravitational equations and the problem of motion}.
\newblock {\em Annals Math.}, 39:65--100, 1938.

\bibitem{Blanchet:2013haa}
Luc Blanchet.
\newblock {Gravitational Radiation from Post-Newtonian Sources and Inspiralling
  Compact Binaries}.
\newblock {\em Living Rev. Rel.}, 17:2, 2014.

\bibitem{Ohta:1973je}
T.~Ohta, H.~Okamura, T.~Kimura, and K.~Hiida.
\newblock {Physically acceptable solution of einstein's equation for many-body
  system}.
\newblock {\em Prog. Theor. Phys.}, 50:492--514, 1973.

\bibitem{Ohta:1974kp}
Tadayuki Ohta, Hiroshi Okamura, Toshiei Kimura, and Kichiro Hiida.
\newblock {Coordinate Condition and Higher Order Gravitational Potential in
  Canonical Formalism}.
\newblock {\em Prog. Theor. Phys.}, 51:1598, 1974.

\bibitem{Ohta:1974pq}
T.~Ohta, H.~Okamura, K.~Hiida, and T.~Kimura.
\newblock {Higher order gravitational potential for many-body system}.
\newblock {\em Prog. Theor. Phys.}, 51:1220--1238, 1974.

\bibitem{Damour:1981bh}
Thibault Damour and Nathalie Deruelle.
\newblock {Radiation Reaction and Angular Momentum Loss in Small Angle
  Gravitational Scattering}.
\newblock {\em Phys. Lett.}, A87:81, 1981.

\bibitem{Damour:1985mt}
Thibault Damour and Gerhard Sch\"afer.
\newblock {Lagrangians for$n$ point masses at the second post-Newtonian
  approximation of general relativity}.
\newblock {\em Gen. Rel. Grav.}, 17:879--905, 1985.

\bibitem{Jaranowski:1997ky}
Piotr Jaranowski and Gerhard Schaefer.
\newblock {Third postNewtonian higher order ADM Hamilton dynamics for two-body
  point mass systems}.
\newblock {\em Phys. Rev. D}, 57:7274--7291, 1998.
\newblock [Erratum: Phys.Rev.D 63, 029902 (2001)].

\bibitem{Jaranowski:1999ye}
Piotr Jaranowski and Gerhard Schaefer.
\newblock {The Binary black hole problem at the third postNewtonian
  approximation in the orbital motion: Static part}.
\newblock {\em Phys. Rev. D}, 60:124003, 1999.

\bibitem{Jaranowski:1999qd}
Piotr Jaranowski and Gerhard Schaefer.
\newblock {The Binary black hole dynamics at the third postNewtonian order in
  the orbital motion}.
\newblock {\em Annalen Phys.}, 9:378--383, 2000.

\bibitem{Damour:2000we}
Thibault Damour, Piotr Jaranowski, and Gerhard Schäfer.
\newblock {On the determination of the last stable orbit for circular general
  relativistic binaries at the third post-Newtonian approximation}.
\newblock {\em Phys. Rev.}, D62:084011, 2000.

\bibitem{Blanchet:2000nv}
Luc Blanchet and Guillaume Faye.
\newblock {Equations of motion of point particle binaries at the third
  postNewtonian order}.
\newblock {\em Phys. Lett. A}, 271:58, 2000.

\bibitem{Blanchet:2000ub}
Luc Blanchet and Guillaume Faye.
\newblock {General relativistic dynamics of compact binaries at the third
  postNewtonian order}.
\newblock {\em Phys. Rev. D}, 63:062005, 2001.

\bibitem{Itoh:2003fy}
Yousuke Itoh and Toshifumi Futamase.
\newblock {New derivation of a third postNewtonian equation of motion for
  relativistic compact binaries without ambiguity}.
\newblock {\em Phys. Rev. D}, 68:121501, 2003.

\bibitem{Foffa:2011ub}
Stefano Foffa and Riccardo Sturani.
\newblock {Effective field theory calculation of conservative binary dynamics
  at third post-Newtonian order}.
\newblock {\em Phys. Rev. D}, 84:044031, 2011.

\bibitem{Jaranowski:2015lha}
Piotr Jaranowski and Gerhard Schäfer.
\newblock {Derivation of local-in-time fourth post-Newtonian ADM Hamiltonian
  for spinless compact binaries}.
\newblock {\em Phys. Rev.}, D92(12):124043, 2015.

\bibitem{Damour:2014jta}
Thibault Damour, Piotr Jaranowski, and Gerhard Schäfer.
\newblock {Nonlocal-in-time action for the fourth post-Newtonian conservative
  dynamics of two-body systems}.
\newblock {\em Phys. Rev.}, D89(6):064058, 2014.

\bibitem{Damour:2016abl}
Thibault Damour, Piotr Jaranowski, and Gerhard Schäfer.
\newblock {Conservative dynamics of two-body systems at the fourth
  post-Newtonian approximation of general relativity}.
\newblock {\em Phys. Rev.}, D93(8):084014, 2016.

\bibitem{Bernard:2015njp}
Laura Bernard, Luc Blanchet, Alejandro Bohé, Guillaume Faye, and Sylvain
  Marsat.
\newblock {Fokker action of nonspinning compact binaries at the fourth
  post-Newtonian approximation}.
\newblock {\em Phys. Rev.}, D93(8):084037, 2016.

\bibitem{Bernard:2016wrg}
Laura Bernard, Luc Blanchet, Alejandro Bohé, Guillaume Faye, and Sylvain
  Marsat.
\newblock {Energy and periastron advance of compact binaries on circular orbits
  at the fourth post-Newtonian order}.
\newblock {\em Phys. Rev.}, D95(4):044026, 2017.

\bibitem{Bernard:2017bvn}
Laura Bernard, Luc Blanchet, Alejandro Bohé, Guillaume Faye, and Sylvain
  Marsat.
\newblock {Dimensional regularization of the IR divergences in the Fokker
  action of point-particle binaries at the fourth post-Newtonian order}.
\newblock {\em Phys. Rev.}, D96(10):104043, 2017.

\bibitem{Marchand:2017pir}
Tanguy Marchand, Laura Bernard, Luc Blanchet, and Guillaume Faye.
\newblock {Ambiguity-Free Completion of the Equations of Motion of Compact
  Binary Systems at the Fourth Post-Newtonian Order}.
\newblock {\em Phys. Rev. D}, 97(4):044023, 2018.

\bibitem{Foffa:2019rdf}
Stefano Foffa and Riccardo Sturani.
\newblock {Conservative dynamics of binary systems to fourth Post-Newtonian
  order in the EFT approach I: Regularized Lagrangian}.
\newblock 2019.

\bibitem{Foffa:2019yfl}
Stefano Foffa, Rafael~A. Porto, Ira Rothstein, and Riccardo Sturani.
\newblock {Conservative dynamics of binary systems to fourth Post-Newtonian
  order in the EFT approach II: Renormalized Lagrangian}.
\newblock 2019.

\bibitem{Goldberger:2006bd}
Walter~D. Goldberger and Ira~Z. Rothstein.
\newblock {Towers of Gravitational Theories}.
\newblock {\em Gen. Rel. Grav.}, 38:1537--1546, 2006.

\bibitem{Blanchet:1985sp}
Luc Blanchet and Thibault Damour.
\newblock {Radiative gravitational fields in general relativity I. general
  structure of the field outside the source}.
\newblock {\em Phil. Trans. Roy. Soc. Lond. A}, 320:379--430, 1986.

\bibitem{Blanchet:1989ki}
L.~Blanchet and T.~Damour.
\newblock {Postnewtonian Generation of Gravitational Waves}.
\newblock {\em Ann. Inst. H. Poincare Phys. Theor.}, 50:377--408, 1989.

\bibitem{Damour:1990gj}
T.~Damour and Bala~R. Iyer.
\newblock {Multipole analysis for electromagnetism and linearized gravity with
  irreducible cartesian tensors}.
\newblock {\em Phys. Rev. D}, 43:3259--3272, 1991.

\bibitem{Sathyaprakash:2009xs}
B.S. Sathyaprakash and B.F. Schutz.
\newblock {Physics, Astrophysics and Cosmology with Gravitational Waves}.
\newblock {\em Living Rev. Rel.}, 12:2, 2009.

\bibitem{Iyer:1993xi}
Bala~R. Iyer and C.~M. Will.
\newblock {PostNewtonian gravitational radiation reaction for two-body
  systems}.
\newblock {\em Phys. Rev. Lett.}, 70:113--116, 1993.

\bibitem{Iyer:1995rn}
Bala~R. Iyer and C.~M. Will.
\newblock {PostNewtonian gravitational radiation reaction for two-body systems:
  Nonspinning bodies}.
\newblock {\em Phys. Rev. D}, 52:6882--6893, 1995.

\bibitem{Gopakumar:1997ng}
A.~Gopakumar, Bala~R. Iyer, and Sai Iyer.
\newblock {Second postNewtonian gravitational radiation reaction for two-body
  systems: Nonspinning bodies}.
\newblock {\em Phys. Rev. D}, 55:6030--6053, 1997.
\newblock [Erratum: Phys.Rev.D 57, 6562 (1998)].

\bibitem{Marchand:2020fpt}
Tanguy Marchand, Quentin Henry, Fran\c{c}ois Larrouturou, Sylvain Marsat,
  Guillaume Faye, and Luc Blanchet.
\newblock {The mass quadrupole moment of compact binary systems at the fourth
  post-Newtonian order}.
\newblock {\em Class. Quant. Grav.}, 37(21):215006, 2020.

\bibitem{Damour:2007nc}
Thibault Damour, Piotr Jaranowski, and Gerhard Schaefer.
\newblock {Hamiltonian of two spinning compact bodies with next-to-leading
  order gravitational spin-orbit coupling}.
\newblock {\em Phys. Rev. D}, 77:064032, 2008.

\bibitem{Arnowitt:1962hi}
R.~L. Arnowitt, S.~Deser, and C.~W. Misner.
\newblock {The dynamics of general relativity}.
\newblock In Louis Witten, editor, {\em {Gravitation: an introduction to
  current research}}, chapter~7, pages 227--265. {John Wiley \& Sons}, New
  York, London, 1962.

\bibitem{Damour:2008qf}
Thibault Damour, Piotr Jaranowski, and Gerhard Schäfer.
\newblock {Effective one body approach to the dynamics of two spinning black
  holes with next-to-leading order spin-orbit coupling}.
\newblock {\em Phys. Rev.}, D78:024009, 2008.

\bibitem{deSitter:1916zza}
W.~de~Sitter.
\newblock {Einstein's theory of gravitation and its astronomical consequences,
  First Paper}.
\newblock {\em Mon. Not. Roy. Astron. Soc.}, 76:699--728, 1916.

\bibitem{Everitt:2011hp}
C.~W.~F. Everitt et~al.
\newblock {Gravity Probe B: Final Results of a Space Experiment to Test General
  Relativity}.
\newblock {\em Phys. Rev. Lett.}, 106:221101, 2011.

\bibitem{Tagoshi:2000zg}
Hideyuki Tagoshi, Akira Ohashi, and Benjamin~J. Owen.
\newblock {Gravitational field and equations of motion of spinning compact
  binaries to 2.5 postNewtonian order}.
\newblock {\em Phys. Rev. D}, 63:044006, 2001.

\bibitem{Faye:2006gx}
Guillaume Faye, Luc Blanchet, and Alessandra Buonanno.
\newblock {Higher-order spin effects in the dynamics of compact binaries. I.
  Equations of motion}.
\newblock {\em Phys. Rev. D}, 74:104033, 2006.

\bibitem{Porto:2010tr}
Rafael~A. Porto.
\newblock {Next to leading order spin-orbit effects in the motion of
  inspiralling compact binaries}.
\newblock {\em Class. Quant. Grav.}, 27:205001, 2010.

\bibitem{Levi:2010zu}
Michele Levi.
\newblock {Next to Leading Order gravitational Spin-Orbit coupling in an
  Effective Field Theory approach}.
\newblock {\em Phys. Rev. D}, 82:104004, 2010.

\bibitem{Hartung:2011te}
Johannes Hartung and Jan Steinhoff.
\newblock {Next-to-next-to-leading order post-Newtonian spin-orbit Hamiltonian
  for self-gravitating binaries}.
\newblock {\em Ann. Phys. (Berlin)}, 523:783--790, 2011.

\bibitem{Hartung:2013dza}
Johannes Hartung, Jan Steinhoff, and Gerhard Schafer.
\newblock {Next-to-next-to-leading order post-Newtonian linear-in-spin binary
  Hamiltonians}.
\newblock {\em Annalen Phys.}, 525:359--394, 2013.

\bibitem{Marsat:2012fn}
Sylvain Marsat, Alejandro Bohe, Guillaume Faye, and Luc Blanchet.
\newblock {Next-to-next-to-leading order spin-orbit effects in the equations of
  motion of compact binary systems}.
\newblock {\em Class. Quant. Grav.}, 30:055007, 2013.

\bibitem{Levi:2015uxa}
Michele Levi and Jan Steinhoff.
\newblock {Next-to-next-to-leading order gravitational spin-orbit coupling via
  the effective field theory for spinning objects in the post-Newtonian
  scheme}.
\newblock {\em JCAP}, 1601:011, 2016.

\bibitem{Bini:2019nra}
Donato Bini, Thibault Damour, and Andrea Geralico.
\newblock {Novel approach to binary dynamics: application to the fifth
  post-Newtonian level}.
\newblock {\em Phys. Rev. Lett.}, 123:231104, 2019.

\bibitem{Levi:2020kvb}
Michèle Levi, Andrew~J. McLeod, and Matthew von Hippel.
\newblock {NNNLO gravitational spin-orbit coupling at the quartic order in G}.
\newblock 2020.

\bibitem{Blanchet:2011zv}
Luc Blanchet, Alessandra Buonanno, and Guillaume Faye.
\newblock {Tail-induced spin-orbit effect in the gravitational radiation of
  compact binaries}.
\newblock {\em Phys. Rev. D}, 84:064041, 2011.

\bibitem{Marsat:2013caa}
Sylvain Marsat, Alejandro Boh\'e, Luc Blanchet, and Alessandra Buonanno.
\newblock {Next-to-leading tail-induced spin\textendash{}orbit effects in the
  gravitational radiation flux of compact binaries}.
\newblock {\em Class. Quant. Grav.}, 31:025023, 2014.

\bibitem{Barker:1975ae}
B.~M. Barker and R.~F. O'Connell.
\newblock {Gravitational Two-Body Problem with Arbitrary Masses, Spins, and
  Quadrupole Moments}.
\newblock {\em Phys. Rev. D}, 12:329--335, 1975.

\bibitem{DEath:1975wqz}
Peter~D. D'Eath.
\newblock {Interaction of two black holes in the slow-motion limit}.
\newblock {\em Phys. Rev. D}, 12:2183--2199, 1975.

\bibitem{Thorne:1984mz}
Kip~S. Thorne and James~B. Hartle.
\newblock {Laws of motion and precession for black holes and other bodies}.
\newblock {\em Phys. Rev. D}, 31:1815--1837, 1984.

\bibitem{Poisson:1997ha}
Eric Poisson.
\newblock {Gravitational waves from inspiraling compact binaries: The
  Quadrupole moment term}.
\newblock {\em Phys. Rev.}, D57:5287--5290, 1998.

\bibitem{Steinhoff:2007mb}
Jan Steinhoff, Steven Hergt, and Gerhard Sch{\"a}fer.
\newblock {On the next-to-leading order gravitational spin(1)-spin(2)
  dynamics}.
\newblock {\em Phys. Rev.}, D77:081501, 2008.

\bibitem{Porto:2008jj}
Rafael~A Porto and Ira~Z. Rothstein.
\newblock {Next to Leading Order Spin(1)Spin(1) Effects in the Motion of
  Inspiralling Compact Binaries}.
\newblock {\em Phys. Rev.}, D78:044013, 2008.
\newblock [Erratum: Phys. Rev. \textbf{D81}, 029905 (2010)].

\bibitem{Porto:2008tb}
Rafael~A. Porto and Ira~Z. Rothstein.
\newblock {Spin(1)Spin(2) Effects in the Motion of Inspiralling Compact
  Binaries at Third Order in the Post-Newtonian Expansion}.
\newblock {\em Phys. Rev.}, D78:044012, 2008.
\newblock [Erratum: Phys. Rev.D81,029904(2010)].

\bibitem{Levi:2008nh}
Michele Levi.
\newblock {Next to Leading Order gravitational Spin1-Spin2 coupling with
  Kaluza-Klein reduction}.
\newblock {\em Phys. Rev. D}, 82:064029, 2010.

\bibitem{Steinhoff:2008ji}
Jan Steinhoff, Steven Hergt, and Gerhard Sch{\"a}fer.
\newblock {Spin-squared Hamiltonian of next-to-leading order gravitational
  interaction}.
\newblock {\em Phys.\ Rev.\ D}, 78:101503, 2008.

\bibitem{Hergt:2010pa}
Steven Hergt, Jan Steinhoff, and Gerhard Schäfer.
\newblock {Reduced Hamiltonian for next-to-leading order Spin-Squared Dynamics
  of General Compact Binaries}.
\newblock {\em Class. Quant. Grav.}, 27:135007, 2010.

\bibitem{Hergt:2011ik}
Steven Hergt, Jan Steinhoff, and Gerhard Schaefer.
\newblock {Elimination of the spin supplementary condition in the effective
  field theory approach to the post-Newtonian approximation}.
\newblock {\em Annals Phys.}, 327:1494--1537, 2012.

\bibitem{Bohe:2015ana}
Alejandro Boh\'e, Guillaume Faye, Sylvain Marsat, and Edward~K. Porter.
\newblock {Quadratic-in-spin effects in the orbital dynamics and
  gravitational-wave energy flux of compact binaries at the 3PN order}.
\newblock {\em Class. Quant. Grav.}, 32(19):195010, 2015.

\bibitem{Levi:2011eq}
Michele Levi.
\newblock {Binary dynamics from spin1-spin2 coupling at fourth post-Newtonian
  order}.
\newblock {\em Phys. Rev. D}, 85:064043, 2012.

\bibitem{Hartung:2011ea}
Johannes Hartung and Jan Steinhoff.
\newblock {Next-to-next-to-leading order post-Newtonian spin(1)-spin(2)
  Hamiltonian for self-gravitating binaries}.
\newblock {\em Annalen Phys.}, 523:919--924, 2011.

\bibitem{Levi:2015ixa}
Michele Levi and Jan Steinhoff.
\newblock {Next-to-next-to-leading order gravitational spin-squared potential
  via the effective field theory for spinning objects in the post-Newtonian
  scheme}.
\newblock {\em JCAP}, 1601:008, 2016.

\bibitem{Levi:2014sba}
Michele Levi and Jan Steinhoff.
\newblock {Equivalence of ADM Hamiltonian and Effective Field Theory approaches
  at next-to-next-to-leading order spin1-spin2 coupling of binary inspirals}.
\newblock {\em JCAP}, 12:003, 2014.

\bibitem{Levi:2020uwu}
Mich\`ele Levi, Andrew~J. Mcleod, and Matthew Von~Hippel.
\newblock {NNNLO gravitational quadratic-in-spin interactions at the quartic
  order in G}.
\newblock 3 2020.

\bibitem{Levi:2016ofk}
Michele Levi and Jan Steinhoff.
\newblock {Complete conservative dynamics for inspiralling compact binaries
  with spins at fourth post-Newtonian order}.
\newblock 7 2016.

\bibitem{Khalil:2020mmr}
Mohammed Khalil, Jan Steinhoff, Justin Vines, and Alessandra Buonanno.
\newblock {Fourth post-Newtonian effective-one-body Hamiltonians with generic
  spins}.
\newblock {\em Phys. Rev. D}, 101(10):104034, 2020.

\bibitem{Levi:2019kgk}
Michèle Levi, Stavros Mougiakakos, and Mariana Vieira.
\newblock {Gravitational cubic-in-spin interaction at the next-to-leading
  post-Newtonian order}.
\newblock 2019.

\bibitem{Levi:2020lfn}
Michèle Levi and Fei Teng.
\newblock {NLO gravitational quartic-in-spin interaction}.
\newblock 8 2020.

\bibitem{Kosower:2018adc}
David~A. Kosower, Ben Maybee, and Donal O'Connell.
\newblock {Amplitudes, Observables, and Classical Scattering}.
\newblock 2018.

\bibitem{Damour:2017zjx}
Thibault Damour.
\newblock {High-energy gravitational scattering and the general relativistic
  two-body problem}.
\newblock {\em Phys. Rev.}, D97(4):044038, 2018.

\bibitem{Vines:2018gqi}
Justin Vines, Jan Steinhoff, and Alessandra Buonanno.
\newblock {Spinning-black-hole scattering and the test-black-hole limit at
  second post-Minkowskian order}.
\newblock 2018.

\bibitem{Damour:2019lcq}
Thibault Damour.
\newblock {Classical and quantum scattering in post-Minkowskian gravity}.
\newblock {\em Phys. Rev. D}, 102(2):024060, 2020.

\bibitem{Long:2021ufh}
Oliver Long and Leor Barack.
\newblock {Time-domain metric reconstruction for hyperbolic scattering}.
\newblock 5 2021.

\bibitem{Bini:2017xzy}
Donato Bini and Thibault Damour.
\newblock {Gravitational spin-orbit coupling in binary systems,
  post-Minkowskian approximation and effective one-body theory}.
\newblock {\em Phys. Rev.}, D96(10):104038, 2017.

\bibitem{Vines:2017hyw}
Justin Vines.
\newblock {Scattering of two spinning black holes in post-Minkowskian gravity,
  to all orders in spin, and effective-one-body mappings}.
\newblock {\em Class. Quant. Grav.}, 35(8):084002, 2018.

\bibitem{Damour:2016gwp}
Thibault Damour.
\newblock {Gravitational scattering, post-Minkowskian approximation and
  Effective One-Body theory}.
\newblock {\em Phys. Rev.}, D94(10):104015, 2016.

\bibitem{Damour:2014afa}
Thibault Damour, Federico Guercilena, Ian Hinder, Seth Hopper, Alessandro
  Nagar, and Luciano Rezzolla.
\newblock {Strong-Field Scattering of Two Black Holes: Numerics Versus
  Analytics}.
\newblock {\em Phys. Rev. D}, 89(8):081503, 2014.

\bibitem{Bini:2017wfr}
Donato Bini and Thibault Damour.
\newblock {Gravitational scattering of two black holes at the fourth
  post-Newtonian approximation}.
\newblock {\em Phys. Rev.}, D96(6):064021, 2017.

\bibitem{Kalin:2019rwq}
Gregor Kälin and Rafael~A. Porto.
\newblock {From Boundary Data to Bound States}.
\newblock {\em JHEP}, 01:072, 2020.

\bibitem{Kalin:2019inp}
Gregor Kälin and Rafael~A. Porto.
\newblock {From Boundary Data to Bound States II: Scattering Angle to Dynamical
  Invariants (with Twist)}.
\newblock {\em JHEP}, 02:120, 2020.

\bibitem{Westpfahl:1985}
K.~{Westpfahl}.
\newblock {High-Speed Scattering of Charged and Uncharged Particles in General
  Relativity}.
\newblock {\em Fortschritte der Physik}, 33:417--493, 1985.

\bibitem{Cheung:2018wkq}
Clifford Cheung, Ira~Z. Rothstein, and Mikhail~P. Solon.
\newblock {From Scattering Amplitudes to Classical Potentials in the
  Post-Minkowskian Expansion}.
\newblock {\em Phys. Rev. Lett.}, 121(25):251101, 2018.

\bibitem{Cheung:2020gyp}
Clifford Cheung and Mikhail~P. Solon.
\newblock {Classical gravitational scattering at $ \mathcal{O} $(G$^{3}$) from
  Feynman diagrams}.
\newblock {\em JHEP}, 06:144, 2020.

\bibitem{Bern:2019crd}
Zvi Bern, Clifford Cheung, Radu Roiban, Chia-Hsien Shen, Mikhail~P. Solon, and
  Mao Zeng.
\newblock {Black Hole Binary Dynamics from the Double Copy and Effective
  Theory}.
\newblock {\em JHEP}, 10:206, 2019.

\bibitem{Bern:2019nnu}
Zvi Bern, Clifford Cheung, Radu Roiban, Chia-Hsien Shen, Mikhail~P. Solon, and
  Mao Zeng.
\newblock {Scattering Amplitudes and the Conservative Hamiltonian for Binary
  Systems at Third Post-Minkowskian Order}.
\newblock {\em Phys. Rev. Lett.}, 122(20):201603, 2019.

\bibitem{Bern:2020buy}
Zvi Bern, Andres Luna, Radu Roiban, Chia-Hsien Shen, and Mao Zeng.
\newblock {Spinning Black Hole Binary Dynamics, Scattering Amplitudes and
  Effective Field Theory}.
\newblock 5 2020.

\bibitem{Bern:2021dqo}
Zvi Bern, Julio Parra-Martinez, Radu Roiban, Michael~S. Ruf, Chia-Hsien Shen,
  Mikhail~P. Solon, and Mao Zeng.
\newblock {Scattering Amplitudes and Conservative Binary Dynamics at ${\cal
  O}(G^4)$}.
\newblock 1 2021.

\bibitem{Bjerrum-Bohr:2018xdl}
N.~E.~J. Bjerrum-Bohr, Poul~H. Damgaard, Guido Festuccia, Ludovic Planté, and
  Pierre Vanhove.
\newblock {General Relativity from Scattering Amplitudes}.
\newblock {\em Phys. Rev. Lett.}, 121(17):171601, 2018.

\bibitem{Bjerrum-Bohr:2019kec}
N.E.J. Bjerrum-Bohr, Andrea Cristofoli, and Poul~H. Damgaard.
\newblock {Post-Minkowskian Scattering Angle in Einstein Gravity}.
\newblock {\em JHEP}, 08:038, 2020.

\bibitem{Kalin:2020fhe}
Gregor K\"alin, Zhengwen Liu, and Rafael~A. Porto.
\newblock {Conservative Dynamics of Binary Systems to Third Post-Minkowskian
  Order from the Effective Field Theory Approach}.
\newblock 7 2020.

\bibitem{Mogull:2020sak}
Gustav Mogull, Jan Plefka, and Jan Steinhoff.
\newblock {Classical black hole scattering from a worldline quantum field
  theory}.
\newblock {\em JHEP}, 02:048, 2021.

\bibitem{Jakobsen:2021smu}
Gustav~Uhre Jakobsen, Gustav Mogull, Jan Plefka, and Jan Steinhoff.
\newblock {Classical Gravitational Bremsstrahlung from a Worldline Quantum
  Field Theory}.
\newblock 1 2021.

\bibitem{Damour:2020tta}
Thibault Damour.
\newblock {Radiative contribution to classical gravitational scattering at the
  third order in $G$}.
\newblock {\em Phys. Rev. D}, 102(12):124008, 2020.

\bibitem{Kosmopoulos:2021zoq}
Dimitrios Kosmopoulos and Andres Luna.
\newblock {Quadratic-in-Spin Hamiltonian at $\mathcal{O}(G^2)$ from Scattering
  Amplitudes}.
\newblock 2 2021.

\bibitem{Liu:2021zxr}
Zhengwen Liu, Rafael~A. Porto, and Zixin Yang.
\newblock {Spin Effects in the Effective Field Theory Approach to
  Post-Minkowskian Conservative Dynamics}.
\newblock 2 2021.

\bibitem{Kalin:2020lmz}
Gregor K\"alin, Zhengwen Liu, and Rafael~A. Porto.
\newblock {Conservative Tidal Effects in Compact Binary Systems to
  Next-to-Leading Post-Minkowskian Order}.
\newblock {\em Phys. Rev. D}, 102:124025, 2020.

\bibitem{Bern:1994cg}
Zvi Bern, Lance~J. Dixon, David~C. Dunbar, and David~A. Kosower.
\newblock {Fusing gauge theory tree amplitudes into loop amplitudes}.
\newblock {\em Nucl. Phys.}, B435:59--101, 1995.

\bibitem{Bern:1994zx}
Zvi Bern, Lance~J. Dixon, David~C. Dunbar, and David~A. Kosower.
\newblock {One loop n point gauge theory amplitudes, unitarity and collinear
  limits}.
\newblock {\em Nucl. Phys.}, B425:217--260, 1994.

\bibitem{Bern:2008qj}
Z.~Bern, J.~J.~M. Carrasco, and Henrik Johansson.
\newblock {New Relations for Gauge-Theory Amplitudes}.
\newblock {\em Phys. Rev.}, D78:085011, 2008.

\bibitem{Bern:2010ue}
Zvi Bern, John Joseph~M. Carrasco, and Henrik Johansson.
\newblock {Perturbative Quantum Gravity as a Double Copy of Gauge Theory}.
\newblock {\em Phys. Rev. Lett.}, 105:061602, 2010.

\bibitem{Amaro-Seoane:2014ela}
Pau Amaro-Seoane, Jonathan~R. Gair, Adam Pound, Scott~A. Hughes, and Carlos~F.
  Sopuerta.
\newblock {Research Update on Extreme-Mass-Ratio Inspirals}.
\newblock {\em J. Phys. Conf. Ser.}, 610(1):012002, 2015.

\bibitem{Babak:2017tow}
Stanislav Babak, Jonathan Gair, Alberto Sesana, Enrico Barausse, Carlos~F.
  Sopuerta, Christopher P.~L. Berry, Emanuele Berti, Pau Amaro-Seoane, Antoine
  Petiteau, and Antoine Klein.
\newblock {Science with the space-based interferometer LISA. V: Extreme
  mass-ratio inspirals}.
\newblock {\em Phys. Rev. D}, 95(10):103012, 2017.

\bibitem{LeTiec:2011bk}
Alexandre Le~Tiec, Abdul~H. Mroue, Leor Barack, Alessandra Buonanno, Harald~P.
  Pfeiffer, Norichika Sago, and Andrea Taracchini.
\newblock {Periastron Advance in Black Hole Binaries}.
\newblock {\em Phys. Rev. Lett.}, 107:141101, 2011.

\bibitem{Barausse:2011dq}
Enrico Barausse, Alessandra Buonanno, and Alexandre Le~Tiec.
\newblock {The complete non-spinning effective-one-body metric at linear order
  in the mass ratio}.
\newblock {\em Phys. Rev.}, D85:064010, 2012.

\bibitem{LeTiec:2011dp}
Alexandre Le~Tiec, Enrico Barausse, and Alessandra Buonanno.
\newblock {Gravitational Self-Force Correction to the Binding Energy of Compact
  Binary Systems}.
\newblock {\em Phys. Rev. Lett.}, 108:131103, 2012.

\bibitem{Rifat:2019ltp}
Nur E.~M. Rifat, Scott~E. Field, Gaurav Khanna, and Vijay Varma.
\newblock {Surrogate model for gravitational wave signals from comparable and
  large-mass-ratio black hole binaries}.
\newblock {\em Phys. Rev. D}, 101(8):081502, 2020.

\bibitem{vandeMeent:2020xgc}
Maarten van~de Meent and Harald~P. Pfeiffer.
\newblock {Intermediate mass-ratio black hole binaries: Applicability of small
  mass-ratio perturbation theory}.
\newblock {\em Phys. Rev. Lett.}, 125(18):181101, 2020.

\bibitem{Warburton:2021kwk}
Niels Warburton, Adam Pound, Barry Wardell, Jeremy Miller, and Leanne Durkan.
\newblock {Gravitational-wave energy flux for compact binaries through second
  order in the mass ratio}.
\newblock 7 2021.

\bibitem{Quinn:1996am}
Theodore~C. Quinn and Robert~M. Wald.
\newblock {An Axiomatic approach to electromagnetic and gravitational radiation
  reaction of particles in curved space-time}.
\newblock {\em Phys. Rev.}, D56:3381--3394, 1997.

\bibitem{Mino:1996nk}
Yasushi Mino, Misao Sasaki, and Takahiro Tanaka.
\newblock {Gravitational radiation reaction to a particle motion}.
\newblock {\em Phys. Rev.}, D55:3457--3476, 1997.

\bibitem{Barack:2009ux}
Leor Barack.
\newblock {Gravitational self force in extreme mass-ratio inspirals}.
\newblock {\em Class. Quant. Grav.}, 26:213001, 2009.

\bibitem{Pound:2015tma}
Adam Pound.
\newblock {Motion of small objects in curved spacetimes: An introduction to
  gravitational self-force}.
\newblock {\em Fund. Theor. Phys.}, 179:399--486, 2015.

\bibitem{Poisson:2011nh}
Eric Poisson, Adam Pound, and Ian Vega.
\newblock {The Motion of point particles in curved spacetime}.
\newblock {\em Living Rev. Rel.}, 14:7, 2011.

\bibitem{DEath:1975jps}
Peter~D. D'Eath.
\newblock {Dynamics of a small black hole in a background universe}.
\newblock {\em Phys. Rev. D}, 11:1387--1403, 1975.

\bibitem{Kates:1980zz}
Ronald~E. Kates.
\newblock {Motion of a small body through an external field in general
  relativity calculated by matched asymptotic expansions}.
\newblock {\em Phys. Rev. D}, 22:1853--1870, 1980.

\bibitem{Barack:2018yvs}
Leor Barack and Adam Pound.
\newblock {Self-force and radiation reaction in general relativity}.
\newblock {\em Rept. Prog. Phys.}, 82(1):016904, 2019.

\bibitem{Detweiler:2002mi}
Steven~L. Detweiler and Bernard~F. Whiting.
\newblock {Selfforce via a Green's function decomposition}.
\newblock {\em Phys. Rev. D}, 67:024025, 2003.

\bibitem{Barack:1999wf}
Leor Barack and Amos Ori.
\newblock {Mode sum regularization approach for the selfforce in black hole
  space-time}.
\newblock {\em Phys. Rev. D}, 61:061502, 2000.

\bibitem{Barack:2001bw}
Leor Barack.
\newblock {Gravitational selfforce by mode sum regularization}.
\newblock {\em Phys. Rev. D}, 64:084021, 2001.

\bibitem{Barack:2007jh}
Leor Barack and Darren~A. Golbourn.
\newblock {Scalar-field perturbations from a particle orbiting a black hole
  using numerical evolution in 2+1 dimensions}.
\newblock {\em Phys. Rev. D}, 76:044020, 2007.

\bibitem{Barack:2007we}
Leor Barack, Darren~A. Golbourn, and Norichika Sago.
\newblock {m-Mode Regularization Scheme for the Self Force in Kerr Spacetime}.
\newblock {\em Phys. Rev. D}, 76:124036, 2007.

\bibitem{Barack:2007tm}
Leor Barack and Norichika Sago.
\newblock {Gravitational self force on a particle in circular orbit around a
  Schwarzschild black hole}.
\newblock {\em Phys. Rev.}, D75:064021, 2007.

\bibitem{Shah:2010bi}
Abhay~G. Shah, Tobias~S. Keidl, John~L. Friedman, Dong-Hoon Kim, and Larry~R.
  Price.
\newblock {Conservative, gravitational self-force for a particle in circular
  orbit around a Schwarzschild black hole in a Radiation Gauge}.
\newblock {\em Phys. Rev. D}, 83:064018, 2011.

\bibitem{Barack:2010tm}
Leor Barack and Norichika Sago.
\newblock {Gravitational self-force on a particle in eccentric orbit around a
  Schwarzschild black hole}.
\newblock {\em Phys. Rev.}, D81:084021, 2010.

\bibitem{Shah:2012gu}
Abhay~G. Shah, John~L. Friedman, and Tobias~S. Keidl.
\newblock {EMRI corrections to the angular velocity and redshift factor of a
  mass in circular orbit about a Kerr black hole}.
\newblock {\em Phys. Rev. D}, 86:084059, 2012.

\bibitem{vandeMeent:2016pee}
Maarten van~de Meent.
\newblock {Gravitational self-force on eccentric equatorial orbits around a
  Kerr black hole}.
\newblock {\em Phys. Rev.}, D94(4):044034, 2016.

\bibitem{vandeMeent:2017bcc}
Maarten van~de Meent.
\newblock {Gravitational self-force on generic bound geodesics in Kerr
  spacetime}.
\newblock {\em Phys. Rev.}, D97(10):104033, 2018.

\bibitem{Cohen:1974cm}
J.~M. Cohen and L.~S. Kegeles.
\newblock {Electromagnetic fields in curved spaces - a constructive procedure}.
\newblock {\em Phys. Rev. D}, 10:1070--1084, 1974.

\bibitem{Chrzanowski:1975wv}
P.~L. Chrzanowski.
\newblock {Vector Potential and Metric Perturbations of a Rotating Black Hole}.
\newblock {\em Phys. Rev. D}, 11:2042--2062, 1975.

\bibitem{Wald:1978vm}
Robert~M. Wald.
\newblock {Construction of Solutions of Gravitational, Electromagnetic, Or
  Other Perturbation Equations from Solutions of Decoupled Equations}.
\newblock {\em Phys. Rev. Lett.}, 41:203--206, 1978.

\bibitem{Ori:2002uv}
Amos Ori.
\newblock {Reconstruction of inhomogeneous metric perturbations and
  electromagnetic four potential in Kerr space-time}.
\newblock {\em Phys. Rev. D}, 67:124010, 2003.

\bibitem{Lousto:2002em}
Carlos~O. Lousto and Bernard~F. Whiting.
\newblock {Reconstruction of black hole metric perturbations from Weyl
  curvature}.
\newblock {\em Phys. Rev. D}, 66:024026, 2002.

\bibitem{Keidl:2010pm}
Tobias~S. Keidl, Abhay~G. Shah, John~L. Friedman, Dong-Hoon Kim, and Larry~R.
  Price.
\newblock {Gravitational Self-force in a Radiation Gauge}.
\newblock {\em Phys. Rev. D}, 82(12):124012, 2010.
\newblock [Erratum: Phys.Rev.D 90, 109902 (2014)].

\bibitem{Pound:2013faa}
Adam Pound, Cesar Merlin, and Leor Barack.
\newblock {Gravitational self-force from radiation-gauge metric perturbations}.
\newblock {\em Phys. Rev. D}, 89(2):024009, 2014.

\bibitem{Merlin:2014qda}
Cesar Merlin and Abhay~G. Shah.
\newblock {Self-force from reconstructed metric perturbations: numerical
  implementation in Schwarzschild spacetime}.
\newblock {\em Phys. Rev. D}, 91(2):024005, 2015.

\bibitem{vandeMeent:2015lxa}
Maarten van~de Meent and Abhay~G. Shah.
\newblock {Metric perturbations produced by eccentric equatorial orbits around
  a Kerr black hole}.
\newblock {\em Phys. Rev.}, D92(6):064025, 2015.

\bibitem{Merlin:2016boc}
Cesar Merlin, Amos Ori, Leor Barack, Adam Pound, and Maarten van~de Meent.
\newblock {Completion of metric reconstruction for a particle orbiting a Kerr
  black hole}.
\newblock {\em Phys. Rev. D}, 94(10):104066, 2016.

\bibitem{Pound:2019lzj}
Adam Pound, Barry Wardell, Niels Warburton, and Jeremy Miller.
\newblock {Second-order self-force calculation of the gravitational binding
  energy in compact binaries}.
\newblock {\em Phys. Rev. Lett.}, 124(2):021101, 2020.

\bibitem{Pound:2012nt}
Adam Pound.
\newblock {Second-order gravitational self-force}.
\newblock {\em Phys. Rev. Lett.}, 109:051101, 2012.

\bibitem{Pound:2012dk}
Adam Pound.
\newblock {Nonlinear gravitational self-force. I. Field outside a small body}.
\newblock {\em Phys. Rev. D}, 86:084019, 2012.

\bibitem{Pound:2014koa}
Adam Pound.
\newblock {Conservative effect of the second-order gravitational self-force on
  quasicircular orbits in Schwarzschild spacetime}.
\newblock {\em Phys. Rev. D}, 90(8):084039, 2014.

\bibitem{Pound:2015wva}
Adam Pound.
\newblock {Second-order perturbation theory: problems on large scales}.
\newblock {\em Phys. Rev. D}, 92(10):104047, 2015.

\bibitem{Pound:2014xva}
Adam Pound and Jeremy Miller.
\newblock {Practical, covariant puncture for second-order self-force
  calculations}.
\newblock {\em Phys. Rev. D}, 89(10):104020, 2014.

\bibitem{Miller:2016hjv}
Jeremy Miller, Barry Wardell, and Adam Pound.
\newblock {Second-order perturbation theory: the problem of infinite mode
  coupling}.
\newblock {\em Phys. Rev. D}, 94(10):104018, 2016.

\bibitem{Pound:2017psq}
Adam Pound.
\newblock {Nonlinear gravitational self-force: second-order equation of
  motion}.
\newblock {\em Phys. Rev. D}, 95(10):104056, 2017.

\bibitem{Warburton:2011fk}
Niels Warburton, Sarp Akcay, Leor Barack, Jonathan~R. Gair, and Norichika Sago.
\newblock {Evolution of inspiral orbits around a Schwarzschild black hole}.
\newblock {\em Phys. Rev.}, D85:061501, 2012.

\bibitem{Osburn:2015duj}
Thomas Osburn, Niels Warburton, and Charles~R. Evans.
\newblock {Highly eccentric inspirals into a black hole}.
\newblock {\em Phys. Rev.}, D93(6):064024, 2016.

\bibitem{Miller:2020bft}
Jeremy Miller and Adam Pound.
\newblock {Two-timescale evolution of extreme-mass-ratio inspirals: waveform
  generation scheme for quasicircular orbits in Schwarzschild spacetime}.
\newblock {\em Phys. Rev. D}, 103(6):064048, 2021.

\bibitem{vandeMeent:2018rms}
Maarten van~de Meent and Niels Warburton.
\newblock {Fast Self-forced Inspirals}.
\newblock {\em Class. Quant. Grav.}, 35(14):144003, 2018.

\bibitem{Hughes:2021exa}
Scott~A. Hughes, Niels Warburton, Gaurav Khanna, Alvin J.~K. Chua, and
  Michael~L. Katz.
\newblock {Adiabatic waveforms for extreme mass-ratio inspirals via multivoice
  decomposition in time and frequency}.
\newblock 2 2021.

\bibitem{Chua:2020stf}
Alvin J.~K. Chua, Michael~L. Katz, Niels Warburton, and Scott~A. Hughes.
\newblock {Rapid generation of fully relativistic extreme-mass-ratio-inspiral
  waveform templates for LISA data analysis}.
\newblock {\em Phys. Rev. Lett.}, 126(5):051102, 2021.

\bibitem{Katz:2021yft}
Michael~L. Katz, Alvin J.~K. Chua, Lorenzo Speri, Niels Warburton, and Scott~A.
  Hughes.
\newblock {FastEMRIWaveforms: New tools for millihertz gravitational-wave data
  analysis}.
\newblock 4 2021.

\bibitem{Tiec:2014lba}
Alexandre Le~Tiec.
\newblock {The Overlap of Numerical Relativity, Perturbation Theory and
  Post-Newtonian Theory in the Binary Black Hole Problem}.
\newblock {\em Int. J. Mod. Phys.}, D23(10):1430022, 2014.

\bibitem{Bini:2014zxa}
Donato Bini and Thibault Damour.
\newblock {Gravitational self-force corrections to two-body tidal interactions
  and the effective one-body formalism}.
\newblock {\em Phys. Rev.}, D90(12):124037, 2014.

\bibitem{Akcay:2015pjz}
Sarp Akcay and Maarten van~de Meent.
\newblock {Numerical computation of the effective-one-body potential $q$ using
  self-force results}.
\newblock {\em Phys. Rev.}, D93(6):064063, 2016.

\bibitem{Bini:2015bfb}
Donato Bini, Thibault Damour, and Andrea Geralico.
\newblock {Confirming and improving post-Newtonian and effective-one-body
  results from self-force computations along eccentric orbits around a
  Schwarzschild black hole}.
\newblock {\em Phys. Rev.}, D93(6):064023, 2016.

\bibitem{Regge:1957td}
Tullio Regge and John~A. Wheeler.
\newblock {Stability of a Schwarzschild singularity}.
\newblock {\em Phys. Rev.}, 108:1063--1069, 1957.

\bibitem{Zerilli:1971wd}
F.~J. Zerilli.
\newblock {Gravitational field of a particle falling in a schwarzschild
  geometry analyzed in tensor harmonics}.
\newblock {\em Phys. Rev. D}, 2:2141--2160, 1970.

\bibitem{Teukolsky:1973ha}
Saul~A. Teukolsky.
\newblock {Perturbations of a rotating black hole. 1. Fundamental equations for
  gravitational electromagnetic and neutrino field perturbations}.
\newblock {\em Astrophys. J.}, 185:635--647, 1973.

\bibitem{Mano:1996mf}
Shuhei Mano, Hisao Suzuki, and Eiichi Takasugi.
\newblock {Analytic solutions of the Regge-Wheeler equation and the
  postMinkowskian expansion}.
\newblock {\em Prog. Theor. Phys.}, 96:549--566, 1996.

\bibitem{Buonanno:1998gg}
A.~Buonanno and T.~Damour.
\newblock {Effective one-body approach to general relativistic two-body
  dynamics}.
\newblock {\em Phys. Rev.}, D59:084006, 1999.

\bibitem{Buonanno:2000ef}
Alessandra Buonanno and Thibault Damour.
\newblock {Transition from inspiral to plunge in binary black hole
  coalescences}.
\newblock {\em Phys. Rev.}, D62:064015, 2000.

\bibitem{Pan:2011gk}
Yi~Pan, Alessandra Buonanno, Michael Boyle, Luisa~T. Buchman, Lawrence~E.
  Kidder, Harald~P. Pfeiffer, and Mark~A. Scheel.
\newblock {Inspiral-merger-ringdown multipolar waveforms of nonspinning
  black-hole binaries using the effective-one-body formalism}.
\newblock {\em Phys. Rev.}, D84:124052, 2011.

\bibitem{Taracchini:2012ig}
Andrea Taracchini, Yi~Pan, Alessandra Buonanno, Enrico Barausse, Michael Boyle,
  Tony Chu, Geoffrey Lovelace, Harald~P. Pfeiffer, and Mark~A. Scheel.
\newblock {Prototype effective-one-body model for nonprecessing spinning
  inspiral-merger-ringdown waveforms}.
\newblock {\em Phys. Rev.}, D86:024011, 2012.

\bibitem{Taracchini:2013rva}
Andrea Taracchini et~al.
\newblock {Effective-one-body model for black-hole binaries with generic mass
  ratios and spins}.
\newblock {\em Phys. Rev.}, D89(6):061502, 2014.

\bibitem{Bohe:2016gbl}
Alejandro Bohé et~al.
\newblock {Improved effective-one-body model of spinning, nonprecessing binary
  black holes for the era of gravitational-wave astrophysics with advanced
  detectors}.
\newblock {\em Phys. Rev.}, D95(4):044028, 2017.

\bibitem{Damour:2015isa}
Thibault Damour, Piotr Jaranowski, and Gerhard Schäfer.
\newblock {Fourth post-Newtonian effective one-body dynamics}.
\newblock {\em Phys. Rev.}, D91(8):084024, 2015.

\bibitem{Bini:2020nsb}
Donato Bini, Thibault Damour, and Andrea Geralico.
\newblock {Sixth post-Newtonian local-in-time dynamics of binary systems}.
\newblock {\em Phys. Rev. D}, 102(2):024061, 2020.

\bibitem{Akcay:2012ea}
Sarp Akcay, Leor Barack, Thibault Damour, and Norichika Sago.
\newblock {Gravitational self-force and the effective-one-body formalism
  between the innermost stable circular orbit and the light ring}.
\newblock {\em Phys. Rev.}, D86:104041, 2012.

\bibitem{Barausse:2009aa}
Enrico Barausse, Etienne Racine, and Alessandra Buonanno.
\newblock {Hamiltonian of a spinning test-particle in curved spacetime}.
\newblock {\em Phys. Rev.}, D80:104025, 2009.
\newblock [Erratum: Phys. Rev. \textbf{D85}, 069904 (2012)].

\bibitem{Barausse:2009xi}
Enrico Barausse and Alessandra Buonanno.
\newblock {An Improved effective-one-body Hamiltonian for spinning black-hole
  binaries}.
\newblock {\em Phys. Rev.}, D81:084024, 2010.

\bibitem{Balmelli:2013zna}
Simone Balmelli and Philippe Jetzer.
\newblock {Effective-one-body Hamiltonian with next-to-leading order spin-spin
  coupling for two nonprecessing black holes with aligned spins}.
\newblock {\em Phys. Rev. D}, 87(12):124036, 2013.
\newblock [Erratum: Phys.Rev.D 90, 089905 (2014)].

\bibitem{Balmelli:2015zsa}
Simone Balmelli and Thibault Damour.
\newblock {New effective-one-body Hamiltonian with next-to-leading order
  spin-spin coupling}.
\newblock {\em Phys. Rev. D}, 92(12):124022, 2015.

\bibitem{Damour:2001tu}
Thibault Damour.
\newblock {Coalescence of two spinning black holes: an effective one-body
  approach}.
\newblock {\em Phys.\ Rev.\ D}, 64:124013, 2001.

\bibitem{Nagar:2011fx}
Alessandro Nagar.
\newblock {Effective one body Hamiltonian of two spinning black-holes with
  next-to-next-to-leading order spin-orbit coupling}.
\newblock {\em Phys. Rev.}, D84:084028, 2011.
\newblock [Erratum: Phys. Rev.D88,no.8,089901(2013)].

\bibitem{Barausse:2011ys}
Enrico Barausse and Alessandra Buonanno.
\newblock {Extending the effective-one-body Hamiltonian of black-hole binaries
  to include next-to-next-to-leading spin-orbit couplings}.
\newblock {\em Phys. Rev.}, D84:104027, 2011.

\bibitem{Damour:2007xr}
Thibault Damour and Alessandro Nagar.
\newblock {Faithful effective-one-body waveforms of small-mass-ratio coalescing
  black-hole binaries}.
\newblock {\em Phys. Rev.}, D76:064028, 2007.

\bibitem{Damour:2008gu}
Thibault Damour, Bala~R. Iyer, and Alessandro Nagar.
\newblock {Improved resummation of post-Newtonian multipolar waveforms from
  circularized compact binaries}.
\newblock {\em Phys. Rev.}, D79:064004, 2009.

\bibitem{Pan:2010hz}
Yi~Pan, Alessandra Buonanno, Ryuichi Fujita, Etienne Racine, and Hideyuki
  Tagoshi.
\newblock {Post-Newtonian factorized multipolar waveforms for spinning,
  non-precessing black-hole binaries}.
\newblock {\em Phys. Rev.}, D83:064003, 2011.
\newblock [Erratum: Phys. Rev.D87,no.10,109901(2013)].

\bibitem{Pretorius:2005gq}
Frans Pretorius.
\newblock {Evolution of binary black hole spacetimes}.
\newblock {\em Phys. Rev. Lett.}, 95:121101, 2005.

\bibitem{Campanelli:2005dd}
Manuela Campanelli, C.~O. Lousto, P.~Marronetti, and Y.~Zlochower.
\newblock {Accurate evolutions of orbiting black-hole binaries without
  excision}.
\newblock {\em Phys. Rev. Lett.}, 96:111101, 2006.

\bibitem{Baker:2005vv}
John~G. Baker, Joan Centrella, Dae-Il Choi, Michael Koppitz, and James van
  Meter.
\newblock {Gravitational wave extraction from an inspiraling configuration of
  merging black holes}.
\newblock {\em Phys. Rev. Lett.}, 96:111102, 2006.

\bibitem{Baker:2006ha}
John~G. Baker, James~R. van Meter, Sean~T. McWilliams, Joan Centrella, and
  Bernard~J. Kelly.
\newblock {Consistency of post-Newtonian waveforms with numerical relativity}.
\newblock {\em Phys. Rev. Lett.}, 99:181101, 2007.

\bibitem{Buonanno:2006ui}
Alessandra Buonanno, Gregory~B. Cook, and Frans Pretorius.
\newblock {Inspiral, merger and ring-down of equal-mass black-hole binaries}.
\newblock {\em Phys. Rev. D}, 75:124018, 2007.

\bibitem{Hannam:2007ik}
Mark Hannam, Sascha Husa, Ulrich Sperhake, Bernd Bruegmann, and Jose~A.
  Gonzalez.
\newblock {Where post-Newtonian and numerical-relativity waveforms meet}.
\newblock {\em Phys. Rev. D}, 77:044020, 2008.

\bibitem{Berti:2007fi}
Emanuele Berti, Vitor Cardoso, Jose~A. Gonzalez, Ulrich Sperhake, Mark Hannam,
  Sascha Husa, and Bernd Bruegmann.
\newblock {Inspiral, merger and ringdown of unequal mass black hole binaries: A
  Multipolar analysis}.
\newblock {\em Phys. Rev. D}, 76:064034, 2007.

\bibitem{Gonzalez:2008bi}
Jose~A. Gonzalez, Ulrich Sperhake, and Bernd Bruegmann.
\newblock {Black-hole binary simulations: The Mass ratio 10:1}.
\newblock {\em Phys. Rev. D}, 79:124006, 2009.

\bibitem{Buchman:2012dw}
Luisa~T. Buchman, Harald~P. Pfeiffer, Mark~A. Scheel, and Bela Szilagyi.
\newblock {Simulations of non-equal mass black hole binaries with spectral
  methods}.
\newblock {\em Phys. Rev. D}, 86:084033, 2012.

\bibitem{Berti:2007nw}
E.~Berti, V.~Cardoso, J.~A. Gonzalez, U.~Sperhake, and Bernd Bruegmann.
\newblock {Multipolar analysis of spinning binaries}.
\newblock {\em Class. Quant. Grav.}, 25:114035, 2008.

\bibitem{Hannam:2007wf}
Mark Hannam, Sascha Husa, Bernd Bruegmann, and Achamveedu Gopakumar.
\newblock {Comparison between numerical-relativity and post-Newtonian waveforms
  from spinning binaries: The Orbital hang-up case}.
\newblock {\em Phys. Rev. D}, 78:104007, 2008.

\bibitem{Chu:2009md}
Tony Chu, Harald~P. Pfeiffer, and Mark~A. Scheel.
\newblock {High accuracy simulations of black hole binaries: Spins anti-aligned
  with the orbital angular momentum}.
\newblock {\em Phys. Rev. D}, 80:124051, 2009.

\bibitem{Szilagyi:2009qz}
Bela Szilagyi, Lee Lindblom, and Mark~A. Scheel.
\newblock {Simulations of Binary Black Hole Mergers Using Spectral Methods}.
\newblock {\em Phys. Rev. D}, 80:124010, 2009.

\bibitem{Campanelli:2008nk}
Manuela Campanelli, Carlos~O. Lousto, Hiroyuki Nakano, and Yosef Zlochower.
\newblock {Comparison of Numerical and Post-Newtonian Waveforms for Generic
  Precessing Black-Hole Binaries}.
\newblock {\em Phys. Rev. D}, 79:084010, 2009.

\bibitem{Centrella:2010mx}
Joan Centrella, John~G. Baker, Bernard~J. Kelly, and James~R. van Meter.
\newblock {Black-hole binaries, gravitational waves, and numerical relativity}.
\newblock {\em Rev. Mod. Phys.}, 82:3069, 2010.

\bibitem{Baumgarte:2010ndz}
Thomas~W. Baumgarte and Stuart~L. Shapiro.
\newblock {\em {Numerical Relativity: Solving Einstein's Equations on the
  Computer}}.
\newblock Cambridge University Press, 2010.

\bibitem{Sperhake:2011xk}
Ulrich Sperhake, Emanuele Berti, and Vitor Cardoso.
\newblock {Numerical simulations of black-hole binaries and gravitational wave
  emission}.
\newblock {\em Comptes Rendus Physique}, 14:306--317, 2013.

\bibitem{SXS}
\url{https://data.black-holes.org/waveforms}.

\bibitem{Ossokine:2017dge}
Serguei Ossokine, Tim Dietrich, Evan Foley, Reza Katebi, and Geoffrey Lovelace.
\newblock {Assessing the Energetics of Spinning Binary Black Hole Systems}.
\newblock {\em Phys. Rev.}, D98(10):104057, 2018.

\bibitem{Babak:2016tgq}
Stanislav Babak, Andrea Taracchini, and Alessandra Buonanno.
\newblock {Validating the effective-one-body model of spinning, precessing
  binary black holes against numerical relativity}.
\newblock {\em Phys.\ Rev.\ D}, 95(2):024010, 2017.

\bibitem{Nagar:2018plt}
Alessandro Nagar, Francesco Messina, Piero Rettegno, Donato Bini, Thibault
  Damour, Andrea Geralico, Sarp Akcay, and Sebastiano Bernuzzi.
\newblock {Nonlinear-in-spin effects in effective-one-body waveform models of
  spin-aligned, inspiralling, neutron star binaries}.
\newblock {\em Phys. Rev.}, D99(4):044007, 2019.

\bibitem{Nagar:2018zoe}
Alessandro Nagar et~al.
\newblock {Time-domain effective-one-body gravitational waveforms for
  coalescing compact binaries with nonprecessing spins, tides and self-spin
  effects}.
\newblock {\em Phys. Rev.}, D98(10):104052, 2018.

\bibitem{Hannam:2013oca}
Mark Hannam, Patricia Schmidt, Alejandro Bohé, Leïla Haegel, Sascha Husa,
  Frank Ohme, Geraint Pratten, and Michael Pürrer.
\newblock {Simple Model of Complete Precessing Black-Hole-Binary Gravitational
  Waveforms}.
\newblock {\em Phys. Rev. Lett.}, 113(15):151101, 2014.

\bibitem{Husa:2015iqa}
Sascha Husa, Sebastian Khan, Mark Hannam, Michael Pürrer, Frank Ohme, Xisco
  Jiménez~Forteza, and Alejandro Bohé.
\newblock {Frequency-domain gravitational waves from nonprecessing black-hole
  binaries. I. New numerical waveforms and anatomy of the signal}.
\newblock {\em Phys. Rev.}, D93(4):044006, 2016.

\bibitem{Khan:2015jqa}
Sebastian Khan, Sascha Husa, Mark Hannam, Frank Ohme, Michael Pürrer, Xisco
  Jiménez~Forteza, and Alejandro Bohé.
\newblock {Frequency-domain gravitational waves from nonprecessing black-hole
  binaries. II. A phenomenological model for the advanced detector era}.
\newblock {\em Phys. Rev.}, D93(4):044007, 2016.

\bibitem{Khan:2019kot}
Sebastian Khan, Frank Ohme, Katerina Chatziioannou, and Mark Hannam.
\newblock {Including higher order multipoles in gravitational-wave models for
  precessing binary black holes}.
\newblock {\em Phys. Rev.}, D101(2):024056, 2020.

\bibitem{Garcia-Quiros:2020qpx}
Cecilio García-Quirós, Marta Colleoni, Sascha Husa, Héctor Estellés,
  Geraint Pratten, Antoni Ramos-Buades, Maite Mateu-Lucena, and Rafel Jaume.
\newblock {IMRPhenomXHM: A multi-mode frequency-domain model for the
  gravitational wave signal from non-precessing black-hole binaries}.
\newblock 1 2020.

\bibitem{Pratten:2020fqn}
Geraint Pratten, Sascha Husa, Cecilio Garcia-Quiros, Marta Colleoni, Antoni
  Ramos-Buades, Hector Estelles, and Rafel Jaume.
\newblock {Setting the cornerstone for the IMRPhenomX family of models for
  gravitational waves from compact binaries: The dominant harmonic for
  non-precessing quasi-circular black holes}.
\newblock 1 2020.

\bibitem{Pratten:2020ceb}
Geraint Pratten et~al.
\newblock {Let's twist again: computationally efficient models for the dominant
  and sub-dominant harmonic modes of precessing binary black holes}.
\newblock 4 2020.

\bibitem{LIGOScientific:2018mvr}
B.~P. Abbott et~al.
\newblock {GWTC-1: A Gravitational-Wave Transient Catalog of Compact Binary
  Mergers Observed by LIGO and Virgo during the First and Second Observing
  Runs}.
\newblock {\em Phys. Rev.}, X9(3):031040, 2019.

\bibitem{Bohe:2012mr}
Alejandro Bohe, Sylvain Marsat, Guillaume Faye, and Luc Blanchet.
\newblock {Next-to-next-to-leading order spin-orbit effects in the near-zone
  metric and precession equations of compact binaries}.
\newblock {\em Class. Quant. Grav.}, 30:075017, 2013.

\bibitem{Dolan:2014pja}
Sam~R. Dolan, Patrick Nolan, Adrian~C. Ottewill, Niels Warburton, and Barry
  Wardell.
\newblock {Tidal invariants for compact binaries on quasicircular orbits}.
\newblock {\em Phys. Rev.}, D91(2):023009, 2015.

\bibitem{Harte:2011ku}
Abraham~I. Harte.
\newblock {Mechanics of extended masses in general relativity}.
\newblock {\em Class. Quant. Grav.}, 29:055012, 2012.

\bibitem{Dolan:2013roa}
Sam~R. Dolan, Niels Warburton, Abraham~I. Harte, Alexandre Le~Tiec, Barry
  Wardell, and Leor Barack.
\newblock {Gravitational self-torque and spin precession in compact binaries}.
\newblock {\em Phys. Rev.}, D89(6):064011, 2014.

\bibitem{Akcay:2016dku}
Sarp Akcay, David Dempsey, and Sam~R. Dolan.
\newblock {Spin\textendash{}orbit precession for eccentric black hole binaries
  at first order in the mass ratio}.
\newblock {\em Class. Quant. Grav.}, 34(8):084001, 2017.

\bibitem{Kavanagh:2017wot}
Chris Kavanagh, Donato Bini, Thibault Damour, Seth Hopper, Adrian~C. Ottewill,
  and Barry Wardell.
\newblock {Spin-orbit precession along eccentric orbits for extreme mass ratio
  black hole binaries and its effective-one-body transcription}.
\newblock {\em Phys. Rev. D}, 96(6):064012, 2017.

\bibitem{Akcay:2017azq}
Sarp Akcay.
\newblock {Self-force correction to geodetic spin precession in Kerr
  spacetime}.
\newblock {\em Phys. Rev. D}, 96(4):044024, 2017.

\bibitem{Bini:2018ylh}
Donato Bini, Thibault Damour, Andrea Geralico, Chris Kavanagh, and Maarten
  van~de Meent.
\newblock {Gravitational self-force corrections to gyroscope precession along
  circular orbits in the Kerr spacetime}.
\newblock {\em Phys. Rev. D}, 98(10):104062, 2018.

\bibitem{Bini:2019lkm}
Donato Bini and Andrea Geralico.
\newblock {New gravitational self-force analytical results for eccentric
  equatorial orbits around a Kerr black hole: gyroscope precession}.
\newblock {\em Phys. Rev.}, D100(10):104003, 2019.

\bibitem{Detweiler:2008ft}
Steven~L. Detweiler.
\newblock {A Consequence of the gravitational self-force for circular orbits of
  the Schwarzschild geometry}.
\newblock {\em Phys. Rev.}, D77:124026, 2008.

\bibitem{Blanchet:2009sd}
Luc Blanchet, Steven~L. Detweiler, Alexandre Le~Tiec, and Bernard~F. Whiting.
\newblock {Post-Newtonian and Numerical Calculations of the Gravitational
  Self-Force for Circular Orbits in the Schwarzschild Geometry}.
\newblock {\em Phys. Rev. D}, 81:064004, 2010.

\bibitem{Blanchet:2010zd}
Luc Blanchet, Steven~L. Detweiler, Alexandre Le~Tiec, and Bernard~F. Whiting.
\newblock {High-Order Post-Newtonian Fit of the Gravitational Self-Force for
  Circular Orbits in the Schwarzschild Geometry}.
\newblock {\em Phys. Rev. D}, 81:084033, 2010.

\bibitem{Bini:2019lcd}
Donato Bini and Andrea Geralico.
\newblock {New gravitational self-force analytical results for eccentric
  equatorial orbits around a Kerr black hole: redshift invariant}.
\newblock {\em Phys. Rev.}, D100:104002, 2019.

\bibitem{Barack:2011ed}
Leor Barack and Norichika Sago.
\newblock {Beyond the geodesic approximation: conservative effects of the
  gravitational self-force in eccentric orbits around a Schwarzschild black
  hole}.
\newblock {\em Phys. Rev.}, D83:084023, 2011.

\bibitem{LeTiec:2011ab}
Alexandre Le~Tiec, Luc Blanchet, and Bernard~F. Whiting.
\newblock {The First Law of Binary Black Hole Mechanics in General Relativity
  and Post-Newtonian Theory}.
\newblock {\em Phys. Rev.}, D85:064039, 2012.

\bibitem{Zimmerman:2016ajr}
Aaron Zimmerman, Adam G.~M. Lewis, and Harald~P. Pfeiffer.
\newblock {Redshift factor and the first law of binary black hole mechanics in
  numerical simulations}.
\newblock {\em Phys. Rev. Lett.}, 117(19):191101, 2016.

\bibitem{Blanchet:2012at}
Luc Blanchet, Alessandra Buonanno, and Alexandre Le~Tiec.
\newblock {First law of mechanics for black hole binaries with spins}.
\newblock {\em Phys. Rev.}, D87(2):024030, 2013.

\bibitem{Tiec:2015cxa}
Alexandre Le~Tiec.
\newblock {First Law of Mechanics for Compact Binaries on Eccentric Orbits}.
\newblock {\em Phys. Rev.}, D92(8):084021, 2015.

\bibitem{Blanchet:2017rcn}
Luc Blanchet and Alexandre Le~Tiec.
\newblock {First Law of Compact Binary Mechanics with Gravitational-Wave
  Tails}.
\newblock {\em Class. Quant. Grav.}, 34(16):164001, 2017.

\bibitem{Fujita:2016igj}
Ryuichi Fujita, Soichiro Isoyama, Alexandre Le~Tiec, Hiroyuki Nakano, Norichika
  Sago, and Takahiro Tanaka.
\newblock {Hamiltonian Formulation of the Conservative Self-Force Dynamics in
  the Kerr Geometry}.
\newblock {\em Class. Quant. Grav.}, 34(13):134001, 2017.

\bibitem{Bardeen:1973gs}
James~M. Bardeen, B.~Carter, and S.~W. Hawking.
\newblock {The Four laws of black hole mechanics}.
\newblock {\em Commun. Math. Phys.}, 31:161--170, 1973.

\bibitem{Damour:2011fu}
Thibault Damour, Alessandro Nagar, Denis Pollney, and Christian Reisswig.
\newblock {Energy versus Angular Momentum in Black Hole Binaries}.
\newblock {\em Phys. Rev. Lett.}, 108:131101, 2012.

\bibitem{Nagar:2015xqa}
Alessandro Nagar, Thibault Damour, Christian Reisswig, and Denis Pollney.
\newblock {Energetics and phasing of nonprecessing spinning coalescing black
  hole binaries}.
\newblock {\em Phys. Rev.}, D93(4):044046, 2016.

\bibitem{Nagar:2020xsk}
Alessandro Nagar, Piero Rettegno, Rossella Gamba, and Sebastiano Bernuzzi.
\newblock {Effective-one-body waveforms from dynamical captures in black hole
  binaries}.
\newblock {\em Phys. Rev. D}, 103(6):064013, 2021.

\bibitem{Siemonsen:2019dsu}
Nils Siemonsen and Justin Vines.
\newblock {Test black holes, scattering amplitudes and perturbations of Kerr
  spacetime}.
\newblock {\em Phys. Rev. D}, 101(6):064066, 2020.

\bibitem{Mathisson:1937zz}
Myron Mathisson.
\newblock {Neue mechanik materieller systemes}.
\newblock {\em Acta Phys.\ Polon.}, 6:163--2900, 1937.

\bibitem{Papapetrou:1951pa}
Achille Papapetrou.
\newblock {Spinning test particles in general relativity. 1.}
\newblock {\em Proc.\ Roy.\ Soc.\ Lond.\ A}, A209:248--258, 1951.

\bibitem{Dixon:1970zza}
W.G. Dixon.
\newblock {Dynamics of extended bodies in general relativity. I. Momentum and
  angular momentum}.
\newblock {\em Proc. Roy. Soc. Lond. A}, 314:499--527, 1970.

\bibitem{Dixon:1979}
W.~G. {Dixon}.
\newblock {Extended bodies in general relativity: their description and
  motion}.
\newblock In J.~{Ehlers}, editor, {\em Isolated Gravitating Systems in General
  Relativity}, pages 156--219, 1979.

\bibitem{Bini:2017pee}
Donato Bini, Andrea Geralico, and Justin Vines.
\newblock {Hyperbolic scattering of spinning particles by a Kerr black hole}.
\newblock {\em Phys. Rev.}, D96(8):084044, 2017.

\bibitem{Tulczyjew:1959}
W.~Tulczyjew.
\newblock {Equations of motion of rotating bodies in general relativity
  theory}.
\newblock {\em Acta Phys. Polon.}, 18:37--55, 1959.
\newblock [Erratum: Acta Phys. Polon. \textbf{18}, 534 (1959)].

\bibitem{Akcay:2015pza}
Sarp Akcay, Alexandre Le~Tiec, Leor Barack, Norichika Sago, and Niels
  Warburton.
\newblock {Comparison Between Self-Force and Post-Newtonian Dynamics: Beyond
  Circular Orbits}.
\newblock {\em Phys. Rev.}, D91(12):124014, 2015.

\bibitem{Darwin1961}
Dawin C.
\newblock {\em Proc. R. Soc. Lond. A.}, 263:39, 1961.

\bibitem{Kavanagh:2016idg}
Chris Kavanagh, Adrian~C. Ottewill, and Barry Wardell.
\newblock {Analytical high-order post-Newtonian expansions for spinning extreme
  mass ratio binaries}.
\newblock {\em Phys. Rev.}, D93(12):124038, 2016.

\bibitem{Bini:2016dvs}
Donato Bini, Thibault Damour, and Andrea Geralico.
\newblock {High post-Newtonian order gravitational self-force analytical
  results for eccentric equatorial orbits around a Kerr black hole}.
\newblock {\em Phys. Rev.}, D93(12):124058, 2016.

\bibitem{Dietrich:2016lyp}
Tim Dietrich, Sebastiano Bernuzzi, Maximiliano Ujevic, and Wolfgang Tichy.
\newblock {Gravitational waves and mass ejecta from binary neutron star
  mergers: Effect of the stars' rotation}.
\newblock {\em Phys. Rev.}, D95(4):044045, 2017.

\bibitem{LIGOScientific:2020stg}
R.~Abbott et~al.
\newblock {GW190412: Observation of a Binary-Black-Hole Coalescence with
  Asymmetric Masses}.
\newblock {\em Phys. Rev. D}, 102(4):043015, 2020.

\bibitem{Damour:2009sf}
Thibault Damour.
\newblock {Gravitational Self Force in a Schwarzschild Background and the
  Effective One Body Formalism}.
\newblock {\em Phys.Rev.}, D81:024017, 2010.

\bibitem{Tiec:2013twa}
Alexandre Le~Tiec et~al.
\newblock {Periastron Advance in Spinning Black Hole Binaries: Gravitational
  Self-Force from Numerical Relativity}.
\newblock {\em Phys. Rev.}, D88(12):124027, 2013.

\bibitem{Pizzella:2016ooh}
G.~Pizzella.
\newblock {Birth and initial developments of experiments with resonant
  detectors searching for gravitational waves}.
\newblock {\em Eur. Phys. J. H}, 41(4-5):267--302, 2016.

\bibitem{Howl:2021giy}
Richard Howl and Ivette Fuentes.
\newblock {Quantum Frequency Interferometry: with applications ranging from
  gravitational wave detection to dark matter searches}.
\newblock 3 2021.

\bibitem{Detweiler:1979wn}
Steven~L. Detweiler.
\newblock {Pulsar timing measurements and the search for gravitational waves}.
\newblock {\em Astrophys. J.}, 234:1100--1104, 1979.

\bibitem{TheLIGOScientific:2014jea}
J.~Aasi et~al.
\newblock {Advanced LIGO}.
\newblock {\em Class. Quant. Grav.}, 32:074001, 2015.

\bibitem{Buikema:2020dlj}
Aaron Buikema et~al.
\newblock {Sensitivity and performance of the Advanced LIGO detectors in the
  third observing run}.
\newblock {\em Phys. Rev. D}, 102(6):062003, 2020.

\bibitem{Davis:2021ecd}
D.~Davis et~al.
\newblock {LIGO Detector Characterization in the Second and Third Observing
  Runs}.
\newblock 1 2021.

\bibitem{Samajdar:2021egv}
Anuradha Samajdar, Justin Janquart, Chris Van Den~Broeck, and Tim Dietrich.
\newblock {Biases in parameter estimation from overlapping gravitational-wave
  signals in the third generation detector era}.
\newblock 2 2021.

\bibitem{Pizzati:2021gzd}
Elia Pizzati, Surabhi Sachdev, Anuradha Gupta, and Bangalore Sathyaprakash.
\newblock {Bayesian inference of overlapping gravitational wave signals}.
\newblock 2 2021.

\bibitem{Himemoto:2021ukb}
Yoshiaki Himemoto, Atsushi Nishizawa, and Atsushi Taruya.
\newblock {Impacts of overlapping gravitational-wave signals on the parameter
  estimation: Toward the search for cosmological backgrounds}.
\newblock 3 2021.

\bibitem{Armano:2009zz}
M.~Armano et~al.
\newblock {LISA Pathfinder: The experiment and the route to LISA}.
\newblock {\em Class. Quant. Grav.}, 26:094001, 2009.

\bibitem{Neyman:1933wgr}
Jerzy Neyman and Egon~Sharpe Pearson.
\newblock {On the Problem of the Most Efficient Tests of Statistical
  Hypotheses}.
\newblock {\em Phil. Trans. Roy. Soc. Lond. A}, 231(694-706):289--337, 1933.

\bibitem{whittle:1957}
P.~Whittle.
\newblock Curve and periodogram smoothing.
\newblock {\em Journal of the Royal Statistical Society: Series B (Statistical
  Methodology)}, 19:38--63, 1957.

\bibitem{Usman:2015kfa}
Samantha~A. Usman et~al.
\newblock {The PyCBC search for gravitational waves from compact binary
  coalescence}.
\newblock {\em Class. Quant. Grav.}, 33(21):215004, 2016.

\bibitem{Owen:1995tm}
Benjamin~J. Owen.
\newblock {Search templates for gravitational waves from inspiraling binaries:
  Choice of template spacing}.
\newblock {\em Phys. Rev. D}, 53:6749--6761, 1996.

\bibitem{DalCanton:2017ala}
Tito Dal~Canton and Ian~W. Harry.
\newblock {Designing a template bank to observe compact binary coalescences in
  Advanced LIGO's second observing run}.
\newblock 5 2017.

\bibitem{Allen:2004gu}
Bruce Allen.
\newblock {${\chi}^{2}$ time-frequency discriminator for gravitational wave
  detection}.
\newblock {\em Phys. Rev. D}, 71:062001, 2005.

\bibitem{Babak:2012zx}
S.~Babak et~al.
\newblock {Searching for gravitational waves from binary coalescence}.
\newblock {\em Phys. Rev. D}, 87(2):024033, 2013.

\bibitem{Veitch:2014wba}
J.~Veitch et~al.
\newblock {Parameter estimation for compact binaries with ground-based
  gravitational-wave observations using the LALInference software library}.
\newblock {\em Phys. Rev. D}, 91(4):042003, 2015.

\bibitem{Ashton:2018jfp}
Gregory Ashton et~al.
\newblock {BILBY: A user-friendly Bayesian inference library for
  gravitational-wave astronomy}.
\newblock {\em Astrophys. J. Suppl.}, 241(2):27, 2019.

\bibitem{ForemanMackey:2012ig}
Daniel Foreman-Mackey, David~W. Hogg, Dustin Lang, and Jonathan Goodman.
\newblock {emcee: The MCMC Hammer}.
\newblock {\em Publ. Astron. Soc. Pac.}, 125:306--312, 2013.

\bibitem{Crowder_2007}
Jeff Crowder and Neil~J. Cornish.
\newblock Solution to the galactic foreground problem for lisa.
\newblock {\em Physical Review D}, 75(4), Feb 2007.

\bibitem{B_aut_2010}
Arkadiusz Błaut, Stanislav Babak, and Andrzej Królak.
\newblock Mock lisa data challenge for the galactic white dwarf binaries.
\newblock {\em Physical Review D}, 81(6), Mar 2010.

\bibitem{Robson:2017ayy}
Travis Robson and Neil Cornish.
\newblock {Impact of galactic foreground characterization on a global analysis
  for the LISA gravitational wave observatory}.
\newblock {\em Class. Quant. Grav.}, 34(24):244002, 2017.

\bibitem{Roebber:2020hso}
Elinore Roebber et~al.
\newblock {Milky Way Satellites Shining Bright in Gravitational Waves}.
\newblock {\em Astrophys. J. Lett.}, 894(2):L15, 2020.

\bibitem{Korol:2020hay}
Valeriya Korol, Vasily Belokurov, Christopher~J. Moore, and Silvia Toonen.
\newblock {Weighing Milky Way Satellites with LISA}.
\newblock 10 2020.

\bibitem{Robson_2019}
Travis Robson, Neil~J Cornish, and Chang Liu.
\newblock The construction and use of lisa sensitivity curves.
\newblock {\em Classical and Quantum Gravity}, 36(10):105011, Apr 2019.

\bibitem{Berti:2004bd}
Emanuele Berti, Alessandra Buonanno, and Clifford~M. Will.
\newblock {Estimating spinning binary parameters and testing alternative
  theories of gravity with LISA}.
\newblock {\em Phys. Rev. D}, 71:084025, 2005.

\bibitem{Allen:2005fk}
Bruce Allen, Warren~G. Anderson, Patrick~R. Brady, Duncan~A. Brown, and
  Jolien~D.E. Creighton.
\newblock {FINDCHIRP: An Algorithm for detection of gravitational waves from
  inspiraling compact binaries}.
\newblock {\em Phys. Rev. D}, 85:122006, 2012.

\bibitem{Gair:2010bx}
Jonathan~R. Gair, Alberto Sesana, Emanuele Berti, and Marta Volonteri.
\newblock {Constraining properties of the black hole population using LISA}.
\newblock {\em Class. Quant. Grav.}, 28:094018, 2011.

\bibitem{Gair:2010yu}
Jonathan~R. Gair, Christopher Tang, and Marta Volonteri.
\newblock {LISA extreme-mass-ratio inspiral events as probes of the black hole
  mass function}.
\newblock {\em Phys. Rev. D}, 81:104014, 2010.

\bibitem{Sesana:2010wy}
Alberto Sesana, Jonathan Gair, Emanuele Berti, and Marta Volonteri.
\newblock {Reconstructing the massive black hole cosmic history through
  gravitational waves}.
\newblock {\em Phys. Rev. D}, 83:044036, 2011.

\bibitem{Greene:2007xw}
Jenny~E. Greene and Luis~C. Ho.
\newblock {The Mass Function of Active Black Holes in the Local Universe}.
\newblock {\em Astrophys. J.}, 667:131--148, 2007.
\newblock [Erratum: Astrophys.J. 704, 1743--1747 (2009)].

\bibitem{TheLIGOScientific:2016pea}
B.~P. Abbott et~al.
\newblock {Binary Black Hole Mergers in the first Advanced LIGO Observing Run}.
\newblock {\em Phys. Rev.}, X6(4):041015, 2016.
\newblock [erratum: Phys. Rev.X8,no.3,039903(2018)].

\bibitem{TheVirgo:2014hva}
F.~Acernese et~al.
\newblock {Advanced Virgo: a second-generation interferometric gravitational
  wave detector}.
\newblock {\em Class. Quant. Grav.}, 32(2):024001, 2015.

\bibitem{Aso:2013eba}
Yoichi Aso, Yuta Michimura, Kentaro Somiya, Masaki Ando, Osamu Miyakawa,
  Takanori Sekiguchi, Daisuke Tatsumi, and Hiroaki Yamamoto.
\newblock {Interferometer design of the KAGRA gravitational wave detector}.
\newblock {\em Phys. Rev.}, D88(4):043007, 2013.

\bibitem{LIGOIndia}
Bala Iyer et~al.
\newblock {LIGO-India, Proposal of the Consortium for Indian Initiative in
  Gravitational-wave Observations}, 2011.
\newblock {LIGO} Document M1100296-v2.

\bibitem{Schafer:2018kuf}
Gerhard Schäfer and Piotr Jaranowski.
\newblock {Hamiltonian formulation of general relativity and post-Newtonian
  dynamics of compact binaries}.
\newblock {\em Living Rev. Rel.}, 21(1):7, 2018.

\bibitem{Futamase:2007zz}
Toshifumi Futamase and Yousuke Itoh.
\newblock {The post-Newtonian approximation for relativistic compact binaries}.
\newblock {\em Living Rev. Rel.}, 10:2, 2007.

\bibitem{Poisson:2014}
E.~Poisson and C.M. Will.
\newblock {\em Gravity: Newtonian, Post-Newtonian, Relativistic}.
\newblock Cambridge University Press, 2014.

\bibitem{Goldberger:2007hy}
Walter~D. Goldberger.
\newblock {Effective field theories and gravitational radiation}.
\newblock In {\em {Les Houches Summer School - Session 86: Particle Physics and
  Cosmology: The Fabric of Spacetime}}, pages 351--353, 355--396, 2007.

\bibitem{Rothstein:2014sra}
Ira~Z. Rothstein.
\newblock {Progress in effective field theory approach to the binary inspiral
  problem}.
\newblock {\em Gen. Rel. Grav.}, 46:1726, 2014.

\bibitem{Mroue:2013xna}
Abdul~H. Mroue et~al.
\newblock {Catalog of 174 Binary Black Hole Simulations for Gravitational Wave
  Astronomy}.
\newblock {\em Phys. Rev. Lett.}, 111(24):241104, 2013.

\bibitem{Chu:2015kft}
Tony Chu, Heather Fong, Prayush Kumar, Harald~P. Pfeiffer, Michael Boyle,
  Daniel~A. Hemberger, Lawrence~E. Kidder, Mark~A. Scheel, and Bela Szilagyi.
\newblock {On the accuracy and precision of numerical waveforms: Effect of
  waveform extraction methodology}.
\newblock {\em Class. Quant. Grav.}, 33(16):165001, 2016.

\bibitem{Pan:2013rra}
Yi~Pan, Alessandra Buonanno, Andrea Taracchini, Lawrence~E. Kidder, Abdul~H.
  Mroué, Harald~P. Pfeiffer, Mark~A. Scheel, and Béla Szilágyi.
\newblock {Inspiral-merger-ringdown waveforms of spinning, precessing
  black-hole binaries in the effective-one-body formalism}.
\newblock {\em Phys. Rev.}, D89(8):084006, 2014.

\bibitem{Abbott:2016nmj}
B.~P. Abbott et~al.
\newblock {GW151226: Observation of Gravitational Waves from a 22-Solar-Mass
  Binary Black Hole Coalescence}.
\newblock {\em Phys. Rev. Lett.}, 116(24):241103, 2016.

\bibitem{TheLIGOScientific:2016wfe}
B.~P. Abbott et~al.
\newblock {Properties of the Binary Black Hole Merger GW150914}.
\newblock {\em Phys. Rev. Lett.}, 116(24):241102, 2016.

\bibitem{TheLIGOScientific:2016src}
B.~P. Abbott et~al.
\newblock {Tests of general relativity with GW150914}.
\newblock {\em Phys. Rev. Lett.}, 116(22):221101, 2016.

\bibitem{Abbott:2016izl}
B.~P. Abbott et~al.
\newblock {An improved analysis of GW150914 using a fully spin-precessing
  waveform model}.
\newblock {\em Phys. Rev.}, X6(4):041014, 2016.

\bibitem{Yunes:2009ef}
Nicolas Yunes, Alessandra Buonanno, Scott~A. Hughes, M.~Coleman~Miller, and
  Yi~Pan.
\newblock {Modeling Extreme Mass Ratio Inspirals within the Effective-One-Body
  Approach}.
\newblock {\em Phys. Rev. Lett.}, 104:091102, 2010.

\bibitem{Damour:2009sm}
Thibault Damour.
\newblock {Gravitational Self Force in a Schwarzschild Background and the
  Effective One Body Formalism}.
\newblock {\em Phys. Rev.}, D81:024017, 2010.

\bibitem{Yunes:2010zj}
Nicolás Yunes, Alessandra Buonanno, Scott~A. Hughes, Yi~Pan, Enrico Barausse,
  M.~Coleman Miller, and William Throwe.
\newblock {Extreme Mass-Ratio Inspirals in the Effective-One-Body Approach:
  Quasi-Circular, Equatorial Orbits around a Spinning Black Hole}.
\newblock {\em Phys. Rev.}, D83:044044, 2011.
\newblock [Erratum: Phys. Rev.D88,no.10,109904(2013)].

\bibitem{Antonelliea:2019}
Andrea Antonelli, Alessandra Buonanno, Jan Steinhoff, Maarten van~de Meent, and
  Justin Vines.
\newblock {\em in preparation}, 2019.

\bibitem{Bertotti:1956}
B.~{Bertotti}.
\newblock {On gravitational motion}.
\newblock {\em Il Nuovo Cimento}, 4:898--906, October 1956.

\bibitem{Bertotti:1960}
B.~{Bertotti} and J.~{Plebanski}.
\newblock {Theory of gravitational perturbations in the fast motion
  approximation}.
\newblock {\em Annals of Physics}, 11:169--200, October 1960.

\bibitem{Rosenblum:1978zr}
A.~Rosenblum.
\newblock {Gravitational Radiation Energy Loss in Scattering Problems and the
  Einstein Quadrupole Formula}.
\newblock {\em Phys. Rev. Lett.}, 41:1003, 1978.
\newblock [Erratum: Phys. Rev. Lett.41,1140(1978)].

\bibitem{Bel:1981be}
LLuis Bel, T.~Damour, N.~Deruelle, J.~Ibanez, and J.~Martin.
\newblock {Poincaré-invariant gravitational field and equations of motion of
  two pointlike objects: The postlinear approximation of general relativity}.
\newblock {\em Gen. Rel. Grav.}, 13:963--1004, 1981.

\bibitem{Portilla:1979xx}
M.~Portilla.
\newblock Momentum and angular momentum of two gravitating particles.
\newblock {\em J. Phys.}, A12:1075--1090, 1979.

\bibitem{Portilla:1980uz}
M.~Portilla.
\newblock Scattering of two gravitating particles: Classical approach.
\newblock {\em J. Phys.}, A13:3677--3683, 1980.

\bibitem{Westpfahl:1979gu}
K.~Westpfahl and M.~Goller.
\newblock {GRAVITATIONAL SCATTERING OF TWO RELATIVISTIC PARTICLES IN POSTLINEAR
  APPROXIMATION}.
\newblock {\em Lett. Nuovo Cim.}, 26:573--576, 1979.

\bibitem{Schafer:1986}
Gerhard {Sch{\"a}fer}.
\newblock {The ADM Hamiltonian at the postlinear approximation.}
\newblock {\em General Relativity and Gravitation}, 18:255--270, March 1986.

\bibitem{Westpfahl:1987}
K.~{Westpfahl}, R.~{Mohles}, and H.~{Simonis}.
\newblock {Energy-momentum conservation for gravitational two-body scattering
  in the post-linear approximation}.
\newblock {\em Classical and Quantum Gravity}, 4:L185--L188, September 1987.

\bibitem{Ledvinka:2008tk}
Tomas Ledvinka, Gerhard Schaefer, and Jiri Bicak.
\newblock {Relativistic Closed-Form Hamiltonian for Many-Body Gravitating
  Systems in the Post-Minkowskian Approximation}.
\newblock {\em Phys. Rev. Lett.}, 100:251101, 2008.

\bibitem{Foffa:2013gja}
Stefano Foffa.
\newblock {Gravitating binaries at 5PN in the post-Minkowskian approximation}.
\newblock {\em Phys. Rev.}, D89(2):024019, 2014.

\bibitem{Bini:2018ywr}
Donato Bini and Thibault Damour.
\newblock {Gravitational spin-orbit coupling in binary systems at the second
  post-Minkowskian approximation}.
\newblock {\em Phys. Rev.}, D98(4):044036, 2018.

\bibitem{Blanchet:2018yvb}
Luc Blanchet and Athanassios~S. Fokas.
\newblock {Equations of motion of self-gravitating $N$-body systems in the
  first post-Minkowskian approximation}.
\newblock {\em Phys. Rev.}, D98(8):084005, 2018.

\bibitem{Bjerrum-Bohr:2013bxa}
N.~E.~J. Bjerrum-Bohr, John~F. Donoghue, and Pierre Vanhove.
\newblock {On-shell Techniques and Universal Results in Quantum Gravity}.
\newblock {\em JHEP}, 02:111, 2014.

\bibitem{Holstein:2004dn}
Barry~R. Holstein and John~F. Donoghue.
\newblock {Classical physics and quantum loops}.
\newblock {\em Phys. Rev. Lett.}, 93:201602, 2004.

\bibitem{Neill:2013wsa}
Duff Neill and Ira~Z. Rothstein.
\newblock {Classical Space-Times from the S Matrix}.
\newblock {\em Nucl. Phys.}, B877:177--189, 2013.

\bibitem{Vaidya:2014kza}
Varun Vaidya.
\newblock {Gravitational spin Hamiltonians from the S matrix}.
\newblock {\em Phys. Rev.}, D91(2):024017, 2015.

\bibitem{Parke:1986gb}
Stephen~J. Parke and T.~R. Taylor.
\newblock {An Amplitude for $n$ Gluon Scattering}.
\newblock {\em Phys. Rev. Lett.}, 56:2459, 1986.

\bibitem{Britto:2004nc}
Ruth Britto, Freddy Cachazo, and Bo~Feng.
\newblock {Generalized unitarity and one-loop amplitudes in N=4
  super-Yang-Mills}.
\newblock {\em Nucl. Phys.}, B725:275--305, 2005.

\bibitem{Britto:2004ap}
Ruth Britto, Freddy Cachazo, and Bo~Feng.
\newblock {New recursion relations for tree amplitudes of gluons}.
\newblock {\em Nucl. Phys.}, B715:499--522, 2005.

\bibitem{Britto:2005fq}
Ruth Britto, Freddy Cachazo, Bo~Feng, and Edward Witten.
\newblock {Direct proof of tree-level recursion relation in Yang-Mills theory}.
\newblock {\em Phys. Rev. Lett.}, 94:181602, 2005.

\bibitem{Bern:2010yg}
Zvi Bern, Tristan Dennen, Yu-tin Huang, and Michael Kiermaier.
\newblock {Gravity as the Square of Gauge Theory}.
\newblock {\em Phys. Rev.}, D82:065003, 2010.

\bibitem{Guevara:2017csg}
Alfredo Guevara.
\newblock {Holomorphic Classical Limit for Spin Effects in Gravitational and
  Electromagnetic Scattering}.
\newblock 2017.

\bibitem{Arkani-Hamed:2017jhn}
Nima Arkani-Hamed, Tzu-Chen Huang, and Yu-tin Huang.
\newblock {Scattering Amplitudes For All Masses and Spins}.
\newblock 2017.

\bibitem{Ochirov:2018uyq}
Alexander Ochirov.
\newblock {Helicity amplitudes for QCD with massive quarks}.
\newblock {\em JHEP}, 04:089, 2018.

\bibitem{Buonanno:2009zt}
Alessandra Buonanno, Bala Iyer, Evan Ochsner, Yi~Pan, and B.~S. Sathyaprakash.
\newblock {Comparison of post-Newtonian templates for compact binary inspiral
  signals in gravitational-wave detectors}.
\newblock {\em Phys. Rev.}, D80:084043, 2009.

\bibitem{Damour:2013tla}
Thibault Damour, Alessandro Nagar, and Lo\"\i{}c Villain.
\newblock {Merger states and final states of black hole coalescences: a
  numerical-relativity-assisted effective-one-body approach}.
\newblock {\em Phys. Rev. D}, 89(2):024031, 2014.

\bibitem{Buonanno:2005xu}
Alessandra Buonanno, Yanbei Chen, and Thibault Damour.
\newblock {Transition from inspiral to plunge in precessing binaries of
  spinning black holes}.
\newblock {\em Phys. Rev.}, D74:104005, 2006.

\bibitem{Jani:2016wkt}
Karan Jani, James Healy, James~A. Clark, Lionel London, Pablo Laguna, and
  Deirdre Shoemaker.
\newblock {Georgia Tech Catalog of Gravitational Waveforms}.
\newblock {\em Class. Quant. Grav.}, 33(20):204001, 2016.

\bibitem{Healy:2017psd}
James Healy, Carlos~O. Lousto, Yosef Zlochower, and Manuela Campanelli.
\newblock {The RIT binary black hole simulations catalog}.
\newblock {\em Class. Quant. Grav.}, 34(22):224001, 2017.

\bibitem{Dietrich:2018phi}
Tim Dietrich, David Radice, Sebastiano Bernuzzi, Francesco Zappa, Albino
  Perego, Bernd Brügmann, Swami~Vivekanandji Chaurasia, Reetika Dudi, Wolfgang
  Tichy, and Maximiliano Ujevic.
\newblock {CoRe database of binary neutron star merger waveforms}.
\newblock {\em Class. Quant. Grav.}, 35(24):24LT01, 2018.

\bibitem{Boyle:2019kee}
Michael Boyle et~al.
\newblock {The SXS Collaboration catalog of binary black hole simulations}.
\newblock {\em Class. Quant. Grav.}, 36(19):195006, 2019.

\bibitem{Abbott:2017vtc}
Benjamin~P. Abbott et~al.
\newblock {GW170104: Observation of a 50-Solar-Mass Binary Black Hole
  Coalescence at Redshift 0.2}.
\newblock {\em Phys. Rev. Lett.}, 118(22):221101, 2017.
\newblock [Erratum: Phys. Rev. Lett.121,no.12,129901(2018)].

\bibitem{Abbott:2017gyy}
B..~P.. Abbott et~al.
\newblock {GW170608: Observation of a 19-solar-mass Binary Black Hole
  Coalescence}.
\newblock {\em Astrophys. J.}, 851(2):L35, 2017.

\bibitem{Evans:2016mbw}
Benjamin~P. Abbott et~al.
\newblock {Exploring the Sensitivity of Next Generation Gravitational Wave
  Detectors}.
\newblock {\em Class. Quant. Grav.}, 34(4):044001, 2017.

\bibitem{Foffa:2019hrb}
Stefano Foffa, Pierpaolo Mastrolia, Riccardo Sturani, Christian Sturm, and
  William~J. Torres~Bobadilla.
\newblock {Static two-body potential at fifth post-Newtonian order}.
\newblock {\em Phys. Rev. Lett.}, 122(24):241605, 2019.

\bibitem{Blumlein:2019zku}
J.~Blümlein, A.~Maier, and P.~Marquard.
\newblock {Five-Loop Static Contribution to the Gravitational Interaction
  Potential of Two Point Masses}.
\newblock 2019.

\bibitem{Damour:2009kr}
Thibault Damour and Alessandro Nagar.
\newblock {An Improved analytical description of inspiralling and coalescing
  black-hole binaries}.
\newblock {\em Phys. Rev.}, D79:081503, 2009.

\bibitem{Hinderer:2016eia}
Tanja Hinderer et~al.
\newblock {Effects of neutron-star dynamic tides on gravitational waveforms
  within the effective-one-body approach}.
\newblock {\em Phys. Rev. Lett.}, 116:181101, 2016.

\bibitem{Damour:2016pm}
Thibault Damour.
\newblock {Gravitational scattering, post-Minkowskian approximation and
  Effective One-Body theory}.
\newblock {\em Phys.Rev.}, D94:104015, 2016.

\bibitem{Vines:2017pm}
Justin Vines.
\newblock {Scattering of two spinning black holes in post-Minkowskian gravity,
  to all orders in spin, and effective-one-body mappings}.
\newblock {\em Class.Quant.Grav.}, 35:084002, 2018.

\bibitem{Hinderer:2008dm}
Tanja Hinderer and Eanna~E. Flanagan.
\newblock {Two timescale analysis of extreme mass ratio inspirals in Kerr. I.
  Orbital Motion}.
\newblock {\em Phys. Rev.}, D78:064028, 2008.

\bibitem{Barack:2009ey}
Leor Barack and Norichika Sago.
\newblock {Gravitational self-force correction to the innermost stable circular
  orbit of a Schwarzschild black hole}.
\newblock {\em Phys. Rev. Lett.}, 102:191101, 2009.

\bibitem{Barack:2010ny}
Leor Barack, Thibault Damour, and Norichika Sago.
\newblock {Precession effect of the gravitational self-force in a Schwarzschild
  spacetime and the effective one-body formalism}.
\newblock {\em Phys. Rev.}, D82:084036, 2010.

\bibitem{vandeMeent:2016hel}
Maarten van~de Meent.
\newblock {Self-force corrections to the periapsis advance around a spinning
  black hole}.
\newblock {\em Phys. Rev. Lett.}, 118(1):011101, 2017.

\bibitem{Bini:2014ica}
Donato Bini and Thibault Damour.
\newblock {Two-body gravitational spin-orbit interaction at linear order in the
  mass ratio}.
\newblock {\em Phys. Rev.}, D90(2):024039, 2014.

\bibitem{Kavanagh:2015lva}
Chris Kavanagh, Adrian~C. Ottewill, and Barry Wardell.
\newblock {Analytical high-order post-Newtonian expansions for extreme mass
  ratio binaries}.
\newblock {\em Phys. Rev.}, D92(8):084025, 2015.

\bibitem{Shah:2013uya}
Abhay~G. Shah, John~L Friedman, and Bernard~F Whiting.
\newblock {Finding high-order analytic post-Newtonian parameters from a
  high-precision numerical self-force calculation}.
\newblock {\em Phys. Rev.}, D89(6):064042, 2014.

\bibitem{Johnson-McDaniel:2015vva}
Nathan~K. Johnson-McDaniel, Abhay~G. Shah, and Bernard~F. Whiting.
\newblock {Experimental mathematics meets gravitational self-force}.
\newblock {\em Phys. Rev.}, D92(4):044007, 2015.

\bibitem{kavanaghetal:2015}
Chris Kavanagh, Adrian~C. Ottewill, and Barry Wardell.
\newblock {Analytical high-order post-Newtonian expansions for extreme mass
  ratio binaries }.
\newblock {\em Phys.Rev.}, D92:024017, 2015.

\bibitem{Hopper:2015icj}
Seth Hopper, Chris Kavanagh, and Adrian~C. Ottewill.
\newblock {Analytic self-force calculations in the post-Newtonian regime:
  eccentric orbits on a Schwarzschild background}.
\newblock {\em Phys. Rev.}, D93(4):044010, 2016.

\bibitem{Bini:2013rfa}
Donato Bini and Thibault Damour.
\newblock {High-order post-Newtonian contributions to the two-body
  gravitational interaction potential from analytical gravitational self-force
  calculations}.
\newblock {\em Phys. Rev.}, D89(6):064063, 2014.

\bibitem{Bini:2014nfa}
Donato Bini and Thibault Damour.
\newblock {Analytic determination of the eight-and-a-half post-Newtonian
  self-force contributions to the two-body gravitational interaction
  potential}.
\newblock {\em Phys. Rev.}, D89(10):104047, 2014.

\bibitem{Bini:2015bla}
Donato Bini and Thibault Damour.
\newblock {Detweiler’s gauge-invariant redshift variable: Analytic
  determination of the nine and nine-and-a-half post-Newtonian self-force
  contributions}.
\newblock {\em Phys. Rev.}, D91:064050, 2015.

\bibitem{Bini:2015xua}
Donato Bini, Thibault Damour, and Andrea Geralico.
\newblock {Spin-dependent two-body interactions from gravitational self-force
  computations}.
\newblock {\em Phys. Rev.}, D92(12):124058, 2015.
\newblock [Erratum: Phys. Rev.D93,no.10,109902(2016)].

\bibitem{Bini:2016qtx}
Donato Bini, Thibault Damour, and andrea Geralico.
\newblock {New gravitational self-force analytical results for eccentric orbits
  around a Schwarzschild black hole}.
\newblock {\em Phys. Rev.}, D93(10):104017, 2016.

\bibitem{Damour:2002vi}
Thibault Damour, Bala~R. Iyer, Piotr Jaranowski, and B.~S. Sathyaprakash.
\newblock {Gravitational waves from black hole binary inspiral and merger: The
  Span of third postNewtonian effective one-body templates}.
\newblock {\em Phys. Rev.}, D67:064028, 2003.

\bibitem{Buonanno:2007pf}
Alessandra Buonanno, Yi~Pan, John~G. Baker, Joan Centrella, Bernard~J. Kelly,
  Sean~T. McWilliams, and James~R. van Meter.
\newblock {Toward faithful templates for non-spinning binary black holes using
  the effective-one-body approach}.
\newblock {\em Phys. Rev.}, D76:104049, 2007.

\bibitem{Bini:2012gu}
Donato Bini, Thibault Damour, and Guillaume Faye.
\newblock {Effective action approach to higher-order relativistic tidal
  interactions in binary systems and their effective one body description}.
\newblock {\em Phys. Rev.}, D85:124034, 2012.

\bibitem{Steinhoff:2016rfi}
Jan Steinhoff, Tanja Hinderer, Alessandra Buonanno, and Andrea Taracchini.
\newblock {Dynamical Tides in General Relativity: Effective Action and
  Effective-One-Body Hamiltonian}.
\newblock {\em Phys. Rev.}, D94(10):104028, 2016.

\bibitem{BHPToolkit}
{Black Hole Perturbation Toolkit}.
\newblock (\href{http://bhptoolkit.org/}{bhptoolkit.org}).

\bibitem{LIGOScientific:2018jsj}
B.~P. Abbott et~al.
\newblock {Binary Black Hole Population Properties Inferred from the First and
  Second Observing Runs of Advanced LIGO and Advanced Virgo}.
\newblock {\em Astrophys. J.}, 882(2):L24, 2019.

\bibitem{Blumlein:2020pog}
J.~Blümlein, A.~Maier, P.~Marquard, and G.~Schäfer.
\newblock {Fourth post-Newtonian Hamiltonian dynamics of two-body systems from
  an effective field theory approach}.
\newblock {\em Nucl. Phys. B}, 955:115041, 2020.

\bibitem{Blumlein:2020znm}
J.~Bl\"umlein, A.~Maier, P.~Marquard, and G.~Sch\"afer.
\newblock {Testing binary dynamics in gravity at the sixth post-Newtonian
  level}.
\newblock {\em Phys. Lett. B}, 807:135496, 2020.

\bibitem{Aoyama:2012wj}
Tatsumi Aoyama, Masashi Hayakawa, Toichiro Kinoshita, and Makiko Nio.
\newblock {Tenth-Order QED Contribution to the Electron g-2 and an Improved
  Value of the Fine Structure Constant}.
\newblock {\em Phys. Rev. Lett.}, 109:111807, 2012.

\bibitem{Odom:2006zz}
Brian~C. Odom, D.~Hanneke, B.~D'Urso, and G.~Gabrielse.
\newblock {New Measurement of the Electron Magnetic Moment Using a One-Electron
  Quantum Cyclotron}.
\newblock {\em Phys. Rev. Lett.}, 97:030801, 2006.
\newblock [Erratum: Phys. Rev. Lett.99,039902(2007)].

\bibitem{Zackay:2019tzo}
Barak Zackay, Tejaswi Venumadhav, Liang Dai, Javier Roulet, and Matias
  Zaldarriaga.
\newblock {Highly spinning and aligned binary black hole merger in the Advanced
  LIGO first observing run}.
\newblock {\em Phys. Rev. D}, 100(2):023007, 2019.

\bibitem{Huang:2020ysn}
Yiwen Huang, Carl-Johan Haster, Salvatore Vitale, Aaron Zimmerman, Javier
  Roulet, Tejaswi Venumadhav, Barak Zackay, Liang Dai, and Matias Zaldarriaga.
\newblock {Source properties of the lowest signal-to-noise-ratio binary black
  hole detections}.
\newblock 2020.

\bibitem{Bini:2015mza}
Donato Bini and Thibault Damour.
\newblock {Analytic determination of high-order post-Newtonian self-force
  contributions to gravitational spin precession}.
\newblock {\em Phys. Rev.}, D91(6):064064, 2015.

\bibitem{Bini:2018aps}
Donato Bini, Thibault Damour, and Andrea Geralico.
\newblock {Spin-orbit precession along eccentric orbits: improving the
  knowledge of self-force corrections and of their effective-one-body
  counterparts}.
\newblock {\em Phys. Rev.}, D97(10):104046, 2018.

\bibitem{LIGOaplus}
National~Science Foundation.
\newblock {The A+ Upgrade to Advanced LIGO}, 2018.
\newblock Retrieved online on 18-04-2020.

\bibitem{Aasi:2013wya}
B.~P. Abbott et~al.
\newblock {Prospects for Observing and Localizing Gravitational-Wave Transients
  with Advanced LIGO, Advanced Virgo and KAGRA}.
\newblock {\em Living Rev. Rel.}, 21(1):3, 2018.

\bibitem{Pati:2000vt}
Michael~E. Pati and Clifford~M. Will.
\newblock {PostNewtonian gravitational radiation and equations of motion via
  direct integration of the relaxed Einstein equations. 1. Foundations}.
\newblock {\em Phys. Rev. D}, 62:124015, 2000.

\bibitem{Porto:2016pyg}
Rafael~A. Porto.
\newblock {The effective field theorist’s approach to gravitational
  dynamics}.
\newblock {\em Phys. Rept.}, 633:1--104, 2016.

\bibitem{Levi:2018nxp}
Michele Levi.
\newblock {Effective Field Theories of Post-Newtonian Gravity: A comprehensive
  review}.
\newblock 2018.

\bibitem{Bernard:2017ktp}
Laura Bernard, Luc Blanchet, Guillaume Faye, and Tanguy Marchand.
\newblock {Center-of-Mass Equations of Motion and Conserved Integrals of
  Compact Binary Systems at the Fourth Post-Newtonian Order}.
\newblock {\em Phys.\ Rev.\ D}, 97(4):044037, 2018.

\bibitem{Bini:2020wpo}
Donato Bini, Thibault Damour, and Andrea Geralico.
\newblock {Binary dynamics at the fifth and fifth-and-a-half post-Newtonian
  orders}.
\newblock {\em Phys. Rev. D}, 102(2):024062, 2020.

\bibitem{Bini:2020hmy}
Donato Bini, Thibault Damour, and Andrea Geralico.
\newblock {Sixth post-Newtonian nonlocal-in-time dynamics of binary systems}.
\newblock {\em Phys. Rev. D}, 102(8):084047, 2020.

\bibitem{Bini:2020uiq}
Donato Bini, Thibault Damour, Andrea Geralico, Stefano Laporta, and Pierpaolo
  Mastrolia.
\newblock {Gravitational dynamics at $O(G^6)$: perturbative gravitational
  scattering meets experimental mathematics}.
\newblock 8 2020.

\bibitem{Levi:2015msa}
Michele Levi and Jan Steinhoff.
\newblock {Spinning gravitating objects in the effective field theory in the
  post-Newtonian scheme}.
\newblock {\em JHEP}, 09:219, 2015.

\bibitem{Levi:2014gsa}
Michele Levi and Jan Steinhoff.
\newblock {Leading order finite size effects with spins for inspiralling
  compact binaries}.
\newblock {\em JHEP}, 06:059, 2015.

\bibitem{Vines:2016qwa}
Justin Vines and Jan Steinhoff.
\newblock {Spin-multipole effects in binary black holes and the test-body
  limit}.
\newblock {\em Phys.\ Rev.\ D}, 97(6):064010, 2018.

\bibitem{Bini:2018zde}
Donato Bini, Thibault Damour, Andrea Geralico, and Chris Kavanagh.
\newblock {Detweiler's redshift invariant for spinning particles along circular
  orbits on a Schwarzschild background}.
\newblock {\em Phys. Rev. D}, 97(10):104022, 2018.

\bibitem{Bini:2020zqy}
Donato Bini, Andrea Geralico, and Jan Steinhoff.
\newblock {Detweiler\textquoteright{}s redshift invariant for extended bodies
  orbiting a Schwarzschild black hole}.
\newblock {\em Phys. Rev. D}, 102(2):024091, 2020.

\bibitem{Blanchet:2011aha}
Luc Blanchet, Steven Detweiler, Alexandre Le~Tiec, and Bernard~F. Whiting.
\newblock {High-Accuracy Comparison between the Post-Newtonian and Self-Force
  Dynamics of Black-Hole Binaries}.
\newblock {\em Fundam. Theor. Phys.}, 162:415--442, 2011.
\newblock [,415(2010)].

\bibitem{Westpfahl:1980mk}
K.~Westpfahl and H.~Hoyler.
\newblock {GRAVITATIONAL BREMSSTRAHLUNG IN POSTLINEAR FAST MOTION
  APPROXIMATION}.
\newblock {\em Lett. Nuovo Cim.}, 27:581--585, 1980.

\bibitem{Aoude:2020onz}
Rafael Aoude, Kays Haddad, and Andreas Helset.
\newblock {On-shell heavy particle effective theories}.
\newblock {\em JHEP}, 05:051, 2020.

\bibitem{Chung:2020rrz}
Ming-Zhi Chung, Yu-tin Huang, Jung-Wook Kim, and Sangmin Lee.
\newblock {Complete Hamiltonian for spinning binary systems at first
  post-Minkowskian order}.
\newblock {\em JHEP}, 05:105, 2020.

\bibitem{Steinhoff:2014kwa}
Jan Steinhoff.
\newblock {Spin and quadrupole contributions to the motion of astrophysical
  binaries}.
\newblock {\em Fund. Theor. Phys.}, 179:615--649, 2015.

\bibitem{fokker1929relativiteitstheorie}
Adriaan~Dani{\"e}l Fokker.
\newblock {\em Relativiteitstheorie}.
\newblock P. Noordhoff, 1929.

\bibitem{Maybee:2019jus}
Ben Maybee, Donal O'Connell, and Justin Vines.
\newblock {Observables and amplitudes for spinning particles and black holes}.
\newblock {\em JHEP}, 12:156, 2019.

\bibitem{Guevara:2019fsj}
Alfredo Guevara, Alexander Ochirov, and Justin Vines.
\newblock {Black-hole scattering with general spin directions from
  minimal-coupling amplitudes}.
\newblock {\em Phys. Rev.}, D100(10):104024, 2019.

\bibitem{Guevara:2018wpp}
Alfredo Guevara, Alexander Ochirov, and Justin Vines.
\newblock {Scattering of Spinning Black Holes from Exponentiated Soft Factors}.
\newblock {\em JHEP}, 09:056, 2019.

\bibitem{pryce1948mass}
Maurice Henry~Lecorney Pryce.
\newblock The mass-centre in the restricted theory of relativity and its
  connexion with the quantum theory of elementary particles.
\newblock {\em Proceedings of the Royal Society of London. Series A.
  Mathematical and Physical Sciences}, 195(1040):62--81, 1948.

\bibitem{newton1949localized}
Theodore~Duddell Newton and Eugene~P Wigner.
\newblock Localized states for elementary systems.
\newblock {\em Reviews of Modern Physics}, 21(3):400, 1949.

\bibitem{Vines:2016unv}
Justin Vines, Daniela Kunst, Jan Steinhoff, and Tanja Hinderer.
\newblock {Canonical Hamiltonian for an extended test body in curved spacetime:
  To quadratic order in spin}.
\newblock {\em Phys. Rev.}, D93(10):103008, 2016.

\bibitem{Siemonsen:2017yux}
Nils Siemonsen, Jan Steinhoff, and Justin Vines.
\newblock {Gravitational waves from spinning binary black holes at the leading
  post-Newtonian orders at all orders in spin}.
\newblock {\em Phys. Rev. D}, 97(12):124046, 2018.

\bibitem{Kabat:1992tb}
Daniel~N. Kabat and Miguel Ortiz.
\newblock {Eikonal quantum gravity and Planckian scattering}.
\newblock {\em Nucl. Phys. B}, 388:570--592, 1992.

\bibitem{Akhoury:2013yua}
Ratindranath Akhoury, Ryo Saotome, and George Sterman.
\newblock {High Energy Scattering in Perturbative Quantum Gravity at Next to
  Leading Power}.
\newblock 8 2013.

\bibitem{Goldstein:2000}
Herbert Goldstein, Charles Poole, and John Safko.
\newblock {\em Classical Mechanics}.
\newblock Addison-Wesley, Boston, 3rd edition, 2000.

\bibitem{Carter:2010}
Brandon {Carter}.
\newblock {Republication of: Black hole equilibrium states Part II. General
  theory of stationary black hole states}.
\newblock {\em General Relativity and Gravitation}, 42:653--744, 2010.

\bibitem{Regimbau:2009rk}
Tania Regimbau and Scott~A. Hughes.
\newblock {Gravitational-wave confusion background from cosmological compact
  binaries: Implications for future terrestrial detectors}.
\newblock {\em Phys. Rev. D}, 79:062002, 2009.

\bibitem{1853026}
Rory Smith et~al.
\newblock {Bayesian inference for gravitational waves from binary neutron star
  mergers in third-generation observatories}.
\newblock 3 2021.

\bibitem{Bonetti_2020}
Matteo Bonetti and Alberto Sesana.
\newblock Gravitational wave background from extreme mass ratio inspirals.
\newblock {\em Physical Review D}, 102(10), Nov 2020.

\bibitem{Littenberg:2020bxy}
Tyson Littenberg, Neil Cornish, Kristen Lackeos, and Travis Robson.
\newblock {Global Analysis of the Gravitational Wave Signal from Galactic
  Binaries}.
\newblock {\em Phys. Rev. D}, 101(12):123021, 2020.

\bibitem{Karnesis:2021tsh}
Nikolaos Karnesis, Stanislav Babak, Mauro Pieroni, Neil Cornish, and Tyson
  Littenberg.
\newblock {Characterization of the stochastic signal originating from compact
  binaries populations as measured by LISA}.
\newblock 3 2021.

\bibitem{wiener1930generalized}
Norbert Wiener et~al.
\newblock Generalized harmonic analysis.
\newblock {\em Acta mathematica}, 55:117--258, 1930.

\bibitem{khintchine1934korrelationstheorie}
Alexander Khintchine.
\newblock Korrelationstheorie der station{\"a}ren stochastischen prozesse.
\newblock {\em Mathematische Annalen}, 109(1):604--615, 1934.

\bibitem{Cornish:2020vtw}
Neil~J. Cornish and Kevin Shuman.
\newblock {Black Hole Hunting with LISA}.
\newblock {\em Phys. Rev. D}, 101(12):124008, 2020.

\bibitem{Marsat:2020rtl}
Sylvain Marsat, John~G. Baker, and Tito Dal~Canton.
\newblock {Exploring the Bayesian parameter estimation of binary black holes
  with LISA}.
\newblock {\em Phys. Rev. D}, 103(8):083011, 2021.

\bibitem{Vallisneri:2013rc}
Michele Vallisneri and Nicolas Yunes.
\newblock {Stealth Bias in Gravitational-Wave Parameter Estimation}.
\newblock {\em Phys. Rev. D}, 87(10):102002, 2013.

\bibitem{regimbau2012mock}
Tania Regimbau, Thomas Dent, Walter Del~Pozzo, Stefanos Giampanis, Tjonnie~GF
  Li, Craig Robinson, Chris Van Den~Broeck, Duncan Meacher, Carl Rodriguez,
  Bangalore~S Sathyaprakash, et~al.
\newblock Mock data challenge for the einstein gravitational-wave telescope.
\newblock {\em Physical Review D}, 86(12):122001, 2012.

\bibitem{Albanesi:2021rby}
Simone Albanesi, Alessandro Nagar, and Sebastiano Bernuzzi.
\newblock {Effective one-body model for extreme-mass-ratio spinning binaries on
  eccentric equatorial orbits: testing radiation reaction and waveform}.
\newblock 4 2021.

\bibitem{shannon1949communication}
Claude~Elwood Shannon.
\newblock Communication in the presence of noise.
\newblock {\em Proceedings of the IRE}, 37(1):10--21, 1949.

\bibitem{wen2005detecting}
Linqing Wen and Jonathan~R Gair.
\newblock Detecting extreme mass ratio inspirals with lisa using
  time--frequency methods.
\newblock {\em Classical and Quantum Gravity}, 22(10):S445, 2005.

\bibitem{abuse_fisher}
Michele Vallisneri.
\newblock Use and abuse of the fisher information matrix in the assessment of
  gravitational-wave parameter-estimation prospects.
\newblock {\em Physical Review D}, 77(4):042001, 2008.

\bibitem{porter2009overview}
Edward~K Porter.
\newblock An overview of lisa data analysis algorithms.
\newblock {\em arXiv preprint arXiv:0910.0373}, 2009.

\bibitem{rodriguez2012verifying}
Carl~L Rodriguez, Ilya Mandel, and Jonathan~R Gair.
\newblock Verifying the no-hair property of massive compact objects with
  intermediate-mass-ratio inspirals in advanced gravitational-wave detectors.
\newblock {\em Physical Review D}, 85(6):062002, 2012.

\bibitem{gair2013testing}
Jonathan~R Gair, Michele Vallisneri, Shane~L Larson, and John~G Baker.
\newblock Testing general relativity with low-frequency, space-based
  gravitational-wave detectors.
\newblock {\em Living Reviews in Relativity}, 16(1):7, 2013.

\bibitem{porter2015fisher}
Edward~K Porter and Neil~J Cornish.
\newblock Fisher versus bayes: A comparison of parameter estimation techniques
  for massive black hole binaries to high redshifts with elisa.
\newblock {\em Physical Review D}, 91(10):104001, 2015.

\bibitem{amaro2018relativistic}
Pau Amaro-Seoane.
\newblock Relativistic dynamics and extreme mass ratio inspirals.
\newblock {\em Living reviews in relativity}, 21(1):4, 2018.

\bibitem{PhysRevD.102.124054}
Ollie Burke, Jonathan~R. Gair, Joan Sim\'on, and Matthew~C. Edwards.
\newblock Constraining the spin parameter of near-extremal black holes using
  lisa.
\newblock {\em Phys. Rev. D}, 102:124054, Dec 2020.

\bibitem{Gupta:2020lxa}
Anuradha Gupta, Sayantani Datta, Shilpa Kastha, Ssohrab Borhanian, K.G. Arun,
  and B.S. Sathyaprakash.
\newblock {Multiparameter tests of general relativity using multiband
  gravitational-wave observations}.
\newblock {\em Phys. Rev. Lett.}, 125(20):201101, 2020.

\bibitem{metropolis1953equation}
Nicholas Metropolis, Arianna~W Rosenbluth, Marshall~N Rosenbluth, Augusta~H
  Teller, and Edward Teller.
\newblock Equation of state calculations by fast computing machines.
\newblock {\em The journal of chemical physics}, 21(6):1087--1092, 1953.

\bibitem{roberts1997weak}
Gareth~O Roberts, Andrew Gelman, Walter~R Gilks, et~al.
\newblock Weak convergence and optimal scaling of random walk metropolis
  algorithms.
\newblock {\em The annals of applied probability}, 7(1):110--120, 1997.

\end{thebibliography}

\end{document}